\documentclass[faculty=fw,department=nat,phddegree=fys,10pt,final]{adsphd}

\title{Computational modeling of biological nanopores}

\author{Kherim}{Willems}

\supervisor{Prof.~dr.~ir.~P.~Van~Dorpe}{}
\supervisor{Prof.~dr.~G.~Maglia}{}
\supervisor{Prof.~dr.~J.~Hofkens}{}
\president{Prof.~dr.~E.~Carlon}{}
\jurymember{Prof.~dr.~ir.~T.~Stakenborg}{}
\jurymember{Prof.~dr.~ir.~M.~Roeffaers}{}
\jurymember{Prof.~dr.~C.~Bartic}{}
\externaljurymember{Prof.~dr.~rer.~nat.~U.~Keyser}{University of Cambridge, UK}

\researchgroup{Experimental Biophotonics}
\website{www.imec.be} 
\email{Kherim.Willems@imec.be} 

\address{Kapeldreef 75}
\addresscity{B-3001 Leuven} 

\date{December 2020}
\copyyear{2020}

\setcustomcoverpage{mycoverpage.tex} 

\newcommand{\attrib}[1]{%
\nopagebreak{\raggedleft\normalsize #1\par}}

\usepackage[cfg=thesis.cfg]{nomencl}   
\renewcommand{\nomname}{List of Symbols}
\newcommand{\myprintnomenclature}{%
  \cleardoublepage%
  \printnomenclature%
  \chaptermark{\nomname}
  \addcontentsline{toc}{chapter}{\nomname} 
}
\makenomenclature%

\usepackage[acronym, indexonlyfirst, nopostdot]{glossaries-extra} 
\newcommand{\glossname}{List of Abbreviations}
\newcommand{\myprintglossary}{%
  \renewcommand{\glossaryname}{\glossname}
  \cleardoublepage%
  \printglossary[title=\glossname]
  \chaptermark{\glossname}
}
\renewcommand{\glossarypreamble}{\footnotesize}
\setabbreviationstyle{long-short}%
\makeglossaries%

\glsaddkey*
{is}
{\glsentrytext{\glslabel}is}
{\glsentryis}
{\Glsentryis}
{\glsis}
{\Glsis}
{\GLSis}

\glsaddkey*
{ic}
{\glsentrytext{\glslabel}ic}
{\glsentryic}
{\Glsentryic}
{\glsic}
{\Glsic}
{\GLSic}

\usepackage[utf8]{inputenc}
\usepackage{csquotes}
\usepackage[
  hyperref=auto,
  mincrossrefs=999,
  backend=biber,
  date=year,
  url=false,
  eprint=false,
  style=numeric,
  sorting=none
]{biblatex}
\usepackage{textcomp} 
\usepackage{pifont}   
\usepackage{booktabs, tabularx, multirow, threeparttable}
\usepackage{amssymb,amsthm}
\usepackage{amsmath}
\usepackage{mathtools}
\usepackage{calligra}
\DeclareMathAlphabet{\mathcalligra}{T1}{calligra}{m}{n}
\usepackage{textgreek}

\usepackage[version=4]{mhchem}
\usepackage{chemfig}

\usepackage{siunitx}
\DeclareSIUnit \Molar {\textsc{M}}
\DeclareSIUnit \atm {\textsc{atm}}

\usepackage{cleveref}
  \creflabelformat{equation}{#2#1#3}
  \crefname{figure}{Fig.}{Figs.}
  \Crefname{figure}{Figure}{Figures}
  \crefname{table}{Tab.}{Tabs.}
  \Crefname{table}{Table}{Tables}
  \crefname{equation}{Eq.}{Eqs.}
  \Crefname{equation}{Equation}{Equations}
  \crefname{section}{Sec.}{Secs.}
  \Crefname{section}{Section}{Sections}

\usepackage[font=scriptsize,labelfont=bf,labelsep=period]{caption}
\usepackage[font=footnotesize,labelfont=bf,labelsep=period,margin=1mm]{subcaption}
\DeclareCaptionLabelFormat{bold}{\textbf{#2}}
\usepackage{epigraph}
\newcommand{\epipos}{80}
\usepackage{tikz}

\usepackage{pdflscape}

\usepackage{listings}
\usepackage{hyphenat}

\usepackage{afterpage}

\usepackage[usenames,dvipsnames]{xcolor}
\usepackage{transparent}

\definecolor{greyblue}{RGB}{201,215,222}

\makeatletter 
\newlength{\figrulesep} 
\makeatother

\DeclareSIUnit \angstrom {\text {Å}}
\DeclareSIUnit \molar {\mole\per\cubic\deci\metre}
\DeclareSIUnit \Molar {\textsc{M}}
\DeclareSIUnit \atm {\textsc{atm}}
\DeclareSIUnit \dalton {\textsc{D\!a}}
\DeclareSIUnit \bp {\textsc{bp}}

\makeatletter
\DeclareSIUnit\elementarycharge{\protect \protect \unhbox \voidb@x \hbox {\protect $\relax e$}}
\makeatother

\DeclareSIUnit \kT {\textit{k}\textsubscript{B} \textit{T}}
\DeclareSIUnit \kTe { \kT\per\elementarycharge }
\DeclareSIUnit \cal { \text{cal} }
\DeclareSIUnit \photon { \text{photon} }
\DeclareSIUnit \kDa { \kilo \dalton }
\DeclareSIUnit \uM  { \micro \Molar }
\DeclareSIUnit \mM  { \milli \Molar }
\DeclareSIUnit \nM  { \nano \Molar }
\DeclareSIUnit \nS  { \nano \siemens }
\DeclareSIUnit \pN  { \pico \newton }
\DeclareSIUnit \dC { \degreeCelsius}
\DeclareSIUnit \mps { \meter \per \second}
\DeclareSIUnit \mmps { \milli \meter \per \second}
\DeclareSIUnit \cnmpnspv { \cubic\nano\meter\per\nano\second\per\volt}

\newcommand{\mSI}[2]{\text{\SI{#1}{#2}}}
\newcommand{\msi}[2]{\text{\si{#1}{#2}}}
\newcommand{\mSIrange}[3]{\text{\SIrange{#1}{#2}{#3}}}
\newcommand{\mSIlist}[2]{\text{\SIlist{#1}{#2}}}

\newcommand{\SIgrid}[2]{%
   \SIlist[list-separator = $\times$, list-final-separator = $\times$, list-units = single]%
   {#1}{#2}
}

\newcommand{\code}[1]{\texttt{#1}}
\newcommand\ta{\textalpha}
\newcommand\tb{\textbeta}

\newcommand{\pdbid}[1]{PDBID:~\href{https://www.rcsb.org/structure/#1}{#1}}

\newcommand\etal{\textit{et al.}}
\newcommand\ie{\textit{i.e.},}
\newcommand\eg{\textit{e.g.},}
\newcommand\cisi{\textit{cis}}
\newcommand\transi{\textit{trans}}
\newcommand\lumen{\textit{lumen}}

\newcommand{\DHFR}[2]{DHFR\textsubscript{#1}#2}
\newcommand{\DHFRt}{DHFR\textsubscript{tag}}
\newcommand{\ires}{\current_{\rm{res}}}
\newcommand{\iresp}{\ires\%}
\newcommand{\iopen}{\current_0}
\newcommand{\iblock}{\current_b}
\newcommand{\ilevel}[1]{\rm{L}_{#1}}

\newcommand\dsum{\displaystyle\sum}
\renewcommand{\vec}[1]{\boldsymbol{#1}}
\newcommand{\rpos}{\vec{r}}
\newcommand{\normvec}{\vec{\hat{n}}}
\newcommand{\identity}{\vec{\rm{I}}}
\newcommand{\stdev}{\sigma}
\newcommand{\pav}[2]{\left< #1 \right>_{\text{#2}}}

\newcommand{\bessel}[1]{I_{#1}}
\newcommand{\pd}[2]{\displaystyle\frac{\partial #1}{\partial #2}}
\newcommand{\ddt}[1]{\frac{d #1}{dt}}
\newcommand\variance{{\rm{Var}}}

\newcommand\stresstensor{\sigma}

\newcommand\hydrostresstensor{\stresstensor^{\rm{H}}_{ij}}
\newcommand\maxwellstresstensorvec{\vec{\stresstensor}^{\rm{E}}}
\newcommand \hydrostresstensorvec{\vec{\stresstensor}^{\rm{H}}}

\newcommand\boltzmann{k_{\rm{B}}}
\newcommand\avogadro{N_{\rm{A}}}
\newcommand\temperature{T}
\newcommand\faraday{\mathcal{F}}
\newcommand\ec{e}
\newcommand\kbt{\boltzmann \temperature}
\newcommand\Vt{\frac{\kbt}{\ec}}

\newcommand\timedim{t}
\newcommand\timestep{\Delta \timedim}

\newcommand\velatom{v}
\newcommand\accelatom{a}

\newcommand\energymd{U}

\newcommand\force{\vec{F}}
\newcommand\forcedens{\vec{f}}
\newcommand\potential{\varphi}
\newcommand\varpotential{V}
\newcommand\concentration{c}
\newcommand\vel{u}
\newcommand\velocity{\vec{u}}
\newcommand\pressure{p}
\newcommand\electricfield{E}
\newcommand\efield{\vec{E}}
\newcommand\displacement{\vec{D}}
\newcommand\walldistance{d_{\rm{w}}}
\newcommand\dipolemoment{\vec{m}}

\newcommand\reynolds{\mathrm{Re}}

\newcommand\barrier{\Delta E}
\newcommand\potbar{\Delta V}
\newcommand\rate{k}
\newcommand\dwelltime{t_{\rm{d}}}
\newcommand\vthresh{V_{\rm{max}}^{\rm{bias}}}
\newcommand\tthresh{t_{\rm{d,max}}}
\newcommand\kinetic{E}
\newcommand\transition{c}
\newcommand\cbulk{\concentration_\text{s}}
\newcommand\vbias{\varpotential_\text{b}}
\newcommand\ep{{\rm{ep}}}
\newcommand\eo{{\rm{eo}}}
\newcommand\st{{\rm{ster}}}
\newcommand\osm{{\rm{eo}}}
\newcommand\dep{{\rm{dep}}}
\newcommand\es{{\rm{es}}}
\newcommand\ext{{\rm{ex}}}

\newcommand\static{{\rm{es}}}
\newcommand\steric{{\rm{st}}}
\newcommand\Eep{E_{\ep}}
\newcommand\Eosm{E_{\osm}}

\newcommand\Eext{E_{\ext}}

\newcommand\forceele{\force_{\rm{E}}}
\newcommand\forcehyd{\force_{\rm{H}}}
\newcommand\forceep{\force_{\ep}}
\newcommand\forceosm{\force_{\osm}}
\newcommand\forceeo{\force_{\eo}}
\newcommand\forcedep{\force_{\dep}}

\newcommand\forcest{\force_{\st}}

\newcommand\forcedrag{\force_{d}}
\newcommand\forcelj{\force_{\rm{LJ}}}
\newcommand\forceentr{\force_{\Delta \entropy}}

\newcommand\forcedensvol{\forcedens_{\rm{V}}}
\newcommand\forcedensedl{\forcedens_{\rm{EDL}}}

\newcommand\cis{{\rm{cis}}}
\newcommand\trans{{\rm{trans}}}
\renewcommand{\min}{{\rm{min}}}
\newcommand\xcis{x^\cis}
\newcommand\xtrans{x^\trans}
\newcommand\xmin{x^\min}

\newcommand\dxcis{\Delta x^\cis}
\newcommand\dxtrans{\Delta x^\trans}
\newcommand\Nnet{N_{\rm{net}}}
\newcommand\Ntag{N_{\rm{tag}}}
\newcommand\Nbody{N_{\rm{body}}}

\newcommand\Neo{N_{\eo}}
\newcommand\Ntot{N_{\rm{tot}}}
\newcommand\ratefc{\rate_{\rm{eff}}^\cis}
\newcommand\rateft{\rate_{\rm{eff}}^\trans}

\newcommand{\onefiguresource}[5]
{\begin{figure}[#5]
   \centering
   \begin{minipage}{0.9\textwidth}
      \centering
      #1
      \caption[#3]{#4}
      \label{#2}
   \end{minipage}
\end{figure}
}

\makeatletter
\newcommand\reaction@[1]{\begin{equation}\ce{#1}\end{equation}}
\newcommand\reaction@nonumber[1]%
{\begin{equation*}\ce{#1}\end{equation*}}
\newcommand\reaction{\@ifstar{\reaction@nonumber}{\reaction@}}
\makeatother

\newcommand{\dconc}{\bar{\concentration}}
\newcommand{\dwall}{\bar{\walldistance}}
\newcommand{\permittivity}{\varepsilon}
\newcommand{\absperm}{\permittivity_0}
\newcommand{\relperm}{\permittivity_r}

\newcommand{\diffusion}{\mathcal{D}}
\newcommand{\mobility}{\mu}
\newcommand{\transportn}{t}
\newcommand{\chargen}{z}
\newcommand{\chargeni}{z_{i}}

\newcommand{\chargeq}{q}

\newcommand{\ionsize}{a}
\newcommand{\density}{\varrho}
\newcommand{\viscosity}{\eta}
\newcommand{\permeability}{P}
\newcommand{\length}{l}
\newcommand{\diameter}{d}

\newcommand{\radiusa}{a}
\newcommand{\particleradius}{\radiusa_{\rm{p}}}

\newcommand{\avionconc}{\langle\concentration\rangle}
\newcommand{\tna}{\transportn_{\ce{Na+}}}
\newcommand{\gna}{\conductance_{\ce{Na+}}}
\newcommand{\gcl}{\conductance_{\ce{Cl-}}}
\newcommand{\pna}{\permeability_{\ce{Na+} / \ce{Cl-}}}
\newcommand{\pcl}{\permeability_{\ce{Cl-} / \ce{Na+}}}
\newcommand{\molarconductivity}{\Lambda}
\newcommand{\specmolarconductivity}{\lambda}

\newcommand\scd{\rho}
\newcommand\scdpore{\scd_{\text{pore}}}
\newcommand\scdion{\scd_{\text{ion}}}
\newcommand\scdfix{\scd^{\rm{f}}}
\newcommand\scdmob{\scd^{\rm{m}}}
\newcommand\flux{\vec{J}}

\newcommand\volumeforce{\forcedens_{\rm ion}}
\newcommand\surfcd{\sigma_s}
\newcommand\current{I}
\newcommand\currentsim{\current_{\rm sim}}

\newcommand\conductance{G}
\newcommand\icr{\alpha}
\newcommand\flowrate{Q_{\text{eo}}}
\newcommand\flowcond{\conductance_{\text{eo}}}
\newcommand\eor{\alpha_{\text{eo}}}

\newcommand\ionstr{\mathcal{I}}
\newcommand\ci{\concentration_{i}}

\newcommand\Na{\ce{Na+}}
\newcommand\Cl{\ce{Cl-}}
\newcommand\cNa{\concentration_{\Na}}
\newcommand\cCl{\concentration_{\Cl}}
\newcommand\Qion{Q_{\rm ion}}
\newcommand\radpot{\left<\potential\right>_\text{rad}}
\newcommand\cylpot{\left<\potential\right>_\text{cyl}}
\newcommand\eqpot{\potential^{0}}
\newcommand\radeqpot{\left<\eqpot\right>_\text{rad}}
\newcommand\radenergy{\left< U_{\text{E},i} \right>_{\text{rad}}}
\newcommand\deltaEt{\Delta E_{\text{B},i}}
\newcommand{\pH}[1]{pH~\num{#1}}
\newcommand{\pHrange}[2]{pH~\numrange{#1}{#2}}
\newcommand{\pHlist}[1]{pH~\numlist{#1}}
\newcommand\pKa{\text{p}K_{\mathrm{a}}}
\newcommand\revpot{V_{\text{r}}}
\newcommand\pavi{\pav{\ci/\cbulk}{PT}}
\newcommand\pavNa{\pav{\cNa/\cbulk}{PT}}
\newcommand\pavCl{\pav{\cCl/\cbulk}{PT}}

\newcommand\bjl{\mathcalligra{l}_{B}}
\newcommand\dbl{\lambda_{D}}
\newcommand\stl{\lambda_{S}}
\newcommand\invdbl{\kappa}
\newcommand\potdiff{\Delta V}

\newcommand\chargedensrelaxtime{\tau_{\rho}}

\newcommand\partialcharge{Q}

\newcommand\atomradius{R}

\newcommand\bfactor{\mathrm{B}}
\newcommand\msd{\left< u_i^2 \right>}

\newcommand{\grid}[4]{$#1 \times #2 \times #3 $~\si{#4}}
\newcommand{\apbsgrid}[4]{$#1 \times #2 \times #3 $~\si{#4}}

\newcommand\ionaccess{\xi}

\newcommand\sliplength{b}

\newcommand\protfrac{f}

\newcommand\gibbs{G}

\newcommand\entropy{S}
\newcommand\microstates{\Omega}

\newcommand\potLJ{U_{\rm{LJ}}}
\newcommand\welldepthLJ{\epsilon_{\rm{LJ}}}
\newcommand\dminLJ{r_{\rm{min}}}

\newcommand\energyelec{\Delta \gibbs^{\es}}

\newcommand\zdna{z_{\mathrm{DNA}}}
\newcommand\zdnac{z_{\mathrm{DNA}}^{\text{\cisi}}}
\newcommand\zdnat{z_{\mathrm{DNA}}^{\text{\transi}}}

\newcommand\probability{\mathcal{P}}
\newcommand\probtrans{\probability_{\rm{transl}}}

\newcommand{\permeabilityratio}[2]{ \permeability_{#1} / \permeability_{#2} }

\newcommand{\sliceline}{\tikz \draw [>-<,>={Straight Barb},>={Straight Barb}] (0,.5) -- (0.5,.5);}

\newcommand\resistance{R}
\newcommand{\Rpore}{\resistance_{\text{p}}}
\newcommand{\Raccess}{\resistance_{\text{a}}}
\newcommand{\dchar}{\tilde{\diameter}}
\newcommand{\potcis}{\potential_{\text{pore}}^{\text{\cisi}}}
\newcommand{\pottrans}{\potential_{\text{pore}}^{\text{\transi}}}

\newcommand{\reprRSC}[1]%
   {Reproduced from Ref.~\cite{#1} with permission from The Royal Society of Chemistry.}
\newcommand{\adapRSC}[1]%
   {Adapted from Ref.~\cite{#1} with permission from The Royal Society of Chemistry.}
\newcommand{\reprACS}[1]%
   {Reproduced with permission from \fullcite{#1} Copyright~\citedate{#1} American Chemical Society.}
\newcommand{\adapACS}[1]%
   {Adapted with permission from \fullcite{#1} Copyright~\citedate{#1} American Chemical Society.}

\newcounter{objective}
\newcommand{\objective}[1]{%
  \refstepcounter{objective}{%
  \par\vspace{.5cm}\noindent \normalsize \bfseries%
  Objective\hspace{1ex}\arabic{objective}\hspace{1em}\textbar}%
  \hspace{1em}\textbf{#1}
}

\newabbreviation{lhc}{LHC}{large hadron collidor}

\newabbreviation{ligo}{LIGO}{laser interferometer gravitational-wave observatory}

\newabbreviation{fet}{FET}{field effect transistor}

\newabbreviation{zmw}{ZMW}{zero-mode waveguide}
\newabbreviation{nsom}{NSOM}{near-field scanning optical microscope}

\newabbreviation{afm}{AFM}{atomic force microscopy}
\newabbreviation{stm}{STM}{scannning tunneling microscopy}

\newabbreviation{ont}{ONT}{Oxford Nanopore Technologies}

\newabbreviation{sam}{SAM}{self-assembled monolayer}

\newabbreviation{pft}{PFT}{pore-forming toxin}
\newabbreviation{bnp}{BNP}{biological nanopore}
\newabbreviation{ssnp}{SSNP}{solid-state nanopore}

\newabbreviation{ahl}{\textalpha{}HL}{\textalpha-hemolysin}

\newglossaryentry{7r-ahl}{
  name={7R-\textalpha{}HL},
  description={\textalpha{}HL variant with the mutations M113R, T115R, T117R, G119R, N121R, N123R, T125R}
}

\newabbreviation{clya}{ClyA}{cytolysin~A}

\newglossaryentry{clya-as}{
  name={ClyA-AS},
  description={\textit{S. typhi} wild-type ClyA variant with the mutations C87A, L99Q, E103G, F166Y, I203V, C285S, K294R, H307Y}
}

\newglossaryentry{clya-r}{
  name={ClyA-R},
  description={ClyA-AS variant with the mutation S110R}
}

\newglossaryentry{clya-rr}{
  name={ClyA-RR},
  description={ClyA-AS variant with the mutations S110R, D64R}
}

\newglossaryentry{clya-rr56}{
  name={ClyA-RR$_{56}$},
  description={ClyA-AS variant with the mutations S110R, Q56R}
}

\newglossaryentry{clya-rr56k}{
  name={ClyA-RR$_{56}$K},
  description={ClyA-AS variant with the mutations S110R, Q56R, Q8K}
}

\newabbreviation{mspa}{MspA}{\textit{Mycobacterium smegmatis} porin~A}
\newabbreviation{csgg}{CsgG}{curli specific gene~G}

\newabbreviation{frac}{FraC}{fragaceatoxin~C}

\newglossaryentry{wtfrac}{
  name={WtFraC},
  description={wild-type FraC pore}
}

\newglossaryentry{refrac}{
  name={ReFraC},
  description={FraC variant with mutations D10R/K159E}
}

\newabbreviation{plyab}{PlyAB}{pleurotolysin~AB}
\newabbreviation{plya}{PlyA}{pleurotolysin~A}
\newabbreviation{plyb}{PlyB}{pleurotolysin~B}

\newglossaryentry{plyab-wt}{
  name={PlyAB-WT},
  description={PlyAB wild-type variant}
}

\newglossaryentry{plyab-e2}{
  name={PlyAB-E2},
  description={PlyAB variant with mutations C62S, C94S (PlyA) and N26D, N107D, G218R, A328T, C441A, A464V (PlyB)}
}

\newglossaryentry{plyab-r}{
  name={PlyAB-R},
  description={PlyAB variant with mutations C62S, C94S (PlyA) and N26D, K255E, E260R, E270R, A328T, C441A, A464V (PlyB)}
}

\newglossaryentry{kcsa}{
  name={KcsA},
  description={potassium channel of \textit{Streptomyces A}}
}

\newglossaryentry{ael}{
  name={AeL},
  description={aerolysin}
}

\newabbreviation{phi29p}{\textPhi29p}{\textPhi29 packaging motor}
\newabbreviation{ompf}{OmpF}{outer membrane porin F}
\newabbreviation{ompg}{OmpG}{outer membrane porin G}
\newabbreviation{fhua}{FhuA}{ferric hydroxamate uptake protein component A}

\newabbreviation{iv}{IV}{current--voltage}

\newglossaryentry{ep}{
  name={EP},
  description={electrophoresis, electrophoretic}
}

\newglossaryentry{eo}{
  name={EO},
  description={electro-osmosis, electro-osmotic}
}

\newglossaryentry{dep}{
  name={DEP},
  description={dielectrophoresis, dielectrophoretic}
}

\newglossaryentry{es}{
  name={ES},
  description={electrostatic}
}

\newabbreviation{eof}{EOF}{electro-osmotic flow}
\newabbreviation{ghk}{GHK}{Goldman-Hodgkin-Katz}
\newabbreviation{icr}{ICR}{ionic current rectification}

\newabbreviation{pdb}{PDB}{protein data bank}
\newabbreviation{tmd}{TMD}{transmembrane domain}
\newabbreviation{sm}{SM}{sphingomyelin}
\newabbreviation{macpf}{MACPF}{membrane attack complex/perforin-like}
\newabbreviation{hth}{HTH}{helix-turn-helix}
\newabbreviation{dna}{DNA}{deoxyribonucleic acid}
\newabbreviation{ssdna}{ssDNA}{single-stranded DNA}
\newabbreviation{dsdna}{dsDNA}{double-stranded DNA}
\newabbreviation{sph}{SPH}{sphingomyelin}
\newabbreviation{dphpc}{DPhPC}{1,2-diphytanoyl-\textit{sn}-glycero-3-phosphocholine}
\newabbreviation{mops}{MOPS}{4-morpholinepropanesulfonic acid}

\newabbreviation{md}{MD}{molecular dynamics}
\newabbreviation{bd}{BD}{Brownian dynamics}
\newabbreviation{mc}{MC}{Monte Carlo}
\newabbreviation{vmd}{VMD}{visual molecular dynamics}
\newabbreviation{namd}{NAMD}{nanoscale molecular dynamics}
\newabbreviation{apbs}{APBS}{adaptive Poisson-Boltzmann solver}
\newabbreviation{com}{COM}{center-of-mass}
\newabbreviation{mdff}{MDFF}{molecular dynamics flexible fitting}
\newabbreviation{tcl}{Tcl}{Tool Command Language}
\newabbreviation{msd}{MSD}{mean-square displacement}
\newabbreviation{rmsd}{RMSD}{root-mean-square deviation}
\newabbreviation{pme}{PME}{particle mesh Ewald}
\newabbreviation{lj}{LJ}{Lennard-Jones}

\newabbreviation{mdh}{\code{mdh}}{multiple Debye-H\"{u}ckel}

\newglossaryentry{pdb2pqr}{
  name={PDB2PQR},
  description={software package to convert PDB into PQR files}
}

\newglossaryentry{propka}{
  name={PROPKA},
  description={software package to predict the $\pKa$-values of protein residues with an empirical approach}
}

\newglossaryentry{delphipka}{
  name={DelPhiPKa},
  description={software package to predict the $\pKa$-values of protein residues with an electrostatic approach}
}

\newglossaryentry{charmm36}{
  name={CHARMM36},
  description={chemistry at Harvard macromolecular mechanics force field version 36}
}

\newglossaryentry{parse}{
  name={PARSE},
  description={parameters for solvation energy force field}
}

\newabbreviation{pcr}{PCR}{polymerase chain reaction}
\newabbreviation{page}{PAGE}{polyacrylamide gel electrophoresis}
\newabbreviation{cryo-em}{cryo-EM}{cryogenic electron microscopy}
\newabbreviation{cd}{CD}{circular dichroism}
\newabbreviation{ddm}{DDM}{\textit{n}-dodecyl \textbeta-\textsc{D}-maltoside}
\newabbreviation{rbc}{RBC}{red blood cells}
\newabbreviation{fret}{FRET}{F\"{o}rster resonance energy transfer}
\newabbreviation{nmr}{NMR}{nuclear magnetic resonance}
\newabbreviation{edl}{EDL}{electrical double layer}
\newabbreviation{vdw}{VDW}{van der Waals}
\newabbreviation{dhfr}{DHFR}{dihydrofolate reductase}
\newabbreviation{bsa}{BSA}{bovine serum albumin}
\newabbreviation{edta}{EDTA}{Ethylenediaminetetraacetic acid}
\newabbreviation{iptg}{IPTG}{isopropyl \textbeta-\textsc{D}-1-thiogalactopyranoside}

\newabbreviation{nadph}{NADPH}{nicotinamide adenine dinucleotide phosphate}
\newabbreviation{abel}{ABEL}{anti-Brownian electrokinetic}
\newabbreviation{nta}{NTA}{nitrilotriacetic acid}
\newabbreviation{mtx}{MTX}{methotrexate}

\newabbreviation{pb}{PB}{Poisson-Boltzmann}
\newabbreviation{np}{NP}{Nernst-Planck}
\newabbreviation{ns}{NS}{Navier-Stokes}
\newabbreviation{pnp}{PNP}{Poisson-Nernst-Planck}
\newabbreviation{pnp-ns}{PNP-NS}{Poisson-Nernst-Planck and Navier-Stokes}
\newabbreviation{epnp-ns}{ePNP-NS}{extended Poisson-Nernst-Planck and Navier-Stokes}

\newabbreviation{pe}{PE}{Poisson's equation}
\newabbreviation{pbe}{PBE}{Poisson-Boltzmann equation}
\newabbreviation{npe}{NPE}{Nernst-Planck equation}
\newabbreviation{nse}{NSE}{Navier-Stokes equations}
\newabbreviation{pnpe}{PNPE}{Poisson-Nernst-Planck equations}
\newabbreviation{pnp-nse}{PNP-NSE}{Poisson-Nernst-Planck and Navier-Stokes equations}
\newabbreviation{smnpe}{smNPE}{size-modified Nernst-Planck equation}

\newabbreviation{fea}{FEA}{finite element analysis}
\newabbreviation{bc}{BC}{boundary condition}

\newabbreviation{fem}{FEM}{finite element method}
\newabbreviation{fdm}{FDM}{finite difference method}
\newabbreviation{fvm}{FVM}{finite volume method}

\newglossaryentry{nve}{
  name={NVE},
  description={microcanonical ensemble, constant particle number, volume and energy}
}

\newglossaryentry{nvt}{
  name={NVT},
  description={canonical ensemble, constant particle number, volume and temperature}
}

\newglossaryentry{npt}{
  name={NpT},
  description={isothermal-isobaric ensemble, constant particle number, pressure and temperature}
}

\newabbreviation{qcp}{QCP}{quaternion characteristic polynomial}

\newabbreviation{ode}{ODE}{ordinary differential equation}
\newabbreviation{pde}{PDE}{partial differential equation}

\nomenclature[I1]{$\bessel{1}$}{first order modified Bessel function}
\nomenclature[I2]{$\bessel{2}$}{second order modified Bessel function}
\nomenclature[rpos]{$\vec{r}$}{position vector}
\nomenclature[identity]{$\identity$}{identity matrix}
\nomenclature[r]{$r$}{position between two atoms, see~\cref{eq:nanopores_LJ_force}}

\nomenclature[zdna]{$\zdna$}{z-position of the {DNA} strand's center-of-mass}

\nomenclature[kb]{$\boltzmann$}{Boltzmann constant}
\nomenclature[ec]{$\ec$}{elementary charge}
\nomenclature[NA]{$\avogadro$}{Avogadro's constant}
\nomenclature[faraday]{$\faraday$}{Faraday constant}
\nomenclature[epsilon0]{$\absperm$}{permittivity of free space}

\nomenclature[epsilon]{$\permittivity$}{permittivity}
\nomenclature[epsilonr]{$\relperm$}{relative permittivity}
\nomenclature[epsilonrf]{$\permittivity_{r,\text{f}}(\avionconc)$}{concentration dependent relative permittivity of the electroolyte, see~\cref{eq:epnp-ns_relperm_concentration}}
\nomenclature[epsilonrf0]{$\permittivity_{r,\text{f}}^0$}{relative permittivity of the electrolyte at infinite dilution, see~\cref{eq:epnp-ns_relperm_concentration}}
\nomenclature[epsilonrfc]{$\permittivity_{r,\text{f}}^c(\avionconc)$}{concentration dependent empirical scaling fuction of the electrolyte relative permittivity, see~\cref{eq:epnp-ns_relperm_concentration,eq:epnpns_concentration_solv_epsr}}

\nomenclature[T]{$\temperature$}{temperature}

\nomenclature[beta]{$\beta$}{inverse thermal energy, see~\cref{eq:nanopores_electrostatic_energy}}

\nomenclature[Vi]{$V_i$}{repulsive potential, see~\cref{eq:nanopores_electrostatic_energy}}

\nomenclature[u]{$\vec{\vel}$}{fluid velocity field}
\nomenclature[E]{$\efield$}{electric field}
\nomenclature[Ez]{$\electricfield_z$}{z-component of the electric field}

\nomenclature[phi]{$\potential$}{electric potential}
\nomenclature[phirescis]{$\potcis$}{electric potential at the \cisi{} entry of the nanopore, see~\cref{eq:nanopores_potcis}}
\nomenclature[phirestrans]{$\pottrans$}{electric potential at the \transi{} entry of the nanopore, see~\cref{eq:nanopores_pottrans}}

\nomenclature[phicyl]{$\cylpot$}{cylindrically averaged electrostatic potential}
\nomenclature[phirad]{$\radpot$}{radially averaged electrostatic potential, see~\cref{eq:epnp-ns_radpot}}
\nomenclature[Urad]{$\radenergy$}{radially averaged electrostatic energy for a monovalent ion, see~\cref{eq:epnp-ns_radenergy}}

\nomenclature[Qeo]{$\flowrate$}{volumetric flowrate, see~\cref{eq:nanopores_eof_rate,eq:flowrate}}
\nomenclature[Geo]{$\flowcond$}{electro-osmotic conductance, see~\cref{sec:eof}}
\nomenclature[alphaeo]{$\eor$}{electro-osmotic conductance rectification, see~\cref{sec:eof}}

\nomenclature[eta]{$\viscosity$}{fluid dynamic viscosity}
\nomenclature[etacd]{$\viscosity(\avionconc,\walldistance)$}{concentration and wall distance dependent dynamic viscosity of the electrolyte, see~\cref{eq:epnp-ns_viscosity_density}}
\nomenclature[eta0]{$\viscosity^0$}{electrolyte dynamic viscosity at infinite dilution, see~\cref{eq:epnp-ns_viscosity_density}}
\nomenclature[etac]{$\viscosity^c(\avionconc)$}{concentration dependent empirical scaling fuction of the electrolyte dynamic viscosity, see~\cref{eq:epnp-ns_viscosity_density,eq:epnpns_concentration_solv_eta}}
\nomenclature[etad]{$\viscosity^w(\walldistance)$}{wall distance dependent empirical scaling fuction of the electrolyte dynamic viscosity, see~\cref{eq:epnp-ns_viscosity_density,eq:epnpns_distance_eta}}

\nomenclature[rhodens]{$\density$}{fluid density}
\nomenclature[rhodensc]{$\density(\avionconc)$}{concentration dependent electrolyte density, see~\cref{eq:epnp-ns_viscosity_density}}
\nomenclature[rhodens0]{$\density^0$}{electrolyte density at infinite dilution, see~\cref{eq:epnp-ns_viscosity_density}}
\nomenclature[rhodenscf]{$\density^c(\avionconc)$}{oncentration dependent empirical scaling fuction of the electrolyte density, see~\cref{eq:epnp-ns_viscosity_density,eq:epnpns_concentration_solv_rho}}

\nomenclature[pressure]{$\pressure$}{pressure}

\nomenclature[mdip]{$\dipolemoment$}{dipole moment, see~\cref{eq:nanopores_dep_force}}

\nomenclature[forcedens]{$\forcedens$}{force density}
\nomenclature[forcedensedl]{$\forcedensedl$}{force density on the electrical double layer, see~\cref{eq:nanopores_edl_force}}
\nomenclature[forcedensvol]{$\forcedensvol$}{volume force density, see~\cref{eq:nanopores_navier_stokes}}
\nomenclature[forcedension]{$\volumeforce$}{volume force density on the electrical double layer, same as $\forcedensedl$, see~\cref{eq:ion_force_density}}

\nomenclature[forcev]{$\force$}{force vector}
\nomenclature[forcevep]{$\forceep$}{electrophoretic (Lorentz) force, see~\cref{eq:nanopores_lorentz_force}}
\nomenclature[forcevdep]{$\forcedep$}{dielectrophoretic force, see~\cref{eq:nanopores_dep_force}}
\nomenclature[forcevE]{$\forceele$}{electric force, see~\cref{eq:nanopores_electric_force}}
\nomenclature[forcevd]{$\forcedrag$}{drag force, see~\cref{eq:nanopores_drag_force}}
\nomenclature[forcevH]{$\forcehyd$}{hydrodynamic force, see~\cref{eq:nanopores_hydrodynamic_force}}
\nomenclature[forcevesm]{$\force_{\es,m}$}{electrostatic force on atom $m$, see~\cref{eq:nanopores_electrostatic_force}}
\nomenclature[forcevLJ]{$\forcelj$}{Lennard-Jones force, see~\cref{eq:nanopores_LJ_force}}
\nomenclature[forcevDeltaS]{$\forceentr$}{entropic force, see~\cref{eq:nanopore_entropic_force}}

\nomenclature[sigmae]{$\maxwellstresstensorvec$}{Maxwell stress tensor, see~\cref{eq:nanopores_maxwell_stresstensor}}
\nomenclature[sigmah]{$\hydrostresstensorvec$}{hydrodynamic stress tensor, see~\cref{eq:nanopores_hydrodynamic_stresstensor}}

\nomenclature[DeltaGelec]{$\energyelec$}{electrostatic energy, see~\cref{eq:nanopores_electrostatic_energy}}

\nomenclature[ULJ]{$\potLJ$}{Lennard-Jones potential, see~\cref{eq:nanopores_electrostatic_energy}}
\nomenclature[epsilonLJ]{$\welldepthLJ$}{well depth of the Lennard-Jones potential, see~\cref{eq:nanopores_LJ_force}}
\nomenclature[rmin]{$\dminLJ$}{position of energy minimum in the Lennard-Jones potential, see~\cref{eq:nanopores_LJ_force}}

\nomenclature[S]{$\entropy$}{entropy, see~\cref{eq:nanopores_entropy}}
\nomenclature[Omega]{$\microstates$}{number of available microstates, see~\cref{eq:nanopores_entropy}}

\nomenclature[DeltaS]{$\Delta \entropy$}{entropy difference, see~\cref{eq:nanopore_entropic_energy}}

\nomenclature[ULJ]{$\energymd$}{interaction potential, see~\cref{eq:nanopores_md_force,eq:nanopores_md_energy}}
\nomenclature[ULJb]{$\energymd_{\text{b}}$}{bonded energy term of $\energymd$}
\nomenclature[ULJnb]{$\energymd_{\text{nb}}$}{non-bonded energy term of $\energymd$}

\nomenclature[Q]{$\partialcharge$}{atomic partial charge, see~\cref{eq:pka_partial_charge}}
\nomenclature[Qi]{$\partialcharge_i$}{partial charge of atom $i$, see~\cref{eq:electrostatics_scdf}}
\nomenclature[q]{$\chargeq$}{charge}

\nomenclature[lb]{$\bjl$}{Bjerrum length, see~\cref{eq:nanopores_bjerrum_length}}
\nomenclature[lambdaD]{$\dbl$}{Debye length, see~\cref{eq:nanopores_debye_length}}
\nomenclature[lambdaS]{$\stl$}{Stern layer thickness, see~\cref{eq:nanopores_relaxation_time}}
\nomenclature[kappa]{$\invdbl$}{inverse Debye length, see~\cref{eq:nanopores_pbe_linear}}

\nomenclature[concentration]{$\concentration$}{chemical concentration}
\nomenclature[concentrations]{$\cbulk$}{bulk salt concentration}
\nomenclature[concentrationi]{$\ci$}{concentration of ion species $i$}
\nomenclature[concentrationi0]{$\ci^0$}{bulk concentration of ion species $i$}
\nomenclature[concentrationav]{$\avionconc$}{average ion concentration, see~\cref{sec:epnp-ns:electrostatic_field}}

\nomenclature[zn]{$\chargen$}{charge number}
\nomenclature[zni]{$\chargeni$}{charge number of ion species $i$}

\nomenclature[fHA]{$\protfrac_{\mathrm{HA}}$}{protonated fraction, see~\cref{eq:pka_protonation}}

\nomenclature[t]{$\timedim$}{time}
\nomenclature[J]{$\flux$}{chemical species flux, see~\cref{eq:nanopores_nernst_planck}}

\nomenclature[Diffusion]{$\diffusion$}{diffusion coefficient, see~\cref{eq:nanopores_nernst_planck}}
\nomenclature[Diffusioni]{$\diffusion_{i}(\avionconc,\walldistance)$}{concentration and wall distance dependent diffusion coefficient of ion $i$, see~\cref{eq:epnp-ns_diffusion_and_mobility}}
\nomenclature[Diffusioni0]{$\diffusion_{i}^0$}{diffusion coefficient of ion $i$ at infinite dilution, see~\cref{eq:epnp-ns_diffusion_and_mobility}}
\nomenclature[Diffusionic]{$\diffusion_{i}^c(\avionconc)$}{concentration dependent empirical scaling fuction of the diffusion coefficient of ion $i$, see~\cref{eq:epnp-ns_diffusion_and_mobility,eq:concentration_d_mu_tna}}
\nomenclature[Diffusionid]{$\diffusion_{i}^w(\walldistance)$}{wall distance dependent empirical scaling fuction of the diffusion coefficient of ion $i$, see~\cref{eq:epnp-ns_diffusion_and_mobility,eq:epnpns_distance_d}}

\nomenclature[mu]{$\mobility$}{electrophoretic mobility, see~\cref{eq:nanopores_nernst_planck}}
\nomenclature[mui]{$\mobility_{i}(\avionconc,\walldistance)$}{concentration and wall distance dependent electrophoretic mobility of ion $i$, see~\cref{eq:epnp-ns_diffusion_and_mobility}}
\nomenclature[mui0]{$\mobility_{i}^0$}{electrophoretic mobility of ion $i$ at infinite dilution, see~\cref{eq:epnp-ns_diffusion_and_mobility}}
\nomenclature[muic]{$\mobility_{i}^c(\avionconc)$}{concentration dependent empirical scaling fuction of the electrophoretic mobility of ion $i$, see~\cref{eq:epnp-ns_diffusion_and_mobility,eq:concentration_d_mu_tna}}
\nomenclature[muid]{$\mobility_{i}^w(\walldistance)$}{wall distance dependent empirical scaling fuction of the electrophoretic mobility of ion $i$, see~\cref{eq:epnp-ns_diffusion_and_mobility,eq:epnpns_distance_d}}

\nomenclature[betai]{$\vec{\beta_{i}}$}{steric flux vector of ion $i$, see~\cref{eq:sm-nernst-planck,eq:epnp-ns_steric_flux_vector}}

\nomenclature[ai]{$\ionsize_{i}$}{steric size of ion $i$, \cref{eq:epnp-ns_steric_flux_vector}}
\nomenclature[a0]{$\ionsize_{0}$}{steric size of a water molecule, \cref{eq:epnp-ns_steric_flux_vector}}

\nomenclature[Jw]{$\flux_{w}$}{water flux through a nanopore, see~\cref{eq:water_flux_permeability}}
\nomenclature[Nw]{$N_{w}$}{number of water molecules per ion, see~\cref{eq:water_flux_permeability}}

\nomenclature[Rpore]{$\Rpore$}{ionic resistance of the nanopore, see~\cref{sec:np:potential}}
\nomenclature[Raccess]{$\Raccess$}{ionic access resistance of the nanopore, see~\cref{sec:np:potential}}

\nomenclature[length]{$\length$}{pore length}
\nomenclature[diameter]{$\diameter$}{pore diameter}
\nomenclature[dchar]{$\dchar$}{characteristic length of the nanopore system, see~\cref{eq:nanopores_characteristic_length}}

\nomenclature[a]{$\radiusa$}{nanopore radius}
\nomenclature[ap]{$\particleradius$}{particle radius}

\nomenclature[I]{$\ionstr$}{ionic strength}
\nomenclature[rhocharge]{$\scd$}{space charge density, see~\cref{eq:electrostatics_scd}}
\nomenclature[rhochargeion]{$\scdion$}{ion space charge density, see~\cref{eq:nanopores_ion_scd,eq:scdion}}
\nomenclature[rhochargepore]{$\scdpore$}{space charge density of the nanopore, typically the same as $\scdfix$, see~\cref{eq:scdpore}}
\nomenclature[rhochargemob]{$\scdmob$}{mobile space charge density, typically the same as $\scdion$, see~\cref{eq:electrostatics_scdm,eq:electrostatics_scdm_salt}}
\nomenclature[rhochargefix]{$\scdfix$}{fixed space charge density, see~\cref{eq:electrostatics_scdf}}

\nomenclature[Vdelta]{$\potdiff$}{electric potential difference, see~\cref{sec:np:potential}}
\nomenclature[Vbias]{$\vbias$}{applied bias potential, typically the same as $\potdiff$}

\nomenclature[sigma]{$\sigma$}{electrolyte conductivity, see~\cref{eq:nanopores_ohms_law}}

\nomenclature[I]{$\current$}{ionic current through the nanopore, see~\cref{eq:nanopores_ohms_law}}
\nomenclature[Sz]{$S(z)$}{cross-sectional area of the nanopore as a function of the axial coordinate $z$,
see~\cref{eq:nanopores_current_cross_section}}

\nomenclature[zeta]{$\ionaccess$}{ion accessibility coefficient, see~\cref{eq:electrostatics_scdm_salt}}

\nomenclature[PermNa]{$\pna$}{sodium permeability ratio}
\nomenclature[PermCl]{$\pcl$}{chloride permeability ratio}
\nomenclature[transpNa]{$\tna$}{sodium transport number, see~\cref{eq:tna}}

\nomenclature[Vw]{$V_{w}$}{volume of a single water molecule, see~\cref{eq:water_flux_to_velocity}}

\nomenclature[prob]{$\probability$}{probability, see~\cref{eq:electrostatics_probability}}
\nomenclature[probtrans]{$\probtrans$}{translocation probability, see~\cref{eq:ptrans}}

\nomenclature[Vthresh]{$\vthresh$}{threshold voltage, see~\cref{eq:threshold_voltage_simple,eq:threshold_voltage_complex}}

\nomenclature[Nbody]{$\Nbody$}{number of elementary charges in the body of \DHFRt, see~\cref{sec:trapping:phenomenology}}
\nomenclature[Ntag]{$\Ntag$}{number of elementary charges in the tag of \DHFRt, see~\cref{sec:trapping:phenomenology}}
\nomenclature[Nnet]{$\Nnet$}{number of elementary charges in \DHFRt, see~\cref{sec:trapping:phenomenology}}
\nomenclature[Neo]{$\Neo$}{equivalent electro-osmotic charge number, see~\cref{eq:external-barrier}}

\nomenclature[forceepbody]{$\forceep^{\rm{body}}$}{electrophoretic force on the body of \DHFRt, see~\cref{sec:trapping:phenomenology}}
\nomenclature[forceeptag]{$\forceep^{\rm{tag}}$}{electrophoretic force on the tag of \DHFRt, see~\cref{sec:trapping:phenomenology}}
\nomenclature[forceeo]{$\forceeo$}{electro-osmotic force on \DHFRt, see~\cref{sec:trapping:phenomenology}}
\nomenclature[forcest]{$\forcest$}{steric force on \DHFRt, see~\cref{sec:trapping:phenomenology}}

\nomenclature[DeltaE]{$\barrier$}{energy barrier}
\nomenclature[DeltaEes]{$\barrier_{\rm{es}}$}{electrostatic energy barrier, see~\cref{eq:decomposition}}
\nomenclature[DeltaEst]{$\barrier_{\steric}$}{steric energy barrier, see~\cref{eq:decomposition}}
\nomenclature[DeltaEex]{$\barrier_{\ext}$}{external energy barrier, see~\cref{eq:decomposition}}
\nomenclature[DeltaEstat]{$\barrier_\static$}{static energy barrier, see~\cref{eq:static-barrier}}

\nomenclature[Deltax]{$\Delta x$}{barrier distance, see~\cref{eq:external-barrier}}

\nomenclature[DeltaVtag]{$\potbar_{\rm{tag}}$}{electrostatic potential at the tag of \DHFRt, see~\cref{eq:static-barrier}}

\nomenclature[k]{$\rate$}{observed escape rate, see~\cref{eq:double_barrier}}

\nomenclature[taud]{$\dwelltime$}{dwelltime, see~\cref{eq:double_barrier_simple,eq:double_barrier}}
\nomenclature[taurho]{$\chargedensrelaxtime$}{surface charge density relaxation time, see~\cref{eq:nanopores_relaxation_time}}

\nomenclature[Iopen]{$\iopen$}{open pore ionic current}
\nomenclature[Iblock]{$\iblock$}{blocked pore ionic current}
\nomenclature[Isim]{$\currentsim$}{simulated ionic current, see~\cref{eq:currentsim}}

\nomenclature[Bi]{$\bfactor_i$}{temperature factor of atom $i$, see~\cref{eq:bfactor}}

\nomenclature[rhomol]{$\rho_\text{mol}$}{molecular density, see~\cref{eq:denspore}}

\nomenclature[G]{$\conductance$}{nanopore ionic conductance}
\nomenclature[GNa]{$\gna$}{nanopore conductance carries by sodium ion}
\nomenclature[GCl]{$\gcl$}{nanopore conductance carries by chloride ion}

\nomenclature[alphaicr]{$\icr$}{ionic current rectification, see~\cref{eq:icr}}

\nomenclature[dw]{$\walldistance$}{distance from the nanopore wall}

\begin{document}


\makefrontcoverXII

\maketitle

\thispagestyle{empty}

\pagebreak
\hspace{0pt}
\vfill

\settowidth{\versewidth}{the effect that no book or poem is}
\begin{verse}[\versewidth]
After ten standard months I was done,\\
acknowledging the ancient aphorism to\\
the effect that no book or poem is\\
ever finished, merely abandoned.
\end{verse}
\attrib{\textit{`Martin Silenus' in `Hyperion' by Dan Simmons}}

\vfill
\hspace{0pt}
\pagebreak

\cleardoublepage


\frontmatter 

\includepreface{preface}
\includeabstract{abstract}
\includeabstractnl{abstractnl}

\myprintglossary

\myprintnomenclature

\tableofcontents
\listoffigures
\listoftables


\mainmatter 

\cleardoublepage

%
  \graphicspath{{\chaptersdir/nanopores/image/}}%
\chapter{Biological nanopores}
\label{ch:nanopores}

\epigraphhead[\epipos]{%
\epigraph{%
  ``There is a place in our minds now that tools---made things, things of value---occupy, and there was a time
  when our ancestors' minds had no such place. Nothing was ever for keeps, and there was no tomorrow for the
  tool-makers.''
}{%
  \textit{`Ruth Emerson' in `The Doors of Eden' by Adrian Tchaikovsky}
}
}


%
%
\definecolor{shadecolor}{gray}{0.85}
\begin{shaded}
Parts of this chapter were adapted from:
\begin{itemize}
  \item K. Willems*, V. Van Meervelt*, C. Wloka and G. Maglia.
        \textit{Phil. Trans. R. Soc. B} \textbf{372}, 20160230 (2017) 
\end{itemize}
*equal contributions
\newpage
\end{shaded}
This chapter serves as a comprehensive introduction to the primary concepts required to understand the
objectives and relevance of the work performed in this dissertation. In particular, the reader will be
familiarized with biological nanopores as single-molecule sensors, the physical mechanisms that govern
nanopore sensing, and the current approaches for the modeling of nanopores.
The text and figures of this introductory chapter were entirely written and created by me.
The section
``Confining proteins within biological nanopores''
was adapted from ref.~\cite{Willems-VanMeervelt-2017}.

\clearpage

\section{Use of tools}

Some 5 million years ago, our evolutionary ancestors became bipedal in order to better observe and understand
the world around them. Up until the \textit{Homo habilis} introduced the \emph{use of tools}, these ancients
depended solely on their senses to provide them with information, and their hands to shape their environment.
Arguably, this extensive tool usage also set in motion a process of rapid encephalization (\ie~increase of the
brain-to-body mass ratio), further improving the human clade's capability to process, interpret and
communicate information. Even though many other species are active tool-users, to date, only the \textit{Homo}
genus appears to employ them to fabricate more complex tools. This behavior has been observed as far back as
2.6~million years, but the advances brought forth by the scientific and industrial revolutions truly ushered
in a new era of complexity, which was topped only by the invention of the modern computer and the internet,
which have led us into the present day's information age.

Just as we shaped our tools, so have they shaped us. They have impacted virtually all aspects of our society
and are still doing so at an increasing rate. This is particularly true in the sciences, which, in their quest
to find new ways to probe, analyze, quantify, and understand the world around us, have been both tremendous
benefactors and beneficiaries. To this end, we have succeeded in extending our senses far beyond their natural
capabilities with a host of external `sensors', which, quoting the Cambridge dictionary, are
\begin{quote}
  devices that are used to record that something is present or that there are changes in something.
\end{quote}
Contemporary technical capabilities aside, both the scale of the sensor and the nature of the `something' it
is supposed to detect, are constrained solely by our ingenuity and imagination. For example, we have built
both gigantic machines---a \SI{27}{\kilo\meter}-long particle accelerator for studying the physical basis of
gravity~\cite{ATLAS-2012}, or a \SI{4}{\kilo\meter}-long laser interferometer for detecting gravitational
waves produced during the merging of two black holes~\cite{Abbott-2016}---as well as tiny
devices---sub-millimeter-sized implantable sensors for monitoring physiological functions~\cite{Dong-2019}, or
\SI{20}{\nm}-wide \glspl{fet} capable of quantifying biomarkers from blood samples~\cite{Krivitsky-2016}.

This dissertation should be framed in the latter category. Specifically, we have used physical, computational,
and mathematical tools---in the form of experiments, numerical simulations, and physical models,
respectively---to improve our fundamental understanding of minuscule sensors, approximately \num{10000} times
smaller than the width of a human hair: \emph{nanopores}.

\section{Sensing at the single-molecule level}

Before delving into the ``what'', ``how'', and ``why'' of nanopores, let us take a step back and put them into
a broader context. Nanopores, and particularly biological ones, belong to the class of single-molecule
biosensors. Here, the `bio' reflects that one or more biological components (\eg~tissues or cells, or be parts
thereof, such as organelles, enzymes, antibodies, or \gls{dna} molecules) comprise an integral part of the
sensing process, usually as the element that interacts (specifically) with the analyte molecule of
interest~\cite{Banica-2012}. The sensor element can be any type of physicochemical detector (\eg~optical,
electrochemical, mechanical, etc.) that quantifies the interactions between the biological element and the
analyte molecule. In this section, we will introduce the concept of single-molecule sensing, together with a
non-exhaustive selection of some of the most prominent single-molecule techniques.

\subsection{From bulk to single-molecule}

Classical `bulk' (bio)sensors typically measure the ensemble-average of a quantity of interest, meaning that
the final signal originates from a great many (read: from \numrange{10}{e21}) molecules at the same time. For
many applications, such as the enzyme-linked immunosorbent assay or the quantitative polymerase chain
reaction, this information is typically more than sufficient. However, considering that there are many ways to
obtain the same average value, the very nature of the ensemble-average also hides the property of interest's
true distribution and may obscure rare or transient events. For example, when measuring the activity of an
enzyme in bulk, does the signal stem from a single, `average' sub-population, or are there multiple
sub-populations active at the same time---some working slowly and others faster, balancing each other to yield
the observed ensemble-average? One way to answer this question is to interrogate the individual enzymes
themselves, at the so-called `single-molecule' level. Whereas bulk experiments can be considered deterministic
and continuous due to the sheer number of molecules involved in every measurement, the process of
single-molecule sensing is inherently stochastic, meaning that a property can only be quantified by
`sampling' it multiple times, up until a given statistically relevance is reached and the true distribution is
revealed. Besides the study of enzyme kinetics, single-molecule measurements empower several super-resolution
imaging techniques---for which the 2014 Nobel Prize in Chemistry was awarded~\cite{Weiss-2014}---and they lie
at the heart of the third-generation \gls{dna} sequencing technologies, enabling faster and longer reads,
direct detection of epigenetic markers, improved portability, and lower cost~\cite{Schadt-2010}.

Regardless of its precise application or mode of action, any single-molecule technique faces two fundamental
challenges compared to bulk sensors: (1) the amplification of a signal emanating from a single molecule to a
macroscopically measurable signal, and (2) the reduction of the sensing volume such that it contains only a
single analyte molecule. Given that the majority of molecules are intrinsically nanoscopic (\ie~with
characteristic length scales between \SIrange[range-phrase={ and }]{1}{10}{\nm}), it follows that, without
some means of amplification, the signal it generates is equally small, and hence demanding to detect reliably.
The second challenge, on the other hand, is not related to the size of the analyte molecule \textit{per se},
but rather to its concentration. The reason is that, for the duration of at least a single measurement, the
detection volume may contain only a single molecule (and the same one at that). For example, the
inter-molecular distances inside solutions with analyte concentrations of \SI{1}{\nM}, \SI{1}{\uM}, and
\SI{1}{\mM} are, on average, \SIlist{1500;150;15}{\nm}, respectively. Hence, achieving the small sensing
volumes required for physiologically relevant concentrations (\si{\uM} to \si{\mM}) may be difficult to
achieve~\cite{Zhu-2012}.

\subsection{Fluorescence spectroscopy}

One phenomenon that lends itself particularly well to single-molecule techniques is fluorescence: the
emission of light by a molecule (\ie~fluorophore) after absorbing energy from an external electro-magnetic
radiation source. Because some energy is lost to the environment during the fluorescence process, the emitted
light is often redshifted compared to its absorbed counterpart, allowing the two to be easily separated from
one another. Moreover, considering that a single fluorophore can emit \numrange{1}{100}~million photons every
second, they have intrinsically high signal-to-noise ratios. However, because the diffraction poses a physical
lower limit on how strongly light can be confined, the reduction of the sensing volume is not so easily
solved. For state-of-the-art confocal microscopes, this limit is equal to roughly half the wavelength of the
light used (\ie~\SIrange{\approx150}{350}{\nm} for visible light), which mandates the use of very low
analyte concentrations (\SI{\le1}{\nM}). Luckily, several solutions to this problem have been developed over
the years, notably, those based on near-field techniques. For example, the exponentially decaying
(`evanescent') near-fields inside nanoscale metal apertures called
\glspl{zmw}~\cite{Levene-2003,Eid-2009,Zhu-2012}, or protruding from the tip of a
\gls{nsom}~\cite{Ambrose-1994,Hosaka-2001}, allow the confinement of light to volumes of
$50\times50\times50$~\si{\cubic\nm}. Even stronger confinement, down to a few \si{\cubic\nm}, can be achieved
with specialized metal structures that generate plasmonic `hot-spots'~\cite{Xin-Lu-2019}. Finally, the
combination of the above-mentioned techniques with another distance-dependent phenomenon such as \gls{fret},
which is the non-radiative transfer of energy between a donor and acceptor fluorophore when they are closer
than \SI{\approx10}{\nm}~\cite{Roy-2008}, allows one to both limit the observation volume and study
conformational dynamics at the nanoscale~\cite{Kim-2013}. A major drawback of fluorescence-based techniques is
that contrary to label-free methods, they often necessitate attaching a fluorescent dye molecule to the
target molecule. This adds additional complexity to the sensing process and may affect the properties of the
analyte. Perhaps more importantly, the inevitable photobleaching of the typical organic fluorophore
(\ie~seconds) imposes strict limitations on the length that any individual molecule can be observed, and hence
restricts its use in the study of relatively fast processes.

\subsection{Force spectroscopy}

A different approach is taken by force spectroscopy techniques, where a large `probe' is monitored with a
highly sensitive force-feedback system to measure the properties of individual molecules when they are placed
under some form of mechanical strain~\cite{Neuman-2008}. The type of probe depends on the technique in
question. In \gls{afm}, it consists of a very sharp tip that is either scanned across a surface to determine
its topology (hence the term microscopy) or attached (non)-covalently to the molecule of interest such that
it can exert a precise pulling force on it. In optical and magnetic tweezers, analyte molecules are attached
to respectively dielectric and magnetic probe particles and trapped within optical or magnetic field
gradients. Because the displacement of the particle correlates with the exerted force, the latter can be
easily deduced by monitoring the particle's position over time. Hence, these techniques transduce nanoscale
motions and forces using much larger probes and reduce the sensing volume \textit{via} covalent
attachment to the probe or the surface. Force spectroscopy has been used to investigate the (un)folding of
proteins~\cite{Bustamante-2020,Jagannathan-2013} and protein domains~\cite{Rief-1997,Kellermayer-1997}, and to
elucidate protein-\gls{dna} interactions~\cite{Abbondanzieri-2005,Shlyakhtenko-2007,Vanderlinden-2019} and
dynamics~\cite{Lyubchenko-2018,Brouns-2018}. Major drawbacks of force spectroscopy techniques are their
limited throughput (all), the complexity of the instrumentation (all), and the high laser intensities required
(optical tweezers).

\subsection{Ionic current spectroscopy (\ie~nanopores)}
\begin{figure*}[b]
  \centering
  \medskip
  \includegraphics[scale=1]{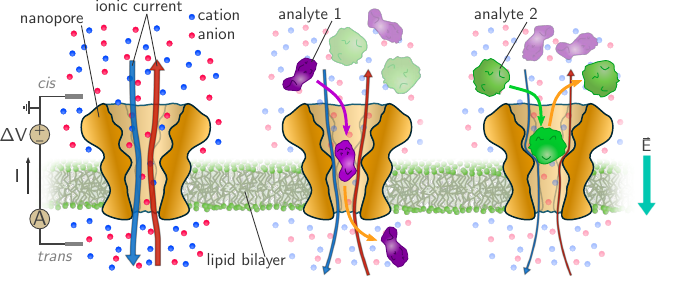}
  \\
  \includegraphics[scale=1]{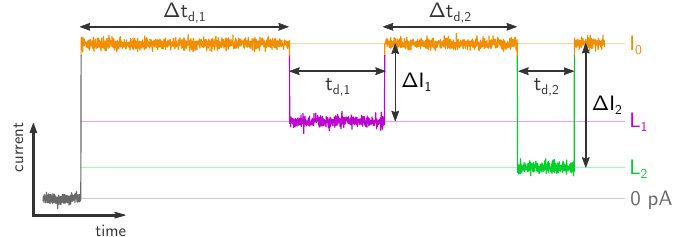}
\caption[Concept of single-molecule sensing with nanopores]{%
  \textbf{Concept of single-molecule sensing with nanopores.}
  (Left panel)
  When a net potential difference ($\Delta V$) is applied across two electrolyte reservoirs (\cisi{} and
  \transi{}) separated by a lipid bilayer containing a single nanopore, the electric field drives cations
  (blue) and anions (red) through the pore, resulting in a measurable `open pore' current ($\iopen$).
  (Middle panel)
  After a given time, $\Delta t_{\rm{d,1}}$, the entry of analyte 1 molecule (purple) into the water-filled
  channel of the pore from the \cisi{}-side transiently blocks the ionic current until the molecule exits
  into the \transi{} reservoir (\ie~it `translocates'). This gives rise to the $\ilevel{1}$ current level,
  which has a characteristic `dwell time' $t_{\rm{d,1}}$.
  (Right panel)
  Similarly, entry of the larger analyte 2 molecule (green) into pore from the \cisi{} side, after inter-event
  time $\Delta t_{\rm{d,2}}$, leads to the deeper current blockade level $\ilevel{2}$ with a dwell time
  $t_{\rm{d,2}}$. Because of its size, however, it cannot pass through the narrowest point of the channel and
  hence will either remain trapped inside the pore or eventually exit back into the \cisi{} reservoir.
  Whereas the duration and depth of the block yield information on the properties of the molecule, the
  inter-event time correlates with its concentration.
  }\label{fig:nanopores_concept}
\end{figure*}

This brings us back to the technique at hand, nanopore-based sensing~\cite{Howorka-2009,Wang-2018}.
Conceptually, single-molecule nanopore measurements (\cref{fig:nanopores_concept}) are perhaps the easiest to
understand of all the techniques discussed in this section. A nanopore device consists of a nanoscopic
aperture (\SIrange{1}{100}{\nm} in diameter) in an otherwise impermeable membrane. This opening serves as the
sole liquid connection between two reservoirs filled with a saline solution (\ie~the electrolyte, typical
ionic strengths vary between \SI{10}{\mM}~and~\SI{3}{\Molar}). When an electrical potential difference is
applied between the two reservoirs (\ie~the \cisi{} and \transi{} sides), a strong directional electric field
is formed within the nanopore that sets into motion the charged ions contained within. This opposing stream of
cations and anions towards their respective electrode polarities results in a steady-state ionic current. For
example, in a \SI{0.15}{\Molar} \ce{KCl} solution, the conductivity of a typical biological nanopore (\eg~with
a diameter of \SI{3}{\nm} and a length of \SI{10}{\nm}) would be approximately \SI{1}{\nS}
(see~\cref{sec:np:potential}). Applying a bias voltage of \SI{0.1}{\volt} across such a nanopore would result
in a current of \SI{100}{\pA}, which is equivalent to the passage of \num{600} million ions through the pore
every second. Hence, the large number of ions enables the real-time measurement of the ionic current at high
sampling frequencies, typically in the \SIrange{1}{50}{\kHz} range~\cite{Maglia-2010}, though experiments up
to \SI{1}{\mega\hertz} have been reported~\cite{Rosenstein-2012}.\footnotemark%
\footnotetext{
Note that the actual sampling limit is set by the signal-to-noise ratio. Noise in nanopores has many
contributions from a diverse set of physical phenomena, and includes flicker, shot, thermal current,
dielectric, and capacitive noise. A detailed discussion falls beyond the scope of this dissertation, and the
interested reader is referred to the comprehensive overview of current noise in nanopores published recently
by Fragasso \etal{}~\cite{Fragasso-2020}.
}
If the characteristic diameter of the nanopore is comparable to that of the molecule of interest (\ie~the
analyte), its entry into the pore will temporarily and significantly disrupt the regular `open pore' ionic
current, resulting in a so-called `blocked-pore' signal. Here, the depth of the current blockade and its
duration yield information on the properties of the molecule, whereas the inter-event time correlates with its
concentration. For any given analyte molecule, both the magnitude and the duration of a current blockade
depend on the exact type of nanopore (biological or solid-state, geometry, material, etc.), and the conditions
of the experiment (ionic strength, pH, bias voltage, etc.), making them notoriously difficult to predict and
interpret unambiguously. Typical blocked-pore current values range from \SIrange{10}{95}{\percent} of the open
pore current~\cite{Fragasso-2020,Zernia-2020,Howorka-2009}. The regular dwell times vary from
\SI{10}{\us}~to~\SI{100}{\ms} (\eg~for \gls{dna}~\cite{Maglia-2010} or protein~\cite{Watanabe-2017}
translocation) but can increase to seconds or even minutes in specific cases (\eg~proteins trapped within a
nanopore~\cite{Soskine-2012}). Finally, the fact that the interior dimensions of the pore are typically such
that only a single molecule can fit inside, means that nanopores are intrinsically well-suited as
single-molecule sensors.

Over the past 24 years, the primary driving force behind the development of nanopores has been, and perhaps
still is, single-molecule \gls{dna} sequencing~\cite{Deamer-2016}. This application has now been successfully
commercialized by the company \gls{ont}~\cite{ONT-2020,Jain-2018}. This has spurred researchers to deploy
nanopores for sensing a wide variety of analytes~\cite{Wang-2018} with both organic chemistries---such as
nucleic acids~\cite{Kasianowicz-1996,Meller-2000,Stoddart-2009,Manrao-2012},
proteins~\cite{Mohammad-2008,Firnkes-2010,Spiering-2011,RodriguezLarrea-2013}, and
polymers~\cite{Robertson-2007,Baaken-2011}---and inorganic ones---such as
ions~\cite{Bezrukov-1993,Kasianowicz-1995,Kasianowicz-1999,Ali-2011,Roozbahani-2020} and metallic
nanoparticles~\cite{Astier-2009,Angevine-2014,Campos-2018}. Evidently, the application space is equally broad,
ranging from proteomics~\cite{Yusko-2017,Houghtaling-2019} and protein
sequencing~\cite{Restrepo-Perez-2018,Huang-2019}, to single-molecule
enzymology~\cite{Willems-VanMeervelt-2017,Ho-2015,Wloka-2017,Harrington-2019,Galenkamp-2020} and
metabolomics~\cite{VanMeervelt-2017,Zernia-2020}.

Not only are nanopores extremely versatile, their inherently small size and relatively simple read-out
mechanism (\ie~a current amplifier) makes them amenable to both miniaturization and parallelization. Both of
which are exemplified by \gls{ont}'s \code{MinION} device---a handheld, third-generation sequencer powered by
\num{512} individual nanopores capable of providing \SI{30}{\giga\bp} (10 times the human genome) of real-time
sequencing data~\cite{ONT-2020}. Next to these advantages, nanopores are also readily combined with the other
single-molecule techniques discussed above, providing both complementary and confirmatory information. Examples
from literature include \gls{afm}~\cite{Aramesh-2019}, optical
tweezers~\cite{Keyser-2006,vanDorp-2009,Hall-2009,Galla-2014}, magnetic tweezers~\cite{Peng-2009},
fluorescence microscopy~\cite{McNally-2010,Anderson-2014,Assad-2014,Huang-2015},
\glspl{zmw}~\cite{Auger-2014,Larkin-2017,Spitzberg-2019}, and plasmonic
structures~\cite{Im-2010,Chen-2018,Verschueren-2018,Garoli-2019}.

\begin{figure*}[b]
  \centering
  \medskip
  \begin{subfigure}[b]{115mm}
    \centering
    \caption{}\vspace{-2.5mm}\label{fig:nanopores_types_biological}
    \includegraphics[scale=1]{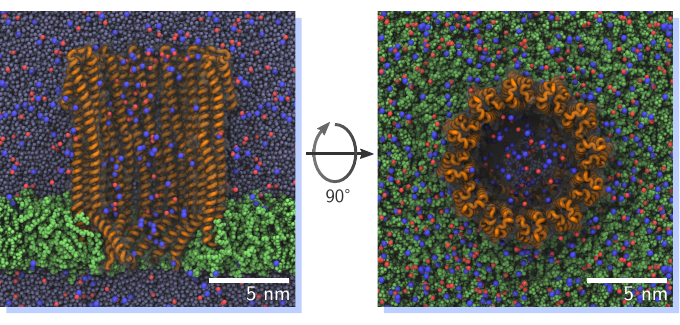}
  \end{subfigure}
  \\ \vspace{-1mm}
  \begin{subfigure}[b]{115mm}
    \centering
    \caption{}\vspace{-2.5mm}\label{fig:nanopores_types_solidstate}
    \includegraphics[scale=1]{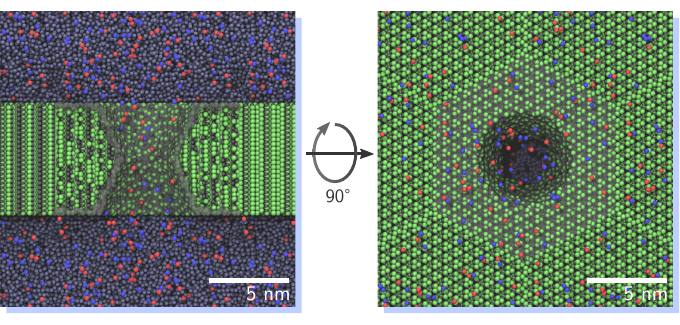}
  \end{subfigure}
\caption[Types of nanopores.]{%
  \textbf{Types of nanopores.}
  Cross-sectional (left) and top (right) views for a fully atomistic representation of 
  (\subref{fig:nanopores_types_biological})
  a Cytolysin A (ClyA) biological nanopore (orange) embedded in a lipid bilayer (green), and
  (\subref{fig:nanopores_types_solidstate})
  a double-conical solid-state nanopore drilled through a \SI{7}{\nm}-thick \ce{Si3N4} membrane (green and
  gray spheres).
  Both pores are surrounded by a \SI{0.15}{\Molar} electrolyte solution containing cations (blue spheres),
  anions (red spheres), and water molecules (light blue spheres). Note that for clarity, only the water
  molecules behind the pore are shown, represented by their oxygen atom.
  Molecular structures were prepared and rendered using \gls{namd}~\cite{Phillips-2005} and
  \gls{vmd}~\cite{Humphrey-1996,Stone-1998}.
  }\label{fig:nanopores_types}
\end{figure*}
\section{Types of nanopores}
\label{sec:np:types}

Depending on their constituent material, nanopores are typically classified into two distinct groups: the
protein-based \glspl{bnp}~\cite{Willems-VanMeervelt-2017} (\cref{fig:nanopores_types_biological}), and the
semiconductor-based \glspl{ssnp}~\cite{Dekker-2007} (\cref{fig:nanopores_types_solidstate}). This division
runs deeper than just the material, however: whereas \glspl{bnp} are formed through \emph{bottom-up}
self-assembly by inserting themselves into the semi-fluid membrane material (\eg~a lipid
bilayer)~\cite{Howorka-2017}, \glspl{ssnp} are fabricated by permanently removing material from the membrane
material itself with \emph{top-down}, micromachining techniques~\cite{Dekker-2007}. In the next sections we
will describe both types in more detail.

\subsection{Biological nanopores}

The majority of the \glspl{bnp} in use today are in fact `repurposed' \glspl{pft}, water-soluble proteins
excreted by pathogenic bacteria that oligomerize and self-assemble onto the membranes of other cells to
permeabilize them, disrupting vital ionic gradients, leaking nutrients into the environment, and potentially
killing the target cell~\cite{Peraro-2015}. \Glspl{pft} belong to either the \ta- or \tb-\gls{pft} class,
depending on whether the architecture of their \gls{tmd} is based on \ta-helices or \tb-sheets, respectively.
In \ta-\glspl{pft}, the \gls{tmd} is composed of amphipathic \ta-helices, typically arranged in a barrel
configuration with titled sidewalls. The hollow interior of the pore is lined with hydrophilic amino acids and
its exterior surface is covered with hydrophobic residues. This double nature allows the toxin to form a
stable, water-filled channel into an otherwise water-impermeable membrane. \Gls{clya} from \textit{E. coli} or
\textit{S. typhi}~\cite{Mueller-2009} (also known as HlyE) and \gls{frac} from \textit{Actinia
fragacea}~\cite{Tanaka-2015} are the most well-studied nanopores from this class. The \gls{tmd} of
\tb-\glspl{pft} consists of an amphipathic \tb-barrel fold, which is typically a perfect cylinder. The most
well-known \tb-\glspl{pft} that are also used as nanopores are \gls{ahl} from \textit{Staphylococcus
aureus}~\cite{Song-1996}, aerolysin from \textit{Aeromonas hydrophila}~\cite{Iacovache-2016}, and the recently
characterized \gls{plyab} from \textit{Pleurotus ostreatus}~\cite{Lukoyanova-Kondos-2015}. Notable non-toxin
membrane channels with a \tb-barrel fold include \gls{mspa} outer-membrane protein~\cite{Faller-2004} and the
\gls{csgg} amyloid secretion channel from \textit{E. coli}~\cite{Goyal-2014}---both of which have been
successfully employed for nanopore \gls{dna} sequencing~\cite{Manrao-2012,Brown-2016}---and
\gls{fhua}~\cite{Locher-1998}, \gls{ompf}~\cite{Yamashita-2008} and \gls{ompg}~\cite{Subbarao-2006} proteins
from \textit{E. coli}. Besides native membrane proteins, also other channel-like proteins have been engineered
to behave like nanopores. For example, the \gls{phi29p}~\cite{Xu-2019}, an essential component of the
\gls{dna} packaging mechanism of the \textPhi29 phage, was modified to insert (reversibly) into lipid bilayers
and shown to translocate \gls{dsdna}~\cite{Wendell-2009}. Not all membrane-inserting nanopores are
protein-based though, as researchers employed \gls{dna} origami~\cite{Rothemund-2006} to design and create
nanopores---from scratch---using nucleic acids as the primary building
block~\cite{Bell-2011,Langecker-2012,Burns-2013,Bell-2014,Gopfrich-2016,Gopfrich-2019}.

\subsection{Solid-state nanopores}

The methodologies for fabrication \glspl{ssnp} are derived from techniques used in the semiconductor industry
to create the nanometer-scale transistors found in virtually any contemporary electronic device. Generally,
the fabrication of a \gls{ssnp} starts with the deposition of a thin dielectric layer (typically
\SIrange{5}{50}{\nm}) on top of a flat silicon substrate. Because this layer will form both the membrane and
the nanopore, it must have both high mechanical strength and high chemical resistance. Hence, suitable
materials include \ce{Si3N4}~\cite{Li-2001,Storm-2003}, \ce{SiO2}~\cite{Storm-2005},
\ce{Al2O3}~\cite{Venkatesan-2009}, \ce{HfO2}~\cite{Larkin-2013} or even \ce{Si}
itself~\cite{Malachowski-2013}. Next, a part of the silicon at the backside of the substrate is removed using
wet or dry etching, leaving a small window that contains the so-called `free-standing membrane'. The \gls{ssnp}
is then formed by creating an aperture into this membrane, which can be achieved by either \textit{ex-situ}
drilling with a focused ion~\cite{Li-2001} or electron~\cite{Storm-2003} beams, or by electron-beam
lithography patterning in combination with wet/dry etching~\cite{Nam-2009}. With controlled dielectric
breakdown, \glspl{ssnp} can also be fabricated \textit{in situ}, where an intact membrane is mounted into the
measurement cell and locally etched using a high transmembrane voltage~\cite{Kwok-2014}. By the
(non-)conformal deposition of oxides~\cite{Chen-2004}, metals~\cite{Li-2013d,Auger-2014,Spitzberg-2019} or
organic molecules~\cite{Wanunu-2007,Yusko-2011,Wei-2012,Rotem-2012} inside the pores, their size, surface
properties and biocompatibility can be manipulated with relative ease~\cite{Eggenberger-2019}. Additionally,
by transferring or growing quasi-2D materials on top of large, `classical' nanopores, researchers have been
able to fabricate ultra-thin (\ie~monoatomic) nanopores from graphene~\cite{Fischbein-2008} and
\ce{MoS2}~\cite{Feng-2015b} for both \gls{dna}~\cite{Feng-2015,Merchant-2010,Song-2011} and
protein~\cite{Shan-2013} detection. Next to these complex semiconductor-based methods, the nanoscale apertures
obtained by pulling glass pipettes provide a powerful, although intrinsically unscalable, approach for
fabricating \glspl{ssnp}~\cite{Wang-2006}. They allow the low-noise detection of
\gls{dna}~\cite{Steinbock-2013} and proteins~\cite{Li-2013B} and can even be integrated with a field-effect
transistor~\cite{Ren-2020}.

\subsection{Biological or solid-state: choosing is losing?}

Because they are made of either proteins or \gls{dna}, \glspl{bnp} can be produced using well-established
bacterial expression systems, making them relatively inexpensive to work with. Using molecular biology
techniques such as site-directed mutagenesis, this same proteinaceous nature also provides a straightforward
method for tailoring the surface properties of \glspl{bnp} with atomic
precision~\cite{Howorka-2001,RinconRestrepo-2011}. Even though \glspl{bnp} typically have superior
signal-to-noise ratios~\cite{Fragasso-2020}, the fragility of the lipid bilayer severely limits the lifetime
of a \gls{bnp}. This prevents the storage of any individual \gls{bnp} for longer than a few hours, its usage
at harsh experimental conditions (\eg~elevated temperatures). The stochastic nature of the formation process
dictates that neither the time nor location of the nanopore's insertion into the lipid bilayer can be easily
controlled, making experiments often exceedingly tedious. The top-down approach taken with \glspl{ssnp}, on
the other hand, provides significantly more flexibility regarding the nanopore's size and material properties.
Their solid-state nature makes them both physically and chemically robust, enabling them to withstand even
harsh treatments (\eg~piranha or \ce{O2}-plasma cleaning). However, the expertise and equipment required to
reproducibly fabricate \glspl{ssnp} (\ie~a cleanroom with micromachining tools) can be prohibitively
expensive. Without a proper protective layer, such as \ce{HfO2}, the size of a \ce{Si3N4} pore is known to
`drift' due to slow chemical etching, either during measurements or when improperly stored~\cite{Chou-2020}. The
destructive nature of the fabrication process (\eg~\textit{via} ion beam milling, plasma etching, or
dielectric breakdown) makes precise control of the pore's shape challenging to achieve~\cite{vandenHout-2010}.
Finally, without an appropriate, bio-compatible coating, the intrinsic `stickiness' of their surfaces leads to
the physisorption of most biomolecules~\cite{Eggenberger-2019,Awasthi-2020}. This results in the denaturing
of analytes of interest at best, and the rapid clogging of the pore at worst~\cite{Yusko-2011}. Currently, the
biological pores still trump their artificial counterparts both in terms of performance and applications.
However, the field is rapidly working to close this gap, and the intrinsic scalability (\eg~parallelization),
the possibilities of integration with other devices (\eg~transistors), and the mass-manufacturability of
\glspl{ssnp} remain intriguing advantages. As with many competing concepts, perhaps the optimal solution may
lie somewhere in the middle, with the so-called `hybrid nanopores' in which one attempts to combine the best
of both worlds~\cite{Hall-2010,Im-2010,Cai-2018}.

\clearpage

\section{Biological nanopores of interest}
\label{sec:np:interest}

As is evident from \cref{sec:np:types}, there are many \glspl{bnp} worthy of an in-depth description of their
structural characteristics and pore formation mechanism. Nevertheless, we will limit ourselves here to those
that are directly relevant to this dissertation. We will start with \glsfirst{ahl}, the first nanopore used
for the detection of biomolecules, and hence also the one with the largest body of experimental, theoretical,
and computational work. Next, we will discuss \glsfirst{clya}, \glsfirst{frac} and \glsfirst{plyab}, the three
nanopores that will be the subject of interest within the results chapters.

\subsection[Alpha-Hemolysin (aHL)]{\ta-Hemolysin (\ta{}HL)}
\label{sec:np:ahl}
\begin{figure*}[p]
  \centering
  \medskip

  \begin{minipage}[t]{53mm}
    \begin{subfigure}[b]{53mm}
      \centering
      \caption{}\vspace{-8.5mm}\hspace{1.5mm}\label{fig:nanopores_ahl_pore_side}
      \includegraphics[scale=1]{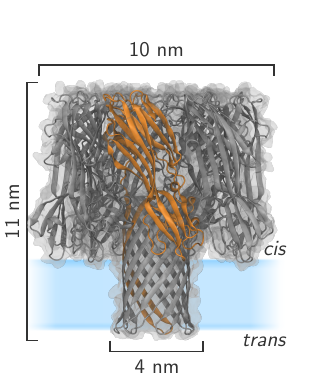}
    \end{subfigure}
    \vspace{5mm} \\
    \begin{subfigure}[b]{53mm}
      \centering
      \caption{}\vspace{-8.5mm}\hspace{1.5mm}\label{fig:nanopores_ahl_pore_top}
      \includegraphics[scale=1]{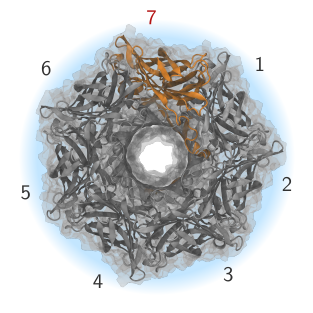}
    \end{subfigure}
  \end{minipage}
  \begin{minipage}{65mm}
    \begin{subfigure}[t]{53mm}
      \centering
      \caption{}\vspace{-8.5mm}\hspace{1.5mm}\label{fig:nanopores_ahl_pore_section}
      \includegraphics[scale=1]{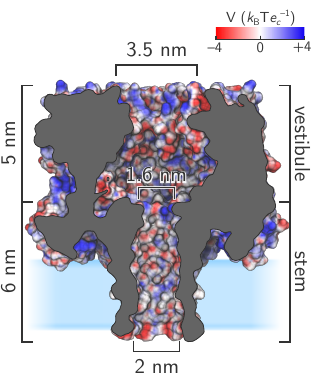}
    \end{subfigure}
    \vspace{5mm} \\
    \begin{subfigure}[b]{32mm}
      \caption{}\vspace{-8.5mm}\hspace{1.5mm}\label{fig:nanopores_ahl_monomer}
      \includegraphics[scale=1]{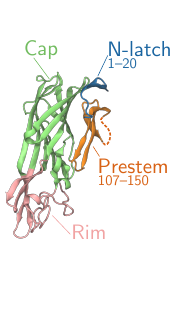}
    \end{subfigure}
    \begin{subfigure}[b]{32mm}
      \centering
      \caption{}\vspace{-8.5mm}\hspace{1.5mm}\label{fig:nanopores_ahl_protomer}
      \includegraphics[scale=1]{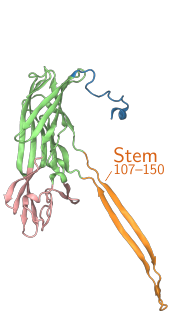}
    \end{subfigure}
  \end{minipage}

\caption[Structure of the \ta-hemolysin (\ta HL) porin]{%
  \textbf{Structure of the \ta-hemolysin (\ta HL) porin.}
  (\subref{fig:nanopores_ahl_pore_side})
  Side view and 
  (\subref{fig:nanopores_ahl_pore_top})
  top view of the heptameric \glsfirst{ahl} nanopore (\pdbid{7AHL}~\cite{Song-1996}), with a single subunit
  (protomer) highlighted in orange. The location of the lipid bilayer is indicated in blue, together with the
  \cisi{}  (`extracellular') and \transi{} (`intracellular') sides of the membrane. The pore is shaped as
  \SI{11}{\nm}-tall mushroom shaped with a cap (\ie~the extracellular part) and stem (\ie~the transmembrane
  \tb-barrel) widths of \SI{10}{\nm} and \SI{4}{\nm}, respectively.
  (\subref{fig:nanopores_ahl_pore_section})
  Electrostatically colored cross-section of the interior molecular surface of \gls{ahl}. The potential at
  physiological conditions was calculated using \gls{apbs}~\cite{Baker-2001,Baker-2005}. Indicated are the
  sizes of the \cisi{} vestibule, the transmembrane stem domain, and the \SI{1.6}{\nm} wide central
  constriction.
  (\subref{fig:nanopores_ahl_monomer})
  Crystal structure of the \SI{33}{\kDa} water-soluble \gls{ahl} monomer (\pdbid{4YHD}~\cite{Sugawara-2015})
  compared to that of
  (\subref{fig:nanopores_ahl_protomer})
  the protomer (\pdbid{7AHL}~\cite{Song-1996}) in the final pore. While the extracellular cap and the
  membrane-binding rim domains remain virtually unaltered during the pore formation process, the N-terminal
  latch and (pre)stem regions undergo considerable conformational changes. 
  All images were prepared and rendered using \gls{vmd}~\cite{Humphrey-1996,Stone-1998}.
  }\label{fig:nanopores_ahl}
\end{figure*}

The \gls{ahl} protein (\cref{fig:nanopores_ahl}) is a \tb-\gls{pft} secreted by \textit{Staphylococcus aureus}
in its post-exponential and stationary growth phases~\cite{Bhakdi-1991}. It shows a high affinity for
sphingomyelin-cholesterol microdomains~\cite{Menestrina-2001,Valeva-2006}, but also readily binds
non-specifically to other membranes at higher concentrations~\cite{Hildebrand-1991}. Its high chemical and
structural stability (\eg~temperature, pH, urea, mutations), together with the early availability of its
crystal structure (1996)~\cite{Song-1996}, have made \gls{ahl} the workhorse of choice in the nanopore field
for many years.

\subsubsection{Pore structure, shape, and charge distribution.}

The \gls{ahl} pore (\pdbid{7AHL}~\cite{Song-1996}) is composed out of seven identical subunits of 293~amino
acids each, that form a \SI{232}{\kDa} mushroom-shaped protein complex with a height of \SI{11}{\nm} and a
width of \SI{10}{\nm} (\cref{fig:nanopores_ahl_pore_side,fig:nanopores_ahl_pore_top}). Its secondary structure
elements are predominantly \tb-sheets (\SI{\approx50}{\percent}), exemplified by the \SI{4}{\nm}-wide
\tb-barrel transmembrane region (the `stem') and the \tb-sandwich folds of its extracellular domains (the
`cap' and `rim'), supplemented with \SI{\approx35}{\percent} of non-conventional structure~\cite{Song-1996}.
The interior of \gls{ahl} (\cref{fig:nanopores_ahl_pore_section}) consists of a vestibule compartment at the
\cisi{} side, with an entry diameter of \SI{3.5}{\nm} and a height of \SI{5}{\nm}, followed by a
\SI{6}{\nm}-long, \SI{2}{\nm}-wide cylindrical channel. Separating both compartments is a narrow constriction
\SI{1.6}{\nm} in diameter, formed by residues Glu111 and Lys147. The electrostatic surface potential
(see~coloring of \cref{fig:nanopores_ahl_pore_section}) reveals that \gls{ahl}'s internal surface charge
distribution is balanced, with neither positive nor negative charges dominating. However, due to the slight
excess of positive charges at the \cisi{} entry and the constriction, wild type \gls{ahl} pores are found to
be somewhat anion selective~\cite{Menestrina-1986}.

\subsubsection{Oligomerization stoichiometry.}

Up until the release of the crystal structure, it was believed that the \gls{ahl} pore complex was a 6-mer
(\ie~hexameric), instead of the now widely accepted 7-mer (\ie~heptameric)~\cite{Song-1996}. Even though it is
likely that \gls{ahl} also forms hexamers, the majority of pores (\SI{>90}{\percent}) formed under typical
conditions are heptameric~\cite{Menestrina-1986}, as supported by scientific evidence from a wide variety of
techniques such as crystallography~\cite{Song-1996,Galdiero-2004}, single-molecule fluorescence
photobleaching~\cite{Das-2007} and particle tracking~\cite{Thompson-2011}, and \gls{md}
simulations~\cite{Aksimentiev-2005,Bhattacharya-2011,Basdevant-2019}.

\subsubsection{Monomer \textit{versus} protomer.}

When comparing the structures of the \gls{ahl} water-soluble monomer (\cref{fig:nanopores_ahl_monomer},
\pdbid{4YHD}~\cite{Sugawara-2015}) with that of the protomer (\cref{fig:nanopores_ahl_protomer},
\pdbid{7AHL}~\cite{Song-1996}), it is evident that the majority of the secondary structure is maintained
during pore formation. Major conformational changes are observed for the stem domain, which unfolds to form
the transmembrane \tb-barrel, and within the triangle region, which connects it to the core of the pore.
Additionally, the first 20 N-terminal amino acids (\ie~the `N-latch') move away from their monomer positions
to (1) make contacts with the cap domain of its neighboring protomer and (2) destabilize the prestem domain.
Indeed, an intact N-latch was found to be essential for both stable oligomerization and high hemolytic
activity~\cite{Song-1996}. The \tb-sandwich folds of the extracellular `cap' and membrane-binding `rim'
domains remain nearly identical before and after pore formation.

\subsubsection{Mechanism of pore formation.}

After secretion by the host cell, the highly water-soluble \gls{ahl} monomers diffusive freely in the solution
until they encounter, and bind to, the membrane of a target cell, mediated by both electrostatic and
hydrophobic interactions. It has been shown that \gls{ahl} has both specific and aspecific membrane-binding
modes~\cite{Hildebrand-1991}, with the former likely mediated by sphingomyelin-cholesterol
microdomains~\cite{Valeva-2006}. Once bound, the monomers will collide and reversibly oligomerize,
self-assembling into a heptameric `prepore' that finally transitions (quasi-irreversibly) into the final pore
by inserting its stem domains into the lipid bilayer. Single-molecule fluorescence particle tracking
experiments have revealed this process to be extremely rapid, with single monomers transitioning into
heptameric pores typically within \SI{5}{\ms}, and seemingly without the formation of intermediate
states~\cite{Thompson-2011}. The ability of \gls{ahl} (and other \glspl{pft}) to keep the concentration of
intermediates low is thought to limit assembly errors~\cite{Lee-2016b}, mainly by preventing the formation of
higher-order oligomers. At best, these may form pores with sub-optimal stoichiometries, at worst, they may
lead to non-functional aggregates, both of which are outcomes that reduce the efficacy of the
toxin~\cite{Fahie-2013,Subburaj-2015}. Energetically speaking, oligomerization is stabilized by both entropic
and enthalpic means, as in the final heptameric pore each protomer buries approximately \SI{33}{\percent} of
its solvent-accessible area, establishes 850 van der Waals contacts, and forms 120 salt bridges and hydrogen
bonds~\cite{Song-1996}. The prepore-to-pore transition is initiated by the destabilization of the prestem due to
steric hindrance with its own N-latch and the neighboring prestem~\cite{Sugawara-2015}. Specifically, residues
Asp13--Gly15 of the N-latch compete with residue Tyr118 of the prestem for binding to D45 in the cap domain,
which causes all the prestem regions to release from their respective cores and assemble into the final stem
domain. This is likely to be a two-step process, in which first the extramembrane part of the \tb-barrel is
formed, followed by the transmembrane section~\cite{Sugawara-2015}.

\subsection{Cytolysin A (ClyA)}
\label{sec:np:clya}

\begin{figure*}[p]
  \centering
  \medskip
  \begin{minipage}[t]{51mm}
    \begin{subfigure}[b]{51mm}
      \centering
      \caption{}\vspace{-8.5mm}\hspace{1.5mm}\label{fig:nanopores_clya_pore_side}
      \includegraphics[scale=1]{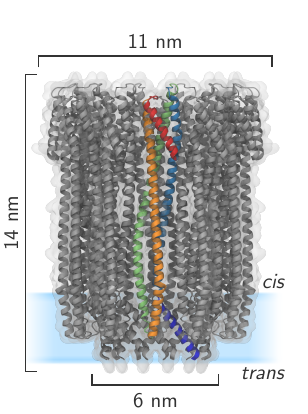}
    \end{subfigure}
    \vspace{5mm} \\
    \begin{subfigure}[b]{51mm}
      \centering
      \caption{}\vspace{-8.5mm}\hspace{1.5mm}\label{fig:nanopores_clya_pore_top}
      \includegraphics[scale=1]{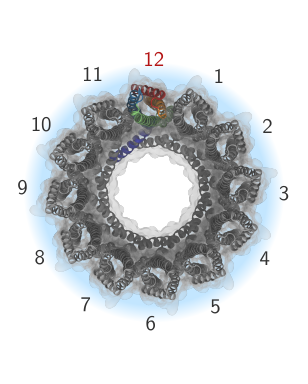}
    \end{subfigure}
  \end{minipage}
  \begin{minipage}{66mm}
    \begin{subfigure}[t]{60mm}
      \centering
      \caption{}\vspace{-8.5mm}\hspace{1.5mm}\label{fig:nanopores_clya_pore_section}
      \includegraphics[scale=1]{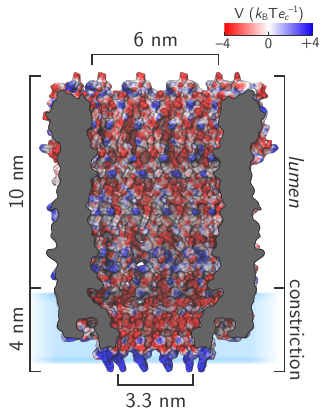}
    \end{subfigure}
    \vspace{5mm} \\
    \begin{subfigure}[b]{37mm}
      \caption{}\vspace{-8.5mm}\hspace{1.5mm}\label{fig:nanopores_clya_monomer}
      \includegraphics[scale=1]{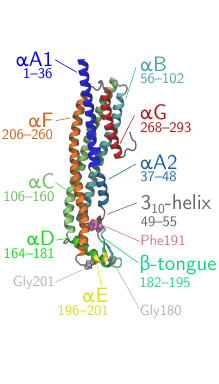}
    \end{subfigure}
    \begin{subfigure}[b]{28mm}
      \centering
      \caption{}\vspace{-8.5mm}\hspace{1.5mm}\label{fig:nanopores_clya_protomer}
      \includegraphics[scale=1]{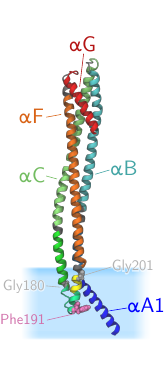}
    \end{subfigure}
  \end{minipage}
\caption[Structure of the cytolysin A (ClyA) porin]{%
  \textbf{Structure of the cytolysin A (ClyA) porin.}
  (\subref{fig:nanopores_clya_pore_side})
  Side view and 
  (\subref{fig:nanopores_clya_pore_top})
  top view of the dodecameric \glsfirst{clya} nanopore (\pdbid{6MRT}~\cite{Peng-2019}), with a single
  subunit (protomer) highlighted in color and the location of the lipid bilayer is indicated in blue.
  (\subref{fig:nanopores_clya_pore_section})
  Cross-section view of the interior molecular surface of the pore, colored according to its estimated
  electrostatic potential in physiological conditions as calculated by
  \gls{apbs}~\cite{Baker-2001,Baker-2005}.
  (\subref{fig:nanopores_clya_monomer})
  The \SI{34}{\kDa} water-soluble \gls{clya} monomer (\pdbid{1QOY}~\cite{Wallace-2000}) consists of a core
  bundle of \ta-helices (\ta A1, \ta A2, \ta B, \ta C, \ta F), onto which a hydrophobic \tb-tongue
  motif---flanked by two short \ta-helices (\ta D and \ta E)---and a C-terminal \ta-helix (\ta G) are packed.
  The two glycine residues in the \tb-tongue confer it with a high degree of flexibility, while the removal of
  Phe191 from its hydrophobic pocket triggers pore formation.
  (\subref{fig:nanopores_clya_protomer})
  A single \gls{clya} protomer (\pdbid{6MRT}~\cite{Peng-2019}), with the same coloring as the monomer, showing
  the insertion of the \ta-tongue into the lipid bilayer and the near \SI{14}{\nm} conformational change
  required for the \ta A1 helix to form the transmembrane part of the pore.
  All images were prepared and rendered using \gls{vmd}~\cite{Humphrey-1996,Stone-1998}.
  }\label{fig:nanopores_clya}
\end{figure*}

In 2009, Mueller \etal{} resolved the crystal structure of \gls{clya}
(\cref{fig:nanopores_clya})~\cite{Mueller-2009}, a large \ta-\gls{pft}, secreted by various \textit{S.
enterica} and \textit{E. coli} strains, with a high affinity for mammalian cell membranes. The large size of
its \lumen{}---compared to \gls{ahl} and \gls{mspa}, the most commonly used protein nanopores at the
time---makes \gls{clya} particularly well equipped to fully capture and study larger biopolymers, such as
proteins, in their native three-dimensional configuration
(\ie~folded)~\cite{Soskine-2013,Soskine-Biesemans-2015}.

\subsubsection{Pore structure, shape, and charge distribution.}

The \SI{408}{\kDa} \gls{clya} pore (\pdbid{6MRT}~\cite{Peng-2019}) consists of 12 identical subunits
(protomers), arranged in a ring-like quaternary structure to form a cylindrical complex roughly \SI{14}{\nm}
in height and \SI{11}{\nm} in diameter (\cref{fig:nanopores_clya_pore_side,fig:nanopores_clya_pore_top}). The
pore walls in contact with the solvent are formed by a core bundle of four tightly packed \ta-helices per
protomer, and the transmembrane region is formed by the iris-like arrangement of the amphipathic N-terminal
\ta-helices. Virtually all helices contribute to interprotomer contacts, resulting in the burying of
\SI{\approx2400}{\square\angstrom} surface area, and the formation of 25~H-bonds and 13~salt bridges per
protomer-protomer interface. The interior shape of \gls{clya} can be described as a set of two hollow
cylinders placed on top of each other: a large \cisi{} chamber (`\lumen{}') with an inner diameter of
\SI{\approx5.5}{\nm} and a height of \SI{10}{\nm} and a smaller \transi{} chamber (`\textit{constriction}')
with a width of \SI{\approx3.6}{\nm} and a height of \SI{4}{\nm} (\cref{fig:nanopores_clya_pore_section}).
\Gls{clya} has an excess negative charge (\SI{-60}{\ec} at \pH{7.5}), most of which riddle the interior walls
of the pore. This results in a negative electrostatic potential inside both the \lumen{} and the constriction
of the pore (\cref{fig:nanopores_clya_pore_section}) and explains the observed cation
selectivity~\cite{Soskine-2012,Franceschini-2016}.

\subsubsection{Oligomerization stoichiometry.}

Next to the typical 12-subunit pore (dodecamer, `Type I'), \gls{clya} can also form 13-mers (tridecamers,
`Type II') and 14-mers (tetradecamers, `Type III')~\cite{Soskine-2013,Peng-2019}, with constriction diameters
of \SIrange{3.3}{4.0}{\nm}, \SIrange{3.7}{4.4}{\nm} and \SIrange{4.2}{5.2}{\nm}, respectively. Aside from
their increased aperture size, the overall structure and assembly mechanism of these pores is identical to
that of the dodecamer, and hence the following discussion will be limited to the Type I pore. Interested
readers are referred to the work of Peng and coworkers~\cite{Peng-2019}, who recently obtained high-resolution
structures of Type II and III pores using \gls{cryo-em}.

\subsubsection{Monomer \textit{versus} protomer.}

The water-soluble \gls{clya} monomer (\cref{fig:nanopores_clya_monomer}) consists of predominantly
\ta-helices, with the core of the protein formed by a bundle of four long \ta-helices (\ta A, \ta B, \ta C and
\ta F). Packed at the top and bottom of this bundle are respectively the C-terminal helix (\ta G) and a
hydrophobic \tb-hairpin motif flanked by two short \ta-helices (\ta{}D and
\ta{}E)~\cite{Wallace-2000,Mueller-2009}. Comparing this structure to that of the protomer
(\cref{fig:nanopores_clya_protomer}), reveals that while three of the core helices are merely straightened and
elongated (at the expense of the \tb-hairpin and the \ta{}A2, \ta{}D and \ta{}E helices), the \ta{}A1 helix
must move to the opposite side of the monomer, a distance of \SI{\approx14}{\nm}! The mechanism that enables
these large conformational changes is similar to that of a spring-loaded lock, which is opened upon membrane
binding.

\subsubsection{Mechanism of pore formation.}

Initial binding of \gls{clya} monomers to the membrane (\textit{via} both specific and non-specific
interactions) is mediated by the partially solvent-exposed, hydrophobic residues of the \tb-hairpin motif.
Using Gly180 and Gly201 as hinges, the \tb-hairpin swings outwards to insert itself into the lipid bilayer.
This removes a stabilizing aromatic residue (Phe190) from its hydrophobic pocket and unlatches a two-helix
spring-loaded mechanism. This results in 1) the straightening and extension of three of the core helices (\ta
B, \ta C and \ta F), and 2) a \SI{180}{\degree} swinging of the \ta A helix (relative to \ta B, \ta C and \ta
F). The amphipathic \ta-A1 helix, which will form the \gls{tmd}, now rests on top of the membrane and is
connected to the rest of the protein \textit{via} a \gls{hth} motif, with Pro36 acting as a hinge. The
subsequent oligomerization of the membrane-bound monomers further packs and buckles the \ta A1 helices,
effectively pushing them downwards and allowing them to `wedge' a hole into the lipid bilayer. In contrast
with \gls{ahl}, \gls{clya} pore formation takes place on a timescale of minutes rather than
milliseconds~\cite{Benke-2015}. Initial far-UV \gls{cd} measurements of the \gls{clya} oligomerization process
in the presence of the mild, non-ionic detergent \gls{ddm} suggested a two-step process, in which the soluble
monomers undergo a rapid transition ($t_{1/2} \approx \SI{80}{\second}$) to a molten globule-like
intermediate, followed by a slow assembly  ($t_{1/2} \approx \SI{1000}{\second}$) of the membrane-bound
protomers into the full dodecameric pore~\cite{Eifler-2006}. However, subsequent investigation of the
oligomerization kinetics using \gls{rbc} hemolysis assays~\cite{Vaidyanathan-2014} and single-molecule
\gls{fret}~\cite{Benke-2015} revealed that the intermediate state is actually an off-pathway by-product that
slows down the pore formation~\cite{Roderer-2017}. Nevertheless, in the long term, most of the monomers will
convert to protomers rather than the molten globule intermediate, and intermolecular collisions between
protomers result in the formation of linear mixtures of oligomers (`protomer
elongation'~\cite{Roderer-2017}). Because virtually all linear oligomer species contribute to this process,
the effective protomer concentration does not decrease, making it both rapid and efficient. Formation of the
full pore then occurs rapidly \textit{via} ring closure when a pair of oligomers---with a total protomer count
of 12---collide and associate. Even though all linear oligomer species contribute to the formation of the
dodecameric pore, the relatively high abundance of the 5-, 6- and 7-mers dictate that \SI{>50}{\percent} of
all full pores are formed by these oligomers~\cite{Benke-2015}.

\subsection{Fragaceatoxin C (FraC)}
\label{sec:np:frac}

\begin{figure*}[p]
  \centering
  \medskip
  \begin{minipage}[t]{58mm}
    \begin{subfigure}[t]{58mm}
      \centering
      \caption{}\vspace{-8.5mm}\hspace{1.5mm}\label{fig:nanopores_frac_pore_side}
      \includegraphics[scale=1]{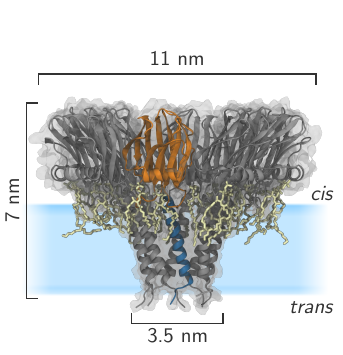}
    \end{subfigure}
    \vspace{5mm} \\
    \begin{subfigure}[t]{58mm}
      \centering
      \caption{}\vspace{-8.5mm}\hspace{1.5mm}\label{fig:nanopores_frac_pore_top}
      \includegraphics[scale=1]{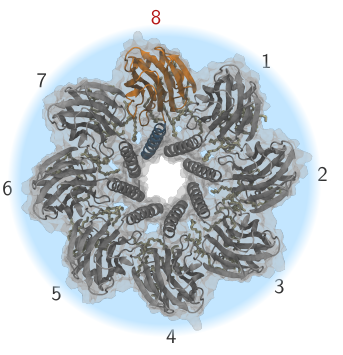}
    \end{subfigure}
  \end{minipage}
  \begin{minipage}[t]{58mm}
    \begin{subfigure}[t]{58mm}
      \centering
      \caption{}\vspace{-8.5mm}\hspace{1.5mm}\label{fig:nanopores_frac_pore_section}
      \includegraphics[scale=1]{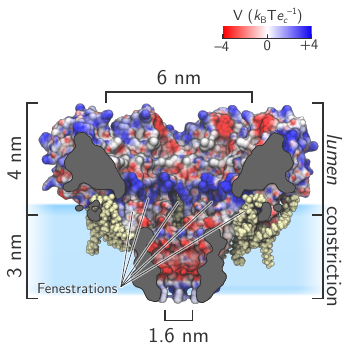}
    \end{subfigure}
    \vspace{5mm} \\
    \begin{subfigure}[t]{28mm}
      \caption{}\vspace{-8.5mm}\hspace{1.5mm}\label{fig:nanopores_frac_monomer}
      \includegraphics[scale=1]{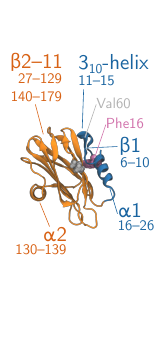}
    \end{subfigure}
    \begin{subfigure}[t]{28mm}
      \centering
      \caption{}\vspace{-8.5mm}\hspace{1.5mm}\label{fig:nanopores_frac_protomer}
      \includegraphics[scale=1]{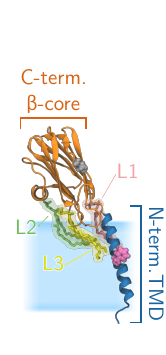}
    \end{subfigure}
  \end{minipage}
\caption[Structure of the fragaceatoxin C (FraC) porin]{%
  \textbf{Structure of the fragaceatoxin C (FraC) porin.}
  (\subref{fig:nanopores_frac_pore_side})
  Side and 
  (\subref{fig:nanopores_frac_pore_top})
  top views of the octameric \SI{176}{\kDa} \gls{frac} porin crystal structure
  (\pdbid{4TSY}~\cite{Tanaka-2015}) with a single protein subunit highlighted in color and the co-crystallized
  sphingomyelin lipids indicated in yellow. The overall shape is that of a funnel with a height of
  \SI{7}{\nm}, and top and bottom diameters of \SIlist{11;3.5}{\nm}, respectively.
  (\subref{fig:nanopores_frac_pore_section})
  Cross-sectional view of the internal molecular surface of \gls{frac}, colored according to the local
  electrostatic potential at physiological conditions as calculated by
  \gls{apbs}~\cite{Baker-2001,Baker-2005}. \Gls{frac} has inner diameters of \SIlist{6;1.6}{\nm} on the
  \cisi{} and \transi{} sides, respectively. Note the fenestrations between the subunits, which are partly
  filled by permanently bound lipids and hence expose the core of the membrane bilayer to the solvent.
  (\subref{fig:nanopores_frac_monomer})
  The crystal structure of the \SI{20}{\kDa} \gls{frac} monomer (\pdbid{3VWI}~\cite{Tanaka-2015}) shows that
  in its water-soluble form, the N-terminal \gls{tmd} (blue) is packed tightly against the \tb-sheet rich core
  domain (orange). Displacement of Phe16 (mauve) from its hydrophobic pocket by Val60 (gray) triggers pore
  formation.
  (\subref{fig:nanopores_frac_protomer})
  The N-terminal \ta-helix of each \gls{frac} protomer (\pdbid{4TSY}~\cite{Tanaka-2015}) first elongates and
  then flips \SI{180}{\degree} to span the entire membrane. The conformation of \tb-core remains virtually
  identical to that of the monomer. Where the L2 and L3 lipids are bound to a single subunit, the L1 lipid is
  wedged in between two protomers and contributes structurally to the solvent-exposed pore walls.
  Images were rendered using \gls{vmd}~\cite{Humphrey-1996,Stone-1998}.
  }\label{fig:nanopores_frac}
\end{figure*}

\Gls{frac} (\cref{fig:nanopores_frac}) is an \ta-\gls{pft} belonging to the family of the actinoporins and is
produced by the sea anemone \textit{Actinia fragacea}. Its X-ray structure was resolved in 2015 by Tanaka
\etal{}~\cite{Tanaka-2015}, and surprisingly contained several co-crystallized \gls{sm} lipids per subunit.
This indicates that \gls{frac} does not merely exhibit a high degree of specificity towards
\gls{sm}-containing membranes, but that these lipids comprise an essential structural element of the pore
itself. Recently, \gls{frac} has become popular for the analysis of peptides~\cite{Huang-2017,Huang-2019}
and amino acids~\cite{Restrepo-Perez-2019a}, making it a promising candidate for protein sequencing
applications~\cite{Restrepo-Perez-2018}.

\subsubsection{Pore structure, shape, and charge distribution.}

The full \SI{176}{\kDa} \Gls{frac} porin (\pdbid{4TSY}~\cite{Tanaka-2015}) consists of eight identical protein
subunits and three \gls{sm} lipids per protomer, bound at highly specific locations
(\cref{fig:nanopores_frac_pore_side,fig:nanopores_frac_pore_top}). It is shaped like a funnel with a height of
\SI{7}{\nm} and \cisi{} and \transi{} outer diameters of \SIlist{11;3.5}{\nm}, respectively. The extracellular
(\cisi) part of the pore comprises the bulk of each protomer and is rich in \tb-sheets (\tb-core). The
\gls{tmd} is formed by the iris-like arrangement of the N-terminal \ta-helices of each subunit, which are long
enough (\SI{3.5}{\nm}) to span the entire lipid bilayer (\SI{\approx2.8}{\nm}). The protomer-protomer
interface buries a surface area of \SI{777}{\square\angstrom}, to which both the \tb-core and the \gls{tmd}
\ta-helices contribute and which is stabilized by protein-lipid H-bonding~\cite{Tanaka-2015}. Close inspection
of the pore walls reveals the presence of eight lateral fenestrations, one between each protomer, that expose
the hydrophobic core of the membrane to the aqueous solvent (\cref{fig:nanopores_frac_pore_section}). These
openings are at least partially occluded by the co-crystallized \gls{sm} lipids, which, as such, contribute
structurally to the pore. Even though the precise biological function of these fenestrations remains unclear,
it is likely that they contribute significantly to the toxicity of sea anemone venom by 1) facilitating the
diffusion of its small hydrophobic components directly into the core of the lipid bilayer, and 2) disrupting
the leaflet composition of the membrane by catalyzing the flip-flop movement of lipids~\cite{Tanaka-2015}. The
interior of \gls{frac} is funnel-shaped, with a \cisi{} diameter of \SI{6}{\nm} and a \trans{} diameter of
only \SI{1.6}{\nm}, the latter also being the pore's narrowest location. Consistent with the observation of
its cation selectivity~\cite{Garcia-Ortega-2011,Wloka-2016}, the interior walls of the pore are predominantly
negatively charged, particularly at the \transi{} constriction (\cref{fig:nanopores_frac_pore_section}). The
extracellular regions of the pore near the lipid headgroups exhibit a strong positive potential, however, which
likely aids in the initial adhesion of \gls{frac} to the membrane~\cite{Tanaka-2015}.

\subsubsection{Oligomerization stoichiometry.}

The lytically active \gls{frac} pore can be obtained as an 8-mer (octamer, `Type~I'), 7-mer (heptamer
`Type~II') and 6-mer (hexamer, `Type~III')~\cite{Huang-2019}, with constriction diameters of
\SIlist{1.6;1.1;0.84}{\nm}, respectively. A non-lytic 9-mer (nonamer) has also been crystallized
(\pdbid{3LIM}~\cite{Mechaly-2011}), and lower oligomeric states also appear to be capable of forming
pores~\cite{Rojko-2016}.

\subsubsection{Monomer \textit{versus} protomer.}

In the water-soluble monomer of the \gls{frac} pore (\pdbid{3VWI}~\cite{Tanaka-2015}), the N-terminal region
(residues 1--29; contains a short \tb-sheet, a $3_{10}$ helix and an \ta-helix) is packed closely against the
C-terminal \tb-core (residues 30--179; contains 11 \tb-sheets in a \tb-sandwich fold and one \ta-helix) and
held in place by the interaction between Phe16 and a hydrophobic cavity in the \tb-core
(\cref{fig:nanopores_frac_monomer}). The \tb-core remains virtually unchanged in the protomer structure
(\pdbid{4TSY}~\cite{Tanaka-2015}), whereas the N-terminal domain forms a much longer \ta-helix that spans the
entire lipid bilayer (\cref{fig:nanopores_frac_protomer}). Out of the three lipids (`L1', `L2', and `L3') that
are bound at specific locations to the \tb-core, only the L1 position can be considered to be of high affinity
and specific only to \gls{sm}, whereas L2 and L3 are of lower affinity and hence more
promiscuous~\cite{Tanaka-2015}. 

\subsubsection{Mechanism of pore formation.}

Even though the precise mechanism with which actinoporins form pores is still under debate, several
experiments have already provided significant insights. Upon initial attachment of the monomer to the
membrane, the binding of a single \gls{sm} lipid to the L1 pocket induces dimerization~\cite{Tanaka-2015}. The
resulting small conformational change (at residues 14--17) displaces Phe16 from its hydrophobic pocket in
favor of Val60, and partially unfolds the N-terminal domain. Because dimerization is only possible in the
presence of \gls{sm}, this lipid acts as both a receptor and an assembly co-factor that contributes to the
final structure of the pore. Sequential addition of other dimeric units \textit{via} random collisions, which
can occur frequently in the small \gls{sm} lipid raft domains, give rise to higher oligomeric complexes that
further destabilize the N-terminal region. Given the limited space available in the \lumen{} of the final pore
(\SI{\approx2}{\nm} diameter), it is likely that the insertion of the N-terminal \ta-helix occurs in a
non-concerted manner prior to full ring closure~\cite{Cosentino-2016}. It is also possible that a pre-pore
structure is formed, similar to the non-lytic \gls{frac} nonamer~\cite{Mechaly-2011}, with the amphipathic
N-termini pushing through the membrane in a concerted fashion~\cite{Tanaka-2015,Rojko-2016} much like the
\gls{clya} porin. Evidence favors the first mechanism, however, given that a sole N-terminal helix of many
actinoporins can exist at the lipid-water interface in a so-called `protein-lipid pore'~\cite{Cosentino-2016}.

\clearpage
\subsection{Pleurotolysin AB (PlyAB)}
\label{sec:np:plyab}

\begin{figure*}[p]
  \centering
  \medskip
  \begin{minipage}[t]{56mm}
    \begin{subfigure}[t]{56mm}
      \centering
      \caption{}\vspace{-8.5mm}\hspace{1.5mm}\label{fig:nanopores_plyab_pore_side}
      \includegraphics[scale=1]{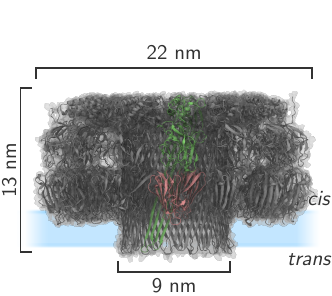}
    \end{subfigure}
    \vspace{5mm} \\
    \begin{subfigure}[t]{56mm}
      \centering
      \caption{}\vspace{-8.5mm}\hspace{1.5mm}\label{fig:nanopores_plyab_pore_top}
      \includegraphics[scale=1]{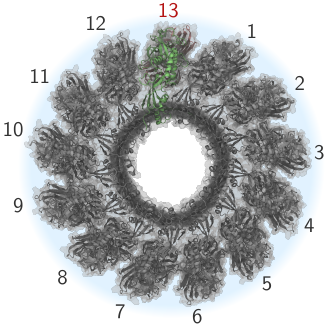}
    \end{subfigure}
  \end{minipage}
  \begin{minipage}[t]{60mm}
    \begin{subfigure}[t]{56mm}
      \centering
      \caption{}\vspace{-8.5mm}\hspace{1.5mm}\label{fig:nanopores_plyab_pore_section}
      \includegraphics[scale=1]{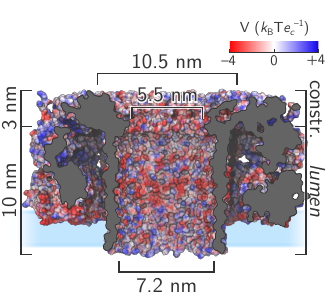}
    \end{subfigure}
    \vspace{5mm} \\ 
    \begin{subfigure}[t]{35mm}
      \caption{}\vspace{-8.5mm}\hspace{1.5mm}\label{fig:nanopores_plyab_monomer}
      \includegraphics[scale=1]{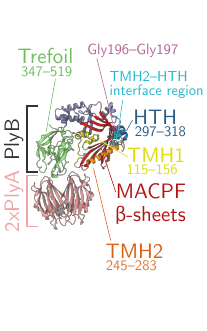}
    \end{subfigure}
    \begin{subfigure}[t]{24mm}
      \centering
      \caption{}\vspace{-8.5mm}\hspace{1.5mm}\label{fig:nanopores_plyab_protomer}
      \includegraphics[scale=1]{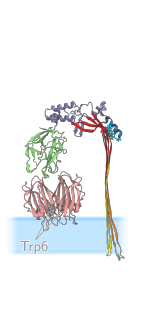}
    \end{subfigure}
  \end{minipage}
\caption[Structure of the pleurotolysin AB (PlyAB) porin]{%
  \textbf{Structure of the pleurotolysin AB (PlyAB) porin.}
  (\subref{fig:nanopores_plyab_pore_side})
  Side and
  (\subref{fig:nanopores_plyab_pore_top})
  top view of the \SI{1067}{\kDa} \gls{plyab} \tb-\gls{pft} porin
  (\pdbid{4V2T}~\cite{Lukoyanova-Kondos-2015}), formed by 13 identical subunits which are each composed of two
  \gls{plya} (pink) and one \gls{plyb} (green) protein chains. The full pore is \SI{13}{\nm} high with
  external diameters of \SI{22}{\nm} and \SI{9}{\nm} at the \cisi{} and \transi{} sides of the  membrane. 
  (\subref{fig:nanopores_plyab_pore_section})
  The cross-section of \gls{plyab}'s molecular surface, colored according to the local electrostatic potential
  at physiological salt concentration (\gls{apbs}~\cite{Baker-2001,Baker-2005}), shows that the interior of
  the pore is predominantly negative. The \lumen{} of the pore is divided by a \SI{5.5}{\nm} wide constriction
  into a conical \cisi{} chamber of \SI{3}{\nm} height and \SI{10.5}{\nm} diameter, and a cylindrical
  \transi{} chamber of \SI{10}{\nm} height and \SI{7.2}{\nm} diameter. 
  (\subref{fig:nanopores_plyab_monomer})
  Cartoon view of the water-soluble monomer configurations of \gls{plya}
  (\pdbid{4OEB}~\cite{Lukoyanova-Kondos-2015}) and \gls{plyb} (\pdbid{4OEJ}~\cite{Lukoyanova-Kondos-2015}),
  aligned according to their position in a single protomer of the full pore (using
  \gls{vmd}~\cite{Humphrey-1996}). Important regions and residues are highlighted in color.
  (\subref{fig:nanopores_plyab_protomer})
  A single \gls{plyab} protomer subunit (\pdbid{4V2T}~\cite{Lukoyanova-Kondos-2015}), showing the extension of
  the transmembrane \ta-helices into \tb-sheets.
  Images were rendered using \gls{vmd}~\cite{Humphrey-1996,Stone-1998}.
  }\label{fig:nanopores_plyab}
\end{figure*}

\Gls{plyab} (\cref{fig:nanopores_plyab}) is a \tb-\gls{pft} secreted by the fungus \textit{Pleurotus
ostreatus} and belongs to the family of \gls{macpf} proteins. Lukoyanova and Kondos \etal{} proposed a
plausible pore structure based on the \gls{cryo-em} image analysis and molecular modeling with a molecular
weight in excess of \SI{1000}{\kDa}~\cite{Lukoyanova-Kondos-2015}, making it one of the largest \glspl{pft}
with a known structure to date.

\subsubsection{Pore structure, shape, and charge distribution.}

The \gls{plyab} pore (\pdbid{4V2T}~\cite{Lukoyanova-Kondos-2015}) consists of 13 identical `subunits', each of
which is composed of a lipid-binding \gls{plya} dimer and a single pore-forming \gls{plyb} protein chain
(\cref{fig:nanopores_plyab_pore_side,fig:nanopores_plyab_pore_top}). Similar to \gls{ahl}, the entire pore
complex is mushroom-shaped, with a height of \SI{13}{\nm}, a head diameter of \SI{22}{\nm}, and a stem
diameter of \SI{9}{\nm}. When traversing the channel of \gls{plyab} (\cref{fig:nanopores_plyab_pore_section})
from the \cisi{}-side, the \SI{10.5}{\nm}-wide pore entry quickly begins to narrow down to a diameter of
\SI{5.5}{\nm} at a depth of \SI{3}{\nm}. Once past this constriction, the pore opens up into the pore
\lumen{}, a \SI{10}{\nm}-long, \SI{7.2}{\nm}-wide cylindrical chamber formed by the large \tb-barrel
\gls{tmd}. The interior walls of wild-type \gls{plyab} are predominantly negatively charged, particularly at
the constriction and in the middle of the \lumen{} (see~surface coloring in
\cref{fig:nanopores_plyab_pore_section}). Likely, this negatively charged interior enables \gls{plyab} to
employ charge-based facilitated diffusion to preferentially deliver small cationic proteins, such as
granzyme~B, over neutral and anionic ones~\cite{Stewart-2014,Reboul-2016}.

\subsubsection{Oligomerization stoichiometry.}

As with the other pores discussed above, \gls{plyab} can form pores with different stoichiometries.
\Gls{cryo-em} analysis revealed a diverse population of 14- (\SI{5}{\percent}), 13- (\SI{75}{\percent}), 12-
(\SI{15}{\percent}) and 11-mers (\SI{5}{\percent})~\cite{Lukoyanova-Kondos-2015}.

\subsubsection{Monomer \textit{versus} protomer.}

The water-soluble \gls{plya} crystal structure (\pdbid{4OEB}~\cite{Lukoyanova-Kondos-2015}) contains a
membrane-binding \tb-sandwich fold (\cref{fig:nanopores_plyab_monomer})---similar to that of the actinoporin
family (\eg~\gls{frac}, \cref{fig:nanopores_frac_monomer})---but lacks the typical N-terminal transmembrane
region. The water-soluble \gls{plyb} structure (\pdbid{4OEJ}~\cite{Lukoyanova-Kondos-2015}) can be subdivided
in a \tb-rich globular trefoil domain at the C-terminus, and a \gls{macpf} domain at the N-terminus. At its
core, the latter contains a four-stranded bent and twisted \tb-sheet (typical for the \gls{macpf}
superfamily), a flexible \gls{hth}-motif and two \ta-helical regions ({TMH1} and {TMH2}). In the protomer
structure (\pdbid{4VT2}~\cite{Lukoyanova-Kondos-2015}), the {TMH1} and {TMH2} regions become fully unwound
into \tb-hairpins to become part of the 52-sheet membrane-spanning \tb-barrel
(\cref{fig:nanopores_plyab_protomer}).

\subsubsection{Mechanism of pore formation.}

Pore formation is initiated by the binding of the water-soluble, potentially pre-dimerized, \gls{plya}
components to the target membrane. Each \gls{plya} dimer will then recruit a larger \gls{plyb} subunit to form
the membrane-bound \gls{plyab} monomer. High-speed imaging of perforin pore assembly has revealed that these
protomers loosely (but irreversibly) assemble into prepore oligomers of up to 8 subunits, which combine
further into larger prepores~\cite{Leung-2017}. When they are sufficiently large, these prepores will close
their ring structure, followed by a set of large conformational changes that ends with the formation of a
large \gls{tmd} \tb-barrel. The \gls{cryo-em} and mutagenesis study performed by
Lukoyanova~\etal{}~\cite{Lukoyanova-Kondos-2015} revealed that the \tb-barrel formation is triggered by the
disruption of the interaction between the \gls{hth}-motif and the {TMH2} helix during oligomerization. This
causes the opening of the central {MACPF} \tb-sheets, resulting in the gradual cooperative refolding of the
{TMH1} and {TMH2} helices towards the membrane and the top-down zippering of the
\tb-barrel~\cite{Reboul-2016}.

\clearpage
\section{A physical perspective on nanopore sensing}
\label{sec:np:physical_perspective}

The apparent simplicity of the principles behind nanopore sensing, together with its scalability and high
sensitivity towards even the smallest of changes within them, have made nanopores into one the most successful
and broadly applied single-molecule sensors. Nanopores have been used as detectors for small molecules, down
to individual ions, as well as biological polymers or particles, such as \gls{dna}, peptides, proteins, and
virions. Equally broad is the number of strategies that have been deployed to improve the selectivity and
sensitivity of nanopores: from the conjugation of binding sites, peptides, aptamers, or even entire proteins,
within the pore or to one of its entries, to the modification of the analyte molecules themselves. An in-depth
overview of the literature describing these strategies can be found in the Ph.D. dissertation of Veerle Van
Meervelt~\cite{VanMeervelt-2017-PhD}. Regardless of the precise details of the sensing approach, before any
analyte molecule can be detected, it must first be attracted towards the pore and subsequently captured for
a period sufficiently long to be sampled. In the following sections, we will describe the origin of
the forces at play in nanopores and clarify their importance during all stages of the sensing process. We will
start this section by discussing a phenomenon that mediates many of these interactions: the \gls{edl}.

\subsection{The electrical double layer within a nanopore}
\label{sec:np:edl}
\begin{figure*}[ptb]
  \centering

  \begin{subfigure}[t]{115mm}
    \centering
    \caption{}\vspace{-2.5mm}\label{fig:nanopores_edl_overview}
    \includegraphics[scale=1]{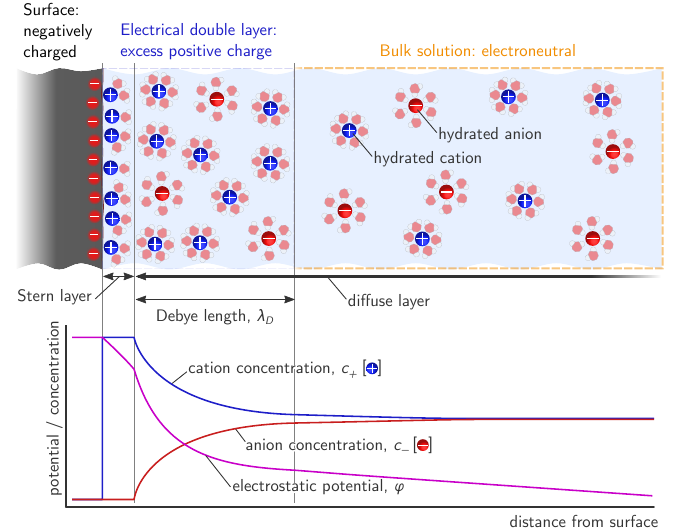}
  \end{subfigure}
  \\
  \begin{subfigure}[t]{115mm}
    \centering
    \caption{}\vspace{-2.5mm}\label{fig:nanopores_edl_confined}
    \includegraphics[scale=1]{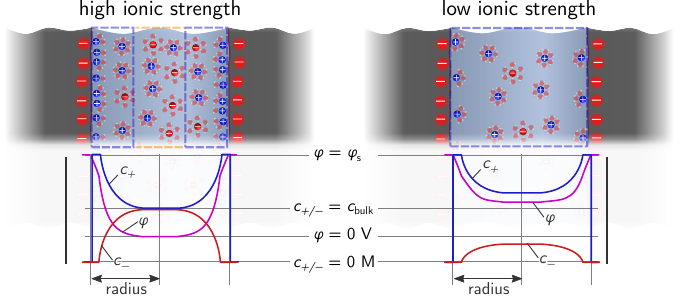}
  \end{subfigure}

\caption[Schematic overview of the structure of the electrical double layer]{%
  \textbf{Schematic overview of the structure of the electrical double layer (EDL).}
  (\subref{fig:nanopores_edl_overview})
  When a charged surface is brought into contact with an electrolyte solution, mobile ions of opposite charge
  (\ie~the counterions) will accumulate near the surface, whereas same-charge ions (\ie~the co-ions) will
  deplete. This region of excess charge in the fluid is called the \glsfirst{edl} and has a characteristic
  thickness equal to the Debye length ($\dbl$, see~\cref{eq:nanopores_debye_length}). Counterions in the Stern
  layer are semi-permanently bound to the surface and partially lose their hydration shell. Because the
  surface electrostatic potential drops exponentially inside the electrical double layer, its influence
  becomes negligible after a few Debye lengths, rapidly resulting in an electroneutral bulk ionic
  distribution.
  (\subref{fig:nanopores_edl_confined})
  Effect of geometrical confinement on the composition of the \gls{edl} within a nanopore at high (left) and
  low (right) ionic strengths. At high salt concentrations, $\dbl$ is short enough to allow the center of the
  pore to remain electroneutral. At low salt concentrations, on the other hand, $\dbl$ is the same or larger
  than the radius of the pore, the \glspl{edl} begin to overlap and the center of the pore also contains an
  excess of counterions. Regardless of the salt concentration, the \gls{edl} takes up a significant portion of
  the total pore volume and hence plays an important role in all nanopore transport phenomena.
  }\label{fig:nanopores_edl}
\end{figure*}

Virtually all intermolecular interactions are either directly (\eg~charge-charge, H-bonds) or indirectly
(\eg~\gls{vdw} forces) electrostatic in nature. Indeed, most surfaces, including those of nanopores and large
analyte molecules, contain ionizable groups that become charged when brought into contact with an aqueous
solution. Hence, it is instrumental to reflect on both their magnitude and the length scales at which they
exert their influence. At the level of interactions between two individual charges, a first characteristic
length scale for electrostatics is the Bjerrum length, $\bjl$, which corresponds to the distance at which the
electrostatic interaction energy between two charged species, $\energyelec (\bjl)$, becomes similar to their
thermal energy, $\boltzmann \temperature$. In other words, it represents the distance below which the
electrostatic interactions start to dominate over thermal effects. For electrolyte solutions containing two
ion species, it can be expressed as~\cite{Bocquet-2010}
\begin{equation}\label{eq:nanopores_bjerrum_length}
  \bjl = \dfrac{\ec^2}{4 \pi \absperm \relperm \boltzmann \temperature}
  \text{ ,}
\end{equation}
with $\ec$ the elementary charge, $\absperm$ the vacuum permittivity, $\relperm$ the relative permittivity of
the solution, $\boltzmann$ the Boltzmann constant, and $\temperature$ the temperature. For a typical
monovalent salt at room temperature $\bjl = \mSI{0.7}{\nm}$, a value that comes close to the radius of many
(biological) nanopores. Hence, it is evident that the transport of ions and other charged molecules through
nanopores will be influenced by electrostatic interactions. However, the typical electrolyte contains many
highly mobile ions in proximity to each other: for an electrolyte with a physiological salt
concentration (\eg~\SI{0.15}{\Molar}), a volume of \grid{10}{10}{10}{\cubic\nm} contains 180~ions, which
corresponds to a mean ion-ion spacing of only \SI{\approx1.8}{\nm}. Consequently, even though thermal effects
dominate at distances $>\bjl$ for individual ions, the extent to which local perturbations can propagate to
longer length scales will depend on the (local) density of all the ions in the solution. This phenomenon leads to
the formation of the \glsfirst{edl} (\cref{fig:nanopores_edl_overview}), which describes the rearrangement of
the ion density in response to a fixed surface charge density or potential. The \gls{edl} is a diffuse layer
that contains an excess of counter-ions (\ie~opposing the wall charge) and a depletion of co-ions (\ie~same as
the wall charge), whose width is inversely proportional to both $\bjl$ and the ionic strength of the
electrolyte, $\ionstr$. A characteristic value is given by the so-called Debye length, $\dbl$, which, for an
electrolyte with $N$ ion species, is given by~\cite{Bocquet-2010}
\begin{equation}\label{eq:nanopores_debye_length}
  \dbl = \left(
          \dfrac{ \permittivity \boltzmann \temperature }%
                { \ec^2 \avogadro \dsum_i^N \ci^0 \chargeni^2 } \right)^{1/2}
       \equiv \left( 8 \pi \bjl \avogadro \ionstr \right)^{-1/2}
  \text{ ,}
\end{equation}
with $\ci^0$ the molar concentration of ion $i$ (the $0$ superscript indicates that it is the `bulk'
concentration), $\chargeni$ its charge number (\ie~valence), and $\avogadro$ the Avogadro constant. Being the
characteristic length scale of the \gls{edl}, $\dbl$ corresponds to the thickness of the diffuse layer next to
the charged wall. Additionally, it represents the length scale over which the electrolyte screens the bulk
medium from the surface electrostatic potential. For salt concentrations of \SIlist{0.01;0.1;1}{\Molar},
$\dbl$ amounts to \SIlist{3.1;0.97;0.31}{\nm}, respectively. The local violation of electroneutrality, caused
by the imbalance between cations and anions within the diffuse layer, creates a macroscopic zone with a
permanent non-zero volume charge density onto which a tangential electric field can exert a Coulombic force.
Hence, the \gls{edl} will not only mediate electrostatic interactions between the nanopore and the analyte
molecules, but it will also result in a drag force on the liquid close to the nanopore walls that gives rise to a
net directional flux of water known as the \gls{eof}. Additionally, in a nanopore with a radius of
\SI{1}{\nm}, the \gls{edl} will begin to overlap with itself even at moderate (\eg~\SI{0.1}{\Molar}) ionic
strengths (\cref{fig:nanopores_edl_confined}). This leads to phenomena such as ion concentration
depletion/enrichment~\cite{Plecis-2005}, surface conductance~\cite{Stein-2004},
permselectivty~\cite{Plecis-2005} or pre-concentration~\cite{Pu-2004}---all of which dramatically impact the
transport properties of these pores.

We conclude this section with a brief discussion of the \glsfirst{pbe}, the most widely used approach to
describe the potential and ion distribution in the \gls{edl}. The \gls{pbe} relates the electrostatic
potential---expressed using Poisson's equation---to the electrochemical thermodynamic equilibrium of an ionic
solution---expressed through Boltzmann statistics---and reads~\cite{Gouy-1910,Chapman-1913,Baker-2005}
\begin{equation}\label{eq:nanopores_pbe}
  \nabla^2 \potential(\vec{r}) = - \dfrac{\scd(\vec{r})}{\absperm \relperm}
  = - \dfrac{1}{\absperm \relperm}
  \sum_i \left[ \avogadro \ci^0 \chargen_i \ec
         \exp \left( \dfrac{-\chargen_i \ec \potential(\vec{r}) }{\kbt} \right)
        \right]
  \text{ ,}
\end{equation}
where $\potential(\vec{r})$ is the electrostatic potential at position $\vec{r}$ and $\scd(\vec{r})$ is the
charge density. Linearizing the exponential term in \cref{eq:nanopores_pbe} with a truncated Taylor expansion
(valid for $\left| \chargeni \ec \potential \right| (\kbt)^{-1} < 1$) yields the linearized \gls{pbe}
\begin{equation}\label{eq:nanopores_pbe_linear}
  \nabla^2 \potential(\vec{r})
  = \dfrac{\ec^2}{\absperm \relperm \kbt} \sum_i \left[ \avogadro \ci^0 \chargeni \right]
      \potential(\vec{r})
  = \invdbl^2 \potential(\vec{r})
  \text{ ,}
\end{equation}
where $\invdbl = \dbl^{-1}$ provides a clear connection to the Debye length described above. Because it is
derived using mean-field theory, the \gls{pbe} represents ions as densities of point charges in a
structureless medium and ignores any ion-ion correlations~\cite{Bocquet-2010}. Particular care must be taken
for systems with very high surface potentials (more than a few \si{\kTe}), or for phenomena where specific
ion-ion or water-water interactions must be considered~\cite{Collins-2012}. Nevertheless, \gls{pb} theory
captures many of the essential physical properties of electrical double layers. While meaningful
analytical solutions to this partial differential equation require considerable simplifications and
linearization, the \gls{pbe} can be solved with relative ease for complex (bio)molecular
systems using numerical methods~\cite{Baker-2001,Baker-2005}.

\subsection{On the timescale of ion and analyte dynamics}
\label{sec:np:dynamics}

The measured nanopore signal results from the complex interplay between a few very large molecules (\ie~the
pore and the analyte) and a great many small particles (\ie~water and ions). Hence, before delving further
into the physics of nanopores, it is instructive to reflect on the timescales over which the behavior of ions
and analyte molecules within the pore can be safely averaged. Because the diffusivity of a molecule is roughly
proportional to its size, ions will move \numrange{10}{100}-fold faster relative to the typical analyte
molecule (\ie~\gls{dna} or proteins). This suggests that ions will be able to respond quasi-instantly to a
change in the pore--analyte configuration, such as a different position, orientation, or applied bias voltage.
More concretely, the characteristic timescale at which the \gls{edl} can adapt to change is given by
$\chargedensrelaxtime$, the surface charge density relaxation time~\cite{Bazant-2004}
\begin{equation}\label{eq:nanopores_relaxation_time}
  \chargedensrelaxtime \sim \dfrac{\dbl L}{\diffusion} 
                            - \dfrac{\dbl^2}{\diffusion}
                            - \dfrac{\dbl \stl}{\diffusion}
  \text{ ,}
\end{equation}
with $\diffusion$ the average ion diffusion coefficient, $L$ the characteristic system size, and $\stl$ the
Stern layer thickness. The value of $\chargedensrelaxtime$ for a nanopore lies in the order of
\SIrange{0.1}{10}{\ns}. For example, in a system with $L = \mSI{10}{\nm}$ (the length of the typical
\gls{bnp}), $\diffusion = \mSI{2}{\square\nm\per\ns}$ (\eg~\ce{K+} and \ce{Cl-} ions), $\dbl = \mSI{1}{\nm}$
(\ie~physiological ionic strength), and $\stl = \mSI{0.1}{\nm}$ (the average ionic radius),
$\chargedensrelaxtime = \mSI{4.5}{\ns}$. In comparison, the dwell time of a large analyte molecule at any
given position within a biological nanopore ranges from \SI{1}{\us}~to~\SI{1}{\second}. For example, at
\SI{+100}{\mV}, \gls{dsdna} translocates through \gls{clya} at a rate of \SI{\approx1}{\bp} every
\SI{7000}{\ns} (\SI{290}{\bp} in \SI{2}{\ms})~\cite{Franceschini-2013}. Hence, given the three
orders-of-magnitude separation in timescale of these two phenomena, it is reasonable to assume that the ions
respond instantaneously to any change in the nanopore--analyte system. In other words, the ionic
atmosphere can always be considered to be at equilibrium with respect to the pore and the analyte. As we shall
see in the results chapters, this assumption will allow us to calculate meaningful energies and forces with
`steady-state' solutions, rather than necessitating the use of complex time-dependent analysis.

\subsection{The electric field: the potentials they are a-changin'}
\label{sec:np:potential}

\begin{figure*}[b]
  \centering

  \begin{subfigure}[t]{15mm}
    \centering
    \caption{}\vspace{-2.5mm}\label{fig:nanopores_efield_circuit}
    \includegraphics[scale=1]{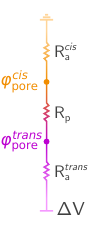}
  \end{subfigure}
  \hspace{-3mm}
  \begin{subfigure}[t]{25mm}
    \centering
    \caption{}\vspace{-2.5mm}\label{fig:nanopores_efield_cartoon}
    \includegraphics[scale=1]{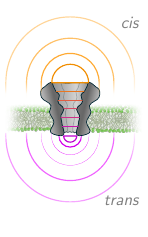}
  \end{subfigure}
  \hspace{-2.5mm}
  \begin{subfigure}[t]{35mm}
    \centering
    \caption{}\vspace{-2.5mm}\label{fig:nanopores_efield_potential}
    \includegraphics[scale=1]{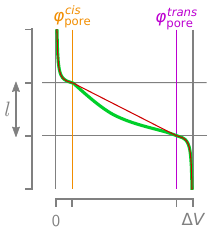}
  \end{subfigure}
  \hspace{-0.5mm}
  \begin{subfigure}[t]{35mm}
    \centering
    \caption{}\vspace{-2.5mm}\label{fig:nanopores_efield_efield}
    \includegraphics[scale=1]{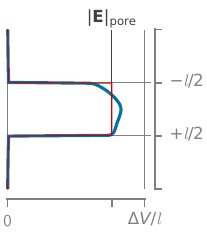}
  \end{subfigure}

\caption[Equiv. circuit model for the elec. potential and field in a nanopore]{%
  \textbf{Equivalent circuit model for the electric potential and field in a nanopore.}
  (\subref{fig:nanopores_efield_circuit})
  The potential distribution within a nanopore system can be approximated by pore resistance ($\Rpore$),
  flanked by two access resistances ($\Raccess = \Raccess^{\textit{cis}} =
  \Raccess^{\textit{trans}}$)~\cite{Kowalczyk-2011}. Because of the access resistance, the potential
  difference between the \cisi{} and \transi{} electrodes is not exactly, $\potdiff$, but is given more
  accurately by $\potcis - \pottrans$.
  (\subref{fig:nanopores_efield_cartoon})
  Cartoon model showing the equipotential planes outside and within the pore.
  (\subref{fig:nanopores_efield_potential})
  Potential and 
  (\subref{fig:nanopores_efield_efield})
  electric field profiles
  along the central axis of the pore for a perfectly cylindrical pore (thin red lines,
  \cref{eq:nanopores_potential_profile,eq:nanopores_efield_profile}), or modified to approximately reflect the
  influence of the geometry profile of the cartoon pore (thick green and blue lines). Due to the access
  resistance, a portion of $\potdiff$ decays outside of the pore near its entries, which helps with the
  capture of analytes. However, the bulk of the potential change occurs within the pore itself, which results
  in a drastic increase in the electric field strength. Note that most biological nanopores are not perfect
  cylinders and hence will have non-uniform electric fields along their length.
  Figure adapted with permission from Ref.~\cite{Chinappi-2015}. Copyrighted by the American Physical Society.
  }\label{fig:nanopores_efield}
\end{figure*}

Due to their high ionic resistivity, the majority of the potential difference ($\potdiff$) applied between the
\cisi{} and \transi{} electrolyte compartments of a nanopore system will change within, and around, the pore.
The overall profile of the electric potential---and by extension the electric field---can be approximated with
the help of an equivalent circuit model~\cite{Wanunu-2009,Grosberg-2010,Kowalczyk-2011}, where the pore is
represented by a resistor ($\Rpore$) and is surrounded by an access resistance ($\Raccess$) on either side
(\cref{fig:nanopores_efield_circuit,fig:nanopores_efield_cartoon}). $\Raccess$ represents the transition from
the bulk electrolyte to the confinement of the pore. Note that even though such a model does not consider the
electrostatics due to the pore's fixed charge distribution, it has proven to be instrumental in the
development of analytically tractable models for describing the capture dynamics of \gls{dna} in both
solid-state~\cite{Wanunu-2009,Grosberg-2010,Muthukumar-2010} and biological
nanopores~\cite{Chinappi-2015,Nomidis-2018}, or for estimating the conductance of nanopores with complex
shapes~\cite{Wanunu-2009,Kowalczyk-2011}.

Assuming that the pore is a perfect cylinder with length $\length$ and diameter $\diameter$ ($\Rpore = 4
\length / (\pi \sigma \diameter^2)$~\cite{Grosberg-2010}), and the pore entries act as planar disk electrodes
with diameter $\diameter$ ($\Raccess = 1 / (2\sigma\diameter)$~\cite{Hall-1975}), the ionic current $\current$
flowing through the pore is given by Ohm's law~\cite{Kowalczyk-2011}
\begin{equation}\label{eq:nanopores_ohms_law}
  \current = \dfrac{\potdiff}{\Rpore + 2 \Raccess} 
           = \potdiff \sigma \left( \dfrac{4 \length}{\pi \diameter^2} + \dfrac{1}{\diameter} \right)^{-1}
           \equiv \potdiff \sigma 2 \pi \dchar
  \text{ ,}
\end{equation}
with $\potdiff$ the potential difference between the \cisi{} and \transi{} electrodes and $\sigma$ the
conductivity of the electrolyte. Note that while constant in the bulk solution, $\sigma$ can be, and usually
is, location-dependent within the confined environment of a nanopore~\cite{Chinappi-2015}. As defined in
Nomidis~\etal{}~\cite{Nomidis-2018}, $\dchar$ represents the characteristic length of the system
\begin{equation}\label{eq:nanopores_characteristic_length}
  \dchar = \dfrac{1}{2\pi} \left[\dfrac{4 \length}{\pi\diameter^2} + \dfrac{1}{\diameter} \right]^{-1}
  \text{ .}
\end{equation}
Because $\current$ is constant, it can also be computed using the differential form of Ohm's
law~\cite{Chinappi-2015}, allowing us to formulate the spatial derivative of the potential
\begin{equation}\label{eq:nanopores_current_cross_section}
  \dfrac{d \potential(z)}{dz} = \dfrac{\current}{\sigma S(z)}
  \text{ ,}
\end{equation}
where $z$ is either the radial distance from the pore entry or the location along the length of the pore,
$S(z)$ the cross-sectional area through which the current is measured, and $\potential(z)$ the electrostatic
potential. In most nanopores, the diameter of a pore can vary significantly along its length, but for a
perfectly cylindrical pore $S(z) = \pi \diameter^2 / 4$, which does not depend on $z$. Assuming spherical
symmetry, outside of the pore $S(z) = \pi z^2$. Applying Kirchhoff's circuit laws to
\cref{eq:nanopores_ohms_law,eq:nanopores_current_cross_section} yields the potential drops due the access
resistance of the \cisi{} and \transi{} entries of the pore, given by
\begin{align}\label{eq:nanopores_potcis}
  \potcis   ={}& \potdiff \dfrac{\dchar}{\diameter}
  \text{ ,}
\end{align}
and
\begin{align}\label{eq:nanopores_pottrans}
  \pottrans ={}& \potdiff \dfrac{\dchar}{\diameter} \left[ \pi + \dfrac{8 \length}{\diameter} \right]
  \text{ ,}
\end{align}
respectively. Finally, an axial profile of the potential ($\potential (z)$,
\cref{fig:nanopores_efield_potential}) can be obtained by (somewhat arbitrarily) positioning the disk
electrode at their respective \cisi{} ($z = -\length / 2$) and \transi{} pore entries ($z = \length /
2$)~\cite{Chinappi-2015}, and integrating \cref{eq:nanopores_current_cross_section} with the boundary
conditions $\potential(-\infty) = 0$ and $\potential(+\infty) = \potdiff$,
\begin{equation}\label{eq:nanopores_potential_profile}
  \dfrac{\potential (z)}{\potdiff} =
  \begin{dcases}
    - \dfrac{\dchar}{2z + \length - \diameter / \pi}
    & \quad z < -\dfrac{\length}{2} \text{ ,} \\
    \dfrac{z}{\length} \left[ \dfrac{\pi \diameter}{4 \length} + 1 \right]^{-1} + \dfrac{1}{2}
    & \quad -\dfrac{\length}{2} \le z \le \dfrac{\length}{2} \text{ ,}\\
    1 - \dfrac{\dchar}{2z - \length + \diameter / \pi}
    & \quad z > \dfrac{\length}{2} \text{ .}
  \end{dcases}
\end{equation}
Because the potential in the reservoir is inversely proportional to the distance from the pore entry, it
decays rapidly (\ie~over a few \si{\nm}) to its bulk reservoir value. Within the pore, $\potential (z)$ is a
simple linear function of $z$, at least for a perfectly cylindrical nanopore. However, most actual nanopores
are far from perfect cylinders, and their diameters can easily vary by a factor of two or more within a few
nanometers along the channel's length, leading to deviations from linearity. Additionally, the fixed charge
distributions lining the nanopore walls can also significantly influence the overall electrostatic profile,
which will be the main subject matter of \cref{ch:electrostatics}. The electric field $\efield (z)$
(\cref{fig:nanopores_efield_efield}) is then simply given by the gradient of $\potential (z)$,
\begin{equation}\label{eq:nanopores_efield_profile}
  \efield (z) = \nabla \potential (z) =
  \begin{dcases}
    \potdiff \dfrac{2 \dchar}{ \left( 2z - \length + \diameter / \pi \right)^2}
    & \quad z < -\dfrac{\length}{2} \text{ ,} \\
    \dfrac{\potdiff}{\length} \left[ \dfrac{\pi \diameter}{4 \length} + 1 \right]^{-1}
    = \left| \efield \right|_{\text{pore}}
    & \quad -\dfrac{\length}{2} \le z \le \dfrac{\length}{2} \text{ ,}\\
    \potdiff \dfrac{2 \dchar}{ \left( 2z + \length - \diameter / \pi \right)^2 }
    & \quad z > \dfrac{\length}{2} \text{ .} \\
  \end{dcases}
\end{equation}
Even though in this system, $\efield (z)$ is constant within the pore ($\left| \efield
\right|_{\text{pore}}$), even relatively small changes in diameter along the nanopore's length will result in
large differences in the local electric field strength. This is because $\efield (z)$ depends on the local
cross-sectional area (\ie~quadratic with respect to diameter, see~\cref{eq:nanopores_current_cross_section}),
with narrower sections resulting in higher field strengths.

\subsection{The electro-osmotic flow: ions pushing water}
\label{sec:np:eof}

\begin{figure*}[b]
  \centering

  \begin{subfigure}[t]{35mm}
    \centering
    \caption{}\vspace{0mm}\label{fig:nanopores_eof_principle}
    \includegraphics[scale=1]{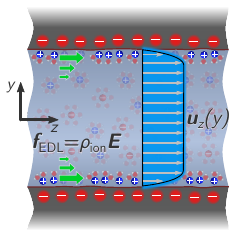}
  \end{subfigure}
  \hspace{10mm}
  \begin{subfigure}[t]{25mm}
    \centering
    \caption{}\vspace{-2.5mm}\label{fig:nanopores_eof_cartoon}
    \includegraphics[scale=1]{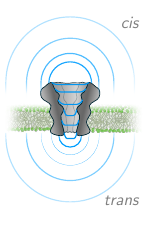}
  \end{subfigure}
  \hspace{-0.5mm}
  \begin{subfigure}[t]{40mm}
    \centering
    \caption{}\vspace{-2.5mm}\label{fig:nanopores_eof_zprofile}
    \includegraphics[scale=1]{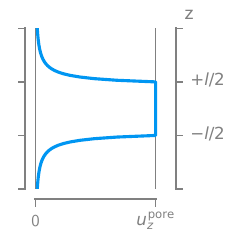}
  \end{subfigure}

\caption[Electro-osmotic flow in a nanopore]{%
  \textbf{Electro-osmotic flow in a nanopore.}
  (\subref{fig:nanopores_eof_principle})
  When an external electric field ($\efield_z$) is applied along a charged nanochannel, a Coulombic force
  density ($\forcedensedl$) is exerted on the excess charge density ($\scdion$) present in the \gls{edl} near
  the wall. The resulting unidirectional movement of these excess counterions drags on the surrounding water
  molecules, giving rise to an \glsfirst{eof} with a net water velocity in the z-direction ($\velocity_z(y)$).
  Note that while in hydrophilic nanopores, $\velocity_z(y)\approx0$ at the wall, a non-zero velocity will be
  observed in hydrophobic nanopores~\cite{Bocquet-2010,Manghi-2018}.
  (\subref{fig:nanopores_eof_cartoon})
  Cartoon model showing the flow velocity profiles outside, and within, a nanopore.
  (\subref{fig:nanopores_eof_zprofile})
  Approximate profile of the average \gls{eof} velocity, $\velocity (z)$, along the central axis of the pore
  for a perfectly cylindrical pore (\cref{eq:nanopores_eof_velocity}). Similar to the electric field, the
  decay of $\velocity (z)$ is inversely proportional to the square of the distance from the pore.
  }\label{fig:nanopores_eof}
\end{figure*}

As touched upon in~\cref{sec:np:edl}, when a tangential electric field, $\efield$, is applied across the
\gls{edl}, the non-zero charge density, $\scdion$, within the fluid will be subject to a Coulombic force
density (\cref{fig:nanopores_eof_principle})
\begin{equation}\label{eq:nanopores_edl_force}
  \forcedensedl = \scdion \efield
  \text{ ,}
\end{equation}
where
\begin{equation}\label{eq:nanopores_ion_scd}
  \scdion = \faraday \dsum_i^N \chargeni \ci
  \text{ ,}
\end{equation}
with $\faraday = \ec \avogadro$ the Faraday constant, and $\ci$ the \emph{local} concentration of ion $i$.
Even though, at an atomic scale, $\forcedensedl$ is exerted on the ions rather than the fluid as a whole, a
portion of the force---which is typically assumed to be all of it---is transferred to the water molecules through
friction. In turn, the unidirectional movement of the \gls{edl} will drag along the rest of the fluid, charged
or not, giving rise to a unidirectional fluid flow---the \glsfirst{eof}~\cite{Bocquet-2010}.

As with the electrical field, tractable analytical formulas that accurately describe the \gls{eof} generated
by (biological) nanopores are not possible due to their complex geometries and charge distributions.
Nevertheless, considerable insights can be gained by simplifying the pore geometry and its surface charge
density. For low surface charge potentials (\ie~$\left| \ec \potential\right| / (\kbt) \le 1$), the
volumetric flowrate ($\flowrate$) of water molecules through an infinitely long, cylindrical nanopore with
radius $\radiusa$ is given by~\cite{Laohakunakorn-2015}
\begin{equation}\label{eq:nanopores_eof_rate}
  \flowrate = - \dfrac{\pi \radiusa^2}{\viscosity \invdbl} 
  \dfrac{ \bessel{2} ( \invdbl \radiusa ) }{ \bessel{1} (\invdbl \radiusa) }
  \electricfield_z \surfcd
  \equiv - \dfrac{\pi \radiusa^2}{\viscosity \invdbl} H(\invdbl \radiusa) \electricfield_z \surfcd
  \text{ ,}
\end{equation}
where $\viscosity$ the fluid viscosity, $\invdbl = (\dbl)^{-1}$ the inverse Debye length
(see~\cref{eq:nanopores_debye_length}), $\electricfield_z$ the electric field along the pore axis, $\surfcd$
the surface charge density of the pore, and $\bessel{n}$ the $n$th-order modified Bessel function of the first
kind. Note that the minus sign indicates that the flow direction will be against the electrical field for
positive surface charge densities (\ie~a negative \gls{edl}) and with it for negative surface charge densities
(\ie~a positive \gls{edl}). $H(\invdbl \radiusa)$ can be considered as an ionic strength `correction factor'
that accounts for the overlapping of the \gls{edl} within the pore. Hence, for conditions where $\invdbl
\radiusa \gg 1$ (\eg~at high ionic strengths or for pores with a large radius), $H(\invdbl \radiusa) \approx
1$ and \cref{eq:nanopores_eof_rate} simplifies to
\begin{equation}\label{eq:nanopores_eof_rate_simple}
  \flowrate = - \dfrac{\pi \radiusa^2}{\viscosity \invdbl} \electricfield_z \surfcd
  \text{ .}
\end{equation}
As with the ionic current (\cref{eq:nanopores_current_cross_section}), $\flowrate$ is a linear function of the
electric field magnitude, but it is inversely proportional to the square root of the ionic strength
($\flowrate \propto \kappa^{-1} \propto \ionstr^{-1/2}$). This means that $\flowrate$ rapidly approaches
infinity at very low the ion concentrations  ($\flowrate \to \infty$ for $\ionstr \to 0$), a non-physical
result that arises from the infinite cylinder assumption~\cite{Mao-2014}. For realistic pores of finite
length, the maximum flow rate is restricted by `end effects' (akin to the access resistance for ion flow), and
analytical models that address this are available~\cite{Sherwood-2014}.

For an incompressible fluid such as water, the conservation of mass dictates that the number of water
molecules passing through any given cross-section must be equal. Thus, at location $z$ along the central axis
of the pore, the average fluid velocity $\velocity(z)$ within a cross-section $S(z)$ for a perfectly
cylindrical pore with radius $\radiusa(z)$ (\cref{fig:nanopores_eof_zprofile}) can be estimated using
\begin{equation}\label{eq:nanopores_eof_velocity}
  \velocity (z) = \dfrac{\flowrate}{S(z)} =
  \begin{dcases}
    \dfrac{\flowrate}{2 \pi \left(z - b \right)^2}
    & \quad z < -\dfrac{\length}{2} \text{ ,} \\
    \dfrac{\flowrate}{\pi \radiusa(z)^2}
    & \quad -\dfrac{\length}{2} \le z \le \dfrac{\length}{2} \text{ ,}\\
    \dfrac{\flowrate}{2 \pi \left(z + b \right)^2}
    & \quad z > \dfrac{\length}{2} \text{ ,}
  \end{dcases}
\end{equation}
where the offset $b = 2^{-1/2} \radiusa + \length / 2$ was included to match the velocities at the pore
entries. Similar to the electric field, $\velocity$ scales with the cross-sectional area and axial changes in
pore diameter quickly result in large changes in flow velocity.

It is worth mentioning that the water flow will also depend on the degree of interaction between the water
molecules and the atoms in the nanopore walls, a phenomenon that is often expressed in fluid dynamics by the
`slip length' $\sliplength$~\cite{Bocquet-2010}. Walls with a high degree of friction---such as the rough,
hydrophilic surfaces of proteins~\cite{Zhang-2014,Wong-Ekkabut-2016b,Pronk-2014}---interact strongly with the
water molecules, and yield a fluid velocity close to zero at the liquid-solid interface ($\sliplength = 0$,
the `no-slip' boundary condition). In contrast, the water flow through frictionless pores---such as the
smooth, hydrophobic surfaces within carbon nanotubes~\cite{Ye-2011,Manghi-2018,Bocquet-2020}---is not slowed
down to zero at the liquid-solid interface ($\sliplength \to \infty$, the `slip' boundary condition). In most
nanopores, the appropriate value for $\sliplength$ will lie somewhere in between these extremes, although for
the hydrophilic channels of \glspl{bnp}, $\sliplength$ can be expected to be closer to $0$ than to
$\infty$~\cite{Bocquet-2010,Manghi-2018}.

\subsection{External forces: (di)electrophoresis and electro-osmosis}
\begin{figure*}[b]
  \centering

  \begin{subfigure}[t]{60mm}
    \centering
    \caption{}\vspace{-2.5mm}\label{fig:nanopores_forces_ext_ep_eo}
    \includegraphics[scale=1]{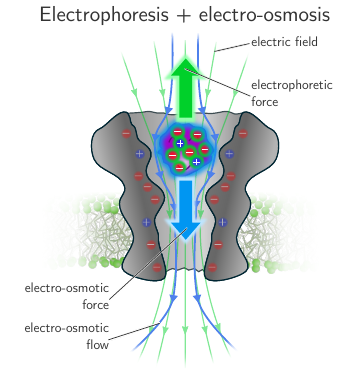}
  \end{subfigure}
  \hspace{-5mm}
  \begin{subfigure}[t]{60mm}
    \centering
    \caption{}\vspace{-2.5mm}\label{fig:nanopores_forces_ext_dep}
    \includegraphics[scale=1]{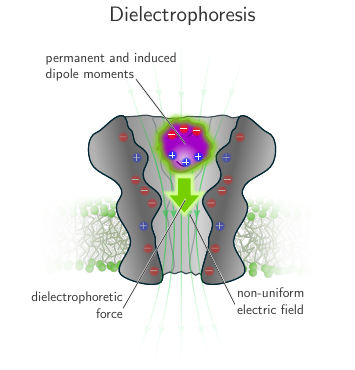}
  \end{subfigure}

\caption[Forces acting on a particle due to the external electric field]{%
  \textbf{Forces acting on a particle due to the external electric field.}
  (\subref{fig:nanopores_forces_ext_ep_eo})
  Because the electric field is strongest within the pore, both the electrophoretic and electro-osmotic forces
  are at their strongest during particle translocation. Note that, depending on the charge of the particle and
  the direction of the \gls{eof} (\ie~the charge of the pore walls), these forces can either reinforce or
  oppose one another.
  (\subref{fig:nanopores_forces_ext_dep})
  At locations where the electric field is not uniform (\ie~$\nabla \efield \neq 0$), such as near the pore
  entries or near a constriction within the pore, dielectric particles with a permanent or induced dipole
  moment will be subjected to a dielectrophoretic force, which pushes them towards increasing electric field
  magnitudes.
  }\label{fig:nanopores_forces_ext}
\end{figure*}

The externally applied bias voltage creates a strong electric field around and within the pore. This field
induces several distinct forces that influence the movement of analyte molecules towards and within the pore
(\cref{fig:nanopores_forces_ext}): the \glsdisp{ep}{electrophoretic~(EP)}, \glsdisp{eo}{electro-osmotic~(EO)}
and \glsdisp{dep}{dielectrophoretic~(DEP)} forces. The former two are the strongest, and hence also the most
well-known. For proteins and \gls{dna} molecules, typical magnitudes of these forces within the pore fall in
the range of \SIrange{1}{100}{\pN}~\cite{Keyser-2006,vanDorp-2009}. Besides attracting analytes and inducing
their translocation, external forces may also impact the structure of the molecules themselves. Electric
field-induced forces within solid-state nanopores have been shown to stretch~\cite{Freedman-2013} or even
unfold~\cite{Freedman-2011} proteins, allowing one to probe their mechanical properties during
translocation~\cite{Waduge-2017,Zhou-2020b}. Whereas the \gls{ep} force is dominant for molecules with high
charge densities (such as \gls{dna})~\cite{vanDorp-2009}, the force balance for other biomolecules
(\eg~proteins) is often more evenly matched, or even in favor of the \gls{eo}
force~\cite{Soskine-2013,Zhang-2020}. The \gls{dep} force is typically much weaker, given that it acts on the
(induced) dipole moment of a particle and is active only in places with a strong electric field gradient
(\ie~around the pore entries or near a constriction within the pore). Hence, while dielectrophoresis can often
be ignored without repercussions, in specific cases it plays an important role in the capture and
translocation processes of analyte molecules~\cite{Freedman-2016,Asandei-2016,Chinappi-2020}, and hence its
existence should be acknowledged.

\subsubsection{Electrophoretic and dielectrophoretic forces: sailing the field lines.}

For a charged particle in an electric field, the \gls{ep} force, $\forceep$ is given by the electric component
of the Lorentz force (\ie~Coulomb's law)~\cite{Lu-2012}
\begin{equation}\label{eq:nanopores_lorentz_force}
  \forceep = \oint_{V} \scd \efield dV = \chargeq \efield
  \text{ ,}
\end{equation}
with $\scd$ the charge density, $V$ the volume, and $\chargeq$ the net charge of the particle. In principle,
the integral form of \cref{eq:nanopores_lorentz_force} allows one the evaluate $\forceep$ for any dielectric
particle within arbitrary electric field distributions. The integrated form, on the other hand, is technically
only valid for a charged particle in a homogeneous electric field (\ie~far away from the pore). But, as we
shall see in \cref{ch:trapping}, it can be a reasonable approximation even for more complex systems
(\eg~within the pore). The \gls{dep} force, $\forcedep$, is proportional to the spatial change of the electric
field and the dipole moment of the particle~\cite{Hoelzel-2020}
\begin{equation}\label{eq:nanopores_dep_force}
  \forcedep = \left( \dipolemoment \cdot \nabla \right) \efield
  \text{ ,}
\end{equation}
with $\dipolemoment$ the dipole moment of the particle. Note that the $\dipolemoment$ contains both a
permanent contribution, due to the non-uniform charge distribution within the
particle~\cite{Hoelzel-2020,VanMeervelt-2017}, and an induced component, resulting from the particle's
effective polarizability~\cite{Minerick-2015}. The latter depends on the complex permittivity of both the particle
and the surrounding electrolyte (\ie~the Clausius--Mossotti relation), and it is influenced strongly by the
frequency of the electric field and not just the strength of its gradient. A more in-depth discussion on the state
of theory and experiment of \gls{dep} for small biomolecules such as proteins can be found in
Ref.~\cite{Hoelzel-2020}.

Both \gls{ep} and \gls{dep} contributions can be accounted for in a single framework if the force is
calculated \textit{via} integration of the electrostatic Maxwell stress tensor, $\maxwellstresstensorvec$,
over the surface ($\Gamma$) of the particle~\cite{Ai-2011}
\begin{equation}\label{eq:nanopores_electric_force}
  \forceele = \oint_{\Gamma} \left( \maxwellstresstensorvec \cdot \normvec \right) d\Gamma
  \text{ ,}
\end{equation}
where
\begin{equation}\label{eq:nanopores_maxwell_stresstensor}
  \maxwellstresstensorvec = \permittivity \efield \efield
                            - \dfrac{1}{2} \permittivity \left( \efield \cdot \efield \right) \identity
  \text{ ,}
\end{equation}
with $\permittivity$ the permittivity, $\efield$ the electric field vector, $\normvec$ the normal vector of
the surface, and $\identity$ the identity tensor vector. Note that the full expression of
$\maxwellstresstensorvec$ should include several magnetic field terms, but since magnetism does not (yet?)
contribute significantly to the forces landscape within nanopores, these are set to zero. Because
\cref{eq:nanopores_electric_force} makes no assumptions regarding the shape of the electric field or particle,
its integral nature allows it to be evaluated (numerically) for arbitrary particle shapes, charge densities,
and electric field distributions, if those are available (\eg~through simulations).

\subsubsection{Electro-osmotic force: such a drag.}

In the first order, the \gls{eo} force acting on a particle can be calculated using the well-known Stokes' law
\begin{equation}\label{eq:nanopores_drag_force}
  \forcedrag = - 6 \pi \viscosity \particleradius \velocity
  \text{ ,}
\end{equation}
where $\forcedrag$ is the drag force on the particle with radius $\particleradius$, $\viscosity$ the
electrolyte viscosity, and $\velocity$ the velocity vector of the fluid. Note that
\cref{eq:nanopores_drag_force} assumes that (1) the flow is laminar (which is typically the case for nanopore
systems\footnotemark), %
\footnotetext{%
In a laminar flow, the fluid moves in smooth `layers' next to each other, with little to no intermixing
besides diffusion. It can be recognized by its low Reynolds number
\begin{equation}
\reynolds = \dfrac{\density \vel \diameter}{\viscosity} \text{ ,}\notag
\end{equation}
with $\diameter$ the nanopore diameter, and $\density$, $\vel$, and $\viscosity$ the fluid density, velocity,
and viscosity, respectively. For example, for a nanopore with $\diameter = \mSI{5}{\nm}$, and a moderate
flow velocity of $\vel = \mSI{0.1}{\meter\per\second}$, $\reynolds\approx\num{5e-4} \ll 1$~\cite{Schoch-2008}.
}%
(2) the particle is spherical with a perfectly smooth surface, (3) the surrounding environment (\ie~density
and viscosity) is homogeneous, and (4) the particle does not perturb the flow itself. Evidently, given that
most biomolecules are not smooth spheres, this is often compensated for by using a so-called Stokes radius (or
hydrodynamic radius), rather than the true particle radius~\cite{Ortega-2011}. This makes $\forcedrag$ useful as a
good first-order approximation for computing the electro-osmotic force, particularly outside of the pore.
Within the pore, however, its validity is less clear given that the size of the particle is often similar to
the pore diameter. Hence, its presence is expected to influence the flow itself. Moreover, as we shall discuss
in \cref{ch:epnpns}, the electrolyte properties within the pore are not uniform, particularly in proximity to
the nanopore~\cite{Qiao-Aluru-2003,Vo-2016,Hsu-2017,Ye-2011} and particle~\cite{Pronk-2014,Makarov-1998}
surfaces, leading to further deviations.

As with the (di)electrophoretic force, a more general framework involves the integration of a stress
tensor---in this case, the hydrodynamic stress tensor $\hydrostresstensorvec$---across the entire particle
surface ($\Gamma$)~\cite{Ghosal-2019}
\begin{equation}\label{eq:nanopores_hydrodynamic_force}
  \forcehyd = \oint_{\Gamma} \left( \hydrostresstensorvec \cdot \normvec \right) d\Gamma
  \text{ ,}
\end{equation}
where
\begin{align}\label{eq:nanopores_hydrodynamic_stresstensor}
  \hydrostresstensorvec =
  \pressure\identity - \viscosity\left[\nabla\velocity + \left(\nabla\velocity \right)^\mathsf{T}\right]
  \text{ ,}
\end{align}
with $\pressure$ the pressure, $\viscosity$ the viscosity, $\velocity$ the velocity vector, and $\normvec$ the
surface normal vector. The hydrodynamic force, $\forcehyd$, does not only include the viscous force (\ie~the
interaction between the particle and the moving fluid molecules), it also takes into account any pressure
differences that may arise around the particle prior to, or during
translocation~\cite{Hoogerheide-2014,Wilson-2018}. Hence, if a full, self-consistent solution of the flow and
pressure fields around a particle is available, the numerical integration of
\cref{eq:nanopores_hydrodynamic_force} will yield the most accurate results~\cite{Galla-2014}.

\subsection{Intrinsic forces: electrostatics, steric hindrance, and entropy}

\begin{figure*}[b]
  \centering

  \begin{subfigure}[t]{60mm}
    \centering
    \caption{}\vspace{-2.5mm}\label{fig:nanopores_forces_int_es}
    \includegraphics[scale=1]{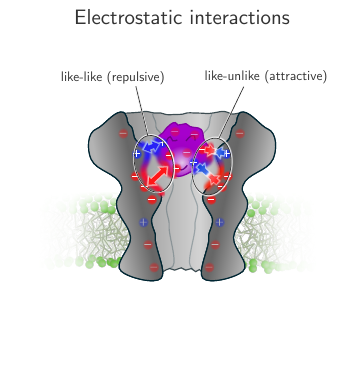}
  \end{subfigure}
  \hspace{-5mm}
  \begin{subfigure}[t]{60mm}
    \centering
    \caption{}\vspace{-2.5mm}\label{fig:nanopores_forces_int_steric_entropic}
    \includegraphics[scale=1]{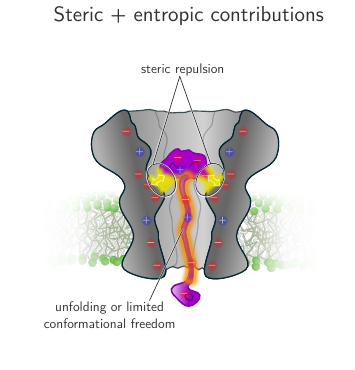}
  \end{subfigure}

\caption[Forces acting on a particle due to the intrinsic interactions]{%
  \textbf{Forces acting on a particle due to the intrinsic interactions.}
  (\subref{fig:nanopores_forces_int_es})
  The intrinsic electrostatic interactions between charged atoms on the translocating particle and the
  nanopore walls can yield repulsive (like-like charges) or attractive (like-unlike charges) forces. Unlike
  their externally applied counterpart, the electric field generated by these forces are highly local, and
  strongly mediated by the ionic strength.
  (\subref{fig:nanopores_forces_int_steric_entropic})
  If the diameter of the particle closely matches that of the pore, the confinement leads to steric clashes
  (\ie~repulsive intramolecular interactions) between the particle and the nanopore walls. In the case of
  proteins, this may prevent them from moving through the pore in their native state, resulting in unfolded
  translocation, or force them to assume a specific conformation or orientation---both of which induce an
  entropic penalty.  
  }\label{fig:nanopores_forces_int}
\end{figure*}

Next to the forces originating from the externally applied electric field, there are also several forces that
stem from the `intrinsic' interactions between the analyte molecules and the pore itself
(\cref{fig:nanopores_forces_int}). The nature of these forces can either be enthalpic (\ie~due to
intermolecular forces such as ion-ion interactions, hydrogen bonding, \gls{vdw} forces, etc.), or entropic
(\ie~due to a change in the microscopic degrees of freedom during the translocation process). To facilitate
the discussion, we will roughly subdivide the enthalpic forces into an electrostatic
(\cref{fig:nanopores_forces_int_es}) and a steric (\cref{fig:nanopores_forces_int_steric_entropic}, yellow
glow) component, with the latter being a `clumped' term that encompasses all intermolecular forces besides
electrostatics. For the entropic forces (\cref{fig:nanopores_forces_int_steric_entropic}, orange glow), we
will limit ourselves to a qualitative description of its impact on unfolded polymer translocation.

\subsubsection{Electrostatic forces: ionically mediated love and hate.}

Electrostatic forces arise when two charged atoms interact with one another, either in a repulsive (for
like-like charges) or attractive (for like-unlike charges) manner (\cref{fig:nanopores_forces_int_es}). For a
static charge distribution in a pure dielectric environment, Coulomb's law provides an excellent estimation
for these forces. However, in an electrolyte, the situation is significantly more complex due to the presence
of mobile charges (\ie~ions) that form an \gls{edl}, minimizing the overall energy of the system and strongly
mediating the distance at which fixed charges can still influence one another. By providing a proper
description of the \gls{edl}, the \gls{pb} theory discussed in~\cref{sec:np:edl} also provides us with a
method to calculate the total electrostatic energy, $\energyelec$, which for a solvated molecular system with
a volume $\Omega$, amounts to~\cite{Gilson-1993,Im-1998,Baker-2005}
\begin{align}\label{eq:nanopores_electrostatic_energy}
  \energyelec = 
  \dfrac{1}{4 \pi} \int_{\Omega}
  \bigg[
    \scd^{\rm{f}} \potential
    - \permittivity \dfrac{\left(\nabla \potential \right)^2}{2}
    - \dsum_{i}^{N} \ci^0 \pd{ e^{- \beta V_i} }{\rpos}
                    \left( e^{ - \beta \chargen_i \ec \potential } - 1 \right)
  \bigg] \, d\rpos
  \text{ ,}
\end{align}
with $\beta = (\kbt)^{-1}$, $\scd^{\rm{f}}(\rpos)$ the fixed charge density (\ie~from the pore or the
analyte), $\permittivity(\rpos)$ the spatial distribution of the permittivity and $\potential(\rpos)$ the
electrostatic potential. For every ion $i$ in the electrolyte, $\ci^0$ is the bulk concentration, $\chargen_i$
is the charge number, and $V_i(\rpos)$ is a repulsive potential that excludes it from non-solvent accessible
areas (\eg~within the biomolecules). Because force is the (negative) change in energy over distance, the
electrostatic force on atom $m$, $\force_{\es,m}$, can be calculated by differentiating $\energyelec$ with
respect to the atomic coordinates, $\rpos_m$, yielding~\cite{Gilson-1993,Im-1998,Baker-2005}
\begin{align}\label{eq:nanopores_electrostatic_force}
  \force_{\es,m} ={}& - \pd{\energyelec}{\rpos_m} \notag \\
    ={}& - \dfrac{1}{4 \pi} \int_{\Omega} \bigg[ 
      \underbrace{ \pd{\scd^{\rm{f}}}{\rpos_m} \potential }_{ \text{RF} }
      - \underbrace{ \dfrac{\left(\nabla \potential \right)^2}{2} 
      \pd{\permittivity}{\rpos_m} }_{ \text{DB} }
      - \underbrace{ \dsum_{i}^{N}
          \ci^0 \pd{ e^{- \beta V_i} }{\rpos_m}
          \left( e^{ - \beta \chargen_i \ec \potential } - 1 \right) }_{ \text{IB} }
          \bigg] \, d\rpos
  \text{ .}
\end{align}
As indicated, the components of the integral in \cref{eq:nanopores_electrostatic_force} can be split into
three distinct contributions. The `RF' or `reaction field' term expresses the response of the potential to the
spatial variations of the fixed charge density. It is mathematically equivalent to Coulomb's law
(\cref{eq:nanopores_lorentz_force})~\cite{Im-1998}, and hence yields the bare (\ie~free-space) force between
charges. The `DB' or `dielectric boundary' term arises from spatial variations of the permittivity between the
solute ($\relperm \approx \text{\numrange{2}{20}}$~\cite{Li-2013}) and the solvent ($\relperm \approx 80$) in
a charged, non-uniform environment. In essence, these differences of polarizability induce a net surface
charge at the interface, onto which the electric field will exert a non-negligible force. The DB force is
particularly important when moving charges into a confined environment, such as ion
channels~\cite{Nadler-2003}. The `IB' or `ionic boundary' term is proportional to the ionic concentration and
accounts for the contribution of the \gls{edl} to the force. Note that the ionic boundary term will always
diminish the contributions of the RF force, whether those are positive or negative. This can intuitively be
understood as the screening effect of the \gls{edl}: higher ionic strengths lead to more screening and hence
lower energetic contributions between fixed charges.

Given that a typical nanopore measurement is performed at physiologically relevant ionic strengths
(\eg~\SIrange{50}{500}{\mM}), and that both the pore and the analyte often carry a significant number of
charges, all the electrostatic force components described above are likely to be non-negligible. Moreover, its
manipulation can be used to dramatically alter the behavior of translocating molecules. Examples include
enhancing the translocation of \gls{dna}~\cite{Maglia-2008}, improving the capture of
peptides~\cite{Asandei-2015b,Asandei-2016}, or mediating the recognition between two protein
isoforms~\cite{Fahie-2015b}. The electrostatic force acting on analyte molecules translocating through
\glspl{bnp}---in the form of the electrostatic energy landscape---will be discussed in detail in
\cref{ch:electrostatics} (\gls{dna} through \gls{frac} and \gls{clya}) and \cref{ch:trapping} (proteins
through \gls{clya}).

\subsubsection{Steric forces: come close, but not too close.}

Next to the long-range electrostatic interactions, the close confinement of an analyte molecule within the
interior of the pore is bound to lead to steric clashes between the atoms of the analyte and those of the
pore~\cite{Buchsbaum-2013}. For two (neutral) atoms at distances below \SI{\approx1}{\nm}, the \glsfirst{vdw}
forces start to dominate the interatomic potential. These originate from transient fluctuations in the
electron densities of the atoms, leading to Keesom (permanent--permanent dipole interaction), Debye
(permanent--induced dipole interaction), and London dispersion (fluctuating dipole--induced dipole
interaction) forces. The total \gls{vdw} force is typically first attractive, up until the electron clouds of
the atoms begin the overlap, and the Pauli exclusion principle induces a strong repulsive force. The \gls{lj}
potential, $\potLJ$, is perhaps the most famous approach that captures the essential properties of the
\gls{vdw} interaction~\cite{Paquet-2015}. Spatially differentiating $\potLJ$ yields the \gls{lj} force 
\begin{equation}\label{eq:nanopores_LJ_force}
  \forcelj = - \nabla \cdot \potLJ( r ) = - \welldepthLJ \nabla \cdot \bigg[ 
    \underbrace{ \left( \dfrac{ \dminLJ }{ r } \right)^{12} }_{\text{repulsive}}
    -
    \underbrace{ 2 \left( \dfrac{ \dminLJ }{ r } \right)^{6} }_{\text{attractive}}
  \bigg]
  \text{ ,}
\end{equation}
with $r$ the distance between the two atoms, $\welldepthLJ$ the depth of the potential well, located at a distance
$\dminLJ$. The repulsive (`12') term ensures that $\potLJ$ becomes highly repulsive at very short distances.
Even though the mathematical form has no physical meaning, it adequately reproduces the repulsive force caused
by the overlapping of electron orbitals (\ie~Pauli repulsion). The minus sign in front of the second (`6')
term indicates that is attractive, and its mathematical form can be related back to the underlying physics.
Because the attractive term acts at longer distances, the resulting energy landscape contains an energy
minimum, $\welldepthLJ$, when the atoms are at $r = \dminLJ$. In other words, $\dminLJ$ represents the tipping
point after which the interatomic force becomes repulsive rather than attractive. Looking at the {CHARMM36}
force field parameters~\cite{Huang-2016}, typical \gls{lj} values for $\dminLJ$ and $\welldepthLJ$ within
proteins are \SI{4.1}{\angstrom} and \SI{-0.11}{\kbt} (carbon), \SI{3.6}{\angstrom} and \SI{-0.31}{\kbt}
(nitrogen), \SI{3.5}{\angstrom} and \SI{-0.22}{\kbt} (oxygen) and \SI{2.3}{\angstrom} and \SI{-0.06}{\kbt}
(hydrogen). Do note that the \gls{lj} potential is merely an approximation (\eg~ it is not directional) and
exact solutions require expensive quantum mechanical calculations~\cite{Paquet-2015}.

Given their weak nature and limited range, the \gls{vdw} forces acting on individual atoms are easily overcome
by thermal energy. However, the simultaneous close contact of hundreds of atoms can easily induce a very large
repulsive force (\ie~a high steric energy barrier) that will influence the translocation dynamics of analyte
molecules. This could easily occur when a protein is squeezed through a narrow constriction within a nanopore
(\cref{fig:nanopores_forces_int_steric_entropic}, yellow glow). In~\cref{ch:trapping}, such a steric barrier
will be crucial for explaining the escape dynamics of a protein trapped in \gls{clya}.

\subsubsection{Entropic forces: threading the needle.}

Up until now, all the discussed forces are enthalpic in nature---they originate from the need of the analyte
molecules to minimize their internal free energy. However, the confinement of proteins and nucleic acids
within a nanoscale cage~\cite{Liu-2015b}, or their (forced) unfolding during nanopore
translocations~\cite{Muthukumar-2010,Cressiot-2012,Cressiot-2015}, induce an additional energy barrier that is
of entropic nature. This means that certain transitions can only occur from a limited subset of analyte/pore
conformations. We will limit the discussion to the threading of a polymer through a nanopore, though the
enthalpic and entropic effects of folded protein confinement within a biological nanopore will be discussed in
more detail in \cref{sec:np:confinement}.

The entropy, $\entropy$, of a thermodynamic system with the number of available microstates, $\microstates$, is
given by Boltzmann's equation~\cite{Neumann-1980}
\begin{equation}\label{eq:nanopores_entropy}
  \entropy = \boltzmann \ln \microstates \text{ .}
\end{equation}
The energy barrier resulting from the entropy difference between a nanopore-analyte configuration where all
microstates are able to translocate (``0''), and one where only a limited subset can traverse the pore (``1'')
is given by
\begin{equation}\label{eq:nanopore_entropic_energy}
  - \temperature \Delta \entropy = - \temperature \left( \entropy_1 - \entropy_0\right)
  = - \boltzmann \temperature \ln \dfrac{\microstates_1}{\microstates_0}
  \text{ ,}
\end{equation}
with $\microstates_n$ the number of available microstates in configuration $n$. Several intuitive insights can
be gained from \cref{eq:nanopore_entropic_energy}: (1) if $\microstates_1 = \microstates_0$ (\eg~all
conformational configuration can translocate), $- \temperature \Delta \entropy = 0$, and the free energy
landscape is controlled solely by enthalpic forces, and (2) if $\microstates_1 < \microstates_0$ (\eg~a
polymer translocating polymer that can only enter the pore with one of its ends), $- \temperature \Delta
\entropy > 0$, indicating that a reduction in the number of suitable microstates indeed results in an energy
barrier, and \textit{vice versa} for $\microstates_1 > \microstates_0$ (\eg~when exiting the confines of the
pore). Concretely, microstate ratios of $\microstates_1 / \microstates_0 =
\text{\numlist{0.5;0.1;0.01;0.001}}$ would induce energy barriers of \SIlist{0.69;2.3;4.6;6.9}{\kbt},
respectively.

Note that, as before, we can define an entropic force as the gradient of the entropic
energy~\cite{Neumann-1980}
\begin{equation}\label{eq:nanopore_entropic_force}
  \forceentr = - \nabla \cdot (- \temperature \Delta \entropy )  
             = \boltzmann \temperature \nabla \ln \dfrac{\microstates_1}{\microstates_0}
  \text{ .}
\end{equation}
The magnitude of $\forceentr$ can be difficult to estimate experimentally, as it is often challenging to
differentiate it from the steric forces~\cite{Buchsbaum-2013}. Nevertheless, quantitative insights can be
obtained by either analytically estimating (the ratio of) the number of microstates at the relevant points in
space~\cite{Tian-2003,Muthukumar-2010,Cressiot-2015}, or by measuring the temperature dependence of the total
force exerted on the analyte experimentally~\cite{Meller-2002,Payet-2015} or through
simulations~\cite{Tian-2003,Matysiak-2006,Vaitheeswaran-2014,Luo-2017}.

\subsection{Analyte capture}
\begin{figure*}[b]
  \centering

  \begin{subfigure}[t]{115mm}
    \centering
    \caption{}\vspace{-2.5mm}\label{fig:nanopores_capture_overview}
    \includegraphics[scale=1]{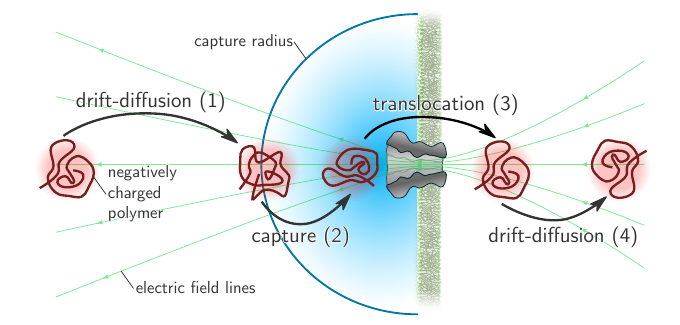}
  \end{subfigure}
  \\
  \begin{subfigure}[t]{115mm}
    \centering
    \caption{}\vspace{-2.5mm}\label{fig:nanopores_capture_energy}
    \includegraphics[scale=1]{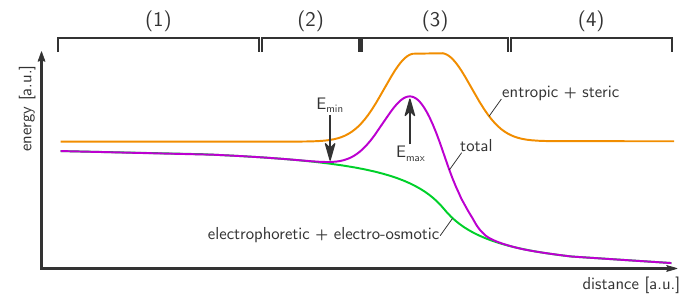}
  \end{subfigure}

\caption[Polymer capture process by a nanopore]{%
  \textbf{Polymer capture process by a nanopore.}
  (\subref{fig:nanopores_capture_overview})
  Schematic overview of the different steps of the capture process, accompanied by
  (\subref{fig:nanopores_capture_energy})
  a sketch of the electrophoretic, entropic, and total energy profiles.
  (1) Far away from the nanopore, the weak electric field causes charged polymer analyte molecules---such as
  \gls{dna} or proteins---to slowly diffuse or drift towards the pore. (2) Once inside the capture radius, the
  electric field becomes strong enough to readily overcome the random diffusive motion, and it is pinned at an
  energy minimum close to the entry of the pore. (3) The limited set of conformations with which the molecule
  can translocate through the pore (\eg~\gls{dna} must enter with one of its ends and then fully unfold),
  incurs an entropic energy penalty prior to, or during, the translocation process. (4) After fully
  translocating, the molecule is again free to drift away from the pore and diffuse in the reservoir.
  Adapted with permission from~\cite{Nomidis-2018}.
  }\label{fig:nanopores_capture}
\end{figure*}

The capture process of a polymer molecule by a nanopore is outlined in~\cref{fig:nanopores_capture}. It is
important to note that while the magnitude of the electric field inside the exterior reservoirs is negligible
compared to the one within nanopore, is not equal to zero. This means that charged molecules in the reservoir
are subject to both diffusion and electrophoretic drift
(\cref{fig:nanopores_capture_overview})~\cite{Muthukumar-2010}. Given that the electric field is inversely
proportional to the square of the distance from the pore
(see~\cref{eq:nanopores_efield_profile})~\cite{Grosberg-2010}, it becomes particularly relevant in close
proximity to the entries of the pore, where the electric field can become significantly stronger than the
thermal energy. Similarly, the \gls{eof} is non-negligible outside of the nanopore~\cite{Wong-2007} and
contributes significantly to the attraction or repulsion of \gls{dna}~\cite{Firnkes-2010} and
proteins~\cite{Soskine-2012,Soskine-2013} towards the pore. Finally, if a polymer requires a specific
configuration in order to translocate (\ie~unfold)---as is typically the case for the analysis of
\gls{dna}~\cite{Muthukumar-2010}, but not folded proteins~\cite{Yusko-2011,Soskine-2012,Plesa-2013}---it must
also overcome a significant entropic energy barrier before it can commence the translocation process
(\cref{fig:nanopores_capture_energy})~\cite{Muthukumar-2010}.

However, because the focus of this dissertation lies with the transport of molecules through (biological)
nanopores after they have been captured, the interested reader is referred to the seminal papers on polymer
capture by nanopores by Wong and Muthukumar (electro-osmotic flow)~\cite{Wong-2007}, Grosberg and Rabin
(electrophoresis)~\cite{Grosberg-2010}, and Muthukumar (entropic energy barrier)~\cite{Muthukumar-2010}. Of
note as well is the work of Nomidis~\etal~\cite{Nomidis-2018}, whose theoretical and experimental analysis
found that the capture rate of both \gls{ssdna} and \gls{dsdna} by \gls{clya} was limited by the entropic
barrier at the entry of the pore, rather than their arrival due to drift and diffusion.

\subsection{Confining proteins within biological nanopores}
\label{sec:np:confinement}

Given the magnitude of the forces involved and the degree of nanoscale confinement, it is worthwhile to
reflect on the structural impact the translocation process may have on the analyte molecule itself. This is
particularly relevant for proteins (\eg~enzymes), where strong pulling forces can cause deformation or
unfolding, and strong confinement can alter their function. In the following section, we will briefly discuss
these items in the context of a protein confined within a biological nanopore~\cite{Galenkamp-2020}.

\subsubsection{The role of electrophoresis and electro-osmosis.}

The role of the applied potential across the pore---and the resulting electrophoretic and electro-osmotic
forces---on the folding of proteins inside a (biological) nanopore is unclear. To estimate their magnitude,
using the \gls{clya} pore as an example (\cref{fig:nanopores_clya}), we can estimate the electric field within
the \lumen{} of \gls{clya} by describing it as two cylindrical electrolyte chambers in series. The strength of
the electrical field along the central nanopore axis inside the wide \cisi{} chamber of dodecameric \gls{clya}
can then be estimated to be approximately \SI{3}{\mV\per\nm} for a bias voltage of
\SI{-50}{\mV}.\footnotemark%
\footnotetext{
As calculated with Ohm's law for \SI{50}{\mV} bias voltage from the resistance of the \cisi{} chamber
($R_{\text{\cisi}} \approx \mSI{2.6e8}{\ohm}$) in series with the \transi{} chamber ($R_{\text{\transi}}
\approx \mSI{2.2e8}{\ohm}$), resulting in the corresponding voltage drops of $\Delta V_{\text{\cisi}} \approx
\mSI{27}{\mV}$ (over a distance $l_{\text{\cisi}} \approx \mSI{10}{\nm}$), and $\Delta V_{\text{\transi}}
\approx \mSI{23}{\mV}$ (over a distance $l_{\text{\transi}} \approx \mSI{3}{\nm}$),
respectively~\cite{Soskine-2013}.
}
Thus, if an immobilized protein carries a significant net charge (\eg~\SI{10}{\ec}) it would be subject to a
Coulomb force of approximately \SI{4}{\pN} (\cref{eq:nanopores_lorentz_force}). Forces of this magnitude have
been shown before to partially unfold or deform proteins by \gls{afm}~\cite{Best-2001}. However, the Pelta
group showed that within the \SIrange{50}{250}{\mV} voltage range, the fractional residual current (defined as
the blocked pore current divided by the open pore current) of maltose binding protein translocating through a
solid-state nanopore remains constant, as expected for a protein that remains folded~\cite{Talaga-2009}. Using
\gls{clya} and sampling several proteins (human thrombin, {AlkB} and \gls{dhfr}), no changes in fractional
residual current up to \SI{-200}{\mV} were observed, confirming that the protein normally should not unfold
under moderately applied potentials~\cite{Soskine-2013}. However, the net force experienced by a protein
immobilized inside a nanopore will also depend on the magnitude and direction of the electro-osmotic flow. The
negative charges lining the interior walls of \gls{clya} make the pore highly cation-selective and induce a
strong electro-osmotic flow~\cite{Franceschini-2016}, enabling the capture of proteins even against the
electrical field~\cite{Soskine-2012} (see~\cref{ch:trapping,ch:transport}). While the precise magnitude of the
water velocity inside \gls{clya} is currently unknown, experimental data using \gls{ahl}
nanopores~\cite{Paula-1999} and simulations performed with solid-state~\cite{vanDorp-2009,Luan-2008} and
\gls{ahl}~\cite{Aksimentiev-2005,Pederson-2015} nanopores, revealed velocities in the order of
\SI{100}{\mmps}. Assuming a similar water velocity inside our \gls{bnp} and applying Stokes' law
(\cref{eq:nanopores_drag_force}), a static and uncharged sphere of \SI{5}{\nm} in diameter would endure a
force of approximately \SI{4}{\pN}. Thus, under negatively applied potentials (\transi{}) the electrophoretic
and electro-osmotic forces acting on negatively charged proteins would oppose and cancel each other out,
resulting in less stress on the protein and a more diffusion-dominated environment~\cite{Firnkes-2010}. For
positively charged molecules, these two forces will add and increase the net force, resulting in additional
stress on the captured protein.

\subsubsection{The role of confined water.}

Another important consideration to make when confining proteins in nanopores, for example when measuring
enzymatic reactions, is the role of the nanopore walls and the relatively strong electrostatic forces inside
the nanopore. As described in \cref{sec:np:clya}, the \gls{clya} protein contains a plethora of negatively
charged residues lining its interior walls (\cref{fig:nanopores_clya_pore_section}). Such a negatively charged
nanocavity is reminiscent of the well-studied \textit{E. coli} {GroEL/GroES} chaperone~\cite{Xu-1997}.
Although the precise mechanism by which {GroEL} aids protein refolding remains
controversial~\cite{England-2008,England-2008b,Motojima-2012,Weber-2013}, it is likely that the confinement
prevents misfolded proteins from aggregating. Another explanation could be that the rich electrostatic
environment inside the {GroEL} cavity increases the water density and hence enhances folding due to an
enhanced hydrophobic effect compared with bulk~\cite{England-2008,England-2008b}. The diffusivity and the
structuring of the water molecules confined inside {GroEL} in the absence of substrate protein are likely to
be similar to that in bulk solution~\cite{Franck-2014}. When there is a protein confined inside the chaperonin
cavity, however, these properties might change depending on the interaction strength between the protein and
the internal walls of the chaperonin~\cite{Weber-2013}. Interestingly, the folding pathways inside the
chaperonin chamber have been proposed to be similar~\cite{Horst-2007} or different~\cite{Jewett-2004} compared
with bulk~\cite{Apetri-2008}. Regardless of the precise folding mechanism and the behavior of the confined
solvent, a charged, hydrophilic cage appears to promote the native state of proteins. Therefore, no
detrimental effects on the conformation of proteins trapped inside \gls{clya} are to be expected.

%
%
%
%

\section{Approaches for computational modeling of nanopores}
\label{sec:np:modeling}

Barring the use of excessive simplifications, our ability to describe the behavior of any arrangement of atoms
and molecules with tractable analytical approaches quickly disappears with increasing system complexity. It is
for this reason that scientists resort to simulations, where the equations that describe underlying the
physical phenomena are solved numerically, to a certain accuracy and typically with a computer, rather than
analytically, for an appropriate, non-simplified model of the molecular system. In the final section of this
chapter, we aim to familiarize the reader with some of the most commonly used computational approaches for
modeling complex molecular problems.


%
\begin{figure*}[b]
  \centering

  \includegraphics[scale=1]{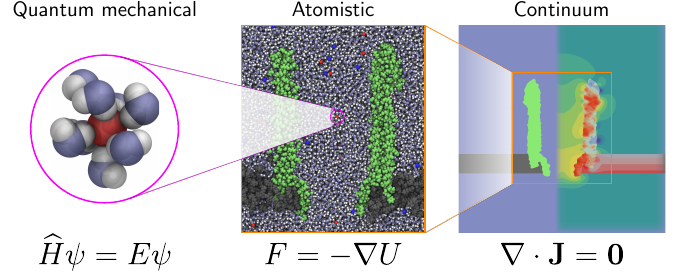}

\caption[Hierarchy of simulation methodologies]{%
  \textbf{Hierarchy of simulation methodologies.}
  Quantum mechanical methods (left) have the highest precision, as they model the nuclei and electron
  distributions of atoms explicitly. However, they also have the highest computational cost and are often only
  employed to parameterize the interaction potentials and partial charges used for the atomistic modeling
  methods (middle, \eg~molecular dynamics). Further averaging of the interaction potential over multiple atoms
  gives rise to continuum or `mean-field' methods (right), where the discreteness of the atoms is replaced by
  a structureless medium with material properties.
  }\label{fig:nanopores_simulations_types}
\end{figure*}

\subsection{Atomistic, `discrete' modeling}

\subsubsection{The interaction potential.}

One of the most commonly used simulation approaches to study the behavior of individual atoms in complex
(macro)molecular systems, is to model the atoms (or a combination of atoms) as discrete, hard spheres, whose
motions are governed by an \emph{intermolecular interaction potential}, $\energymd$
(\cref{fig:nanopores_simulations_types}, middle). Roughly speaking, computational approaches that make use of
such an interaction potential can be subdivided into three classes: \gls{mc} simulations, \glsfirst{md},
and stochastic dynamics~\cite{Paquet-2015}. Whereas the \gls{mc} methods employ random sampling to map out the
properties for a given conformational state of a molecular assembly, \gls{md} can be used to track the
time-evolution of this system from one state to the next. In stochastic dynamics, the realism of the
(intrinsically approximate) interaction potential is improved by the explicit inclusion of a friction
term (to account for solvent viscosity) and an additional random energy term (to account for the occasional
perturbation by a high-velocity collision). In the context of a molecular system, these inclusions lead to the
well-known Langevin dynamics, which, in the overdamped regime (\ie~no inertia) it is known as \glsfirst{bd}.
Because we will only discuss \gls{md} in detail, the interested reader is referred to the comprehensive
comparison compiled by Paquet and Viktor in Ref.~\cite{Paquet-2015}.

\subsubsection{Molecular dynamics: atoms pushing atoms.}

In \gls{md}, Newton's equations of motion are solved numerically for every atom as a function of time,
yielding a so-called `trajectory' that contains the time-dependent positions and velocities for every atom
(\ie~a set of microstates). Concretely, the \gls{md} algorithm computes the position, $\rpos(\timedim)$,
velocity, $\velatom(\timedim)$, and acceleration, $\accelatom(\timedim)$, of each atom for a small
timestep\footnotemark, $\timestep$, for a given set of starting positions and velocities at $\timedim = 0$.%
\footnotetext{
To avoid discretization errors, $\timestep$ must necessarily be smaller than the fastest vibrational frequency
in the system (typically hydrogen bond vibration), leading to the use of time steps in the order of
\SI{1}{\fs} (\SI{e-15}{\second}). This imposes a practical limit of \SIrange{0.1}{1}{\us} on the length
trajectories of large molecular assemblies. For atomistic systems with \num{\approx e6}~atoms (\eg~a large
biological nanopore), the upper limit  for long-timescale trajectories (\ie~computational times within the
duration of the typical doctoral program), using current algorithms and supercomputers, is \SI{\approx
1}{\ms}~\cite{Vendruscolo-2011,Phillips-2020}.
}
The net force on every atom, $\force$, is computed from the gradient of the potential energy functional
(\ie~the interaction potential), which consists of a summation of `bonded' (\ie~between atoms sharing a
covalent bond: bond length, angle, and dihedral) and `non-bonded' (\ie~between all atoms: \gls{vdw} and
electrostatics) interactions. Hence,
\begin{equation}\label{eq:nanopores_md_force}
  \force (\rpos) = - \nabla \energymd (\rpos)
  \text{ ,}
\end{equation}
with
\begin{equation}\label{eq:nanopores_md_energy}
  \energymd (\rpos) = \dsum \energymd_{\text{b}} (\rpos) + \dsum \energymd_{\text{nb}} (\rpos)
  \text{ ,}
\end{equation}
where $\energymd_{\text{b}}$ and $\energymd_{\text{nb}}$ are the bonded and the non-bonded energy terms,
respectively. Because the calculation $\energymd_{\text{nb}}$ involves the summations over all atoms of the
system, it is particularly computationally taxing ($O(N^2)$ scaling). However, the use of approximations
(\eg~distance cut-offs) or clever algorithmic implementations (\eg~neighbor lists) can significantly reduce
the computational cost ($O(N\log N)$ or even $O(N)$ scaling~\cite{Eastman-2010}).

The parameterization of the functions that describe the (non)-bonded interactions are typically based on
\textit{ab initio} quantum-mechanical simulations (\cref{fig:nanopores_simulations_types}, left), combined
with, and verified by, experimental data when possible. The compilation of parameters for a given set of
(bio)molecules is called a molecular mechanics `force-field', of which many variants exist---each one
optimized for a specific set of applications or conditions. In the case of macromolecular simulations
involving proteins, nucleic acids, and ions, the {AMBER}~\cite{Ponder-2003}, {CHARMM}~\cite{Huang-2016} and
{GROMOS}~\cite{Oostenbrink-2004} force fields are the most widely used.

Depending on whether one chooses to keep constant the total energy or the temperature of the system, the
time-average of the \gls{md} trajectory will correspond to respectively the canonical (\gls{nvt}: number,
volume, and temperature), or microcanonical (\gls{nve}: number, volume, and energy) ensemble. The
velocity-Verlet algorithm is one of the most well-known approaches for \gls{nve} \gls{md} simulations, as it
conserves the total energy of the system with a high degree of accuracy~\cite{Swope-1982}. In contrast,
\gls{nvt} simulations mandate the addition or removal of energy from the system to keep the temperature fixed.
This can be achieved by adjusting the velocities of atoms with an additional `thermostat' algorithm such as
Langevin dynamics~\cite{Bussi-2008}, or the velocity-rescaling~\cite{Heyes-1983},
Nos\'{e}-Hoover~\cite{Nose-1984,Hoover-1985}, or Andersen~\cite{Andersen-1980} thermostats.

Even though \gls{md} simulations are among the most realistic simulation methodologies for atomistic
systems, the high computational costs associated with both running the simulations and analyzing the
resulting trajectories still limit its scalability~\cite{Vendruscolo-2011,Phillips-2020}. Additionally, great
care must be taken when setting up the simulation domain (\ie~boundary conditions), the choice of force-field,
and the simulation parameters, as each of these can lead to nonphysical results, particularly in confined
spaces~\cite{Wong-ekkabut-2016a}.

\subsubsection{Molecular dynamics applied to biological nanopores.}

Due to its widespread experimental use, and the early availability of its crystal structure, the \gls{ahl}
nanopore has been the primary subject of interest for \gls{md} studies on \glspl{bnp} over the past
20~years~\cite{Aksimentiev-2005,DeBiase-2016,Basdevant-2019}, most notably by the research groups of Aleksei
Aksimentiev and Mauro Chinappi. The former was the first to use \gls{md} to provide a complete `image' of
\gls{ahl} in terms of electrostatic potential distribution, ion concentrations, ionic conductance, and
electro-osmotic permeability~\cite{Aksimentiev-2005}. Since then, many more \gls{md} studies have been
performed on \gls{ahl}, including the translocation of nucleic acids~\cite{Wells-2007}, and unfolded
proteins~\cite{DiMarino-2015}, the cation-type dependence of the ionic current
rectification~\cite{Bhattacharya-2011}, the magnitude of the electro-osmotic flow~\cite{Bonome-2017}, and the
recognition mechanisms of nucleobases~\cite{Manara-2015b,DeBiase-2016} and amino acids~\cite{DiMuccio-2019}.

Beyond \gls{ahl}, the transport properties of several other \glspl{bnp} have been scrutinized with \gls{md}
simulations. These include the \gls{mspa}~\cite{Bhattacharya-2012,Manara-2015,Bhattacharya-2016,Zhou-2020},
\gls{ael}~\cite{Cao-2018,Ouldali-2020,Zhou-2020}, \gls{frac}~\cite{Zhao-2019}, and
\gls{clya}~\cite{Mandal-2016,Wilson-2019,Li-2020} pores, and recent work by Zhou~\etal{} provides a
comprehensive overview of the nanofluidic properties of the \gls{mspa}, \gls{ahl}, \gls{csgg}, and \gls{ael}
nanopores~\cite{Zhou-2020}.



\subsection{Continuum, `mean-field' modeling}

\subsubsection{The mean-field approximation: averaging the interaction potential.}

The high computational cost associated with the vast amount of degrees of freedom in atomic systems enforces
harsh limits on both the attainable system size and the maximal duration of a simulation. Even though this
computational burden can be reduced by `coarse-graining' (\ie~combining multiple real atoms into a single
pseudoatom with pseudo interaction potentials), the fundamental scaling problem remains unaltered.
Fortunately, the stochasticity of individual atoms tends to rapidly average out in groups of atoms in a
predictable manner, allowing their behavior to be condensed into a single material property that replaces the
unique interaction potentials of a group of atoms with a single effective interaction potential, described by
a structureless medium or continuum (\cref{fig:nanopores_simulations_types}, right). From an energetic
point of view, this is equivalent to finding the best approximation for the total energy of a system of
interacting particles that does not include any explicit interaction terms, also known as the \emph{mean-field
approximation}. Importantly, the mean-field approximation reduces the many-body problem presented by explicit
molecular simulations to a much more scalable one-body problem. Well-known examples of material properties
include the relative permittivity, the viscosity of a liquid, or the diffusion coefficient of an ion. Even
though mean-field theories excel at upward scaling (\ie~macroscopic system sizes and timescales), they must be
used with caution when scaling downwards (\ie~microscopic system sizes and timescales), as the averaged
property will no longer adequately represent the true physical behavior. For the case of electrostatics, the
review by Collins provides an exhaustive set of reasons why continuum methods are unrealistic at the
nanoscale, with the lack of explicit ion-water interactions being the main culprit~\cite{Collins-2012}.
Nevertheless, the mean-field approximation has been used extensively, and successfully, for the (qualitative)
simulation of ion channels~\cite{Im-2002,Furini-2006,Liu-2015},
\glspl{bnp}~\cite{Simakov-2010,Pederson-2015,Aguilella-Arzo-2017,Simakov-2018} and
\glspl{ssnp}~\cite{Cervera-2005,White-2008,Chaudhry-2014,Laohakunakorn-2015}, providing meaningful physical
insights at a fraction of the computational cost of a \gls{md} simulation.

\subsubsection{Molecular electrostatics and transport using continuum methods.}

In \cref{sec:np:edl}, we have already introduced a continuum representation for modeling the electrostatics in
electrolytes: the \glsfirst{pbe} (see~\cref{eq:nanopores_pbe}). Whereas \gls{pb} theory provides a good
approximation for electrostatics at equilibrium, it contains no terms that consider the dynamics of a system
(\eg~diffusion coefficients), making it incapable of modeling nonequilibrium processes, such as the flux of
ions, water, or analyte molecules through a nanopore under an applied bias voltage. A set of equations that
can tackle this problem are the \gls{npe}, a mass conservation equation that extends Fick's law of diffusion
with electrostatic and convective forces, in combination with \gls{pe} to model the electrostatics, and the
\gls{nse} to compute the fluid velocity and pressure distribution. When properly coupled, the \gls{pnp-nse}
provide a self-consistent framework with which one can model the transport of ions and water through
nanopores. In \cref{ch:epnpns} we will introduce a set of corrections for \glsdisp{pnp-ns}{PNP-NS} theory to
improve their validity at the nanoscale and enable accurate, quantitative predictions for the nanofluidic
properties of biological nanopores.

Let us start with the \gls{npe}, which can be used to describe the flux of a charged chemical species in a
fluid medium. It is given by~\cite{Lu-2011}
\begin{equation}\label{eq:nanopores_nernst_planck}
  \underbrace{ \pd{\concentration}{\timedim} }_{\text{variation}}
  = - \nabla \cdot \flux = \nabla \cdot (
    \underbrace{ \diffusion \nabla \concentration }_{\text{diffusion}}
    +
    \underbrace { \chargen \mobility \concentration \nabla \potential }_{\text{migration}}
    -
    \underbrace{\velocity \concentration}_{\text{convection}}
    )
  \text{ ,}
\end{equation}
with $\flux$ the flux, $\concentration$ the concentration, $\diffusion$ the diffusion coefficient, $\chargen$
the charge number, and $\mobility$ the electrophoretic mobility of the charged molecule, and $\potential$ the
electric potential, and $\velocity$ the fluid velocity. As indicated, the \gls{npe} consists of
diffusive, (electro)migratory, and convective flux terms. The diffusive flux makes molecules move along their
concentration gradient (from high to low concentration), and it is proportional to the diffusivity of each
individual molecule. Similarly, the electromigratory flux drives the movement of charged molecules with
(positive charge) or against (negative charge) the electric potential gradient (\ie~the electric field),
proportional to their mobility. Note that the Einstein relation,
$\mobility=\frac{\ec}{\kbt}\chargen\diffusion$, which dictates that the diffusion coefficient is
directly proportional to the electric mobility, only holds at infinite dilution (\ie~$\concentration \to 0$),
and hence should not be used at finite concentrations (even though it often is). Finally, the convective flux
represents the molecules that are dragged along by the movement of the fluid itself, irrespective of their
charge or diffusivity. Note that each chemical species present in the system is described by its own
\gls{npe}, each fully independent from the other. This lack of (steric) interactions between the species can
lead to nonphysical results (\ie~very high ion concentrations in the \gls{edl}), indicating that the \gls{npe}
is only applicable to very dilute solutions. However, as we shall see in~\cref{ch:epnpns,ch:transport}, the
inclusion of coupled, size-dependent terms can mitigate these problems~\cite{Lu-2011}.

As with the \gls{pbe}, the spatial distribution of $\potential$ can be obtained from \gls{pe}
\begin{equation}\label{eq:nanopores_poisson}
  \underbrace { \nabla \cdot \left(\absperm \relperm \nabla \potential \right) }_{\text{field divergence}}
  = \underbrace{ - \scd }_{\text{charge source}}
  \text{ ,}
\end{equation}
where $\scd$ is the charge distribution within the system. In the case of a charged analyte molecule within a
nanopore, $\scd = \scdfix + \scdion$, with the first term representing the fixed ionized groups on the pore
and the analyte, and $\scdion$ the ionic charge imbalance within the \gls{edl}
(see~\cref{eq:nanopores_ion_scd}). In essence, \cref{eq:nanopores_poisson} states that, within a system with a
given permittivity distribution, the electric field must adapt itself, spatially, in such a way (\ie~the field
divergence) as to perfectly cancel out the imposed charge distribution(s). Note that the inclusion of
$\scdion$ in \cref{eq:nanopores_poisson} couples it back to the \gls{npe}.

In continuum mechanics, the velocity of a viscous fluid (\eg~water) can be described using the famous
\gls{nse}, which comprise a mathematical representation for the conservation of mass and momentum. For an
incompressible flow (\ie~water flow through a nanopore), the momentum balance of the \gls{nse} reads
\begin{equation}\label{eq:nanopores_navier_stokes}
  \underbrace{ \density \pd{\velocity}{\timedim} }_{\text{variation}}
  +
  \underbrace{ \density \left( \velocity \cdot \nabla \right) \velocity }_{\text{convection}}
  +
  \underbrace{ \viscosity \nabla \cdot \left(
    \nabla \velocity + \left(\nabla\velocity \right)^\mathsf{T} \right) }_{\text{diffusion}}
  =
  \underbrace{\nabla \cdot \left(\pressure \identity \right)}_{\text{int. source}}
  +
  \underbrace{ \forcedensvol }_{\text{ext. source}}
  \text{ ,}
\end{equation}
and its mass balance is given by the continuity equation
\begin{equation}\label{eq:nanopores_velocity_continuity}
  \density \nabla \cdot \left( \velocity \right) = 0
  \text{ ,}
\end{equation}
with $\density$ the fluid density, $\velocity$ the fluid velocity, $\pressure$ the pressure, and
$\forcedensvol$ a volume force density acting on the fluid. The first two terms of
\cref{eq:nanopores_navier_stokes} represent the local inertia of the fluid, with the first one (variation)
being the time-dependent change of the velocity (zero at steady state), and the second one (convection)
expressing the inertia due to the intrinsic velocity of the fluid. The third (diffusion) and fourth (internal
source) terms result from the divergence of the hydrodynamic stress tensor
(see~\cref{eq:nanopores_hydrodynamic_stresstensor}). Intuitively, the diffusion term can be interpreted as
follows: if the fluid velocities at two neighboring locations differ, the viscous friction between them will
result in a transfer of the momentum from the high- to low-velocity location, analogous to the diffusion of a
chemical species from high to low concentration. The internal source term stems from the pressure differences
within a volume of fluid, causing it to flow from high to low pressures. Finally, the last term represents all
forces that act upon the fluid from external forces, such as gravity or an electric field
(\ie~$\forcedensedl$, see~\cref{eq:nanopores_edl_force}), providing a way to couple the \gls{nse} back to both
the \gls{npe} and the \gls{pe}.

The radical simplification of replacing all interatomic interactions with an average mean-field brings with it a
massive reduction in computational cost, enabling continuum methods to investigate large problems over long
timescales. When numerically solving \glspl{pde}, such as the \gls{pnp-nse}, the total computational domain is
discretized into a set of sub-domains, each holding the local solution for the variables of interest. The
nature of this discretization, and how it is linked algebraically to the \glspl{pde} themselves, depends on
the numerical algorithm, of which the \glsdisp{fdm}{finite difference~(FDM)}, \glsdisp{fem}{finite
element~(FEM)}, and \glsdisp{fvm}{finite volume~(FVM)} methods are the most widely used. Note that only the
\gls{fvm} is truly conservative (\ie~mass cannot (dis)appear out of nowhere), as its formulation intrinsically
ensures that, for any given volume element, the inbound and outbound fluxes are identical. Nevertheless, the
\gls{fdm} and \gls{fem} are very popular, as their mathematical elegance makes them straightforward to
implement for a variety of numerical problems, whereas using fine-grained discretization can mitigate problems
with conservation.

\subsubsection{Continuum modeling of biological nanopores.}

Initial work on predicting the permeation of ions and water through biological transmembrane channels was
focused towards ion channels~\cite{Eisenberg-1996,Chen-1997,Corry-2003}. As was the case with \gls{md}, to
date, virtually all the continuum modeling studies of \glspl{bnp} have been performed with \gls{ahl} as the
subject of
interest~\cite{Noskov-2004,Cozmuta-2005,OKeeffe-2007,Simakov-2010,Pederson-2015,Simakov-2018,Aguilella-Arzo-2020},
though its use in the \gls{ssnp} field is significantly more
prevalent~\cite{Daiguji-2004,Cervera-2005,White-2008,Lu-2012,Chaudhry-2014,Laohakunakorn-2015,Hulings-2018,Rigo-2019,Melnikov-2020}.
However, despite the widespread use of continuum transport models, their ability to quantitatively predict
ionic currents---in their unmodified form---is typically poor~\cite{Corry-2000,Collins-2012}. To remedy this,
researchers have sought ways to mitigate the shortcomings of {PNP}(-{NS}) theory by including steric
ion-ion interactions~\cite{Kilic-2007,Lu-2011,Liu-2020}, the non-uniform spatial distribution of the
diffusivities~\cite{Cozmuta-2005,Furini-2006,Simakov-2010,}, and the concentration dependencies of electrolyte
properties~\cite{Baldessari-2008-1,Burger-2012,Chen-2016}. Such corrections can be based on theory,
experiment, or extracted from other simulations (\eg~\gls{md}). Conversely, the field distributions (electric
field and electro-osmotic flow) can also be used as input for \gls{bd} simulations to stochastically map
the capture dynamics of individual particles~\cite{Pederson-2015,Hulings-2018}. Given that \cref{ch:epnpns} of
this dissertation is devoted to building an improved \gls{pnp-ns} framework for the predictive modeling of
biological nanopores, a more detailed discussion of this topic can be found there. Additionally,
in~\cref{ch:transport} we have applied this improved framework to \gls{clya}, in an extensive study of its
nanofluidic properties.


%
%

\section{In summary}

The field of single-molecule sensing has grown tremendously over the past two decades. Nanopores have
established themselves as a powerful, label-free single-molecule sensing technology, providing both
confirmatory and complementary data relative to the traditional techniques of force spectroscopy and
fluorescence microscopy. Whereas initial research was driven by nanopore-based \gls{dna} sequencing, a goal
that was recently successfully attained and commercialized by Oxford Nanopore Technologies, the application
space of nanopore sensing has now expanded considerably: from proteomics~\cite{Yusko-2017,Houghtaling-2019}
and protein sequencing~\cite{Restrepo-Perez-2018}, to metabolomics~\cite{Zernia-2020} and single-molecule
enzymology~\cite{Galenkamp-2020,Willems-VanMeervelt-2017}. This expanding set of applications has also come
with an equally diverse set of nanopores in terms of size, shape, and material, well beyond the classical
\ta-hemolysin and silicon nitride variants. The choice of biological nanopores now includes pores with large
interiors for whole-protein sensing, such as cytolysin A~\cite{Soskine-2012} or pleurotolysin
AB~\cite{Huang-2020}, or small funnel-shaped ones for small-molecule sensing, such as fragaceatoxin
C~\cite{Huang-2017,Restrepo-Perez-2019a}. Likewise, their solid-state counterparts have advanced
significantly, from quasi-2D pores in graphene~\cite{Fischbein-2008} or \ce{MoS2}~\cite{Feng-2015b} for
ultimate constriction sensitivity, to the integration of field-effect transistors~\cite{Ren-2020} that provide
a new mode of sensing, to sub-angstrom pores that provide novel insights into confined molecular
transport~\cite{Rigo-2019}. However, as the arsenal of available nanopores types and the complexity of their
applications have grown, so has the need to better understand their physical working principles. Even though a
plethora of mathematical frameworks and simulation methodologies currently exist~\cite{Maffeo-2012}, most
notably molecular dynamics and Poisson-Nernst-Planck theory, the intricate physics of nanoscale transport
within nanopores is pushing their capabilities to the limit and is testing the boundaries of their
validity~\cite{Collins-2012}. Nevertheless, it is here that modeling approaches---by providing a physical
understanding of existing phenomena and a guiding hand for novel experiments---can help to drive the nanopore
field forward.

\cleardoublepage

%

%
  \graphicspath{{\chaptersdir/aims/image/}}%
\chapter{Objectives}
\label{ch:objectives}

\epigraphhead[\epipos]{%
\epigraph{%
  ``You could find out most things, if you knew the right questions to ask. Even if you didn't, you could
  still find out a lot.''
}{%
  \textit{`Gurgeh' in `The Player of Games' by Iain M. Banks}
}}

Given their introduction in \cref{ch:nanopores}, it should stand without a doubt that nanopores---and biological
ones in particular---have become powerful single-molecule sensing tools. Among the key strengths of nanopores
are their ability to interrogate individual molecules label-free, at high sampling frequencies, and with long
observation times. Whereas over the past 25 years nanopore research was driven by genomics, researchers are
now recognizing these advantages may be useful also in other fields, including proteomics, metabolomics,
single-molecule enzymology~\cite{Willems-VanMeervelt-2017}, and biosensing. This expanding application space
also presents an increasing challenge: translating the `two-dimensional' nanopore current into a meaningful
set of information. Even though complementary experimental techniques may alleviate the problem, it is here
that modeling approaches---be it analytical or computational---can bestow researchers with a thorough
understanding of the complete system itself, rather than its individual components. Specifically, by
elucidating the physical mechanisms that govern the interactions between the nanopore and the analyte
molecules, they may aid in the unambiguous interpretation of the current signals, provide insights into the
prevailing conditions within the pore, and detail guidelines for tailoring the properties of nanopores. In
this dissertation, the focus lies on the development of analytical and computational methodologies
(\textbf{Objective~1}) that explain the physical mechanisms behind the experimentally observed behavior of
nanopore-analyte systems (\textbf{Objectives~2 and 3}), or that can be used as a `computational microscope' to
study all the properties of the nanopore itself (\textbf{Objective~4}).

\clearpage
\objective{Develop methodologies for accurate modeling of biological nanopores}

Due to their proteinaceous nature, the primary computational tool for studying biological nanopores are
\glsfirst{md} simulations, where every atom is modeled explicitly. Unfortunately, the wealth of information
accessible through \gls{md} simulations comes at a heavy computational cost: often necessitating the use of
months of supercomputer time. Even though we will still make use of \gls{md} to construct realistic and
well-equilibrated homology models of various biological nanopores, most simulations in this work will be based
on continuum, rather than discrete, representations of the nanopore systems. In~\cref{ch:electrostatics} we will
show how the \glsfirst{pb} equation can be employed to quantify the equilibrium (\ie~without an external bias
voltage) electrostatic interactions between a nanopore and the surrounding electrolyte on the one hand, and
its analyte molecules on the other. Nanopores are governed by more than just electrostatics, however, and thus
in~\cref{ch:epnpns} we develop the \gls{epnp-ns} equations: a simulation framework that introduces several
self-consistent corrections to the electrolyte properties in an attempt to mitigate the shortcomings of the
classical {PNP-NS} equations at the nanoscale and beyond infinite dilution. Because the \gls{epnp-ns}
equations can be solved for continuum systems, they aim to provide a fast yet accurate means to model the
transport of ions and the flow of water through a nanopore. In~\cref{ch:trapping} we will also use an analytic
model---which intrinsically necessitates a reduction to the most essential components---to gain valuable
insights.

\objective{Investigate the equilibrium electrostatics of biological nanopores}

The interior walls of most biological nanopores are typically lined with charged amino acids. Hence, it is
not surprising that electrostatic interactions often play a determining role in the overall behavior of a pore.
In~\cref{ch:electrostatics}, we use the \gls{pb} equations to calculate the electrostatic potential within the
several variants of the \glsfirst{plyab}~\cite{Huang-2020}, \glsfirst{frac}~\cite{Wloka-2016,Huang-2017} and
\glsfirst{clya}~\cite{Franceschini-2016} nanopores and investigate the effect of ionic strength and pH on
their emergent properties, such as the \glsfirst{eof}. Additionally, we map out the electrostatic free
energies associated with \gls{ssdna} and \gls{dsdna} translocation through variants of the \gls{frac} and
\gls{clya} pores, respectively, and link the observed differences with published experimental findings. A
similar methodology will be used in \cref{ch:trapping} for quantifying the electrostatics energy associated
with the trapping of a protein within \gls{clya}~\cite{Soskine-Biesemans-2015}, which is the focus of
\textbf{Objective~3}.

\clearpage

\objective{Elucidate the trapping behavior of a protein inside a biological nanopore}

In their pioneering work with \gls{clya}, Soskine and Biesemans~\etal{} showed that the dwell time of the
\glsfirst{dhfr} enzyme (\SI{\approx19}{\kDa}) within \gls{clya} could be increased by orders of magnitude by
(1) fusing a positively charged polypeptide tag to the C-terminus of the protein (`\DHFRt'), and (2) allowing
it to bind to \gls{mtx}, a small negatively charged inhibitor
(\SI{454}{\dalton})~\cite{Soskine-Biesemans-2015}. Moreover, the dwell time of \DHFRt was found to depend
strongly and non-monotonously on the magnitude applied bias voltage~\cite{Biesemans-2015}. The physical
mechanisms behind this behavior are elucidated in~\cref{ch:trapping}, using a combination of experiments,
electrostatic energy calculations, and an analytical `double energy barrier' model. The latter captures the
essential physics governing the escape of \DHFRt{} from either the \cisi{} or \transi{} sides of \gls{clya},
and the fitting to an extensive set of experimental data will yield quantitative values for the magnitude of the
electrophoretic, electro-osmotic, electrostatic, and steric forces acting on proteins captured by \gls{clya}.

\objective{Mapping the transport properties of a biological nanopore}

In their inspiring 2005 publication, Aksimentiev \etal{} made of use \SI{\approx100}{\ns} of all-atom \gls{md}
simulations to map out the electrostatic potential, electro-osmotic flow, and ionic concentrations within the
\gls{ahl} nanopore~\cite{Aksimentiev-2005}. Even though the available computational power has increased
\num{\approx1000}-fold over the past 15~years, \gls{md} simulations remain prohibitively expensive compared to
continuum approaches, which are approximately 1000-fold faster at obtaining the same information. Moreover, to
a large extent, they have benefited from the same advances in computational power and algorithms. Hence,
in~\cref{ch:transport} we apply the \gls{epnp-ns} framework developed in~\cref{ch:epnpns} to a 2D-axisymmetric
model of \gls{clya} nanopore to show that continuum simulation can provide the same or more information as
\gls{md} simulations---at a fraction of the (computational) cost. By simulating the \gls{clya} over a wide
range of ionic strengths (\SIrange{0.005}{5}{\Molar}~\ce{NaCl}) and bias voltages
(\SIrange{-200}{+200}{\mV}), we will be able to gauge the accuracy of the
\gls{epnp-ns} equations for predicting nanoscale conductances. Furthermore, it will allow us to paint a
quantitative picture of nanopore properties that are difficult to access experimentally, including ion
selectivity, ion concentrations, (non)equilibrium electrostatic potentials, and the electro-osmotic flow.

\cleardoublepage

%

%
  \graphicspath{{\chaptersdir/electrostatics/image/}}%
\chapter{Equilibrium electrostatics of biological nanopores}
\label{ch:electrostatics}

\epigraphhead[\epipos]{%
\epigraph{%
  ``Available energy is the main object at stake in the struggle for existence and the evolution of the
  world.''
}{%
  \textit{`Ludwig Eduard Boltzmann'}
}}
\definecolor{shadecolor}{gray}{0.85}
\begin{shaded}
Parts of this chapter were adapted from:
\begin{itemize}
  \item K. Willems*, D. Rui\'{c}*, A. Biesemans*, N. S. Galenkamp, P. Van Dorpe and G. Maglia.
        \textit{ACS Nano} \textbf{13 (9)}, 9980--9992 (2019) 
  \item G. Huang, K. Willems, M. Soskine, C. Wloka and G. Maglia.
        \textit{Nat. Commun.} \textbf{8 (1)}, 1--9 (2017) 
  \item L. Franceschini,  T. Brouns, K. Willems, E. Carlon and G. Maglia.
        \textit{ACS Nano} \textbf{10 (9)}, 8394--8402 (2017) 
  \item G. Huang, K. Willems, M. Bartelds, P. Van Dorpe, M. Soskine and G. Maglia.
        \textit{Nano Lett.} \textbf{20 (50)}, 3819--3827 (2020) 
  \item M. Bayoumi, S. Nomidis, K. Willems, E. Carlon and G. Maglia.
        \textit{Nano Lett.}, Accepted (2020) 
\end{itemize}
*equal contributions
\newpage
\end{shaded}

In this chapter, the electrostatic properties of the \glsfirst{plyab}, \glsfirst{frac} and \glsfirst{clya}
nanopores are investigated \textit{in silico} using the \glsfirst{pbe}, corroborating results published in
previous studies. We demonstrate the extent to which their electrostatic potentials can be manipulated by
modification of specific amino acids of the pore on the one hand, and by altering the ionic strength and the
pH of the electrolyte on the other. Additionally, we use electrostatic energy computations to analyze the
translocation of \gls{ssdna} and \gls{dsdna} through respectively \gls{frac} and \gls{clya} nanopores. \\
The text and figures of this chapter represent entirely my own work. Several parts were adapted from
co-authored papers.
The section
``Construction of the PlyAB homology models''
was adapted from ref.~\cite{Huang-2020}.
The sections
``Computing the electrostatic potentials within nanopores''
and
``Computing nanopore-particle electrostatic interaction energies''
were adapted from ref.~\cite{Willems-Ruic-Biesemans-2019}.
The paragraph
``Partial protonation: partial charges at pH-values close to the $\pKa$.''
and the section
``A link between the electrostatic potential, ion-selectivity and electro-osmotic flow.''
were adapted from ref.~\cite{Huang-2017}.
The section
``All charges are important at physiological ionic strength.''
was adapted from ref.~\cite{Franceschini-2016}.
The section
``Electrostatic confinement of {dsDNA} within {ClyA}''
was adapted from ref.~\cite{Bayoumi-2020}.
%

%
%
%
%

%
\section{Introduction}
\label{sec:elec:intro}

As one of the four fundamental forces of nature, electromagnetism dominates the behavior and properties of
virtually all physical phenomena beyond the size of the atomic nucleus (aside from gravity). At the
macroscale, electromagnetism powers almost all of our technology, whereas at the nanoscale it powers life
itself: it is involved in all chemical phenomena, from the properties of salty solutions to intra- and
intermolecular interactions and chemical reactions. Here, our interest lies in the electrostatic component
of the electromagnetic force (\ie~the static interaction between fixed and mobile charges) and the
electrostatic potential and energies that result from it.

For a biological nanopore, aside from its geometry, the electrostatic interactions resulting from the fixed
charge distribution lining its interior walls dominate the transport properties for both ions and many larger
molecules. Direct interactions between the charges of the nanopore and the ions in the surrounding electrolyte
lead to observable phenomena such as ion selectivity (by attracting counter-ions and repelling co-ions),
ionic current rectification (in case of a conical geometry) and electro-osmotic flow (see~\cref{sec:np:eof}).
Indirect interactions (\ie~mediated through the external electric field) can result in voltage- or
pH-dependent gating. For example, the direct electrostatic interactions between ions and the precisely
positioned charges within the sub-nanometer aperture of the BK ion channel confers it not only with a
selectivity towards ions of the opposite charge, but even allows distinguishing between ions of the same
charge and valence. In larger pores, the predominant charges on the wall either promote or hinder the
transport of respectively opposing- and same-charge ions \textit{via} electrostatic interactions. When coupled
with an asymmetric geometry, charged walls typically result in a higher conductance at opposing bias voltages
of the same magnitude, a phenomenon called ionic current
rectification~\cite{White-2008,Lin-2015,Hsu-2017b,Valisko-2019}. Charges that are not directly exposed to the
electrolyte may also respond to changes in the electrostatic field by inducing conformational changes. These
typically result in either an opening or closure of the channel, as is the case in the \gls{kcsa} ion
channel~\cite{Kopec-2019}. This behavior is not limited to ion channels however and can even be engineered in
larger pores. The \gls{ahl} variant \gls{7r-ahl}---in which 49 ($7\times7$) positively charged residues were
added in the stem region---rapidly closes and opens under negative and positive bias voltages,
respectively~\cite{Maglia-2009}.

In this chapter, we have investigated the electrostatic properties of the several experimentally relevant
variants of the biological nanopores \glsfirst{plyab} (\SI{5.5}{\nm} diameter), \glsfirst{frac} (\SI{1.6}{\nm}
diameter), and \glsfirst{clya} (\SI{3.3}{\nm} diameter) by numerically solving the \glsfirst{pbe} for their
atomistic models. After describing our computational methods, we begin with an analysis of the electrostatic
potential distribution within the wild-type \gls{plyab}, \gls{plyab-e2} and \gls{plyab-r}, at
\SIlist{0.3;1.0}{\Molar} ionic strength. These two pores have oppositely charged constrictions and showed
different protein capture characteristics in experiments. These could be attributed to a significantly reduced
\gls{eof} in \gls{plyab-r}~\cite{Huang-2020} and are supported by our computed electrostatic potentials.

Next, we switch to the much smaller \gls{wtfrac} (wild-type) and \gls{refrac} pores, which also have oppositely
charged constrictions and were used in two separate studies. In the first~\cite{Huang-2017}, the opposing
response of the \glsfirst{eof} strength to lowering of the electrolyte pH from \numrange{7.5}{4.5} revealed
that the capture of peptides by \gls{frac} is driven primarily by the \gls{eof}. In the
second~\cite{Wloka-2016}, it was shown that, unlike its wild-type counterpart, \gls{refrac} was able to
translocate both \gls{ssdna} and \gls{dsdna}. We show that the pH-dependency of the \gls{eof} agrees with our
computations and that the energy barrier for \gls{ssdna} translocation is significantly lowered for
\gls{refrac}.

Lastly, we turn to \gls{clya}, for which we analyzed several variants of \gls{clya-as} at ionic strengths of
\SIlist{0.15;2.5}{\Molar}. These mutants showed a varying ability to translocate
\gls{dsdna}~\cite{Franceschini-2016}, a result corroborated by our electrostatic potential and \gls{dsdna}
translocation electrostatic energy calculations. Finally, we show that the strong repulsion between the
negatively charged constriction and the \gls{dna} confines the latter to the center of the
pore~\cite{Bayoumi-2020}.

\clearpage
\section{Computational methods}
\label{sec:elec:methods}

\begin{figure*}[b]
  \centering

  \includegraphics[scale=1]{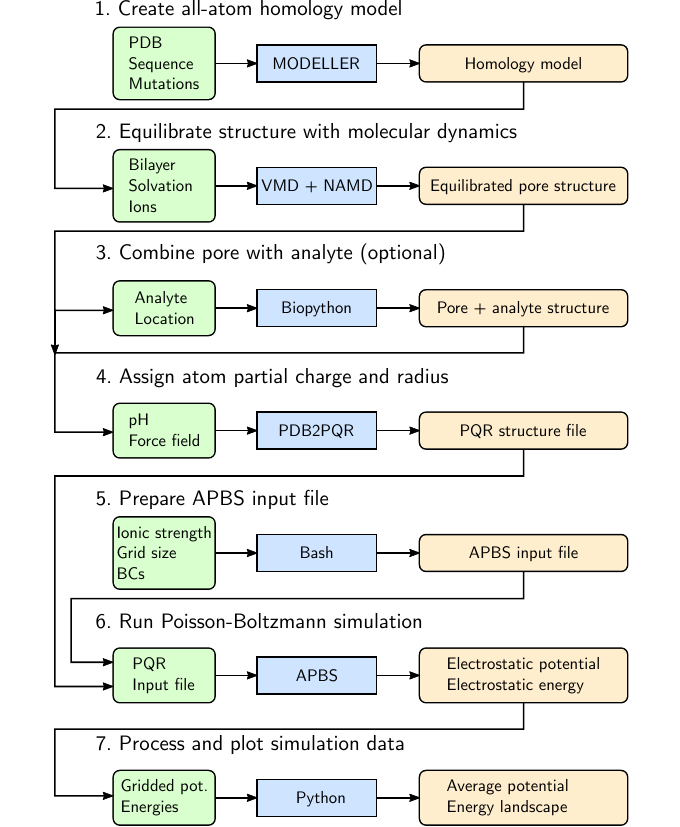}

\caption[Diagram of the electrostatic simulation methodology]{%
  \textbf{Diagram of the electrostatic simulation methodology.}
  Overview of the molecular modeling, electrostatic simulation, and data analysis steps performed within this
  chapter.
  }\label{fig:electrostatics_overview_diagram}
\end{figure*}

For this chapter, we have made use of a variety of simulation methods, existing software programs, and custom
analysis scripts. For reference, a schematic overview describing the general workflow can be found
in~\cref{fig:electrostatics_overview_diagram}.

\clearpage

\subsection{Molecular modeling}
\label{sec:elec:methods:molec}

\subsubsection{Construction of the PlyAB homology models.}
%

\paragraph{Note on the \gls{plyab}, \gls{plya} and \gls{plyb} structures.}
As described in \cref{sec:np:plyab}, the \Gls{plyab} nanopore consist of 39 individual protein chains (26
\gls{plya} and 13 \gls{plyb} chains), arranged in an \ce{(BA2)13} configuration with a 13-fold symmetry
(\cref{fig:nanopores_plyab}). Even though a \gls{cryo-em} structure of the full pore complex is available
(\pdbid{4V2T}~\cite{Lukoyanova-Kondos-2015}), it only contains the C\ta{} atoms of the entire protein
backbone. Fortunately, the same authors also released full-atom crystal structures of the \gls{plya}
(\pdbid{4OEB}~\cite{Lukoyanova-Kondos-2015}) and \gls{plyab} (\pdbid{4OEJ}~\cite{Lukoyanova-Kondos-2015})
monomers in their water-soluble conformations, which can be used, together with the \gls{cryo-em} map and the
C\ta{} atoms of the full pore, to create a full atom model of \gls{plyab} \textit{via} constrained homology
modeling and \gls{mdff}.

\paragraph{Construction of an artificial full atom `prepore' complex.}
First, an artificial `prepore' full atom structure was constructed by aligning 13 copies of the soluble
\gls{plyb} monomer and 26 copies of the \gls{plya} monomer to the C\ta{} atoms of the \gls{cryo-em} structure
(with the PyMOL \code{align} command~\cite{PyMOL}), resulting in \gls{rmsd} values of \SI{0.126}{\angstrom}
(\gls{plya} to subunit A1), \SI{0.081}{\angstrom} (\gls{plya} to subunit A2) and \SI{1.371}{\angstrom}
(\gls{plyb} to subunit B). Hence, the backbone conformations of the water-soluble and pore for both \gls{plya}
and \gls{plyb} match very closely, with the exception of the transmembrane region of \gls{plyb} (\ie~residues
\numrange{105}{171} and \numrange{229}{296}).

\paragraph{Homology modeling of the full-atom pore structure.}
The prepore complex is a better start for building a full-atom homology model than the \gls{cryo-em}
structure, given that the former contains realistic locations for all of the side-chains in the \gls{plya}
subunits, and for the bulk of the residues in \gls{plyb} protomers (\ie~everything except for the pore-forming
region). The full sequence of \gls{plyab}, together with the positional information given by the all the
C\ta{} atoms in the \gls{cryo-em} structure and that of the pre-pore, barring those residues involved in the
transmembrane region, were then used as input for MODELLER \code{automodel} algorithm~\cite{Sali-1993}.
MODELLER produced five full-atom homology models, of which the best one (\ie~the one with the lowest DOPE
score) was fitted to the PlyAB cryo-EM map using simulated annealing molecular dynamics optimization as
implemented in the Flex-EM~\cite{Topf-2008} extension of MODELLER. This entailed a \gls{md} simulated
annealing, constrained by the \gls{cryo-em} density map and a maximum shift of \SI{0.39}{\angstrom}, of 100
steps at \SIlist{150;250;500;1000}{\kelvin}, followed by 200 steps at \SIlist{800;500;250;150;50;0}{\kelvin}
and two conjugate gradients energy minimizations of 200 steps with and without density constraints. Further
refinement of the structure was carried out using symmetry constrained \gls{mdff} (implicit solvent) to the
\gls{cryo-em} density map with \gls{namd}~\cite{Phillips-2005,McGreevy-2014} for \SI{5}{\ns}, followed by
5~cycles of simulated annealing (\ie~heating from \SIrange{25}{350}{\kelvin} and cooling from
\SIrange{350}{300}{\kelvin} in steps of \SI{10}{ps} per \SI{25}{\kelvin}) and a \SI{5}{\ns} equilibration at
\SI{300}{\kelvin} with light harmonic constraints (force constant $k =
\mSI{0.1}{\kilo\cal\per\mole\per\square\angstrom}$) on the C\ta{} atoms.

\paragraph{Construction of the \gls{plyab-e2} and \gls{plyab-r} mutants.}
The homology models of \gls{plyab-e2} (\gls{plya}: C62S, C94S; \gls{plyb}: N26D, N107D, G218R, A328T, C441A,
A464V) and \gls{plyab-r} (\gls{plya}: C62S, C94S; \gls{plyb}: N26D, K255E, E260R, E270R, A328T, C441A, A464V)
were created from the homology model of the wild-type structure with a custom \gls{tcl} script and \gls{vmd}'s
\code{psfgen} plugin~\cite{Humphrey-1996}. Finally, the energy of each mutant was briefly minimized with an
\gls{md} run (implicit solvent, CHARMM36 force field~\cite{Best-2012}) of \SI{100}{\ps} using \gls{namd}, with
harmonic constraints on the C\ta{} atoms (force constant $k =
\mSI{1}{\kilo\cal\per\mole\per\square\angstrom}$).

\subsubsection{Construction of the FraC homology models.}

\paragraph{On the \gls{frac} crystal structure.}
Next to the protein chains, the crystal structure of \gls{frac} (\pdbid{4TSY}~\cite{Tanaka-2015}) also
contains several fragments of the lipid \gls{sph} that form a structural part of the pore itself
(\cref{sec:np:frac,fig:nanopores_frac}). Given that their charged headgroups are positioned outside of the
\lumen{} of the pore, we do not expect these lipids to contribute significantly to its electrostatics.
Therefore, we opted not to include these in our homology models and simulations.

\paragraph{Construction of the \gls{wtfrac} and \gls{refrac} mutants.}
Similarly to the \gls{plyab} mutants, the \gls{refrac} mutant (D10R, K159E) was constructed from the crystal
structure using \gls{vmd}'s \code{psfgen} plugin~\cite{Humphrey-1996}. \Gls{wtfrac} required no additional
mutations, but its missing atoms were also added with \code{psfgen}. Both structures were minimized with an
\gls{md} run (implicit solvent, CHARMM36 force field~\cite{Best-2012}) of \SI{100}{\ps} using \gls{namd}, with
harmonic constraints on the C\ta{} atoms (force constant $k=\mSI{1}{\kilo\cal\per\mole\per\square\angstrom}$).

\subsubsection{Construction of the ClyA homology models.}

\paragraph{On the \gls{clya} crystal structure.}
The dodecameric \textit{E. coli} \gls{clya} crystal structure (\pdbid{2WCD}~\cite{Mueller-2009}) contains all
residues, aside from short fragments at the N- (residues \numrange{1}{7}) and C-termini (residues
\numrange{293}{303}) of each chain. Those at the C-terminus are located outside of the \lumen{} of pore and
hence not expected to contribute significantly to the electrostatics. The missing residues at the N-terminus
are more important however, given their location at the \transi{} entry (\cref{fig:nanopores_clya}). Hence, we
opted to add a single negatively charged amino acid (Glu 7) in our models.

\paragraph{Construction of the \gls{clya-as} mutant.}
As with the previous pore, the wild-type crystal structure was processed with \code{psfgen} to add any missing
atoms and converted to \gls{clya-as} by introducing the relevant mutations (Q8K, N15S, Q38K, A57E, T67V,
C87A, A90V, A95S, L99Q, E103G, K118R, L119I, I124V, T125K, V136T, F166Y, K172R, V185I, K212N, K214R, S217T,
T224S, N227A, T244A, E276G, C285S, and K290Q). Next, a 200-by-\SI{200}{\angstrom} \gls{dphpc} lipid bilayer
patch was created with the CHARMM-GUI membrane builder~\cite{Jo-2008,Lee-2015}, and any lipid residues whose
headgroups were located either inside the pore \lumen{} or within the protein core were physically removed, followed by
a \SI{100}{\ps} steered \gls{md} run in \gls{namd}, using the mass-density of \gls{clya} as a virtual
repulsive force, to push any lipid tails out of the protein core. The resulting bilayer was then merged with
the \gls{clya-as} structure, solvated with 213364 {TIP3P} water molecules (\gls{vmd}'s \code{solvate} plugin)
and neutralized to \SI{0.15}{\Molar} with 674 \ce{Na+} and 602 \ce{Cl-} ions (\gls{vmd}'s \code{autoionize}
plugin). The resulting system was then equilibrated with \gls{md} (\gls{npt} ensemble with a Langevin dynamics
thermostat and a Nos\'{e}-Hoover Langevin barostat, {CHARMM36} force field parameters~\cite{Best-2012}) for
\SI{5}{\ns} at \SI{298.15}{\kelvin} while constraining the proteins C\ta{} atoms harmonically (force constant
$k = \mSI{1}{\kilo\cal\per\mole\per\square\angstrom}$).

\paragraph{Construction of the \gls{clya-r}, \gls{clya-rr}, \gls{clya-rr56} and \gls{clya-rr56k} mutants.}
The other \gls{clya} mutants were constructed from the \gls{clya-as} system described above by introducing
the relevant mutations with \code{psfgen}: \gls{clya-r} (S110R), \gls{clya-rr} (S110R/D64R), \gls{clya-rr56}
(S110R/Q56R) and \gls{clya-rr56k} (S110R/Q56R/Q8K). The resulting systems were equilibrated with \gls{md} for
\SI{1}{\ns} using the same conditions are those described above for \gls{clya-as} and the coordinates of the
pores isolated in individual files.

\subsection{Electrostatic modeling}
\label{sec:elec:methods:elec}

\subsubsection{Computing the electrostatic potentials within nanopores.}
%

\begin{figure*}[p]
  \centering
  \medskip
  \begin{minipage}[t]{50mm}
    \centering
    \begin{subfigure}[t]{50mm}
      \centering
      \caption{}\vspace{-5mm}\hspace{1.5mm}\label{fig:apbs_focussing_plyab_pqr}
      \includegraphics[scale=1]{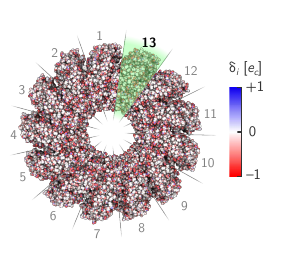}
    \end{subfigure}
    \\ \vspace{2mm}
    \centering
    \begin{subfigure}[t]{50mm}
      \centering
      \caption{}\vspace{-5mm}\hspace{1.5mm}\label{fig:apbs_focussing_frac_pqr}
      \includegraphics[scale=1]{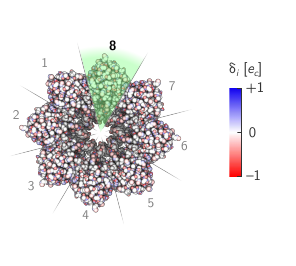}
    \end{subfigure}
    \\ \vspace{2mm}
    \centering
    \begin{subfigure}[t]{50mm}
      \centering
      \caption{}\vspace{-5mm}\hspace{1.5mm}\label{fig:apbs_focussing_clya_pqr}
      \includegraphics[scale=1]{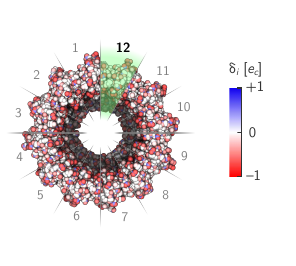}
    \end{subfigure}
  \end{minipage}
  \hspace{2mm}
  \begin{minipage}[t]{55mm}
    \centering
    \begin{subfigure}[t]{55mm}
      \centering
      \caption{}\vspace{-5mm}\hspace{1.5mm}\label{fig:apbs_focussing_plyab_apbs_setup}
      \includegraphics[scale=1]{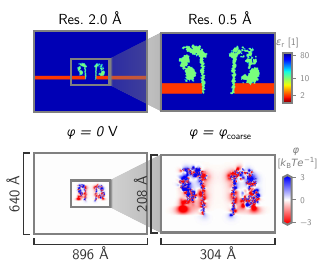}
    \end{subfigure}
    \\ \vspace{2mm}
    \begin{subfigure}[t]{55mm}
      \centering
      \caption{}\vspace{-5mm}\hspace{1.5mm}\label{fig:apbs_focussing_frac_apbs_setup}
      \includegraphics[scale=1]{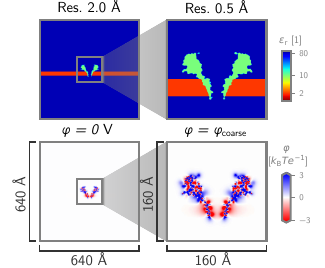}
    \end{subfigure}
    \\ \vspace{2mm}
    \begin{subfigure}[t]{55mm}
      \centering
      \caption{}\vspace{-5mm}\hspace{1.5mm}\label{fig:apbs_focussing_clya_apbs_setup}
      \includegraphics[scale=1]{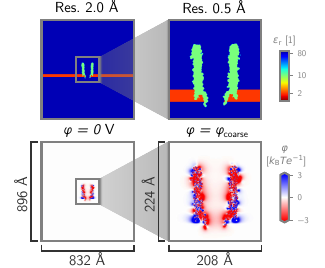}
    \end{subfigure}
  \end{minipage}

\caption[{APBS} simulation setup]{%
  \textbf{{APBS} simulation setup.}
  Top-side views of the {PQR} files of  
  (\subref{fig:apbs_focussing_plyab_pqr})
  \gls{plyab},
  (\subref{fig:apbs_focussing_frac_pqr})
  \gls{frac}, and
  (\subref{fig:apbs_focussing_clya_pqr})
  \gls{clya}. Molecules are represented as \gls{vdw} spheres and colored according to their partial charge
  $\partialcharge$. The area of a single subunit is highlighted with a green wedge.
  Axial cross-sections of the relative permittivity (top) and calculated electrostatic potential (bottom)
  grids for
  (\subref{fig:apbs_focussing_plyab_apbs_setup})
  \gls{plyab},
  (\subref{fig:apbs_focussing_frac_apbs_setup})
  \gls{frac}, and
  (\subref{fig:apbs_focussing_clya_apbs_setup})
  \gls{clya}. The \gls{apbs} focussing approach uses two sequential calculations, `coarse' and `fine', to
  obtain a realistic potential distribution.
  Molecular representations were rendered using \gls{vmd}~\cite{Humphrey-1996,Stone-1998}.
  }\label{fig:apbs_focussing}
\end{figure*}

\paragraph{The Poisson-Boltzmann equation.}
As described in \cref{sec:np:edl}, the electrostatic potential distribution of a protein dissolved in an
electrolyte medium can be estimated by solving the \glsfirst{pbe}~\cite{Baker-2001,Baker-2005}
\begin{equation}\label{eq:pbe}
  - \nabla \cdot \left[ \permittivity (\rpos) \nabla \potential (\rpos) \right] = \scd (\rpos)
  \text{ ,}
\end{equation}
with $\rpos$ the location vector, $\potential$ the electrostatic potential (normalized over the thermal
voltage \si{\kT\per\ec}), $\permittivity$ the local permittivity, and $\scd$ the charge density. For a
molecular system in an electrolyte, $\scd$ can be split into two parts
\begin{equation}\label{eq:electrostatics_scd}
  \scd (\rpos) = \scdfix(\rpos) + \scdmob(\rpos)
  \text{ ,}
\end{equation}
with $\scdfix$ the fixed charge density---resulting from the distribution of to the atomic partial
charges---and $\scdmob$ the mobile charge density---resulting from the distribution of charged ions in the
electrolyte. For $M$ atomic partial charges $\scdfix$ is given by~\cite{Baker-2001,Baker-2005}
\begin{equation}\label{eq:electrostatics_scdf}
  \scdfix (\rpos) = \dfrac{4\pi\ec^2}{\kbt} \sum_{i=1}^{M} \partialcharge_i \delta(\rpos - \vec{r_i})
\end{equation}
with $\ec$ the elementary charge, $\kbt$ the thermal energy, and $\delta$ the Dirac delta function.
$\partialcharge_i$ and $\vec{r_i}$ represent the atomic partial charge and the location of atom $i$,
respectively. The charge density due to $N$ mobile charge species can be expressed
as~\cite{Baker-2001,Baker-2005}
\begin{equation}\label{eq:electrostatics_scdm}
  \scdmob(\rpos) =
    \dfrac{4\pi\ec^2}{\kbt}
      \sum_{i=1}^{N} \ci^0 \chargeni
      \exp{\left[- \chargeni \potential(\rpos) - V_i(\rpos) \right]}
\end{equation}
with $\ci^0$ the bulk concentration, $\chargeni$ the charge number, and $V_i(\rpos)$ the steric potential of
ion species $i$. For a monovalent salt such as \ce{NaCl} this expression can be reduced to
\begin{equation}\label{eq:electrostatics_scdm_salt}
    \scdmob (\rpos) = - \ionaccess^{-2} (\rpos) \sinh \left[ \potential (\rpos) \right]
\end{equation}
with $\ionaccess$ a coefficient that includes both the ion accessibility and the bulk ion concentration.

\paragraph{Solving the \gls{pbe} with \gls{apbs} and \gls{pdb2pqr}.}
The \gls{pbe} solver \gls{apbs}~\cite{Jurrus-2018,Baker-2001} requires several inputs regarding the nanopore
(\ie~geometry, fixed charge distribution, relative permittivity) and the electrolyte medium (\ie~ion
concentration, ion size, and relative permittivity) and the computational system itself (\ie~grid size, grid
spacing, boundary conditions). These are provided by the user in two input files: (1) a configuration file
that contains all user-defined properties and (2) a ``PQR'' file that contains a coordinate, radius, and
partial charge for every atom in the system. The PQR format is similar to the PDB format and can be generated
from a PDB file using the \gls{pdb2pqr} software package~\cite{Jurrus-2018,Dolinsky-2004,Dolinsky-2007}. We
used the atom radius and partial charge parameters of the \gls{charmm36} force field~\cite{Huang-2013} for all
calculations except the \gls{frac} only pore, where we used the \gls{parse} force
field~\cite{Sitkoff-1994,Sitkoff-1996}. During conversion, \gls{pdb2pqr} assigns a protonation states to all
titratable residues at any user-specified pH value using its \gls{propka} submodule~\cite{Li-2005}.

To improve their accuracy, \gls{apbs} simulations consist of a initial `coarse' calculation (\ie~with grid
lengths \SI{>1}{\angstrom}) on a large grid size and with Dirichlet boundary conditions ($\potential = 0$ or
$\potential = \potential^{\code{mdh}}$). This is followed by a second `fine' calculation (\ie~with grid
lengths \SI{\approx0.5}{\angstrom}) on a smaller grid size that uses the potentials obtained in the coarse
simulation as a Dirichlet boundary condition ($\potential = \potential^{\mathrm{coarse}}$). The \code{mdh}
superscript refers to the multiple Debye-H\"{u}ckel approach, where approximate values are obtained
analytically with a Debye-H\"{u}ckel model for multiple, non-interacting spheres containing point
charges~\cite{Baker-2001,Baker-2005,Stone-2010}. The full list of simulation parameters can be found in
\cref{tab:pdb2pqr_apbs_parameters} and the simulation setups in \cref{fig:apbs_focussing}.

\afterpage{
\begin{landscape}
  \begin{threeparttable}[p]
    \footnotesize
    \centering

    \captionsetup{width=18cm}
    \caption[Summary of the {PDB2PQR} and {APBS} input parameters]%
            {Summary of the {PDB2PQR} and {APBS} input parameters.}
    \label{tab:pdb2pqr_apbs_parameters}

    \renewcommand{\arraystretch}{1.2}
    \scriptsize
  
    \begin{tabularx}{18cm}{%
        >{\hsize=1cm}l%
        >{\hsize=0.5cm}X%
        >{\hsize=0.5cm}X%
        >{\hsize=2.3cm}X%
        >{\hsize=2.3cm}X%
        >{\hsize=2.3cm}X%
        >{\hsize=2.3cm}X%
        >{\hsize=2.3cm}X%
      }
      \toprule
      \multicolumn{2}{c}{Parameter} & & \multicolumn{5}{c}{System} \\
      \cmidrule(r){1-2}\cmidrule(l){4-8}
      Name/Symbol & Unit & Dom.\tnote{a}
        & PlyAB & FraC & FraC\,+\,ssDNA & ClyA &  ClyA\,+\,dsDNA \\
      \midrule
      \multicolumn{8}{l}{\textbf{PDB2PQR}} \\[1mm]
      pH\tnote{b} & 1 
        & n.a.
        & 7.5 & 4.5, 7.5 & 7.5 & 7.5 & 7.5 \\
      Force-field & n.a. 
        & n.a.
        & CHARMM & PARSE & CHARMM & CHARMM & CHARMM \\
      \midrule
      \multicolumn{8}{l}{\textbf{APBS}} \\[1mm]
      $\temperature$ & \si{\kelvin}
        & G      
        & 298.15  & 298.15 & 298.15 & 298.15 & 298.15  \\
      $\permittivity_{r,\text{electrolyte}}$ & 1 
        & E
        & 80 & 80 & 78.15 & 80 & 78.15 \\
      $\permittivity_{r,\text{protein}}$ & 1
        & P
        & 10 & 10 & 10 & 10 & 10 \\
      $\permittivity_{r,\text{bilayer}}$ & 1
        & B
        & 2  & 2 & n.a. & 2 & n.a. \\
      $\ionsize_{\pm}$ & \si{\angstrom}
        & E
        & 4 & 4 & 4 & 4 & 4 \\
      $\chargen_{\pm}$ & 1
        & E
        & $\pm1$ & $\pm1$ & $\pm1$ & $\pm1$ & $\pm1$\\
      $\concentration_{\pm}$ & \si{\Molar}
        & E
        & 0.3, 1.0 & 1.0 & 0.15, 2.5 & 1.0 & 0.15, 2.0, 2.5 \\
      $g_{x}^{\mathrm{c}} \times g_{y}^{\mathrm{c}} \times g_{z}^{\mathrm{c}}$ & \si{\angstrom}
        & G
        & \apbsgrid{898}{898}{642}{} 
        & \apbsgrid{642}{642}{642}{} 
        & \apbsgrid{400}{400}{900}{} 
        & \apbsgrid{834}{837}{898}{} 
        & \apbsgrid{400}{400}{1100}{} \\ 
      $\Delta g_{x}^{\mathrm{c}} \times \Delta g_{y}^{\mathrm{c}} \times \Delta g_{z}^{\mathrm{c}}$ & \si{\angstrom}
        & G
        & \apbsgrid{2.00}{2.00}{2.00}{} 
        & \apbsgrid{2.00}{2.00}{2.00}{} 
        & \apbsgrid{1.25}{1.25}{0.74}{} 
        & \apbsgrid{2.00}{2.00}{2.00}{} 
        & \apbsgrid{1.38}{1.38}{0.82}{} \\ 
      $g_{x}^{\mathrm{f}} \times g_{y}^{\mathrm{f}} \times g_{z}^{\mathrm{f}}$ & \si{\angstrom}
        & G
        & \apbsgrid{225}{225}{209}{} 
        & \apbsgrid{161}{161}{161}{} 
        & \apbsgrid{150}{150}{600}{} 
        & \apbsgrid{209}{209}{225}{} 
        & \apbsgrid{150}{150}{700}{} \\ 
      $\Delta g_{x}^{\mathrm{f}} \times \Delta g_{y}^{\mathrm{f}} \times \Delta g_{z}^{\mathrm{f}}$ & \si{\angstrom}
        & G
        & \apbsgrid{0.50}{0.50}{0.50}{} 
        & \apbsgrid{0.50}{0.50}{0.50}{} 
        & \apbsgrid{0.47}{0.47}{0.49}{} 
        & \apbsgrid{0.50}{0.50}{0.50}{} 
        & \apbsgrid{0.52}{0.52}{0.52}{} \\ 
      $\potential_{\mathrm{boundary}}^{\mathrm{coarse}}$ & \si{\volt}
        & G
        & $\potential = 0$
        & $\potential = 0$
        & $\potential = \potential^{\code{mdh}}$
        & $\potential = 0$
        & $\potential = \potential^{\code{mdh}}$ \\
      $\potential_{\mathrm{boundary}}^{\mathrm{fine}}$ & \si{\volt}
        & G
        & $\potential = \potential^{\mathrm{coarse}}$
        & $\potential = \potential^{\mathrm{coarse}}$
        & $\potential = \potential^{\mathrm{coarse}}$
        & $\potential = \potential^{\mathrm{coarse}}$
        & $\potential = \potential^{\mathrm{coarse}}$ \\
      \midrule
      \multicolumn{8}{l}{\textbf{\code{draw\_membrane2}}} \\[1mm]
      $d_{\textrm{bilayer}}$ & \si{\angstrom}
        & B
        & 27 & 27 & 27 & n.a. & n.a. \\
      $r_{\textrm{bilayer}}^{\textrm{top}}$ & \si{\angstrom}
        & B
        & 43 & 23 & 37 & n.a. & n.a. \\
      $r_{\textrm{bilayer}}^{\textrm{bottom}}$ & \si{\angstrom}
        & B
        & 43 & 12 & 18 & n.a. & n.a. \\
      \bottomrule
    \end{tabularx}
  
    \begin{tablenotes}
     \item[a] Domains: global (G), electrolyte (E), protein/DNA (P), lipid bilayer (B);
     \item[b] pH values used by \gls{propka} during \gls{pdb2pqr} runtime.
    \end{tablenotes}
  
  \end{threeparttable}
\end{landscape}
}

\paragraph{Adding a lipid bilayer to APBS calculations.}
Being membrane proteins, active biological nanopores are embedded in a lipid bilayer and hence not surrounded
by electrolyte medium everywhere. From an electrostatic point of view, a lipid bilayer can be modeled as an
uncharged and ion-impermeable dielectric slab with a low permittivity: this is indeed very different from an
ion-containing electrolyte medium with high permittivity. Hence, it is not surprising that the presence of a
lipid bilayer influences the distribution of the electrostatic potential~\cite{Homeyer-2015}. Even though
\Gls{apbs} does not natively support the inclusion of a lipid bilayer, the software package does include an
external program (\code{draw\_membrane2}) that allows adding one to the permittivity, ion-accessibility, and
charge maps produced by \gls{apbs} in the form of a dielectric slab with a conical hole at its center. The
relevant parameters (thickness $d_{\textrm{bilayer}}$, top cone radius $r_{\textrm{bilayer}}^{\textrm{top}}$
and bottom cone radius $r_{\textrm{bilayer}}^{\textrm{bottom}}$) are listed in
\cref{tab:pdb2pqr_apbs_parameters}. The use of \code{draw\_membrane2} is a very I/O intensive task, as it
requires \gls{apbs} to write out a set of 5 3D grids (typically \SI{>2}{\gibi\byte} each), for every
individual simulation (\ie~both the coarse and the fine runs) to disk. Next, they are loaded into memory by
\code{draw\_membrane2}, which adds in the bilayer and finally writes them out to the disk again, where they
can be accessed by \gls{apbs}. When running many simulations in parallel, this entire process requires both
large amounts of disk space and very high read/write speeds. Particularly the latter results in a significant
increase of the simulation time: from \SIrange{20}{200}{\min}. Even though the presence of a bilayer has an
impact on the electrostatic energy~\cite{Bonthuis-2006}, its inclusion in \gls{apbs} comes at a significant
computational cost and its absence does not influence the qualitative outcome of our energy comparisons.
Hence, this approach was only used when computing the electrostatic potential inside the pore-only systems
(\num{\pm20} runs), and the bilayer was omitted entirely for all energy-based calculations of particle
translocation (\num{>500} runs).

\paragraph{Partial protonation: partial charges at pH-values close to the p$\boldsymbol{K}_{\mathbf{a}}$.}
%
%
When the pH of the solution is close to the $\pKa$ of a titratable residue (\ie~within \num{\pm2}~pH units),
the residue will, on average, be neither fully protonated nor deprotonated. The protonated fraction of an acid
($\protfrac_{\mathrm{HA}}$) is given by 
\begin{align}\label{eq:pka_protonation}
  \protfrac_{\mathrm{HA}} = \left[ 1 + 10^{\mathrm{pH} - \pKa} \right]^{-1}
  \text{ .}
\end{align}
Hence, when $\mathrm{pH} = \pKa$, the residue will be protonated \SI{50}{\percent} of the time and
deprotonated for other \SI{50}{\percent}. This also means that, on average, the partial charge of the residue
($\partialcharge$) lies somewhere between \SIrange{-1}{+1}{\ec}:
\begin{align}\label{eq:pka_partial_charge}
  \partialcharge = \partialcharge_{\mathrm{HA}} \protfrac_{\mathrm{HA}}
                    + \partialcharge_{\mathrm{A}^{-}} \left( 1 - \protfrac_{\mathrm{HA}} \right)
  \text{ ,}
\end{align}
with $\partialcharge_{\mathrm{HA}}$ and $\partialcharge_{\mathrm{A}^{-}}$ the partial charge of the fully
protonated and deprotonated residue, respectively. \gls{pdb2pqr} does not take into account partial
(de)protonation, and assigns partial charges in a discrete, all-or-nothing manner. This mean that acid
residues (ASP, GLU, TYR, C-terminus) have a charge of \SI{-1}{\ec} if $\mathrm{pH} > \pKa$, and \SI{0}{\ec}
otherwise. Likewise, basic residues (ARG, LYS, HIS, N-terminus) have a charge of \SI{+1}{\ec} if $\mathrm{pH}
< \pKa$, and \SI{0}{\ec} otherwise. Because the side-chain $\pKa$'s of (free) amino acids are either below 6
or above 9, this is a reasonable approximation for simulations at pH values close to 7. However, it does not
adequately represent the equilibrium situation for lower (or higher) pH values. For example, \pH{4.5} is very
close to the $\pKa$ values of glutamate (\num{4.4}) and aspartate (4.0) residues, and most of these residues
would hence be fractionally charged. In addition, the proximity of many charged residues tends to shift their
effective $\pKa$'s up or down, depending on whether the de(protonated) state is energetically favorable or
not, which may even cause residues to be partially (de)protonated at neutral pH.

To calculate the correct equilibrium electrostatic potentials of biological nanopores at lower pH values, we
extended the Python code of \gls{pdb2pqr} (`\code{fractional-PDB2PQR}') with the ability to assign partial
protonation states. The adapted program uses as input (1) a fully protonated PQR molecule (obtained by running
\gls{pdb2pqr} at \pH{1}), (2) the per-residue $\pKa$ values (calculated by either
\gls{propka}~\cite{Olsson-2011} or \gls{delphipka}~\cite{Wang-2015}), (3) a user-defined pH value and (4) the
force-field of choice. This approach was used to compute the electrostatics of the \gls{frac} variants at
\pH{4.5} and \num{7.5}~\cite{Huang-2017}.

\subsubsection{Computing nanopore-particle electrostatic interaction energies.}
\label{sec:elec:methods:elec:energy}
%

%
\begin{figure*}[b]
  \centering
  \medskip
  \includegraphics[scale=1]{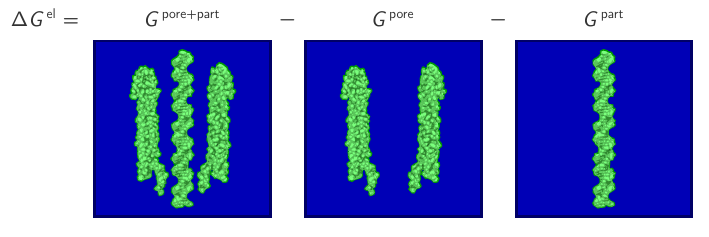}

\caption[Computing net electrostatic energies with {APBS}]{%
  \textbf{Computing net electrostatic energies with {APBS}.}
  The net electrostatic energy difference $\energyelec$ for placing a particle, illustrated here as a
  \gls{dsdna} molecule, inside a nanopore can be computed from the energy of the pore+particle system
  $\gibbs^{\rm{pore+part}}$ (left) by subtracting the energies of the empty pore $\gibbs^{\rm{pore}}$
  (middle) and particle only $\gibbs^{\rm{part}}$ (right) systems.
  }\label{fig:apbs_energy_scheme}
\end{figure*}

When using the \gls{pbe} to evaluate the electrostatics of a system, its electrostatic energy is given
by~\cite{Baker-2005}
\begin{equation}\label{eq:pbe_energy}
  \gibbs (\potential, \permittivity) = \int_\Omega \left[
    \scdfix(\rpos) \potential
    - \dfrac{\permittivity}{2}\left( \nabla \potential \right)^2
    - \ionaccess^{-2} \left( \cosh \potential - 1 \right)
  \right] \,dr
  \text{ ,}
\end{equation}
and hence depends not only on the electrostatic potential of the system (\textit{via} $\potential$), but also
on the choice of relative permittivities (\textit{via} $\permittivity$) and ionic properties (\textit{via}
$\ionaccess$). \Cref{eq:pbe_energy} can be evaluated directly by \gls{apbs} by including the \code{calcenergy}
keyword in the input file, which enables the rapid evaluation of the electrostatic energies of complex
biomolecular systems. For example, the free electrostatic energy change caused by placing another molecule
(\ie~`particle') inside a nanopore ($\energyelec$) can be computed using~\cite{Homeyer-2015}
\begin{equation}\label{eq:electrostatic_energy}
  \energyelec = \gibbs^{\rm{pore+part}} - \gibbs^{\rm{pore}} - \gibbs^{\rm{part}}
\end{equation}
with $\gibbs^{\rm{pore+part}}$, $\gibbs^{\rm{pore}}$ and $\gibbs^{\rm{part}}$ the electrostatic free energies
of systems containing respectively both the nanopore and the particle molecule, the empty nanopore and the
particle molecule alone. A sketch of these different systems is given in \cref{fig:apbs_energy_scheme}. This
approach was used to map out the electrostatic energy landscape of \gls{ssdna} and \gls{dsdna} translocation
through respectively the \gls{frac} and \gls{clya} nanopores in this chapter, and of a tagged \gls{dhfr}
molecule through \gls{clya} in \cref{ch:trapping}.

\clearpage

\section{Results and discussion}
\label{sec:elec:results}

\subsection{Electrostatics of the PlyAB nanopore}
\label{sec:elec:plyab}

\subsubsection{Size matters: large proteins require large nanopores.}

\begin{figure*}[p]
  \centering
  \medskip

  \begin{subfigure}[t]{120mm}
    \centering
    \caption{}\vspace{-5mm}\hspace{1.5mm}\label{fig:plyab_mutants_residues}
    \includegraphics[scale=1]{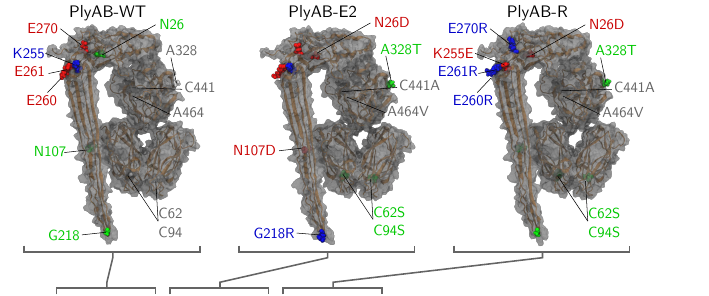}
  \end{subfigure}
  \\ \vspace{-2mm}
  \begin{subfigure}[t]{120mm}
    \centering
    \caption{}\vspace{-2.5mm}\hspace{1.5mm}\label{fig:plyab_mutants_potential_0300}
    \includegraphics[scale=1]{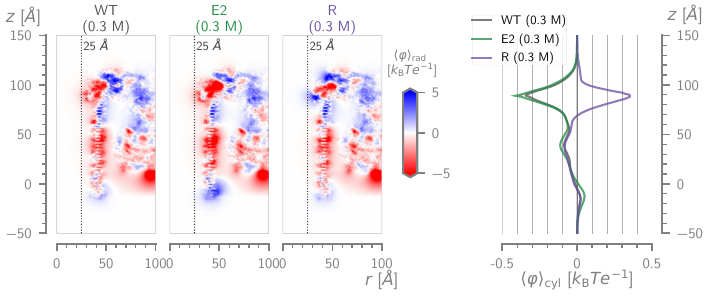}
  \end{subfigure}
  \\ \vspace{0mm}
  \begin{subfigure}[t]{120mm}
    \centering
    \caption{}\vspace{-2.5mm}\hspace{1.5mm}\label{fig:plyab_mutants_potential_1000}
    \includegraphics[scale=1]{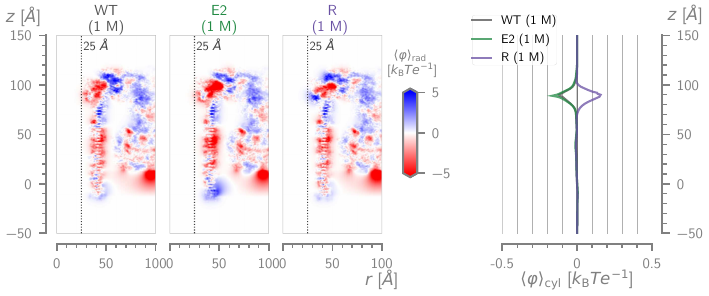}
  \end{subfigure}

\caption[Electrostatic potential inside {PlyAB} mutants]{%
  \textbf{Electrostatic potential inside {PlyAB} mutants.}
  (\subref{fig:plyab_mutants_residues})
  Cartoon representation of a single subunit of the \gls{plyab-wt} (left), \gls{plyab-e2} (middle), and
  \gls{plyab-r} (right), which were investigated for their ability to capture and analyze large protein in
  Ref.~\cite{Huang-2020}. The labels indicate the mutations w.r.t. the wild-type \gls{plyab}.
  Radially averaged cross-sections ($\radpot$, left) and \SI{25}{\angstrom}-radius cylindrically averaged line
  plots ($\cylpot$, right) of the electrostatic potentials at
  (\subref{fig:plyab_mutants_potential_0300})
  \SI{0.3}{\Molar}
  (\subref{fig:plyab_mutants_potential_1000})
  \SI{1.0}{\Molar}.
  Computations were performed using \gls{apbs}~\cite{Baker-2001,Baker-2005}.
  Molecular representations were rendered using \gls{vmd}~\cite{Humphrey-1996,Stone-1998}.
  }\label{fig:plyab_mutants_potential}
\end{figure*}

To analyze folded proteins with biological nanopores, they must interact in a characteristic manner with the
pore for a time period that is sufficiently long to be properly sampled~\cite{Willems-VanMeervelt-2017}. For
small nanopores such as \gls{ahl} or \gls{mspa}, this can be achieved by tethering the protein in close
proximity to the entry of pore~\cite{Movileanu-2000,Fahie-2015,Ho-2015,Laszlo-2016,Thakur-2019}. Larger
nanopores, such as \gls{clya}, allow for the immobilization of the entire protein within the \lumen{} of the
nanopore
itself~\cite{Soskine-2012,Soskine-2013,Soskine-Biesemans-2015,Biesemans-2015,Wloka-2016,VanMeervelt-2017,Galenkamp-2018,Galenkamp-2020},
which typically requires no modifications to the analyte proteins, even though better signals can sometimes be
obtained with engineered versions~\cite{Soskine-Biesemans-2015,Galenkamp-2020}. The latter approach does
mandate that the protein physically fits inside the pore, which puts a strict upper size limit. Given that
many eukaryotic proteins (average molecular weight of \SI{\approx50}{\kDa}~\cite{Kozlowski-2016}) exceed
\SI{5}{\nm} in diameter, only \gls{clya} has been used successfully so far. Compounding the lack of suitable
nanopores is that, for more than two decades, nanopore research was focused primarily on \gls{dna} analysis,
resulting in a wide selection of biological nanopores with narrow diameters (\SIrange{\approx1}{2}{\nm})
entries. With a \lumen{} of \SI{\approx7.2}{\nm} in diameter, the \gls{plyab} nanopore (see
\cref{sec:np:plyab,fig:nanopores_plyab}) is the ideal candidate for analyzing larger globular proteins. Huang
\etal. employed directed evolution to improve the electrophysiological stability of the wild-type pore,
resulting in the variants \gls{plyab-e2} (\gls{plya}: C62S, C94S; \gls{plyb}: N26D, N107D, G218R, A328T,
C441A, A464V; \cref{fig:plyab_mutants_residues}, middle) and \gls{plyab-r} (\gls{plya}: C62S, C94S; \gls{plyb}
N26D, K255E, E260R, E270R, A328T, C441A, A464V; \cref{fig:plyab_mutants_residues}, right)~\cite{Huang-2020}.
Notably, \gls{plyab-r} showed significantly altered behavior when capturing negatively charged proteins. In
\gls{plyab-e2}, blockades are only observed for the relatively small (\SI{24}{\kDa}) \tb-casein when it was
added at the \transi{} side (at positive bias voltages), and not for the larger (\SI{66.5}{\kDa}) \gls{bsa}.
\gls{plyab-r}, on the other hand, exhibited blockades for both \tb-casein and \gls{bsa}, and from either the
\cisi{} (at positive bias voltages) or \transi{} (at positive bias voltages) side. In the following section, we
will describe the electrostatics of these \gls{plyab} variants, discuss the implications of their differences,
and finally link them to the experimentally observed differences.

\subsubsection{Charge reversals induce potential reversals.}

The interior of \gls{plyab-wt} is dominated by negatively charged residues along the entire length of the
pore, with the exception of a small positive patch at the \transi{} entry (\cref{fig:plyab_mutants_residues},
left). The mutations of \gls{plyab-e2} (\cref{fig:plyab_mutants_residues}, middle) introduce only one
additional negative charge in the center of the \transi{} \lumen{} (N107D) and a single positive charge at the
\transi{} entry (G218R). \Gls{plyab-r} (\cref{fig:plyab_mutants_residues}, right) hosts three
negative-to-positive (E260R, E261R, E270R) and one positive-to-negative (K255E) charge reversals inside the
\cisi{} constriction. At an ionic strength of \SI{0.3}{\Molar}, the effect of these mutations is reflected
clearly in the electrostatic potential distributions within these pores
(\cref{fig:plyab_mutants_potential_0300}). Whereas the radially averaged potential, $\radpot$, represents the
average potential within a 2D-cross-section of the pore (\cref{fig:plyab_mutants_potential_0300}, left), the
cylindrically averaged potential, $\cylpot$, is the average potential along the length of the pore and within
the radius of the narrowest part of the channel, set here to \SI{25}{\angstrom}
(\cref{fig:plyab_mutants_potential_0300}, right). From an electrostatic point-of-view, \gls{plyab-e2} is
virtually identical to the wild-type, with the exception of a slightly more negative constriction (peak
$\cylpot$ changes from \SIrange{-0.35}{-0.4}{\kTe} at $z = \mSI{90}{\angstrom}$) and a lightly more positive
\transi{} entry (peak $\cylpot$ changes from \SIrange{+0.25}{+0.5}{\kTe} at $z = \mSI{-15}{\angstrom}$) due to
the N26D and G218R mutations, respectively. The many charge reversals in \gls{plyab-r}, on the other hand,
also reversed the potential inside the constriction to be strongly positive (peak $\cylpot$ changes from
\SIrange{-0.35}{+0.35}{\kTe} at $z = \mSI{90}{\angstrom}$). At higher ionic strengths (\SI{1}{\Molar},
\cref{fig:plyab_mutants_potential_1000}), the additional screening diminishes the magnitude of the peak
$\cylpot$ values in the constriction, and virtually abolishes any differences in the \transi{} \lumen{} by
dropping $\cylpot$ close to zero.

\subsubsection{An electro-osmotic tug of war.}

In a (large) nanopore, the electro-osmotic flow is caused by force exerted on the \gls{edl} by the external
electric field, which results in a unidirectional movement of the \gls{edl} that drags on the fluid in the
rest of the pore~\cite{Qiao-Aluru-2003,Mao-2014,Tagliazucchi-2015,Bonome-2017}. The electrostatic potential
along the walls of the wild-type and E2 variants of \gls{plyab} is predominantly negative, which results in a
uniform, positively charged \gls{edl}. Hence, these pores are expected to generate a significant
electro-osmotic flow from \cisi{} to \transi{} at negative voltages (and \textit{vice versa} at positive
voltages). In \gls{plyab-r}, the negative potential inside the \transi{} \lumen{} is reversed abruptly to a
high positive value at the \cisi{} constriction. Indeed, the \gls{edl} inside the constriction of
\gls{plyab-r} is negatively charged and will exert the opposite force on the fluid as the \gls{edl} in the
\transi{} \lumen{}. The resulting direction of the electro-osmotic flow will depend on whether the `tug of
war' game is won by the \cisi{} or \transi{} side of the pore. Regardless of the final direction of the
\gls{eof} in \gls{plyab-r}, this competition will significantly reduce the magnitude of the \gls{eof} compared
to \gls{plyab-e2}, which may cause the electrophoretic force to trump the electro-osmotic competition, giving
rise to the capture behavior observed experimentally~\cite{Huang-2020}.


\subsection{The {pH}-dependent electrostatics of the FraC nanopore}
\label{sec:elec:frac}

\subsubsection{{WtFraC} and {ReFraC}: minor mutations with a major impact.}
\begin{figure*}[t]
  \centering
  \medskip

  \begin{subfigure}[t]{65mm}
    \centering
    \caption{}\vspace{-5mm}\hspace{1.5mm}\label{fig:frac_mutants_pka_residues}
    \includegraphics[scale=1]{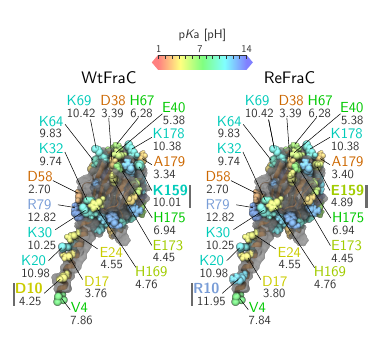}
  \end{subfigure}
  \begin{subfigure}[t]{50mm}
    \centering
    \caption{}\vspace{-2.5mm}\hspace{1.5mm}\label{fig:frac_mutants_pka_scatter}
    \includegraphics[scale=1]{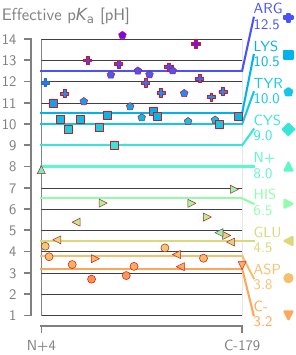}
  \end{subfigure}

\caption[Effective $\pKa$ values of {WtFraC} and {ReFraC}]{%
  \textbf{Effective p$\boldsymbol{K}_{\mathbf{a}}$ values of {WtFraC} and {ReFraC}.}
  (\subref{fig:frac_mutants_pka_residues})
  Molecular representation the most salient titratable residues of \gls{wtfrac} (left) and \gls{refrac}
  (D10R/K159E, right), colored according to their subunit-averaged $\pKa$ values, as computed by
  \gls{propka}~\cite{Olsson-2011,Sondergaard-2011}.
  (\subref{fig:frac_mutants_pka_scatter})
  Scatter plot of the chain-averaged $\pKa$ values of all titratable residues of \gls{wtfrac} (red outlines)
  and \gls{refrac} (blue outlines), as a function of their residue number. Because most $\pKa$ values are
  identical between these two variants, only residues 10 and 159 were plotted for \gls{refrac}. The solid
  lines indicate the baseline $\pKa$ values of the given amino acids used by
  \gls{propka}~\cite{Olsson-2011,Sondergaard-2011}.
  Molecular structures were rendered using \gls{vmd}~\cite{Humphrey-1996,Stone-1998}.
  }\label{fig:frac_mutants_pka}
\end{figure*}

As with other nanopores, the narrowest location in \gls{frac}'s hydrophilic channel is expected to dominate
its nanofluidic properties. Hence, it is no surprise that \gls{refrac} (D10R/K159E,
\cref{fig:frac_mutants_pka_residues})~\cite{Wloka-2016}, which contains a negative-to-positive charge reversal
mutation at the bottom of its constriction, showed a much lower conductance (cf.~$\conductance =
\mSI{1.91(17)}{\nS}$ (\gls{wtfrac}) and $\conductance = \mSI{1.19(12)}{\nS}$ (\gls{refrac}), at \SI{+50}{mV}
and \SI{1}{\Molar} \ce{NaCl}~\cite{Wloka-2016}) and a reversed ion selectivity
(cf.~$\permeabilityratio{\ce{K+}}{\ce{Cl-}} = 3.6$ (\gls{wtfrac}), $\permeabilityratio{\ce{K+}}{\ce{Cl-}} =
0.57$ (\gls{refrac}), for \SI{0.467}{\Molar} \ce{KCl} (\cisi) $\|$ \SI{1.960}{\Molar} \ce{KCl} (\transi) at
\pH{7.5}~\cite{Huang-2017}). Contrary to its wild-type counterpart, \gls{refrac} was able to induce \gls{dna}
translocation, allowing it to differentiate between homopolymeric stretches of A, C, and T
polynucleotides~\cite{Wloka-2016}. Finally, in a subsequent study, it was shown that \gls{wtfrac}, but not
\gls{refrac}, exhibited different peptide capture dynamics at low pH values~\cite{Huang-2017}. In this section
we aim to rationalize these findings, using electrostatic simulations as a guide. First, the difference in
electrostatic potential between these two variants is investigated when changing the electrolyte pH from
\numrange{7.5}{4.5}. Second, we look into their ability to translocate \gls{dna} by computing the
electrostatic energy landscape of a translocating piece of \gls{ssdna}. For reference, a detailed discussion
of the structure and general properties of \gls{frac} can be found in \cref{sec:np:frac}.

\subsubsection{{WtFraC} and {ReFraC} titrate differently at low pH.}

Before proceeding to the electrostatic simulations at \pHlist{7.5;4.5}, it is instructive to analyze the
effective $\pKa$ values of all titratable residues in \gls{wtfrac} and \gls{refrac}
(\cref{fig:frac_mutants_pka}). These values can be calculated self-consistently by software packages such as
\gls{propka}~\cite{Olsson-2011,Sondergaard-2011} or \gls{delphipka}~\cite{Wang-2015}. Of all the titratable
side-chains that face the interior channel wall, only specific residues will be influenced if the pH is
lowered to \num{4.5} (\cref{fig:frac_mutants_pka_residues}). For \gls{wtfrac} these comprise D10 ($\pKa =
4.25$), D17 ($\pKa = 3.76$), E24 ($\pKa = 4.55$), E40 ($\pKa = 5.38$), H67 ($\pKa = 6.28$), H169 ($\pKa =
4.76$), E173 ($\pKa = 4.45$) and H175 ($\pKa = 6.94$); for \gls{refrac} they consist of D17 ($\pKa = 3.80$),
E24 ($\pKa = 4.55$), E40 ($\pKa = 5.38$), H67 ($\pKa = 6.28$), H169 ($\pKa = 4.76$), E173 ($\pKa = 4.45$) and
H175 ($\pKa = 6.94$). Note that these lists are virtually identical, except for the absence of residue 10 in
the case of \gls{refrac}, since it was mutated to an arginine residue ($\pKa = 11.95$). This suggests that the
electrostatic influence of residue 10 will remain unchanged for \gls{refrac} at \pH{4.5}
($\protfrac_{\mathrm{HA}} = 1.00$), whereas the D10 in \gls{wtfrac} will be in a protonated state for a
significantly part of the time ($\protfrac_{\mathrm{HA}} = 0.36$). Inside the confined space of a nanopore
\lumen{}, the close proximity of the like-charged and oppositely-charged residues respectively destabilizes
and stabilizes their (de)protonated states~\cite{Olsson-2011}. This can be observed as significant $\pKa$
shifts relative to the typical values found inside proteins ($\Delta \pKa = \pKa^{\textrm{effective}} -
\pKa^{\textrm{baseline}}$). In \cref{fig:frac_mutants_pka_scatter}, we plotted the chain-averaged $\pKa$
values of all titratable residues in \gls{wtfrac} and \gls{refrac} together with their conventional baseline
values, as used by \gls{propka}. These values are also summarized \cref{tab:frac_pka_summary}. The effective
$\pKa$ of most charged residues---acidic and basic---are lower compared to their baseline values. This
suggests that the deprotonated states of the \num{\approx10} acidic residues (Glu + Asp) are stabilized, at
the cost of the destabilization of their \num{\approx20} basic counterparts (Lys + Arg). Given this 2:1 ratio,
this is a logical result.

\begin{table}[t]
  \centering
\begin{threeparttable}
  \footnotesize
  \centering

  \caption{Summary of the WtFraC's and ReFraC's $\pKa$ values.}
  \label{tab:frac_pka_summary}

  \renewcommand{\arraystretch}{1.2}
  \scriptsize

  \begin{tabularx}{9.5cm}{lSSSSSSS}
    \toprule
    \multicolumn{2}{c}{Residue}
    & \multicolumn{2}{c}{$\pKa^{\textrm{effective}}$}
    & \multicolumn{2}{c}{$\Delta \pKa$\tnote{*}}
    & \multicolumn{2}{c}{Count} \\
    \cmidrule(r){1-2} \cmidrule(lr){3-4} \cmidrule(lr){5-6} \cmidrule(lr){7-8}
    {Name} & {$\pKa^{\text{baseline}}$} & {Wt} & {Re} & {Wt} & {Re} & {Wt} & {Re} \\
    \midrule
    C-  & 3.20  & 3.34  & 3.40  & +0.14 & +0.20 & 1   & 1   \\
    ASP & 3.80  & 3.52  & 3.42  & -0.28 & -0.38 & 8   & 7   \\
    GLU & 4.50  & 4.31  & 4.40  & -0.19 & -0.10 & 5   & 6   \\
    HIS & 6.50  & 5.59  & 5.60  & -0.91 & -0.90 & 6   & 6   \\
    N+  & 8.00  & 7.86  & 7.84  & -0.14 & -0.16 & 1   & 1   \\
    TYR & 10.00 & 11.61 & 11.61 & +1.61 & +1.61 & 11  & 11  \\
    LYS & 10.50 & 10.16 & 10.18 & -0.34 & -0.32 & 11  & 10  \\
    ARG & 12.50 & 12.21 & 12.19 & -0.29 & -0.31 & 10  & 11  \\
    \bottomrule
  \end{tabularx}

  \begin{tablenotes}
    \item[*] $\pKa$-shift $\Delta \pKa = \pKa^{\textrm{effective}} - \pKa^{\textrm{baseline}}$.
  \end{tablenotes}

\end{threeparttable}
\end{table}
\begin{figure*}[b]
  \centering
  \medskip
  \begin{subfigure}[t]{120mm}
    \centering
    \caption{}\vspace{-2.5mm}\hspace{1.5mm}\label{fig:frac_mutants_partialcharge_residues}
    \includegraphics[scale=1]{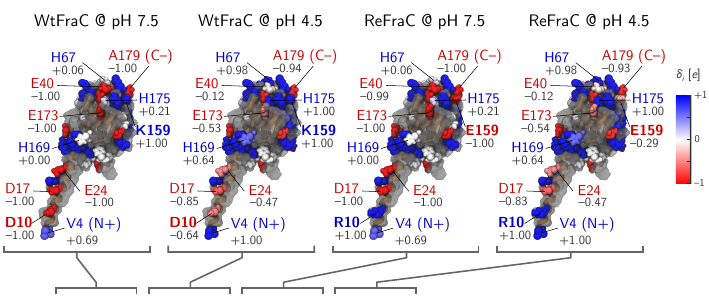}
  \end{subfigure}
  \\ \vspace{-2mm}
  \begin{subfigure}[t]{120mm}
    \centering
    \caption{}\vspace{-2.5mm}\hspace{1.5mm}\label{fig:frac_mutants_potential_1000}
    \includegraphics[scale=1]{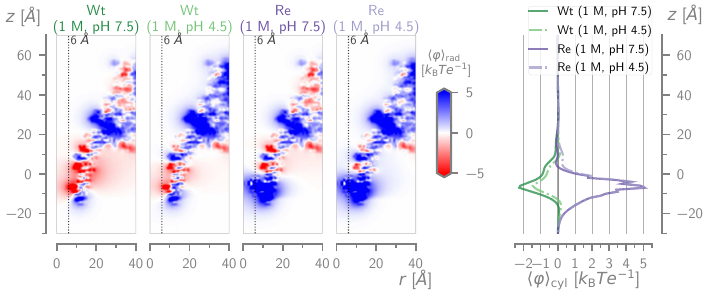}
  \end{subfigure}

\caption[Electrostatic potential distribution inside {WtFraC} and {ReFraC}]{%
  \textbf{Electrostatic potential distribution inside {WtFraC} and {ReFraC}.}
  (\subref{fig:frac_mutants_partialcharge_residues})
  Molecular models of the partial charges of the most important titratable residues of \gls{wtfrac} and
  \gls{refrac} at \pHlist{7.5;4.5}, as computed with \cref{eq:pka_partial_charge}.
  (\subref{fig:frac_mutants_potential_1000})
  Radially averaged cross-sections ($\radpot$, left) and \SI{6}{\angstrom}-radius cylindrically averaged line
  plots ($\cylpot$, right) of the electrostatic potentials at at \SI{1}{\Molar} ionic strength and for
  \pHlist{7.5;4.5}, computed from the fractionally charged PQR files with \gls{apbs}.
  Computations were performed using \gls{apbs}~\cite{Baker-2001,Baker-2005}.
  Molecular representations were rendered with \gls{vmd}~\cite{Humphrey-1996,Stone-1998}.
  }\label{fig:frac_mutants_potential}
\end{figure*}

\subsubsection{Partial charges of WtFraC and ReFraC at pH 4.5 and 7.5.}

Whereas most residues have a unitary charge at \pH{7.5}, this is no longer the case at \pH{4.5}. As discussed
in the methods of this chapter (\cref{sec:elec:methods:elec}) and corroborated by the effective $\pKa$
analysis in the previous paragraph, most of \gls{frac}'s acidic residues will not be fully (de)protonated
at \pH{4.5}. Translated to electrostatic simulations, this means that at \pH{4.5}, these residues will carry,
on average, a charge between \SIrange{-1}{0}{\ec} (\cref{fig:frac_mutants_partialcharge_residues}), depending
on their exact $\pKa$ value (\cref{fig:frac_mutants_pka,tab:frac_pka_summary}). For \gls{wtfrac}, the most
important negatively charged residues are D10 (\SIrange{-1}{-0.64}{\ec}), D17 (\SIrange{-1}{-0.85}{\ec}), E24
(\SIrange{-1}{-0.47}{\ec}), E40 (\SIrange{-1}{-0.12}{\ec}), and E173 (\SIrange{-1}{-0.53}{\ec}). In the case
of \gls{refrac}, these are D17 (\SIrange{-1}{-0.85}{\ec}), E24 (\SIrange{-1}{-0.47}{\ec}), E40
(\SIrange{-1}{-0.12}{\ec}), E173 (\SIrange{-1}{-0.54}{\ec}), and E159 (\SIrange{-1}{-0.29}{\ec}). Notice that
all basic residues (aside from histidine) do not change their protonation state when going from
\pHrange{7.5}{4.5}. This includes R10 in \gls{refrac}, which, in contrast to D10 in \gls{wtfrac}, keeps its
full charge at low pH.

\subsubsection{WtFraC and ReFraC: electrostatic opposites.}

As evidenced by the radially averaged cross-sections and the cylindrical averages of \gls{wtfrac}'s and
\gls{refrac}'s electrostatic potentials at \pH{7.5} and \SI{1}{\Molar} ionic strength
(\cref{fig:frac_mutants_potential_1000}), the D10R mutation fully reverses the electrostatic potential inside
the constriction from negative to positive. Because arginine is significantly bulkier compared to glutamate,
the charged groups are packed closer together inside the constriction, resulting in an increase of the peak
cylindrically averaged potential ($\cylpot$) from \SI{\approx-2.5}{\kTe} in \gls{wtfrac} to
\SI{\approx+5}{\kTe} in \gls{refrac}. The central part of \gls{frac} (at $z \approx
\mSI{30}{\angstrom}$) is strongly positive, whereas the rest of the \cisi{} side ($z > \mSI{30}{\angstrom}$)
is negative.

Lowering the pH to \num{4.5} has a large effect on the constriction of \gls{wtfrac}, which becomes
significantly less negative. The $\cylpot$ amplitude inside the constriction reduces by \SI{40}{\percent},
from \SIrange{-2.5}{-1.5}{\kTe}. In contrast, maximum $\cylpot$ value inside the constriction of \gls{refrac}
remains steady at \SI{+5}{\kTe}. Finally, at low pH, the interior walls of the larger
\cisi{} side ($z > \mSI{30}{\angstrom}$) also reverse their potential, which becomes predominantly positive
instead of negative in both \gls{frac} variants.

\subsubsection{A link between the electrostatic potential, ion-selectivity, and electro-osmotic flow.}
%
%

The electrostatic potentials described in the previous section corroborate the experimentally determined ion
selectivities at \pHlist{7.5;4.5}~\cite{Huang-2017}. Whereas \gls{wtfrac} becomes less cation-selective, as
evidenced by the \SI{\approx42}{\percent} drop of its potassium permeability ratio
($\permeabilityratio{\ce{K+}}{\ce{Cl-}}$) from \num{3.64(37)} (\pH{7.5}) to \num{2.11(23)} (\pH{4.5}),
\gls{refrac} becomes more anion-selective, with a \SI{\approx60}{\percent} increase of its chloride
permeability ratio ($\permeabilityratio{\ce{Cl-}}{\ce{K+}}$) from \num{1.75(12)} (\pH{7.5}) to \num{2.78(62)}
(\pH{4.5}). Overall, these findings are in agreement with changes observed in the electrostatic potential of
these pores, but the increase in anion selectivity for \gls{refrac} at \pH{4.5} indicates that all charges
along the wall of a nanopore contribute to the ion selectivity, not just those at its narrowest location. If
we assume the preferential transport of a given ion and its hydration shell results in the net directional
transport of water through a nanopore, it follows that the \gls{eof} moves in the direction of the most
permeable ion and the net number of water molecules flowing through the pore ($\flux_{w}$) can be estimated
using~\cite{Piguet-2014}
\begin{equation}\label{eq:water_flux_permeability}
  \flux_{w} = N_{w} \dfrac{\current(\vbias)}{\ec} 
    \left[ \dfrac%
        {1 - \permeabilityratio{\ce{K+}}{\ce{Cl-}}}%
        {1 + \permeabilityratio{\ce{K+}}{\ce{Cl-}}}%
    \right]
    \text{ ,}
\end{equation}
with $N_{w}$ the average number of water molecules per ion (\ie~the hydration shell), $\current$ the ionic
current through the pore at a given bias voltage, $\vbias$, and $\ec$ the elementary charge. Note that
\cref{eq:water_flux_permeability} likely underestimates the true water flux, as it does not take the
additional drag on the fluid---due to the unidirectional movement of the electrical double layer---into
account (see~\cref{sec:np:eof}).

Assuming an hydration shell of 10 water molecules (\ie~$N_{w} = 10$)~\cite{Boukhet-2016,Piguet-2014} and a
bias voltage of \SI{-50}{\mV}, \gls{wtfrac} transports water molecules from \cisi{} to \transi{} at rates of
\SI{6.1e9}{\per\second} (\pH{7.5}, $\current = \mSI{-171}{\pA}$) and \SI{2.5e9}{\per\second} (\pH{4.5},
$\current = \mSI{-111}{\pA}$). A reduction of \SI{\approx59}{\percent} from \pHrange{7.5}{4.5}. Because
\Gls{refrac} is anion-selective, water will flow in the opposite direction, or from \cisi{} to \transi{} at
positive potentials. Using \gls{refrac}'s current at \SI{+50}{\mV} in \cref{eq:water_flux_permeability} yields
rates of \SI{1.4e9}{\per\second} (\pH{7.5}, $\current = \mSI{80}{\pA}$) and \SI{2.1e9}{\per\second} (\pH{4.5},
$\current = \mSI{71}{\pA}$). An increase of \SI{\approx51}{\percent} from \pHrange{7.5}{4.5}. The flow rates
can be translated to velocities using
\begin{equation}\label{eq:water_flux_to_velocity}
  \velocity = \dfrac{\flux_{w} V_{w}}{\pi R^2}
  \text{ ,}
\end{equation}
with $\velocity$ the velocity, $V_{w}$ the volume occupied by a single water molecule
(\SI{\approx30}{\cubic\angstrom}) and $R$ the local radius of the pore. Assuming a radius of \SI{1}{\nm}
(\eg~in the center of \gls{frac}'s channel), the rates above become velocities of \SI{58}{\mmps} (Wt @
\pH{7.5}), \SI{24}{\mmps} (Wt @ \pH{4.5}), \SI{13}{\mmps} (Re @ \pH{7.5}), and \SI{20}{\mmps} (Re @ \pH{4.5}).
These velocities are comparable to literature values for pores of similar
size~\cite{Boukhet-2016,Piguet-2014,Pederson-2015}.

Overall, these data confirm the experimental finding from Huang \etal{} (see~\cite{Huang-2017}), that, at
least for larger proteins such as chymotrypsin, the \gls{eof} drives their interaction with \gls{frac}. The
observed reduction and increase in capture rates for respectively \gls{wtfrac} (at negative bias) and
\gls{refrac} (at positive bias) when lowering the pH from \numrange{7.5}{4.5}, is in agreement with the
respective trends of their electro-osmotic flow velocities at these pH values.

\begin{figure*}[p]
  \centering
  \medskip
  \begin{subfigure}[t]{120mm}
    \centering
    \caption{}\vspace{-2.5mm}\hspace{1.5mm}\label{fig:frac_mutants_bead-ssdna_1000_energy}
    \includegraphics[scale=1]{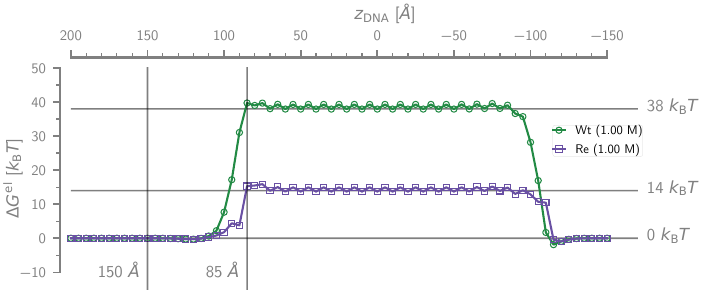}
  \end{subfigure}
  \\ \vspace{3.38mm}
  \begin{subfigure}[t]{120mm}
    \centering
    \caption{}\vspace{-10mm}\hspace{1.5mm}\label{fig:frac_mutants_bead-ssdna_1000_potential}
    \includegraphics[scale=1]{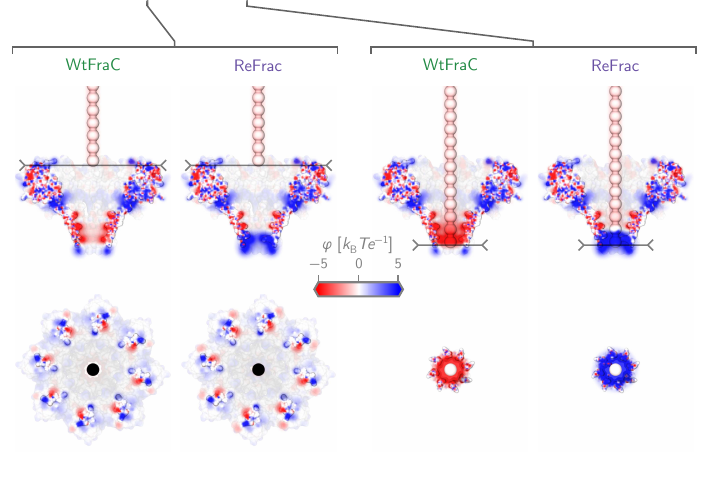}
  \end{subfigure}

\caption[Electrost. energy of ssDNA transl. through {WtFraC} and {ReFraC}]{%
  \textbf{Electrostatic energy of ssDNA translocation through {WtFraC} and {ReFraC}.}
  (\subref{fig:frac_mutants_bead-ssdna_1000_energy})
  Electrostatic energy landscape ($\energyelec$) for the translocation of \gls{ssdna}, represented here by a
  string of \SI{1}{\nm} diameter beads with \SI{-1}{\ec} charge each, through \gls{wtfrac} and \gls{refrac} at
  \SI{1.0}{\Molar} ionic strength. Due to its positively charged constriction, the energy barrier for
  \gls{refrac} is only \SI{14}{\kT}, almost a third of the \SI{38}{\kT} observed for \gls{wtfrac}.
  (\subref{fig:frac_mutants_bead-ssdna_1000_potential})
  Vertical (top) and horizontal (bottom) cross-sections of the electrostatic potential ($\potential$) inside
  \gls{wtfrac} and \gls{refrac} as the \gls{ssdna} enters the pore from the \cisi{}-side ($\zdna =
  \mSI{150}{\angstrom}$, left) and when it reaches the constriction ($\zdna = \mSI{85}{\angstrom}$, left). The
  locations of the horizontal slices are indicated using a gray line (\sliceline).
  Computations were performed using \gls{apbs}~\cite{Baker-2001,Baker-2005}.
  Molecular representations were rendered with \gls{vmd}~\cite{Humphrey-1996,Stone-1998}.
  }\label{fig:frac_mutants_bead-ssdna_1000}
\end{figure*}

\subsection[Energetics of {ssDNA} transl. through {WtFraC} and {ReFraC}]%
           {Energetics of {ssDNA} translocation through {WtFraC} and {ReFraC}}
\label{sec:elec:frac:dna}

\subsubsection{Calculating electrostatic energies using a {ssDNA} bead model.}

When observing the contrary electrostatic potentials within the constriction of \gls{wtfrac} and \gls{refrac}
(\cref{fig:frac_mutants_bead-ssdna_1000_potential}), it is unsurprising that only the latter is able to
translocate \gls{dna}. As evinced by the rotaxane experiments performed by Wloka \etal{}, \gls{refrac} is able
to translocate both \gls{ssdna} and \gls{dsdna}~\cite{Wloka-2016}. The fact that the diameter of \gls{dsdna}
(\SI{22}{\angstrom}) is almost twice the width of constriction (\SI{12}{\angstrom}) implies that the
N-terminal \ta-helices of \gls{refrac}'s are pushed apart during \gls{dsdna} translocation. Due to the static
nature of our simulations, we cannot model this without resorting to \gls{md} simulations and are thus
limited to \gls{ssdna} (\SI{\approx10}{\angstrom} diameter). However, the short persistence length of
\gls{ssdna} (\SI{13}{\angstrom}~\cite{Tinland-1997}) makes it difficult to model atomistically with a
reasonable conformation. Hence, we opted to represent it as a simple linear string of 20 beads with a radius
of \SI{5}{\angstrom} and a charge of \SI{-1}{\ec}, spaced \SI{10}{\angstrom} apart, mimicking a fully
stretched conformation of \gls{ssdna}. By placing this bead model at different heights (from $200 \ge \zdna
\ge \mSI{-150}{\angstrom}$) along the center of \gls{wtfrac} and \gls{refrac}, we were able to map the net
electrostatic energy ($\energyelec$, \cref{eq:electrostatic_energy}) caused by the translocation of a
\gls{ssdna} molecule through these pores at \SI{1}{\Molar} ionic strength
(\cref{fig:frac_mutants_bead-ssdna_1000}). 

\subsubsection{Reversing the electrostatic potential at the constriction significantly lowers the energy barrier.}

When translocating the \gls{ssdna} model from \cisi{} to \transi{}
(\cref{fig:frac_mutants_bead-ssdna_1000_energy}), there is virtually no change in $\energyelec$ observed
between the \gls{dna} entering the pore ($\zdna = \mSI{150}{\angstrom}$,
\cref{fig:frac_mutants_bead-ssdna_1000_potential}, left) up until the moment it reaches the top of the
constriction ($\zdna = \mSI{110}{\angstrom}$). Entering the constriction, however, is accompanied by a steep
increase of $\energyelec$ between $ \mSI{110}{\angstrom} > \zdna > \mSI{85}{\angstrom}$
(\cref{fig:frac_mutants_bead-ssdna_1000_potential}, right) from \SI{\approx0}{\kT} to \SI{38}{\kT} and
\SI{14}{\kT} for \gls{wtfrac} and \gls{refrac}, respectively. Both energy changes are positive, and hence
energetically unfavorable. Even though the interactions between the positively charged R10 residues in
\gls{refrac} and the negatively charged beads in the \gls{ssdna} model reduce the energy barrier in the
constriction by approximately a third, it is still insufficient to make \gls{ssdna} translocation
electrostatically favorable. However, the electrostatic energy also includes the effect of induced image
charges due to changes in permittivity (see~\cref{eq:nanopores_electrostatic_force}). It is therefore likely
that the displacement of the electrolyte (high permittivity) by the \gls{dna} strand (low permittivity) in the
confined area of the constriction results in a repulsive dielectric boundary force, which would contribute
significantly to total energy cost. Once inside the constriction, the energy level remains stable at their
respective maxima until drops rapidly to \SI{0}{\kT} once the \gls{ssdna} exits from the pore at the \transi{}
entry ($\mSI{-90}{\angstrom} > \zdna > \mSI{-120}{\angstrom}$).

The electrostatic energy barriers computed above for \gls{ssdna} translocation through \gls{wtfrac}
(\SI{38}{\kT}) and \gls{refrac} (\SI{15}{\kT}) are very similar in magnitude to those determined
experimentally by Payet~\etal{} for the \gls{ahl} nanopore (\SI{35}{\kT})~\cite{Payet-2015}. In the case of
\gls{ahl}, however, the electrostatic \gls{ssdna}--pore interactions were found to be attractive rather than
repulsive and provided a dominant contribution to determining the translocation speed of the \gls{dna} strand
through the pore. Hence, it is reasonable to assume that the electrostatics within \gls{frac} nanopores are of
similar importance.

\subsubsection{Overcoming the energy barriers inside the constriction.}

For the piece of \gls{ssdna} to translocate (at positive bias potentials), it must overcome the energy barrier
presented by the constriction: \SI{\approx38}{\kT} for \gls{wtfrac} and \SI{\approx14}{\kT} for \gls{refrac}.
In the case of \gls{wtfrac}, the energy barrier is not only more than twice as high as the one in
\gls{refrac}, the effective force exerted on the \gls{dna} at any given bias voltage will also be less due to
the opposing \gls{eof}. Nevertheless, if we ignore the \gls{eof}, and assume that, in order for it to
translocate, the \gls{dna} only uses the electrophoretic force exerted on it as the first base moves from the
\cisi{} entry ($\zdnac = \mSI{150}{\angstrom}$, \cref{fig:frac_mutants_bead-ssdna_1000_potential}, left) to
the bottom of \transi{} constriction ($\zdnat = \mSI{85}{\angstrom}$,
\cref{fig:frac_mutants_bead-ssdna_1000_potential}, right). The electrophoretic energy ($\Delta
\gibbs^{\textrm{ep}}$) gained by moving a piece of \gls{ssdna} from positions $z_1$ to $z_2$ along the length
of the pore is given by
\begin{align}\label{eq:ep_energy}
  \Delta \gibbs^{\textrm{ep}} ={}& - \int_{z_1}^{z_2} \forceep (z) \,dz \nonumber \\
                              ={}& - \int_{z_1}^{z_2} \chargeq (z) \efield (z) \,dz
  \text{ ,}
\end{align}
with $\forceep (z)$ the electrophoretic force, $\chargeq (z)$ the amount of charge inside the pore, and the
$\efield (z)$ the electric field, all a function of \gls{dna} position $z$. Assuming a uniform decay of the
applied potential $\vbias$ along the length of the pore ($\length$), $\efield$ can be simplified to
\begin{equation}\label{eq:linear_efield}
  \efield (z) = \dfrac{\vbias}{\length}
  \text{ .}
\end{equation}
Note that even though \cref{eq:linear_efield} does not take into account the non-uniform geometry of the pore,
nor its access resistance (cf.~\cref{sec:np:potential,eq:nanopores_potential_profile}), it should be adequate
for our back-of-the-envelope calculations. Using a linear function, $\chargeq (z)$ can be approximated by
\begin{equation}\label{eq:ssdna_charge_in_pore}
  \chargeq (z) = 
        \dfrac{ n_2 - n_1 } { z_2 - z_1 } 
        \left( z - z_1 \right) \chargeq_{\textrm{base}}
  \text{ ,}
\end{equation}
with $n_1$ and $n_2$ the number of bases inside the pore when the $z$ is at positions $z_1$ and $z_2$,
respectively, and $q_{\textrm{base}}$ the net charge of a single base.

Finally, plugging \cref{eq:linear_efield,eq:ssdna_charge_in_pore} in \cref{eq:ep_energy} and integrating
yields
\begin{align}\label{eq:ep_energy_dna}
  \Delta \gibbs^{\textrm{ep}} %
     ={}& - \int_{z_1}^{z_2} %
          \left(  n_2 - n_1 \right) %
          \dfrac{ z - z_1 }{z_2 - z_1}
          \chargeq_{\textrm{base}} \dfrac{\vbias}{\length} \,dz \nonumber \\
     ={}& - 0.5 \left(  n_2 - n_1 \right)  \chargeq_{\textrm{base}}
            \dfrac{\vbias}{\length} \left(  z_2 - z_1 \right)
  \text{ ,}
\end{align}
Using the parameters $n_1 = 0$, $n_2 = 7$, $z_1 = \mSI{150}{\angstrom}$, $z_2 = \mSI{85}{\angstrom}$,
$q_{\textrm{base}} = \mSI{-1}{\ec}$, $\vbias = \mSI{+100}{\mV}$ and $\length = \mSI{65}{\angstrom}$, in
\cref{eq:ep_energy_dna} yields $\Delta \gibbs^{\textrm{ep}} = \mSI{-13.6}{\kT}$. Indeed, this value is very
close to the barrier height of \SI{14}{\kT} observed for \gls{refrac}, but it is far from sufficient for
crossing the \SI{38}{\kT} barrier presented by \gls{wtfrac}. Additionally, because the \gls{eof} works with
the $\forceep$ in \gls{refrac}, but against it in \gls{wtfrac}, the magnitude of $ \left| \Delta
\gibbs^{\textrm{ep}} \right| = \mSI{13.6}{\kT}$ is an underestimation for \gls{refrac}, but an overestimation
for \gls{wtfrac}, which would reduce its ability to translocate \gls{ssdna} even further.

\subsection{Electrostatics of the ClyA nanopore}
\label{sec:elec:clya}

\subsubsection{Precise manipulation of charges enables {DNA} translocation at physiological ionic strengths.}

\begin{figure*}[p]
  \centering
  \medskip

  \begin{subfigure}[t]{120mm}
    \centering
    \caption{}\vspace{-5mm}\hspace{1.5mm}\label{fig:clya_mutants_residues}
    \includegraphics[scale=1]{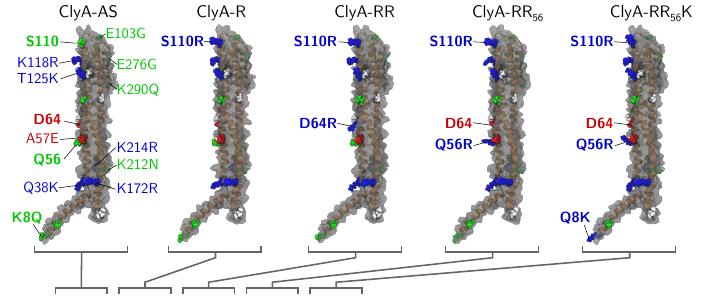}
  \end{subfigure}
  \\ \vspace{-2mm}
  \begin{subfigure}[t]{120mm}
    \centering
    \caption{}\vspace{-2.5mm}\hspace{1.5mm}\label{fig:clya_mutants_potential_0150}
    \includegraphics[scale=1]{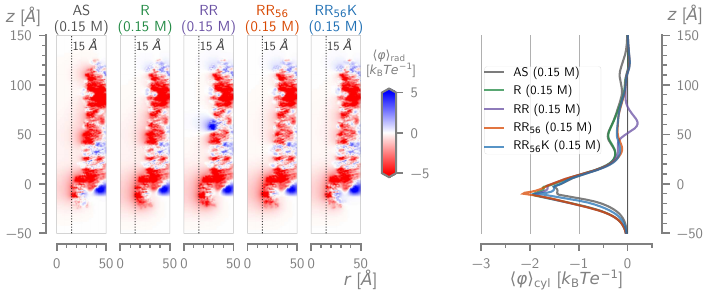}
  \end{subfigure}
  \\ \vspace{0mm}
  \begin{subfigure}[t]{120mm}
    \centering
    \caption{}\vspace{-2.5mm}\hspace{1.5mm}\label{fig:clya_mutants_potential_2500}
    \includegraphics[scale=1]{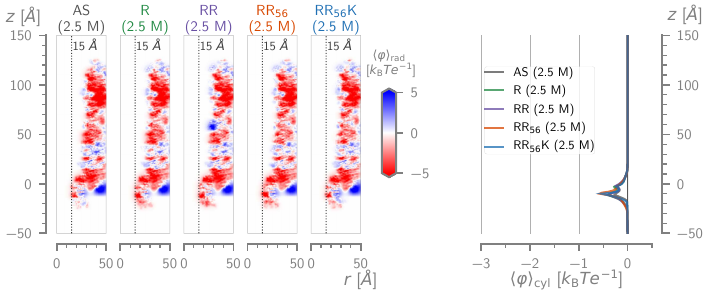}
  \end{subfigure}

\caption[Electrost. potential distribution inside several {ClyA} variants]{%
  \textbf{Electrostatic potential distribution inside several {ClyA} variants.}
  (\subref{fig:clya_mutants_residues})
  Cartoon representation of a single subunit of the ClyA-AS, -R, -RR, RR$_{56}$ and -RR$_{56}$K pores, all of
  which were investigated for their ability to translocate DNA under physiological ionic strength in
  ref.~\cite{Franceschini-2016}. The labels indicate the mutations w.r.t. the \textit{S. typhi} wild-type (for
  AS), or w.r.t. \gls{clya-as} (for the others).
  Radially averaged cross-sections ($\radpot$, left) and \SI{15}{\angstrom}-radius cylindrically averaged line
  plots ($\cylpot$, right) of the electrostatic potentials at
  (\subref{fig:clya_mutants_potential_0150})
  \SI{0.15}{\Molar}
  (\subref{fig:clya_mutants_potential_2500})
  \SI{2.5}{\Molar}.
  Computations were performed using \gls{apbs}~\cite{Baker-2001,Baker-2005}.
  Molecular representations were rendered using \gls{vmd}~\cite{Humphrey-1996,Stone-1998}.
  }\label{fig:clya_mutants_potential}
\end{figure*}

As indicated in \cref{sec:np:clya}, the interior walls of \gls{clya} are lined predominantly with negatively
charged residues (cf.~\cref{fig:nanopores_clya_pore_section}), which prevents the pore from translocating
\gls{dna} at salt concentrations lower than \SI{\approx1}{\Molar}~\cite{Franceschini-2013,Franceschini-2016}.
This concentration is far from the physiological ionic strengths (\ie~\SI{0.15}{\Molar}) required for the
(optimal) functioning of many \gls{dna}-processing enzymes. In an extensive mutagenesis study by Franceschini
\etal, more than 20 variants of \gls{clya-as} were tested for their ability to translocate \gls{dsdna} at an
ionic strength of \SI{0.15}{\Molar}~\cite{Franceschini-2016}. These mutants introduced additional positive
charges at key locations throughout the interior walls of the pore (either at the \cisi{} entry, \lumen{} or
\transi{} constriction), typically by the mutagenesis of a neutral (\ie~net \SI{+1}{\ec}) or negatively
charged (\ie~net \SI{+2}{\ec}) residue to arginine or lysine. Interestingly, only the S110R/D64R double mutant
(\gls{clya-rr}) was able to translocate \gls{dsdna}, and only from the \cisi{} side. Because analyzing all 20
variants would be too computationally intensive, we have limited our analysis here to several mutants
representative of the observed electrostatic phenomena: \gls{clya-as}, \gls{clya-r} (S110R), \gls{clya-rr}
(S110R/D64R), \gls{clya-rr56} (S110R/Q56R), and \gls{clya-rr56k} (S110R/Q56R/Q8K). The locations of their
mutations are indicated in \cref{fig:clya_mutants_residues}.

\subsubsection{All charges are important at physiological ionic strength.}
%
%

When observing the evolution of the electrostatic potential at an ionic strength of \SI{0.15}{\Molar}
(\cref{fig:clya_mutants_potential_0150,tab:clya_potential_summary}) from \cisi{} to \transi{}, the effect of
the different mutations is clearly visible. The mutation S110R is located at the \cisi{} entry of \gls{clya}
($z \approx \mSI{120}{\angstrom}$), and it switches the potential from negative (\SI{-0.12}{\kTe}) in
\gls{clya-as} to a slightly positive value (\SIrange{+0.02}{0.04}{\kTe}) in the other mutants. Even though
this is a small change, it means that \gls{dna} will now be attracted towards, instead of repelled from, the
\cisi{} entry of the pore. This may help with aligning the \gls{dna} strand with the central axis of the pore,
priming it for entry into the \lumen{}. Next up are mutations D64R and Q56R, both of which are located in the
middle of the \lumen{} ($z \approx \mSI{60}{\angstrom}$). However, whereas D64R introduces two positive
charges, Q56R only yields a single one. This is clearly reflected in the electrostatic potential, where for
Q56R it remains negative (from \SIrange{-0.32}{-0.20}{\kTe}), but D64R manages to push it to be significantly
positive (from \SIrange{-0.32}{+0.20}{\kTe}). This large change towards a positive potential observed inside
the \lumen{} of \gls{clya-rr} may allow the \gls{dna} to penetrate sufficiently deep into the pore and hence to
build up enough force from the external electric field to overcome the high negative potential at the
\transi{} constriction. At the bottom of the \lumen{} ($z \approx \mSI{30}{\angstrom}$), the electrostatic
potential falls rapidly to high negative values. The influence of the Q8K mutation, which is located in at
bottom of the constriction ($z \approx \mSI{-10}{\angstrom}$), is not clearly visible when comparing
\gls{clya-as} with \gls{clya-rr56k} since the potential decreases from \SIrange{-1.44}{-1.84}{\kTe}.
Nevertheless, relative to the other mutants, \gls{clya-rr56k} does have the least negative constriction. The
discrepancies between the peak values inside the constrictions are likely caused by minor conformational
changes in the side chains the residues that make up \gls{clya}'s \gls{tmd}.\footnotemark%
\footnotetext{
  Because all \gls{apbs} simulations for each pore were performed with a single set of atomic coordinates,
  each the result of its own (random) energy minimization, not all of the non-mutated side-chains are necessarily
  at exactly the same position across all variants. This is particularly likely in case a mutated amino acid
  is nearby.
}
Given that they are located at the protein's N-terminus, they enjoy much more conformational freedom that the
other mutated residues, and hence may have moved more during the \gls{md} equilibration phase of the model
creation (see \cref{sec:elec:methods:molec}). Regardless of its conformation, the Q8K mutation does not
significantly impact the negative potential inside the constriction and as such is not expected to impact
\gls{dna} translocation.

\subsubsection{The constriction dominates at high ionic strengths.}

At \SI{2.5}{\Molar}, the overall electrostatic potential inside the pore drops to virtually \SI{0}{\kTe}
everywhere, except inside the constriction
(\cref{fig:clya_mutants_potential_2500,tab:clya_potential_summary}). The magnitude of the negative potential
is significantly reduced for all mutants (\SI{\approx-0.5}{\kTe}) and its influence starts only inside the
constriction itself ($z \approx \mSI{0}{\angstrom}$). This means that at high salt concentrations, the
\gls{dna} can enter into the pore and move down to its constriction unopposed, after which only a small
barrier must be overcome to complete the translocation.

\begin{table}
  \footnotesize
  \centering

  \captionsetup{width=8.5cm}
  \caption[Electrostatic potential at key locations for several {ClyA} variants]%
          {Electrostatic potential at key locations for several {ClyA} variants.}
  \label{tab:clya_potential_summary}

  \renewcommand{\arraystretch}{1.2}
  \scriptsize

  \begin{tabularx}{8.5cm}{lSSSSSS}
    \toprule
     & \multicolumn{6}{c}{ {Electrostatic potential ($\cylpot$) [\si{\kTe}]} } \\
    \cmidrule(l){2-7}
     & \multicolumn{2}{c}{$z = \mSI{120}{\angstrom}$}
     & \multicolumn{2}{c}{$z = \mSI{60}{\angstrom}$}
     & \multicolumn{2}{c}{$z = \mSI{-10}{\angstrom}$} \\
    \cmidrule(r){2-3} \cmidrule(lr){4-5} \cmidrule(lr){6-7} 
    {Variant}
      & \SI{0.15}{\Molar} & \SI{2.5}{\Molar}
      & \SI{0.15}{\Molar} & \SI{2.5}{\Molar}
      & \SI{0.15}{\Molar} & \SI{2.5}{\Molar} \\
    \midrule
    AS         & -0.12 & -0.00 & -0.32 & -0.00 & -1.44 & -0.25 \\
    R          & +0.02 & +0.00 & -0.31 & -0.00 & -1.91 & -0.36 \\
    RR         & +0.03 & +0.00 & +0.20 & +0.00 & -1.88 & -0.36 \\
    RR$_{56}$  & +0.04 & +0.00 & -0.20 & -0.00 & -2.14 & -0.57 \\
    RR$_{56}$K & +0.04 & +0.00 & -0.20 & +0.00 & -1.84 & -0.53 \\
    \bottomrule
  \end{tabularx}
\end{table}
\subsection{Energetics of {dsDNA} translocation through {ClyA}}
\label{sec:elec:clya:dna}

\subsubsection{Computing the energy of DNA translocation with a full-atom {dsDNA} model.}

\begin{figure*}[p]
  \centering
  \medskip
  \begin{subfigure}[t]{40mm}
    \centering
    \caption{}\vspace{-5mm}\hspace{1.5mm}\label{fig:clya_mutants_dsdna_2500_energy}
    \includegraphics[scale=1]{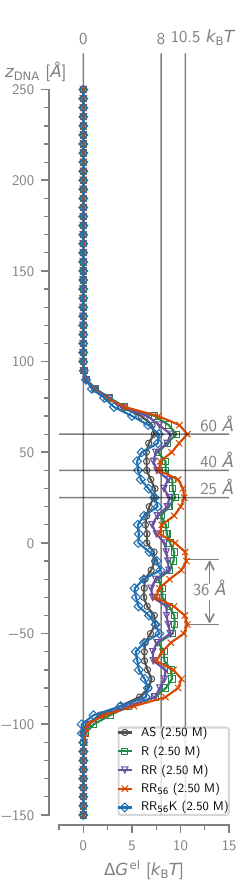}
  \end{subfigure}
  \hspace{-3.7mm}
  \begin{subfigure}[t]{81.5mm}
    \centering
    \caption{}\vspace{-5mm}\hspace{1.5mm}\label{fig:clya_mutants_dsdna_2500_potential}
    \includegraphics[scale=1]{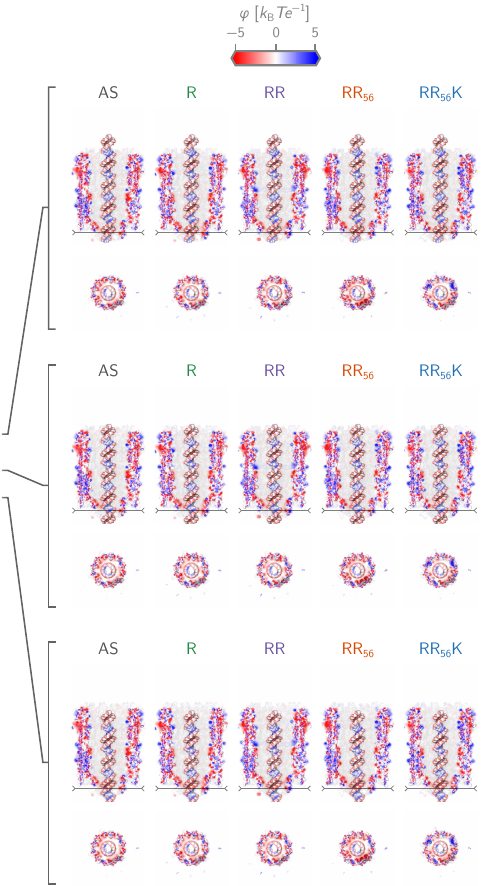}
  \end{subfigure}

\caption[Electrost. energy of {dsDNA} transl. through {ClyA} at high ionic strength]{%
  \textbf{Electrostatic energy of {dsDNA} translocation through {ClyA} at high ionic strength.}
  (\subref{fig:clya_mutants_dsdna_2500_energy})
  Net electrostatic energy ($\energyelec$, \cref{eq:electrostatic_energy}), at high ionic strength
  (\SI{2.5}{\Molar}), for a \SI{51}{\bp} piece of \gls{dsdna} (\SI{\approx175}{\angstrom}) as a function of
  its $z$-position ($z_{\mathrm{DNA}}$) inside the \gls{clya} variants AS, R, RR, RR$_{56}$ and RR$_{56}$K.
  Once inside the \transi{} constriction, $\energyelec$ fluctuates periodically with a magnitude of
  \SI{\approx2.5}{\kT} every \SI{36}{\angstrom}, which is close to the pitch of \SI{34}{\angstrom} of B-DNA.
  (\subref{fig:clya_mutants_dsdna_2500_potential})
  Vertical and horizontal cross-sections of the electrostatic potentials ($\potential$) inside the \gls{clya}
  variants as the \gls{dsdna} traverses the \transi{} constriction, showing the first maximum
  (\SI{60}{\angstrom}, top), first minimum (\SI{40}{\angstrom}, middle) and second maximum
  (\SI{25}{\angstrom}, bottom). The locations of the horizontal slices are indicated using a gray line
  (\sliceline).
  Computations were performed using \gls{apbs}~\cite{Baker-2001,Baker-2005}.
  Molecular representations were rendered with \gls{vmd}~\cite{Humphrey-1996,Stone-1998}.
  }\label{fig:clya_mutants_dsdna_2500}
\end{figure*}

Using the same methodology that was used with \gls{ssdna} translocation through \gls{frac} in
\cref{sec:elec:frac}, we used a full atom \gls{dsdna} model to map out the electrostatic energy landscape of
\gls{dna} translocation through the \gls{clya} variants discussed above at \SI{0.15}{\Molar}
(\cref{fig:clya_mutants_dsdna_0150}) and \SI{2.5}{\Molar} (\cref{fig:clya_mutants_dsdna_2500}) ionic
strengths. Here, a bead model is not needed since the long persistence length of \gls{dsdna}
(\SI{390}{\angstrom}~\cite{Gross-2011}) effectively fixes the conformation of a \SI{51}{\bp} strand
(\SI{\approx175}{\angstrom} long) to that of a rigid rod. Additionally, \gls{clya}'s \transi{} constriction is
sufficiently wide (\SI{33}{\angstrom}) to accommodate the full width of a \gls{dsdna} strand
(\SI{22}{\angstrom}).

\subsubsection{DNA translocation at physiological ionic strengths is a fragile balance of energies.}

\begin{figure*}[p]
  \centering
  \medskip
  \begin{subfigure}[t]{40mm}
    \centering
    \caption{}\vspace{-5mm}\hspace{1.5mm}\label{fig:clya_mutants_dsdna_0150_energy}
    \includegraphics[scale=1]{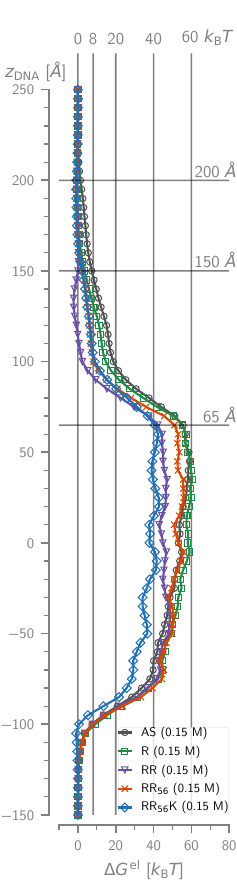}
  \end{subfigure}
  \hspace{-3.7mm}
  \begin{subfigure}[t]{81.5mm}
    \centering
    \caption{}\vspace{-5mm}\hspace{1.5mm}\label{fig:clya_mutants_dsdna_0150_potential}
    \includegraphics[scale=1]{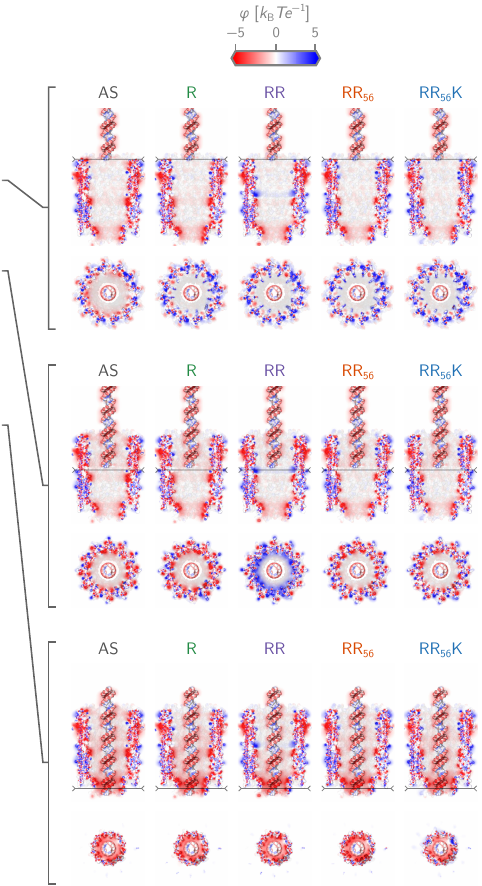}
  \end{subfigure}

\caption[Electrost. energy of {dsDNA} transl. through {ClyA} at physiol. ionic strength]{%
  \textbf{Electrostatic energy of {dsDNA} translocation through {ClyA} at physiological ionic strength.}
  (\subref{fig:clya_mutants_dsdna_0150_energy})
  Net electrostatic energy ($\energyelec$, \cref{eq:electrostatic_energy}), at physiological ionic strength
  (\SI{150}{\mM}), for a \SI{51}{\bp} piece of \gls{dsdna} (\SI{\approx175}{\angstrom}) as a function of its
  $z$-position ($z_{\mathrm{DNA}}$) inside the \gls{clya} variants AS, R, RR, RR$_{56}$ and RR$_{56}$K. The
  mutation S110R eliminates the \SI{8}{\kT} energy penalty for \gls{dsdna} to enter \lumen{} of \gls{clya}
  from the \cisi{} side up until \SI{150}{\angstrom}. In contrast to D64R, which briefly promotes the
  downwards motion up of \gls{dsdna} starting from \SI{150}{\angstrom}, Q56R only manages to slow the rise of
  the energy barrier. Once inside the constriction, the Q8K mutation results in an energy drop from
  \SIrange{\approx40}{60}{\kT}.
  (\subref{fig:clya_mutants_dsdna_0150_potential})
  Vertical and horizontal cross-sections of the electrostatic potential ($\potential$) inside the \gls{clya}
  variants as the \gls{dsdna} begins to enter the \cisi{} entrance (\SI{200}{\angstrom}, top), reaches the
  center of the \lumen{} (\SI{150}{\angstrom}, middle) and fully entered the \transi{} constriction
  (\SI{65}{\angstrom}, bottom). The locations of the horizontal slices are indicated using a gray line
  (\sliceline).
  Computations were performed using \gls{apbs}~\cite{Baker-2001,Baker-2005}.
  Molecular representations were rendered with \gls{vmd}~\cite{Humphrey-1996,Stone-1998}.
  }\label{fig:clya_mutants_dsdna_0150}
\end{figure*}

Upon entry into the pore from the \cisi{} side ($\zdna = \mSI{200}{\angstrom}$,
\cref{fig:clya_mutants_dsdna_0150_potential}, top), the S110R mutation interacts favorably with the \gls{dna}
strand, preventing the gradual \SI{+8}{\kT} rise in energy observed when \gls{clya-as} when the \gls{dna}
moves towards the location of the  D64R (or Q56R) mutation in the middle of the lumen ($\zdna =
\mSI{150}{\angstrom}$, \cref{fig:clya_mutants_dsdna_0150_potential}, middle). When moving further down, the
Q56R mutation in \gls{clya-rr56} and \gls{clya-rr56k} manages to limit the energy increase compared to
\gls{clya-as} and \gls{clya-r}, but the D64R mutation in \gls{clya-rr} lowers the energy by
\SI{\approx5}{\kT}. Once the \gls{dna} begins to enter the constriction (from $\zdna = \mSI{100}{\angstrom}$),
however, the energy rises dramatically for all variants by an additional \SI{\approx40}{\kT} until it reaches
a maximum when the \gls{dna} starts to exit the pore from the \transi{} side ($\zdna = \mSI{200}{\angstrom}$,
\cref{fig:clya_mutants_dsdna_0150_potential}, bottom). The influence of the Q8K mutation is also clearly
visible, with \gls{clya-rr56k} having a significantly lower energy barrier (\SI{\approx10}{\kT}) compared to
\gls{clya-rr56}. As the \gls{dna} moves through the constriction, its energy levels appear to fluctuate by a
few \si{\kT} with a period of \SI{\approx30}{\angstrom}. This phenomenon is even more prominent at high salt
concentrations and will hence be discussed in the next section.


As we shall see in the next chapter, at positive bias voltages the \gls{eof} in \gls{clya} flows from
\transi{} to \cisi{} and hence exerts a significant amount of upward force on the \gls{dna}, pushing it back
towards \cisi{} entry. If the electro-osmotic force is equal to half the electrophoretic force---a
conservative estimate---any energy gains due to the electrophoretic force will be cut in half. In other words,
this would equate to an effective charge of \SI{-1}{\ec\per\bp} instead of \SI{-2}{\ec\per\bp}. In order to
successfully translocate, the \gls{dsdna} strand must reduce its energy sufficiently to overcome the
\SI{\approx60}{\kT} energy barrier presented by the constriction.  Moving the \gls{dna} strand from the
\cisi{} entry to top of the energy barrier (\cref{eq:ep_energy_dna}, $n_1 = 0$, $n_2 = 41$, $z_1 =
\mSI{200}{\angstrom}$, $z_2 = \mSI{65}{\angstrom}$, $q_{\textrm{base}} = \mSI{-1}{\ec}$, $\vbias =
\mSI{+100}{\mV}$) yields an energy reduction of $\Delta \gibbs^{\textrm{ep}} = \mSI{-76}{\kT}$, which should
indeed be sufficient to overcome the barrier. However, for the \gls{dna} to reduce its energy by such amounts,
it must be able to physically reach that location in the first place. The S110R mutation helps here by
removing the initial \SI{+8}{\kT} energy increase between $200 > \zdna > \mSI{150}{\angstrom}$ observed only
in \gls{clya-as}. Past this point, the Q56R mutation reduces the rate at which \gls{dna} gains energy compared
to \gls{clya-as} and \gls{clya-r}, but not to the same extent as D64R in \gls{clya-rr}, where moving downwards
briefly reduces the energy. Likely, at this point in the translocation process, the force balance is
precarious in all mutants. Hence, where the downwards force in \gls{clya-rr} tips the energy balance in favor
of the downwards movement of the \gls{dna}, the opposite is true for the other mutants. In other words, it
appears that the D64R mutation enables the \gls{dna} to move further down into the \lumen{} than the Q56R
mutation, allowing it to sufficiently reduce its energy and eventually overcome the barrier at the
constriction. Note that this analysis presents the situation where the \gls{dsdna} is aligned perfectly with
the entry of the pore, which will often not be the case in reality. Additionally, outside of the pore, the
relative magnitude of the electro-osmotic force is likely to be significantly stronger than its
electrophoretic counterpart, which would further reduce the chances of \gls{dna} to enter \gls{clya} in the
first place.

\subsubsection{High ionic strength smooths out and lowers energy barriers.} 

At an ionic strength of \SI{2.5}{\Molar}, no large differences are observed between energy landscapes of the
different \gls{clya} mutants (\cref{fig:clya_mutants_dsdna_2500_energy}). As the \gls{dsdna} enters the pore
from \cisi{} side, its $\energyelec$ remains flat until the it reaches the constriction ($200 > \zdna >
\mSI{100}{\angstrom}$), after which it rises rapidly by \SI{\approx+10}{\kT} inside the constriction. As
discussed in the previous paragraph, at $\zdna = \mSI{100}{\angstrom}$ the \gls{dna} strand will have
decreased its energy by $\Delta \gibbs^{\textrm{ep}} = \mSI{-76}{\kT}$, which exceeds the energy barrier at
the constriction by almost an order of magnitude. This is in full agreement with the experimental observations
of \gls{dsdna} translocation through \gls{clya-as} at higher ionic
strengths~\cite{Franceschini-2013,Franceschini-2016}.

\subsubsection{The chirality of \gls{dsdna} may induce rotation.}

As the \gls{dsdna} traverses the constriction, its energy is observed to fluctuate in a sinusoidal manner with
an amplitude of \SI{2.5}{\kT} and a period of \SI{36}{\angstrom} (\cref{fig:clya_mutants_dsdna_2500_energy}).
Likely, the peaks (\cref{fig:clya_mutants_dsdna_2500_potential}, top and bottom) and valleys
(\cref{fig:clya_mutants_dsdna_2500_potential}, middle) correlate with the precise number of negative charges
present inside the constriction, which in turn corresponds to how much of the B-\gls{dna}'s minor groove
(\ie~high charge density) or major groove (\ie~low charge density) occupies the constriction. Given that, from
a geometric point of view, moving a B-\gls{dna} strand up or down is identical to rotating it around its
length axis, it is likely that a \SI{2.5}{\kT} energy reduction is sufficient to make the \gls{dna} rotate as it
translocates through \gls{clya}.

\subsection{Electrostatic confinement of {dsDNA} within {ClyA}}
%

\begin{figure*}[t]
  \centering
  \medskip
  \begin{subfigure}[t]{60mm}
    \centering
    \caption{}\vspace{-5mm}\hspace{1.5mm}\label{fig:clya_dsdna_constriction_energy_heatmap}
    \includegraphics[scale=1]{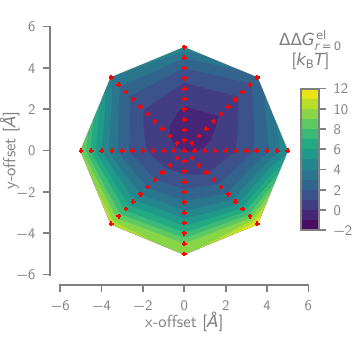}
  \end{subfigure}
  \hspace{-3mm}
  \begin{subfigure}[t]{60mm}
    \centering
    \caption{}\vspace{-5mm}\hspace{1.5mm}\label{fig:clya_dsdna_constriction_energy_averaged}
    \includegraphics[scale=1]{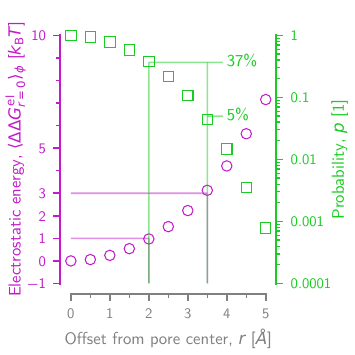}
  \end{subfigure}

\caption[Electrost. confinement of {dsDNA} within the constr. of {ClyA-AS}]{%
  \textbf{Electrostatic confinement of {dsDNA} within the constriction of {ClyA-AS}.}
  (\subref{fig:clya_dsdna_constriction_energy_heatmap})
  Contour plot of the electrostatic energy cost/gain ($\Delta \energyelec_{r=0}$) of moving a piece of
  \gls{dsdna} away from the axial center of \gls{clya-as}. The red crosses indicate the location of data
  points.
  (\subref{fig:clya_dsdna_constriction_energy_averaged})
  Energy costs averaged over all angles ($\langle \Delta \energyelec_{r=0} \rangle_{\phi}$), together with the
  corresponding Boltzmann probability distribution $\probability$.
  Computations were performed using \gls{apbs}~\cite{Baker-2001,Baker-2005}.
  Figure adapted with permission from~\cite{Bayoumi-2020}.
  }\label{fig:clya_dsdna_constriction}
\end{figure*}

Given the electrostatic repulsion between \gls{clya-as}'s constriction and the \gls{dsdna}, we expect that the
latter will be pushed to remain in the center of the pore. To quantify the extent of this confinement, we
computed the electrostatic energy cost of moving a piece of dsDNA away from the center of the pore (\ie~
closer to the walls of \gls{clya}). This information allowed the use of a narrower constriction within the
coarse-grained \gls{md} simulations presented in Bayoumi \etal{}~\cite{Bayoumi-2020}, and was crucial to
properly model the behavior of the \gls{ssdna}- and \gls{dsdna}-rotaxanes within \gls{clya}.

\subsubsection{System setup and energy calculation.}

The same piece of \SI{51}{\bp} \gls{dsdna} from previous sections was placed inside \gls{clya-as} at a fixed
height ($\zdna = \mSI{60}{\angstrom}$). Next, the \gls{dna} was moved away from the center of the
constriction ($x = 0$, $y = 0$), using cylindrical coordinates $\rpos_{\mathrm{DNA}} = (r, \phi)$ such that
$x = r \cos{\phi}$ and $y = r \sin{\phi}$:
\begin{equation}
  \rpos_{\mathrm{DNA}} = %
  \begin{cases}
    r: [0,5] \msi{\angstrom} &\mbox{with } \Delta r    = \mSI{0.5}{\angstrom} \\
    r: [0,360] \msi{\degree} &\mbox{with } \Delta \phi = \mSI{45}{\degree} \\
  \end{cases}
  \text{ ,}
\end{equation}
resulting in 177 unique systems (\ie~1 reference system with only the pore, 88 reference systems with only the
DNA and 88 systems containing the pore and the DNA). The electrostatic potentials and energies at an ionic
strength of \SI{2}{\Molar} where computed as before with \gls{apbs}, using two sequential focusing
calculations (coarse grid: \apbsgrid{500}{500}{700}{\angstrom} size with
\apbsgrid{1.42}{1.42}{0.87}{\angstrom} spacing; fine grid: \apbsgrid{175}{175}{400}{\angstrom} size with
\apbsgrid{0.50}{0.50}{0.50}{\angstrom} spacing). The energy cost (or gain) for moving the \gls{dna} away from
the center of the pore ($r = 0$) was then calculated using
\begin{align}\label{eq:electrostatics_deltadelta_energy}
  \Delta \energyelec_{r=0} (r, \phi) = \energyelec (r, \phi) - \energyelec (0,0)
  \text{  .}
\end{align}

\subsubsection{The energy minimum is not at the center and effective pore size}

The contour plot of $\Delta \energyelec_{r=0}$ (\cref{fig:clya_dsdna_constriction_energy_heatmap}) reveals
that the energetic minimum for a piece of \gls{dsdna} does not lie at the center of the pore but is slightly
offset, approximately \SI{1}{\angstrom} in both $x$- and $y$-directions. Likely, this is a consequence of the
chirality of the \gls{dna} and the precise location of the major and minor grooves within the constriction of
\gls{clya}, similarly to the fluctuations observed during translocation
(\cref{fig:clya_mutants_dsdna_0150_energy,fig:clya_mutants_dsdna_2500_energy}). 


When one wants to perform long \gls{md} simulations of \gls{dna} dynamics within \gls{clya} at reasonable
computational times, it can be useful to use coarse-grained models of both the pore and the \gls{dna},
which do not (explicitly) include the electrostatic interaction between them. Hence, to compensate for the
lack of repulsion between the \gls{dna} strand and the walls of the pore, one approach could be to
artificially increase the confinement by reducing the diameter of the latter to an `effective' pore size.
The angle-averaged relative energy $\langle \Delta \energyelec_{r=0} \rangle_{\phi}$
(\cref{fig:clya_dsdna_constriction_energy_averaged}) reveals a quasi exponential increase of energy required to
move a mere \SI{5}{\angstrom} away from the center of the pore: from \SIrange{0}{7}{\kT}. Even a small offset confers a
significant electrostatic energy penalty, with shifts of \SI{2.0}{\angstrom} and \SI{3.5}{\angstrom} giving
rise to increases of \SI{1}{\kT} and \SI{3}{\kT}, respectively. Assuming Boltzmann statistics, the probability
$\probability$ of the \gls{dna} to be present at a certain offset from the center of the pore
(\cref{fig:clya_dsdna_constriction_energy_averaged}) is given by 
\begin{align}\label{eq:electrostatics_probability}
  \probability = \exp{
    \left( \dfrac{\langle \Delta \energyelec_{r=0} \rangle_{\phi}}{\boltzmann\temperature} \right)
    }
    \text{  .}
\end{align}
This means that at any given time, \SI{63}{\percent} and \SI{95}{\percent} of \gls{dna} molecules will be
within radius of \SI{2.0}{\angstrom} and \SI{3.5}{\angstrom} from the center of the pore, respectively. Hence,
when placing a cut-off at \SI{3}{\kT}, the electrostatic energy confines the DNA to a circular region with a
radius of \SI{14.5}{\angstrom} ($\mSI{11}{\angstrom} + \mSI{3.5}{\angstrom}$), yielding an effective
constriction size of \SI{29}{\angstrom}, instead of the traditional \SI{33}{\angstrom}.

\section{Conclusion}
\label{sec:elec:conclusion}

The electrostatic interactions between nanopores and analyte molecules can play a determining role in the
properties of nanopores. In this chapter, I have investigated the influence of ionic strength and solution pH
on the distribution of the equilibrium electrostatic potential (\ie~without an external electric field) within
several variants of the \gls{plyab}, \gls{frac} and \gls{clya} nanopores. This enabled us to provide insights
into experimentally observed properties such as the ion-selectivity and postulate the influence on second-order
effects such as the electro-osmotic flow. Self-consistent computation of the electrostatic interaction
energies between a translocating \gls{dna} strand and the \gls{frac} and \gls{clya} nanopores, provided
further insights into the \gls{dna} translocation mechanisms of these pores, confirming experimental
observations.

The large (interior) size of the \gls{plyab} nanopore makes it particularly well-suited for studying large
proteins in their native state. However, experiments on the protein capture of three \gls{plyab}
variants---namely wild-type, E2, and R---showed that only \gls{plyab-r} yielded blockades that provide
sufficient information for protein identification~\cite{Huang-2020}. Analysis of the electrostatics of these
three pores revealed the \lumen{} of all three pores was predominantly negative, but that \gls{plyab-r}
displayed a positive potential at its constriction, rather than a negative one. As expected, increasing the
ionic strength from \SIrange{0.3}{1.0}{\Molar} significantly reduced the influence of the potential within the
center of the pore. As a consequence of the competing forces on the \gls{edl} in \gls{plyab-r}, we propose
that its \gls{eof} is significantly lower compared to \gls{plyab-e2} (and \gls{plyab-wt}), explaining the
experimentally observed changes in protein blockade duration and frequency.

Next, we investigated the much smaller \gls{frac} nanopore, comparing again the electrostatics of the
wild-type pore, \gls{wtfrac}, with a variant, \gls{refrac}, that contained a single negative-to-positive
charge reversal at its constriction. Trapping experiments showed these pores captured proteins at opposing
bias voltages, and that the event frequency of \gls{wtfrac}, but not \gls{refrac}, was severely affected when
lowering the electrolyte pH from \pHrange{4.5}{7.5}~\cite{Huang-2017}. The computational electrostatic
analysis, which took into account the partial protonation of all titratable residues, revealed two key
conclusions that support the experimental data. (1) The single charge reversal inside \gls{refrac} was
sufficient to fully switch the electrostatic potential within the pore \lumen{} from negative to positive,
indicating that also the direction of the is \gls{eof} reversed. (2) Whereas the lowering of the pH to 4.5
also reduced the magnitude of the potential in \gls{wtfrac} by \SI{40}{\percent}, it remained similar for
\gls{refrac}. Originally, the \gls{refrac} pore was developed to enable \gls{frac} to be used for \gls{dna}
analysis~\cite{Wloka-2016}. To quantify the differences in the ability of both pores to translocate \gls{dna},
we computed the electrostatic energy cost of moving a piece of \gls{ssdna} across each pore. Surprisingly,
even though the high energy barrier observed in \gls{wtfrac} was reduced by \SI{63}{\percent} in \gls{refrac},
the energy barrier was still there, despite the favorable electrostatic interactions between the positively
charged constriction and the negatively charged \gls{dna}. Likely, it is caused by a repulsive dielectric
boundary force (see~\cref{eq:nanopores_electrostatic_force}), which results from the displacement of the high
permittivity electrolyte by a low permittivity \gls{dna} strand. Nevertheless, a rough estimation indicates
that the force exerted by the electric field on the \gls{dna} is sufficiently high to overcome the
barrier in the case of \gls{refrac}, but not \gls{wtfrac}, in agreement with the experiments.

In the last section, we investigated the electrostatics of five \gls{clya} variants: AS, R, RR, RR$_{56}$, and
RR$_{56}$K. Single-channel electrophysiology experiments demonstrated previously that, even though all pores
could translocate \gls{dsdna} fragments at high ionic strengths (\eg~\SI{2.5}{\Molar}), only \gls{clya-rr} had
this ability under physiological conditions (\eg~\SI{0.15}{\Molar})~\cite{Franceschini-2016}. As expected, at
high salt concentrations, the complex potential landscape within all pores was found to be virtually the same
(\ie~close to 0), with only a slightly negative constriction remaining. However, the analysis of the
electrostatic potential differences within these pores at \SI{0.15}{\Molar} showed that the S110R mutation
(located at the \cisi{} entry) neutralized the repulsive potential at the mouth of the pore, perhaps allowing
\gls{dna} to be captured more easily. More importantly, we found that only \gls{clya-rr}'s D64R mutation, a
charge reversal positioned in the center of the \lumen{}, was able to produce a positive potential within an
otherwise highly negative environment. Subsequent energy calculations of a piece of translocating \gls{dsdna}
at physiological conditions show that favorable interactions present in \gls{clya-rr} result in a lowering of
the energy as it nears the high energy barrier (\SI{\approx60}{\kT}) located at the constriction. This
suggests that, by allowing it to enter so deeply into the \lumen{} of the pore, only \gls{clya-rr} lets the
\gls{dna} strand build up a sufficient amount of force to punch through the barrier of the constriction. At high ionic
strengths, the energy landscape appeared to fluctuate by \SI{\approx2.5}{\kT} every \SI{36}{\angstrom},
suggesting that it might rotate during translocation. Further calculations with off-center \gls{dsdna} strands
showed that the high electrostatic repulsion within the constriction effectively confines it to the center,
reducing the effective size of the pore.

It is clear that, by studying the electrostatics of a nanopore, one can already gain significant insights into
its physics. Nevertheless, as we shall see in the following chapters, equilibrium electrostatics comprise only
a part of a nanopore's properties. Hence, developing a full understanding requires a thorough quantitative
analysis of the \emph{nonequilibrium} effects (\ie~with an external electric field), such as electrophoresis,
electro-osmosis, and steric hindrance.

\cleardoublepage

%

%
  \graphicspath{{\chaptersdir/trapping/image/}}%
\chapter{Trapping of a single protein inside a nanopore}
\label{ch:trapping}
\epigraphhead[\epipos]{%
\epigraph{%
  ``Magic is organizing chaos. And while oceans of mystery remain, we have deduced that this requires two
  things. Balance and control.''
}{%
  \textit{`Tissaia De Vries'}
}
}
\definecolor{shadecolor}{gray}{0.85}
\begin{shaded}
This chapter was published as:
\begin{itemize}
  \item K. Willems*, D. Rui\'{c}*, A. Biesemans*, N. S. Galenkamp, P. Van Dorpe and G. Maglia.
        \textit{ACS Nano} \textbf{13 (9)}, 9980--9992 (2019) 
        \\
        *equal contributions
\end{itemize}
\newpage
\end{shaded}
%
%

%

In this chapter, we used experimental, computational, and theoretical methods to investigate the trapping
behavior of a small protein, \glsfirst{dhfr}, within the \glsfirst{clya} nanopore. It builds upon the
electrostatic energy methodology developed in \cref{ch:electrostatics} to parameterize and verify a
mathematical model that captures the essential physics of the experimentally observed voltage- and
charge-dependent dwell times of a tagged \gls{dhfr} molecule in \gls{clya}. \\
To the extent possible, the main and supplementary texts and figures of the original manuscript were
integrated into this chapter, and the remainder of the supplementary information can be found in
\cref{ch:trapping_appendix}. The text and figures of this chapter represent entirely my own work. All
experimental work was performed by Dr.~Annemie Biesemans, and the trapping model was derived by
Dr.~rer.~nat.~Dino Rui\'{c} (see~\cref{sec:trapping_appendix:escape_rates}).
\adapACS{Willems-Ruic-Biesemans-2019}
Note that further permissions related to the material excerpted should be directed to the ACS.
%
%
%

%
\section{Abstract}
\label{sec:trapping:abstract}

The ability to confine and study single molecules has enabled important advances in natural and applied
sciences. Recently, we have shown that unlabeled proteins can be confined inside the biological nanopore
\glsfirst{clya} and conformational changes monitored by ionic current recordings. However, trapping small
proteins remains a challenge. Here we describe a system where steric, electrostatic, electrophoretic, and
electro-osmotic forces are exploited to immobilize a small protein, \glsfirst{dhfr}, inside \gls{clya}.
Assisted by electrostatic simulations, we show that the dwell time of \gls{dhfr} inside \gls{clya} can be
increased by orders of magnitude (from milliseconds to seconds) by manipulation of the \gls{dhfr} charge
distribution. Further, we describe a physical model that includes a double energy barrier and the main
electrophoretic components for trapping \gls{dhfr} inside the nanopore. Simultaneous fits to the voltage
dependence of the dwell times allowed direct estimates of the \cisi{} and \transi{} translocation
probabilities, the mean dwell time, and the force exerted by the electro-osmotic flow on the protein
(\SI{\approx9}{\pN} at \SI{-50}{\mV}) to be retrieved. The observed binding of \gls{nadph} to the trapped
\gls{dhfr} molecules suggested that the engineered proteins remained folded and functional inside \gls{clya}.
Contact-free confinement of single proteins inside nanopores can be employed for the manipulation and
localized delivery of individual proteins and will have further applications in single-molecule analyte
sensing and enzymology studies.

\section{Introduction}
\label{sec:trapping:intro}

Sensors capable of the label-free interrogation of proteins at the single-molecule level have applications in
biosensing, biophysics, and enzymology~\cite{Gooding-2016,Xie-2001,Willems-VanMeervelt-2017}. In particular,
the ability to observe the behavior of individual proteins allows one to directly retrieve the rates of
kinetic processes and provides a wealth of mechanistic, energetic, and structural information, which are not
readily obtained from statistically averaged ensemble (bulk) measurements~\cite{Gooding-2016}. To achieve
single-molecular sensitivity at high signal-to-noise ratios, the observational volume of the sensor should be
similar in size to the object of interest (\ie~zeptoliter-range for a protein with a radius of \SI{2.5}{\nm}).
Moreover, many kinetic processes have relatively long time scales (\eg~\SIrange{e-3}{e0}{\second}) which in
turn necessitate long observational times to obtain a statistically relevant number of events. Hence, the
protein must also remain inside the observational volume for seconds or minutes, a feat that is only possible
if the protein is either physically immobilized or trapped in a local energetic minimum that is significantly
deeper than the thermal energy~\cite{Krishnan-2010,Myers-2015}.

To counteract the random thermal motion of nanoscale objects in solution several optical, microfluidic, and
nanofluidic methodologies have been developed over the years. The optical trapping of nanoscale objects
(\SI{<50}{\nm} radius) requires the sub-diffraction limited confinement of light~\cite{Neuman-2004,Baker-2017,
Bradac-2018} which can be achieved with photonic~\cite{Yang-2009,Mandal-2010} or
plasmonic~\cite{Juan-2009,Chen-2011,
Pang-2011,Bergeron-2013,Kotnala-2014,Kerman-2015,Chen-2018,Verschueren-2019}, nanostructures. Although optical
techniques have been shown to be capable of trapping proteins with a radius of
\SI{\approx2.3}{\nm}~\cite{Kotnala-2014}, the high optical intensities required, and the solid-state nature of
the devices tend to not only trap the proteins but also unfold them, limiting the scope of their
applicability~\cite{Pang-2011, Verschueren-2019}.

Microfluidic techniques might offer softer alternatives for the immobilization of single molecules. The
\gls{abel} trap makes use of optical tracking to electrophoretically counteract the Brownian motion of
individual dielectric particles~\cite{Cohen-2005,Cohen-2006,Goldsmith-2010, Goldsmith-2011}, enabling the
trapping of proteins down to \SI{\approx2.9}{\nm} radius~\cite{Goldsmith-2011} and even single
fluorophores~\cite{Fields-2011}. Because this technique uses fluorescence microscopy to track the movement of
their targets, the observational time window is ultimately limited by the photobleaching of the
dye~\cite{Cohen-2005,Goldsmith-2010}.

Nanopores, which are nanometer-sized apertures in a membrane separating two electrolyte reservoirs, have been
used extensively to study single
molecules~\cite{Ma-2009,Laszlo-2016,Willems-VanMeervelt-2017,Varongchayakul-2018}. In nanopore analyses, an
electric field is applied across the membrane and information about a molecule passing through the pore is
collected by monitoring the modulations of the ionic charge current. As proteins typically transit the pore at
high velocities (\SIrange{\approx e-3}{e-2}{\meter\per\second})~\cite{Fologea-2007,Plesa-2013,Li-2013B,
Larkin-2014}, the dwell time (\ie~the duration a molecule of interest spends inside the observable volume) is
on the order of \SIrange{e-6}{e-3}{\second}. These timescales have proven sufficient for obtaining structural
information such as protein size, shape, charge, dipole moment, and
rigidity~\cite{Fologea-2007,Larkin-2014,Waduge-2017,Hu-2018,Houghtaling-2019}, but they are too brief to
efficiently study the enzymatic cycle of the majority of human enzymes (turnover numbers between
\SIrange{e-3}{e3}{\per\second}).

\begin{figure*}[t]
  \begin{minipage}{6cm}
    \begin{subfigure}[t]{5.5cm}
      \centering
      \caption{}\vspace{-3mm}\label{fig:trapping_concept_clya}
      \includegraphics[scale=1]{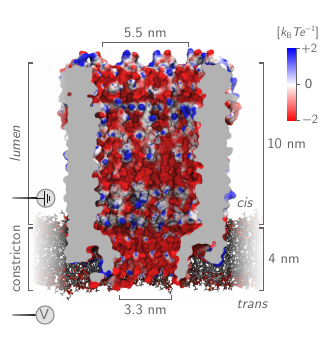}
    \end{subfigure}
  \end{minipage}
  \begin{minipage}{6cm}
    \begin{subfigure}[t]{5.5cm}
      \centering
      \caption{}\vspace{-3mm}\label{fig:trapping_concept_dhfr}
      \includegraphics[scale=1]{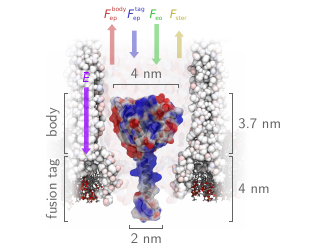}
    \end{subfigure}
    \\
    \begin{subfigure}[t]{5.5cm}
      \centering
      \caption{}\vspace{-3mm}\label{fig:trapping_concept_sequence}
      \includegraphics[scale=1]{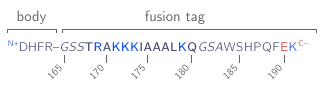}
    \end{subfigure}
  \end{minipage}

\caption[Trapping of proteins inside the {ClyA-AS} nanopore]{%
  \textbf{Trapping of proteins inside the {ClyA-AS} nanopore.}
  (\subref{fig:trapping_concept_clya})
  Surface representation of a type I \gls{clya-as} nanopore embedded in a planar lipid bilayer (derived
  through homology modeling from \pdbid{2WCD}~\cite{Mueller-2009}, see~\cref{sec:elec:methods:molec}), and
  colored according to its electrostatic potential in \SI{150}{\mM} \ce{NaCl} (calculated by
  \gls{apbs}~\cite{Baker-2001,Dolinsky-2004,Dolinsky-2007}, see~\cref{sec:elec:methods:elec}).
  (\subref{fig:trapping_concept_dhfr})
  Depiction of a single \glsfirst{dhfr} molecule extended with a positively charged C-terminal polypeptide tag
  (\DHFR{4}{S}) inside a \gls{clya-as} nanopore. The secondary structure of the tag (primarily
  \textalpha-helical) was predicted by the {PEP-FOLD} server~\cite{Thevenet-2012,Shen-2014}. At negative
  applied bias voltages relative to \transi{}, the electric field ($\protect\overrightarrow{E}$) is expected
  to pull the negatively charged body of \gls{dhfr} upward ($\forceep^{\rm{body}}$) and the positively charged
  fusion tag downward ($\forceep^{\rm{tag}}$), while the electro-osmotic flow pushes the entire protein
  downward ($\forceeo$). Lastly, as the body of \gls{dhfr} is larger than the diameter of the \transi{}
  constriction, the force required to overcome the steric hindrance ($\forcest$) during full
  \cisi{}-to-\transi{} translocation is expected to be significant.
  (\subref{fig:trapping_concept_sequence})
  Sequence of \DHFR{4}{S} fusion tag with its positive and negative residues colored blue and red,
  respectively. The sequence of the Strep-tag starts at residue 183, and the GSS and GSA linkers are shown in
  italicized font. Note that, at \pH{7.5}, the C- and N-termini contribute one negative charge to the body and
  one positive charge to the tag, respectively.
  Images were rendered using \gls{vmd}~\cite{Humphrey-1996,Stone-1998}.
  }\label{fig:clya_dhfr_trapping_concept}
\end{figure*}

To increase the observation window of proteins by nanopores, researchers have made extensive use of
non-covalent interactions. By coating solid-state nanopores with \gls{nta} receptors, the dwell time of
His-tagged proteins could be prolonged up to six orders of magnitude~\cite{Wei-2012}. In another account, the
diffusion coefficient of several proteins was reduced 10-fold \textit{via} tethering to a lipid bilayer coated
nanopore~\cite{Yusko-2011,Yusko-2017}. The decoration of biological nanopores with thrombin-specific aptamers
enabled the investigation of the binding kinetics of thrombin to its aptamer~\cite{Rotem-2012} and the
selective detection in the presence of a 100-fold excess of non-cognate proteins~\cite{Soskine-2012}.
Electrophoretic translocation of protein-DNA complexes through small nanopores (\SI{<3}{\nm} diameter)
typically results in the temporary trapping of the entire complex, which has allowed for the study of
polymerase enzymes~\cite{Lieberman-2010,Derrington-2015} and DNA-binding
proteins~\cite{Squires-2015,Yang-2018}. Although promising, none of these approaches could efficiently control
the trapping of the protein inside the nanopore or allow for the observation of enzyme kinetics or
ligand-induced conformational changes.

The energetic landscape of a protein translocating through a nanopore stems directly from the electrostatic,
electrophoretic, electro-osmotic, and steric forces exerted on it~\cite{Muthukumar-2014}. Given the relatively
high motility of proteins, the creation of a long-lasting (\SIrange{10}{100}{\second}), contact-free trap
within a spatial region of a few nanometers mandates the presence of a deep potential energy well within the
nanopore~\cite{Movileanu-2005}. Such a potential profile was achieved by Luchian and co-workers, who showed
that the dwell time of a polypeptide inside the \gls{ahl} pore could be significantly increased by
manipulating the strength of the electro-osmotic flow~\cite{Mereuta-2014,Asandei-2016} or by the placement of
oppositely charged amino acids at the polypeptide's termini~\cite{Asandei-2015}. In a similar approach, a
single barnase enzyme was trapped inside \gls{ahl} \textit{via} the addition of a positively charged
N-terminal tag~\cite{Mohammad-2008}.

Previous work in the Maglia group on protein analysis with nanopores was centered around the biological
nanopore \glsfirst{clya}---a protein with a highly negatively charged interior whose shape can best be
described by a large (\SI{\approx 5.5}{\nm} diameter, \SI{\approx 10}{\nm} height, \cisi{} \lumen{}) and a
small (\SI{\approx 3.3}{\nm} diameter, \SI{\approx 4}{\nm} height, \transi{} constriction) cylinder stacked on
top of each other (\cref{fig:trapping_concept_clya})~\cite{Soskine-2012,Soskine-2013}. Upon capture from the
\cisi{} side of the pore, certain proteins exhibited exceptionally long dwell times inside \gls{clya} from
seconds up to tens of
minutes~\cite{Soskine-2012,Soskine-2013,Soskine-Biesemans-2015,Wloka-2017,VanMeervelt-2017,Galenkamp-2018},
enabling the monitoring of conformational changes~\cite{VanMeervelt-2014, Galenkamp-2018,Biesemans-2015} and
even of the orientation~\cite{VanMeervelt-2014} of the proteins inside the nanopore. A subset of the
investigated proteins, such as lysozyme, Dendra2\_M159A and \glsfirst{dhfr}, resided inside the nanopore
\lumen{} only for hundreds of microseconds and hence could not be
studied~\cite{Soskine-2012,Soskine-Biesemans-2015}. It was observed that the size of the nanopore plays a
crucial role in the effectiveness of protein trapping, as a mere \SI{<10}{\percent} increase of \gls{clya}'s
diameter (\ie~by using \gls{clya} nanopores with a higher oligomeric state) is enough to reduce the dwell time
of proteins by almost three orders of magnitude~\cite{Soskine-2013}. Next to pore size, the charge
distribution of proteins can significantly affect their dwell time inside a nanopore. For example, the binding
of the negatively charged (\SI{-2}{\ec}) inhibitor \gls{mtx} to a modified \gls{dhfr} molecule with positively
charged fusion tag at the C-terminus (\DHFRt{}) increased the dwell time of the protein inside the \gls{clya}
nanopore from \SI{\approx3}{\ms} to \SI{\approx3}{\second} at \SI{-90}{\mV}~\cite{Soskine-Biesemans-2015}.

In this work the immobilization of individual \textit{Escherichia coli} \gls{dhfr} molecules
(\cref{fig:trapping_concept_dhfr}) inside the \gls{clya} biological nanopore (specifically type I
\gls{clya}-AS~\cite{Soskine-2013}, \cref{fig:trapping_concept_clya}) is investigated in detail. Using
nanoscale protein electrostatic simulations as a guideline, our results show that the dwell time of
\DHFR{4}{S}---a molecule identical to the above-mentioned \DHFRt{} aside from the insertion of a single
alanine residue its fusion tag (A174\_A175insA, \cref{fig:trapping_concept_sequence})---inside \gls{clya} can
be increased several orders of magnitude by manipulating the distribution of positive and negative charges on
its surface. To elucidate the physical origin of the trapping mechanism, a double energy barrier model was
developed which---by fitting the voltage dependency of the dwell times for various \gls{dhfr} mutants---yields
direct estimates of the \cisi{} and \transi{} translocation rates and the magnitude of the force exerted by the
electro-osmotic flow on \gls{dhfr}\@. Our method provides an efficient means to increase the dwell time of the
\gls{dhfr} protein inside the \gls{clya} nanopore and suggests a general mechanism to tune the dwell time of
other proteins, which we believe hold significant value for single-molecule sensing and analysis applications.

\section{Results and discussion}
\label{sec:trapping:results_discussion}

\subsection{Phenomenology of {DHFR} trapped inside {ClyA}}
\label{sec:trapping:phenomenology}

To effectively study the enzymes at the single-molecular level, one must be able to collect a statistically
significant number (\ie~typically hundreds) of catalytic cycles from the same enzyme. In the case of the
\textit{E.~coli} \gls{dhfr}, which has a turnover number of \SI{\approx0.08}{\second}~\cite{Kohen-2015}, this
means that the protein must remain trapped inside the pore for tens of seconds. However, as detailed above,
such long dwell times were only achieved for \gls{dhfr} by adding a positively charged polypeptide tag to the
C-terminus of \gls{dhfr}, together with the binding of the negatively charged inhibitor
\gls{mtx}~\cite{Soskine-Biesemans-2015}. Although these long dwell times are encouraging, the requirement for
\gls{mtx} excludes the study of the full enzymatic cycle. Hence, using these previous findings as a starting
point we aim to find out how to prolong the dwell time of a tagged \gls{dhfr} molecule inside the
\gls{clya-as} nanopore without the use of \gls{mtx} and to understand the fundamental physical mechanisms that
determine the escape of \gls{dhfr} from the pore.

The structure of \DHFR{4}{S}, the tagged \gls{dhfr} molecule used as a starting point in this work, can be
roughly divided into a `body', which encompasses the enzyme itself and has a net negative charge,
$\Nbody=\mSI{-10}{\ec}$, and a `tag', which comprises the C-terminal polypeptide extension and bears a net
positive charge, $\Ntag=\mSI{+4}{\ec}$ (\cref{fig:trapping_concept_dhfr,fig:trapping_concept_sequence}). To
capture a tagged \gls{dhfr} molecule, an electric field oriented from \cisi{} to \transi{} (\ie~negative bias
voltage) must be applied across the nanopore which gives rise to an electro-osmotic flow pushing the protein
into the pore ($\forceeo$). The electrophoretic force on the body ($\forceep^{\rm{body}}$) strongly opposes
this electro-osmotic force, but is significantly weakened by the electrophoretic force on the tag
($\forceep^{\rm{tag}}$), allowing the protein to be
captured~\cite{Soskine-2012,Soskine-Biesemans-2015,Biesemans-2015}. Because the body and tag of the \gls{dhfr}
molecules bear a significant amount of opposing charges, it is likely that the molecule will align itself with
the electric field, where the tag is oriented toward the \transi{} side. In this configuration, the body sits
in the \gls{clya} \lumen{} and the tag is located in or near the narrow constriction. Because the body
(\SI{\approx4}{\nm}) is larger than the diameter of the constriction (\SI{3.3}{\nm}), the steric hindrance
between the body and the pore is expected to strongly disfavor full translocation to the \transi{} reservoir,
giving rise to an apparent `steric hindrance' force ($\forcest$). Finally, Poisson-Boltzmann electrostatic
calculations showed that the negatively charged interior of \gls{clya-as} creates a negative electrostatic
potential within both the \lumen{} (\SI{\approx-0.3}{\kTe}) and the constriction (\SI{\approx-1}{\kTe}) of the
pore~\cite{Franceschini-2016}, which will result in unfavorable and favorable interactions with the body and
the tag, respectively.

\begin{figure*}[p]
  \centering
  \begin{minipage}{6cm}
    \begin{subfigure}[t]{5.5cm}
      \caption{}\vspace{-3mm}\label{fig:trapping_apbs_model}
      \includegraphics[scale=1]{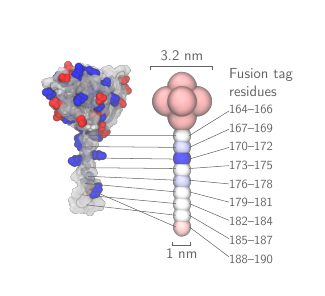}
    \end{subfigure}
    \\
    \begin{subfigure}[t]{5.5cm}
      \caption{}\vspace{-3mm}\label{fig:trapping_apbs_energy}
      \includegraphics[scale=1]{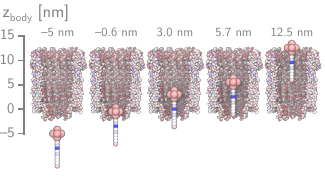}
      \includegraphics[scale=1]{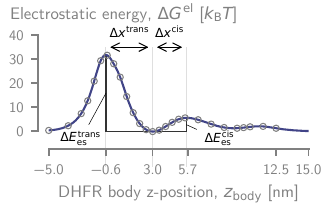}
    \end{subfigure}
  \end{minipage}
  \hspace{-5mm}
  \begin{minipage}{6cm}
    \begin{subfigure}[t]{5.5cm}
      \caption{}\vspace{-3mm}\label{fig:trapping_apbs_barrier_body}
      \includegraphics[scale=1]{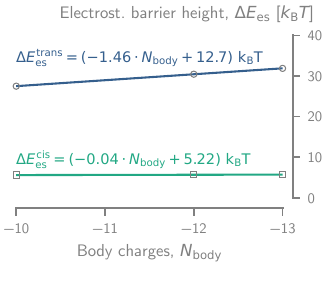}
    \end{subfigure}
    \\
    \begin{subfigure}[t]{5.5cm}
      \caption{}\vspace{-3mm}\label{fig:trapping_apbs_barrier_tag}
      \includegraphics[scale=1]{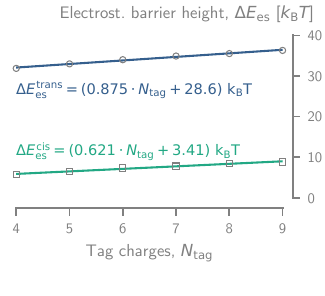}
    \end{subfigure}
  \end{minipage}
  \caption[Energy landscape of \DHFR{4}{S} inside {ClyA-AS}]{%
    \textbf{Energy landscape of \DHFR{4}{S} inside {ClyA-AS}.}
    (\subref{fig:trapping_apbs_model})
    Coarse-grained model of \DHFR{4}{S} used in the electrostatic energy calculations in \gls{apbs}\@. The
    body of \gls{dhfr} consists of seven negatively charged (\SI{-1.43}{\ec}) beads (\SI{1.6}{\nm} diameter)
    in a spherical configuration (\SI{0.8}{\nm} spacing), whereas the tail is represented by a linear string
    of beads (\SI{1}{\nm} diameter, \SI{0.6}{\nm} spacing), each holding the net charge of three amino acids.
    (\subref{fig:trapping_apbs_energy})
    Electrostatic energy ($\energyelec$) resulting from a series of \gls{apbs} energy calculations where the
    coarse-grained \DHFR{4}{S} bead model is moved along the central axis of the pore. The distances $\dxcis$
    and $\dxtrans$ refer to the distances between the energy minimum near the bottom of the \lumen{}
    ($z_{\rm{body}} = \mSI{3}{\nm}$) and the maximum at, respectively, \cisi{} ($z_{\rm{body}} =
    \mSI{5.7}{\nm}$) and \transi{} ($z_{\rm{body}} = \mSI{-0.6}{\nm}$)
    (\subref{fig:trapping_apbs_barrier_body})
    Although every additional negative charge to the body of \gls{dhfr} increases the \transi{} electrostatic
    barrier by \SI{1.46}{\kT}, it has virtually no effect on the \cisi{} barrier, which increases only by
    \SI{0.04}{\kT} per charge.
    (\subref{fig:trapping_apbs_barrier_tag})
    The addition of a single positive charge to \gls{dhfr}'s tag similarly affects the height of the \transi{}
    and \cisi{}, with increases of \SI{0.875}{\kT} and \SI{0.621}{\kT} per charge, respectively.
  }\label{fig:apbs_simulation_results}
\end{figure*}

\subsection{Energy landscape of {DHFR} in {ClyA}}

To increase the dwell time of \gls{dhfr}---and to generalize our findings for other proteins---it is necessary
to understand how the forces exerted on \gls{dhfr} inside the pore behave as a function of the experimental
conditions (\ie~charge distribution and applied bias). In the absence of specific high-affinity interactions,
\gls{dhfr}'s trapping behavior should be chiefly determined by its electrostatic interactions with the pore,
whereas the external electrophoretic and electro-osmotic forces can be viewed as modifications thereof. Hence,
we will start by investigating the molecule's electrostatic energy landscape within \gls{clya} in equilibrium
where the externally applied electric field vanishes.

To this end, we used \gls{apbs}~\cite{Baker-2001,Dolinsky-2004,Dolinsky-2007,Li-2005} to compute the
electrostatic energy ($\energyelec$) of a simplified bead-like tagged \gls{dhfr} molecule model as it moves
through the pore (\cref{fig:trapping_apbs_model,fig:trapping_apbs_energy}), similar to the approach used in
\cref{ch:electrostatics} (\cref{sec:elec:methods:elec:energy}) to investigate \gls{ssdna}
(\cref{sec:elec:frac:dna}) and \gls{dsdna} (\cref{sec:elec:clya:dna}) translocation through \gls{frac} and
\gls{clya}, respectively. The tagged \gls{dhfr} molecule was reduced to a coarse-grained `bead' model
(\cref{fig:trapping_apbs_model}), where the bulk of the protein (body) was defined by seven negatively charged
beads ($r = \SI{0.8}{\nm}$, $Q_i = \SI{-1.7143}{\ec}$) in a spherical configuration (\SI{0.8}{\nm} spacing);
and the C-terminal fusion tag (tail) was represented by nine smaller beads ($r = \SI{0.5}{\nm}$,
$\partialcharge{}_i = \num{-3}$ to \SI{+3}{\ec} depending on the amount of charges in their corresponding
amino acids) each representing three amino acids in an alpha-helix (\SI{0.6}{\nm}   spacing). Our reasons for
this simplification were two-fold: (1) the high degree of axial symmetry in the bead model resulted in a free
energy that was independent of the precise orientation of \gls{dhfr}, significantly reducing the number of
required computations; and (2) the reduced body size of the coarse-grained model compared to the full atom
model allowed for the placement of \gls{dhfr} along the entire length of the pore without nonphysical overlaps
between the atoms of \gls{clya} and \gls{dhfr} inside the \transi{} constriction, resulting in more realistic
free energies. This allows the body to pass the constriction without necessitating conformational changes,
which cannot be modeled using \gls{apbs}. Hence, this also means that the magnitude of the maxima of the
electrostatic energy landscape, which occur when the charges of the bead model come close to those of the
pore, should be viewed as indicative and not absolute.

Nevertheless, the energy profile of \DHFRt{} (\cref{fig:trapping_apbs_energy}) clearly shows that there is a
significant electrostatic barrier, $\barrier^{\trans}_{\rm{es}}$, to overcome when the body of the \gls{dhfr}
moves through the constriction of the pore. Moreover, we observed a second smaller electrostatic barrier,
$\barrier^{\cis}_{\rm{es}}$, toward the \cisi{} side so that an energetic minimum exists inside \gls{clya} in
which the molecule can reside. The size difference between these two barriers clearly suggests that in the
absence of an external force (\ie~at \SI{0}{\mV} bias) the molecule will exit toward \cisi{} with overwhelming
probability. Nevertheless, the multiple distinct blockade current levels observed experimentally
(see~\cref{fig:trapping_traces}) suggest the presence of multiple energy minima within the pore, of which the
equilibrium energy landscape might provide a potential physical location.

To estimate how the charges on \gls{dhfr} impact its dwell time, we modified the number of charges in the body
from \SIrange{-10}{-13}{\ec} and recomputed the energy landscape using \gls{apbs}
(\cref{fig:trapping_apbs_barrier_body}). We found that the electrostatic energy barrier toward \cisi{} was
largely unaffected (\SI{0.04}{\kT} increase per negative charge) while the barrier for \transi{} exit
increased significantly (\SI{1.46}{\kT} increase per negative charge). The latter is a reflection of the
highly negatively charged and narrow \transi{} constriction of \gls{clya}.

Contrary to the body of \gls{dhfr}, the modification of the charge in the tag from \num{+4} to \SI{+9}{\ec}
influenced the heights of both the \cisi{} and the \transi{} barrier similarly, with increases of
\SI{0.621}{\kT} and \SI{0.875}{\kT} per positive elementary charge, respectively
(\cref{fig:trapping_apbs_barrier_tag}). This behavior can be explained by the fact that, at \gls{dhfr}'s
equilibrium position within the pore, the positively charged tag resides in the highly negatively
electrostatic well present in the \transi{} constriction of the nanopore (\cref{fig:trapping_apbs_energy}).
Moving the molecule from this position in either direction requires this Coulombic attraction to be overcome
which is directly proportional to the number of charges on the tag, irrespective of whether the molecule moves
toward \cisi{} or toward \transi{}.

Note that when an external electric field is applied, the electrophoretic and electro-osmotic forces must be
taken into account. If their net balance is positive (\ie~a net force toward \cisi{}) or negative (\ie~a net
force toward \transi{}), the electrostatic landscape will be tilted upward and downward, respectively
(see~\cref{fig:trapping_apbs_external_biased}). The capture of highly negative charged (\SI{-11}{\ec}) wild
type \gls{dhfr} molecules against the electric field~\cite{Soskine-Biesemans-2015} strongly indicates that the
electro-osmosis outweighs electrophoresis and the energy landscape will be shifted downward at \transi{}
(see~\cref{sec:trapping:biased_landscape}), resulting in higher and lower barrier heights at \cisi{} and
\transi{}, respectively. This effectively deepens the energy minimum, which should manifest as an increase of
\gls{dhfr}'s dwell time.

\subsection{Dwell time measurements}

The entry of a single protein into \gls{clya} results in a temporary reduction of the ionic current from the
`open pore' ($\iopen$) to a characteristic `blocked pore' ($\iblock$) level. Previously, we revealed that the
\gls{dhfr} protein shows a main current blockade with $\iresp = \iblock / \iopen \approx \SI{70}{\percent}$
(see~\cref{fig:trapping_traces}). However, occasionally deeper blocks are observed which most likely represent
the transient visit of \gls{dhfr} to multiple locations inside the nanopore. Here we assume that the dwell
time ($\dwelltime$) is simply given by the time from the initial capture to the final release where the
current level returns to the open pore current.

After gathering sufficient statistics for the dwell time events, we computed the expectation value of
$\dwelltime$ by taking the arithmetic mean of all dwell time events. This is because the chance for an escape
can be modeled as the probability of overcoming a potential barrier whose distribution function is exponential
(see~\cref{sec:trapping_appendix:escape_rates}). Note that even if the molecule transitions through multiple
meta-states with individual rates connecting each of them before it exits, the expectation value is still
given by the arithmetic mean (see~\cref{eq:tau_arithmetic_mean}).

\begin{table}[t]
  \begin{threeparttable}
    \centering
    \footnotesize

    \captionsetup{width=12cm}
    \caption[Mutations and charges of all {DHFR} variants]%
            {Mutations and charges of all {DHFR} variants.}
    \label{tab:dhfr_variants}

    \renewcommand{\arraystretch}{1.15}
    \scriptsize
    
    \begin{tabularx}{12cm}{Xllccc}
      \toprule
        & \multicolumn{2}{c}{Mutations\tnote{a}} &
      \multicolumn{3}{c}{$\chargeq_{\rm{DHFR}}$\tnote{b} [\si{\ec}]} \\
      \cmidrule(r){2-3}  \cmidrule(l){4-6}
      Name  & Body & Tag & Body & Tag & Total \\
      \midrule
      \DHFR{4}{S}   &  --- & ---
                    & \num{-10} & \num{+4} & \num{-6} \\
      \DHFR{4}{I}   & V88E P89E & ---
                    & \num{-12} & \num{+4} & \num{-8} \\
      \DHFR{4}{C}   & A82E A83E & ---
                    & \num{-12} & \num{+4} & \num{-8} \\
      \DHFR{4}{O1}  & E71Q & ---
                    & \num{-12} & \num{+4} & \num{-8} \\
      \DHFR{4}{O2}  & T68E R71E & ---
                    & \num{-13} & \num{+4} & \num{-9} \\
      \midrule
      \DHFR{5}{O1}  & E71Q      & A175K
                    & \num{-12} & \num{+5} & \num{-7} \\
      \DHFR{7}{O1}  & E71Q      & A175K A174K A176K
                    & \num{-12} & \num{+7} & \num{-5} \\
      \DHFR{5}{O2}  & T68E R71E & A175K
                    & \num{-13} & \num{+5} & \num{-8} \\
      \DHFR{6}{O2}  & T68E R71E & A175K A174K
                    & \num{-13} & \num{+6} & \num{-7} \\
      \DHFR{7}{O2}  & T68E R71E & A175K A174K A176K
                    & \num{-13} & \num{+7} & \num{-6} \\
      \DHFR{8}{O2}  & T68E R71E & A175K A174K A176K A169K
                    & \num{-13} & \num{+8} & \num{-5} \\
      \DHFR{9}{O2}  & T68E R71E & A175K A174K A176K A169K L177K
                    & \num{-13} & \num{+9} & \num{-4} \\
      \bottomrule
    \end{tabularx}

    \begin{tablenotes}
      \item[a] W.r.t. \DHFR{4}{S}: body residues \numrange{1}{163} and tag residues \numrange{164}{190};
      \item[b] Net charge of \gls{dhfr} at \pH{7.5}.
    \end{tablenotes}

  \end{threeparttable}
\end{table}

We observed before that the dwell time of tagged \gls{dhfr} molecules depends strongly on the applied
bias~\cite{Biesemans-2015}. That is, exponentially rising with voltage until a certain bias---which we will
refer to as the \emph{threshold voltage}---followed by an exponential fall. This behavior has also been
observed for charged peptides in \gls{ahl}~\cite{Movileanu-2005} and is typical for a decay of a bound state
into multiple final states, such as an escape to either \cisi{} or \transi{}
(see~\cref{sec:trapping_appendix:escape_rates}). Therefore, the dwell time of the molecule, as a function of
bias voltage, $\vbias$, can be expressed as the inverse of the sum of two escape rates~\cite{Movileanu-2005}
\begin{align}\label{eq:double_barrier_simple}
  \dfrac{1}{\dwelltime} = \rate ={}& \rate^\cis + \rate^\trans \\
    ={}&
    \rate^\cis_0 \exp \left( -\dfrac{\alpha^\cis \ec \vbias}{\kbt} \right) +
    \rate^\trans_0 \exp \left( \dfrac{\alpha^\trans \ec \vbias}{\kbt} \right)
  \text{ ,}
\end{align}
where $\rate^{\cis/\trans}$ are the molecule's escape rates toward \cisi{} and \transi{}, respectively. These
can be further decomposed into attempt frequencies $\rate^{\cis/\trans}_0$ and bias-dependent barriers in the
exponentials. Although this equation can help to qualitatively describe the experimental data, the reduction
of the entire protein-nanopore system to four parameters does not allow for their physical interpretation.

\begin{figure*}[p]
  \centering

	\begin{subfigure}[t]{4cm}
		\centering
		\caption{}\vspace{-5mm}\label{fig:trapping_dwell_times_body_model}
		\includegraphics[scale=1]{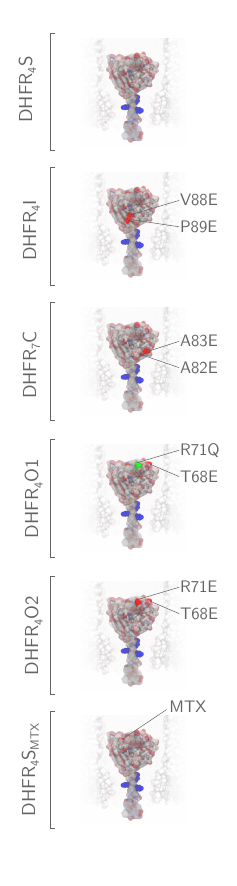}
  \end{subfigure}
	\begin{subfigure}[t]{6.5cm}
		\centering
		\caption{}\vspace{-3mm}\label{fig:trapping_dwell_times_body_data}
    \includegraphics[scale=1]{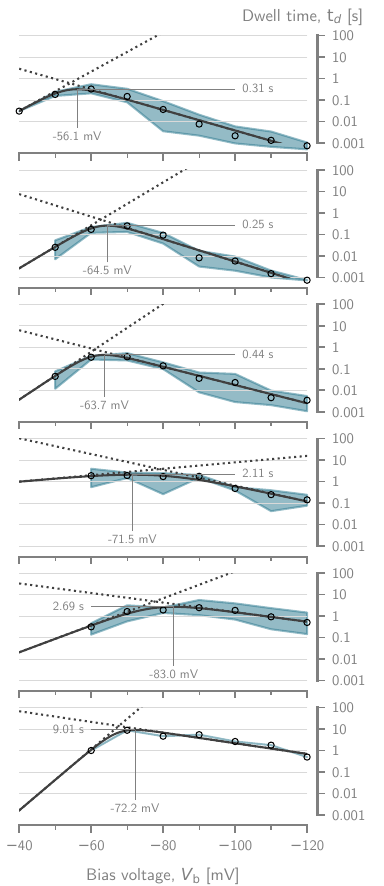}
  \end{subfigure}

	\caption[Effect of the body charge on the dwell time of tagged {DHFR}]{%
    \textbf{Effect of the body charge on the dwell time of tagged {DHFR}\@.}
    (\subref{fig:trapping_dwell_times_body_model})
    Surface representation of the five tested \DHFR{4}{X} body charge mutants and the \DHFR{4}{S}+MTX
    complex. The mutated residues are indicated for each variant. The positive charges in the fusion tag are
    colored blue. From top to bottom: \DHFR{4}{S}, \DHFR{4}{I}, \DHFR{4}{C}, \DHFR{4}{O1}, \DHFR{4}{O2}, and
    \DHFR{4}{S}$_{\text{MTX}}$.
    (\subref{fig:trapping_dwell_times_body_data})
    Voltage dependence of the average dwell time ($\dwelltime$) inside \gls{clya-as} for \gls{dhfr} mutants in
    (\subref{fig:trapping_dwell_times_body_model}). The solid lines represent the voltage dependency predicted
    by fitting the double barrier model given by \cref{eq:double_barrier_simple} to the data
    (see~\cref{tab:fitting_parameters_simple}). The dotted lines represent the dwell times due to the \cisi{}
    (low to high) and \transi{} (high to low) barriers. The threshold voltages at the maximum dwell time were
    estimated by inserting the fitting parameters into \cref{eq:threshold_voltage_simple}. The error envelope
    represents the minimum and maximum values obtained from repeats at the same condition. All measurements
    were performed at \SI{\approx28}{\celsius} in aqueous buffer at \pH{7.5} containing \SI{150}{\mM}
    \ce{NaCl}, \SI{15}{\mM} \ce{Tris-HCl}. Current traces were sampled at \SI{10}{\kilo\hertz} and filtered
    using a low-pass Bessel filter with a \SI{2}{\kilo\hertz} cutoff.
  }\label{fig:trapping_dwell_times_body}
\end{figure*}

\subsection[Engineering {DHFR}'s dwell time by manip. of its charge]%
           {Engineering {DHFR}'s dwell time by manipulation of its charge}

The results from the \gls{apbs} simulations, together with the previous work with \DHFRt{} and
\gls{mtx}~\cite{Soskine-Biesemans-2015}, suggest that the dwell time of \gls{dhfr} in \gls{clya} can be
increased by the manipulation of its charge distribution. To achieve the increase in dwell time without the
need for \gls{mtx}, several non-conserved amino acids on the surface of \DHFR{4}{S} were identified and
mutated to negatively charged glutamate residues, resulting in the molecules \DHFR{4}{I}, \DHFR{4}{C},
\DHFR{4}{O1}, and \DHFR{4}{O2} (\cref{tab:dhfr_variants} and \cref{fig:trapping_dwell_times_body_model}).
These mutations modify the number of charges in the body compared to \DHFR{4}{S} and their charges are also in
different locations. For convenience, this series of mutations will be referred to as the \emph{body charge
variations} from here on out.

We performed ionic current measurements for all body charge variations for a wide range of bias voltages
(\SIrange{-40}{-120}{\mV}, see~\cref{fig:trapping_traces,fig:trapping_blockades}) and extracted the dwell
times as shown in \cref{fig:trapping_dwell_times_body_data}. All body charge variations showed the same
increase in the dwell time at low electric fields and decreased at high fields. However, we observed
differences in the threshold voltage and the magnitude of the maximum dwell time. These differences cannot
simply be explained by the total number of charges as \DHFR{4}{I} and \DHFR{4}{C} have the same charge as
\DHFR{4}{O1} but their dwell times are 10-fold lower (\cref{fig:trapping_dwell_times_body_data}). This result
implies that the location of the body charge on \gls{dhfr} plays an important role.

\begin{figure*}[p]
  \centering

	\begin{subfigure}[t]{4.5cm}
		\centering
		\caption{}\vspace{-5mm}\label{fig:trapping_dwell_times_tag_model}
    \includegraphics[scale=1]{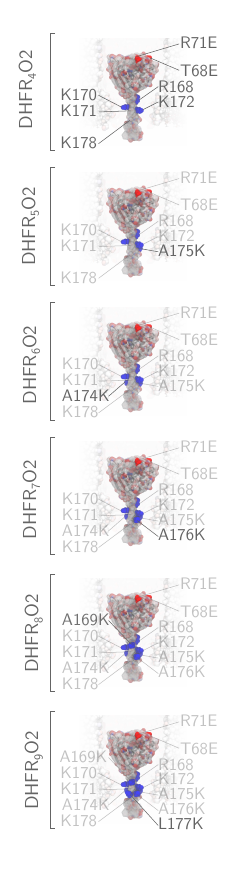}
  \end{subfigure}
  \begin{subfigure}[t]{6.5cm}
		\centering
		\caption{}\vspace{-3mm}\label{fig:trapping_dwell_times_tag_data}
    \includegraphics[scale=1]{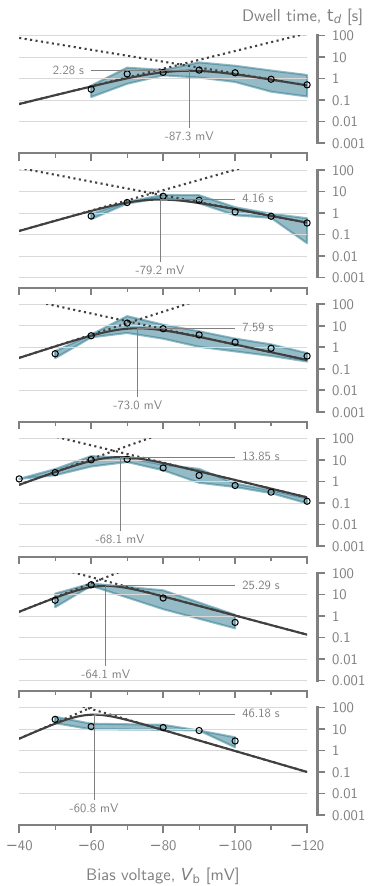}
  \end{subfigure}

	\caption[Effect of the tag charge on the dwell time of \DHFR{$\Ntag$}{O2}]{%
    \textbf{Effect of the tag charge on the dwell time of \DHFR{$\Ntag$}{O2}.}
    (\subref{fig:trapping_dwell_times_tag_model})
    Surface representations of all \DHFR{$\Ntag$}{O2} mutants going from $\Ntag=4$ (top) to $\Ntag=9$
    (bottom). The positively charged residues in the tag have been annotated and highlighted in blue.
    (\subref{fig:trapping_dwell_times_tag_data})
    Voltage dependencies of the mean dwell time ($\dwelltime$) for the mutant on the left-hand side, fitted
    with the double barrier model of \cref{eq:double_barrier_complex}. The annotated threshold voltages were
    computed by ESI \cref{eq:threshold_voltage_complex}. Solid lines represent the double barrier dwell time,
    and the dotted lines show the dwell times due to the \cisi{} (low to high) and \transi{} (high to low)
    barriers. Fitting parameters can be found in \cref{tab:fitting_params_complex}. The error envelope
    represents the minimum and maximum values obtained from repeats at the same condition. Experimental
    conditions are the same as those in \cref{fig:trapping_dwell_times_body}.
  }\label{fig:trapping_dwell_times_tag}
\end{figure*}

Additional body mutations could potentially compromise the catalytic cycle of \gls{dhfr}\@. Hence, we
proceeded by systematically increasing the number of positive charges to the fusion tag ($\Ntag$) of
\DHFR{4}{O2}, the variant that exhibited the longest dwell time, \textit{via} lysine substitution from
\SIrange{+4}{+9}{\ec} (\cref{tab:dhfr_variants,fig:trapping_dwell_times_tag}). The resulting
\DHFR{$\Ntag$}{O2} mutants will be referred to as the \emph{tag charge variations}.

Subsequent characterization of their dwell times revealed that the addition of positive charges to the tag
significantly increased \gls{dhfr}'s dwell time (\cref{fig:trapping_dwell_times_tag_data}). We observed a
similar increase for \DHFR{4}{O1} variants with \num{+5} and \num{+7} tag charge numbers
(see~\cref{fig:trapping_model_comparison_o1_simple}). This behavior is consistent with the tag being trapped
electrostatically inside the negatively charged \transi{}
constriction~\cite{Franceschini-2016,Movileanu-2005,Asandei-2015,Asandei-2016} and it suggests that the tag
plays a crucial role in the trapping of \gls{dhfr}, which was already observed in previous
work~\cite{Soskine-Biesemans-2015}.

\subsection{Binding of {NADPH} reveals that {DHFR} remains folded inside the pore}

\begin{figure*}[!t]
  \centering
  \begin{subfigure}[t]{6.25cm}
    \centering
    \caption{}\vspace{-3mm}\label{fig:trapping_nadph_trace}
    \includegraphics[scale=1]{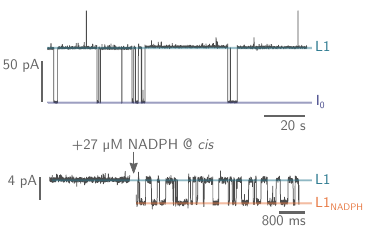}
  \end{subfigure}
  \begin{subfigure}[t]{5cm}
    \centering
    \caption{}\vspace{-3mm}\label{fig:trapping_nadph_ires}
    \includegraphics[scale=1]{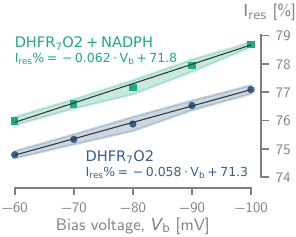}
  \end{subfigure}

  \caption[Binding of NADPH to \DHFR{7}{O2}.]{%
    \textbf{Binding of NADPH to \DHFR{7}{O2}.}
    (\subref{fig:trapping_nadph_trace})
    Top: Typical current trace after the addition of \SI{50}{\nM} \DHFR{7}{O2} to a single \gls{clya-as}
    nanopore added to the \cisi{} reservoir at \SI{-60}{\mV} applied potential. The open-pore current
    (`$\iopen$') and the blocked pore levels (`L1') are highlighted. Bottom: Current trace showing the blocked
    pore current of a single \DHFR{7}{O2} molecule (\SI{50}{\nM}, \cisi) at \SI{-60}{\mV} applied potential
    before (left) and after (right) the addition of \SI{27}{\uM} \gls{nadph} to the \transi{} compartment.
    \Gls{nadph} binding to confined \gls{dhfr} molecule is reflected by current enhancements from the unbound
    `L1' to the \gls{nadph}-bound `L1\textsubscript{NADPH}' current levels, and showed association
    ($\rate_{\rm{on}}$) and dissociation ($\rate_{\rm{off}}$) rate constants of
    \SI{2.03\pm0.58e6}{\per\Molar\per\second} and \SI{71.2\pm20.4}{\per\second}, respectively
    (see~\cref{tab:nadph_rates}).
    (\subref{fig:trapping_nadph_ires})
    Dependence of the $\iresp$ on the applied potential for \DHFR{7}{O2} and \DHFR{7}{O2} bound to
    \gls{nadph}\@. All current traces were collected in \SI{250}{\mM} \ce{NaCl} and \SI{15}{\mM}
    \ce{Tris-HCl}, \pH{7.5}, at \SI{23}{\celsius}, by applying a Bessel low-pass filter with a
    \SI{2}{\kilo\hertz} cutoff and sampled at \SI{10}{\kilo\hertz}.
  }\label{fig:trapping_nadph}

\end{figure*}

To verify that our \gls{dhfr} variants remained folded inside the nanopore, we measured and analyzed the
binding of \gls{nadph} to the enzyme. The addition of the \gls{nadph} co-factor to the \transi{} solution of
nanopore-entrapped \gls{dhfr} molecules induced reversible ionic current enhancements that reflect the binding
and unbinding of the co-factor to the protein (\cref{fig:trapping_nadph_trace} and
\cref{fig:trapping_nadph_o2_traces,tab:nadph_rates}).

Not all \gls{dhfr} variants were found to be suitable for \gls{nadph}-binding analysis: \DHFR{5}{O2} did not
dwell long enough inside \gls{clya-as} at \SI{-60}{\mV} ($\dwelltime = \SI{0.32\pm0.17}{\second}$) to allow a
detailed characterization of \gls{nadph} binding, whereas \gls{nadph}-binding events to \DHFR{8}{O2} were too
noisy for a proper determination of $\rate_{\rm{on}}$ and $\rate_{\rm{off}}$. No \gls{nadph} binding events to
\DHFR{9}{O2} could be observed.  \gls{nadph}-binding events to the other \gls{dhfr} variants (\DHFR{5}{O2},
\DHFR{6}{O2}, and \DHFR{7}{O2}) showed similar values for $\rate_{\rm{on}}$, $\rate_{\rm{off}}$, and event
amplitude (\cref{tab:nadph_rates}), suggesting that the binding of \gls{nadph} to \gls{dhfr} inside the
\gls{clya-as} nanopore is not affected by the number of positive charges in the C-terminal fusion tag.
Possibly, the inability of \DHFR{8}{O2} and \DHFR{9}{O2} to bind the substrate is due to the lodging of
\gls{dhfr} closer to the \transi{} constriction.

Work with solid-state nanopores also previously reported that electric fields inside a nanopore may unfold
proteins during translocation~\cite{Talaga-2009}, suggesting that the high degree of charge separation between
the body and tag of \gls{dhfr} might destabilize its structure. To further investigate the effect of the
applied potential on the protein structure, we analyzed the dependency of the residual current on the applied
potential (\cref{fig:trapping_nadph}). We found that the residual current of both the apo-\gls{dhfr} and the
ligand-bound enzyme increased by \SI{\approx2.5}{\percent} from \SIrange{-60}{-100}{\mV}. A voltage-dependent
change in residual current is compatible with a force-induced stretching of the enzyme. However,
single-molecule force spectroscopy experiments showed that \gls{nadph} binding increases the force required to
unfold the protein by more than 3-fold from \num{27} to \SI{98}{\pN}~\cite{Ainavarapu-2005}. As the change of
residual current over the potential was identical for both apo- and ligand-bound \gls{dhfr}
(\cref{fig:trapping_nadph}), a likely explanation is that, rather than stretching \gls{dhfr}, the applied bias
changes the position of \gls{dhfr} within the nanopore. Hence, our data suggest that, as previously reported
for several other proteins~\cite{VanMeervelt-2017,Galenkamp-2018}, the protein remains folded at different
applied bias voltages.

\subsection{Double barrier model for the trapping of {DHFR}}

Puzzled by the strong dependence of the dwell time on the tag charge, we set out to understand the underlying
trapping mechanism by building a quantitative model. To this end, we will focus on the data set of the dwell
time of \DHFR{$\Ntag$}{O2} shown in \cref{fig:trapping_dwell_times_tag_data}.

We propose a double barrier model that describes the trapping of the molecule as a combination of escape rates
toward \cisi{} and toward \transi{} (see~\cref{sec:trapping_appendix:double_barrier}). Similar to
\cref{eq:double_barrier_simple}, the dwell time is defined in terms of the rate $\rate$ which in turn is given
by the sum of the rate for \cisi{} exit and the rate for \transi{} exit. However, now we define the rates in
terms of energy barriers:
\begin{align}\label{eq:double_barrier}
    \frac{1}{\dwelltime} = \rate =
        \rate_0 \exp \left( - \dfrac{\barrier^{\cis}}{\kbt} \right)
        + \rate_0 \exp \left( - \dfrac{\barrier^{\trans}}{\kbt} \right)
    \text{ ,}
\end{align}
where $\rate_0$ is the attempt rate and $\barrier^{\cis/\trans}$ are the energy barriers the molecule has to
overcome in order to escape toward \cisi{} and \transi{}, respectively. These can be readily decomposed into
\emph{steric}, \emph{electrostatic}, and \emph{external} contributions:
\begin{subequations}\label{eq:decomposition}
\begin{align}
  \barrier^{\cis} ={}&
    \barrier^{\cis}_{\steric,0}
    + \barrier^{\cis}_\static
    + \barrier^{\cis}_\ext \text{ , and,} \\
  \barrier^{\trans} ={}&
    \barrier^{\trans}_{\steric,0}
    + \barrier^{\trans}_\static
    + \barrier^{\trans}_\ext
  \text{ .}
\end{align}
\end{subequations}
The steric components $\barrier^{\cis/\trans}_{\steric,0}$ are defined as those interactions of the molecule
with the nanopore that are not electrostatic in nature, such as size- or conformation-related effects as
\gls{dhfr} translocates through the narrow constriction toward \transi{}.

Supported by the \gls{apbs} simulations (\cref{fig:trapping_apbs_energy}) and the corresponding barrier
height to tag charge dependency analyses (\cref{fig:trapping_apbs_barrier_tag}), we infer that the
electrostatic components $\barrier^{\cis/\trans}_{\static}$ can be further decomposed as
\begin{subequations}\label{eq:static-barrier}
\begin{align}
	\barrier^{\cis}_\static ={}&
			\barrier^{\cis}_{\static,0}
			+ \Ntag \ec \potbar_{\rm{tag}}^{\cis},\\
	\barrier^{\trans}_\static ={}&
			\barrier^{\trans}_{\static,0}
      + \Ntag \ec \potbar_{\rm{tag}}^{\trans}
  \text{ ,}
\end{align}
\end{subequations}
where $\potbar_{\rm{tag}}^{\cis/\trans}$ are the electrostatic potentials associated with the tag charge
$\Ntag$ for the \cisi{} and \transi{} barriers (\ie~the change in barrier height per additional charge in
$\Ntag$) and $\barrier^{\cis/\trans}_{\static,0}$ are two constant terms that combine all electrostatic
interactions between the protein and the pore that do not depend on $\Ntag$ (\eg~body charge related
interactions with the electric fields in the nanopore).

\begin{table}[t]
  \centering
  \begin{threeparttable}
    \footnotesize
    \centering

    \captionsetup{width=12cm}
    \caption[Fitting parameters for \DHFR{$\Ntag$}{O2}]%
            {Fitting parameters for \DHFR{$\Ntag$}{O2}.}
    \label{tab:fitting_params_complex}

    \renewcommand{\arraystretch}{1.5}
    \scriptsize
  
    \begin{tabularx}{12cm}{XXll}
      \toprule
      Parameter   & Description & Type  & Value\tnote{a} \\
      \midrule
      $\vbias$
        & Applied bias voltage
        & independent & \SIrange{40}{120}{\mV} \\
      $\Ntag$
        & Tag charge number
        & independent & \SIrange{4}{9}{} \\
      $\Nbody$
        & Body charge number
        & fixed       & \SI{-13}{} \\
      $\Neo$
        & Equivalent osmotic charge number
        & dependent   & \SI{15.5\pm0.9}{} \\
      $L$
        & Nanopore length
        & fixed       & \SI{14}{\nm} \\
      $\dxtrans$
        & Distance to \transi{} barrier\tnote{b}
        & fixed       & \SI{3.5}{\nm} \\
      $\dxcis$
        & Distance to \cisi{} barrier\tnote{b}
        & dependent   & \SI{5.21\pm1.32}{\nm} \\
      $\potbar_{\rm{tag}}^{\trans}$
        & Change of $\barrier^{\trans}_\static$ with tag charge
        & dependent   & \SI{0.860\pm0.078}{\kbt\per\ec} \\
      $\potbar_{\rm{tag}}^{\cis}$
        & Change of $\barrier^{\cis}_\static$ with tag charge
        & dependent   & \SI{0.218\pm0.167}{\kbt\per\ec} \\
      $\ln (\rateft/\si{\hertz})$
        & Effective attempt rate for the \transi{} barrier
        & dependent   & \num{-3.44\pm1.24} (\SI{3.21e-2}{\hertz}) \\
      $\ln (\ratefc/\si{\hertz})$
        & Effective attempt rate for the \cisi{} barrier
        & dependent   & \num{7.39\pm1.02} (\SI{1.62e3}{\hertz}) \\
      \bottomrule
    \end{tabularx}
  
    \begin{tablenotes}
      \item[a] Errors are confidence intervals for one standard deviation.
      \item[b] Relative to the energetic minimum inside the pore. 
    \end{tablenotes}
  
    \sisetup{table-format=2.4}
  \end{threeparttable}
  \end{table}

The external forces acting on a protein trapped inside \gls{clya} under applied bias voltages manifest in the
barrier contribution $\barrier^{\cis/\trans}_\ext$. They comprise an \emph{electrophoretic} component
$\barrier^{\cis/\trans}_\ep$ and an \emph{electro-osmotic} component $\barrier^{\cis/\trans}_\eo$. The former
results from the strong electric field (\SI{\approx3.5e6}{\volt\per\meter} at \SI{-50}{\mV}) and the nonzero
net charge on the molecule, whereas the latter springs from the force exerted by \gls{clya}'s electro-osmotic
flow, which is strong enough to allow the capture of negatively charged proteins even in opposition to the
electrophoretic force~\cite{Soskine-2012,Soskine-2013,Soskine-Biesemans-2015,Biesemans-2015}. It is assumed
that the bias potential changes linearly over the length of the pore, and hence the external energy barriers
are given by (see~\cref{sec:trapping_appendix:escape_rates})
\begin{subequations}\label{eq:external-barrier}
\begin{align}
  \barrier^{\cis}_\ext ={}&
    \barrier^{\cis}_\ep + \barrier^{\cis}_\eo =
    -(\Nnet + \Neo) \ec \dfrac{\dxcis}{L} \vbias \text{ , and,}\\
  \barrier^{\trans}_\ext ={}&
    \barrier^{\trans}_\ep + \barrier^{\trans}_\eo =
    +(\Nnet + \Neo) \ec \dfrac{\dxtrans}{L} \vbias
    \text{ ,}
\end{align}
\end{subequations}
where $\Nnet = \Nbody + \Ntag$ is the total number of charges on \gls{dhfr}\@, $L$ is the length of the
nanopore (\SI{14}{\nm}), and $\vbias$ is the negative applied bias.  The strength of the electro-osmotic force
is defined by the \emph{equivalent osmotic charge number} $\Neo$---the number of charges that must be added to
\gls{dhfr} to create an equal electrophoretic force on the molecule. Defining the electro-osmotic force in
terms of an equivalent osmotic charge number reveals its complete analogy to an electrophoretic force, which
has the benefit that the magnitudes of both forces can be readily compared. Moreover, the equivalent osmotic
charge number is an invariant related solely to the size and shape of the molecule.

\begin{figure*}[p]
  \centering
  \begin{subfigure}[t]{27.5mm}
		\centering
		\caption{}\vspace{-3mm}\label{fig:trapping_apbs_external_model}
    \includegraphics[scale=1]{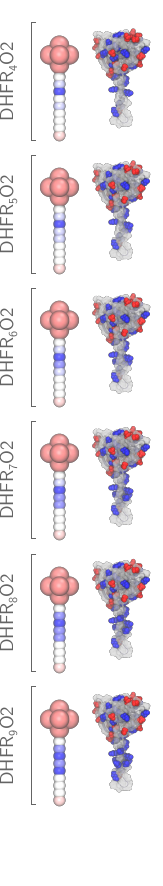}
  \end{subfigure}
  \begin{subfigure}[t]{42.5mm}
		\centering
    \caption{}\vspace{-3mm}%
    \label{fig:trapping_apbs_external_equilibrium}
    \includegraphics[scale=1]{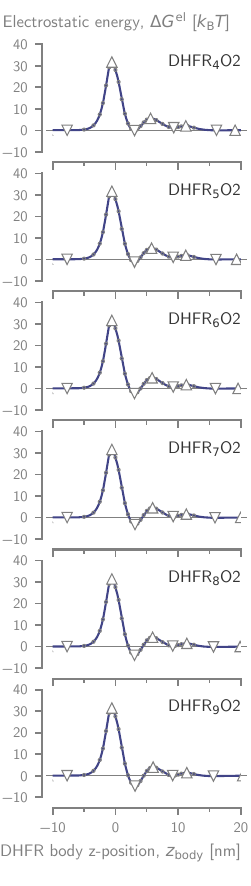}
  \end{subfigure}
  \begin{subfigure}[t]{42.5mm}
		\centering
    \caption{}\vspace{-3mm}%
    \label{fig:trapping_apbs_external_biased}
    \includegraphics[scale=1]{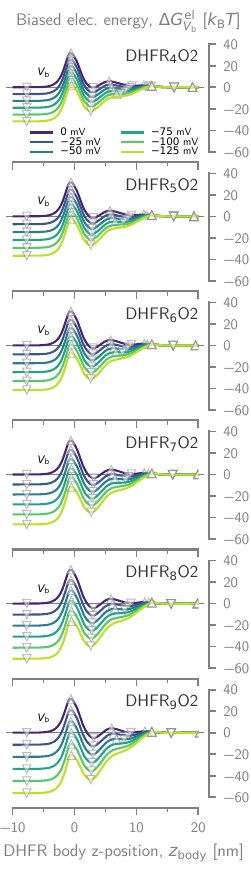}
  \end{subfigure}
  \caption[Electrostatic landscape of a \DHFRt{} bead model in {ClyA-AS}]%
    {%
      \textbf{Electrostatic landscape of a \DHFRt{} bead model in {ClyA-AS}.}
      (\subref{fig:trapping_apbs_external_model})
      Simplified bead model of all \DHFR{$\Ntag$}{O2} in comparison with their corresponding full-atom
      homology model.
      (\subref{fig:trapping_apbs_external_equilibrium})
      The simulated electrostatic energy landscapes of all \DHFR{$\Ntag$}{O2} variants. Increasing the number
      of positive charges in the tag deepens the energetic minimum at $z \approx \mSI{3}{\nm}$.
      (\subref{fig:trapping_apbs_external_biased})
      Approximation of the `tilting' of the energy landscapes from
      (\subref{fig:trapping_apbs_external_equilibrium}) by an applied bias voltage, which acts on the total
      effective charge of \gls{dhfr} through a constant electric field (\ie~linear potential drop) inside the
      pore (\cref{eq:external_energy}). Because the body is negative, increasing the number of positive
      charges in the tag decreases the electrophoretic force countering the electro-osmotic force, resulting
      in a higher degree of tilting. Downward and upward-facing triangles indicate the presence of a local minimum
      and maximum, respectively.
  }\label{fig:trapping_apbs_external}
\end{figure*}

The quantities $\Delta x^{\cis/\trans}$ are defined as the distances from the electrostatic energy minimum to
the \cisi{} and \transi{} barriers, which depend on the energetic landscape of \gls{clya} and on the precise
location of residence of \gls{dhfr} within the pore. To estimate these values, we can use the APBS simulations
(\cref{fig:trapping_apbs_energy}) from which we can read off that $\dxtrans \approx \SI{3.5}{\nm}$. The
\cisi{} distance is more difficult to define as the \cisi{} electrostatic barrier is much shallower. Without
external fields, it has a distance of about $\SI{\approx2.7}{\nm}$, but as we shall see in
\cref{fig:trapping_apbs_external_biased}, when the energy landscape is tilted by an external force, the
barrier that needs to be overcome is actually located at the \cisi{} entrance of the pore. In practice,
$\dxcis$ will need to be adjusted to a value between these two possibilities to give an adequate estimate and
hence will be left as a fitting parameter.

Inserting \cref{eq:decomposition,eq:static-barrier,eq:external-barrier} into \cref{eq:double_barrier} yields
the final dwell time model:
\begin{equation}\label{eq:double_barrier_complex}
	\begin{split}
    \frac{1}{\dwelltime} = \rate ={}&
	\ratefc \exp{%
		\left(
			-\dfrac{%
				\Ntag \ec \potbar_{\rm{tag}}^{\cis}
				- (\Nnet + \Neo) \ec \dfrac{\dxcis}{L} \vbias}{\kbt}
		\right)
		} \\
	&\enspace +
	\rateft \exp{%
		\left(
			-\dfrac{%
				\Ntag \ec \potbar_{\rm{tag}}^{\trans}
				+ (\Nnet + \Neo) \ec \dfrac{\dxtrans}{L} \vbias}{\kbt}
		\right)
    }
    \text{ ,}
	\end{split}
\end{equation}
where the static terms are absorbed into the prefactor to form the effective \cisi{} and \transi{} barrier
attempt rates $\rate_{\rm{eff}}^{\cis/\trans}$. The formulation of \cref{eq:double_barrier_complex} offers a
compact description of the most salient features of the molecule-nanopore system, and it enables us to
describe the dwell time of \gls{dhfr} inside \gls{clya} quantitatively as a function of the physical
properties of the system. Fitting this model to all \DHFR{$\Ntag$}{O2} data simultaneously---with both
$\vbias$ and $\Ntag$ as independent variables---leads to the fitting values in
\cref{tab:fitting_params_complex} and the plots in \cref{fig:trapping_dwell_times_tag_data}, which show
excellent accuracy considering the simplicity of our model. This is a strong indication that we captured the
essence of the trapping mechanism within our model.

\subsection{Effect of tag charge and bias voltage on the energy landscape}
\label{sec:trapping:biased_landscape}

Our equilibrium electrostatics simulations showed the existence of an electrostatic energy minimum
(\cref{fig:trapping_apbs_external_equilibrium}) at the bottom of \gls{clya}'s \lumen{} ($z = \SI{3}{\nm}$),
flanked by two maxima, located at the \transi{} constriction ($z = \SI{-0.6}{\nm}$), and at the middle of the
\cisi{} \lumen{} ($z = \SI{5.7}{\nm}$). Hence, when \gls{dhfr} resides at the electrostatic potential minimum
inside the nanopore, the largest electrostatic barrier is given by the narrower and negatively charged
\transi{} constriction while the barrier at the \cisi{} side is much shallower. Under the influence of an
external force (\ie~the bias voltage), however, the entire energy landscape will become `tilted', reducing
some barriers and enhancing others. The biased energy $\energyelec_{\vbias}$ is thus
\begin{align}\label{eq:external_energy}
  \energyelec_{\vbias} =  \energyelec + \Eext
  \text{ ,}
\end{align}
where $\Eext$ is the energy of the \gls{dhfr} molecule due to the external bias voltage. Assuming the bias
voltage changes linearly from \cisi{} to \transi{}, $\Eext$ can be approximated as
\begin{align}\label{eq:external_energy_model}
  \Eext =
  \begin{cases}
      \Ntot \dfrac{\vbias}{\SI{14}{\nm}} (z - \SI{11}{\nm})
              & \text{for}\; -3 < z < \SI{11}{\nm} \text{ ,} \\
      0,      & \text{for}\;  z >= \SI{11}{\nm} \text{ ,}\\
      \Ntot \vbias, & \text{for}\;  z <  \SI{-3}{\nm} \text{ ,}
  \end{cases}
\end{align}
with $\Ntot = \Nbody + \Ntag + \Neo$ the effective charge of charge of \gls{dhfr}. In the case of
\DHFR{$\Ntag$}{O2} (\cref{fig:trapping_apbs_external_model}), the net charge of the protein is negative
($\Nbody = -13$) and the electro-osmotic flow exerts an opposing force ($\Neo = 15.5$). This means that
increasing the number of positive charges in the tag decreases the net electrophoretic force and thus leads to
a more pronounced tilting of the entire energy landscape, deepening the electrostatic minimum
(cf.~\cref{fig:trapping_apbs_external_equilibrium,fig:trapping_apbs_external_biased}). This increases the
barrier heights at both the \cisi{} and \transi{} sides similarly, which results in the lowering of both the
\cisi{} and \transi{} escape rates and hence longer dwell times. Because the \cisi{} energy barrier in the
\lumen{} is relatively shallow, it disappears at moderate applied voltages (\SI{>-50}{\mV}). This indicates
that, under a negative applied bias voltage, the location of the \cisi{} barrier is voltage-dependent and
hence the distance $\dxcis$ is not well defined. The \cisi{} barrier at low and high biases it is located at
respectively the middle of the \lumen{} ($z \approx \mSI{5.7}{\nm}$) and the \cisi{} entry ($z \approx
\mSIrange{10}{13}{\nm}$).

\begin{figure*}[b]
  \centering

  \begin{subfigure}[t]{5.5cm}
    \centering
    \caption{}\vspace{-3mm}\label{fig:trapping_translocation_voltage}
    \includegraphics[scale=1]{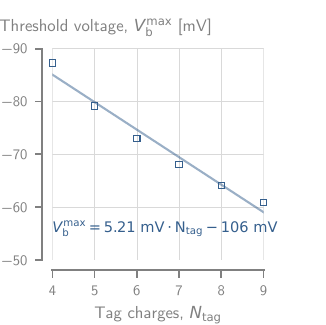}
  \end{subfigure}
  \begin{subfigure}[t]{5.5cm}
    \centering
    \caption{}\vspace{-3mm}\label{fig:trapping_translocation_probability}
    \includegraphics[scale=1]{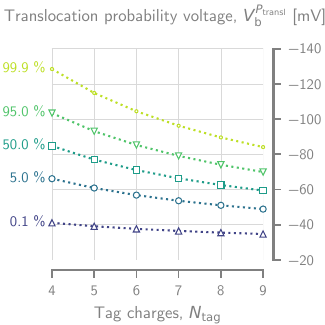}
  \end{subfigure}

  \caption[Tag charge depend. of the threshold voltage and transl. probability]{%
    \textbf{Tag charge dependence of the threshold voltage and translocation probability.}
    (\subref{fig:trapping_translocation_voltage})
    Every additional positive charge in the fusion tag of the \DHFR{$\Ntag$}{O2} variants increases the
    threshold voltage (\cref{eq:threshold_voltage_complex}) by \SI{\approx5.21}{\mV}. The solid line is a
    linear fit to the data.
    (\subref{fig:trapping_translocation_probability})
    Translocation probability voltage $\vbias^{\probability_{\rm{transl}}}$ plotted against tag charge for
    $\probability_{\rm{transl}} = \mSIlist{0.1;5;50;95;99.9}{\percent}$, shows that variants with high tag
    charge require less bias voltage to fully translocate the pore. Values were obtained through interpolation
    from \cref{eq:ptrans}, using the parameters in \cref{tab:fitting_params_complex}.
  }\label{fig:trapping_translocation}
\end{figure*}
\begin{table}
  \centering
  \begin{threeparttable}[t]
    \centering
    \captionsetup{width=12cm}
    \caption[Summary of all threshold voltages and their dwell times]%
            {Summary of all threshold voltages and their dwell times.}
    \label{tab:threshold_voltages_and_dwelltimes}
    \renewcommand{\arraystretch}{1.2}
    \footnotesize
    \begin{tabularx}{12cm}{Xllll}
      \toprule
                    & \multicolumn{2}{c}{Simple model\tnote{a}}
                    & \multicolumn{2}{c}{Complex model\tnote{b}} \\
      \cmidrule(r){2-3}\cmidrule(l){4-5}
      {DHFR} variant & $\vthresh$ [mV] & $\tthresh$ [s]
                    & $\vthresh$ [mV] & $\tthresh$ [s] \\
      \midrule
      \DHFR{4}{S}   & \num{56.1} & \num{0.31} & --- & --- \\
      \DHFR{4}{I}   & \num{65.5} & \num{0.25} & --- & --- \\
      \DHFR{4}{C}   & \num{63.7} & \num{0.44} & --- & --- \\
      \DHFR{4}{O1}  & \num{71.5} & \num{2.11} & \num{75.6} & \num{2.28} \\
      \DHFR{5}{O1}  & \num{70.7} & \num{3.14} & \num{69.8} & \num{4.16} \\
      \DHFR{7}{O1}  & \num{65.4} & \num{15.6} & \num{61.6} & \num{13.9} \\
      \DHFR{4}{O2}  & \num{83.0} & \num{2.69} & \num{87.3} & \num{2.28} \\
      \DHFR{5}{O2}  & \num{78.5} & \num{5.68} & \num{79.2} & \num{4.16} \\
      \DHFR{6}{O2}  & \num{70.8} & \num{11.8} & \num{73.0} & \num{7.59} \\
      \DHFR{7}{O2}  & \num{65.4} & \num{10.8} & \num{68.1} & \num{13.9} \\
      \DHFR{8}{O2}  & \num{63.6} & \num{35.1} & \num{64.1} & \num{25.3} \\
      \DHFR{9}{O2}  & \num{59.8} & \num{80.2} & \num{60.4} & \num{46.2} \\
      \bottomrule
    \end{tabularx}
    \begin{tablenotes}
      \item[a] Estimated using \cref{eq:threshold_voltage_simple} after fitting of
      \cref{eq:double_barrier_simple} the individual mean dwell times of each mutant.
      \item[b] Estimated using \cref{eq:threshold_voltage_complex} after fitting of \cref{eq:double_barrier}
      to all \DHFR{$\Ntag$}{O2} mean dwell time data.
    \end{tablenotes}
  \end{threeparttable}
\end{table}

\subsection{Characteristics of the trapping}

As the double barrier model of \cref{eq:double_barrier_complex} is derived from the underlying physical
interactions of the molecule with the nanopore and with the externally applied field, the fitted parameters of
\cref{tab:fitting_params_complex} are physically relevant quantities that describe the characteristics of the
system.

The sizes of the electrostatic barriers $\potbar_{\rm{tag}}^{\cis/\trans}$ that the tag charges experience are
in direct relation to the gradients of the barrier sizes computed using the \gls{apbs} model
(\cref{fig:trapping_apbs_barrier_tag}). We find that the change of the \transi{} barrier with respect to tag
charge, $\potbar_{\rm{tag}}^{\trans} = \text{\SI{0.860}{\kbt\per\ec}}$, is in excellent agreement with the
simulated gradient of \SI{0.875}{\Vt}. The change observed for the \cisi{} barrier, $\potbar_{\rm tag}^{\cis}
= \text{\SI{0.218}{\kbt\per\ec}}$, is approximately 3-fold smaller compared to its \gls{apbs} value of
\SI{0.621}{\kbt\per\ec}. This deviation likely results from the shallowness of the \cisi{} barrier, causing it
to disappear when the energy landscape is tilted under an applied bias voltage
(see~\cref{fig:trapping_apbs_external_biased}). This gives rise to a \cisi{} barrier that lies at a location
further away from the electrostatic minimum located inside the \transi{} constriction, effectively limiting
the influence of the tag charge number on the barrier height. This claim is further corroborated by the
finding that the fitted value of $\dxcis \approx \SI{5.2}{\nm}$, which is almost twice the distance predicted
by the \gls{apbs} simulations and moves that \cisi{} barrier much closer to the \cisi{} entry.

One of the key insights we obtain from our model is the ability to directly extract information on the
strength of the osmotic flow. However, let us first observe that the equivalent electro-osmotic charge number
$\Neo\approx15.5$ is much bigger than the net charge of all tag charge variations, $\Ntag + \Nbody = -
9,\ldots, -4$, and also has the opposite sign. This is in agreement with the earlier assumption that the
electro-osmotic force is strong enough to overcome the opposing electrophoretic force and is hence responsible
for the capture of the molecule~\cite{Soskine-2012,Soskine-Biesemans-2015}. At a bias of $\vbias =
\SI{-50}{\mV}$ the electro-osmotic force exerted onto the \gls{dhfr} molecule (\cref{eq:osmoticforce}) is
\begin{equation}
  \forceeo = \ec \Neo \frac{\vbias}{L} \approx \SI{9}{\pN}
  \text{ .}
\end{equation}
The magnitude of this force is in line with those found experimentally for
DNA~\cite{Keyser-2006,vanDorp-2009,Lu-2012} and proteins~\cite{Oukhaled-2011} in solid-state nanopores.

We found the threshold voltages (obtained from the fitted model, see~\cref{eq:threshold_voltage_complex}) to
be roughly linearly dependent on the number of tag charges, with a decrease of \SI{\approx5}{\mV} per
additional positive charge (\cref{fig:trapping_translocation_voltage}). This effect is caused by the increase
of the net external force on the molecule with increasing tag charge, resulting in a simultaneous lowering of
the \transi{} barrier and raising of the \cisi{} barrier. The threshold voltages and their corresponding
dwell times for both the simple and complex double barrier models are listed in
\cref{tab:threshold_voltages_and_dwelltimes}.

Another important finding of our model is that the \gls{dhfr} variations are essentially trapped by the
electrostatic forces of the pore on the tag. This can be seen from the direct exponential dependence of the
electrostatics on the tag charge as shown in \cref{eq:static-barrier}. If the molecule was trapped as a whole
between two barriers, we would rather see a dependence on the net charge on the molecule. Indeed, we verified
that such a net charge dependence cannot be fitted to the data. This suggests that the tag acts as an anchor
which is located in the electrostatic minimum created by the \transi{} constriction
(\cref{fig:trapping_apbs_energy}).

We can also determine the probability of a full translocation of \gls{dhfr} using
\begin{align}\label{eq:ptrans}
  \probtrans = \dfrac{\rate^{\trans}}{\rate^{\cis} + \rate^{\trans}}
  \text{ ,}
\end{align}
where $k^\cis$ and $k^\trans$ can be computed using the individual components given by
\cref{eq:double_barrier_complex} and the parameters in \cref{tab:fitting_params_complex}. At zero bias and
zero tag charge, we find that only \SI{0.002}{\percent} of \gls{dhfr} molecules would exit to the \transi{}
side, indicating that in the absence of an electrophoretic driving force a \cisi{} exit is much more likely
than a \transi{} exit. This is in agreement with our expectations because \gls{dhfr}'s size leads to a
significant steric hindrance when it tries to translocate through the nanopore constriction.

Finally, from \cref{eq:ptrans} we can compute the voltage $\vbias^{\probtrans}$ required to obtain a given
translocation probability (\cref{fig:trapping_translocation_probability}). The number of tag charges
significantly lowers the voltage required to achieve full translocation, for example, $\vbias^{\probtrans}$
for $\probtrans = \SI{99.9}{\percent}$ decreases from \SI{\approx-130}{\mV} to \SI{\approx-85}{\mV} going from
$\Ntag = +4$ to \num{+9}. This effect is mainly due to the lowering of the \transi{} barrier height, as the
\cisi{} escape probability voltage (\SI{0.01}{\percent} line in \cref{fig:trapping_translocation_probability})
only changes from \SI{\approx-40}{\mV} to \SI{\approx-35}{\mV} going from \num{+4} to \num{+9} tag charges.
Hence, the higher the tag charge number, the stronger the net external force which pushes the molecule through
the \transi{} constriction of the pore.

\section{Conclusion}
\label{sec:trapping:conclusion}
We showed previously that neutral or weakly charged proteins larger than the \transi{} constriction
(\SI{>3.3}{\nm}) of \gls{clya} can be trapped inside the nanopore for a relatively long duration (seconds to
minutes) and that their behavior can be sampled by ionic current recordings~\cite{Soskine-2013,
Soskine-2012,Soskine-Biesemans-2015,Biesemans-2015,VanMeervelt-2014,VanMeervelt-2017,Wloka-2017}. In contrast,
small proteins rapidly translocate through the nanopore due to the strong electro-osmotic flow and highly
negatively charged proteins remain inside \gls{clya} only briefly or they do not enter at
all~\cite{Soskine-2012}.

In this work, we use \gls{dhfr} as a model molecule to enhance and investigate the trapping of small and
negatively charged proteins inside the \gls{clya} nanopore~\cite{Biesemans-2015}. \gls{dhfr}
(\SIrange{3.5}{4}{\nm}) is slightly too large to pass through the \transi{} constriction, and its negatively
charged body ($\Nbody = -13$) only allowed trapping the protein inside the nanopore for a few milliseconds.
The introduction of a positively charged C-terminal fusion tag partially counterbalanced the electrophoretic
force and introduced an electrostatic trap in the \transi{} constriction of \gls{clya} that increased the
\gls{dhfr} dwell time up to minutes.

The \gls{dhfr} mutants showed a biphasic voltage dependency which was explained by using a physical model
containing a double energy barrier to account for the exit on either side of the nanopore. The model contained
steric, electrostatic, electrophoretic, and electro-osmotic components and it allowed us to describe the
complex voltage-dependent data for the different \gls{dhfr} constructs. Furthermore, fitting to experimental
data of a series of \DHFR{$\Ntag$}{O2} constructs, in which the positive charge of the tag was systematically
increased, enabled us to deduce meaningful values for \gls{dhfr}'s intrinsic \cisi{} and \transi{}
translocation probabilities, as well as an estimate of the force exerted by the electro-osmotic flow on the
protein of \SI{0.178}{\pico\newton\per\milli\volt} (\eg~\SI{9}{\pN} at \SI{-50}{\mV},
\cref{tab:fitting_params_complex}). We also showed that the \gls{apbs} simulation results of a simple bead
model for the molecule are directly related to the independently fitted parameters of the double barrier
model. In conclusion, this means that it should be possible to predict the dwell times of similar experiments
by obtaining parameters directly from these types of \gls{apbs} simulations. Interestingly, some \DHFRt{}
variants showed several well-defined current blockade levels, indicating the presence of several stable energy
minima within the pore. Even though we did not investigate this phenomenon in detail, we did observe the
presence of multiple minima in the electrostatic energy landscape of \DHFRt{}. These findings suggest that the
distinct current blockades observed experimentally might correspond to different physical locations of the
protein along the length of \gls{clya}.

The double barrier model of \cref{eq:double_barrier_complex} in its current form does not adequately describe
mutations that modify the body charge distribution of \gls{dhfr}\@. This is most likely because body charge
variations close to the electrostatically trapped tag will impact the height of the barriers more strongly
than modifications on the far end of the tag. Although a model accounting for this effect could be made, it
would also make the double barrier model significantly more complex without providing any significant
advantages over a more comprehensive atomistic simulation. A more detailed discussion can be found in
\cref{sec:trapping_appendix:body_charge_variations}.

Inside the \lumen{} of \gls{clya}, proteins can bind to their specific substrates at all applied
potentials tested (up to \SI{-100}{\mV}), indicating that the electrostatic potential inside the nanopore and
the electrostatic potential originating from the inner surface of the nanopore did not unfold the protein.
Therefore, our results indicate that \gls{clya} nanopores can be used as nanoscale test tubes to investigate
enzyme function at the single-molecule level. Compared to the wide variety of single-molecule techniques based
on fluorescence, nanopore recordings are label-free, which has the advantage of allowing long observation
times.

The electrophoretic trapping of proteins inside nanopores is likely to have practical applications. For
example, arrays of biological or solid-state nanopores will allow the precise alignment of proteins on a
surface. In addition, proteins immobilized inside glass nanopipettes atop a scanning ion conductance
microscope~\cite{Bruckbauer-2007,Babakinejad-2013} can be manipulated with nanometer-scale precision, which
might be used, for instance, for the localized delivery of proteins. Furthermore, ionic current measurements
through the nanopore can be used for the detection of analyte binding to an immobilized protein, which has
applications in single-molecule protein studies and small analyte sensing.

\section{Materials and methods}
\label{sec:trapping:methods}

\subsection{Electrostatic energy landscape computation}
The electrostatic energy landscape of a coarse-grained \gls{dhfr} molecule translocating through a full-atom
\gls{clya-as} model was computed using the \gls{apbs}~\cite{Baker-2001}, using the approach described in
\cref{sec:elec:methods:elec:energy}. In summary, a full atom model of \gls{clya-as}~\cite{Franceschini-2016}
was prepared \textit{via} homology modeling with MODELLER software package~\cite{Sali-1993} from the wild-type
\gls{clya} crystal structure (\pdbid{2WCD}~\cite{Mueller-2009}) and its energy was further minimized using the
\gls{vmd}~\cite{Humphrey-1996} and \gls{namd} programs~\cite{Phillips-2005}. A coarse-grained bead model of
\gls{dhfr} was placed at various locations along the central axis of the pore using custom \code{Python} code
and the \code{Biopython} package~\cite{Cock-2009}. The bead model of \DHFRt{} consisted of a `body' of seven
negatively charged beads  (\SI{-1.43}{\ec}) beads (\SI{1.6}{\nm} diameter) in a spherical configuration and a
`tail' of nine smaller beads (\SI{1}{\nm} diameter, \SI{0.6}{\nm} spacing) in a linear configuration with
varying charge (depending on the net charge of the three amino acids they represent). Each atom in the
resulting \gls{clya}-\gls{dhfr} complexes was subsequently assigned a radius and partial charge (according to
the CHARMM36 force-field~\cite{Huang-2013}) with the PDB2PQR program~\cite{Dolinsky-2004,Dolinsky-2007}, and
the electrostatic was energy computed with \gls{apbs}. The net electrostatic energy cost or gain of placing a
\gls{dhfr} molecule (\ie~$\energyelec$) along the central z-axis of \gls{clya} (from
$z_{\rm{body}}=\SI{-12.5}{\nm}$ to $\SI{27.5}{\nm}$ relative to the center of the bilayer, with steps of
\SI{0.5}{\nm} inside the pore) was computed by \cref{eq:electrostatic_energy}. To this end, all systems were
solved using the non-linear \gls{pbe} (\code{npbe}) in two steps with the automatic solver (\code{mg-auto}):
(1) a coarse calculation in a box of \SIgrid{40;40;110}{\nm} with grid lengths of
\SIgrid{0.138;0.138;0.122}{\nm} and multiple Debye-H\"{u}ckel boundary conditions (\code{bcfl mdh}), followed
by (2) a finer focusing calculation in a box of \SIgrid{15;15;70}{\nm} with grid lengths of
\SIgrid{0.052;0.052;0.052}{\nm} that used the values of the coarse calculation at its boundaries. The
monovalent salt concentration was set to \SI{0.150}{\Molar} with a radius of \SI{0.2}{\nm} for both ions. The
solvent and solute relative permittivities were set to \num{78.15} and \num{10}, respectively~\cite{Li-2013}.
Both the charge density and the ion accessibility maps were constructed using cubic B-spline discretization
(\code{chgm spl2} and \code{srfm spl2}).

\subsection{Protein mutagenesis, overexpression, and purification}

All \gls{dhfr} variants were constructed, overexpressed and purified using standard molecular biology
techniques~\cite{Soskine-Biesemans-2015,Biesemans-2015}, as described in full detail in the supplementary
information (see~\cref{sec:trapping_appendix:dhfr_cloning}). Briefly, the \DHFR{4}{S} DNA construct was built
from the {pT7-SC1} plasmid containing the \DHFRt{} construct (see ref.~\cite{Soskine-Biesemans-2015}) by
inserting an additional alanine residue at position 175 (located in the fusion tag) with site-directed
mutagenesis. All other variants were derived---again using site-directed mutagenesis---either directly from
\DHFR{4}{S} or from a variant thereof. The plasmids of each \gls{dhfr} variant were used to transform E.
cloni\textsuperscript{\textregistered} EXPRESS BL21(DE3) cells (Lucigen, Middleton, USA), and the \gls{dhfr}
proteins they encode were overexpressed overnight at \SI{25}{\celsius} in a liquid culture. After the
bacterial cells were harvested by centrifugation, the overexpressed proteins were released into the solution
through lysis---using a combination of at least a single freeze-thaw cycle, incubation with lysozyme, and
probe tip sonification. Finally, the \gls{dhfr} proteins were purified from the lysate with affinity
chromatography with Strep-Tactin\textsuperscript{\textregistered} Sepharose\textsuperscript{\textregistered}
(IBA Lifesciences, Goettingen, Germany), aliquoted, and stored at \SI{-20}{\celsius} until further use.

\subsection{ClyA-AS overexpression, purification and oligomerization}

\gls{clya-as} oligomers were prepared as described previously~\cite{Soskine-2013}, and full details can be
found in \cref{sec:trapping_appendix:dhfr_cloning}. Briefly, the \gls{clya-as} monomers were overexpressed and
purified in a manner similar to that for \gls{dhfr}, with the largest difference being the use of
\ce{Ni-NTA}-based affinity chromatography. After purification, \gls{clya-as} monomers were oligomerized in
\SI{0.5}{\percent} \textbeta-dodecylmaltoside (GLYCON Biochemicals GmbH, Luckenwalde, Germany) at
\SI{37}{\celsius} for \SI{30}{\minute}. The type I oligomer (12-mer) was isolated by gel extraction from a
blue native \gls{page}.

\subsection{Electrical recordings in planar lipid bilayers}

Electrical recordings of individual \gls{clya-as} nanopores were carried out using a typical planar lipid
bilayer setup with an AxoPatch 200B (Axon Instruments, San Jose, USA) patch-clamp
amplifier~\cite{Maglia-2010,Soskine-2012}. Briefly, a black lipid membrane consisting of
1,2-diphytanoyl-sn-glycero-3-phosphocholine (Avanti Polar Lipids, Alabaster, USA), was formed inside a
\SI{\approx 100}{\micro\meter} diameter aperture in a thin polytetrafluoroethylene film (Goodfellow Cambridge
Limited, Huntingdon, England) separating two electrolyte compartments. Single nanopores were then made to
insert into the \cisi{}-side chamber (grounded) by addition of \SIrange{0.01}{0.1}{\nano\gram} of
preoligomerized \gls{clya-as} to the buffered electrolyte (\SI{150}{\mM} \ce{NaCl}, \SI{15}{\mM} \ce{Tris-HCl}
\pH{7.5}). All ionic currents were sampled at \SI{10}{\kilo\hertz} and filtered with a \SI{2}{\kilo\hertz}
low-pass Bessel filter. A more detailed description can be found in
\cref{sec:trapping_appendix:nanopore_experiments}.

\subsection{Dwell time analysis and model fitting}

The dwell times of the \gls{dhfr} protein blocks were extracted from single-nanopore channel recordings using
the `single-channel search' algorithm of the pCLAMP 10.5 (Molecular Devices, San Jose, USA) software suite.
The process was monitored manually, and any events shorter than \SI{1}{\ms} were discarded. We processed the
dwell time data, and fitted the double barrier model to it, using a custom Python code employing the
\code{NumPy}~\cite{vanderWalt-2011}, \code{pandas}~\cite{McKinney-2010}, and \code{lmfit}~\cite{Newville-2014}
packages. Fitting of the exponential models to the data was performed using non-linear least-squares
minimization with the Levenberg-Marquardt algorithm as implemented by \code{lmfit}. Note that to improve the
robustness and quality of the fitted parameters, we moved the exponential prefactors into the exponential and
we fitted the natural logarithm of the equations to the natural logarithm of the dwell time data.

\subsection{Analytical expression for the threshold voltages}

An analytic expression for the threshold voltage $\vthresh$---the bias voltage at maximum dwell time---can
found as the bias voltage for which $\dfrac{d\rate}{d\vbias} = 0$. For the simple double barrier model given
by \cref{eq:double_barrier_simple} this becomes
\begin{align}\label{eq:threshold_voltage_simple}
	\vthresh = \dfrac{ \left[
	\frac{%
	\log \left( \rateft / \ratefc \right)+
	\log \left( -N_{\rm{eq}}^\trans / N_{\rm{eq}}^\cis \right)
	}{%
		N_{\rm{eq}}^\trans - N_{\rm{eq}}^\cis
  }\right]}%
  { \ec / \kbt }
  \text{ ,}
\end{align}
whereas for the more complex model given by \cref{eq:double_barrier_complex} it becomes
\begin{equation}\label{eq:threshold_voltage_complex}
	\vthresh = - \dfrac{\left[
	\frac{%
	\log \left( \rateft / \ratefc \right)
	+ \log \left( \dxtrans / \dxtrans \right)
	+ \left( \potbar_{\rm{tag}}^{\cis} - \potbar_{\rm{tag}}^{\trans} \right)\Ntag
	}{%
		\left( \Nnet + \Neo \right) \left(\dxcis + \dxtrans \right) / L
  }\right]}%
  { \ec / \kbt }
  \text{ .}
\end{equation}
%

\cleardoublepage


  \graphicspath{{\chaptersdir/epnpns/image/}}%
\chapter{An improved PNP-NS framework}
\label{ch:epnpns}

\epigraphhead[\epipos]{%
\epigraph{%
  ``The sciences are monuments devoted to the public good; each citizen owes to them a tribute proportional to
  his talents. While the great men, carried to the summit of the edifice, draw and put up the higher floors,
  the ordinary artists scattered in the lower floors, or hidden in the obscurity of the foundations, must only
  seek to improve what cleverer hands have created.''
}{%
  \textit{`Charles-Augustin de Coulomb'}
}}

\definecolor{shadecolor}{gray}{0.85}
\begin{shaded}
This chapter was published as:
\begin{itemize}
  \item K. Willems*, D. Rui\'{c}, F. L. R. Lucas, U. Barman, N. Verellen, G. Maglia and P. Van Dorpe.
        \textit{Nanoscale} \textbf{12}, 16775--16795 (2020) 
\end{itemize}
\newpage
\end{shaded}

In this chapter, we introduce the \glsfirst{epnp-ns} equations, a continuum simulation framework aimed toward
the accurate continuum simulation of ion and water transport through biological nanopores. The \gls{epnp-ns}
equations improve upon the regular {PNP-NS} equations by the inclusion of empirical corrections that account for
the influence of the ionic strength and the proximity of protein walls on the properties of the electrolyte.
To improve readability and to allow for better integration of the supplementary information into the main
chapter text, the original manuscript was split into \cref{ch:epnpns,ch:transport}, detailing the theory of
the framework and its application to the \glsfirst{clya} nanopore, respectively. The remainder of the
supplementary information can be found in \cref{ch:epnpns_appendix}. The text and figures of this chapter
represent entirely my own work.
\adapRSC{Willems-2020}
%
%

\section{Abstract}
\label{sec:epnpns:abstract}

Despite the broad success of biological nanopores as powerful instruments for the analysis of proteins and
nucleic acids at the single-molecule level, a fast simulation methodology to accurately model their
nanofluidic properties is currently unavailable. This limits the rational engineering of nanopore traits and
makes the unambiguous interpretation of experimental results challenging. Here, we present a continuum
approach that can faithfully reproduce the experimentally measured ionic conductance of the biological
nanopore \glsfirst{clya} over a wide range of ionic strengths and bias potentials. Our model consists of the
\glsfirst{epnp-ns} equations and a computationally efficient 2D-axisymmetric representation for the geometry
and charge distribution of the nanopore. Importantly, the \gls{epnp-ns} equations achieve this accuracy by
self-consistently considering the finite size of the ions and the influence of both the ionic strength and the
nanoscopic scale of the pore on the local properties of the electrolyte. These comprise the mobility and
diffusivity of the ions, and the density, viscosity, and relative permittivity of the solvent. Crucially, by
applying our methodology to \gls{clya}, a biological nanopore used for single-molecule enzymology studies, we
could directly quantify several nanofluidic characteristics difficult to determine experimentally. These
include the ion selectivity, the ion concentration distributions, the electrostatic potential landscape, the
magnitude of the electro-osmotic flow field, and the internal pressure distribution. Hence, this work provides
a means to obtain fundamental new insights into the nanofluidic properties of biological nanopores and paves
the way toward their rational engineering.

\section{Introduction}
\label{sec:epnp-ns:intro}

The transport of ions and molecules through nanoscale geometries is a field of intense study that uses
experimental~\cite{Kasianowicz-1996,Meller-2000,Smeets-2006,Stefureac-2006,Wanunu-2009,Buchsbaum-2013,
Franceschini-2016,Willems-Ruic-Biesemans-2019} theoretical~\cite{Bonthuis-2006,Muthukumar-2010,
Sparreboom-2010,Bocquet-2010,Muthukumar-2014,Liu-2020} and computational methods~\cite{Im-2002,Daiguji-2004,
Aksimentiev-2005,Maffeo-2012,Modi-2012,Thomas-2014,Wang-2014,Kim-2015,Bonome-2017,Basdevant-2019,Cao-2019}.
A primary driving force behind this research is the development of nanopores as label-free, stochastic
sensors at the ultimate analytical limit (\ie~single molecule)~\cite{Bayley-2001,Dekker-2007,Venkatesan-2011,
Zhang-2016}. These detectors have applications ranging from the analysis of biopolymers such as
DNA~\cite{Deamer-2016,Kasianowicz-1996,Meller-2000,Maglia-2008,Butler-2008,Stoddart-2009,Manrao-2012,
Franceschini-2013,Jain-2018} or proteins~\cite{Soskine-2012,Yusko-2011,Yusko-2017,Houghtaling-2019,
Restrepo-Perez-2018,Talaga-2009,Rodriguez-Larrea-2013,Nivala-2013,Kennedy-2016,Howorka-2020},
to the detection and quantification of biomarkers~\cite{Chen-2013,Niedzwiecki-2013,VanMeervelt-2014,
Huang-2017,Liu-2018,Galenkamp-2018}, or the fundamental study of chemical or enzymatic reactions at the
single molecular level~\cite{Willems-VanMeervelt-2017,Lieberman-2010,Nivala-2013,Ho-2015,Laszlo-2017,
Harrington-2019,Li-2020,Galenkamp-2020}.

Nanopores are typically operated in the resistive-pulse mode, where the changes in their ionic conductance are
monitored over time~\cite{Bayley-2001,Dekker-2007,Maglia-2010,Venkatesan-2011}. Experimentally, this is
achieved by placing the nanopore between two isolated electrolyte compartments and applying a constant DC (or
AC) voltage across them. The $\approx10^8$ difference in ionic resistance of the nanopore
(\SI{\approx1}{\giga\ohm}) compared to the typical electrolyte reservoir
(\SI{\approx10}{\ohm})~\cite{Maglia-2010}, causes the full potential change to occur within (and around) the
pore. The resulting electric field ($10^6$--$10^7$~\si{\V\per\m}) electrophoretically drives a given number of
ions (\ie~the `open-pore' conductance) and water molecules (\ie~the electro-osmotic flow) through the
pore~\cite{Wong-2007,Mao-2014,Haywood-2014,Laohakunakorn-2015}. Similarly, analyte molecules such as DNA and
proteins are subject to the same Coulombic (electrophoretic) and hydrodynamic (electro-osmotic)
forces~\cite{Wong-2007,Grosberg-2010,Muthukumar-2010,Muthukumar-2014}. This guides them towards---and often
through---the nanopore, resulting in a temporary disruption of the baseline ionic conductance. Naively
speaking, for any given analyte molecule, the rate of these resistive pulses is proportional to its
concentration in the reservoir, whereas their duration, magnitude, and intra-pulse fluctuations contain a
wealth of information on its physical properties (\eg~size, shape, and
charge)~\cite{Howorka-2009,Ying-2019,Lu-2020}. Hence, the unique `fingerprints' provided by the
(sub-)nanometer-sized sensing volume of a nanopore can reveal information that is typically inaccessible to
bulk measurements, such as hidden intermediate states, dynamic non-covalent interactions, and the presence of
subpopulations~\cite{Ying-2019,Lu-2020}. Notably, nanopores have been used to create ultra-long reads of
individual DNA strands~\cite{Jain-2018}, to determine the shape, volume, and dipole moment of
proteins~\cite{Yusko-2017,Houghtaling-2019}, and to investigate the kinetics of single enzymes tethered
to~\cite{Derrington-2015,Harrington-2019} or trapped within~\cite{Li-2020,Galenkamp-2020} the pore. Because
each current blockade is modulated by the complex interactions between the translocating molecule and the
nanopore itself, they will depend on the properties of both. As a result, despite the successful applications
mentioned above, the unambiguous interpretation of the ionic current signal remains a notoriously difficult
task if a full understanding of the nanofluidic phenomena that underlie them is not available.

The computational approaches most widely used to study nanofluidic transport in ion channels or biological
nanopores comprise discrete methods such as \gls{md}~\cite{Lynden-Bell-1996,Allen-1999,
Aksimentiev-2005,Luan-2008,Bhattacharya-2011,Zhang-2014,DiMarino-2015,Belkin-2016,Basdevant-2019,Cao-2019} and
\gls{bd}~\cite{Schirmer-1999,Im-2002,Noskov-2004,Millar-2008,Egwolf-2010,DeBiase-2015,Pederson-2015}, and
mean-field (continuum) methods based on solving the \gls{pb}~\cite{Grochowski-2008, Baldessari-2008-1} and
\gls{pnp}~\cite{Eisenberg-1996,Gillespie-2002,Simakov-2010} equations. The latter two can be coupled with the
\gls{ns} equation to include electro-osmotically or pressure-driven fluid flow~\cite{Lu-2012,Pederson-2015}.
Due to their explicit atomic or particle nature, \gls{md} and \gls{bd} simulations are considered to yield the
most accurate results. However, the large computational cost of simulating a complete biological nanopore
system (100K--1M atoms) for hundreds of nanoseconds still necessitates the use of
supercomputers~\cite{Aksimentiev-2005,Bhattacharya-2011,Wilson-2019,Cao-2019}. The PNP(-NS) equations, on the
other hand, are of particular interest due to their low computational demands and analytical tractability. In
a continuum approach, the simulated system is subdivided into several `structureless' domains, the behavior of
which is parameterized by material properties such as the relative permittivity, diffusion coefficient,
electrophoretic mobility, viscosity, or density. Because these properties can only emerge from the collective
behavior or interactions between small groups of atoms (\ie~the mean-field approximation), great care must be
taken when using them to compute fluxes and fields at the nanoscale, where computational elements may only
contain a few molecules~\cite{Corry-2000,Collins-2012}. Nevertheless, even though the PNP equations have been
used extensively for the qualitative simulation of ion channels~\cite{Im-2002,Furini-2006,Liu-2015},
biological nanopores~\cite{Simakov-2010,Pederson-2015, Aguilella-Arzo-2017,Simakov-2018} and their solid-state
counterparts~\cite{Cervera-2005,White-2008, Chaudhry-2014,Laohakunakorn-2015}, the extent to which they are
quantitatively accurate is often challenged~\cite{Corry-2000,Collins-2012,Maffeo-2012,Thomas-2014,Kim-2015}.
To remedy the shortcomings of PNP and NS theory, a number of modifications have been proposed over the years.
These include, among others, (1) steric ion-ion interactions, (2) the effect of protein-ion/water interactions
on their `motility' (\ie~diffusivity and electrophoretic mobility), (3) the concentration dependencies of ion
motility, and solvent relative permittivity, viscosity, and density.

The steric ion-ion interactions can be accounted for by computing the excess in chemical potential
($\mu_{i}^\text{ex}$) resulting from the finite size of the ions~\cite{Eisenberg-1996,Bazant-2009,
Daiguji-2010}, or by taking the dielectric-self energy of the ions into account~\cite{Corry-2003,
Bonthuis-2006}. Gillespie \etal{} combined PNP and density functional theory---where $\mu_{i}^\text{ex}$ was
split up into ideal, hard-sphere, and electrostatic components---to successfully predict the selectivity and
current of ion channels~\cite{Gillespie-2002}. In the Poisson-Nernst-Planck-Fermi model developed by Liu and
Eisenberg, steric effects are included by treating water molecules as discrete particles, described using
Fermi distributions, and by implementing a mean-field version of the van der Waals
potential~\cite{Liu-2013,Liu-2015,Liu-2020}. In another approach, Kilic \etal{} derived a set of modified PNP
equations based on the free energy functional of Borukhov's modified \gls{pb} model~\cite{Borukhov-1997}
and observed significantly more realistic concentrations for high surface potentials compared to the classical
PNP equations~\cite{Kilic-2007}. To allow for non-identical ion sizes and more than two ion species, this
model was later extended by Lu \etal{}, who used it to probe the effect of finite ion size on the rate
coefficients of enzymes~\cite{Lu-2011}.

The interaction of ions or small molecules such as water with the heavy atoms of proteins or DNA results in a
strong reduction of their motility, as observed in \gls{md}
simulations~\cite{Makarov-1998,Pronk-2014,Wilson-2019}. Since these effects happen only at distances
\SI{\le1}{\nm}, they can usually be neglected for macroscopic simulations. However, in small nanopores
(\SI{\le10}{\nm} radius), they comprise a significant fraction of the total nanopore radius and hence must be
taken into account~\cite{Noskov-2004,Simakov-2010,Pederson-2015, McMullen-2017}. In continuum simulations,
this can be achieved with the use of positional-dependent ion diffusion coefficients. An example
implementation is the `soft-repulsion PNP' developed by Simakov and
Kurnikova~\cite{Simakov-2010,Simakov-2018}, who used it to predict the ionic conductance of the \glsfirst{ahl}
nanopore. Similar reductions in ion diffusion coefficients have been proposed to improve PNP theory's
estimations of the ionic conductance of ion channels~\cite{Furini-2006,Liu-2015, DeBiase-2015}. The motility
of water molecules is expressed by the NS equations as the fluid's viscosity. Hence, as also observed in
\gls{md} simulations for water molecules near proteins~\cite{Pronk-2014} and confined in hydrophilic
nanopores~\cite{Qiao-Aluru-2003,Vo-2016,Hsu-2017}, the water-solid interaction leads to a viscosity several
times higher compared to the bulk values. Note that this is valid for hydrophilic interfaces only, as the lack
of interaction with hydrophobic interfaces, such as carbon nanotubes, leads to a lower
viscosity~\cite{Ye-2011}.

It is well known that the self-diffusion coefficient $\diffusion_{i}$ and electrophoretic mobility
$\mobility_{i}$ of an ion $i$ depends on the local concentrations of all the ions in the
electrolyte~\cite{ContrerasAburto-2013-1}. Their values typically decrease with increasing salt concentration,
and should not be treated as constants. Moreover, even though the Nernst-Einstein relation
($\mobility_{i}=\diffusion_{i}/\kbt$) is strictly speaking only valid at infinite dilution and a good
approximation at low concentrations (\SI{<10}{\mM}), it significantly overestimates the ionic mobility at
higher salt concentrations~\cite{Mills-1989,Panopoulos-1986,ContrerasAburto-2013-1,ContrerasAburto-2013-2}. In
an empirical approach, Baldessari and Santiago formulated an ionic-strength dependency of the ionic mobility
based on the activity coefficient of the salt~\cite{Baldessari-2008-1} and showed excellent correspondence
between the experimental and simulated ionic conductance of long nanochannels over a wide concentration
range~\cite{Baldessari-2008-2}. Alternatively, Burger \etal{} used a microscopic lattice-based model to derive
a set of PNP equations with non-linear, ion density-dependent mobilities and diffusion coefficients that
provided significantly more realistic results for ion channels~\cite{Burger-2012}. Note that other electrolyte
properties, such as the viscosity~\cite{Hai-Lang-1996}, density~\cite{Hai-Lang-1996} and relative
permittivity~\cite{Gavish-2016}, may also significantly affect the ion and water flux. To better compute the
charge flux in ion channels, Chen derived a new PNP framework~\cite{Chen-2016} that includes water-ion
interactions in the form of a concentration-dependent relative permittivity and an additional ion-water
interaction energy term.

To the best of our knowledge, no attempt has been made to consolidate all of the corrections discussed above
into a single framework. Hence, we propose the \glsfirst{epnp-ns} equations, which improve the predictive
power of the traditional \gls{pnp-ns} equations at the nanoscale and beyond infinite dilution. Our
\gls{epnp-ns} framework takes into account the finite size of the ions using a size-modified PNP
theory~\cite{Borukhov-1997,Lu-2011}, and implements spatial-dependencies for the solvent
viscosity~\cite{Pronk-2014,Vo-2016,Hsu-2017}, the ion diffusion coefficients and their
mobilities~\cite{Makarov-1998,Noskov-2004,Pederson-2015}. It also includes self-consistent
concentration-dependent properties---based on empirical fits to experimental data---for all ions in terms of
diffusion coefficients and mobilities~\cite{Baldessari-2008-1,Mills-1989}, and for the solvent in terms of
density, viscosity~\cite{Hai-Lang-1996} and relative permittivity~\cite{Gavish-2016}. In
\cref{sec:epnp-ns:model} we detail the governing equations, followed by their parameterization with
experimental data from literature in \cref{sec:epnp-ns:parameterization}.

\section{Mathematical model}
\label{sec:epnp-ns:model}

The use of continuum or mean-field representations for both the nanopore and the electrolyte enables us to
efficiently compute the steady-state ion and water fluxes under almost any condition. The dynamic behavior of
our complete system is described by the coupled Poisson, Nernst-Planck, and Navier-Stokes equations, a set of
partial differential equations that describe the electrostatic field, the total ionic flux, and the fluid flow,
respectively~\cite{Eisenberg-1996,Cervera-2005,Lu-2012}. To improve upon the quantitative accuracy of the
\gls{pnp-ns} equations for nanopore simulations, we developed an extended version of these equations
({\gls{epnp-ns}}) and implemented it in the commercial finite element solver {COMSOL} Multiphysics (v5.4,
COMSOL Inc., Burlington, MA, USA). 

\subsection{Electrostatic field}
\label{sec:epnp-ns:electrostatic_field}

We will make use of Poisson's equation to evaluate the electric potential
\begin{align}
  \label{eq:poisson}
  \nabla \cdot \left(\absperm \relperm \nabla \potential \right) = -\left( \scdpore + \scdion \right)
  \text{ ,}
\end{align}
with $\potential$ the electric potential, $\absperm$ the vacuum permittivity
(\SI{8.85419e-12}{\farad\per\meter}) and $\relperm$ local relative permittivity. The pore's fixed charge
distribution, $\scdpore$,  can be derived directly from the full atom model of the pore
(see~\cref{eq:scdpore}). The ionic charge density in the fluid is given by
\begin{align}\label{eq:scdion}
  \scdion = \faraday\sum_{i}\chargen_{i}\concentration_{i}
  \text{ ,}
\end{align}
with $\faraday$ Faraday's constant (\SI{96485.33}{\coulomb\per\mole}), and $\concentration_{i}$ the ion
concentration and $\chargen_{i}$ ion charge number of ion $i$. To account for the concentration dependence of
the electrolyte's relative permittivity, we replaced $\relperm$ inside the electrolyte with the expression
\begin{align}\label{eq:epnp-ns_relperm_concentration}
  \permittivity_{r,\text{f}}(\avionconc) =
        \permittivity_{r,\text{f}}^0 \permittivity_{r,\text{f}}^c(\avionconc)
  \text{ ,}
\end{align}
with $\avionconc=\frac{1}{n}\sum_{i}^{n}\concentration_{i}$ the average ion concentration,
$\permittivity_{r,\text{f}}^0$ the relative permittivity at infinite dilution and
$\permittivity_{r,\text{f}}^c$ a concentration-dependent empirical function parameterized with experimental
data
(\cref{eq:epnpns_concentration_solv_epsr,fig:epnpns_concentration_solv_epsr,tab:corrections_equations,tab:corrections_parameters}).

\subsection{Ionic flux}

The total ionic flux $\flux_{i}$ of each ion $i$ is given by the size-modified Nernst-Planck
equation~\cite{Lu-2011}, and can be expressed as the sum of diffusive, electrophoretic, convective, and steric
fluxes
\begin{align}
  \label{eq:sm-nernst-planck}
  \flux_{i} = -\left[
    \diffusion_{i} \nabla \concentration_{i}
    + \chargen_{i} \mobility_{i} \concentration_{i} \nabla \potential
    - \velocity \concentration_{i}
    + \diffusion_{i} \vec{\beta_{i}} \concentration_{i} \right]
  \text{ ,}
\end{align}
where $\vec{\beta_{i}}$ is the steric flux vector
\begin{align}\label{eq:epnp-ns_steric_flux_vector}
  \vec{\beta_{i}} =
      \frac{ \ionsize_{i}^3 / \ionsize_{0}^3 \dsum_{j} \avogadro \ionsize_{j}^3 \nabla \concentration_{j} }
          { 1 - \dsum_{j} \avogadro \ionsize_{j}^3 \concentration_{j} }
  \text{ ,}
\end{align}
and at steady-state
\begin{align}
  \dfrac{\partial \concentration_{i}}{\partial \timedim} ={}& - \nabla \cdot \flux_{i} = 0
  \text{ ,}
\end{align}
with $\diffusion_{i}$ the ion diffusion coefficient, $\concentration_{i}$ the ion concentration,
$\chargen_{i}$ the ion charge number, $\mobility_{i}$ the electrophoretic mobility of ion $i$. $\potential$ is
the electrostatic potential, $\velocity$ the fluid velocity, and $\avogadro$ Avogadro's constant
(\SI{6.022e23}{\per\mole}). $\ionsize_{i}$ and $\ionsize_{0}$ are steric cubic diameters of ions
and water molecules, respectively. Because currently there are no experimentally verified values available for
$\ionsize_{i}$ and $\ionsize_{0}$, we set them to \SI{0.5}{\nm} (max. \SI{13.3}{\Molar}) and \SI{0.311}{\nm}
(max. \SI{55.2}{\Molar}), respectively~\cite{Bazant-2009}.

The reduction of the ionic motility at increasing salt concentrations and in proximity to the nanopore walls
was implemented self-consistently by replacing $\diffusion_{i}$ and $\mobility_{i}$ with the expressions
\begin{align}\label{eq:epnp-ns_diffusion_and_mobility}
  \diffusion_{i}(\avionconc,\walldistance) ={}&
      \diffusion_{i}^0 \diffusion_{i}^c(\avionconc) \diffusion_{i}^w(\walldistance)  \\
  \mobility_{i}(\avionconc,\walldistance) ={}&
      \mobility_{i}^0 \mobility_{i}^c(\avionconc) \mobility_{i}^w(\walldistance)
  \text{ ,}
\end{align}
where $\diffusion_{i}^0$ and $\mobility_{i}^0$ represent the values at infinite dilution. The concentration
dependent factors $\diffusion_{i}^c(\avionconc)$ and $\mobility_{i}^c(\avionconc)$ are empirical functions
fitted to experimental data (between \SIrange{0}{5}{\Molar} \ce{NaCl}) of respectively the ion self-diffusion
coefficients~\cite{Mills-1989} (\cref{eq:concentration_d_mu_tna,fig:epnpns_concentration_d}) and the
electrophoretic mobilities~\cite{Bianchi-1989,Currie-1960,Goldsack-1976,DellaMonica-1979}
(\cref{eq:concentration_d_mu_tna,fig:epnpns_concentration_mu}). Likewise, the factors
$\diffusion_{i}^w(\walldistance)$ and $\mobility_{i}^w(\walldistance)$ are empirical functions that introduce
a spatial dependency on the distance from the nanopore wall $\walldistance$, and were parameterized by fitting
to molecular dynamics data
(\cref{eq:epnpns_distance_d,fig:epnpns_distance_d,tab:corrections_equations,tab:corrections_parameters})~\cite{Noskov-2004,Simakov-2010,Makarov-1998,Wilson-2019}.

Based on the observation that the diffusivity of nanometer- to micrometer-sized particles reduces
significantly when confined in pores and slits of comparable dimensions~\cite{Renkin-1954,Deen-1987,
Dechadilok-2006,Muthukumar-2014,Kannam-2017}, Simakov \etal{}~\cite{Simakov-2010} and Pederson
\etal{}~\cite{Pederson-2015} reduced the ion motilities inside the pore as a function of the ratio between the
ion and the nanopore radii. We chose not to include this correction into our model, as extrapolating its
applicability for ions with a hydrodynamic radius comparable to the size of the solvent molecules is
questionable~\cite{Anderson-1972,Deen-1987}.

\subsection{Fluid flow}

As derived by Axelsson \etal{}~\cite{Axelsson-2015}, the fluid flow and pressure field inside an
incompressible fluid with a variable density and variable viscosity is given by the Navier-Stokes equations:
\begin{align}
  \label{eq:navier-stokes-variable}
  \dfrac{\partial}{\partial \timedim} \left( \density \velocity \right) +
  \left( \velocity \cdot \nabla \right) \left( \density\velocity \right)
  + \nabla \cdot \hydrostresstensor = \volumeforce
  \text{ ,}
\end{align}
where
\begin{align}
  \hydrostresstensor =
  \pressure\identity - \viscosity\left[\nabla\velocity+\left(\nabla\velocity \right)^\mathsf{T}\right]
  \text{ ,}
\end{align}
together with the continuity equations for the fluid density
\begin{align}
  \label{eq:continuity-density}
  \dfrac{\partial \density}{\partial \timedim} + \velocity \cdot \nabla \density  = 0
  \text{ ,}
\end{align}
and the divergence constraint for the momentum
\begin{align}
  \nabla \cdot \left( \density\velocity \right) - \velocity \cdot \nabla \density ={}& 0
  \text{ ,}
\end{align}
with $\velocity$ the fluid velocity, $\density$ the fluid density, $\hydrostresstensor$ the hydrodynamic
stress tensor, $\viscosity$ the viscosity, and $\pressure$ the pressure. The external body force density
$\volumeforce$ that acts on the fluid is given by
\begin{align}\label{eq:ion_force_density}
  \volumeforce = \scdion \efield
  \text{ ,}
\end{align}
with $\efield = - \nabla \potential$ the electric field vector. At steady-state, the partial derivatives with
respect to time in \cref{eq:navier-stokes-variable,eq:continuity-density} become equal to zero:
\begin{align}
  \dfrac{\partial}{\partial \timedim} \left( \density \velocity \right) ={}& 0 \\
  \dfrac{\partial \density}{\partial \timedim} ={}& 0
  \text{ .}
\end{align}
As with the previous equations, we introduced a concentration and wall distance dependency for $\viscosity$
and a concentration dependency for the $\density$ by replacing their constant values by
\begin{align}\label{eq:epnp-ns_viscosity_density}
  \viscosity(\avionconc,\walldistance) ={}&
    \viscosity^0 \viscosity^c(\avionconc) \viscosity^w(\walldistance) \\
  \density(\avionconc) ={}&
    \density^0 \density^c(\avionconc)
  \text{ ,}
\end{align}
where $\viscosity^0$ and $\density^0$ are the values at infinite dilution (\ie~pure water). The empirical
functions $\viscosity^c(\avionconc)$, $\density^c(\avionconc)$ and $\viscosity^w(\walldistance)$ were
parameterized \textit{via} fitting to experimental~\cite{Hai-Lang-1996}
(\cref{eq:epnpns_concentration_solv_eta,eq:epnpns_concentration_solv_rho,fig:epnpns_concentration_solv_eta,fig:epnpns_concentration_solv_rho})
and molecular dynamics~\cite{Pronk-2014} (\cref{eq:epnpns_distance_eta,fig:epnpns_distance_eta}) data obtained
from literature, as summarized in \cref{tab:corrections_equations,tab:corrections_parameters}.

\subsubsection{From {ePNP-NS} to {PNP-NS}}

The \gls{epnp-ns} equations revert into the regular \gls{pnp-ns} equations by disabling the steric flux vector
($\vec{\beta}=0$) and all concentration ($\permittivity_{r,\text{f}}^c=1$, $\diffusion_{i}^c=1$,
$\mobility_{i}^c=1$, $\viscosity^c=1$, $\density^c=1$) and wall distance functions ($\diffusion_{i}^w=1$,
$\mobility_{i}^w=1$, $\viscosity^w=1$).

\begin{table}[p]
  \begin{threeparttable}
    \centering
    \captionsetup{width=12cm}
    \caption{Parameters and fitting equations used in the {ePNP-NS} equations.}
    \label{tab:corrections_equations}
    \footnotesize
    \renewcommand{\arraystretch}{1.2}
    \begin{tabularx}{12cm}{>{\raggedright\hsize=2.5cm}X >{\hsize=1.5cm}l >{\hsize=5cm}X >{\hsize=2cm}l}
      \toprule
    
      Name & Symbol\tnote{a} & Infinite dilution value\tnote{b}/Function\tnote{c} & Reference \\
    
      \midrule
    
      \multirow{4}{1.5cm}{Relative permittivity}
        & $\permittivity_{r,\text{p}}$ & \num{20} & \cite{Li-2013} \\
        & $\permittivity_{r,\text{m}}$ & \num{3.2} & \cite{Gramse-2013} \\
        & $\permittivity_{r,\text{f}}^0$ & \num{78.15} & \cite{Gavish-2016} \\
        & $\permittivity_{r,\text{f}}^c(\dconc)$ & $1 - \left(1 -	\dfrac{P_1}{P_0}\right) L \left(
          \dfrac{3P_2}{P_0 - P_1} \dconc \right)$ & \cite{Gavish-2016} \vspace{0.5cm} \\
      
      \multirow{4}{1.5cm}{Ion self-diffusion coefficient}
        & $\diffusion_{\Na}^0$ & \SI{1.334e-9}{\square\meter\per\second} & \cite{Mills-1989} \\
        & $\diffusion_{\Cl}^0$ & \SI{2.032e-9}{\square\meter\per\second} & \cite{Mills-1989} \\
        & $\diffusion_{i}^c(\dconc)$ & $\left( 1 + P_1\dconc^{0.5} + P_2\dconc + P_3\dconc^{1.5} + P_4\dconc^2
          \right)^{-1}$ & This work \\
        & $\diffusion_{i}^w(\dwall)$ & $1-\exp{\left(-P_1(\dwall+P_2)\right)}$ &
          \cite{Makarov-1998,Simakov-2010} \vspace{0.25cm} \vspace{0.5cm} \\
      
      \multirow{4}{1.5cm}{Ion electro\hyp{}phoretic mobility} & $\mobility_{\Na}^0$ &
        \SI{5.192e-4}{\square\meter\per\second\per\volt} & \cite{Bianchi-1989} \\
        & $\mobility_{\Cl}^0$ & \SI{7.909e-4}{\square\meter\per\second\per\volt} & \cite{Bianchi-1989} \\
        & $\mobility_{i}^c(\dconc)$ & $\left( 1 + P_1\dconc^{0.5} + P_2\dconc + P_3\dconc^{1.5} + P_4\dconc^2
          \right)^{-1}$ & This work \\
        & $\mobility_{i}^w(\dwall)$ & $1-\exp{\left(-P_1(\dwall+P_2)\right)}$ &
          \cite{Makarov-1998,Simakov-2010} \vspace{0.5cm} \\
    
      \multirow{2}{1.5cm}{Ion transport number}
        & $\transportn_{\Na}^0$ & 0.396 & \cite{Bianchi-1989} \\
        & $\transportn_{\Na}^c(\dconc)$ & $\left( 1 + P_1\dconc^{0.5} + P_2\dconc + P_3\dconc^{1.5} +
          P_4\dconc^2 \right)^{-1}$ & This work \vspace{0.5cm} \\
    
      \multirow{3}{1.5cm}{Dynamic viscosity}
        & $\viscosity^0$ & \SI{8.904e-4}{\pascal\second} & \cite{Hai-Lang-1996} \\
        & $\viscosity^c(\dconc)$ & $ 1 + P_1 \dconc^{0.5} + P_2 \dconc + P_3 \dconc^2 + P_4 \dconc^{3.5}$ &
          This work \\
        & $\viscosity^w(\dwall)$ & $1 + \exp{ \left( -P_1(\dwall-P_2) \right) }$ & \cite{Pronk-2014}
          \vspace{0.5cm} \\
    
      \multirow{2}{1.5cm}{Fluid density}
        & $\density^0$ & \SI{997}{\kilogram\per\cubic\meter} & \cite{Hai-Lang-1996} \\
        & $\density^c(\dconc)$ & $1 + P_1 \dconc + P_2 \dconc^2$ & This work \\
      \bottomrule
    \end{tabularx}
    \begin{tablenotes}
      \item[a] Dependencies on either $\dconc = \avionconc/1$~M (dimensionless average ion concentration) and
      $\dwall = \walldistance/1$~nm (dimensionless distance from the nanopore wall);
      \item[b] Values at infinite dilution for a system temperature of \SI{298.15}{\kelvin};
      \item[c] These functions are empirical and hence have no physical meaning. $L$ is the Langevin function
      $L (x) = \coth(x) - 1/x$. The values of the fitting parameters $P_x$ of each property can be found in
      \cref{tab:corrections_parameters} and graphs of the fits in
      \cref{fig:epnpns_concentration_d,fig:epnpns_concentration_mu,fig:epnpns_concentration_solv,fig:epnpns_distance}.
    \end{tablenotes}
  \end{threeparttable}
\end{table}

\section{Empirical parameterization}
\label{sec:epnp-ns:parameterization}

The properties of the ions and water molecules in the electrolyte depend strongly on the salt concentration
($\concentration$) and the distance from the protein wall ($\walldistance$). To obtain reasonable estimates of
these values for any given \ce{NaCl} concentration or wall distance, we fitted a series of empirical
polynomial functions, using non-linear regression with the \code{lmfit} Python package~\cite{Newville-2014},
to literature data: either bulk electrolyte experimental data for the concentration dependencies and molecular
dynamics data for the wall distance dependencies. In the next sections, we will describe the fits in detail,
and the best-fit parameters are summarized in~\cref{tab:corrections_parameters}.

\subsection{Concentration-dependent fitting to bulk electrolyte data}

\subsubsection{Diffusivity, mobility, and transport number.}
The \ce{NaCl} concentration dependency data of ion self-diffusion coefficients, the ion electrophoretic
mobilities and the cation transport number were fitted to
\begin{align}\label{eq:concentration_d_mu_tna}
  X_{i} (\concentration) = P_0 X_{i}^c (\concentration) =
  P_0 \left[ 1
           + P_1 \concentration^{0.5}
           + P_2 \concentration
           + P_3 \concentration^{1.5}
           + P_4 \concentration^2 
      \right]^{-1}
  \text{ ,}
\end{align}
with $X$ either $\diffusion$ (diffusion), $\mobility$ (mobility) or $\transportn$ (transport number) and $i$
either \Na{} or \Cl{}. $P_0$ is the value at infinite dilution (\ie~$\concentration = \SI{0}{\Molar}$) and
fixed during fitting of parameters $P_1$, $P_2$, $P_3$ and $P_4$.

\begin{figure*}[!b]
  \centering

  \begin{subfigure}[t]{11cm}
    \centering
    \caption{}\vspace{-1mm}\label{fig:epnpns_concentration_d_sodium}
    \includegraphics[scale=1]{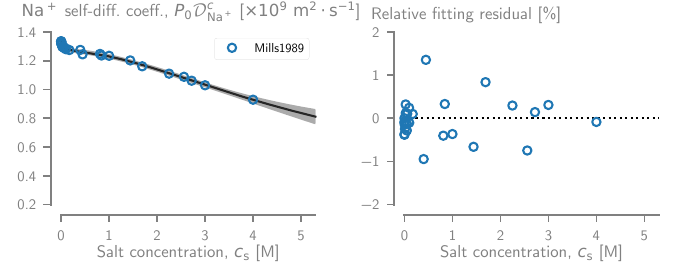}
  \end{subfigure}
  \\
  \begin{subfigure}[t]{11cm}
    \centering
    \caption{}\vspace{-1mm}\label{fig:epnpns_concentration_d_chloride}
    \includegraphics[scale=1]{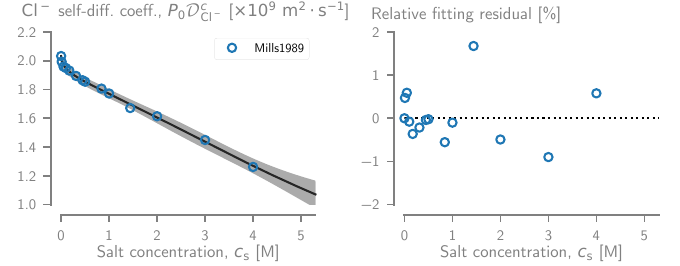}
  \end{subfigure}

  \caption[Concentration depend. of ion self-diffusion coefficients in \ce{NaCl}]%
    {%
      \textbf{Concentration dependency of ion self-diffusion coefficients in \ce{NaCl}.}
      (\subref{fig:epnpns_concentration_d_sodium})
      \Na{} and
      (\subref{fig:epnpns_concentration_d_chloride})
      \Cl{} self-diffusion coefficients~\cite{Mills-1989} as a function of the bulk \ce{NaCl} concentration
      (left) and the relative residuals after fitting (right) of \cref{eq:concentration_d_mu_tna}. Circles
      represent the experimental data and solid lines the fitted equation with the gray shading as the
      $3\sigma$ confidence interval.
  }\label{fig:epnpns_concentration_d}
\end{figure*}
\begin{figure*}[p]
  \centering

  \begin{subfigure}[t]{11cm}
    \centering
    \caption{}\vspace{-1mm}\label{fig:epnpns_concentration_mu_tsodium}
    \includegraphics[scale=1]{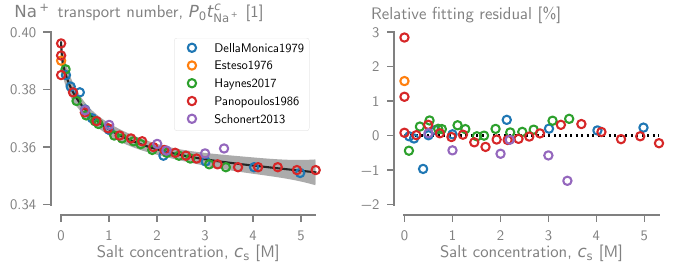}
  \end{subfigure}
  \\
  \begin{subfigure}[t]{11cm}
    \centering
    \caption{}\vspace{-1mm}\label{fig:epnpns_concentration_mu_sodium}
    \includegraphics[scale=1]{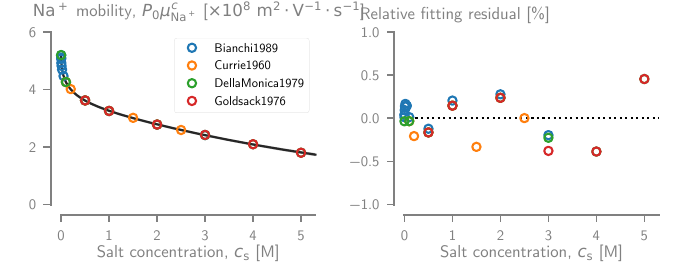}
  \end{subfigure}
  \\
  \begin{subfigure}[t]{11cm}
    \centering
    \caption{}\vspace{-1mm}\label{fig:epnpns_concentration_mu_chloride}
    \includegraphics[scale=1]{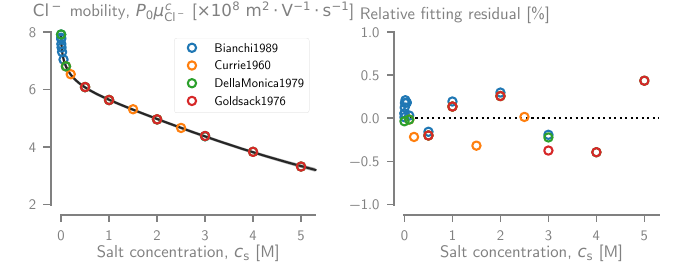}
  \end{subfigure}

  \caption%
    [Concentration depend. of the \ce{Na+} transport nb. and the ion electroph. mob. in \ce{NaCl}]
    {%
      \textbf{Concentration dependency of the cation transport number and the ion electrophoretic mobilities
      in \ce{NaCl}.}
      (\subref{fig:epnpns_concentration_mu_tsodium})
      \Na{} transport numbers
      and~\cite{Esteso-1976,Haynes-2017,DellaMonica-1979,Panopoulos-1986,Schonert-2013},
      (\subref{fig:epnpns_concentration_mu_sodium})
      \Na{} and
      (\subref{fig:epnpns_concentration_mu_chloride})
      \Cl{} electrophoretic mobilities~\cite{Bianchi-1989,Currie-1960,Goldsack-1976,DellaMonica-1979} as a
      function of the bulk \ce{NaCl} concentration (left) and the relative residuals after fitting (right) of
      \cref{eq:concentration_d_mu_tna}. Circles represent the experimental data and solid lines the fitted
      equation with the gray shading as the $3\sigma$ confidence interval.
  }\label{fig:epnpns_concentration_mu}
\end{figure*}

\paragraph{Ion self-diffusion coefficients.}
The data for the fitting of the self-diffusion coefficients of the \Na{} ($\diffusion_{\Na}^c (c)$,
\cref{fig:epnpns_concentration_d_sodium}) and \Cl{} ($\diffusion_{\Cl}^c (c)$,
\cref{fig:epnpns_concentration_d_chloride}) ranged from \SIrange{0}{4}{\Molar} and was taken from the
compilation of Mills~\cite{Mills-1989}. Fitting of \cref{eq:concentration_d_mu_tna} yielded relative residuals
\SI{<\pm2}{\percent} (\cref{fig:epnpns_concentration_d}, left panels), indicating an excellent representation
of the data. Because no experimental data were available beyond \SI{4}{\Molar}, the diffusivities between
\SIrange{4}{5.3}{\Molar} had to be extrapolated, whereas for simulated concentrations \SI{>5.3}{\Molar}, their
respective values at \SI{5.3}{\Molar} were used (\ie~$\diffusion_{\Na} (c > \mSI{5.3}{\Molar}) =
\mSI{0.81e-9}{\meter\squared\per\second}$, $\diffusion_{\Cl} (c > \mSI{5.3}{\Molar}) =
\mSI{1.07e-9}{\meter\squared\per\second}$).

\paragraph{Sodium transport number.}
The literature data for the concentration dependency of the \Na{} transport number ($\transportn_{\Na}^c (c)$,
\cref{fig:epnpns_concentration_mu_tsodium}) was obtained from five separate
sources~\cite{Esteso-1976,Haynes-2017,DellaMonica-1979,Panopoulos-1986,Schonert-2013} and spanned the entire
experimentally accessible \ce{NaCl} concentration range (\SIrange{0}{5.3}{\Molar}). After fitting
\cref{eq:concentration_d_mu_tna}, we found most relative fitting residuals to be \SI{<\pm1}{\percent}
(\cref{fig:epnpns_concentration_mu_tsodium}, left panel).

\paragraph{Ion electrophoretic mobilities.}
The individual data for concentration-dependent ionic mobilities of \Na{} ($\mobility^c_{\Na}$,
\cref{fig:epnpns_concentration_mu_sodium}) and \Cl{} ($\mobility^c_{\Cl}$,
\cref{fig:epnpns_concentration_mu_chloride}) were calculated from literature
values~\cite{Bianchi-1989,Currie-1960,Goldsack-1976,DellaMonica-1979} of the salt's molar conductivity
$\molarconductivity$ (between~\SIrange{0}{5}{\Molar}) using~\cite{ContrerasAburto-2013-1}
\begin{align}\label{eq:conductivity-to-mobility}
  \mobility_{i}(\concentration) = \frac{\specmolarconductivity_{i}(\concentration)}{\chargen_{i}\faraday}
  \quad\text{with}\quad \specmolarconductivity_{i}(\concentration) = \molarconductivity(\concentration)
  \transportn_{i}(\concentration)
  \text{ ,}
\end{align}
where $\specmolarconductivity_{i}(\concentration)$ is the specific molar conductivity of ion $i$, and
$\faraday$ the Faraday constant (\SI{96485}{\coulomb\per\mole}). The transport number for \Na{} was already
computed in the previous paragraph, and the one for \Cl{} is given by $\transportn_{\Cl}^c (c) = 1 -
\transportn_{\Na}^c (c)$. \Cref{eq:concentration_d_mu_tna} was fitted to both data sets individually, and the
relative residuals were found to be \SI{<\pm0.5}{\percent}. Mobilities for simulated concentrations between
\SIlist{5;5.3}{\Molar} were extrapolated, and for concentrations beyond the values were capped to those at
\SI{5.3}{\Molar} (\ie~$\mobility_{\Na} (c > \mSI{5.3}{\Molar}) =
\mSI{1.74e-8}{\meter\squared\per\volt\per\second}$, $\mobility_{\Cl} (c > \mSI{5.3}{\Molar}) =
\mSI{3.21e-8}{\meter\squared\per\volt\per\second}$).

\subsubsection{Viscosity.}

\begin{figure*}[p]
  \centering
  \begin{subfigure}[t]{11cm}
    \centering
    \caption{}\vspace{-1mm}\label{fig:epnpns_concentration_solv_eta}
    \includegraphics[scale=1]{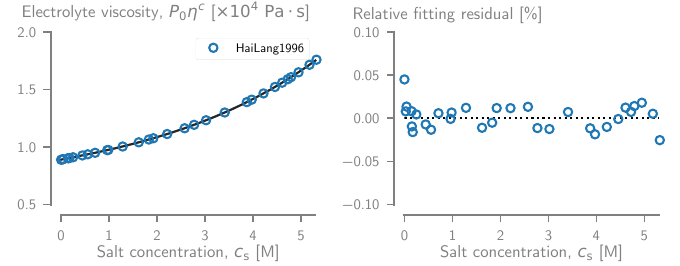}
  \end{subfigure}
  \\
  \begin{subfigure}[t]{11cm}
    \centering
    \caption{}\vspace{-1mm}\label{fig:epnpns_concentration_solv_rho}
    \includegraphics[scale=1]{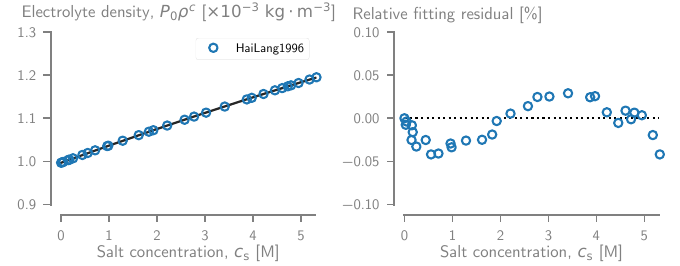}
  \end{subfigure}
  \\
  \begin{subfigure}[t]{11cm}
    \centering
    \caption{}\vspace{-1mm}\label{fig:epnpns_concentration_solv_epsr}
    \includegraphics[scale=1]{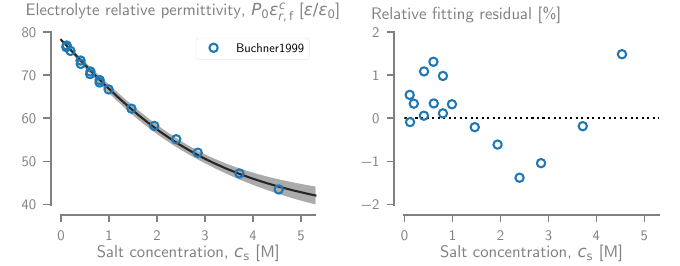}
  \end{subfigure}

  \caption[Concentration depend. of the electrolyte viscosity, density, and rel. perm. in \ce{NaCl}]%
    {%
      \textbf{Concentration dependency of the electrolyte viscosity, density, and relative permittivity in
      \ce{NaCl}.}
      (\subref{fig:epnpns_concentration_solv_eta})
      Viscosity~\cite{Hai-Lang-1996},
      (\subref{fig:epnpns_concentration_solv_rho})
      density~\cite{Hai-Lang-1996} and
      (\subref{fig:epnpns_concentration_solv_epsr})
      relative permittivity~\cite{Buchner-1999,Gavish-2016} as a function of the bulk \ce{NaCl} concentration
      (left) and the relative residuals after fitting (right) to
      \cref{eq:epnpns_concentration_solv_eta,,eq:epnpns_concentration_solv_rho,,eq:epnpns_concentration_solv_epsr},
      respectively. Circles represent the experimental data and solid lines the fitted equation with the gray
      shading as the $3\sigma$ confidence interval.
  }\label{fig:epnpns_concentration_solv}
\end{figure*}

The interpolation function representing the concentration-dependent electrolyte viscosity ($\viscosity
(\concentration)$) is given by
\begin{align}\label{eq:epnpns_concentration_solv_eta}
  \viscosity (\concentration) = P_0 \viscosity^c (\concentration) =
  P_0 \left[ 1
            + P_1 \concentration^{0.5}
            + P_2 \concentration
            + P_3 \concentration^{2}
            + P_4 \concentration^{3.5}
      \right]
  \text{ ,}
\end{align}
with $P_0$ the value at infinite dilution (fixed during fitting), and $P_1$, $P_2$, $P_3$ and $P_4$ fitting
parameters. Experimental data for parameterization of \cref{eq:epnpns_concentration_solv_eta} was taken from a
single literature source~\cite{Hai-Lang-1996} and covers the entire concentration range
(\SIrange{0}{5.3}{\Molar}). The resulting fit to this data, yielding $\viscosity^c (\concentration)$, is shown
in~\cref{fig:epnpns_concentration_solv_eta}. Relative fitting residuals were found to be very low
(\SI{<\pm0.05}{\percent}, \cref{eq:epnpns_concentration_solv_eta}, left panel). As with the previous
parameters, viscosities for concentrations above saturated concentrations were capped to their values at
\SI{5.3}{\Molar} (\ie~$\viscosity (c > \mSI{5.3}{\Molar}) = \mSI{1.75e-4}{\pascal\second}$).

\subsubsection{Density.}

The interpolation function representing the concentration dependency of the electrolyte density is given by
\begin{align}\label{eq:epnpns_concentration_solv_rho}
  \density (\concentration) = P_0 \density^c (\concentration) =
  P_0 \left[ 1
            + P_1 \concentration
            + P_2 \concentration^{2}
      \right]
  \text{ ,}
\end{align}
with $P_0$ the value at infinite dilution (fixed during fitting), and $P_1$ and $P_2$ fitting parameters.
Experimental density data was taken from the source as the viscosity~\cite{Hai-Lang-1996} and the fit,
yielding $\density^c (\concentration)$ (\cref{eq:epnpns_concentration_solv_rho}), also showed very low
relative fitting residuals (\SI{<\pm0.05}{\percent}, \cref{eq:epnpns_concentration_solv_rho}, left panel).
Density values for concentrations above \SI{5.3}{\Molar} were capped as usual (\ie~$\density (c >
\mSI{5.3}{\Molar}) = \mSI{1.19e3}{\kilo\gram\per\meter\cubed}$).

\subsubsection{Relative permittivity.}

For the concentration dependency of the electrolyte relative permittivity, we used the microfield model
developed by Gavish \etal{}~\cite{Gavish-2016}
\begin{align}\label{eq:epnpns_concentration_solv_epsr}
  \relperm (\concentration) = P_0 \permittivity_{r,\text{f}}^c (\concentration) =
  P_0 \left[ 1
              - \left( 1 - \dfrac{P_1}{P_0}\right)
              L \left( \dfrac{3 P_2}{P_0 - P_1} \concentration \right)
      \right]
  \text{ ,}
\end{align}
where $L$ is the Langevin function ($L(x) = \coth{(x)} - 1/x$), $P_0$ the value at infinite dilution, $P_1$
the limiting permittivity ($\permittivity_{r,ms}$) and $P_2$ the total excess polarization of the ions
($\alpha$). Even though we fitted the equation to our literature dataset~\cite{Buchner-1999} ($P_1 =
29.50\pm1.32$ and $P_2 = 11.74\pm0.21$), we opted to make use of the parameters given by Gavish for \ce{NaCl}
at \SI{298.15}{\kelvin} ($P_1 = 30.08$ and $P_2 = 11.5$~\cite{Gavish-2016},
\cref{fig:epnpns_concentration_solv_epsr}). We found the relative fitting residuals
(\cref{fig:epnpns_concentration_solv_epsr}, left panel) to be low (\SI{<\pm2}{percent}). Relative
permittivities for concentrations above saturation were capped as before (\ie~$\relperm (c >
\mSI{5.3}{\Molar}) = \num{42.12}$).

\subsubsection{Note on the concentration dependencies.}

All concentration-dependent parameters use the local ionic strength rather than their individual ion
concentrations. Though valid for electroneutral bulk solutions, this approximation no longer holds inside the
electrical double layer (\ie~near charged surfaces or inside small nanopores), where local electroneutrality
is violated. The main reasons for making this simplification regardless are the lack of non-bulk experimental
data and the absence of a tractable analytical model. Furthermore, we will see in~\cref{ch:transport} that the
current implementation of our concentration-dependent functions will lead to an excellent agreement with the
experimental data in all but the most extreme cases, justifying our choice \textit{a posteriori}.

\subsection[Wall distance-dependent fitting to {MD} data]%
           {Wall distance-dependent fitting to molecular dynamics data}

\subsubsection{Diffusivity and mobility.}
\begin{figure*}[b]
  \centering

  \begin{subfigure}[t]{11cm}
    \centering
    \caption{}\vspace{0mm}\label{fig:epnpns_distance_d}
    \includegraphics[scale=1]{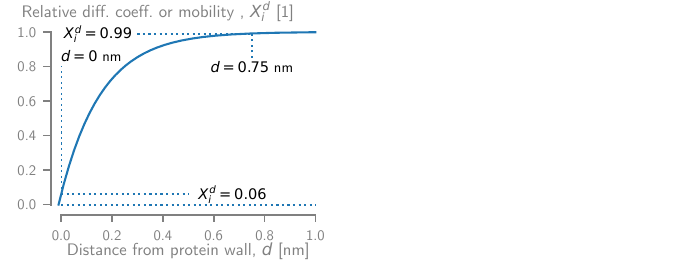}
  \end{subfigure}
  \\
  \begin{subfigure}[t]{11cm}
    \centering
    \caption{}\vspace{0mm}\label{fig:epnpns_distance_eta}
    \includegraphics[scale=1]{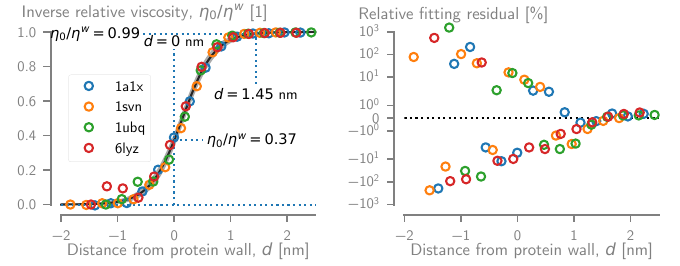}
  \end{subfigure}

  \caption[Wall distance depend. of ion self-diff. coeff., mobilities and electrolyte viscosity]%
    {%
    \textbf{Wall distance dependency of ion self-diffusion coefficients, mobilities, and electrolyte viscosity.}
      (\subref{fig:epnpns_distance_d})
      Relative ion self-diffusion coefficient and
      mobility~\cite{Simakov-2010,Makarov-1998,Pederson-2015,Wilson-2019}, and
      (\subref{fig:epnpns_distance_eta})
      inverse relative viscosity~\cite{Pronk-2014} as a function of the distance from the protein wall (left)
      and the relative residuals after fitting (right) to \cref{eq:epnpns_distance_d,,eq:epnpns_distance_eta},
      respectively. Circles represent the experimental data and solid lines the fitted equation with the gray
      shading as the $3\sigma$ confidence interval.
  }\label{fig:epnpns_distance}
\end{figure*}

The protein wall distance function used for the ion self-diffusion coefficients and the ion electrophoretic
mobilities was taken from Simakov \etal{}~\cite{Simakov-2010}, who fitted it to the molecular dynamics data
from Makarov \etal{}~\cite{Makarov-1998}. The function is a sigmoidal given by
\begin{align}\label{eq:epnpns_distance_d}
  X_{i}^w (\walldistance) = 1 - \exp{-P_1 (\walldistance - P_2)}
  \text{ ,}
\end{align}
with $X$ either $\diffusion$ (diffusion) or $\mobility$ (mobility) and $i$ either \Na{} or \Cl{} and $P_1$
($a$) and $P_2$ ($r_0$) the fitting parameters. For $P_1$, we used the value determined by Simakov \etal{}
(\SI{6.2}{\per\nm}), but for $P_2$ we opted to use a value of \SI{0.01}{\nm} instead of \SI{0.22}{\nm}
proposed by Simakov~\cite{Simakov-2010} and further corroborated by Wilson \etal{}~\cite{Wilson-2019}. We
chose this smaller offset because \cref{eq:epnpns_distance_d} quickly becomes negative beyond the offset
$P_2$, resulting in nonphysical values. Pederson \etal{} solved this by setting an arbitrary minimum positive
value for the diffusion coefficient~\cite{Pederson-2015}, but this would result in a `dead' zone of nearly
\SI{0.2}{\nm} from each wall with a very low diffusivity (\ie~several orders of magnitude below the bulk
value). This would make the current of a pore with a diameter smaller than \SI{0.5}{\nm} effectively zero,
which has been shown experimentally to be not the case~\cite{Rigo-2019}. In addition, $P_2$ represents the
center-to-center distance between the ion and the nearest heavy atom, meaning that approximately half of the
\SI{0.22}{\nm} would already be inside our geometry (\ie~within the van der Waals radius of the heavy atom),
and the other half would be inaccessible to ions due to excluded volume effects. Because we are not modeling
the latter, we opted to assign the full offset given by $P_2$ to the heavy atom, effectively assuming that all
ion-inaccessible space is already taken into account by our geometry. Our choice of parameters leads to
values of $X_{i}^w (\mSI{0}{\nm}) = 0.06$ (\ie~at the wall boundary) and $X_{i}^w (\mSI{0.75}{\nm}) = 0.99$
(\ie~\SI{0.75}{\nm} within the electrolyte, \cref{fig:epnpns_distance_d}), which roughly corresponds to the
values reported by Wilson \etal{}~\cite{Wilson-2019}.
\begin{landscape}
  \hspace{-1.5cm}
  \begin{threeparttable}[p]
    \centering
       \captionsetup{width=19.25cm}
    \caption{Overview of the \ce{NaCl} fitting parameters used for interpolation.}
    \label{tab:corrections_parameters}
    \sisetup{inter-unit-product=\ensuremath{{}\cdot{}}}
    \renewcommand{\arraystretch}{2}
    \scriptsize
    \begin{tabularx}{19.25cm}{@{}
            l l S[table-format=1.3] S[table-format=-1.2(2)e-1] S[table-format=-1.2(2)e-1]
            S[table-format=-1.2(2)e-1] S[table-format=-1.2(2)e-1] S[table-format=>1.2] l @{}}
      \toprule
        & & {Inf. dil.} & \multicolumn{4}{c}{Fitting parameters} & & \\
      \cmidrule(r){3-3} \cmidrule(l){4-7} Property & Eq. & $P_{0}$ & $P_{1}$ & $P_{2}$ & $P_{3}$ & $P_{4}$ &
      $R^{2}$ & References \\
      \midrule
      $\diffusion_{\Na{}}^c(c)$ & \cref{eq:concentration_d_mu_tna} & 1.334 \tnote{a} & 2.02(14)e-1 &
      -3.05(41)e-1 & 2.19(38)e-1 & -3.13(108)e-2 & >0.99 & \cite{Mills-1989} \\
      $\diffusion_{\Cl{}}^c(c)$ & \cref{eq:concentration_d_mu_tna} & 2.032 \tnote{a} & 1.49(30)e-1 &
      -4.94(904)e-2 & 3.40(826)e-2 & 1.43(230)e-2 & >0.99 & \cite{Mills-1989} \\
      $\transportn_{\Na{}}^c(c)$ & \cref{eq:concentration_d_mu_tna} & 0.3963 & 9.38(164)e-2 & 2.86(324)e-3 &
      -1.88(652)e-2 & 4.51(275)e-3 & 0.98 &
      \cite{Esteso-1976,Haynes-2017,DellaMonica-1979,Panopoulos-1986,Schonert-2013}\\
      $\mobility_{\Na{}}^c(c)$ & \cref{eq:concentration_d_mu_tna} & 5.192 \tnote{b} & 7.91(6)e-1 &
      -3.53(17)e-1 & 1.46(15)e-1 & 9.23(389)e-3 & >0.99 &
      \cite{Bianchi-1989,Currie-1960,Goldsack-1976,DellaMonica-1979} \\
      $\mobility_{\Cl{}}^c(c)$ & \cref{eq:concentration_d_mu_tna} & 7.909 \tnote{b} & 6.29(6)e-1 &
      -4.29(17)e-1 & 2.12(14)e-1 & -1.07(37)e-2 & >0.99 &
      \cite{Bianchi-1989,Currie-1960,Goldsack-1976,DellaMonica-1979} \\
      $\viscosity^c(c)$ & \cref{eq:epnpns_concentration_solv_eta} & 0.8904 \tnote{c} & 7.56(27)e-3 &
      7.77(4)e-2 & 1.19(1)e-2 & 5.95(35)e-4 & >0.99 & \cite{Hai-Lang-1996} \\
      $\density^c(c)$ & \cref{eq:epnpns_concentration_solv_rho} & 0.997 \tnote{d} & 4.06(1)e-2 & -6.39(16)e-4
      & & & >0.99 & \cite{Hai-Lang-1996} \\
      $\permittivity_{r,\text{f}}^c(c)$ & \cref{eq:epnpns_concentration_solv_epsr} & 78.15 & 3.08(0)e1 &
      1.15(0)e1 & & & & \cite{Buchner-1999,Gavish-2016} \\
      $\diffusion_{i}^w(d)$ & \cref{eq:epnpns_distance_d} & & 6.2 & 0.01 & & & &
      \cite{Makarov-1998,Simakov-2010,Pederson-2015} \\
      $\mobility_{i}^w(d)$ & \cref{eq:epnpns_distance_d} & & 6.2 & 0.01 & & & &
      \cite{Makarov-1998,Simakov-2010,Pederson-2015} \\
      $\viscosity^w(d)$ & \cref{eq:epnpns_distance_eta} & & 3.36(23) & 1.47(23)e-1 & & & 0.97 &
      \cite{Pronk-2014} \\
      \bottomrule
    \end{tabularx}
    \begin{tablenotes}
      \item[a] Unit and scaling: \SI{e-9}{\square\meter\per\second}
      \item[b] Unit and scaling: \SI{e-8}{\square\meter\per\second\per\volt}
      \item[c] Unit and scaling: \SI{e-4}{\pascal\second}
      \item[d] Unit and scaling: \SI{e3}{\kilo\gram\per\cubic\meter}
    \end{tablenotes}
  \end{threeparttable}
\end{landscape}

\subsubsection{Viscosity.}
Analogously to the diffusivity of ions, the movement of water molecules near a protein wall is also
hampered~\cite{Makarov-1998}, resulting in substantially higher viscosities~\cite{Pronk-2014}. To model this,
we fitted the following logistic curve to the molecular dynamics data from Pronk \etal{}~\cite{Pronk-2014}
\begin{align}\label{eq:epnpns_distance_eta}
  \viscosity^w (\walldistance) = 1 + \exp{ \left( -P_1 (\walldistance - P_2) \right) }
  \text{ ,}
\end{align}
with $P_1$ and $P_2$ the fitting parameters, representing the slope of the sigmoidal and the offset of the
sigmoid's center from 0, respectively. Prior to the fitting, the viscosity data for all proteins were offset
by their hydrodynamic radii, such that $\walldistance = \SI{0}{\nm}$ represents the mean distance from the
wall for each protein, regardless of its size. Additionally, we normalized and inverted the relative
viscosities ($\viscosity_0 / \viscosity^w$) allowing us to fit values between 0 and 1. Our parameters result
in values of $\viscosity_0 / \viscosity^w (\SI{0}{\nm}) = 0.37$ (\ie~at the protein wall) and $\viscosity_0
/ \viscosity^w (\SI{1.45}{\nm}) = 0.99$ (\ie~\SI{1.45}{\nm} within the electrolyte,
\cref{fig:epnpns_distance_eta}). The relative fitting residuals (\cref{fig:epnpns_distance_eta}, left panel)
for $\walldistance \ge \mSI{0}{\nm}$ (the utilized portion of the function) are typically
\SI{<\pm5}{\percent}.

\section{Conclusion}
\label{sec:epnp-ns:conclusion}

In this chapter, we have developed the \glsfirst{epnp-ns} equations, an improved version of the well-known
transport equations that is geared towards the accurate modeling of the transport of ions and water through
biological nanopores. Our \gls{epnp-ns} equations combine many of the improvements already available in the
literature, in addition to several new corrections. These include the finite size of the ions (\ie~steric
repulsion between ions), self-consistent concentration- and positional-dependent parametrization of the ionic
transport coefficients (\ie~diffusion coefficient and mobility), and the electrolyte properties
(\ie~density, viscosity, and relative permittivity).

Most electrolyte properties depend on the salt concentration. Typically, ion self-diffusion coefficients,
ionic conductivities, and relative permittivity decrease with concentrations, whereas density and viscosity
increase. Self-consistent parameterization of the concentration dependencies was based on interpolation
functions, fitted to experimental electrolyte data obtained from literature sources. Note that, given that our
parameters are derived from `bulk' electrolyte data, their accuracy with respect to non-electroneutral
solutions (\eg~the \gls{edl} within a nanopore) is not known. Nevertheless, we expect them to yield
significantly more physical results better compared to infinite dilution values. 

At the nanoscale, the interactions between the nanopore walls on the one hand, and the ions and water
molecules on the other, result in a short-ranged (\SI{<1}{\nm}) reduction of diffusivity and mobility of the
ions~\cite{Makarov-1998}, and an increase of the water viscosity~\cite{Pronk-2014}. We approximated these
distance-dependencies by fitting sigmoid functions to data from \gls{md} simulations, obtained from the
literature. While crude, we expect them to reproduce the physically observed effects adequately.

In \cref{ch:transport}, we will apply our \gls{epnp-ns} framework to a 2D-axisymmetric model of a biological
nanopore, yielding a wealth of information that is both qualitatively and quantitatively accurate.

\cleardoublepage

%

%
  \graphicspath{{\chaptersdir/transport/image/}}%
\chapter{Modeling the transport of ions and water through {ClyA}}
\label{ch:transport}
%


\epigraphhead[\epipos]{%
\epigraph{%
  ``That was one of the odd, persistent defects of simulations: no matter how precise they became, the
  participant remained aware that they were not reality.''
}{%
  \textit{`Sylveste' in `Revelation Space' by Alastair Reynolds}
}}

\definecolor{shadecolor}{gray}{0.85}
\begin{shaded}
This chapter was published as:
\begin{itemize}
  \item K. Willems*, D. Rui\'{c}, F. L. R. Lucas, U. Barman, N. Verellen, G. Maglia and P. Van Dorpe.
        \textit{Nanoscale} \textbf{12}, 16775--16795 (2020) 
\end{itemize}
\newpage
\end{shaded}
%
%

%

In this chapter we apply the \glsfirst{epnp-ns} equations introduced in \cref{ch:epnpns} to the biological
nanopore \glsfirst{clya}. To this end, we set up a 2D-axisymmetric representation of the geometry and charge
distribution of the pore, and numerically solved the \gls{epnp-ns} equations using \glsfirst{fea} for a wide
range of bias voltages and ionic strengths. This allowed us to validate our model against experimental data
and investigate \gls{clya}'s transport properties in detail.
To improve readability and to allow for better integration of the supplementary information into the main
chapter text, the original manuscript was split into \cref{ch:epnpns,ch:transport}, detailing the theory of
the framework and its application to the \gls{clya} nanopore, respectively. The remainder of the supplementary
information can be found in \cref{ch:transport_appendix}. The text and figures of this chapter represent
entirely my own work. All experimental work was performed by Florian L. R. Lucas.
\adapRSC{Willems-2020}
%
%
%

\section{Introduction}
\label{sec:transport:intro}

To validate our new \gls{epnp-ns} framework, we applied it to a 2D-axisymmetric model of
\gls{clya}~\cite{Mueller-2009}, a large protein nanopore~\cite{Soskine-2012} that typically contains 12 or
more identical subunits~\cite{Soskine-2013,Peng-2019}. We choose to model the \gls{clya-as} type I, a
dodecameric variant of the wild type \gls{clya} from \textit{S. Typhii} that was artificially evolved for
improved stability~\cite{Soskine-2013}. \Gls{clya-as} has been extensively used in experimental studies of
proteins~\cite{Soskine-2013,VanMeervelt-2014,Soskine-Biesemans-2015,Biesemans-2015,Wloka-2017,
VanMeervelt-2017,Galenkamp-2018,Willems-Ruic-Biesemans-2019,Zernia-2020,Galenkamp-2020} and
\gls{dna}~\cite{Franceschini-2013,Franceschini-2016,Nomidis-2018} and has even been employed as a targeted
immunotoxin~\cite{Mutter-2018}. This allowed us to gauge the qualitative and quantitative performance of the
\gls{epnp-ns} equations and simultaneously elucidate previously inaccessible details about the environment
inside the pore.

\begin{figure*}[b]
  \centering
  \begin{subfigure}[t]{5.5cm}
    \centering
    \caption{}\hspace{0mm}\label{fig:transport_clya_side}
    \includegraphics[scale=1]{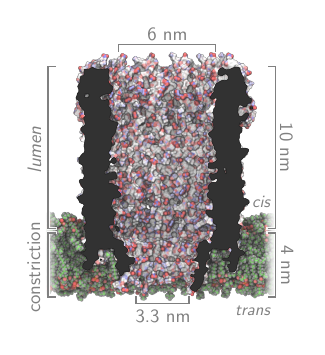}
  \end{subfigure}
  \begin{subfigure}[t]{5.5cm}
    \centering
    \caption{}\hspace{0mm}\label{fig:transport_clya_top}
    \includegraphics[scale=1]{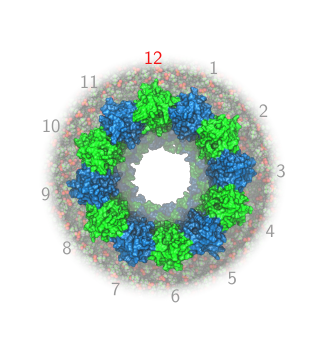}
  \end{subfigure}

  \caption[All-atom model of {ClyA-AS}]
  {%
    \textbf{All-atom models of {ClyA-AS}.}
    (\subref{fig:transport_clya_side})
    Axial cross-sectional and
    (\subref{fig:transport_clya_top})
    top views of the dodecameric nanopore \gls{clya-as}~\cite{Soskine-2013}, derived through homology
    modeling from the \textit{E. coli} \gls{clya} crystal structure (\pdbid{2WCD}~\cite{Mueller-2009}).
    Measurements indicate the average `cylindrical' sizes of the \cisi{} \lumen{} (\SI{6}{\nm} in diameter,
    \SI{10}{\nm} in height) and the \transi{} constriction (\SI{3.3}{\nm} in diameter, \SI{4}{\nm} in height).
    Images were rendered with \glsxtrshort{vmd}~\cite{Humphrey-1996,Stone-1998}.
  }\label{fig:transport_clya}
\end{figure*}

\Gls{clya} is a relatively large protein nanopore that self-assembles on lipid bilayers to form \SI{14}{\nm}
long hydrophilic channels (\cref{fig:transport_clya_side}). The interior of the pore can be divided into
roughly two cylindrical compartments: the \cisi{} \lumen{} (\SI{\approx6}{\nm} diameter, \SI{\approx10}{\nm}
height), and the \transi{} constriction (\SI{\approx3.3}{\nm} diameter, \SI{\approx4}{\nm} height). Because
\gls{clya} consists of 12 identical subunits (\cref{fig:transport_clya_top}), it exhibits a high degree of
radial symmetry, a geometrical feature that can be exploited to obtain meaningful results at a much lower
computational cost~\cite{Cervera-2005,Lu-2012,Pederson-2015}. However, this requires the reduction of the full
3D atomic structure and charge distribution to a realistic 2D-axisymmetric model.

We will start this chapter by describing the process of creating such a 2D-axisymmetric model of a nanopore
from a 3D full atom homology model (\cref{sec:transport:model}). This is followed by the validation of the model
through a direct comparison of simulated and experimentally measured ionic conductances, and a thorough
characterization of the influence of the bulk ionic strength and the applied bias voltage on cation and anion
concentrations inside the pore, the electrostatic potential distribution and magnitude of the electro-osmotic
flow (\cref{sec:transport:results}). Finally, we touch upon our key findings and their impact
(\cref{sec:transport:conclusion}).

\begin{figure*}[p]
  \centering
  \begin{minipage}[t]{6.75cm}
    \begin{subfigure}[t]{6.75cm}
      \centering
      \caption{}\label{fig:transport_model_zoom}
      \includegraphics[scale=1]{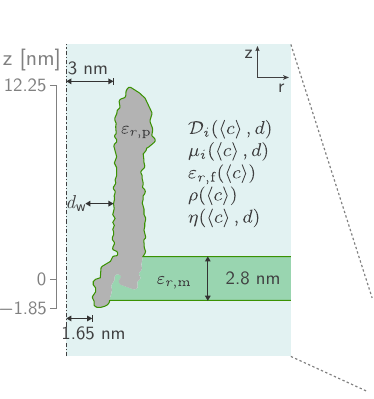}
    \end{subfigure}
    \\
    \begin{minipage}[t]{5.5cm}
      \addtocounter{subfigure}{1}
      \begin{subfigure}[t]{1.5cm}
        \centering
        \caption{}\label{fig:transport_model_outline}
        \vspace{-5mm}
        \includegraphics[scale=1]{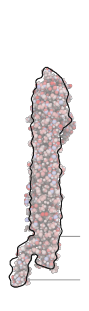}
      \end{subfigure}
      \begin{subfigure}[t]{3cm}
        \centering
        \caption{}\hspace{5mm}\label{fig:transport_model_scd}
        \vspace{-5mm}
        \includegraphics[scale=1]{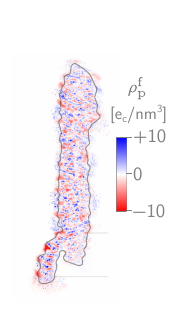}
      \end{subfigure}
    \end{minipage}
  \end{minipage}
  \hspace{-1.0cm}
  \addtocounter{subfigure}{-3}
  \begin{subfigure}[t]{5.65cm}
    \centering
    \caption{}\label{fig:transport_model_full}
    \includegraphics[scale=1]{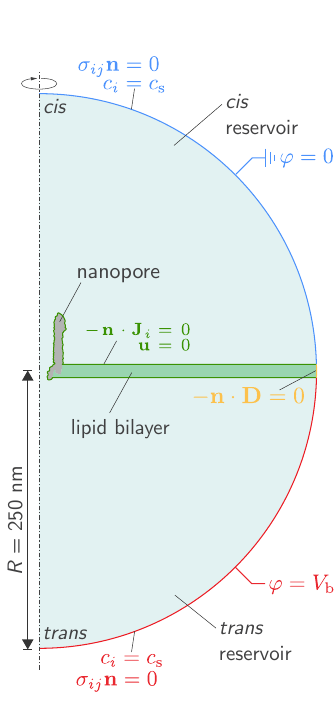}
  \end{subfigure}

  \caption[A 2D-axisymmetric model of {ClyA-AS}]
  {%
    \textbf{A 2D-axisymmetric model of {ClyA-AS}.}
    (\subref{fig:transport_model_zoom}+\subref{fig:transport_model_full})
    The 2D-axisymmetric simulation geometry of \gls{clya-as} (gray) embedded in a lipid bilayer (green) and
    surrounded by a spherical water reservoir (blue). Note that all electrolyte parameters depend on the local
    average ion concentration $\avionconc=\frac{1}{n}\sum_{i}^{n}\concentration_{i}$ and that some are also
    influenced by the distance from the nanopore wall $\walldistance$.
    (\subref{fig:transport_model_outline})
    The 2D-axisymmetric geometry was derived directly from the all-atom model using the radially averaged
    atomic density (see methods for details). Hence, it closely follows the outline of a \ang{30} wedge out of
    the homology model.
    (\subref{fig:transport_model_scd})
    The fixed space charge density ($\scdpore$) map of \gls{clya-as}, obtained by Gaussian projection of each
    atom's partial charge onto a 2D plane (see methods for details).
  }\label{fig:transport_model}
\end{figure*}

\section{A 2D-axisymmetric model of {ClyA}}
\label{sec:transport:model}

\subsection{Simulation system geometry and boundary conditions}
\label{sec:transport:global_geometry}

\subsubsection{Simulation geometry}

The complete system (\cref{fig:transport_model_zoom,fig:transport_model_full}) consists of a large
hemispherical electrolyte reservoir ($R=\SI{250}{\nm}$), split through the middle into a \cisi{} and a
\transi{} compartment by a lipid bilayer ($h=\SI{2.8}{\nm}$), which contains the nanopore at its center. Both
the bilayer and the nanopore are represented by dielectric blocks (see~\cref{tab:corrections_equations} for
parameters) that are impermeable to ions and water.

\subsubsection{Boundary conditions}

The reservoir boundaries were set up, using Dirichlet \glspl{bc}, to act as electrodes: the \cisi{} side was
grounded ($\potential = 0$) and a fixed bias potential was applied along the \transi{} edge ($\potential =
\vbias$). To simulate the presence of an endless reservoir, the ion concentration at both external boundaries
were fixed to the bulk salt concentration ($\concentration_i = \cbulk$) and the unconstrained flow in and out
of the computational domain was enabled by means of a `no normal stress' condition ($\hydrostresstensor\vec{n}
= 0$). The boundary conditions on the edges of the reservoir shared with the nanopore and bilayer were set to
no-flux ($-\vec{n}\cdot\vec{\flux}_{i} = 0$) and no-slip ($\vec{\velocity} = 0$), preventing the flux of ions
through them and mimicking a sticky hydrophilic surface, respectively. Finally, a Neumann boundary condition
was applied at the bilayer's external boundary ($-\vec{n}\cdot\left(\absperm \relperm \nabla \potential
\right) = 0$). More details can be found in~\cref{sec:epnpns_appendix:weak_forms}.

\subsection{From full-atom to a 2D-axisymmetric}
\label{sec:transport:pore_geometry}

\subsubsection{Molecular dynamics simulations.}

To sample the conformational space off the flexible side-chains of the nanopore, we used \gls{md} to
equilibrate the full-atom homology model of \gls{clya-as} type I~\cite{Soskine-2013} from
\cref{ch:electrostatics} (see~\cref{sec:elec:methods:molec}) for \SI{30}{\ns} in an explicit solvent with
harmonic constraints on the protein backbone atoms. The \SI{30}{\ns} production run was performed using a
\gls{nvt} ensemble at \SI{298.15}{\kelvin} and the atomic coordinates saved every \SI{5}{\ps}. Accidental
structural deterioration of the pore was prevented by harmonically restraining the C\ta{} atoms to their
original positions in the crystal structure with a spring constant of \SI{695}{\pN\per\nm} during the entire
\gls{md} run~\cite{Bhattacharya-2011}. All other atoms were allowed to move freely. This allowed us to sample
the conformational landscape of all the \gls{clya} side chains without disrupting the overall structure of the
pore and to generate a well-averaged geometry and charge distribution. The analysis in the next sections were
performed with 50 sets of atomic coordinates---extracted from the final \SI{5}{\ns} of the coordinates of the
\gls{md} production run (\ie~every \SI{100}{\fs})---and aligned by minimizing the \gls{rmsd} between their
backbone atoms (\gls{vmd}'s \code{RMSD tool}~\cite{Humphrey-1996}).

\subsubsection{Side-chain B-factors.}

\begin{figure*}[p]
  \centering

  \begin{subfigure}[t]{11.5cm}
    \centering
    \caption{}\vspace{0mm}\label{fig:transport_bfactor_heatmap}
    \includegraphics[scale=1]{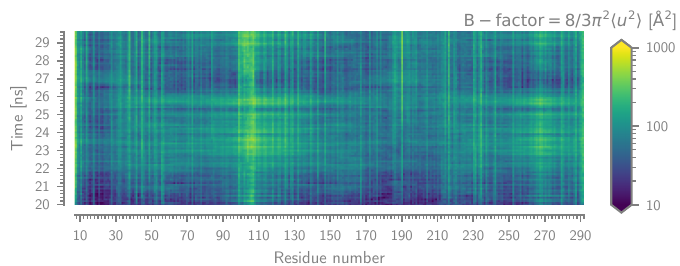}
  \end{subfigure}
  \\
  \begin{subfigure}[t]{11.5cm}
    \centering
    \caption{}\vspace{0mm}\label{fig:transport_bfactor_scatter}
    \includegraphics[scale=1]{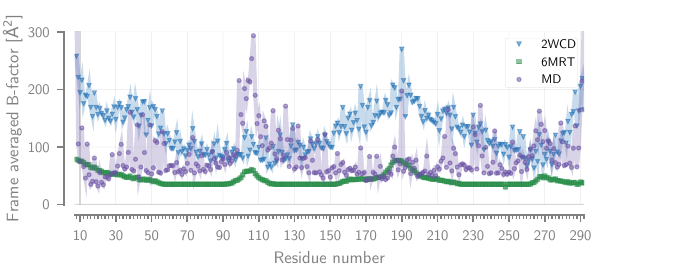}
  \end{subfigure}
  \\
  \begin{subfigure}[t]{11.5cm}
    \centering
    \hspace{-0.5cm}
    \caption{}\vspace{0mm}\label{fig:transport_bfactor_surface}
    \includegraphics[scale=1]{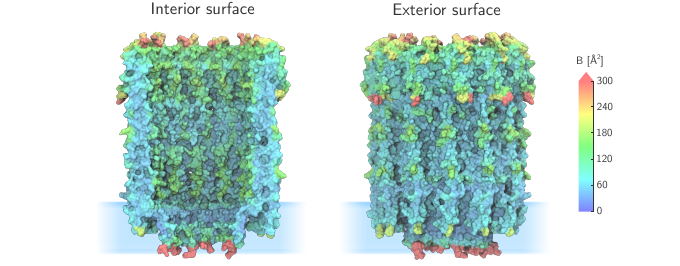}
  \end{subfigure}

  \caption[Per-residue B-factors for the last 10~ns of the {MD} run]%
  {%
    \textbf{Per-residue B-factors for the last 10~ns of the {MD} run.}
    (\subref{fig:transport_bfactor_heatmap})
    Heatmap of the per-residue B-factor ($\bfactor$, \cref{eq:bfactor}) of \gls{clya-as}, averaged over all 12
    monomers, as determined by our \gls{md} simulation.
    (\subref{fig:transport_bfactor_scatter})
    The data from the heatmap averaged over all frames and plotted together with the B-factors given by the
    crystal (\pdbid{2WCD}~\cite{Mueller-2009} and \gls{cryo-em} \pdbid{6MRT}~\cite{Peng-2019}) structures of
    the wild-type \gls{clya} dodecamer.
    (\subref{fig:transport_bfactor_surface})
    Molecular surface plots of the interior (left) and exterior (right) walls of \gls{clya-as}, colored
    according to the chain-averaged per-residue B-factor. Images were rendered using
    \gls{vmd}~\cite{Humphrey-1996}.
  }\label{fig:transport_bfactor}
\end{figure*}

The thermal fluctuations of the side chains can be quantified with the temperature factor or `B-factor'
($\bfactor_i$)~\cite{Bahar-1997}
\begin{align}\label{eq:bfactor}
  \bfactor_i = \dfrac{8}{3} \pi^2 \msd \text{ ,}
\end{align}
which can be computed for each atom $i$ from its \gls{msd} $\msd$. Starting from the last \SI{10}{\ns} of our
\SI{30}{\ns} \gls{md} trajectory, we computed the $\msd$ using the fast \gls{qcp}
algorithm~\cite{Theobald-2005,Liu-2010}, as implemented in the \code{MDAnalysis} Python
package~\cite{MichaudAgrawal-2011}. A heatmap of the resulting B-factors for each residue, averaged over all
12 monomers of \gls{clya}, can be found in~\cref{fig:transport_bfactor_heatmap}. In addition, we plotted the
frame-averaged B-factor (\cref{fig:transport_bfactor_scatter}), allowing us to compare the flexibility
observed in the \gls{md} trajectory with the B-factors given by the dodecamer \gls{clya} crystal
(\pdbid{2WCD}~\cite{Mueller-2009}) and \gls{cryo-em} (\pdbid{6MRT}~\cite{Peng-2019}) structures.
Interestingly, the $\bfactor_i$ of our \gls{md} simulation does not show much agreement with the original
\pdbid{2WCD} crystal structure, but it reproduces the more flexible regions observed in the (newer)
\pdbid{6MRT} \gls{cryo-em} structure (\ie~residue numbers \numrange{7}{20}, \numrange{95}{105},
\numrange{180}{200} and \numrange{260}{280}). Peng and coworkers also started from the \pdbid{2WCD} crystal
structure to generate their refined \gls{clya} model, albeit using a high-resolution \gls{cryo-em} map as a
restraint. Hence, the similarities in flexibility between our model and \pdbid{6MRT} are encouraging that the
\gls{md} relaxes the structure properly. Finally, we also visualized the \gls{md} $\bfactor_i$ on the interior
and exterior molecular surface of \gls{clya-as} (\cref{fig:transport_bfactor_surface}). This shows that the
most flexible regions are the \cisi{} and \transi{} entries of the pore. A cross-section of the final \gls{md}
structure, in comparison with the \gls{clya} dodecamer crystal \pdbid{2WCD}~\cite{Mueller-2009} and
\gls{cryo-em} \pdbid{6MRT}~\cite{Peng-2019} structures can be found
in~\cref{sec:transport_appendix:radius_comparision} (\cref{fig:transport_radius_comparison_surface}).

\subsubsection{Computing an axially symmetric geometry from the atomic density.}

The 2D-axisymmetric geometry of the \gls{clya-as} nanopore (\cref{fig:transport_model_outline}) was derived
directly from its full atom model by means of the molecular density ($\rho_\text{mol}$). For each of the 50
structures mentioned above, we computed the 3D-dimensional molecular density map on a \SI{0.5}{\angstrom}
resolution grid using the Gaussian function~\cite{Li-2013}
\begin{align}\label{eq:denspore}
  \rho_\text{mol} = 1 - \displaystyle\prod_{i} \left[ 1 - 
    \exp{\left(-\dfrac{d_i^2}{(\stdev\atomradius_{i})^2}\right)} \right]
    \text{ ,}
\end{align}
with
\begin{align}
  d_i = \sqrt{(x-x_i)^2 + (y-y_i)^2 + (z-z_i)^2}
  \text{ ,}
\end{align}
where for each atom $i$, $R_i$ is its Van der Waals radius, $d_i$ is the distance of grid coordinates $(x, y,
z)$ from the atom center $(x_i, y_i, z_i)$ and $\stdev = 0.93$ is a width factor. After averaging all 50
density maps in 3D, the resulting map was then radially averaged along the $z$-axis, relative to the center of
the pore, to obtain a 2D-axisymmetric density map. The contour line at \SI{25}{\percent} density was used as
the nanopore simulation geometry, after the manual removal of overlapping and superfluous vertices to improve
the quality of the final computational mesh. This 2D-axisymmetric geometry closely follows the outline of a
\SI{30}{\degree} `wedge' of the full atom structure (\cref{fig:transport_model_outline}).

\subsubsection{Computing an axially symmetric charge density from the atomic charge density.}

The 2D-axially symmetric charge distribution (\cref{fig:transport_model_scd}) was also derived directly from
the 50 sets of aligned nanopore coordinates that were used for the geometry. Inspired by how charges are
represented in the \gls{pme} method~\cite{Aksimentiev-2005}, we computed the fixed charge distribution of the
nanopore $\scdpore(r,z)$ by assuming that an atom $i$ of partial charge $\partialcharge_{i}$, at the location
$(x_i, y_i, z_i)$ in the full 3D atomistic pore model, contributes an amount $\partialcharge_{i}/2\pi r_i$ to
the partial charge at a point $(r_i,z_i)$ with $r_i = \sqrt{x_i^2 + y_i^2}$ in the averaged 2D-axisymmetric
model. This effectively spreads the charge over all angles to achieve axial symmetry. We assumed a Gaussian
distribution of the space charge density of each atom $i$ around its respective 2D-axisymmetric coordinates
$(r_i,z_i)$ such that
\begin{align}\label{eq:scdpore}
  \scdpore(r,z) = \dsum_{i} \dfrac{\ec\partialcharge_{i}}{\pi(\stdev\atomradius_{i})^2}
            \exp{\left(-\dfrac{(r-r_i)^2 + (z-z_i)^2}{(\stdev\atomradius_{i})^2}\right)}
  \text{ ,}
\end{align}
where $\atomradius_{i}$ is the atom radius, $\sigma = 0.5$ is the sharpness factor and $\ec$ is the
elementary charge. To embed $\scdpore$ with sufficient detail, yet efficiently, into a numeric solver, the
spatial coordinates were discretized with a grid spacing of \SI{0.005}{\nm} in the domain of $\scdpore$ and
precomputed values were used during the solver runtime. All partial charges (at \pH{7.5}) and radii were taken
from the CHARMM36 force field~\cite{Best-2012} and assigned using \code{PROPKA}~\cite{Olsson-2011} and
\code{PBD2PQR}~\cite{Jurrus-2018}. Within \code{COMSOL}, the surface charge density $\scdpore(r,z)$ was
imported as a 2D linear interpolation function and converted into a pseudo-3D volumetric charge density by
normalization over the local circumference of each element ($2 \pi r$). Integration of the resulting charge
density over the final mesh yielded a net charge of \SI{-72.9}{\ec}, which is indeed very close to the
\SI{-72}{\ec} of the original atomistic model. The equilibrium 2D-axisymmetric charge map
(\cref{fig:transport_model_scd}) was loaded directly into our solver as a linear interpolation function
($\scdpore$) and applied across all computational domains.

\subsection{On the extension to non-axisymmetric nanopores}

The possibility to use a computationally efficient 2D-axisymmetric model is enabled by \gls{clya}'s high
degree of rotational symmetry (\cref{fig:transport_clya_top}). To successfully extend our \gls{epnp-ns}
framework to non-axisymmetric pores---such as \gls{ompf}~\cite{Yamashita-2008} or
\gls{ompg}~\cite{Subbarao-2006}---and very narrow nanopores---such as \gls{ahl}~\cite{Song-1996},
\gls{frac}~\cite{Tanaka-2015}, or \gls{mspa}~\cite{Faller-2004}---these equations should be solved for a full
3D representation. For narrow pores, the radial averaging of the geometry might remove features essential for
correctly modeling the fluidic properties, such as internal corrugations or opposing charges within the same
radial plane. Even though this will significantly increase the computational cost, it would still be
significantly faster than either \gls{bd} or \gls{md} approaches (\ie~hours per data point instead of days).

Describing a methodology for setting up a 3D model is beyond the scope of this chapter. We expect it to
follow along the same lines as the approach outlined above, with the exception that the actual molecular
surface and charge distribution of the pore (see the methodology used by the
\gls{apbs}~\cite{Baker-2001,Jurrus-2018} for a potential approach) could be used instead of a radially
averaged one. The \gls{epnp-ns} equations themselves would require discretization in Cartesian ($x$,~$y$,~$z$)
rather than cylindrical ($r$,~$\phi$,~$z$) coordinates, but can otherwise remain unchanged.

\section{Results and discussion}
\label{sec:transport:results}

The \gls{iv} relationships of many nanopores, \gls{clya} included, often deviate significantly from Ohm's law.
This is because the ionic flux arises from a complex interplay between the pore's geometry (\eg~size, shape,
and charge distribution), the properties of the surrounding electrolyte (\eg~salt concentration viscosity and
relative permittivity), and the externally applied conditions (\eg~bias voltage, temperature, and pressure). The
ability of a computational model to quantitatively predict the ionic current of a nanopore over a wide range
of bias voltages and salt concentrations strongly indicates that it captures the essential physics governing
the nanofluidic transport. Hence, to validate our model, we experimentally measured the single-channel ionic
conductance of \gls{clya} at a wide range of experimentally relevant salt concentrations ($\cbulk$) and bias
voltages ($\vbias$). We compared these experimental data with the simulated ionic transport properties in
terms of current, conductance, rectification, and ion selectivity, of both the classical \gls{pnp-ns} and the
newly developed \gls{epnp-ns} equations---finding a quantitative match between the experiment and the
simulations.

\begin{figure*}[p]
  \centering
  \includegraphics[scale=1]{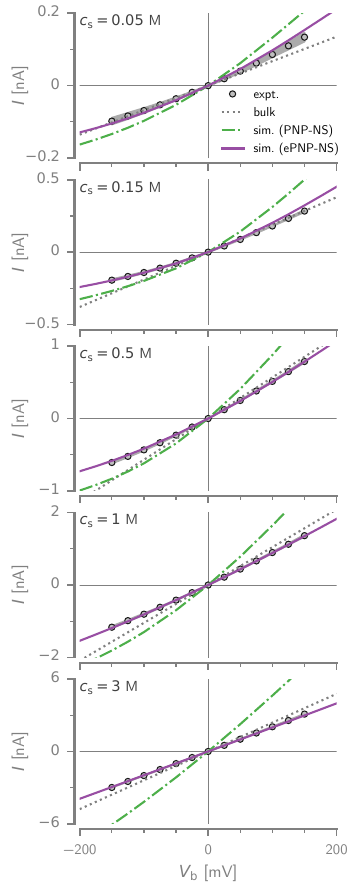}
  \caption[Meas. and sim. ionic current through single {ClyA-AS} nanopores]%
  {%
    \textbf{Measured and simulated ionic current through single {ClyA-AS} nanopores.}
    Comparison between the experimentally (expt.) measured, simple resistor pore model (bulk) and the
    simulated (\gls{pnp-ns} and \gls{epnp-ns}) \gls{iv} curves of \gls{clya-as} at \SI{25\pm1}{\celsius}
    between $\vbias=\mSI{\pm200}{\mV}$, and for $\cbulk=\mSIlist{0.05;0.15;0.5;1;3}{\Molar}$ \ce{NaCl}. The
    simulated currents were computed from the total ionic fluxes using \cref{eq:currentsim}. The bulk current
    was calculated by \cref{eq:bulk_nanopore_current} by modeling \gls{clya} as two series resistors
    (\cref{eq:bulk_nanopore_current}~\cite{Soskine-2013,Kowalczyk-2011}), using the bulk \ce{NaCl}
    conductivity at the given concentrations. The gray envelopes represent the experimental errors (standard
    deviation, $n=3$).
  }\label{fig:transport_validation_ivs}
\end{figure*}

\subsection{Transport of ions through {ClyA}}
\label{sec:iont}

\subsubsection{Ionic current and conductance.}

The ability of our model to reproduce the ionic current of a biological nanopore over a wide range of
experimentally relevant conditions (between $\vbias = \mSIrange{-150}{+150}{\mV}$ and for $\cbulk =
\mSIlist{0.05;0.15;0.5;1;3}{\Molar}$ \ce{NaCl}) can be seen in \cref{fig:transport_validation_ivs}. Here, we
compare IV relationships of \gls{clya-as} as measured experimentally (`expt.'), simulated using our
2D-axisymmetric model (`\gls{pnp-ns}' and `\gls{epnp-ns}') and naively analytically estimated (`bulk') using a
resistor model of the pore (\cref{eq:bulk_nanopore_current})~\cite{Soskine-2013,Kowalczyk-2011}. The simulated
ionic current $\currentsim$ at steady-state was computed by
\begin{align}\label{eq:currentsim}
  \currentsim = \faraday\int_{S}\left(\dsum_{i}\chargen_{i}\normvec\cdot\vec{\flux}_{i}\right)dS
  \text{ ,}
\end{align}
with $\chargen_{i}$ the charge number and $\vec{\flux}_{i}$ the total flux of each ion $i$ across \cisi{}
reservoir boundary $S$, and $\normvec$ the unit vector normal to $S$. The current given by the bulk model is
\begin{align}\label{eq:bulk_nanopore_current}
  \current(\cbulk,\vbias) = 
      \dfrac{\sigma (\cbulk) \pi}{4}
      \left( 
          \dfrac{ l_{\text{cis}} }{ d_{\text{cis}}^2 } +
          \dfrac{ l_{\text{trans}} }{ d_{\text{trans}}^2 }
      \right)^{-1}
      \vbias
      \text{ ,}
\end{align}
with $\sigma$ the (concentration-dependent) bulk electrolyte conductivity. The \cisi{} and \transi{} chambers
have have heights and diameters of $l_{\text{cis}}=\text{\SI{10}{\nm}}$ and
$l_{\text{trans}}=\text{\SI{4}{\nm}}$, and $d_{\text{cis}}=\text{\SI{6}{\nm}}$ and
$d_{\text{trans}}=\text{\SI{3.3}{\nm}}$, respectively~\cite{Soskine-2013}.

\begin{figure*}[p]
  \centering

  \begin{minipage}[t]{3.75cm}
    \begin{subfigure}[t]{3.75cm}
      \centering
      \caption{}\vspace{0mm}\label{fig:transport_ioncurrent_conductance_heatmap}
      \includegraphics[scale=1]{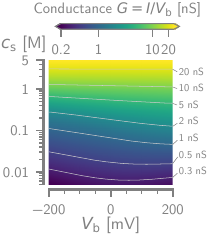}
    \end{subfigure}
    \\
    \begin{subfigure}[t]{3.75cm}
      \centering
      \caption{}\vspace{0mm}\label{fig:transport_ioncurrent_conductance_concentration}
      \includegraphics[scale=1]{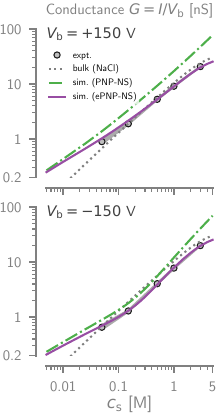}
    \end{subfigure}
  \end{minipage}
  \hspace{-2.5mm}
  \begin{minipage}[t]{3.75cm}
    \begin{subfigure}[t]{3.75cm}
      \centering
      \caption{}\vspace{0mm}\label{fig:transport_ioncurrent_icr_heatmap}
      \includegraphics[scale=1]{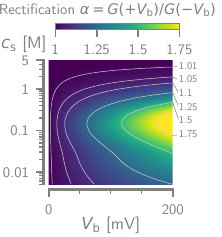}
    \end{subfigure}
    \\
    \begin{subfigure}[t]{3.75cm}
      \centering
      \caption{}\vspace{0mm}\label{fig:transport_ioncurrent_icr_concentration}
      \includegraphics[scale=1]{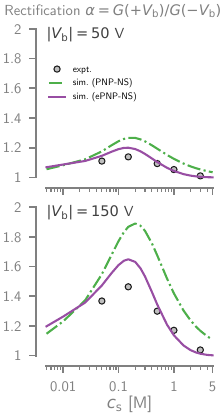}
    \end{subfigure}
  \end{minipage}
  \hspace{-2.5mm}
  \begin{minipage}[t]{3.75cm}
    \begin{subfigure}[t]{3.75cm}
      \centering
      \caption{}\vspace{0mm}\label{fig:transport_ioncurrent_tsodium_heatmap}
      \includegraphics[scale=1]{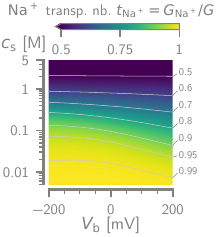}
    \end{subfigure}
    \\
    \begin{subfigure}[t]{3.75cm}
      \centering
      \caption{}\vspace{0mm}\label{fig:transport_ioncurrent_tsodium_concentration}
      \includegraphics[scale=1]{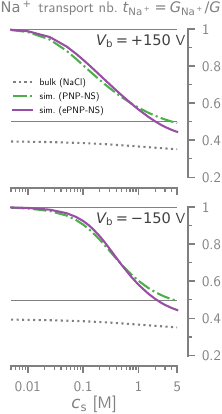}
    \end{subfigure}
  \end{minipage}

  \caption[Meas. and sim. ionic cond., rectification, and \ce{Na+} sel. of single {ClyA-AS} nanopores.]%
  {%
    \textbf{Measured and simulated ionic conductance, rectification, and cation selectivity of single {ClyA-AS}
    nanopores.}
    (\subref{fig:transport_ioncurrent_conductance_heatmap})
    Contour plot of the simulated (\gls{epnp-ns}) ionic conductance $\conductance = \current /\vbias$ as a
    function of $\vbias$ and $\cbulk$.
    (\subref{fig:transport_ioncurrent_conductance_concentration})
    Log-log plots of $\conductance$ as a function of $\cbulk$ at \SI{+150}{\mV} (top) and \SI{-150}{\mV}
    (bottom)---comparing results obtained through experiments, \gls{pnp-ns} and \gls{epnp-ns} simulations, and
    the simple resistor pore model. The gray envelopes represent the experimental errors (standard deviation,
    $n=3$).
    (\subref{fig:transport_ioncurrent_icr_heatmap})
    Contour plot of the ionic current rectification $\icr = \conductance(+\vbias)/\conductance (-\vbias)$,
    computed from the ionic conductances in the \gls{epnp-ns} simulation, as a function of $\vbias$ and
    $\cbulk$.
    (\subref{fig:transport_ioncurrent_icr_concentration})
    Comparison between the simulated ('ePNP-NS' and 'PNP-NS') and measured (expt.) $\icr$ as a function of
    $\cbulk$ for \SIlist{50;150}{\mV}.
    (\subref{fig:transport_ioncurrent_tsodium_heatmap})
    Contour plot of the \Na{} transport number $\tna = \gna / \conductance$, computed from the individual
    ionic conductances in the \gls{epnp-ns} simulation, as a function of $\vbias$ and $\cbulk$. The $\tna$
    expresses the fraction of the ionic current is carried by \Na{} ions (\ie~the cation selectivity).
    (\subref{fig:transport_ioncurrent_tsodium_concentration})
    Simulated (\gls{pnp-ns} and \gls{epnp-ns}) values of $\tna$ as a function of $\cbulk$ for \SI{+150}{\mV}
    (top) and \SI{-150}{\mV} (bottom). Here, the `bulk' line indicates the bulk \ce{NaCl} cation transport
    number, represented by its empirical function $\tna(\cbulk)$ (see~\cref{tab:corrections_equations} and
    \cref{tab:corrections_parameters}). The solid gray line represents $\tna=0.5$.
  }\label{fig:transport_ioncurrent}
\end{figure*}

Whereas the classical \gls{pnp-ns} equations consistently overestimated the ionic current, particularly at
high salt concentrations, the predictions of the \gls{epnp-ns} equations corresponded closely to the measured
values, especially at high ionic strengths ($\cbulk > \mSI{0.5}{\Molar}$). The inability of the classical
\gls{pnp-ns} equations to correctly estimate the current is expected however, as in this regime the model
parameters (\eg~diffusivity, mobility and viscosity, \ldots) already deviate significantly from their
`infinite dilution' values (see~\cref{sec:epnp-ns:parameterization},
\cref{fig:epnpns_concentration_d,fig:epnpns_concentration_mu,fig:epnpns_concentration_solv,fig:epnpns_distance}).
At $\cbulk < \mSI{0.15}{\Molar}$ the \gls{epnp-ns} equations tended to minorly overestimate the ionic current,
but the discrepancies were much smaller than those observed for \gls{pnp-ns}. Further, these ionic strengths
are not usually tested experimentally. Finally, the bulk model managed to capture the currents surprisingly
well at high salt concentrations and positive bias voltages, indicating that, under these conditions, the
distribution of ions inside the pore is similar to the bulk electrolyte. In contrast, this simple model
faltered in the negative voltage regime, indicating that the resistance of the pore is not only determined by
its geometry (\ie~the diameter and length of the `resistor') but that it is strongly modulated by its
electrostatic properties.

The ability of a nanopore to conduct ions can be best expressed by its conductance: $\conductance = \current /
\vbias$. We computed \gls{clya}'s conductance with the \gls{epnp-ns} equations as a function of bias voltage
($\vbias = \mSIrange{-200}{+200}{\mV}$) and bulk \ce{NaCl} concentration ($\cbulk =
\mSIrange{0.005}{5}{\Molar}$), of which a contour plot can be found in
\cref{fig:transport_ioncurrent_conductance_heatmap}. The near horizontal contour lines in the upper part of
the plot show that, at high ionic strengths ($\cbulk > \mSI{1}{\Molar}$), \gls{clya} maintains the same
conductance regardless of the applied bias voltage. This behavior changes at intermediate concentrations
($\mSI{0.1}{\Molar} < \cbulk < \mSI{1}{\Molar}$, typical experimental conditions), where maintaining the same
conductance level with increasing negative bias amplitudes requires increasing salt concentrations. Finally,
at low salt concentrations ($\cbulk < \mSI{0.1}{\Molar}$), the ionic conductance increases when reducing the
negative voltage amplitude but subsequently levels out at positive bias voltages.

The cross-sections of the ionic conductance as a function of concentration at high positive and negative bias
voltages (\cref{fig:transport_ioncurrent_conductance_concentration}), serve to demonstrate the differences
between these respective regimes. At high positive and negative bias voltages ($\vbias=\mSI{\pm150}{\mV}$),
the slopes of the conductance log-log plots with respect to the bulk salt concentration show linear and
bi-linear behavior, respectively. This could be indicative of a different mode of ion conduction of positive
and negative bias voltages, at least for low concentrations ($\cbulk<\mSI{0.15}{\Molar}$). Even though the
bulk model and the \gls{pnp-ns} equations manage to capture the conductance at respectively high and low ionic
strengths, only the \gls{epnp-ns} equations perform well over the entire concentration range, particularly at
experimentally relevant concentrations (\SIrange{0.1}{2}{\Molar})~\cite{Willems-Ruic-Biesemans-2019,
Galenkamp-2020,Franceschini-2013,Franceschini-2016}. Overall, the predictions made using \gls{pnp-ns}
overestimate the conductance over the entire concentration range, but they do converge with those computed
with \gls{epnp-ns} when approaching infinite dilution ($\cbulk<\mSI{0.01}{\Molar}$). A more in-depth
discussion on the effect of the concentration, wall distance and steric corrections on the ionic conductance
can be found in the in~\cref{sec:transport:comparison_corrections} (\cref{fig:transport_comparison_ionic}).

The difference in ionic conduction at opposing bias voltages is also known as the ionic current rectification
(\cref{fig:transport_ioncurrent_icr_heatmap,fig:transport_ioncurrent_icr_concentration}), given by $\icr$:
\begin{align}\label{eq:icr}
  \icr(\vbias) = \conductance(+\vbias) / \conductance(-\vbias)
  \text{ ,}
\end{align}
with $\conductance(+\vbias)$ and $\conductance(-\vbias)$ the conductance of the pore at opposing bias voltages
of the same magnitude. The ionic current rectification is a phenomenon often observed in nanopores that are
both charged, and contain a degree of geometrical asymmetry along the central axis of the
pore~\cite{Constantin-2007,White-2008,Wang-2014}. The latter can be either intrinsic to the geometry of nanopore,
or due to deformations in response to the electric field (\ie~gating or electrostriction). As can be seen in
\cref{fig:transport_ioncurrent_icr_heatmap}, \gls{clya} exhibits a strong degree of rectification, which is to
be expected given its predominantly negatively charged interior and asymmetric \cisi{} (\SI{\approx6}{\nm})
and \transi{} (\SI{\approx3.3}{\nm}) entry diameters (\cref{fig:transport_clya_side}). We found $\icr$ to
increase monotonously with the bias voltage magnitude, at least over the investigated range
(\cref{fig:transport_ioncurrent_icr_concentration}). We found the dependence of $\icr$ on the ionic strength
not to be monotonous, but rather rising rapidly to a peak value at $\cbulk=\mSI{0.15}{\Molar}$, followed by a
gradual decline towards unity at saturating salt concentrations. This concentration is within the transition
zone observed in the conductance at negative bias voltages
(\cref{fig:transport_ioncurrent_conductance_concentration}) and provides further evidence for a change in the
conductive properties of the pore in this regime. Even though the values produced by the \gls{epnp-ns}
simulations do not fully match quantitatively to the experimental results, at least for $\cbulk <
\text{\mSI{0.5}{\Molar}}$, they do (1) reproduce the observed trends qualitatively and (2) exhibit much
smaller errors relative to the \gls{pnp-ns} results (cf.~green and purple lines in
\cref{fig:transport_ioncurrent_icr_concentration}).

The results and comparisons discussed above indicate that \gls{clya}'s conductivity is dominated by the bulk
electrolyte conductivity above physiological salt concentrations ($\cbulk > \mSI{0.15}{\Molar}$). The
breakdown of this simple dependency at lower ionic strengths is particularly evident at negative bias voltages
and is likely caused by the overlapping of the electrical double layer (EDL) inside the pore (\ie~the Debye
length is \SI{\approx1.4}{\nm} at $\cbulk=\mSI{0.05}{\Molar}$). This effectively excludes the co-ions (\Cl)
from the interior of the pore and attracts as many counter-ions (\Na) as needed to screen the fixed charges of
the pore. As a result, \Cl{} ions do not contribute to ionic current and the conductance is dominated by \Na{}
ions attracted by \gls{clya}'s `surface' charges~\cite{Uematsu-2018}. The presence of only a single ion type
inside the pore at low ionic strength may also explain why the \gls{epnp-ns} equations are
more accurate at higher ionic strengths ($\cbulk\ge\mSI{0.15}{\Molar}$). Because our ionic mobilities are
derived from bulk ionic conductances (\ie~for unconfined ions in a locally electroneutral environment) our
mobility model likely begins to break down under conditions where only a single ion type is
present~\cite{Duan-2010}. Another cause of the discrepancies could be a slight narrowing of the nanopore at
low salt concentrations, which cannot be captured by our simulation due to the static nature of its geometry
and charge distributions. Nevertheless, our simplified 2D-axisymmetric model, in conjunction with the
\gls{epnp-ns} equations, can accurately predict the ionic current flowing through \gls{clya} for a wide
range of experimentally relevant ionic strengths and bias voltages. This suggests that our continuum system
can accurately capture the essential physical phenomena that drive the ion and water transport through the
nanopore both \emph{qualitatively} and \emph{quantitatively}. Hence, we expect the distribution of the
resulting properties (\eg~ion concentrations, ion fluxes, electric field, and water velocity) to closely
correspond to their true values.

\subsubsection{Cation selectivity.}
The ion selectivity of a nanopore determines the preference with which it transports one ion type over the
other. Experimentally, it is often determined by placing the pore in a salt gradient (\ie~different salt
concentrations in the \cisi{} and \transi{} reservoirs) and measuring the reversal potential ($\revpot$),
\ie~the bias voltage at which the nanopore current is zero~\cite{Soskine-2013,Franceschini-2016}. The
\gls{ghk} equation can then be used to convert $\revpot$ into the permeability ratio $\pna = \gna / \gcl$.
Here, we represent the \gls{clya}'s ion selectivity
(\cref{fig:transport_ioncurrent_tsodium_heatmap,fig:transport_ioncurrent_tsodium_concentration}) 
by the fraction of the total current that is carried by \Na{} ions: the apparent \Na{} transport number
\begin{align}\label{eq:tna}
  \tna = \dfrac{\pna}{\pna + 1} = \dfrac{\gna}{\gna + \gcl}
  \text{ ,}
\end{align}
with $\pna$ the cation permeability ratio, and $\gna$ and $\gcl$ the cation and anion contributions to the
total conductance. As expected from its negatively charged interior, we found \gls{clya} to be 
cation-selective (\ie~$\tna > 0.5$) for all investigated voltages up to a bulk salt concentration of 
$\cbulk \approx \mSI{2}{\Molar}$ \ce{NaCl} (0.5 contour line in \cref{fig:transport_ioncurrent_tsodium_heatmap}).
Above this concentration, $\tna$ falls to a minimum value of 0.45 at $\cbulk\approx\mSI{5}{\Molar}$,
which is still \num{\approx1.27} times its bulk electrolyte value of 0.35. This shows that---even at saturating
concentrations where the Debye length is \SI{<0.2}{\nm}---\gls{clya} enhances the transport of cations. Below
\SI{2}{\Molar}, the ion selectivity increases logarithmically with decreasing salt concentrations, but it also
becomes more sensitive to the direction and magnitude of the electric field:  with negative bias voltages
yielding higher ion selectivities (\cref{fig:transport_ioncurrent_tsodium_concentration}). For example, to
reach a selectivity of $\tna \approx 0.9$, the salt concentration must fall to \SI{0.05}{\Molar} at
\SI{+150}{\mV}, but only to \SI{0.125}{\Molar} at \SI{-150}{\mV}.

Using the reversal potential method, Franceschini \etal{}~\cite{Franceschini-2016} found \gls{clya}'s ion
selectivity to be $\tna=0.66$ ($\pna=1.9$). This corresponds well to the average between the \cisi{}
($\cbulk=\mSI{1}{\Molar}$, $\tna=0.57$, $\pna=1.3$) and \transi{} ($\cbulk=\mSI{0.15}{\Molar}$, $\tna=0.84$,
$\pna=5.4$) reservoir concentrations used in their experiment. Therefore, this suggests that although
measuring the reversal potential gives valuable insights into the selectivity of ion channels and small
nanopores, it does not describe the ion selectivity under symmetric conditions. In addition, the \gls{ghk}
equation does not consider the ionic flux due to the electro-osmotic flow and assumes that the Nernst-Einstein
relation holds for all used concentrations. These two effects should not be ignored as they contribute
significantly to the total conductance of the pore. Furthermore, because the ion selectivity depends strongly
on the ionic strength and often the applied bias voltage, the measured reversal potential will necessarily be
influenced by the chosen salt gradient and represents the selectivity at an undetermined intermediate
concentration. Tabulated data of the ion selectivity for selected voltages and concentrations can be found in
\cref{tab:ion_selectivities}.

\subsection{Ion concentration distribution}\label{sec:ionc}
\begin{figure*}[p]
  \centering

  \begin{subfigure}[t]{9cm}
    \centering
    \caption{}\vspace{-3mm}\label{fig:transport_concentration_average}
    \includegraphics[scale=1]{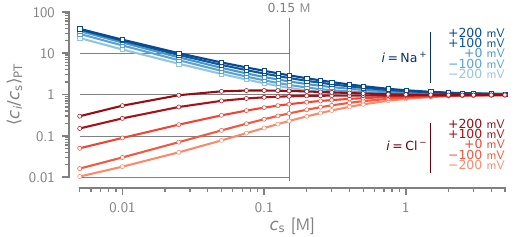}
  \end{subfigure}
  \\
  \begin{subfigure}[t]{11.5cm}
    \centering
    \caption{}\vspace{-3mm}\label{fig:transport_concentration_contours}
    \includegraphics[scale=1]{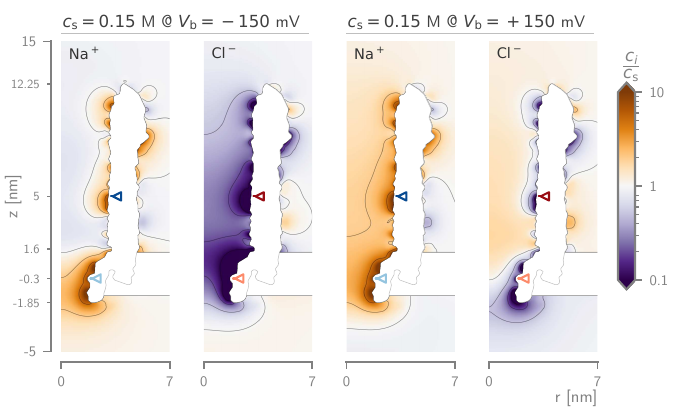}
  \end{subfigure}
  \\
  \begin{subfigure}[t]{11.5cm}
    \centering
    \caption{}\vspace{-3mm}\label{fig:transport_concentration_profiles}
    \includegraphics[scale=1]{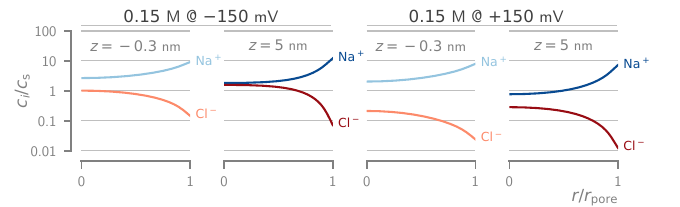}
  \end{subfigure}

  \caption[Ion concentration distribution inside {ClyA-AS}]%
  {%
    \textbf{Ion concentration distribution inside {ClyA-AS}.}
    (\subref{fig:transport_concentration_average})
    Relative \Na{} and \Cl{} concentrations averaged over the entire pore volume ($\pavi$,
    \cref{eq:pore_surface_integral}) as a function of the reservoir salt concentration ($\cbulk =
    \mSIrange{0.005}{5}{\Molar}$) and bias voltage ($\vbias = \mSIrange{-200}{+200}{\mV}$).
    (\subref{fig:transport_concentration_contours})
    Contour plots of the relative ion concentration ($\ci/\cbulk$) for both \Na{} and \Cl{} for $\cbulk =
    \mSI{0.15}{\Molar}$ and at $\vbias = \mSIlist{-150;+150}{\mV}$.
    (\subref{fig:transport_concentration_profiles})
    The relative \Na{} and \Cl{} concentration profiles along the radius of the pore, through the middle of
    the constriction ($z = \mSI{-0.3}{\nm}$) and the \lumen{} ($z = \mSI{5}{\nm}$), as indicated by the arrows
    in (\subref{fig:transport_concentration_contours}).
    }\label{fig:transport_concentration}
\end{figure*}

Following the validation of the model in the previous section, we now proceed by describing the local ionic
concentrations inside \gls{clya}. Detailed knowledge of the ionic environment can be valuable to
experimentalists who seek to trap and study single enzymes with
\gls{clya}~\cite{Soskine-Biesemans-2015,VanMeervelt-2017,Galenkamp-2018}. Moreover, it gives insight into the
origin of the ion current rectification, ion selectivity, and the electro-osmotic flow. Note that the figures
below were obtained from a nanoscale continuum steady-state simulation, they represent a time-averaged
situation on the order of \SIrange{10}{100}{\ns}~\cite{Im-2002}.

\subsubsection{Relative cation and anion concentrations.}

We use the relative ion concentration averaged over the total pore (`PT') volume ($\pavi$), as a measure for
global ionic conditions inside the pore (\cref{fig:transport_concentration_average}). At low ionic strengths
($\cbulk < \mSI{0.05}{\Molar}$), our simulation predicts a strong enhancement of the \Na{} concentration
($\pavNa$) and a clear depletion of the \Cl{} concentration ($\pavCl$) inside the pore relative to the
reservoir concentration, irrespective of the bias voltage. This effect diminishes rapidly with increasing
ionic strengths, which can be explained by the electrolytic screening of the negative charges lining the walls
of \gls{clya} (\ie~the electrical double layer). At low reservoir concentrations ($\cbulk <
\mSI{0.05}{\Molar}$) the number of ions in the bulk is sparse, leading to the attraction and repulsion of
respectively as many \Na{} and \Cl{} ions as the chemical potential allows. As the concentration increases,
the overall availability of ions improves and the extreme concentration differences between the pore and the
bulk are no longer required to offset the fixed charges lining \gls{clya}'s interior walls. For example,
increasing the reservoir concentration at equilibrium ($\vbias = \mSI{0}{\mV}$) from
\SIrange{0.005}{0.05}{\Molar} causes $\pavNa$ to fall an order of magnitude (\numrange{34}{4.4}) and $\pavCl$
to raise an order of magnitude (\numrange{0.05}{0.31}). Even though at physiological ionic strength ($\cbulk =
\mSI{0.15}{\Molar}$) their concentrations still differ significantly from those in the reservoir ($\pavNa =
2.1$ and $\pavCl = 0.58$), they do approach bulk-like values ($1.14 \ge \pavNa \ge 1$ and $0.89 \le \pavCl \le
1$) at higher concentrations ($\cbulk \ge \mSI{1}{\Molar}$).

We also observed a significant difference between their sensitivities to the applied bias voltage,
particularly at low salt concentrations (\cref{fig:transport_concentration_average}, left of
\SI{<0.15}{\Molar} line). Whereas the \Na{} concentration shows only a limited response, the \Cl{}
concentration changes much more dramatically. For example, at $\cbulk = \mSI{0.15}{\Molar}$ and when changing
the bias voltage from \SIrange{-150}{+150}{\mV}, $\pavNa$ rises \num{\approx 1.7}-fold (\numrange{1.6}{2.7})
and $\pavCl$ increases \num{\approx 3.8}-fold (\numrange{0.28}{1.06}). This difference is clearly visualized
by the contour plots of the relative ion concentrations ($\ci/\cbulk$) at $\cbulk = \mSI{0.15}{\Molar}$ and
for $\vbias = \mSIlist{-150;+150}{\mV}$ (\cref{fig:transport_concentration_contours}). They reveal that the
\transi{} constriction ($\num{-1.85} < z < \mSI{1.6}{\nm}$) remains depleted of \Cl{} and enhanced in \Na{}
for both $\vbias = \mSI{-150}{\mV}$ and $\vbias = \mSI{+150}{\mV}$. This is not the case in the \lumen{}
($\num{1.6} < z < \mSI{12.25}{\nm}$), in which the \Na{} concentration is bulk-like for $\vbias <
\mSI{0}{\mV}$ and enhanced for $\vbias > \mSI{0}{\mV}$. Conversely, the number of \Cl{} ions becomes more and
more depleted in the \lumen{} for increasing negative bias magnitudes, and it is virtually bulk-like at higher
positive bias voltages. This is further exemplified by the radial profiles of the ion concentrations
(\cref{fig:transport_concentration_profiles}) through the middle of the constriction ($z = \mSI{-0.3}{\nm}$)
and the \lumen{} ($ z = \mSI{5}{\nm}$), which also clearly shows the extent of the electrical double layer.

\begin{figure*}[p]
  \centering

  \begin{subfigure}[t]{9cm}
    \centering
    \caption{}\vspace{-3mm}\label{fig:transport_charge_total}
    \includegraphics[scale=1]{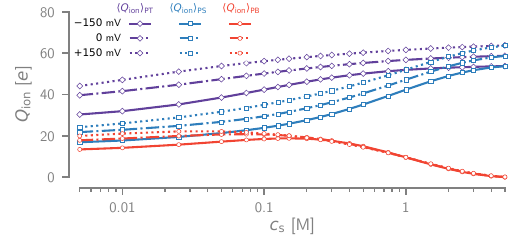}
  \end{subfigure}
  \\
  \begin{subfigure}[t]{11.5cm}
    \centering
    \caption{}\vspace{-3mm}\label{fig:transport_charge_contours}
    \includegraphics[scale=1]{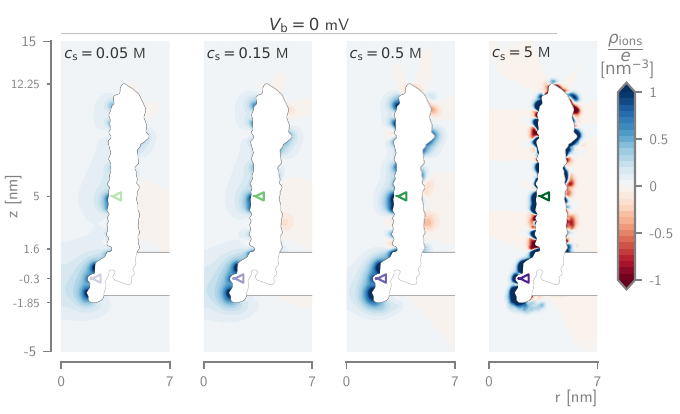}
  \end{subfigure}
  \\
  \begin{subfigure}[t]{11.5cm}
    \centering
    \caption{}\vspace{-3mm}\label{fig:transport_charge_profiles}
    \includegraphics[scale=1]{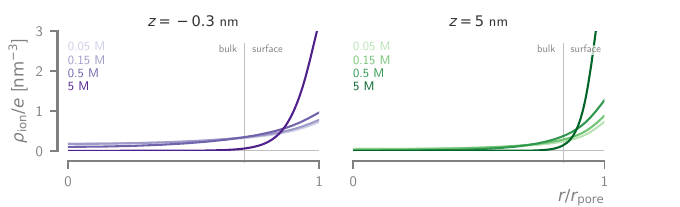}
  \end{subfigure}

  \caption[Ion charge density distribution inside {ClyA-AS}]%
  {%
    \textbf{Ion charge density distribution inside {ClyA-AS}.}
    (\subref{fig:transport_charge_total})
    The average number of ionic charges inside the pore $\pav{\Qion}{\text{PT}}$, is distributed between those
    close to the pore's surface $\pav{\Qion}{\text{PS}}$ (\ie~within \SI{0.5}{\nm} of the wall), and those in
    the `bulk' of the pore's interior $\pav{\Qion}{\text{PB}}$.
    (\subref{fig:transport_charge_contours})
    Cross-section contour plots of the ion space charge density ($\scdion$), expressed as the number of elementary
    charges per \si{\cubic\nano\meter}, at $\vbias = \mSI{0}{\mV}$ and for $\cbulk =
    \mSIlist{0.05;0.15;0.5;5}{\Molar}$.
    (\subref{fig:transport_charge_profiles})
    Radial cross-sections of the $\scdion$ at the center of the constriction ($z = \mSI{-0.3}{\nm}$) and the
    \lumen{} ($z = \mSI{5}{\nm}$) of \gls{clya}. The vertical line represents the division between ions in the
    `bulk' ($\walldistance > \mSI{0.5}{\nm}$) of the pore and those located near its surface ($\walldistance
    \le \mSI{0.5}{\nm}$).
  }\label{fig:transport_charge}
\end{figure*}

\subsubsection{Ion charge density.}

The formation of an electrical double layer inside the pore, and the resulting asymmetry in the cation and
anion concentrations, gives rise to a net charge density inside the pore ($\scdion$, \cref{eq:scdion}). To
investigate the distribution of these charges within \gls{clya}, we divided the total interior volume of the
pore into a `pore surface' (PS) and a `pore bulk' (PB) region. The PB region encompasses a cylindrical volume
at the entry of the pore up until \SI{0.5}{\nm} from the wall ($\walldistance \ge \mSI{0.5}{\nm}$), the
approximate distance from which the wall begins to exert a significant influence on the properties of the
electrolyte (\cref{fig:epnpns_distance}). The PS region includes the remaining volume between the PB domain
and the nanopore wall ($\walldistance < \mSI{0.5}{\nm}$). Integration of $\scdion$ over the PS and PB regions
yields the average number of mobile charges present inside those locations
(\cref{fig:transport_charge_total}): $\pav{\Qion}{\text{PB}}$ and $\pav{\Qion}{\text{PS}}$, respectively.
Although the total number of mobile charges inside the pore, $\pav{\Qion}{\text{PT}} = \pav{\Qion}{\text{PB}}
+ \pav{\Qion}{\text{PS}}$, rises appreciatively with increasing reservoir concentrations, the majority of
these additional charges are confined to the walls of the pore. Up until a reservoir concentration
$\cbulk=\mSI{\approx0.15}{\Molar}$, we found $\pav{\Qion}{\text{PT}}$ to be distributed equally between the
surface (\SI{\approx+27}{\ec}) and bulk (\SI{\approx+22}{\ec}) layers. At high salt concentrations ($\cbulk >
\mSI{1}{\Molar}$), the number of charges in the PS region more than doubles (towards
$\pav{\Qion}{\text{PS}}=\mSI{+58}{\ec}$ at \SI{5}{\Molar}), and those in the PB region diminish (towards
$\pav{\Qion}{\text{PB}} \approx \mSI{0}{\ec}$ at \SI{5}{\Molar}). The bias voltage also influences the total
number of mobile charges in the pore. As can be seen from our simulation results for three different voltages
(\cref{fig:transport_charge_total}), $\pav{\Qion}{\text{PT}}$ is approximately \SIrange{+10}{+15}{\ec} higher
at $\vbias = \mSI{+150}{\mV}$ as compared to $\vbias = \mSI{-150}{\mV}$ for the full range of ion
concentrations. Interestingly, at reservoir concentration \SI{>0.15}{\Molar}, $\pav{\Qion}{\text{PB}}$ becomes
independent of applied voltage and the changes in $\pav{\Qion}{\text{PT}}$ can be attributed to
$\pav{\Qion}{\text{PS}}$.

The cross-section contour plots of $\scdion$ inside \gls{clya} for four different bulk concentrations ($\cbulk
= \mSIlist{0.05;0.15;0.5;5}{\Molar}$) reveal the redistribution of the mobile charges with increasing ionic
strength in more detail. Up until a bulk concentration of $\cbulk = \mSI{\le0.5}{\Molar}$, the EDL inside the
pore overlaps significantly with itself, as evidenced by the net positive charge density found throughout the
interior of the pore (\cref{fig:transport_charge_contours}). Moreover, the absence of \Cl{} ions effectively
prevents the formation of a negatively charged EDL next to the few positively charged residues lining the pore
walls. The situation at high salt concentrations (\eg~\SI{5}{\Molar}) is very different, with almost no charge
density within the PB region of the pore ($\walldistance \ge \mSI{0.5}{\nm}$), but with pockets of highly
charged and alternating positive and negative charge densities close to the nanopore wall
(\cref{fig:transport_charge_contours}, rightmost panel). This sharp confinement is shown clearly by the radial
density profiles (\cref{fig:transport_charge_profiles}) drawn through the constriction ($z = \mSI{-0.3}{\nm}$,
purple triangles) and the \lumen{} ($z = \mSI{5}{\nm}$, green triangles).

It is well known that the activity of an enzyme depends on the composition of the electrolyte that surrounds
it~\cite{Purich-2010-7}. Hence, we expect that the interpretation of kinetic data obtained from enzymes
trapped inside the nanopore~\cite{VanMeervelt-2017,Galenkamp-2018} will benefit from the precise
quantification of the ionic conditions inside the pore, including the concentration difference with the
reservoir but also the significant imbalance between cations and anions~\cite{Warren-1966}.

\subsection{Electrostatic potential and energy}\label{sect:esp}
\begin{figure*}[p]
  \centering

  \begin{subfigure}[t]{2.5cm}
    \centering
    \caption{}\vspace{-3mm}\label{fig:transport_potential_residues}
    \includegraphics[scale=1]{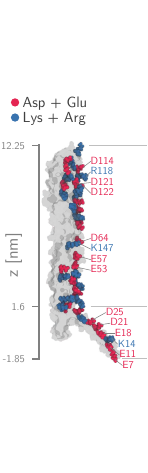}
  \end{subfigure}
  \hspace{-0.5cm}
  \begin{subfigure}[t]{9cm}
    \centering
    \caption{}\vspace{-3mm}\label{fig:transport_potential_contours}
    \includegraphics[scale=1]{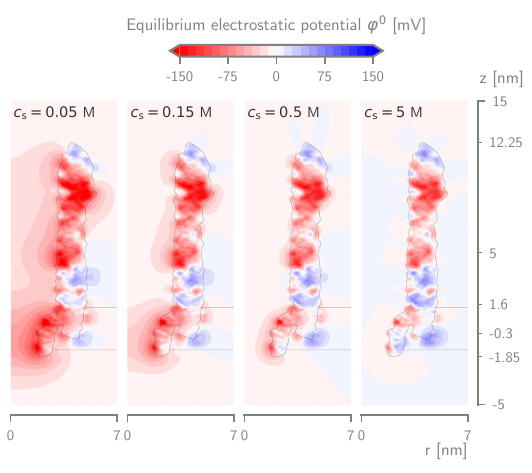}
  \end{subfigure}
  \\
  \begin{subfigure}[t]{11cm}
    \centering
    \caption{}\vspace{-3mm}\label{fig:transport_potential_profiles}
    \includegraphics[scale=1]{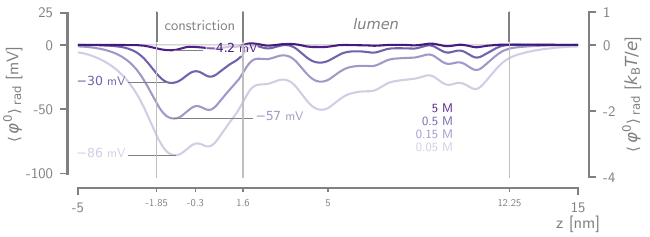}
  \end{subfigure}

  \caption[Equilibrium electrostatic potential inside {ClyA-AS}]%
  {%
    \textbf{Equilibrium electrostatic potential inside {ClyA-AS}}.
    (\subref{fig:transport_potential_residues})
    A single subunit of \gls{clya} in which all amino acids with a net charge and whose side chains face the
    inside of the pore (\ie~those contribute the most to the electrostatic potential) are highlighted.
    Negatively (Asp+Glu) and positively and positively (Lys+Arg) charged residues are colored in red and blue,
    respectively.
    (\subref{fig:transport_potential_contours})
    As a result of these fixed charges, \gls{clya} exhibits a complex electrostatic potential ($\potential$)
    landscape at equilibrium (\ie~at $\vbias = \mSI{0}{\mV}$), and whose values inside the pore we have
    plotted for several key concentrations ($\cbulk = \mSIlist{0.05;0.15;0.5;5}{\Molar}$). Note that even at
    physiological salt concentrations ($\cbulk = \mSI{0.15}{\Molar}$), the negative electrostatic potential
    extends significantly inside the \lumen{} ($\num{1.6} < z < \mSI{12.25}{\nm}$), and even more so inside
    the \transi{} constriction ($\num{1.85} < z < \mSI{1.6}{\nm}$). For the former, localized influential
    negative `hotspots' can be found in the middle ($\num{4} < z < \mSI{6}{\nm}$) and at the \cisi{} entry
    ($\num{10} < z < \mSI{12}{\nm}$).
    (\subref{fig:transport_potential_profiles})
    Radial average of the equilibrium electrostatic potential along the length of the pore ($\radpot$) for the
    same concentrations as in (\subref{fig:transport_potential_contours}). Even though the \lumen{} of the
    pore is almost fully screened for $\cbulk > \mSI{0.5}{\Molar}$, the constriction still retains some of its
    negative influence even at \SI{5}{\Molar}.
  }\label{fig:transport_potential}
\end{figure*}

The electrostatic potential, or rather the spatial change thereof in the form of an electric field, is one of
the primary driving forces within a nanopore. Typically, the potential can be split into an external,
`non-equilibrium' contribution, resulting from the bias voltage applied between the \transi{} and the \cisi{}
reservoirs, and an intrinsic, `equilibrium' component, caused by the fixed charge distribution of the
pore~\cite{Willems-Ruic-Biesemans-2019}. To accurately describe and understand the nanopore transport
processes both contributions to the net electric field inside the pore are essential, as their relative
magnitudes and directions can significantly influence the transport of
ions~\cite{Aksimentiev-2005,Bhattacharya-2011,DeBiase-2015,Basdevant-2019}, water
molecules~\cite{Laohakunakorn-2015,Bhadauria-2017}, and
biopolymers~\cite{Buchsbaum-2013,Muthukumar-2014,Willems-Ruic-Biesemans-2019}.

\subsubsection{A few important charged residues.}

The interior walls of the \gls{clya} nanopore (\cref{fig:transport_potential_residues}) are riddled with
negatively charged amino acids (\ie~aspartate or glutamate), interspaced by a few positively charged residues
(\ie~lysine or arginine). When grouping these charges by proximity, we found three clusters with significantly
more negative than positive residues: inside the \transi{} constriction ($-1.85 < z < \mSI{1.6}{\nm}$; E7,
E11, K14, E18, D21, D25), in the middle of the \cisi{} \lumen{} ($4 < z < \mSI{6}{\nm}$; E53, E57, D64, K147)
and at the top of the pore ($10 < z < \mSI{12}{\nm}$; D114, R118, D121, D122). As we shall see, these clusters
leave strong negative fingerprints in the global electrostatic potential.

\subsubsection{Distribution of the equilibrium electrostatic potential.}

The electrostatic potential at equilibrium ($\eqpot$, \ie~at $\vbias = \mSI{0}{\mV}$) reveals the effect of
\gls{clya}'s fixed charges on the potential inside the pore (\cref{fig:transport_potential_contours}). Due to
electric screening by the mobile charge carriers in the electrolyte, however, the extent of their influence
strongly depends on the bulk ionic strength. The contour plot cross-sections of $\eqpot$ for $\cbulk =
\mSIlist{0.05;0.15;0.5;5}{\Molar}$ (\cref{fig:transport_potential_contours}) and their corresponding radial
averages (\cref{fig:transport_potential_profiles}) demonstrate this effect aptly. The radial average
($\radpot$) represents the mean value along the longitudinal axis of the pore and can be computed using
\begin{align}\label{eq:epnp-ns_radpot}
  \radpot = \dfrac{1}{\pi R(z)^2}\int_{0}^{R(z)}\potential(r,z) \;2 \pi r \; dr
  \text{ ,}
\end{align}
where $R(z)$ is taken as the \gls{clya}'s radius inside the pore ($-1.85 \le z \le \mSI{12.25}{\nm}$), and as
fixed values of \SI{4}{\nm} and \SI{2}{\nm} inside the \cisi{} ($z > \mSI{12.25}{\nm}$) and \transi{} (($z <
\mSI{-1.85}{\nm}$) reservoirs, respectively. Starting from the \cisi{} entry ($z \approx \SI{10}{\nm}$), the
electrostatic potential is dominated by the acidic residues D114, D121, and D122, resulting in a rapid
reduction of $\radeqpot$ upon entering the pore. Next, $\radeqpot$ slowly decreases up until the middle of the
\lumen{} ($z \approx \mSI{5}{\nm}$), where the next set of negative residues, namely E53, E57, and D64, lower
it even further. After a brief increase, $\radeqpot$ attains its maximum amplitude inside the \transi{}
constriction ($z \approx \mSI{0}{\nm}$) due to the close proximity of the amino acids E7, E11, E18, D21 and
D25, and then quickly falls to \num{0} inside the \transi{} reservoir.

At low ionic strengths ($\cbulk < \mSI{0.05}{\Molar}$), the lack of sufficient ionic screening results in
relatively high negative potentials throughout the entire pore. For example, at low concentrations ($\cbulk =
\mSI{0.05}{\Molar}$), the $\radeqpot$ inside the constriction ramps up to a value of \SI{-86}{\mV}
(\SI{-3.35}{\kTe}), which significantly exceeds the single ion thermal voltage $\si{\kTe} = \mSI{25.7}{\mV}$.
Hence, on the one hand, they prohibit anions such as \Cl{} from entering the pore, and on the other hand, they
attract cations such as \Na{} and trap them inside the pore. For intermediate concentrations ($0.05 \le \cbulk
< \mSI{0.5}{\Molar}$) the influence of the negative charges becomes increasingly confined to several
`hotspots' near the nanopore walls, most notably at entry of the pore ($10 < z < \mSI{12}{\nm}$), in the
middle of the \lumen{} ($4 < z < \mSI{6}{\nm}$), and in the constriction ($-1.85 < z < \mSI{1.6}{\nm}$), in
accordance with the charge groups discussed in the previous section. Even though the magnitude of $\radeqpot$
at $\cbulk = \mSI{0.15}{\Molar}$ drops below \SI{1}{\kTe} inside the \lumen{} ($\radeqpot \approx
\mSI{-14}{\mV}$), it remains strongly negative inside the constriction ($\radeqpot \approx \mSI{-47}{\mV}$).
Finally, at high reservoir concentrations ($\cbulk \ge \mSI{0.5}{\Molar}$) the potential is close to
\SI{\approx0}{\mV} over the entire \lumen{} of the pore, with only a small negative potential remaining inside
the constriction. A summary of the most salient $\radeqpot$ values can be found in
\cref{tab:radial_potential}.

\begin{figure}[p]
  \centering

  \begin{subfigure}[t]{11cm}
    \centering
    \caption{}\vspace{-5mm}\label{fig:transport_energy_profiles}
    \includegraphics[scale=1]{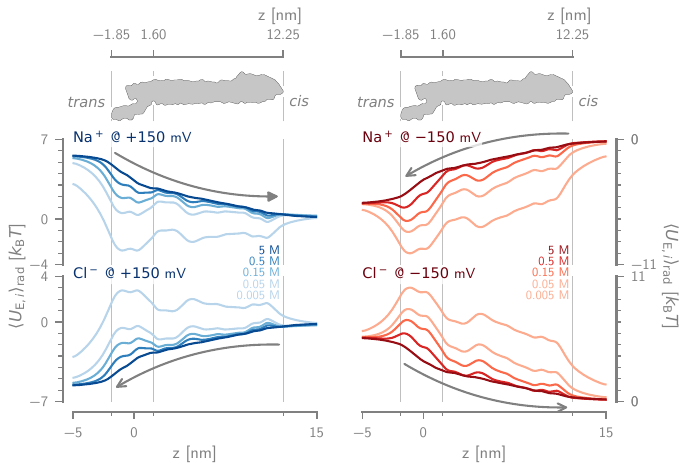}
  \end{subfigure}
  \\
  \begin{subfigure}[t]{9cm}
    \centering
    \caption{}\vspace{-3mm}\label{fig:transport_energy_barriers}
    \includegraphics[scale=1]{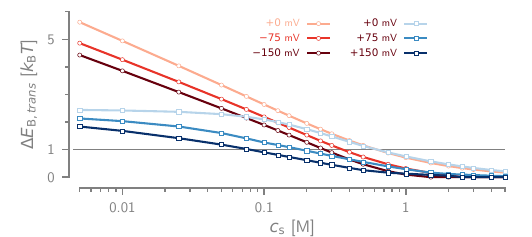}
  \end{subfigure}

  \caption[Non-equilibrium electrostatic energy landscape for single ions]%
  {%
    \textbf{Non-equilibrium electrostatic energy landscape for single ions.}
    (\subref{fig:transport_energy_profiles})
    Radially averaged non-equilibrium electrostatic energy landscape for single ions, $\radenergy =
    \chargen_{i} \ec \radpot$, as calculated directly from the radial electrostatic potential at $\vbias
    = \mSIlist{+150;-150}{\mV}$ for monovalent cations and anions. The gray arrows indicate the direction in
    which the ions must travel in order for them to contribute positively to the ionic current.
    (\subref{fig:transport_energy_barriers})
    Height of the electrostatic energy barrier ($\Delta E_{\text{B},\mathit{trans}}$) at the \transi{}
    constriction as a function of the bulk salt concentration. Note that $\Delta E_{\text{B},\mathit{trans}}$
    is much higher for negative voltages and rises logarithmically at lower concentrations. The divergence
    between \SI{+0}{\mV} and \SI{-0}{\mV} for $\cbulk < \mSI{0.3}{\Molar}$ highlights the difference in
    barrier height when traversing the pore from \cisi{} to \transi{} or \textit{vice versa}.
  }\label{fig:transport_energy}
\end{figure}

\subsubsection{Non-equilibrium electrostatic energy at $\mathbf{\pm150}$~mV.}

To link back the observed ionic conductance properties to the electrostatic potential, we computed
$\radenergy$, the radially averaged electrostatic energy for a monovalent ion
\begin{equation}\label{eq:epnp-ns_radenergy}
  \radenergy = \chargen_{i} \ec \radpot
  \text{ ,}
\end{equation}
at $\vbias = \mSI{\pm150}{\mV}$ for the entire range of simulated ionic strengths
(\cref{fig:transport_energy_profiles}). The resulting plot represents the energy landscape---filled with
barriers (hills) or traps (valleys)---that a positive or negative ion must traverse in order to contribute
positively to (\ie~increase) the ionic current.

At positive bias voltages, cations traverse the pore from \transi{} to \cisi{}
(\cref{fig:transport_energy_profiles}, first plot). Upon entering the negatively charged constriction, their
electrostatic energy drops dramatically, followed by a relatively flat section with a small barrier for entry
in the \lumen{} at $z \approx \mSI{1.6}{\nm}$. At very low ionic strengths ($\cbulk < \mSI{0.05}{\Molar}$),
the energy at \transi{} is significantly lower than the energy of the cation in the \cisi{} compartment
(\eg~$\Delta\radenergy > \mSI{2}{\kT}$ at \SI{0.005}{\Molar}), forcing the ions to accumulate inside the pore.
At higher concentrations ($\cbulk > \mSI{0.05}{\Molar}$), the increased screening smooths out the potential
drop inside the pore, allowing the cations to migrate unhindered across the entire length of the pore. Anions
at $\vbias = \mSI{+150}{\mV}$ travel from \cisi{} to \transi{} (\cref{fig:transport_energy_profiles}, second
plot) and must overcome energy barriers at both sides of the pore. The \cisi{} barrier prevents anions from
entering the pore, but because its magnitude is attenuated strongly with increasing salt concentration
(\SIrange{1.7}{0.5}{\kT} when increasing the reservoir salt concentration from $\cbulk =
\mSIrange{0.005}{0.05}{\Molar}$), it is only relevant at lower ionic strengths ($\cbulk <
\mSI{0.05}{\Molar}$). Once inside the \lumen{}, anions can move relatively unencumbered to the \transi{}
constriction, where they face the second, more significant energy barrier. This prevents them from fully
translocating and causes them to accumulate inside the lumen and explains why we observe higher \Cl{}
concentrations inside the pore at positive bias voltages (\cref{fig:transport_concentration_contours}). As
with the cations, an increase in the ionic strength significantly reduces these hurdles, resulting in a much
smoother landscape for $\cbulk > \mSI{0.15}{\Molar}$.

At negative voltages, cations move through the pore from \cisi{} to \transi{}, with a slow and continuous drop
of the electrostatic energy throughout the \lumen{} of the pore up until the constriction
(\cref{fig:transport_energy_profiles}, third plot). This results in the efficient removal of cations from the
pore \lumen{}, and explains the lower \Na{} concentration observed at positive voltages
(\cref{fig:transport_concentration_average}). To fully exit from the pore, however, cations must overcome a
large energy barrier, which reduces the nanopore's ability to conduct cations compared to positive potentials
and hence contributes to the ion current rectification. The situation for anions at negative bias voltages
(\ie~traveling from \transi{} to \cisi{}) is very different (\cref{fig:transport_energy_profiles}, fourth
plot). Their ability to even enter the pore is severely hampered by an energy barrier of a few \si{\kT} at the
\transi{} constriction. Any anions that do cross this barrier, and those still present in the \lumen{} of
\gls{clya}, will rapidly move towards the \cisi{} entry and exit from the pore due to a continuous drop of
their electrostatic energy. This effectively depletes the entire \lumen{} of anions, which can be observed
from the much lower \Cl{} concentrations at negative voltages
(see~\cref{fig:transport_concentration_average}).

\subsubsection{Concentration and voltage dependencies of the energy barrier at the constriction.}

Many biological nanopores contain constrictions that play crucial roles in shaping their ionic conductance
properties~\cite{Maglia-2008,Franceschini-2016,Huang-2017}. The reason for this is two-fold, (1) the narrowest
part dominates the overall resistance of the pore, and (2) confinement of charged residues results in much
larger electrostatic energy barriers. With its highly negatively charged \transi{} constriction, \gls{clya}'s
affinity for transport of anions is diminished and that for cations is enhanced compared to bulk, even at high
ionic strengths (\cref{fig:transport_ioncurrent_tsodium_concentration})~\cite{Soskine-2013}. To further
elucidate the significance of the \transi{} electrostatic barrier ($\deltaEt$), we quantified its height at
positive and negative voltages as a function of the salt concentration (\cref{fig:transport_energy_barriers}).

Because the application of a bias voltage effectively tilts the energy landscape, it reduces the magnitude of
the energy barriers for both positive and negative potentials, as evidenced by the lowering of the curves with
increasing bias magnitude (\cref{fig:transport_energy_barriers}, light to dark color shading). Likewise,
raising the bulk salt concentration results in a continuous decrease of $\deltaEt$ due to an increase in the
screening of the fixed charges lining the constriction. At moderate to higher reservoir concentrations
($\cbulk > \mSIrange{0.1}{0.5}{\Molar}$, depending on $\vbias$), $\deltaEt$ falls below \SI{1}{\kT} regardless
of the bias voltage, and its effect on the ion transport through the pore is significantly reduced.

The ions under the influence of a positive bias voltage (\ie~\Na{} moving from \transi{} to \cisi{} and \Cl{}
moving from \cisi{} to \transi{}, blue lines in \cref{fig:transport_energy_barriers}) experience a $\deltaEt$
roughly half that of those under a negative voltage (red lines in \cref{fig:transport_energy_barriers}). For
example, increasing the salt concentration from \SIrange{0.005}{0.15}{\Molar}, causes $\deltaEt$ to drop from
\SIrange{1.8}{0.73}{\kT} at $\vbias = \mSI{+150}{\mV}$ and from \SIrange{4.4}{1.5}{\kT} at $\vbias =
\mSI{-150}{\mV}$. These differences in barrier heights are directly reflected by \gls{clya}'s higher degree of
ion selectivity at negative compared to positive bias voltages
(\cref{fig:transport_ioncurrent_tsodium_concentration}).

\subsection{Transport of water through {ClyA}}
\label{sec:eof}

The charged nature of the inner surface of many nanopores gives rise to a net flux of water through the pore,
called the \gls{eof}~\cite{Qiao-Aluru-2003,Thompson-2003,Mao-2014}. The \gls{eof} not only contributes
significantly to the ionic current, but the magnitude of the viscous drag force it exerts on proteins is often
of the same order as the Coulombic electrophoretic force
(EPF)~\cite{vanDorp-2009,Firnkes-2010,Willems-Ruic-Biesemans-2019}. Hence, it strongly influences the capture
and translocation of biomolecules including nucleic acids~\cite{Wong-2007,Luan-2008,Firnkes-2010},
peptides~\cite{Huang-2017,Li-2018,Huang-2019}, and proteins~\cite{Soskine-2012,Soskine-2013,VanMeervelt-2014,
Soskine-Biesemans-2015,Biesemans-2015,Wloka-2017,Galenkamp-2018,Willems-Ruic-Biesemans-2019}. Because the drag
exerted by the \gls{eof} depends primarily on the size and shape of the biomolecule of interest and not on its
charge~\cite{Willems-Ruic-Biesemans-2019}, it can be employed to capture molecules even against the electric
field~\cite{Soskine-2012}. The \gls{eof} is a consequence of the interaction between the fixed charges on the
nanopore walls and mobile charges in the electrolyte and can be described by two closely related mechanisms:
(1) the excess transport of the hydration shell water molecules in one direction due to the pore's ion
selectivity, and (2) the viscous drag exerted by the unidirectional movement of the electrical double layer
inside the pore~\cite{Tagliazucchi-2015,Bonome-2017}. The first mechanism likely dominates in pores with a
diameter close to that of the hydrated ions (\SI{\le1}{\nm}) such as \gls{ahl} or
\gls{frac}~\cite{Huang-2017,Huang-2019}, whereas the second is expected to be stronger for larger pores
(\SI{>1}{\nm}), such as \gls{clya}~\cite{Soskine-2012,Willems-Ruic-Biesemans-2019} or most solid-state
nanopores~\cite{Mao-2014,Laohakunakorn-2015}. In our simulation, the \gls{eof} is generated according to the
second mechanism by coupling the Navier-Stokes and the Poisson-Nernst-Planck equations through a volume
force ($\volumeforce$, \cref{eq:ion_force_density}). This coupling dictates that the electric field exerts a
net force on the fluid if it contains a net ionic charge density---as is the case for the electrical double
layer lining the walls of \gls{clya} (\cref{fig:transport_charge_total}).

\begin{figure*}[p]
  \centering
  
  \begin{minipage}[t]{6cm}
    \begin{subfigure}[t]{6cm}
      \centering
      \caption{}\vspace{-3mm}\label{fig:transport_flow_contour}
      \includegraphics[scale=1]{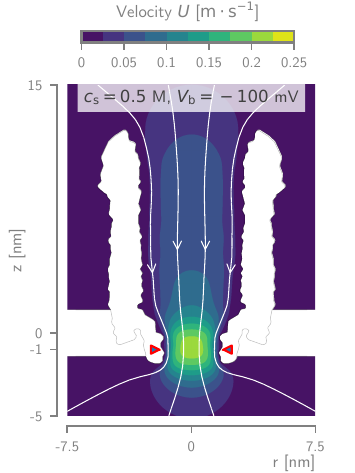}
    \end{subfigure}
    \\
    \begin{subfigure}[t]{5.5cm}
      \centering
      \caption{}\vspace{-3mm}\label{fig:transport_flow_profiles}
      \includegraphics[scale=1]{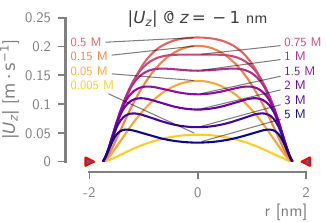}
    \end{subfigure}
  \end{minipage}
  \hspace{0.25cm}
  \begin{minipage}[t]{5cm}
    \begin{subfigure}[t]{5cm}
      \centering
      \caption{}\vspace{-3mm}\label{fig:transport_flow_conductance}
      \includegraphics[scale=1]{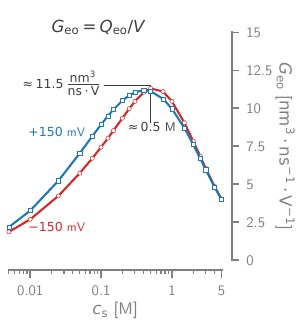}
    \end{subfigure}
    \\
    \begin{subfigure}[t]{5cm}
      \centering
      \caption{}\vspace{-3mm}\label{fig:transport_flow_rectification}
      \includegraphics[scale=1]{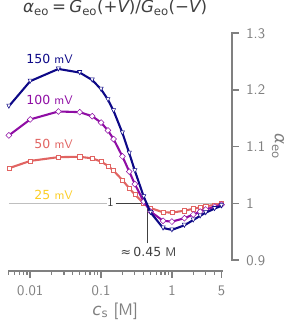}
    \end{subfigure}
  \end{minipage}

  \caption[Concentration and voltage depend. of the {EOF} inside {ClyA-AS}]%
  {%
    \textbf{Concentration and voltage dependency of the electro-osmotic flow inside {ClyA-AS}.}
    (\subref{fig:transport_flow_contour})
    Contour plot of the \gls{eof} velocity magnitude $U$ at \SI{0.5}{\Molar} and \SI{-100}{\mV} bias voltage.
    The arrows on the streamlines indicate the direction of the flow. As observed
    experimentally~\cite{Soskine-2013} and expected from a negatively charged conical nanopore, the \gls{eof}
    follows the direction of the cation (\ie~from \cisi{} to \transi{} under negative bias voltages and
    \textit{vice versa} for positive ones). Note that in contrast to \gls{plyab-r}, whose oppositely charged
    \lumen{} and constriction induce Eddy currents within the pore~\cite{Huang-2020}, the uniform distribution
    of negative charges on the walls of \gls{clya} results in a smooth, unidirectional flow.
    (\subref{fig:transport_flow_profiles})
    Cross-section profiles of the absolute value of the water velocity $\left|U_z\right|$ inside \transi{}
    constriction (at $z = \mSI{-1}{\nm}$) for various salt concentrations at $\vbias = \mSI{-100}{\mV}$.
    Notice that at high salt concentrations ($\cbulk > \mSI{1}{\Molar}$), the velocity profile exhibits two
    `lobes' close to the nanopore walls and hence deviates from the parabolic shape observed at lower ionic
    strengths.
    (\subref{fig:transport_flow_conductance})
    Concentration dependency of the electro-osmotic conductance $\flowcond = \flowrate / \vbias$, with
    $\flowrate$ the total flow rate through the pore (\cref{eq:flowrate}). In the low concentration regime,
    $\flowcond$ increases rapidly between \SIlist{0.005;0.5}{\Molar} after which it decreases logarithmically
    for higher concentrations.
    (\subref{fig:transport_flow_rectification})
    The rectification of the electro-osmotic conductance ($\eor(V) = \flowcond (+V) / \flowcond (-V)$) plotted
    against the bulk salt concentration. The $\eor$ increases with bias voltage and exhibits an inversion
    point at $\cbulk \approx \mSI{0.45}{\Molar}$.
  }\label{fig:transport_flow}

\end{figure*}

\subsubsection{Direction, magnitude, and distribution of the water velocity.}
As expected, given \gls{clya}'s negatively charged interior surface and the resulting positively charged
electrical double layer, the direction of the net water flow inside \gls{clya} follows the electric field:
from \cisi{} to \transi{} at negative bias voltages (\cref{fig:transport_flow_contour}). This corresponds to
the observations and analysis of single-molecule protein capture~\cite{Soskine-2013} and
trapping~\cite{Soskine-Biesemans-2015,Biesemans-2015,Willems-Ruic-Biesemans-2019} experiments using the
\gls{clya-as} nanopore. Along the longitudinal axis ($z$) of the pore, the water velocity is governed by the
conservation of mass, meaning it is lowest in the wide \cisi{} \textit{lumen} and highest in the narrow
\transi{} constriction (\cref{fig:transport_flow_contour}). For example, at $\vbias = \mSI{-100}{\mV}$ and
$\cbulk = \mSI{0.5}{\Molar}$ the velocity at the center of the pore is \SI{\approx0.07}{\mps} in the
\textit{lumen} and \SI{\approx0.21}{\mps} in the constriction.

Along the radial axis ($r$), $\velocity$ has a parabolic profile with the highest value at the center of the
pore and the lowest at the wall due to the no-slip boundary condition (\cref{fig:transport_flow_profiles}).
Such a parabolic profile contrasts the expected `plug flow' for an \gls{eof}, but follows logically from the
overlap of the electrical double layer inside the pore and the resulting uniform volume force---analogous to a
gravity- or pressure-driven Stokes flow. At concentrations higher than \SI{0.5}{\Molar}, however, the
increasing degree of confinement of the double layer---and its charge---to the nanopore walls
(see~\cref{fig:transport_charge_total}, $\pav{\Qion}{s}$) results in a flattening of the central maximum and
hence a plug flow profile. Interestingly, at very high salt concentrations ($\cbulk \ge \mSI{1}{\Molar}$) the
velocity profile in the constriction exhibits a dimple at the center of the pore
(\cref{fig:transport_flow_profiles}). This is the result of a self-induced pressure gradient caused by the
expansion of the \gls{eof} as it exits the pore~\cite{Melnikov-2017}.

\subsubsection{Influence of bulk ionic strength and bias voltage on the electro-osmotic conductance.}
In analogy to the ionic conductance, the amount of water transported by \gls{clya} can be expressed by the
electro-osmotic conductance $\flowcond = \flowrate / \vbias$ (\cref{fig:transport_flow_conductance}). Here,
$\flowrate$ is the net volumetric flow rate of water through the pore and computed by integrating the water
velocity across the reservoir boundary $S$
\begin{align}\label{eq:flowrate}
  \flowrate = \int_{S}\left(\normvec\cdot\vec{\velocity}\right)dS
  \text{ .}
\end{align}
The strength of the \gls{eof} depends strongly and non-monotonically on the bulk ionic strength: $\flowcond$
rapidly increases with ionic strength until a peak value is reached at $\cbulk \approx \mSI{0.5}{\Molar}$,
followed by a gradual logarithmic decline (\cref{fig:transport_flow_conductance}). For example, at $\vbias =
\mSI{-150}{\mV}$, $\flowcond$ first increases from \SI{1.85}{\cnmpnspv} at \SI{0.005}{\Molar} to
\SI{11.3}{\cnmpnspv} at \SI{0.5}{\Molar}, followed by a gradual decline to \SI{4.00}{\cnmpnspv} at
\SI{5}{\Molar}. As with the ionic conductance, more details on the effect of the concentration, wall distance,
and steric corrections on the electro-osmotic conductance can be found
in~\cref{sec:transport:comparison_corrections} (\cref{fig:transport_comparison_water}).

The sensitivity of the \gls{eof} to the magnitude and sign of the bias voltage is given by the electro-osmotic
conductance rectification $\eor(V) = \flowcond (+V) / \flowcond (-V)$
(\cref{fig:transport_flow_rectification}). For all voltage magnitudes, $\eor$ shows a maximum at $\cbulk
\approx \mSI{0.045}{\Molar}$, after which it falls rapidly to reach unity ($\eor = 1$) at approximately
$\cbulk \approx \mSI{0.45}{\Molar}$. A minimum is then reached at \SI{\approx1}{\Molar}, followed by a gradual
approach towards unity at $\cbulk = \mSI{5}{\Molar}$.

\begin{figure*}[t]
  \centering

  \begin{subfigure}[t]{6cm}
    \centering
    \caption{}\vspace{-3mm}\label{fig:transport_pressure_contour}
    \includegraphics[scale=1]{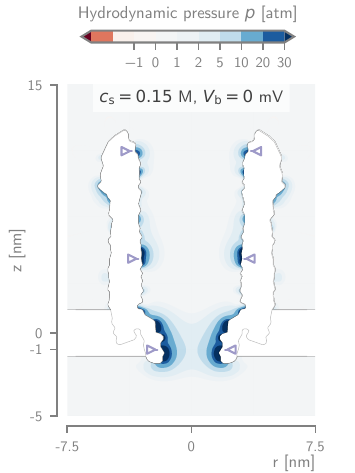}
  \end{subfigure}
  \hspace{-5mm}
  \begin{subfigure}[t]{4.5cm}
    \centering
    \caption{}\vspace{-3mm}\label{fig:transport_pressure_profiles}
    \includegraphics[scale=1]{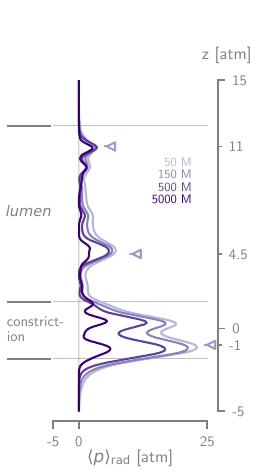}
  \end{subfigure}

  \caption[Pressure distribution inside {ClyA-AS}]
  {%
    \textbf{Pressure distribution inside {ClyA-AS}.}
    (\subref{fig:transport_pressure_contour})
    Contour map of the hydrodynamic pressure $\pressure$ at $\cbulk = \mSI{0.15}{\Molar}$ and
    $\vbias=\mSI{0}{\mV}$, showing that the large variations in \Na{} concentration along the pore wall result
    in osmotic pressure `hotspots' (\SIrange{5}{30}{\atm}) inside the confined fluid.
    (\subref{fig:transport_pressure_profiles})
    The axial pressure profile and averaged along the entire radius of the pore at $\vbias=\SI{0}{\mV}$.
  }\label{fig:transport_pressure}

\end{figure*}

\subsubsection{Pressure distribution inside ClyA.}
The large variations of \Na{} concentration along the walls of \gls{clya}---up to several orders of magnitude
over the course of a few nanometers (see~\cref{fig:transport_concentration_contours})---induce regions of high
`osmotic' pressure with peak values up to \SI{30}{\atm} (\cref{fig:transport_pressure_contour}). The largest
`hotspots' are located at \cisi{} entry of the pore ($z = \mSI{11}{\nm}$), in the middle of the \lumen{} ($z =
\mSI{4.5}{\nm}$) and inside the entire constriction ($z = \mSI{-1}{\nm}$)
(\cref{fig:transport_pressure_profiles}), and their influence extends well towards the center of the pore. Up
until $\cbulk \approx \mSI{0.5}{\Molar}$, increasing the reservoir salt concentration does not seem to
strongly influence the overall magnitude of the pressure spots. Hence, because such large pressure differences
can  exert a significant amount of force on particles translocating through nanopores~\cite{Hoogerheide-2014},
we expect them to play an important role in the detailed trapping dynamics of proteins inside
\gls{clya}~\cite{Soskine-Biesemans-2015,Willems-Ruic-Biesemans-2019}.

\subsection{Effect of the individual corrections on the ionic and water conductances}
\label{sec:transport:comparison_corrections}

\begin{figure*}[p]
  \centering
  \begin{subfigure}[t]{11.5cm}
    \centering
    \caption{}\vspace{0mm}\label{fig:transport_comparison_ionic}
    \includegraphics[scale=1]{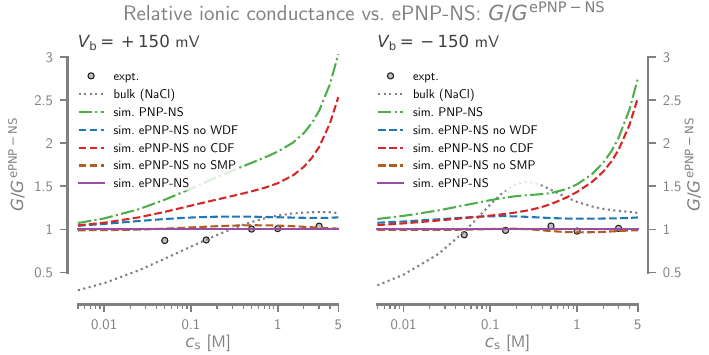}
  \end{subfigure}
  \\
  \begin{subfigure}[t]{11.5cm}
    \centering
    \caption{}\vspace{0mm}\label{fig:transport_comparison_water}
    \includegraphics[scale=1]{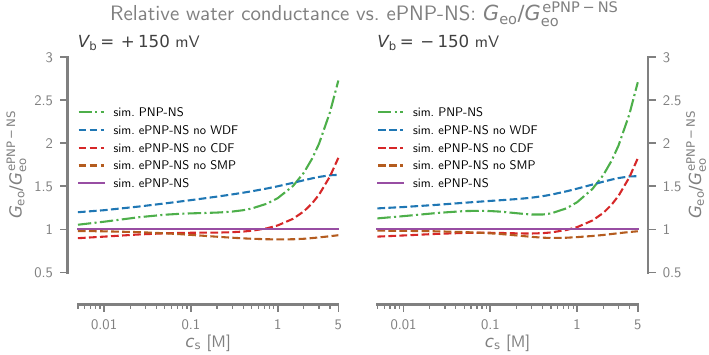}
  \end{subfigure}

  \caption[Effect of individ. corrections on the sim. ionic and water conductance]%
  {%
    \textbf{Effect of individual corrections on the simulated ionic and water conductance.}
    (\subref{fig:transport_comparison_ionic})
    The ionic conductance $\conductance = \current / \vbias$ and
    (\subref{fig:transport_comparison_water})
    the water conductance $\flowcond = \flowrate / \vbias$ of {ClyA-AS} at \SI{+150}{\mV} (left) and
    \SI{-150}{\mV} (right), normalized over the values of the \gls{epnp-ns} equations. Comparison between the
    experimental data (expt.), the simple resistor model (bulk, \cref{eq:bulk_nanopore_current}), classic
    {PNP-NS} ({sim. PNP-NS}), \gls{epnp-ns} without wall distance corrections ({sim. ePNP-NS no WDF}:
    $\diffusion_{i}^w=1$, $\mobility_{i}^w=1$, $\viscosity^w=1$), \gls{epnp-ns} without
    concentration-dependent corrections ({sim. ePNP-NS no CDF}: $\permittivity_{r,\text{f}}^c=1$,
    $\diffusion_{i}^c=1$, $\mobility_{i}^c=1$, $\viscosity^c=1$, $\density^c=1$), \gls{epnp-ns} without steric
    effect ({sim. ePNP-NS no SMP}: $\vec{\beta}=0$) and full with all corrections enabled (\ie~\gls{epnp-ns}).
  }\label{fig:transport_comparison}
\end{figure*}

To estimate the influence to which each group of corrections (\ie~the wall distance, concentration, and steric
effects) on the conductance properties of \gls{clya-as}, we performed a set of simulations for the following
conditions: \gls{epnp-ns} without wall distance corrections (`ePNP-NS no WDF'; $\diffusion_{i}^w=1$,
$\mobility_{i}^w=1$, $\viscosity^w=1$), \gls{epnp-ns} without concentration-dependent corrections (`ePNP-NS no
CDF'; $\permittivity_{r,\text{f}}^c=1$, $\diffusion_{i}^c=1$, $\mobility_{i}^c=1$, $\viscosity^c=1$,
$\density^c=1$), \gls{epnp-ns} without steric effect (`ePNP-NS no SMP'; $\vec{\beta}=0$). For comparison, we
plotted the ionic (\cref{fig:transport_comparison_ionic}) and water (\cref{fig:transport_comparison_water})
conductances of these models, normalized over the values obtained with the full \gls{epnp-ns} equations.

\subsubsection{Effect of disabling the wall distance corrections.}

Disabling the wall distance corrections yields a higher (\SI{\approx10}{\percent} increase) ionic conductance
compared to the full \gls{epnp-ns} equations (\cref{fig:transport_comparison_ionic}, blue curve). In effect,
the implementation of the wall distance corrections results in smaller effective pore sizes.  The influence
on the water flow is stronger, with the reduction of the viscosity near the wall giving rise to a
\SIrange{25}{50}{\percent} increase in the water flow (\cref{fig:transport_comparison_water}, blue curve).

\subsubsection{Effect of disabling the concentration corrections.}

The removal of the concentration-dependent corrections has a large influence on the simulated results. The
ionic conductance increases dramatically: from \SIlist{5;13;29;152}{\percent} at
\SIlist{0.005;0.05;0.5;5}{\Molar}, respectively (\cref{fig:transport_comparison_ionic}, red curve). This is
expected, however, given that the large reduction of both diffusion coefficients and ionic mobilities with
increasing salt concentrations is not taken into account. The influence on the water flow shows only a limited
effect from \SI{0.005}{\Molar} (\SI{9}{\percent} decrease) to \SI{1}{\Molar} (\SI{2}{\percent} increase),
followed by a rapid increase of \SI{82}{\percent} between \SIlist{1;5}{\Molar}
(\cref{fig:transport_comparison_water}, red curve). This is a direct result of the \SI{15}{\percent} increase
of the electrolyte viscosity between \SIrange{1}{5}{\Molar}, compared to the mere \SI{5}{\percent} increase
between \SIrange{0}{1}{\Molar} (\cref{fig:epnpns_concentration_solv_eta}).

\subsubsection{Effect of disabling the steric corrections.}
The steric corrections appear to have little influence on the ionic conductance, with a maximum deviation of
at most \SI{\pm3}{\percent} over the entire concentration range (\cref{fig:transport_comparison_ionic}, brown
curve). The effect on the water flow is larger, with a decrease of \SI{10}{\percent} at \SI{0.5}{\Molar}
(\cref{fig:transport_comparison_water}, brown curve). By placing an upper limit on the total ion concentration
(\SI{\approx13.3}{\Molar} in our case), the steric corrections prevent the excessive screening of fixed charges by
nonphysical ionic strengths (\ie~concentrations that would require fitting more ions into a given space than
what is physically possible). This is an essential mechanic that allows one to realistically model the
electrical double layer near surfaces with high charge densities---as is the case for most biological
nanopores.

\section{Conclusion}
\label{sec:transport:conclusion}

The use of computationally inexpensive continuum models is pervasive in the solid-state nanopore field, but
their application to structurally more complex biological nanopores has been limited to date. We made use
of the radial symmetry of the biological nanopore \gls{clya} to create a 2D-axisymmetric model of the pore
which, in conjunction with the \gls{epnp-ns} equations, is able to accurately describe the ionic current of
\gls{clya} for a wide range of experimentally relevant ionic strengths and bias voltages. Our approach shows
that continuum modeling of biological nanopores is not only feasible but can also be predictive. Our results
describe in great detail the properties of \gls{clya}, such as its true ion selectivity, the differences
between cation and anion concentrations inside the pore, the distribution, and magnitude of the electrostatic
potential, the velocity of the electro-osmotic flow and the presence of highly localized `hotspots' of osmotic
pressure. These findings do not only provide valuable insights into the physical mechanisms of nanoscale
transport, but they also reveal a great deal about the conditions within the pore itself. For example, strong
deviations from bulk values, such as the depletion of anions from the \lumen{} of \gls{clya} at negative bias
voltages, may have a profound impact on the activity and structure of trapped proteins.

The \gls{epnp-ns} framework comprises an empirical approach that significantly improves the quantitative
accuracy for continuum simulations of nanoscale transport phenomena. We expect our framework and its
principles to be transferable to other biological nanopores and nanoscale transport systems. We believe our
model constitutes a powerful and practical tool that can aid with (1) elucidating the link between ionic
current observed during a nanopore experiment and the actual physical phenomenon, (2) describing the
electrophoretic and electro-osmotic properties of any biological nanopore, and (3) guiding the rational design
of new variants of existing nanopores. For example, by further expanding and automating our framework, it
could be used as a probe to gauge---and tailor---the properties of specific nanopore
mutants~\cite{Huang-2020,Cao-2019}, or to study the dynamics of translocating proteins by computing the forces
that act on model particles~\cite{Willems-Ruic-Biesemans-2019}. Both of these applications could play a
crucial role in designing the next generation of (biological) nanopore-based biosensors.

\section{Materials and methods}
\label{sec:transport:methods}

\subsection{{ClyA-AS} expression and purification}

\Gls{clya-as} monomers were expressed, purified, and oligomerized using methods described in detail
elsewhere~\cite{Soskine-2012,Soskine-2013}. Briefly, \textit{E. cloni} {EXPRESS BL21} (DE3) cells (Lucigen
Corporation, Middleton, USA) were transformed with a {pT7-SC1} plasmid containing the \gls{clya-as} gene,
followed by overexpression after induction with \SI{0.5}{\mM} \gls{iptg} (Carl Roth, Karlsruhe, Germany). The
\gls{clya} monomers were purified using \ce{Ni+}-NTA affinity chromatography and oligomerized by incubation in
\SI{0.2}{\percent} \gls{ddm} (Sigma-Aldrich, Zwijndrecht, The Netherlands) for \SI{30}{\minute} at
\SI{37}{\celsius}. Pure \gls{clya-as} {type I} (12-mer) nanopores were obtained using native \gls{page} on a
\SIrange[range-phrase = --]{4}{15}{\percent} gradient gel (Bio-Rad, Veenendaal, The Netherlands) and
subsequent excision of the correct oligomer band.

\subsection{Recording of single-channel current-voltage curves}

Experimental current-voltage curves were measured using single-channel electrophysiology, as detailed
elsewhere~\cite{Maglia-2010,Soskine-2012,Soskine-2013}. First, a black lipid bilayer was formed inside a
\SI{\approx100}{\um} diameter aperture in a thin Teflon film separating two buffered electrolyte compartments.
This was achieved by applying a droplet of \SI{5}{\percent} hexadecane in pentane (Sigma-Aldrich, Zwijndrecht,
The Netherlands) over the aperture and leaving it to dry for \SI{1}{\minute} at \SI{25}{\celsius}. The
buffered electrolyte solution was added to both compartments, topped with \SI{10}{\uL} of
\SI{6.25}{\milli\gram\per\milli\liter} \gls{dphpc} (Avanti Polar Lipids, Alabaster, USA) in pentane. The
pentane was left to evaporate for \SI{2}{\minute} at \SI{25}{\celsius}. A lipid bilayer was formed by lowering
and raising the buffer level over the aperture. Minute amounts (\SI{\approx0.2}{\uL}) of the purified
\gls{clya-as} {type I} oligomer were then added to the grounded \cisi{} reservoir and allowed to insert into
the lipid bilayer. Single-channel current-voltage curves were recorded using a custom pulse protocol of the
\code{Clampex 10.4} (Molecular Devices, San Jose, USA) software package connected to AxoPatch 200B patch-clamp
amplifier (Molecular Devices, San Jose, USA) \textit{via} a Digidata 1440A digitizer (Molecular Devices, San
Jose, USA). Data were acquired at \SI{10}{\kHz} and filtered using a \SI{2}{\kHz} low-pass filter. Measurements
at different ionic strengths were performed at \SI{\approx25}{\celsius} in aqueous \ce{NaCl} (Carl Roth,
Karlsruhe, Germany) solutions, buffered at \pH{7.5} using \SI{10}{\mM} \gls{mops} (Carl Roth, Karlsruhe,
Germany).

\cleardoublepage

%

%
  \graphicspath{{\chaptersdir/conclusion/image/}}%
\chapter{Conclusions and perspectives}
\label{ch:conclusion}

\epigraphhead[\epipos]{%
\epigraph{%
  ``But just because something does not have an ending doesn't mean it doesn't have a conclusion.''
}{%
  \textit{`Ky' in `Use of Weapons' by Iain M. Banks}
}}

In this dissertation, I have demonstrated that computational modeling---in combination with analytical
physical models and systematic experiments---can be a powerful tool to shed light on (1) the physical
mechanisms that drive molecular transport through (biological) nanopores, and (2) the diverse physical
conditions that occur within them. In particular, this work shows that although a great deal of information
can be obtained from equilibrium electrostatic simulations, which effectively mimic the nanopore without an
applied bias voltage, the transport of analyte molecules such as \gls{dna} or proteins is governed by a
complex interplay of extrinsic (\ie~(di)electrophoretic and electro-osmotic) and intrinsic (\ie~electrostatic,
steric, and entropic) forces. It is for this reason that a nonequilibrium simulation framework was developed:
the \glsfirst{epnp-ns} equations. This framework enables one to accurately predict the transport of ions,
water molecules, and analyte molecules through biological nanopores, providing a means to paint a complete
picture of the sensing process. The following sections summarize the key conclusions that can be drawn from
this dissertation, followed by my perspective on the future of nanopore sensing, and the role of computational
modeling therein.

\section{General conclusions}
\label{sec:con:conclusions}
%


\subsubsection{The electrostatics of biological nanopores influence all transport processes.}

The importance of electrostatics within biological nanopores cannot be understated. Electrostatic interactions
have an influential hand in defining the nanofluidic properties of most biological nanopores. Not only do they
determine conductance characteristics such as ion selectivity~\cite{Ramirez-2007} and current
rectification~\cite{Wang-2014}, they are also the fundamental enabler of the
\glsfirst{eof}~\cite{Bocquet-2010}, and finally, they strongly mediate the translocation process of analyte
molecules \textit{via} long-range electrostatic forces~\cite{Maglia-2008,Asandei-2015b,Fahie-2015b}.

In~\cref{ch:electrostatics}, we made use of \glsfirst{pb} theory to investigate the influence of charge
reversal mutations, in combination with the ionic strength and pH of the electrolyte, on the equilibrium
electrostatics (\ie~without an applied bias voltage) of the biological nanopores \glsfirst{plyab},
\glsfirst{frac}, and \glsfirst{clya}. We found that introducing charge reversal mutations
(negative-to-positive) within the constrictions of the \gls{plyab} (\gls{plyab-e2} \textit{versus}
\gls{plyab-r}) and \gls{frac} (\gls{wtfrac} \textit{versus} \gls{refrac}) pores translated itself in the full
reversal of the electrostatic potential at those locations, even at high ionic strengths (\SI{1}{\Molar}).
Given the intimate link between the potential and the \gls{eof} (through the \gls{edl}), we hypothesized that
this could have profound effects on the magnitude of the \gls{eof} for \gls{plyab-r}~\cite{Huang-2020}, and
would even change its directionality for \gls{refrac}~\cite{Huang-2017}, two claims supported by protein
trapping experiments. Additionally, we show that, whereas the magnitude of the electrostatic potential within
the negatively charged wild-type \gls{frac} (\gls{wtfrac}) is strongly reduced upon lowering the pH from
\numrange{7.5}{4.5}, that of the positively charged \gls{refrac} is barely affected. This finding also
confirms the experimental observations~\cite{Huang-2017}. In the case of \gls{clya}, we studied four different
mutants of \gls{clya-as}~\cite{Soskine-2013}: \gls{clya-r} (S110R), \gls{clya-rr} (S110R/D64R),
\gls{clya-rr56} (S110R/Q56R), and \gls{clya-rr56k} (S110R/Q56R/Q8K). Experiments have shown that while all
variants are able to translocate \gls{dsdna} at high salt concentrations
(\SI{>2}{\Molar})~\cite{Franceschini-2013}, only \gls{clya-rr} is able to do so at physiological ionic
strengths (\SI{0.15}{\Molar})~\cite{Franceschini-2016}. This is of importance for applications where the
electrostatic interactions should be as `natural' as possible, such as \gls{dna} mapping or sequencing with
the help of a processive enzyme. We found the S110R mutation at the \cisi{} entry to effectively eliminate the
negative electrostatic potential observed at the \cisi{} entrance of \gls{clya-as}, which in turn is likely to
promote the entry of the \gls{dna} within the \lumen{}. Interestingly, the D64R mutation of \gls{clya-rr}
(\SI{+2}{\ec} per subunit), but not the Q56R mutation of \gls{clya-rr56} (\SI{+1}{\ec} per subunit), resulted
in a reversal of the electrostatic potential in the middle of the \lumen{}, the consequences of which will be
discussed in the next paragraph. Even though the Q8K mutation lowered the magnitude of the negative potential
within the constriction of \gls{clya-rr56k}, it remained significantly higher compared to the rest of the
pore. As expected, at high ionic strengths (\SI{2.5}{\Molar}), however, no differences between the mutants
could be observed, in agreement with experiments~\cite{Franceschini-2013,Franceschini-2016}

To gain a better understanding of the electrostatic energy barriers that arise during the translocation of
analyte molecules, we computed the electrostatic energy costs (or gains) that result from the movement of a
\gls{dna} molecule (in~\cref{ch:electrostatics}) or a protein (in~\cref{ch:trapping}) along the length of the
pore. In the case of \gls{ssdna} translocation through \gls{refrac}, we found that the D10R charge reversal
lowered the energy barrier for entering the constriction of the pore from \SIrange{38}{14}{\kbt}.
Back-of-the-envelope calculations show that even at moderately applied bias potentials (and even ignoring the
\gls{eof}), the barrier in \gls{refrac} can be readily overcome, whereas the one in \gls{wtfrac} cannot. This
is in agreement with the \gls{dna} translocation experiments~\cite{Wloka-2016}. The situation for \gls{dsdna}
translocation through \gls{clya} (at \SI{0.15}{\Molar} ionic strength) is more complex, as the key
charge-reversal mutation, D64R, is not located within the constriction of the pore, but rather in the middle
of the large \cisi{} \lumen{}. We hypothesize that just as the S110R mutation promotes the entry of the
\gls{dna} from the \cisi{} side, so does the D64R mutation encourage the strand to penetrate deeper into the
\lumen{} of the pore by lowering rather than increasing the electrostatic energy. Under (positive) applied
bias voltages, this would allow the \gls{dna} to accumulate sufficient force to overcome \SI{\approx40}{\kbt}
energy barrier within the constriction. At high salt concentrations (\SI{2.5}{\Molar}), the energy landscape
of all variants is similar, and the energy barrier is reduced to a mere \SI{10}{\kbt}, which can be easily
overcome by the electrophoretic force. Interestingly, the energy of the translocating \gls{dna} molecule
appears to fluctuate every \SI{36}{\angstrom} with \SI{\approx2}{\kbt}, suggesting that it might rotate during
translocation. Finally, the strong electrostatic repulsion results in the strong confinement of the
\gls{dsdna} to the center of the pore, even at high salt concentrations.

As an example of the role of electrostatics during protein translocation, we computed the electrostatic
energy of \DHFRt{}, an engineered variant of the \textit{E. coli} \glsfirst{dhfr} enzyme containing a
positively charged C-terminal fusion tag attached to its negative charged core~\cite{Soskine-Biesemans-2015},
as it traverses the \gls{clya} nanopore. Rather than a single barrier at the constriction, the asymmetric
charge distribution of \DHFRt{} gives rise to an energy landscape with an explicit energy minimum, whose depth
is proportional to the number of positive charges in the fusion tag. Hence, this indicates that the protein is
electrostatically trapped within the pore. However, superimposing the externally applied electric field upon
the electrostatic energy of a translocating molecule will `tilt' the energy landscape. In the case of
\DHFRt{}, this tilting caused the magnitude of the \transi{} barrier to decrease, whereas the smaller \cisi{}
barrier was quickly erased and essentially moved to the \cisi{} entry of the pore.

It is clear that the introduction of specific charges, at the right location, can significantly impact the
transport of ions, water molecules, and analytes through biological nanopores. This can be effected either
directly, through electrostatic interactions, or indirectly, through the \gls{eof}, both of which are mediated
by the ionic strength and pH of the electrolyte. Nevertheless, even though equilibrium electrostatic
calculations already provide a wealth of information, applying an external electric field `tilts' the energy
landscape, causing seemingly impassable barriers (at equilibrium) to diminish or even disappear. Hence, a full
understanding of molecular transport through nanopores necessitates the use of nonequilibrium simulations.

\subsubsection{Accurate, nonequilibrium continuum modeling of biological nanopores requires extensive corrections.}

To gain even deeper insights into the nanoscale forces involved in nanopore-based sensing, it is necessary to
go beyond equilibrium electrostatics and approximate analytical models, towards modeling the explicit dynamics
of ion and water transport through nanopores. Whereas \gls{md} simulations are still superior in terms of
accuracy~\cite{Aksimentiev-2005,DeBiase-2016,Basdevant-2019}, the heavy computational cost associated with
simulating the motion of a million individual atoms in \SI{1}{\fs} time steps still limits its usefulness for
high-throughput analysis of long-timescale (\ie~\SI{>1}{\us}) phenomena~\cite{Vendruscolo-2011,Phillips-2020}.
Hence, a fair share of literature has been devoted to the development of continuum models, typically based on
\gls{pnp-ns} theory, of
solid-state~\cite{Daiguji-2004,Cervera-2005,White-2008,Lu-2012,Chaudhry-2014,Laohakunakorn-2015,Hulings-2018,Rigo-2019,Melnikov-2020}
and
biological~\cite{Noskov-2004,Cozmuta-2005,OKeeffe-2007,Simakov-2010,Pederson-2015,Simakov-2018,Aguilella-Arzo-2020}
(\ie~mostly on \gls{ahl}) nanopores. Such models are orders of magnitude more efficient, both in terms of
solution time and the ease of data analysis. However, their applicability for nanoscale transport problems is
often (rightly) questioned~\cite{Corry-2000,Collins-2012}, which has led to the development of modified
continuum transport equations that attempt to compensate for some of the simplifications introduced by
mean-field
theory~\cite{Noskov-2004,Baldessari-2008-1,Daiguji-2010,Simakov-2010,Lu-2011,Burger-2012,Chen-2016,Liu-2020}.
In~\cref{ch:epnpns}, we implemented a novel continuum framework geared specifically towards the modeling of
biological nanopores: the \glsfirst{epnp-ns} equations. It considers steric ion-ion
interactions~\cite{Daiguji-2010,Kilic-2007,Lu-2011,Liu-2020}, the influence of the nanopore walls on the local
ion diffusivity/mobility~\cite{Makarov-1998,Noskov-2004,Pederson-2015,Hulings-2018,Wilson-2019} and the water
viscosity~\cite{Pronk-2014,Vo-2016,Hsu-2017}, as well as the concentration dependencies of ion
diffusivity/mobility~\cite{Baldessari-2008-1,Burger-2012}, and solvent relative permittivity~\cite{Chen-2016},
viscosity and density~\cite{Hai-Lang-1996}. All these dependencies were robustly parameterized using empirical
relations, fitted to experimental or \gls{md} data obtained from literature sources.

Next, in~\cref{ch:transport}, we applied the \gls{epnp-ns} equations to a 2D-axisymmetric model of the
\gls{clya-as} nanopore, whose geometry and charge distribution was derived from series of equilibrated atomic
structures obtained from a \SI{30}{\ns}-long \gls{md} simulation of \gls{clya-as} homology model. The
solutions of the \gls{epnp-ns} equations provided detailed descriptions of the ion concentrations, the
electrostatic potential, the \gls{eof} velocity, and the pressure distributions within the pore, over a wide
range of ionic strengths (\SIrange{0.005}{5}{\Molar}) and applied bias voltages (\SIrange{-200}{+200}{\mV}).
Integration of the cation and anion fluxes through the pore gave direct access to the total ionic current, but
also the true ion selectivity, a property that is difficult to pin down experimentally. Notably, the ionic
conductance of \gls{clya} predicted by the \gls{epnp-ns} equations matched closely with the values obtained
from carefully calibrated single-channel recordings, in contrast to the classical \gls{pnp-ns} equations. The
(cat)ion selectivity, although strongly dependent on the ionic strength and bias voltage, also corresponded to
the permeability ratio derived from the reversal potential measurements. As with the transport of \gls{dna}
(see~\cref{ch:electrostatics}) and proteins (see~\cref{ch:trapping}) through \gls{clya}, we found the
transport of ions to be subject to the electrostatic energy barriers present within the pore. These barriers did
not only influence the conductive properties of the pore but also profoundly impacted the availability of
anions with \gls{clya}'s \lumen{}, particularly at negative bias voltages, a condition that may influence the
structure and activity of enzymes trapped within \gls{clya}. Analysis of the \gls{eof} magnitude showed a
non-monotonic dependence on the ionic strength, with a maximum at \SI{0.5}{\Molar} \ce{NaCl}. This knowledge
may be exploited to fine-tune the force balance exerted on analyte molecules and improve their trapping
efficiency. Interestingly, \gls{clya}'s complex charge distribution and the accompanying non-uniform
\gls{edl}, resulted in confined (\SI{\approx1}{\cubic\nm}) zones of high pressure (\SIrange{5}{30}{\atm})
localized at the \cisi{} entry, the middle of the \lumen{}, and within the \transi{} constriction. The
pressure differences induced by these zones may contribute significantly to the net force exerted on
translocating particles, confirming the importance of using the full hydrodynamic stress tensor to calculate
the force, rather than just the Stokes' drag.

The \gls{epnp-ns} equations, together with a suitable model of the pore, constitute a significant advance in
the continuum simulation of biological nanopores, and the modeling of nanoscale transport phenomena in
general. It is an important step toward building a tool that can help to move the field forward on various
fronts, including elucidating the link between the observed variations in ionic current and the physical
events that cause them, evaluating the transport properties of existing nanopores, and guiding the rational
design of novel characteristics.

\subsubsection{Trapping a protein within {ClyA} involves a complex interplay of forces.}

The ability to retain proteins, and particularly enzymes, for long periods of time within a nanopore, is of
crucial importance for applications such as single-molecule
enzymology~\cite{Willems-VanMeervelt-2017,VanMeervelt-2014,Wloka-2017,Thakur-2019,Galenkamp-2020} and
biomarker sensing~\cite{VanMeervelt-2017,Galenkamp-2018,Zernia-2020}. Previous experiments with \gls{clya}
have shown that, whereas large(r) proteins (that can still enter the pore through the \cisi{} entry) can
remain trapped for seconds to
minutes~\cite{Soskine-2013,Soskine-2012,Soskine-Biesemans-2015,Biesemans-2015,VanMeervelt-2014,VanMeervelt-2017},
smaller proteins (similar in diameter to the \transi{} constriction) translocate through the pore within
milliseconds~\cite{Soskine-2012}. This is too quickly properly sample any conformational changes, enzymatic
reactions, or ligand binding. However, other studies revealed that the dwell time of
peptides~\cite{Movileanu-2005,Asandei-2015,Asandei-2016} within biological nanopores could be lengthened by
orders of magnitude by the introduction of a large permanent dipole moment.

In~\cref{ch:trapping}, a systematic set of experiments was performed with the \DHFRt{} enzyme. Previous
experiments revealed that the binding of the negatively charged inhibitor molecule \glsfirst{mtx} to \DHFRt{}
extended its peak mean dwell time within the pore from \SIrange{0.3}{9}{\second}, a 30-fold
increase~\cite{Soskine-Biesemans-2015}. We successfully eliminated the need for \gls{mtx} (which would prevent
a study of \gls{dhfr}'s enzymatic properties) by (1) introducing three additional negative charges near the
\gls{mtx} binding site, and (2) by increasing the number of positively charged residues in the fusion tag from
\numrange{+4}{+9}. The former raised the dwell time by an order of magnitude to \SI{2}{\second}, whereas the
latter resulted in dwell times of \SI{4}{\second} (\num{+5}), \SI{8}{\second} (\num{+6}), \SI{14}{\second}
(\num{+7}), \SI{25}{\second} (\num{+8}), and \SI{46}{\second} (\num{+9}). All variants were still able to bind
to their cofactor, \ce{NADH}, indicating that neither our modifications, nor the confinement within the pore,
caused structural disruption or unfolding. Notably, the \num{+7} \DHFRt{} variant recently enabled the
real-time detection \gls{dhfr}'s conformational changes of during its enzymatic cycle~\cite{Galenkamp-2020}.

The dwell times of all \DHFRt{} variants exhibited a biphasic relation with the applied bias voltage, where it
first rose exponentially up until a certain threshold voltage, followed by an exponential
decay~\cite{Biesemans-2015}, similar to the escape rate of an \ta-helical peptide trapped within
\gls{ahl}~\cite{Movileanu-2005}. Such a complex behavior was explained by a shift in \gls{dhfr}'s preference
to exit \gls{clya} from the \cisi{} entry at low voltages to the \transi{} entry at high applied biases. To
quantify this process, we developed a double energy barrier physical transport model (\ie~one barrier for each
entry), containing steric, electrostatic, electrophoretic, and electro-osmotic energy components. By
meticulously mapping out these voltage dependencies, together with an extensive set of equilibrium
electrostatic energy calculations for all translocating \DHFRt{} variants, we successfully parameterized this
double barrier model, yielding meaningful values for \gls{dhfr}'s intrinsic \cisi{} and \transi{}
translocation probabilities, as well as an estimate of the force exerted by the \gls{eof} on the protein of
\SI{0.178}{\pico\newton\per\milli\volt} (\eg~\SI{9}{\pN} at \SI{-50}{\mV}).

Interestingly, the electrostatic energy barriers obtained from the electrostatic simulations matched closely
with those obtained from the fitting of the double barrier model, indicating that the simulations closely
represent the true physical situation. Moreover, the magnitude of \gls{eof} force on \gls{dhfr} is equivalent
to a net charge of \SI{+15.5}{\ec}. This value is significantly larger than that of the typical protein
(\SI{99}{\percent} lies between \SI{\pm10}{\ec}~\cite{Requiao-2017}), \emph{quantitatively} confirming that
the \gls{eof} is the dominating force in protein capture by \gls{clya}. Given the general nature of the double
barrier model (\ie~it includes no explicit structural information of the trapped protein besides its charge),
it may be applicable to other proteins and even other pores, providing a solid approach toward rationally
engineering the dwell time of small proteins within nanopores.

\section{Future perspectives}
\label{sec:con:perspectives}

\subsubsection{Technical improvements to the {ePNP-NS} framework.}

Even though the \gls{epnp-ns} framework presented in this dissertation constitutes a solid step forwards for
the accurate and efficient modeling of (biological) nanopores with continuum methods, it is still far from the
being a `user-friendly' tool that can be utilized by nanopore researchers. Currently, the \gls{epnp-ns}
equations have been implemented in the COMSOL Multiphysics software, a versatile, but commercial finite
element solver. In the spirit of `open science', the first (technical) improvement I would endeavor is porting
the current implementation into a free (as in speech, not just beer) and open-source \gls{pde} solver
framework, such as \code{FEniCS} (finite element) or \code{OpenFOAM} (finite volume). Besides making the full
program available, free of charge, to anyone that wishes to use it, this would give full control of the code
to the users, enabling them to utilize, adapt, and contribute as they see fit. Additionally, this would provide
clear, straightforward, and uniform paths toward full integration of both the upstream (molecular modeling and
geometry generation) and downstream (data analysis and visualization) workflows. The next challenge to tackle
would be to implement a true 3D model, rather than the 2D-axisymmetric representation. Even though this would
likely result in a significant increase in computational time, the lack of axisymmetric averaging would remove
a layer of abstraction from the simulation that imposes an inherent limit on the accuracy of the results that
can be obtained with them. Full 3D models could properly capture the complex, corrugated geometries present in
even the most radially symmetric nanopores. As a result, this would also greatly improve the realism of the
nanopore's fixed charge density, as it eliminates overlapping charges, and charges that fall within the
electrolyte rather than the solid dielectric of the pore. Note that the 2D-axisymmetric approach could remain
a valuable tool for situations where speed, rather than accuracy, is of the essence---as would be the case
when screening many nanopore mutants or conditions. A final technical improvement would be to further automate
the simulation pipeline, at the very least for specific and verified pores, to provide a
user-friendly tool that is usable by less `technically-inclined' users.\footnotemark%
\footnotetext{
Note that unlike me, these are usually the scientists that perform useful experiments and breakthroughs with
actual real-world applications. 
}
Perhaps this last enhancement is the most important of all, given that any piece of software can only be
useful if it is being used.

\subsubsection{Scientific improvements to the {ePNP-NS} framework.}

Also from a scientific point-of-view several advances as possible. First, the empirical parameterization that
powers the material properties in the \gls{epnp-ns} should be replaced by a first-principles approach that
truly captures the underlying physics, instead of mimicking the phenomenological results. In other words,
rather than forcing the approximated physics to be more accurate, it is better to improve or remove the
approximations themselves. To date, perhaps the most powerful example of such a first-principles methodology
is the Poisson-Nernst-Planck-Bikerman model developed by Liu and Eisenberg~\cite{Liu-2020}, which, in their
own words ``describes the size, correlation, dielectric, and polarization effects of ions and water in aqueous
electrolytes at equilibrium or nonequilibrium all within a unified framework.'' Nevertheless, despite its
phenomenological approach, the existing \gls{epnp-ns} parameterization remains quantitatively useful and may
serve as a barometer for which approximations one should address first. In the short term, a more attainable
goal is the explicit implementation of the electrolyte pH, through the inclusion of \ce{H+} and \ce{OH-} ions,
in combination with the water equilibrium reaction and buffer molecules. Besides improving the general realism
of the framework, it would enable the self-consistent modeling of the nanopore's fixed charge distribution
\textit{via} the effective $\pKa$ of all ionizable groups. This is not only relevant for biological nanopores
but proves even more instrumental for accurately simulating the solid-liquid interfaces present in solid-state
devices.

\subsubsection{Potential applications of the {ePNP-NS} framework.}

On the application side, there are several low-hanging fruits: (1) simulating nanopore mutants, and (2)
computing the net force exerted on translocating analyte molecules. Recently, we applied our \gls{epnp-ns}
approach to the \gls{plyab-e2} and \gls{plyab-r} pores described in \cref{ch:electrostatics}, and found that
it was indeed able to capture the experimentally observed differences in the \gls{eof} between these two
mutants~\cite{Huang-2020}. The current framework could be applied to numerous other pores and mutants thereof,
to help explain and perhaps even guide experiments. Even more exciting is the self-consistent simulation of
translocating particles of various sizes and charges under nonequilibrium conditions. In addition to the
electrostatic force, these computations would yield the full electrophoretic and hydrodynamic force landscape,
together with the residual current values for each particle position. By subsequently feeding these
data into a Brownian dynamics simulator, which adds a stochastic force to mimic thermal motion, it
would allow one to mimic the actual nanopore sensing process itself~\cite{Pederson-2015,Hulings-2018}. For
example, when using \gls{clya} quantify small molecules (\eg~metabolites) \textit{via} their binding to an
``adaptor protein'' trapped within the nanopore~\cite{Zernia-2020}, the precise origin of the minor current
fluctuations (or lack thereof) is still unclear. By providing a detailed trace of the positional dynamics of
the adaptor protein---with and without its bound ligand---together with the simulated ionic current, our
computational model could significantly better our understanding of such systems. Taking it even further, a 3D
solver could be used to automatically screen all the adaptor proteins available in the \gls{pdb}, assessing
their expected signal strength and yielding a shortlist of candidates to test experimentally. Similarly, the
framework could be employed for the analysis of the proteins themselves (\ie~proteomics). By predicting the
expected residual ionic currents and the dynamics of relevant proteins, simulations can aid with their
identification and classification. Finally, also in the fields of \gls{dna} or protein sequencing, the
\gls{epnp-ns} framework could be of great value. It could be used to optimize the signals of a specific pore
by performing a (semi) automatic screening of mutations within the base-sensing region of the nanopore.
Additionally, simulations of existing \gls{dna} sequencing systems may contribute to the unambiguous
interpretation of the measured signals, which typically requires complex algorithms or specifically trained
neural networks~\cite{Wick-2019}. In short, by providing a mutualistic connection between the experimentalist
and the simulant, the ability to correlate the detailed dynamics of an analyte molecule with the
experimentally observed ionic current signal is a key ingredient for advancing the nanopore field.

\cleardoublepage

%

\appendix
\chapter[SI: Trapping of a single protein inside a nanopore]%
        {Supplementary information: Trapping of a single protein inside a nanopore}
\label{ch:trapping_appendix}

\definecolor{shadecolor}{gray}{0.85}
\begin{shaded}
  \adapACS{Willems-Ruic-Biesemans-2019}
\newpage
\end{shaded}

\section{Escape rates over a potential barrier}
\label{sec:trapping_appendix:escape_rates}

\subsection{Dwell times of bound states}

When measuring the dwell time of molecules in bound states such as when they are trapped in nanopores, we
typically obtain a distribution of dwell times whose appearance yields direct information about the kinetics
of the bound state. In this section, we will discuss how to extract kinematic information from such
distributions and what the distributions of multi-level bound states look like.

\subsubsection{Single bound state}
\label{sec:trapping_appendix:single_bound_state}

The simplest scenario is the decay of a single, bound state. Suppose we measure the duration of the bound state
many times in an experiment. This involves measuring the time trace noting the capture and release times and
then plotting a histogram of events vs.~dwell time in the bound state. Since the energy obtained for the
release of the molecule is thermal, the process is stochastic in nature and with growing event numbers the
histogram approaches the underlying distribution associated with the kinetics of escaping from the bound
state.

For a single bound state with a rate of escape $\rate$, the probability distribution function of remaining in
the bound state is given by the typical exponential distribution
\begin{align*}
    f(t) = \rate \; e^{-\kbt}
    \text{ .}
\end{align*}
The nature of the distribution function $f$ is sometimes confusing as it is often used interchangeably and
without notational clarity as both a probability distribution and an event distribution. In an experiment
where we count events in time series data, we gather statistical data on the event distribution rather than on
the probability distribution. In the simple case of a single, bound state, this event distribution is given by
\begin{align}\label{eq:event_distri}
    F(t) = F_0 \;f(t)
    \text{ ,}
\end{align}
where $F_0$ is the total number of events. Note that since $f$ is a probability distribution, it is
necessarily normalized to unity
\begin{align}
    \langle 1\rangle_f :={}& \int_0^\infty dt\;f(t) = 1
    \text{ ,}
\end{align}
and therefore the event distribution is normalized to the total number of events:
\begin{align}\label{eq:normalization}
    \langle 1\rangle_F :={}& \int_0^\infty dt\;F(t) = F_0
    \text{ .}
\end{align}

Since the event distribution and the probability distribution are so closely related in this case, the rate of
escape $\rate$ can simply be read off of a general exponential fit to the event histogram generated from the
time-series data.

For later reference, let us note that the \emph{expectation value} of the duration of the molecule in the
bound state (\ie~the average dwell time) can be computed from the distribution function as
\begin{align*}
    \tau :={}& \langle t\rangle_f
    = \int_0^\infty dt\; t\; f(t)
    = \int_0^\infty dt\; t\;\rate\; e^{-kt}
    = - t e^{-kt}\Big|_{0}^\infty + \int_0^\infty dt \;e^{-kt}
    = \frac{1}{\rate}
    \text{ .}
\end{align*}
On the other hand, if we want to compute the average dwell time from the event distribution, we need to take
into account that the event distribution is differently normalized:
\begin{align}\label{eq:event_distri_norm}
    \tau ={}& \frac{\langle t\rangle_F}{\langle 1\rangle_F}
    = \frac{F_0\langle t\rangle_f}{F_0\langle 1\rangle_f}
    = \langle t\rangle_f
    \text{ .}
\end{align}

The variance is defined as the squared deviation from the expectation value
\begin{align*}
    \variance(t) = \left\langle (t - \langle t\rangle_f)^2\right\rangle_f
                 = \langle t^2\rangle_f - \langle t\rangle_f^2
    \text{ .}
\end{align*}
In case of an event distribution $F$, we also need to normalize by $\langle 1\rangle_F$ as above.  The
expectation value of the squared time is given by
\begin{align*}
    \langle t^2\rangle_f ={}& \int_0^\infty dt\; t^2 f(t)
        = \int_0^\infty dt\; t^2 \; \rate\; e^{- kt}
        =  - t^2\; e^{-kt}\Big|_0^\infty + \int_0^\infty dt\; 2t\; e^{-kt}\\
        ={}&  \frac{2}{\rate} \langle t\rangle_f = 2\langle t\rangle^2
        \text{ .}
\end{align*}
Therefore we find for the variance of the dwell time of events
\begin{align*}
    \variance(t) = \langle t \rangle^2 = \tau^2 = \frac{1}{\rate^2}
    \text{ .}
\end{align*}

\subsubsection{Multi-level bound states}
\label{sec:trapping_appendix:multi_bound_state}

Now let us take a look at the general $N$-level system. It can be described by the set of coupled \glspl{ode}
\begin{align}\label{eq:multi_ode}
    \ddt{} f_i(t) = \sum_{j=1}^N m_{ij} f_j(t)
    \text{ ,}
\end{align}
where $f_i$ is the probability distribution of the $i$-th state and $N$ is the total number of states. Here we
assume that each bound state can decay into another state \textit{via} the rate $m_{ij}$. If we choose one or
more of the states to represent a free state where the molecule escapes, we can in principle describe trapping
events with an arbitrary number of transitions between meta states before the escape of the molecule. The
rates $m_{ij}$ of \cref{eq:multi_ode} are not entirely unrelated as they need to conserve probability. An
obvious choice is to simply express the decrease of probability of a state as the increase of the other
states:
\begin{align}\label{eq:matrix_elements}
    m_{ij} =
    \begin{cases}
        - \sum_j \rate_{ij}, & \text{if}\; i=j \text{ ,}\\
        \rate_{ji}, & \text{else} \text{ ,}
    \end{cases}
\end{align}
This means that a molecule that disappears in one state has to appear in another one. It cannot simply vanish
nor can it appear out of thin air.

It is known that systems of ODEs as in \cref{eq:multi_ode} describe solutions that can be expressed as the
sums of exponentials of their rates:
\begin{align}\label{eq:ansatz}
    f_i(t)  = \sum_{\ell, m=1}^N a_{i\ell m} e^{- \rate_{\ell m} t}
    \text{ .}
\end{align}
Inserting this ansatz into \cref{eq:multi_ode} tells us how the probability distributions are related:
\begin{align*}
    \ddt{} f_i = \sum_{j} m_{ij} f_j
               = \sum_{j} m_{ij} \sum_{\ell, m} a_{j\ell m} e^{- \rate_{\ell m} t}
    \overset{!}{=} - \sum_{\ell, m} \rate_{\ell m} a_{i\ell m} e^{- \rate_{\ell m} t}
    \text{ .}
\end{align*}
Since all terms are linearly independent, we can compare the coefficients of the exponentials on each side and
find
\begin{align}\label{eq:coeffeq}
    \sum_j m_{ij} a_{j\ell m} = - \rate_{\ell m} a_{i\ell m}
    \text{ .}
\end{align}
Finally using \cref{eq:matrix_elements}, we find
\begin{align}\label{eq:finalcoeffs}
    -\sum_j \rate_{ij} a_{i\ell m} + \sum_{j\ne i} \rate_{ji} a_{j\ell m}
    ={}& - \rate_{\ell m} a_{i\ell m} \nonumber\\
    \Rightarrow\qquad  a_{i\ell m}
    ={}& \frac{\sum_{j\ne i} \rate_{ji} a_{j\ell m}}{\sum_j \rate_{ji} - \rate_{\ell m} }
    \text{ .}
\end{align}
\cref{eq:finalcoeffs} is a manifestation of the probability conservation in terms of a constraint on the
possible values of the coefficients.

This derivation tells us that we can express probability distributions of an arbitrary state in an $N$-level
system by a simple linear combination of all the exponentials with the transition rates present in the
systems. Moreover, the coefficients of the exponential terms in the distribution functions are directly
related to the rates.

\subsubsection{Interpretation of experimental data}

Extracting events from time-series data and plotting a histogram of the dwell time will yield the event
distribution function. In a multi-state system, the observed event may be a consequence of multiple different
states with transition rates between each other. If the experiment or the data analysis technique does not
distinguish between these states but groups them together, what we observe may be the sum of multiple event
distributions.

Event distributions are, as shown \eg~in \cref{eq:event_distri}, mathematically equivalent to probability
distributions, only their interpretation differs. When solving the \gls{ode} of \cref{eq:multi_ode} for event
distributions instead of probability distributions, we can proceed completely analogously.  Only when we apply
the initial conditions, we need to choose a normalization to events as in \cref{eq:event_distri}, rather than
a normalization to probabilities. This is then also reflected in the way that expectation values are
calculated, as was shown in \cref{eq:event_distri_norm}.

In an $N$-level system with event distributions $F_i$, where the measurement or data analysis tools cannot
distinguish between the levels $\alpha,\dots,\beta$, we observe a compound event distribution given by
\begin{align*}
    F_{\alpha,\dots,\beta} = \sum_{i=\alpha}^{\beta} F_i
        = \sum_{i=\alpha}^{\beta} F_{0}f_i
    \text{ .}
\end{align*}
Now $F_{\alpha,\dots,\beta}$ not only contains the sum of all independent event distributions in the states
$\alpha,\dots,\beta$, it also contains all interactions between the states $\alpha,\dots,\beta$ and all other
states. Note that we only need a single overall event normalization $F_0$ since the probabilities of finding
the molecule in one of the many states at $t=0$ is set by the initial conditions chosen when solving the
\gls{ode} of \cref{eq:multi_ode}. All differences between the distributions at time $t$ are then a consequence
of the intrinsic rates between states.

Suppose we measure such a system and now we want to extract the kinetics of the process. The compound event
distribution function can be split up into exponential terms using \cref{eq:ansatz}
\begin{align}\label{eq:compound_distri}
    F_{\alpha,\dots,\beta} =
        \sum_{i=\alpha}^\beta \sum_{\ell, m = 1}^N
            F_{0} a_{i\ell m} e^{- \rate_{\ell m} t}
    =: \sum_{\ell, m=1}^N F_{0} A_{\ell m} \rate_{\ell m} \;e^{-\rate_{\ell m} t}
    \text{ .}
\end{align}
Note that the rate $\rate_{\ell m}$ defined as appearing in the amplitude is to keep the analogy to the
probability distribution of \cref{sec:trapping_appendix:single_bound_state}.  By fitting with a sum of
exponentials, we can simply ignore the results for the coefficients $F_{0}$ and $A_{\ell m}$ and extract the
rates $\rate_{\ell m}$ that are directly related to individual transitions between states.

If we want to compute average dwell times using a compound event distribution function, we find
\begin{align}\label{eq:exp_dwell_time}
    \tau = \frac{\langle t \rangle_{F_{\alpha,\dots,\beta}}}{%
                 \langle 1 \rangle_{F_{\alpha,\dots,\beta}}}
    = \frac{\sum_{\ell, m=1}^N \frac{A_{\ell m} }{\rate_{\ell m}}}{%
        \sum_{\ell, m=1}^N A_{\ell m}}
    \text{ ,}
\end{align}
where we used that
\begin{align*}
    \langle 1 \rangle_{F_{\alpha,\dots,\beta}}
    =  \int_0^\infty dt \sum_{\ell, m=1}^N
        F_{0} A_{\ell m} \rate_{\ell m}\;e^{-\rate_{\ell m} t}
    = F_{0}\sum_{\ell, m=1}^N A_{\ell m}
    \text{ ,}
\end{align*}
and
\begin{align*}
    \langle t \rangle_{F_{\alpha,\dots,\beta}}
    =  \sum_{\ell, m=1}^N
        F_{0} A_{\ell m} \rate_{\ell m}\int_0^\infty dt\;  t\;e^{-\rate_{\ell m} t}
    =  F_{0}\sum_{\ell, m=1}^N \frac{A_{\ell m}}{\rate_{\ell m}}
    \text{ .}
\end{align*}
Note how the event number normalization $F_0$ neatly cancels. This should be obvious because the kinetics of
the bound states cannot depend on whether a hundred or a thousand events are collected.

One useful observation for the calculation of the dwell time of a multi-exponential distribution function is
that the total expectation value is simply the arithmetic mean of the expectation values from the individual
exponentials:
\begin{align}\label{eq:tau_arithmetic_mean}
    \tau = \frac{\langle t \rangle_{F_{\alpha,\dots,\beta}}}{%
                 \langle 1 \rangle_{F_{\alpha,\dots,\beta}}}
    = \frac{\sum_{i=\alpha}^{\beta}\langle t \rangle_{F_i}}{%
            \sum_{i=\alpha}^{\beta}\langle 1 \rangle_{F_i}}
    = \frac{\sum_{i=\alpha}^{\beta}\langle t \rangle_{f_i}}{%
            \sum_{i=\alpha}^{\beta}\langle 1 \rangle_{f_i}}
    = \frac{1}{N_{\alpha,\ldots,\beta}}\sum_{i=\alpha}^{\beta}\tau_i
    \text{ ,}
\end{align}
where $N_{\alpha,\ldots,\beta}$ is the number of bound states of the compound distribution function.

Another type of compounding that is often encountered is that compound distribution functions are collected
from different experiments. The compound distribution function can then be expressed as
\begin{align*}
    \mathcal{F} = \sum_\xi F^{\xi}_{\alpha,\dots,\beta}
    \text{ ,}
\end{align*}
where $\xi$ denotes the experiment and $F^{\xi}_{\alpha,\dots,\beta}$ is the compound distribution function of
experiment $\xi$. Using \cref{eq:compound_distri} we find
\begin{align}\label{eq:exp_sum_distri}
    \mathcal{F} = \sum_\xi\sum_{\ell, m=1}^N F_0^\xi A_{\ell m}\rate^{\xi} _{\ell m}
    \;e^{-\rate^\xi_{\ell m} t}
    \text{ .}
\end{align}
Note that the rates $\rate^\xi_{\ell m}$ can vary systematically between experiments. If, for example, the
temperature is not tightly controlled but measured for each experiment, the temperature dependence of the
rates can in principle be taken into account to remove these kinds of systematic errors. However, for the
present treatment, we will simply assume that the laboratory conditions are controlled tightly enough that we
can assume the rates to be independent of the experiments, \textit{i.e.},
\begin{align}\label{eq:exp_independent}
   \rate^\xi_{\ell m} = \rate_{\ell m}
   \text{ .}
\end{align}
Using \cref{eq:exp_sum_distri,eq:exp_independent} we can therefore express the average dwell time as
\begin{align*}
    \tau = \frac{\langle t\rangle_{\mathcal{F}}}{\langle 1\rangle_{\mathcal{F}}}
    =\frac{\sum_\xi F^\xi_{0}\sum_{\ell, m=1}^N \frac{A_{\ell m} }{\rate_{\ell
        m}}}{\sum_\xi F^\xi_{0}\sum_{\ell, m=1}^N A_{\ell m}}
    = \frac{\sum_{\ell, m=1}^N \frac{A_{\ell m} }{\rate_{\ell m}}}{%
        \sum_{\ell, m=1}^N A_{\ell m}}
    \text{ ,}
\end{align*}
which is identical to \cref{eq:exp_dwell_time}. This means that as long as the experimental conditions are
sufficiently close, we can simply sum up the results of experiments. On the other hand, if we have systematic
differences in rates (\ie~$\rate^\xi_{\ell m}$ varies systematically between experiments) the above treatment
can be easily extended to generalize the computation of average dwell times in the presence of systematic
errors.

Lastly, we want to quantify how event distributions behave when two independent processes are measured within
the same event distribution. An example of this would be that the molecule is captured into a completely
separate state from which it has an escape rate with kinetics that are different from the actual state we want
to measure. This could potentially be a meta state with a negligible transition rate to the main trapping
mechanism or it could simply be a capture in a different orientation.

In that case, we find that the total measured event distribution function is given by
\begin{align}\label{eq:two_population_distribution}
    \mathcal{F} = F^A_{\alpha,\ldots,\beta} + F^B_{\gamma,\ldots,\delta}
    \text{ .}
\end{align}
The main observation, in that case, is that $F^A$ and $F^B$ may contain completely different rates and that the
rates of $F^A$ and $F^B$ are not related as in \cref{eq:finalcoeffs}. A way to distinguish these types of
distribution functions would be to quantify the capture ratio $F^A_0/F^B_0$ in terms of experimental
conditions. Modifying the capture ratio would then yield predictable behavior of the ratio $F^A/F^B$ which can
be used to untangle the two independent physical processes.

\subsection{Escape from a single barrier system}
\label{sec:trapping_appendix:single_barrier_system}

We want to derive an expression for the dwell time of a molecule confined in a potential well with a single
barrier as illustrated in \cref{fig:single_barrier}. This treatment will be useful as a stepping stone to a
double barrier model but it will also tell us why the molecule trapped in a nanopore cannot be described by a
single barrier model.

\onefiguresource{%
\begin{tikzpicture}[scale=1.5]
    \draw[fill=blue!20] (0,-0.3) circle (0.2);

    \draw plot [smooth, tension=0.5]  coordinates {%
        (-1.5,2.5)
        (-1,0.5)
        (0,-0.5)
        (1,0.5)
        (1.5,2)
        (2,1.5)
        (2.5,0.5)
    };

    \draw[->, red,thick] (0,0) -- (0,2.1) node[midway,anchor=east]{$\kinetic$};
    \draw[->, orange,thick] (0,2.1) -- (1.5,2.1) node[midway,anchor=south]{$c$};

    \draw[thick,dotted, teal] (0,-0.5) -- (3,-0.5);
    \draw[thick,dotted, teal] (1.5,2.05) -- (3,2.05);
    \draw[|-|, thick, teal] (3,-0.5) -- (3,2.05)
        node[midway, anchor=west]{$\barrier^\cis$};

    \draw[->, thick, black] (-2,-0.5) --++ (0,1) node[anchor=east]{energy};
    \draw[->, thick, black] (-2,-0.5) --++ (1,0) node[anchor=north]{$x$};

    \draw[green!50!black,densely dashed]
        (-1.5, 1.3) node[anchor=east]{$\Eosm(x)$} --++ (4,1.3) ;

    \draw[magenta!50!black,densely dashed]
        (-1.5, 2)  node[anchor=east]{$\Eep(x)$} --++ (4,-1);

    \draw[black,|-|] (-1.7,3) -- (2.7,3)
        node[midway, anchor=south]{Nanopore} node[midway, anchor=north]{$L$};
    \draw[black, ->] (3,3) --++ (1,0) node[anchor=south]{cis};
    \draw[black, ->] (-2,3) --++ (-1,0) node[anchor=south]{trans};

    \draw[densely dotted] (0,-0.5) --++ (0,-0.5) node[anchor=north]{$\xmin$};
    \draw[densely dotted] (1.55,2) --++ (0,-3) node[anchor=north]{$\xcis$};

\end{tikzpicture}
}{fig:single_barrier}{%
  Escape from a single barrier system.
}{%
  \textbf{Escape from a single barrier system.}
  Molecule (blue circle) trapped in a potential well defined by a nanopore and trying to overcome the energy
  barrier $\barrier^\cis$ by picking up the kinetic energy $\kinetic$ from its environment and then
  propagating over the barrier with an instantaneous transition rate $\transition$. Overlaid are the linear
  energy potentials from the electrophoretic force $\Eep$ and from the force of the electro-osmotic flow
  $\Eosm$.
}{t}

In order to overcome the barrier of \cref{fig:single_barrier}, the molecule has to pick up thermal energy
randomly. The probability of finding a molecule with a kinetic energy of $\kinetic$ in an environment at
thermal equilibrium is given by the usual (normalized) Boltzmann distribution
\begin{align*}
    f(\kinetic) = \frac{1}{\kbt} e^{-\frac{\kinetic}{\kbt}}
    \text{ .}
\end{align*}
Then the scattering rate $\rate$ into some final state can be expressed as
\begin{align}\label{eq:rate}
    \rate = \int_0^\infty d\kinetic\; Z \;c(\kinetic)\;f(\kinetic)
    \text{ ,}
\end{align}
where $Z$ is the density of states and $c$ is the transition rate from an initial state at energy $\kinetic$
into a final state. The density of states is assumed to be energy independent which means that the molecule
can only have a fixed number of internal states, irrespective of the kinetic energy it picks up. Note that the
transition rate $c$ could in principle depend on the internal states of the molecule but we neglected this as
well. We will view the escape rate of the molecule from the potential well as an instantaneous scattering
event that takes it over the barrier.

In order to derive an approximation to this rate, we will assume that the molecule cannot escape (\ie~tunnel)
as long as its kinetic energy is less than the required energy barrier that needs to be overcome. On the other
hand, if the molecule acquires kinetic energy equal to or larger than the barrier, the transition rate $c$ is
assumed to be constant. Thus we have
\begin{align*}
    c(\kinetic) =
        \begin{cases}
            0, & \text{if}\; \kinetic < \barrier^\cis \text{ ,}\\
            c_0, & \text{else} \text{ .}
        \end{cases}
\end{align*}
The escape rate over the barrier in \cref{eq:rate} therefore yields
\begin{align*}
    \rate ={}& Z c_0 \int_{\barrier^\cis}^\infty d\kinetic\;f(\kinetic)\\
    ={}& Z c_0\; e^{-\frac{\barrier^\cis}{\kbt}}\\
    =:& \;\rate^{\cis}_0\; e^{-\frac{\barrier^\cis}{\kbt}}
    \text{ .}
\end{align*}
And the dwell time is the inverse of the escape rate:
\begin{align*}
    \tau = \frac{1}{\rate}
    \text{ .}
\end{align*}
The barrier can be decomposed into steric, electrostatic, and external
contributions:
\begin{align}\label{eq:barrier}
    \barrier^\cis = \barrier^{\cis}_{\steric,0} + \barrier^\cis_\static +
    \barrier^\cis_\ext
    \text{ ,}
\end{align}
where the steric contribution $\barrier^{\cis}_{\steric,0}$ accounts for all interactions of the molecule
within the potential well that are not electrostatic in nature (\textit{e.g.}~size related effects). The
barrier contribution from the electrostatic interactions can be expressed as
\begin{align}\label{eq:barrier_static}
    \barrier^\cis_{\static} =
    \barrier^{\cis}_{\static,0} + \Ntag \ec \potbar_{\rm{tag}}^\cis
    \text{ ,}
\end{align}
where $\Ntag$ is the signed net number of charges on the molecule being trapped in the electrostatic potential
well. Since the molecule is large and charges may be located outside of the potential well, not all charges
modify the barrier. For our present purpose, we will refer to the number of charges present in the well as the tag
charge number, in reference to the term used in \cref{ch:trapping}. Thus, $\potbar_{\rm tag}^\cis$ is the
barrier height change due to a change in the tag charge number $\Ntag$. The electrostatic barrier
contributions that are independent of $\Ntag$ are absorbed into the constant $\barrier^{\cis}_{\static,0}$.

Moreover, $\barrier^\cis_\ext$ stems from the externally applied bias that results in an electrophoretic force
and a force due to the electro-osmotic flow and can therefore be split up as
\begin{align}\label{eq:barrier_ext}
    \barrier^\cis_\ext = \barrier^\cis_\ep + \barrier^\cis_\osm
    \text{ .}
\end{align}
We will assume that both of these external forces are constant throughout the nanopore and therefore the
associated energy potentials are linear in space as is illustrated in \cref{fig:single_barrier}. Thus we can
express the potential energies due to the external forces as
\begin{align*}
    \Eep(x)   ={}& -\forceep \;x + b_{\ep},\\
    \Eosm(x)  ={}& -\forceosm\; x + b_{\osm}
    \text{ ,}
\end{align*}
where $\forceep$ is the constant electrophoretic force and $\forceosm$ is the constant force due to the
electro-osmotic force on the molecule. Note the sign in front of the force since the force points down the
slope of the potential energy.

Using the energy potentials we can compute the effect of the external fields on the energy barrier as
\begin{align*}
    \Delta \Eep^\cis = \Eep(\xcis) - \Eep(\xmin)
    = - \forceep \dxcis
    \text{ ,}
\end{align*}
where $\xcis$ is the location of the barrier and $\xmin$ is the location of the minimum. The distance of the
barrier to the minimum is given by
\begin{align*}
    \dxcis = \left|\xcis - \xmin\right|
    \text{ .}
\end{align*}
The constant energy shift $b_{\ep}$ does not contribute and therefore can be chosen arbitrarily. The
electrophoretic force due to an applied external bias $\vbias$ can be expressed by Coulomb's law
\begin{align}\label{eq:forceep}
    \forceep = \ec \Nnet   \frac{\vbias}{L}
    \text{ ,}
\end{align}
where $\Nnet = \Ntag +\Nbody$ is the signed net number of charges on the molecule, where we defined the body
charge $\Nbody$ in reference to the terms in \cref{ch:trapping}. Moreover, $L$ is the length of the potential
drop from \cisi{} to \transi{} (cf.~\cref{fig:single_barrier}).

Likewise, we can compute the change in barrier height due to the osmotic component as
\begin{align*}
    \barrier^\cis_{\osm} = \Eosm(\xcis) - \Eosm(\xmin)
    = - \forceosm \Delta \xcis
    \text{ ,}
\end{align*}
where the force is linearly dependent on the potential as
\begin{align}\label{eq:osmoticforce}
    \forceosm = \ec \Neo \frac{\vbias}{L}
    \text{ .}
\end{align}
Note that for later convenience we expressed the osmotic force in complete analogy to the electrophoretic
force of \cref{eq:forceep}, where we defined the parameter $\Neo$ that we will refer to as the
\emph{equivalent osmotic charge number}. This number tells us what net charge a molecule would have to have in
order to experience the same force \textit{via} electrophoresis. Thus if $\Nnet = -\Neo$, the net external
force on the molecule vanishes.

Using \cref{eq:barrier,eq:barrier_static,eq:barrier_ext} we find for the total barrier height
\begin{align}\label{eq:cisbarrier}
    \barrier^\cis = \barrier^{\cis}_{\steric,0} + \barrier^{\cis}_{\static,0}
        + \Ntag \ec  \potbar_{\rm{tag}}^\cis
        - \left(\Nnet + \Neo\right) \frac{\Delta \xcis}{L}\ec \vbias
    \text{ .}
\end{align}
The escape rate (\ie~the inverse of the dwell time) is therefore given by
\begin{align}\label{eq:single_barrier_dwell_time}
\begin{split}
    \frac{1}{\tau(\Ntag, \vbias)} ={}& \rate (\Ntag, \vbias)\\
    ={}& \rate^{\cis}_0 \exp\left(-\frac{\barrier^\cis}{\kbt}\right)\\
    ={}& \rate^{\cis}_{\rm{eff}}\; \exp\bigg(
        - \frac{\Ntag \ec  \potbar_{\rm{tag}}^\cis
        - \left(\Nnet + \Neo\right) \frac{\Delta \xcis}{L}\ec \vbias}{\kbt}
    \bigg)
    \text{ ,}
\end{split}
\end{align}
where the constant steric and electrostatic terms $\barrier^{\cis}_{\steric,0}$ and
$\barrier^{\cis}_{\static,0}$, respectively, have been absorbed into the effective attempt rate
$\rate^{\cis}_{\rm{eff}}$.

We immediately see that in a single barrier model the dwell time has to monotonically increase or decrease
with the applied bias.  Data that exhibits a maximum as a function of $\vbias$, such as the dwell time data of
\cref{ch:trapping}, cannot be described by the single barrier model.

\onefiguresource{%
\begin{tikzpicture}
    \draw[fill=blue!20] (0,-0.3) circle (0.2);

    \draw plot [smooth, tension=0.5]  coordinates {%
        (-2.5,1)
        (-2,2)
        (-1.5,2.5)
        (-1,0.5)
        (0,-0.5)
        (1,0.5)
        (1.5,2)
        (2,1.5)
        (2.5,0.5)
    };

    \draw[->, red,thick] (0.1,0) -- (0.1,2.1) ;
    \draw[->, red,thick] (-0.1,0) -- (-0.1,2.6)
        node[midway,anchor=east]{$\kinetic$};
    \draw[->, orange,thick] (0.1,2.1) -- (1.5,2.1)
        node[midway,anchor=south]{$c^\cis$};
    \draw[->, orange,thick] (-0.1,2.6) -- (-1.5,2.6)
        node[midway,anchor=south]{$c^\trans$};

    \draw[thick,dotted, teal] (0,-0.5) -- (3,-0.5);
    \draw[thick,dotted, teal] (1.5,2.05) -- (3,2.05);
    \draw[|-|, thick, teal] (3,-0.5) -- (3,2.05)
        node[midway, anchor=west]{$\barrier^\cis$};

    \draw[thick,dotted, teal] (0,-0.5) -- (-3,-0.5);
    \draw[thick,dotted, teal] (-1.5,2.55) -- (-3,2.55);
    \draw[|-|, thick, teal] (-3,-0.5) -- (-3,2.55)
        node[midway, anchor=east]{$\barrier^\trans$};

    \draw[black,|-|] (-2.5,3.5) -- (2.7,3.5)
        node[midway, anchor=south]{Nanopore} node[midway, anchor=north]{$L$};
    \draw[black, ->] (3,3.5) --++ (1,0) node[anchor=south]{cis};
    \draw[black, ->] (-2.7,3.5) --++ (-1,0) node[anchor=south]{trans};

    \draw[densely dotted] (0,-0.5) --++ (0,-0.5) node[anchor=north]{$\xmin$};
    \draw[densely dotted] (1.55,2) --++ (0,-3) node[anchor=north]{$\xcis$};
    \draw[densely dotted] (-1.6,2.5) --++ (0,-3.5) node[anchor=north]{$\xtrans$};

    \draw[->, thick, black] (-4,-1) --++ (0,1) node[anchor=east]{energy};
    \draw[->, thick, black] (-4,-1) --++ (1,0) node[anchor=north]{$x$};
\end{tikzpicture}
}{fig:second_barrier}{%
  Escape from a double barrier system.
  }{%
  \textbf{Escape from a double barrier system.}
  Molecule (blue ball) trapped in between two energy barriers $\barrier^\cis$ and $\barrier^\trans$. By
  picking up kinetic energy $\kinetic$ from its environment it can transition through either barrier towards
  \cisi{} or \transi{}.
}{b}

\subsection{Escape from a double barrier system}
\label{sec:trapping_appendix:double_barrier}

Now let us assume that there is a second barrier as illustrated in \cref{fig:second_barrier} so that the
molecule can escape towards the \transi{} side with a finite probability. We can define the barrier to the
\transi{} side in complete analogy to the \cisi{} barrier of \cref{eq:cisbarrier} as
\begin{align*}
    \barrier^\trans = \barrier^{\trans}_{\steric,0}
        + \barrier^{\trans}_{\static,0}
        + \Ntag \ec  \potbar_{\rm{tag}}^\trans
        + \left(\Nnet + \Neo\right) \frac{\dxtrans}{L}\ec \vbias
        \text{ .}
\end{align*}
Note, however, the opposite sign in front of the external barrier contribution. This indicated that
the external force points in the same direction as the \transi{} exit, rather than the opposite.

Then the dwell time is given by the inverse of the sum of rates over either barrier
\begin{align}\label{eq:double_barrier_rate}
\begin{split}
    \frac{1}{\tau\left(\Ntag,\vbias\right)} ={}& \rate (\Ntag,\vbias)\\
    ={}&
     \rate_0\exp\left(-\frac{\barrier^\cis}{\kbt}\right)
     + \rate_0\exp\left(-\frac{\barrier^\trans}{\kbt}\right)\\
    ={}& \rate^{\cis}_{\rm{eff}}\; \exp\bigg(
        -\frac{\Ntag \ec  \potbar_{\rm{tag}}^\cis
        - \left(\Nnet + \Neo\right) \frac{\Delta \xcis}{L}\ec \vbias}{\kbt}
    \bigg)\\
    &+ \rate^{\trans}_{\rm{eff}}\; \exp\bigg(
        -\frac{\Ntag \ec  \potbar_{\rm{tag}}^\trans
        + \left(\Nnet + \Neo\right) \frac{\dxtrans}{L}\ec \vbias}{\kbt}
    \bigg)
    \text{ ,}
\end{split}
\end{align}
Again, the parameters $\rate^{\cis/\trans}_{\rm{eff}}$ are effective attempt rates related to the dwell time at
zero tag charge and vanishing applied bias.

The translocation probability is given by the ratio of the \transi{} escape rate to the total escape rate:
\begin{align*}
    \probability ={}& \frac{\rate_0\exp\left(-\frac{\barrier^\trans}{\kbt}\right)}{\rate (\Ntag, \vbias)}
    \text{ .}
\end{align*}
The intrinsic, size-related probability for translocation without any trapping of the tag or applied fields
can be directly computed as the ratio $\rate^{\trans}_{\rm{eff}}/(\rate^{\cis}_{\rm{eff}} +
\rate^{\trans}_{\rm eff})$, which is a small number in the case of a large molecule trying to pass through a
nanopore with a narrow \transi{} constriction as shown in the \cref{ch:trapping}.

\section{Experimentally observed behavior of tagged {DHFR}}

\subsection{Multistate residences of {DHFR} inside {ClyA}}
\begin{table}[b]
  \centering

  \begin{threeparttable}
    \centering
    \captionsetup{width=10cm}
    \caption{$\iresp$ values of the different {DHFR} variants.}
    \label{tab:dhfr_iresp}
    \renewcommand{\arraystretch}{1.0}
    \footnotesize
    \sisetup{table-format=2.4}
    \begin{tabularx}{10cm}{Xllll}
      \toprule
                    &  & \multicolumn{3}{c}{$\iresp$\tnote{a}} \\
                                            \cmidrule{3-5}
      DHFR variant  & $\vbias$ [\si{\mV}] & L1  & L2 & L3   \\
      \midrule
      \DHFR{4}{S}   & \num{-80} & \num{67.4\pm2.1} & \num{46.4\pm0.2} & --- \\
      \DHFR{4}{I}   & \num{-80} & \num{71.3\pm0.6} & --- & --- \\
      \DHFR{4}{C}   & \num{-80} & \num{72.8\pm1.0} & --- & --- \\
      \DHFR{4}{O1}  & \num{-80} & \num{74.1\pm0.4} & \num{57.0\pm1.0}
                                                            & \num{38.7\pm1.1} \\
      \DHFR{4}{O2}  & \num{-80} & \num{74.9\pm0.7} & \num{58.0\pm0.9} & --- \\
      \DHFR{5}{O2}  & \num{-60} & \num{74.0\pm0.3} & \num{57.7\pm0.1} & --- \\
      \DHFR{6}{O2}  & \num{-60} & \num{74.2\pm0.1} & \num{57.7\pm0.1} & --- \\
      \DHFR{7}{O2}  & \num{-60} & \num{73.4\pm0.3} & \num{56.4\pm0.8}
                                                            & \num{39.3\pm2.5} \\
      \DHFR{8}{O2}  & \num{-60} & \num{74.7\pm1.0} & \num{58.4\pm1.1}
                                                            & \num{41.4\pm1.5} \\
      \DHFR{9}{O2}  & \num{-60} & \num{74.0\pm0.1} & \num{57.0\pm0.4}
                                                            & \num{38.9\pm1.8} \\
      \bottomrule
    \end{tabularx}
    \begin{tablenotes}
      \item[a] $\iresp$ values for each \gls{dhfr} variant are based on at least 50 individual \gls{dhfr}
      blockades collected from at least three different single nanopore experiments. Errors are standard
      deviations from the mean.
    \end{tablenotes}
  \end{threeparttable}
\end{table}
\begin{figure*}[p]
  \centering
  \begin{subfigure}[t]{5.75cm}
		\centering
    \caption{}\vspace{-2mm}\label{fig:trapping_traces_body}
    \includegraphics[scale=1]{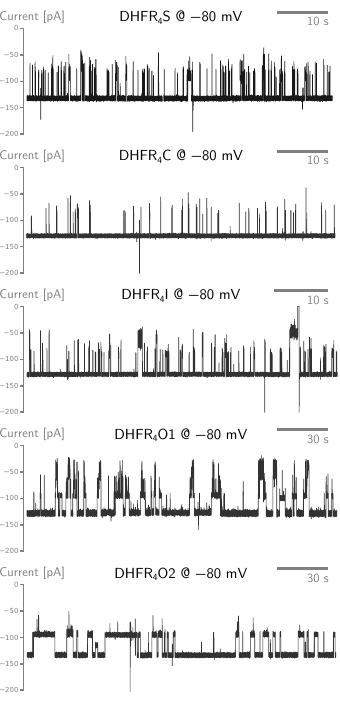}
  \end{subfigure}
  \begin{subfigure}[t]{5.75cm}
		\centering
    \caption{}\vspace{-2mm}\label{fig:trapping_traces_tag}
    \includegraphics[scale=1]{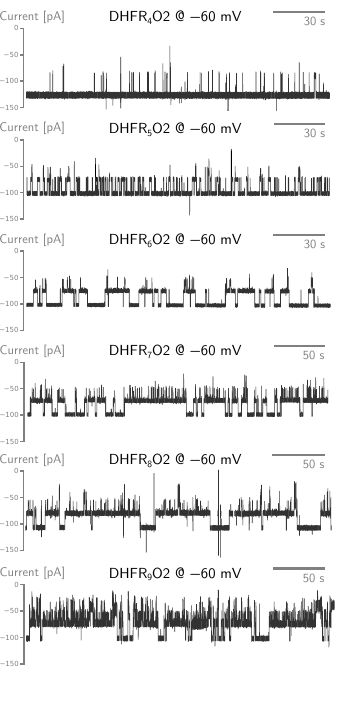}
  \end{subfigure}
  
  \caption[Current blockades of {DHFR} molecules with a fusion tag.]{%
    \textbf{Current blockades of {DHFR} molecules with a fusion tag.}
    (\subref{fig:trapping_traces_body})
    Typical current traces of the various \DHFR{4}{S} body charge variants (\DHFR{4}{C}, \DHFR{4}{I},
    \DHFR{4}{O1} and \DHFR{4}{O2}) at \SI{-80}{\mV} applied bias. The C, I, and O1 mutants have the same net
    charge, but the location of their charges differs, which results in significantly different dwell times.
    (\subref{fig:trapping_traces_tag})
    Typical current recordings for the tag charge variants of \DHFR{4}{O2} (\DHFR{$\Ntag$}{O2}) at
    \SI{-60}{\mV}, revealing the increased dwell time with increasing positive tag charges. Note that most
    \gls{dhfr} variants showed complex multi-level blockades. Therefore, average dwell times ($\dwelltime$)
    were used to guarantee a fair comparison between the different mutants. All current traces were collected
    in \SI{150}{\mM} \ce{NaCl}, \SI{15}{\mM} \ce{Tris-HCl} \pH{7.5} at \SI{28}{\celsius} after adding
    \SI{\approx50}{\nM} of \gls{dhfr} to the \cisi{} side reservoir of a single \gls{clya-as} nanopore.
    Signals were sampled at \SI{10}{\kilo\hertz} and filtered with a \SI{2}{\kilo\hertz} cutoff Bessel
    low-pass filter.
  }\label{fig:trapping_traces}
\end{figure*}

\begin{figure*}[t]
  \centering
  \begin{subfigure}[t]{5.5cm}
    \centering
    \caption{}\vspace{-2mm}\label{fig:trapping_blockades_I}
    \includegraphics[scale=1]{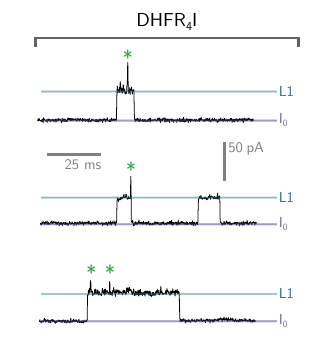}
  \end{subfigure}
  \begin{subfigure}[t]{5.5cm}
    \centering
    \caption{}\vspace{-2mm}\label{fig:trapping_blockades_C}
    \includegraphics[scale=1]{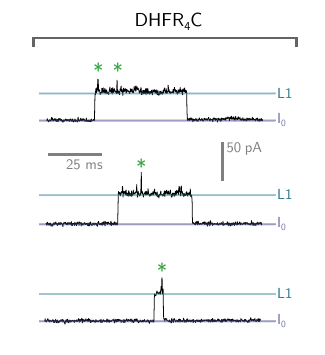}
  \end{subfigure}

  \caption[\DHFR{4}{I} and \DHFR{4}{C} blockades in {ClyA-AS} at \SI{-80}{\mV}.]{%
    \textbf{\DHFR{4}{I} and \DHFR{4}{C} blockades in {ClyA-AS} at \SI{-80}{\mV}.}
    Three individual blockades of (\subref{fig:trapping_blockades_I}) \DHFR{4}{I} and
    (\subref{fig:trapping_blockades_C}) \DHFR{4}{C} in \gls{clya-as} at \SI{-80}{\mV} showing the L1 current
    level (purple line) and short dwelling on a lower current level (green asterisks). The latter is too short
    to be properly sampled at this potential (transitions to this additional current level are observed by
    short, unresolved spikes). The teal line represents the open-pore current $\iopen$. The current traces
    were collected in \SI{150}{\mM} \ce{NaCl}, \SI{15}{\mM} \ce{Tris-HCl} \pH{7.5} at \SI{28}{\celsius}, by
    applying a Bessel low-pass filter with a \SI{2}{\kilo\hertz} cutoff and sampled at \SI{10}{\kilo\hertz}.
  }\label{fig:trapping_blockades}
\end{figure*}

At \SI{-80}{\mV}, in \SI{150}{\mM} \ce{NaCl} \SI{15}{\mM} \ce{Tris-HCl} \pH{7.5}, the current blockades
induced by the \DHFR{4}{S} variants (\cref{fig:trapping_traces,tab:dhfr_iresp}) showed a main current level
($\ilevel{1}$) with relative residual current values ($\iresp$), expressed as a percentage of the open-pore
current ($\iopen$), of \SI{67.4 \pm 2.1}{\percent}, \SI{71.3 \pm 0.6}{\percent}, \SI{72.8 \pm 1.0}{\percent},
\SI{74.1 \pm 0.4}{\percent} and \SI{74.9 \pm 0.7}{\percent} for \DHFR{4}{S}, \DHFR{4}{I}, \DHFR{4}{C},
\DHFR{4}{O1}, \DHFR{4}{O2}, respectively. As observed for other proteins~\cite{Soskine-2012,Soskine-2013,
Soskine-Biesemans-2015,Biesemans-2015}, \DHFR{4}{S}, \DHFR{4}{O1} and \DHFR{4}{O2} also displayed a second
current level ($\ilevel{2}$) with $\iresp$ values of \SI{46.4 \pm 0.2}{\percent} \SI{57.0 \pm 1.0}{\percent}
\SI{58.0 \pm 0.9}{\percent}, respectively (\cref{tab:dhfr_iresp}). \DHFR{4}{I} and \DHFR{4}{C} also
occasionally dwelled on a second current level, however, the dwell time at this level was too short to allow
reliable determination of the $\iresp$ (\cref{fig:trapping_blockades}). \DHFR{4}{O1} often visited a third
current level ($\ilevel{3}$) with $\iresp$ of \SI{38.7\pm 1.1}{\percent}. It is likely that the multiple
current levels observed for the different \gls{dhfr} variants reflect the residence of the protein in
different physical locations inside the nanopore~\cite{Soskine-2012}.

The collected time series data may contain capture events that transition to several meta states before the
molecule escapes the nanopore. For our data analysis, we count the dwell time of an event as
the elapsed time from the capture up until the escape, irrespective of how many transitions to meta states
have been observed. Thus, the complex kinetics of the molecule in the captured state are simply summed over in
the resulting dwell time event histograms (\cref{eq:compound_distri}). For the scope of this work, we make the
assumptions that (1) there is only one dominant capture process involved and that (2) the molecule's escape
from the nanopore is dominated by a single rate, such that the dwell time distribution can then be
approximated by a single exponential. Since a maximum likelihood fit of a single or multi-exponential
distribution function is well represented by the arithmetic mean, we use the arithmetic mean directly as the
expectation value of the entire distribution (\cref{eq:tau_arithmetic_mean}).

\begin{table}[b]
  \centering
  \begin{threeparttable}
    \centering
    \captionsetup{width=10cm}
    \caption[NADPH binding/unbinding kinetics to trapped {DHFR} variants.]{%
      \textbf{NADPH binding/unbinding kinetics to trapped {DHFR} variants.}\tnote{a}}%
    \label{tab:nadph_rates}
    \footnotesize
    \renewcommand{\arraystretch}{1.2}
    \sisetup{table-format=2.4}
    \begin{tabularx}{10cm}{Xlll}
      \toprule
      DHFR variant
        & $\rate_{\rm{on}}$ [\si{\per\second\per\mM}]\tnote{b}
        & $\rate_{\rm{off}}$ [\si{\per\second}]
        & Amplitude [\si{\pA}]  \\
      \midrule
      \DHFR{4}{O2}  & \num{1684\pm225} & \num{60.2\pm23.2} & \num{-1.71\pm0.13} \\
      \DHFR{6}{O2}  & \num{1390\pm397} & \num{55.9\pm5.1} & \num{-1.47\pm0.09} \\
      \DHFR{7}{O2}  & \num{2032\pm578} & \num{71.2\pm20.4} & \num{-1.46\pm0.07} \\
      \bottomrule
    \end{tabularx}
    \begin{tablenotes}
      \item[a] At \SI{-60}{\mV} applied potential.
      \item[b] $\rate_{\rm{on}}$, $\rate_{\rm{off}}$ and amplitude values for each \gls{dhfr} variant are
      based on at least \num{300} \gls{nadph} binding events on more than 15 individual \gls{dhfr} blockades
      collected from three different single nanopore experiments.
    \end{tablenotes}
  \end{threeparttable}
\end{table}

\subsection{Analysis of {NADPH} binding to {DHFR} variants}

\begin{figure*}[b]
  \centering
  \includegraphics[scale=1]{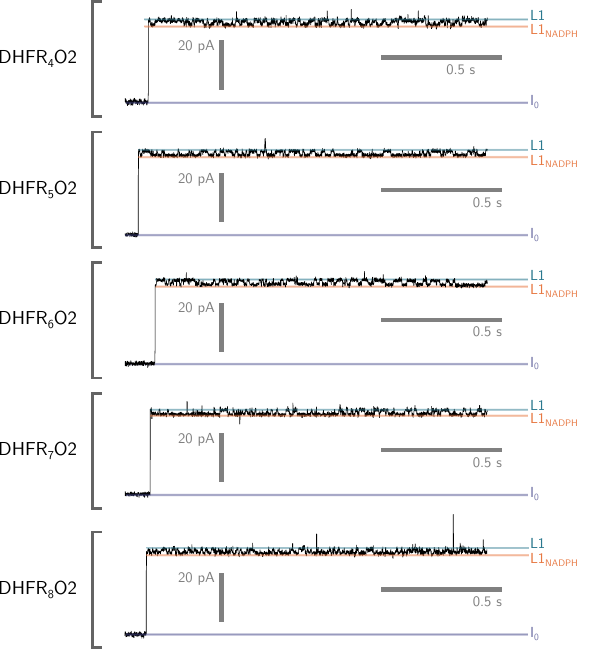}
  \caption[{NADPH} binding to nanopore-confined \DHFR{$\Ntag$}{O2}.]{%
    \textbf{{NADPH} binding to nanopore-confined \DHFR{$\Ntag$}{O2}.}
    Typical current traces of single \DHFR{$\Ntag$}{O2} (\SI{\approx 50}{\nM}, \cisi{}) molecules inside
    \gls{clya-as} at \SI{-80}{\mV} applied potential after addition of \SI{40}{\uM} \gls{nadph} to the trans
    compartment. \gls{nadph} binding to confined \DHFR{$\Ntag$}{O2} is reflected by current enhancements from
    the unbound L1 (purple line) to the \gls{nadph}-bound L1\textsubscript{NADPH} (orange line) current
    levels. The open-pore current $\iopen$ is represented by the teal line. All current traces were collected
    in \SI{150}{\mM} \ce{NaCl}, \SI{15}{\mM} \ce{Tris-HCl} \pH{7.5} at \SI{28}{\celsius}, by applying a Bessel
    low-pass filter with a \SI{2}{\kilo\hertz} cutoff and sampled at \SI{10}{\kilo\hertz}. An additional
    Bessel 8-pole filter with \SI{500}{\hertz} cutoff was digitally applied to the current traces.
    }\label{fig:trapping_nadph_o2_traces}
\end{figure*}

Typical current traces of \gls{nadph} binding to trapped \gls{dhfr} molecules are shown in
\cref{fig:trapping_nadph_o2_traces}. Analysis of the on- ($\rate_{\rm{on}}$) and off-rates
($\rate_{\rm{off}}$), and the event amplitudes of \gls{nadph} binding to \DHFR{4}{O2}, \DHFR{6}{O2} and
\DHFR{7}{O2} entrapped within the nanopore are all similar (\cref{tab:nadph_rates}), suggesting that the
proteins remain folded and active inside \gls{clya}.

\clearpage

\section{Modeling of body charge variations}
\label{sec:trapping_appendix:body_charge_variations}

\subsection{Not all charges on the body are equivalent}

\begin{figure*}[p]
  \centering
  \begin{subfigure}[t]{5.5cm}
		\centering
		\caption{}\vspace{-3mm}\label{fig:trapping_model_comparison_body_complex}
    \includegraphics[scale=1]{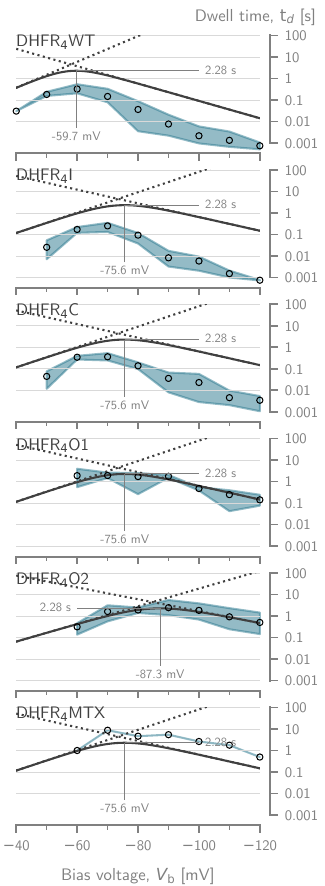}
  \end{subfigure}
  \begin{minipage}[t]{6cm}
    \begin{subfigure}[t]{5.5cm}
      \centering
      \caption{}\vspace{-3mm}\label{fig:trapping_model_comparison_o1_simple}
      \includegraphics[scale=1]{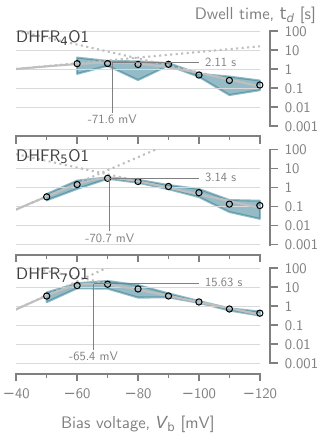}
    \end{subfigure}
    \\
    \begin{subfigure}[t]{5.5cm}
      \centering
      \caption{}\vspace{-3mm}\label{fig:trapping_model_comparison_o1_complex}
      \includegraphics[scale=1]{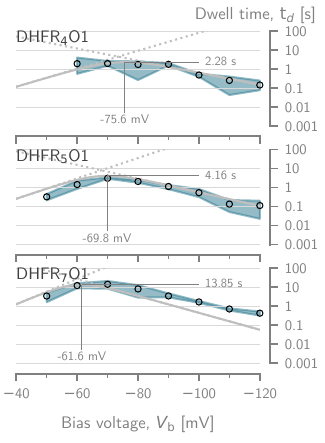}
    \end{subfigure}
  \end{minipage}

  \caption[Effect of body and tag charge on the dwell time of {DHFR}.]
  {%
    \textbf{Effect of body and tag charge on the dwell time of {DHFR}.}
    (\subref{fig:trapping_model_comparison_body_complex}) Predicted dwell times of the body charge variations
    \DHFR{4}{S, -I, -C, -O1 and -O2} by \cref{eq:double_barrier} and using the parameters in
    \cref{tab:fitting_params_complex}. Clearly, the location of the body charge plays an important, but
    uncaptured, role in determining the dwell time of \gls{dhfr}.
    (\subref{fig:trapping_model_comparison_o1_simple})
    and
    (\subref{fig:trapping_model_comparison_o1_complex})
    are the voltage dependencies of the mean dwell time ($\dwelltime$) for several tag charge variants of
    \DHFR{$\Ntag$}{O1}, fitted with the simple barrier model of \cref{eq:double_barrier_simple} and the full
    double barrier model of \cref{eq:double_barrier_complex}, respectively. The annotated threshold voltages
    for (\subref{fig:trapping_model_comparison_o1_simple}) and
    (\subref{fig:trapping_model_comparison_o1_complex}) were computed by respectively
    \cref{eq:threshold_voltage_simple} and \cref{eq:threshold_voltage_complex}. Solid lines represent the
    double barrier dwell time while the dotted lines show the dwell times due the \cisi{} (low to high) and
    \transi{} (high to low) barriers. Fitting parameters can be found in \cref{tab:fitting_parameters_simple}.
    }\label{fig:trapping_model_comparison}
\end{figure*}

The double barrier model of \cref{eq:double_barrier_complex} in its current form cannot adequately account for
the body charge variations of \gls{dhfr}\@. Consider for example the body charge variants \DHFR{4}{I},
\DHFR{4}{C}, and \DHFR{4}{O1} which share the same number of body charges, so that our model would predict the
same dwell time (\cref{fig:trapping_model_comparison_body_complex}). However, the body charges of these
variations are at different locations and as a consequence, they exhibit different dwell times as can be seen
in \cref{fig:trapping_dwell_times_body_data} of \cref{ch:trapping}. The reason for this is that the model
describes the trapping mechanism as a function of the tag charge number $\Ntag$ only, while body charge-related
barrier modifications are absorbed into the constant terms $\barrier^{\cis/\trans}_{\static,0}$ of
\cref{eq:static-barrier}.

If we wanted to modify \cref{eq:static-barrier} such that it can account for changes in the trapping behavior
we would need to account for the location of the body charges as is evident from the dwell time data sets of
\DHFR{4}{I}, \DHFR{4}{C}, and \DHFR{4}{O1}. Such a model would drastically increase in complexity and it is
not clear whether it can still be formulated analytically in a reasonable way. In that case, it may in fact be
more workable to use more refined APBS simulations or even a full molecular dynamics simulation to compute
parameters for the trapping.

\subsection{The distance from the tag matters}

Despite this limitation, the double barrier model is essentially a representation of how the tag is anchored
to the electrostatic minimum in the pore, and hence we can deduce the way the body charge variations will
impact trapping. For example, if we modify the charge on the far end of the body with respect to the tag
location (\textit{i.e.},~far away from the tag). We hypothesize that the barriers responsible for trapping the
tag will not be changed meaningfully---aside from the electrophoretic force which is included in the double
barrier model (\cref{eq:double_barrier_complex}). Indeed, taking the model parameters of
\cref{tab:fitting_params_complex} obtained from the fit to the \DHFR{$\Ntag$}{O2} data set and predicting the
dwell time data of \DHFR{$\Ntag$}{O1}, we find good agreement as shown in
\cref{fig:trapping_model_comparison}.

On the other hand, charges on the body that are close to the tag directly impact the electrostatic energy
landscape and will modify the barrier heights. Such a body charge close to the tag location can be seen as
effectively modifying the net charge on the tag. In that case, we expect that our model fitted to the
\DHFR{$\Ntag$}{O2} data fails to predict the dwell times as can be seen in
\cref{fig:trapping_model_comparison_body_complex}.

\begin{threeparttable}[!b]
  \centering

  \captionsetup{width=11cm}
  \caption[Fitting parameters for the simple double barrier model.]{%
          \textbf{Fitting parameters for the simple double barrier model.}\tnote{a}}
  \label{tab:fitting_parameters_simple}
  \footnotesize
  \renewcommand{\arraystretch}{1.2}

  \begin{tabularx}{11cm}{
    >{\raggedright\hsize=2.5cm}X
    >{\hsize=1cm}l >{\hsize=1cm}l >{\hsize=1cm}l >{\hsize=1cm}l}
    \toprule
                  & \multicolumn{2}{c}{\cisi{} barrier\tnote{b}}
                  & \multicolumn{2}{c}{\transi{} barrier\tnote{b}} \\
    \cmidrule(r){2-3}\cmidrule(l){4-5}
    DHFR variant  & $\ln k^\cis/V_T$ & $\alpha^\cis/V_T$
                  & $\ln k^\trans/V_T$ & $\alpha^\trans/V_T$ \\
    \midrule
    \DHFR{4}{S}   & \num{11.91\pm3.14} & \num{5.38\pm1.83}
                  & \num{-5.45\pm0.86} & \num{2.82\pm0.24} \\
    \DHFR{4}{I}   & \num{15.15\pm4.88} & \num{5.88\pm2.34}
                  & \num{-6.83\pm1.30} & \num{3.08\pm0.34} \\
    \DHFR{4}{C}   & \num{15.87\pm3.52} & \num{6.56\pm1.72}
                  & \num{-5.73\pm0.79} & \num{2.50\pm0.21} \\
    \DHFR{4}{O1}  & \num{1.37\pm3.29}  & \num{0.88\pm1.37}
                  & \num{-8.03\pm2.64} & \num{2.16\pm0.60} \\
    \DHFR{4}{O2}  & \num{9.67\pm2.53}  & \num{3.70\pm1.01}
                  & \num{-5.54\pm1.43} & \num{1.31\pm0.34} \\
    \bottomrule
  \end{tabularx}
  \begin{tablenotes}
    \item[a] Fitting coefficients for \cref{eq:double_barrier_simple} of the \cref{ch:trapping}.
    \item[b] Errors represent one sigma confidence intervals.
  \end{tablenotes}
\end{threeparttable}

\section{Extended materials and methods}

\subsection{Material suppliers}

Unless otherwise specified all chemicals were bought from Sigma-Aldrich (Overijse, Belgium). \Gls{dna} was
purchased from Integrated DNA Technologies (IDT, Leuven, Belgium), enzymes from Fermentas (Merelbeke, Belgium)
and lipids from Avanti Polar Lipids (Alabaster, USA).

\subsection{Cloning of all {DHFR} variants}
\label{sec:trapping_appendix:dhfr_cloning}

\subsubsection{Cloning of DHFR$_{4}$S}

The \DHFR{4}{S} \gls{dna} construct was built from the \DHFRt\ construct~\cite{Soskine-Biesemans-2015} by
inserting an additional alanine residue at position 175 (located in the fusion tag). \DHFRt\ contains two
mutations with respect to wild type \textit{E. coli} \gls{dhfr} (C85A and C152S) and has a C-terminal fusion
tag which possesses five net positive charges and ends with a Strep-tag. To construct \DHFR{4}{S}, the \DHFRt\
circular \gls{dna} template was amplified using the \code{175Ala frwd} (forward) and \code{T7 terminator}
(reverse) primers (\cref{tab:primers}) in the following \gls{pcr} reaction: \SI{\approx200}{\nano\gram} of
template plasmid and \SI{\approx16}{\uM} of forward and reverse primers were mixed in \SI{0.3}{\milli\litre}
final volume of \gls{pcr} mix, which contained \SI{150}{\micro\litre} RED Taq Ready Mix (Sigma-Aldrich). We
performed 34 \gls{pcr} cycles following a pre-incubation step at \SI{98}{\celsius} for \SI{30}{\second}cycling
protocol: denaturation at \SI{98}{\celsius} for \SI{10}{\second}, annealing at \SI{52}{\celsius} for
\SI{30}{\second}, extension at \SI{72}{\celsius} for \SI{60}{\second}; and a final elongation step at
\SI{72}{\celsius} for \SI{10}{\minute}. The resulting \gls{pcr} product was clean-upped using the QIAquick
\gls{pcr} purification kit (Qiagen) and further gel purified using the QIAquick Gel Extraction Kit (Qiagen)
before it was cloned into a pT7 expression plasmid (pT7-SC1)~\cite{Miles-2001} by the MEGAWHOP
procedure~\cite{Miyazaki-2011}: \SI{\approx400}{\nano\gram} of the purified \gls{pcr} product was mixed with
\SI{\approx 200}{\nano\gram} of the \DHFRt\ \gls{dna} template and amplification was carried out with Phire
Hot Start II \gls{dna} polymerase (Finnzymes) in \SI{50}{\micro\litre} final volume (pre-incubation at
\SI{98}{\celsius} for \SI{30}{\second}; then cycling: denaturation at \SI{98}{\celsius} for \SI{5}{\second},
extension at \SI{72}{\celsius} for \SI{90}{\second}, for 30 cycles; followed by a final extension for
\SI{10}{\minute} at \SI{72}{\celsius}). The template \gls{dna} was eliminated by incubation with DpnI (1~FDU)
for \SI{1}{\hour} at \SI{37}{\celsius} and the enzyme was inactivated by incubation at \SI{65}{\celsius} for
\SI{5}{\minute}. Finally, \SI{0.5}{\micro\litre} of the resulting mixture was transformed into
\SI{50}{\micro\litre} of E.~cloni\textsuperscript{\textregistered} 10G electrocompetent cells (Lucigen) by
electroporation. The transformed bacteria were grown overnight at \SI{37}{\celsius} on LB agar plates
supplemented with \SI{100} {\micro\gram\per\milli\litre} ampicillin. The identity of the clones was confirmed
by sequencing. \gls{dna} and protein sequences of \DHFR{4}{S} are listed below.

\begin{footnotesize}
\begin{minipage}[t]{9cm}
\begin{verbatim}
>DHFR4S (DNA)
ATGGCTTCGGCTATGATTTCTCTGATTGCGGCACTGGCTGTCGATCGTGTTATTGGTATG
GAAAACGCTATGCCGTGGAATCTGCCGGCTGATCTGGCGTGGTTTAAACGTAACACTCTG
GACAAGCCGGTCATTATGGGCCGCCATACGTGGGAAAGCATCGGTCGTCCGCTGCCGGGT
CGCAAAAATATTATCCTGAGCAGCCAGCCGGGCACCGATGACCGTGTGACGTGGGTTAAG
AGCGTCGATGAAGCAATTGCGGCGGCAGGCGACGTGCCGGAAATTATGGTTATCGGCGGT
GGCCGCGTTTATGAACAGTTCCTGCCGAAAGCCCAAAAGCTGTACCTGACCCATATCGAT
GCAGAAGTCGAAGGTGATACGCACTTTCCGGACTATGAACCGGATGACTGGGAAAGTGTG
TTCTCCGAATTTCACGACGCCGACGCTCAGAACAGCCACTCATACTCATTCGAAATCCTG
GAACGCCGTGGCAGCAGTACTCGAGCGAAAAAGAAGATTGCGgccGCCCTAAAACAGGGC
AGCGCGTGGAGCCATCCGCAGTTTGAAAAATGATAA
\end{verbatim}
\end{minipage}
\begin{minipage}[t]{3cm}
\begin{verbatim}
>DHFR4S (protein)
MASAMISLIAALAVDRVIGM
ENAMPWNLPADLAWFKRNTL
DKPVIMGRHTWESIGRPLPG
RKNIILSSQPGTDDRVTWVK
SVDEAIAAAGDVPEIMVIGG
GRVYEQFLPKAQKLYLTHID
AEVEGDTHFPDYEPDDWESV
FSEFHDADAQNSHSYSFEIL
ERRGSSTRAKKKIAAALKQG
SAWSHPQFEK**
\end{verbatim}
\end{minipage}
\end{footnotesize}

\subsubsection{Construction of all other variants}

The positions for introduction of negatively charged glutamate residues into \DHFR{4}{S} were chosen after
multiple sequence alignment of \textit{E. coli} \gls{dhfr} (\pdbid{1RH3}, BLAST, 250 results) and
identification of the residues that were located on the opposite end of the molecule than the 4+tag, and which
during evolution had already converted to glutamate in some sequences. In all \DHFR{4}{S} variants described
in this work two native residues of \DHFR{4}{S} were mutated, resulting in \gls{dhfr} constructs with two
(\DHFR{4}{C}; A82E/A83E), (\DHFR{4}{I}; V88E/P89E), (\DHFR{4}{O1}; T68E/R71Q) or three (\DHFR{4}{O2};
T68E/R71E) extra negative charges. To construct the \DHFR{4}{C}, \DHFR{4}{I} and \DHFR{4}{O2} mutants, the
\DHFR{4}{S} gene was amplified using the \code{C-frwd}, \code{I-frwd} or \code{O2-frwd} primer, respectively,
and the \code{T7 terminator} primer. The \gls{pcr} conditions and subsequent purification and cloning steps
were as described above (\cref{sec:trapping_appendix:dhfr_cloning}). The \DHFR{4}{O1} mutant was constructed
starting from the \DHFR{4}{O2} circular \gls{dna} template as described above, using \code{O1-frwd} and
\code{T7 terminator} primers in the first \gls{pcr} amplification step. Following the same strategy,
\DHFR{5}{O1/O2} mutants were constructed using the corresponding \DHFR{4}{O1/O2} circular \gls{dna} template,
\code{6+ frwd} primer and \code{T7 terminator} primer. \DHFR{6}{O1/O2} and \DHFR{7}{O1/O2} mutants were
constructed using the corresponding \DHFR{5}{O1/O2} circular \gls{dna} template, \code{6+ frwd} or \code{7+
frwd} primer, respectively, and \code{T7 terminator} primer. \DHFR{8}{O1/O2} and \DHFR{9}{O1/O2} mutants were
constructed using the corresponding \DHFR{7}{O1/O2} circular \gls{dna} template, \code{8+ frwd} or \code{9+
frwd} primer, respectively, and \code{T7 terminator} primer. All primer sequences are shown in
\cref{tab:primers}.

\begin{table*}[!htb]
  \centering

  \captionsetup{width=11cm}
  \caption{Mutagenesis {DNA} primer sequences.}
  \label{tab:primers}
  \renewcommand{\arraystretch}{1.5}  
  \footnotesize
  \begin{tabularx}{11cm}{l X}
    \toprule
    Primer name  & Primer sequence   \\
    \midrule
    \code{175Ala frwd}
      & GGCAGCAGTACTCGAGCGAAAAAGAAGATTG CGgccGCCCTAAAACAGGGCAGCGCGTGG \\
    \code{T7 terminator}
      & GCTAGTTATTGCTCAGCGG \\
    \code{O2-frwd}
      & CCTGAGCAGCCAGCCGGGCGAAGATGACGA AGTGACGTGGGTTAAGAGCGTCG \\
    \code{I-frwd}
      & GAGCGTCGATGAAGCAATTGAAGAAGCAGG CGACGTGCCGGAAATTATGGTTATCGGCGG \\
    \code{C-frwd}
      & GCAATTGCGGCGGCAGGCGACGAAGAGGAA ATTATGGTTATCGGCGGTGGCCGCG \\
    \code{O1-frwd}
      & CCTGAGCAGCCAGCCGGGCGAAGATGACCA GGTGACGTGGGTTAAGAGCGTCG \\
    \code{5+ frwd}
      & CTCGAGCGAAAAAGAAGATTGCGAAAGCCC TAAAACAGGGCAGCGCGTGGAGCCATCCGC \\
    \code{6+ frwd}
      & CGTGGCAGCAGTACTCGAGCGAAAAAGAAG ATTAAGAAAGCCCTAAAACAGGGCAGCGCG \\
    \code{7+ frwd}
      & CGTGGCAGCAGTACTCGAGCGAAAAAGAAG ATTAAGAAAAAGCTAAAACAGGGCAGCGCG \\
    \code{8+ frwd}
      & GGCAGCAGTACTCGAAAGAAAAAGAAGATT AAGAAAAAGCTAAAACAGGGCAGCGCGTGG \\
    \code{9+ frwd}
      & GGCAGCAGTACTCGAAAGAAAAAGAAGATT AAGAAAAAGAAGAAACAGGGCAGCGCGTGG \\
    \bottomrule
  \end{tabularx}
\end{table*}

\subsection{Protein overexpression and purification}
\label{sec:trapping_appendix:protein_overexpression}

\subsubsection{Strep-tagged {DHFR} mutants}

The pT7-SC1 plasmid containing the \gls{dhfr} gene and the sequence of the Strep-tag at its C-terminus was
transformed into E. cloni\textsuperscript{\textregistered} EXPRESS BL21(DE3) cells (Lucigen, Middleton, USA),
and transformants were selected on LB agar plates supplemented with \SI{100}{\micro\gram\per\milli\litre}
ampicillin after overnight growth at \SI{37}{\celsius}. The resulting colonies were grown at \SI{37}{\celsius}
in 2xYT medium supplemented with \SI{100}{\micro\gram\per\milli\litre} ampicillin until the O.D. at
\SI{600}{\nano\meter} was \num{\approx 0.8} (\SI{200}{rpm} shaking). The \gls{dhfr} expression was
subsequently induced by addition of \SI{0.5}{\mM} \ce{IPTG} (isopropyl \textbeta-D-1-thiogalactopyranoside),
and the temperature was switched to \SI{25}{\celsius} for overnight growth (200~rpm shaking). The next day the
bacteria were harvested by centrifugation at \SI{6000}{g} at \SI{4}{\celsius} for \SI{25}{\minute} and the
resulting pellets were frozen at \SI{-80}{\celsius} until further use.

Bacterial pellets originating from \SI{50}{\milli\litre} culture were resuspended in \SI{30}{\milli\litre}
lysis buffer (\SI{150}{\mM} \ce{NaCl}, \SI{15}{\mM} \ce{Tris-HCl} \pH{7.5}, \SI{1}{\mM} \ce{MgCl2},
\SI{0.2}{units\per\milli\litre} DNase, \SI{10}{\micro\gram\per\milli\litre} lysozyme) and incubated at
\SI{37}{\celsius} for \SI{20}{\minute}. After further disruption of the bacteria by probe sonification the
crude lysate was clarified by centrifugation at \SI{6000}{g} at \SI{4}{\celsius} for \SI{30}{\minute}. The
supernatant was allowed to bind to \SI{\approx 150}{\micro\litre} (bead volume) of
Strep-Tactin\textsuperscript{\textregistered} Sepharose\textsuperscript{\textregistered} (IBA Lifesciences,
Goettingen, Germany) pre-equilibrated with the wash buffer (\SI{150}{\mM} \ce{NaCl}, \SI{15}{\mM}
\ce{Tris-HCl} \pH{7.5}) \textit{via} `end over end' mixing. The resin was then loaded onto a column (Micro Bio
Spin, Bio-Rad) and washed with \num{\approx 20} column volumes of the wash buffer. Elution of \gls{dhfr} from
the column was achieved by addition of \SI{\approx 100}{\micro\litre} of elution buffer (\SI{150}{\mM}
\ce{NaCl}, \SI{15}{\mM} \ce{Tris-HCl} \pH{7.5}, \SI{\approx 15}{\mM} D-Desthiobiotin (IBA)). Proteins were
aliquoted and frozen at \SI{-20}{\celsius} until further use. New aliquots of \gls{dhfr} were thawed prior to
every experiment.

\subsubsection{His-tagged type {I} {ClyA-AS}}

E. cloni\textsuperscript{\textregistered} EXPRESS BL21 (DE3) cells were transformed with the {pT7-SC1} plasmid
containing the \gls{clya-as} gene. \gls{clya-as} contains eight mutations relative to the \textit{S. Typhi}
ClyA-WT: C87A, L99Q, E103G, F166Y, I203V, C285S, K294R and H307Y (the H307Y mutation is in the C-terminal
hexahistidine-tag added for purification)~\cite{Soskine-2013}. Transformants were selected after overnight
growth at \SI{37}{\celsius} on LB agar plates supplemented with \SI{100}{\micro\gram\per\milli\litre}
ampicillin. The resulting colonies were grown at \SI{37}{\celsius} (\SI{200}{rpm} shaking) in 2xYT medium
supplemented with \SI{100}{\micro\gram\per\milli\litre} ampicillin until the O.D. at \SI{600}{\nano\metre} was
\num{\approx 0.8}. \gls{clya-as} expression was then induced by addition of \SI{0.5}{\mM} IPTG, and the
temperature was switched to \SI{25}{\celsius} for overnight growth (\SI{200}{rpm} shaking). The next day the
bacteria were harvested by centrifugation at \SI{6000}{g} for \SI{25}{\minute} at \SI{4}{\celsius} and the
pellets were stored at \SI{-80}{\celsius}.

Pellets containing monomeric \gls{clya-as} arising from \SI{50}{\milli\litre} culture were thawed and
resuspended in \SI{20}{\milli\litre} of wash buffer (\SI{10}{\mM} imidazole, \SI{150}{\mM} \ce{NaCl},
\SI{15}{\mM} \ce{Tris-HCl} \pH{7.5}), supplemented with \SI{1}{\mM} \ce{MgCl2} and
\SI{0.05}{units\per\milli\litre} of DNaseI. After lysis of the bacteria by probe sonication, the crude lysates
were clarified by centrifugation at \SI{6000}{g} for \SI{20}{\minute} at \SI{4}{\celsius} and the supernatant
was mixed with \SI{200}{\micro\litre} of \ce{Ni-NTA} resin (Qiagen, Hilden, Germany) equilibrated in wash
buffer. After \SI{60}{\minute}, the resin was loaded into a Micro Bio Spin column (Bio-Rad, Hercules, USA) and
washed with \SI{\approx 5}{\milli\litre} of the wash buffer. \gls{clya-as} was eluted with approximately
\SI{\approx0.5}{\milli\litre} of wash buffer containing \SI{300}{\mM} imidazole. \gls{clya-as} monomers were
stored at \SI{4}{\celsius} until further use.

\gls{clya-as} monomers were oligomerized by addition of \SI{0.5}{\percent} \textbeta-dodecylmaltoside (DDM,
GLYCON Biochemicals GmbH, Luckenwalde, Germany) and incubation at \SI{37}{\celsius} for \SI{30}{\minute}.
\gls{clya-as} oligomers were separated from monomers by blue native polyacrylamide gel electrophoresis
(BN-PAGE, Bio-Rad, Hercules, USA) using \SIrange{4}{20}{\percent} polyacrylamide gels. The bands corresponding
to Type I \gls{clya-as} were excised from the gel and placed in \SI{150}{\mM} \ce{NaCl}, \SI{15}{\mM}
\ce{Tris-HCl} \pH{7.5}, supplemented with \SI{0.2}{\percent} DDM and \SI{10}{\mM} \gls{edta} to allow
diffusion of the proteins out of the gel.

\subsection{Single nanopore experiments}
\label{sec:trapping_appendix:nanopore_experiments}

\subsubsection{Electrical recordings in planar lipid bilayers}

By convention, the applied potential refers to the potential of the trans electrode in the planar lipid
bilayer set-up. \gls{clya-as} nanopores were inserted into lipid bilayers from the \cisi{} compartment, which
is connected to the ground electrode. The \cisi{} and \transi{} compartments are separated by a
\SI{25}{\micro\meter} thick polytetrafluoroethylene film (Goodfellow Cambridge Limited, Huntingdon, England)
containing an orifice of \SI{\approx 100}{\micro\meter} in diameter. After pre-treatment of the aperture with
\SI{\approx 5}{\micro\litre} of \SI{10}{\percent} hexadecane in pentane, a bilayer was formed by the addition
of \SI{\approx 10}{\micro\litre} of 1,2-diphytanoyl-sn-glycero-3-phosphocholine (DPhPC) in pentane
(\SI{10}{\milli\gram\per\milli\litre}) to both electrophysiology chambers. Typically, the addition of
\SIrange{0.01}{0.1}{\nano\gram} of pre-oligomerised \gls{clya-as} to the \cisi{} compartment
(\SI{0.5}{\milli\litre}) was sufficient to obtain a single channel. Since \gls{clya-as} nanopores displayed a
higher open-pore current at positive than at negative applied potentials, the orientation of the pore could be
easily assessed. All electrical recordings were carried out in \SI{150}{\mM} \ce{NaCl}, \SI{15}{\mM}
\ce{Tris-HCl} \pH{7.5}. The temperature of the recording chamber was maintained at \SI{28}{\celsius} by water
circulating through a metal case in direct contact with the bottom and sides of the chamber.

\subsubsection{Data recording and event analysis}

Electrical signals from planar lipid bilayer recordings were amplified using an Axopatch 200B patch-clamp
amplifier (Axon Instruments, San Jose, USA) and digitized with a Digidata 1440 A/D converter (Axon
Instruments, San Jose, USA). Data were recorded using the Clampex 10.5 software (Molecular Devices, San Jose,
USA) and the subsequent event analysis was carried out with the \code{Clampfit} software (Molecular Devices).
Ionic currents were sampled at \SI{10}{\kilo\hertz} and filtered with a \SI{2}{\kilo\hertz} low-pass Bessel
filter.

Residual current values ($\iresp$) of the different \gls{dhfr} variants were calculated by $\iresp =
\rm{I}_b/\rm{I}_o$, in which $\rm{I}_b$ and $\rm{I}_o$ represent the blocked an open-pore current values,
respectively. $\rm{I}_b$ and $\rm{I}_o$ values were calculated from Gaussian fits to all point current
histograms (\SI{0.1}{\pA} bin size) from at least 3 individual single channels each displaying at least 50
current blockades. The average residence time of the \gls{dhfr} mutants was determined using the `single
channel search' feature in \code{Clampfit}. The detection threshold was set to \SI{\approx 75}{\percent} of
the open-pore current and events shorter than \SI{1}{\ms} were ignored for all mutants except for \DHFR{4}{S},
\DHFR{4}{I} and \DHFR{4}{C}, as they exhibited very short dwell times inside \gls{clya-as}. The process of
event collection was monitored manually.

The average of the mean dwell times obtained from at least three single channels each displaying at least 100
blockades was used to describe the average dwell time ($\dwelltime$) at every potential. For analysis of
\gls{nadph} binding events to \DHFR{$\Ntag$}{O2}, traces were filtered digitally with a 8-pole low-pass Bessel
filter with a \SI{500}{\hertz} cutoff. Current transitions from L1 to L1\textsubscript{NADPH} were analyzed
with the `single channel search' option in \code{Clampfit}. The detection threshold to collect the
\gls{nadph}-induced events was set to \SI{2}{\pA} and events shorter than \SI{0.1}{\ms} were ignored. The
process of event collection was monitored manually. The resulting event dwell times ($\tau_{\rm{off}}$) and
the times between events ($\tau_{\rm{on}}$) were binned together as cumulative distributions and fitted to a
single exponential to retrieve the \gls{nadph}-induced lifetimes ($\tau_{\rm{off}}$) and the
\gls{nadph}-induced inter-event times ($\tau_{\rm{on}}$). The average amplitude of the events was derived from
Gaussian fits to the conventional distributions of the events' amplitudes. Values for off and on rates were
determined as $\rate_{\rm{off}} = 1/\tau_{\rm{off}}$ and $\rate_{\rm{off}} = \left(\tau_{\rm{on}} c
\right)^{-1}$ with $c$ the concentration of \gls{nadph} added to the \transi{} solution. Final values for
$\tau_{\rm{on}}$, $\tau_{\rm{off}}$, $\rate_{\rm{on}}$, $\rate_{\rm{off}}$ and event amplitudes are the
averages derived from three single channel experiments, each analyzing at least 100 binding events on more
than five different \DHFR{$\Ntag$}{O2} blockades.

\cleardoublepage


%
\chapter[SI: An improved PNP-NS framework]%
        {Supplementary information: An improved PNP-NS framework}
\label{ch:epnpns_appendix}

\definecolor{shadecolor}{gray}{0.85}
\begin{shaded}
  \adapRSC{Willems-2020}
\newpage
\end{shaded}

\section{Weak forms of the {ePNP-NS} equations}
\label{sec:epnpns_appendix:weak_forms}

\begin{figure*}[t]
  \centering
  \includegraphics[scale=1]{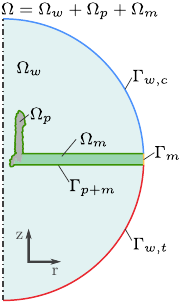}

  \caption[Computational domains and boundaries.]%
  {\textbf{Computational domains and boundaries.}
    The full computational domain ($\Omega$) of the model is subdivided into various subdomains---comprising
    the reservoir ($\Omega_w$), the lipid bilayer ($\Omega_m$) and the pore ($\Omega_p$)---and their limiting
    boundaries---the exterior boundary at the \cisi{} ($\Gamma_{w,c}$) \transi{} and the \transi{}
    ($\Gamma_{w,t}$) sides of the reservoir, the exterior boundary of the membrane ($\Gamma_m$) and the
    interior boundary between the reservoir on the one hand, and the pore and membrane on the other
    ($\Gamma_{p+m}$).
  }\label{fig:epnpns_model_boundaries}
\end{figure*}

To solve partial differential equations with the finite element method, we must derive their weak form.
This is achieved through the multiplication of the equation with an arbitrary test function and integration over
their relevant domains and boundaries (\cref{fig:epnpns_model_boundaries}). The full computational domain of
our model ($\Omega$) is subdivided into domains for the pore ($\Omega_p$), the lipid bilayer ($\Omega_m$), and
the electrolyte reservoir ($\Omega_w$). The relevant boundaries of these domains are also indicated, i.e. the
reservoir's exterior edges at the \cisi{} $\Gamma_{w,c}$) and \transi{} $\Gamma_{w,t}$) sides, the outer edge
of the lipid bilayer ($\Gamma_{m}$) and the interface of the fluid with the nanopore and the bilayer
($\Gamma_{p+m}$).

The following paragraphs detail the weak forms of equations that were used in this
\cref{ch:epnpns,ch:trapping}.

\subsection{Poisson equation}

The global potential distribution is described by the Poisson equation (PE)~\cite{Lu-2012}
\begin{align}
  \label{eq:epnpns_appendix:poisson}
  \nabla \cdot \displacement = -\left( \scdpore + \scdion \right)
  \text{\quad with \quad}
  \displacement = \absperm \relperm \nabla \potential
  \text{ ,}
\end{align}
with $\displacement$ the electrical displacement field, $\potential$ the electric potential, $\absperm$ the
vacuum permittivity (\SI{8.85419e-12}{\farad\per\meter}), and $\relperm$ the local relative permittivity
of the medium. $\scdpore$ and $\scdion$ are the fixed (due to the pore) and mobile (due to the ions) charge
distributions, respectively.

Multiplication of~\cref{eq:epnpns_appendix:poisson} with the potential test function $\psi$ and integration
over the entire model $\Omega=\Omega_w+\Omega_p+\Omega_m$ gives
\begin{align}
  \displaystyle\int_{\Omega}
  \left[
    \nabla \cdot \displacement
  \right]
  \psi \,d\Omega
  ={}&
  - \displaystyle\int_{\Omega} \left[ \scdpore + \scdion \right] \psi \,d\Omega \text{ ,}
\end{align}
which, after applying the Gauss divergence theorem, yields the final weak formulation
\begin{align}
  \displaystyle\int_{\Omega}
  \left[
    \nabla \psi \cdot \displacement
  \right]
  \,d\Omega
  & - \displaystyle\int_{\Gamma_\text{PE}}
  \left[
    \psi \displacement \cdot \vec{n}
  \right]
  \,d\Gamma_\text{PE} \notag \\
  & =
  \displaystyle\int_{\Omega} \left[ \psi \scdpore \right] \,d\Omega
  +
  \displaystyle\int_{\Omega} \left[ \psi \scdion \right] \,d\Omega
  \text{ ,}
\end{align}
with boundaries $\Gamma_\text{PE}=\Gamma_{w,c}+\Gamma_{w,t}+\Gamma_m$ and $\vec{n}$ their normal vector. The
boundary integrals at $\Gamma_{w,c}$ and $\Gamma_{w,t}$ are evaluated using the Dirichlet \glspl{bc}
$\potential=0$ and $\potential=\vbias$, respectively. A zero charge \gls{bc}, $\vec{n} \cdot \displacement =
0$, is used for the integral at $\Gamma_{m}$.

\subsection{Size-modified Nernst-Planck equation}

The total ionic flux $\flux_{i}$ of ion $i$ at steady-state is expressed by the \gls{smnpe}~\cite{Lu-2012}
\begin{equation}\label{eq:smnp}
  \pd{\concentration_{i}}{t} = - \nabla\cdot\flux_{i} = - \nabla \cdot
  \left(
    \diffusion_{i}\nabla\concentration_{i}
    + \chargen_{i}\mobility_{i}\concentration_{i}\nabla\potential
    + \diffusion_{i} \vec{\beta_{i}} \concentration_{i}
    - \velocity\concentration_{i}
  \right)
  \text{ ,}
\end{equation}
where
\begin{equation}\label{eq:steric_vector}
  \vec{\beta_{i}} =
    \frac{\ionsize_{i}^3/\ionsize_{0}^3 \dsum_{j} \avogadro \ionsize_{j}^3 \nabla \concentration_{j}}
        {1 - \dsum_{j} \avogadro \ionsize_{j}^3 \concentration_{j}}
  \text{ ,}
\end{equation}
and with ion diffusion coefficient $\diffusion_{i}$, concentration $\concentration_{i}$, charge number
$\chargen_{i}$, mobility $\mobility_{i}$, electrostatic potential $\potential$, steric saturation factor
$\beta_{i}$ and fluid velocity $\velocity$. $\avogadro$ is Avogadro's constant (\SI{6.022e23}{\per\mole}) and
$\ionsize_{i}$ and $\ionsize_{0}$ are the limiting cubic diameters for ions and water, respectively. Using the
ion concentration test function $d_i$, the weak form of \cref{eq:smnp} becomes
\begin{align}
  \displaystyle\int_{\Omega_w} \left[ \pd{\concentration_{i}}{t} \right] d_{i} \,d\Omega_w =
  \displaystyle\int_{\Omega_w} \left[ - \nabla \cdot \flux_{i} \right] d_{i} \,d\Omega_w
  \text{, }
\end{align}
which can be split into the domain and boundary integrals
\begin{align}
  \displaystyle\int_{\Omega_w} \left[ d_{i} \pd{\concentration_{i}}{t} \right] \,d\Omega_w ={}&
  \displaystyle\int_{\Omega_w} \left[\nabla d_{i} \cdot \flux_{i} \right]\,d\Omega_w
  - \displaystyle\int_{\Gamma_\text{NP}}
  \left[ d_{i} \flux_{i} \cdot \vec{n} \right]\,d\Gamma_\text{NP} \\
  ={}&
  \displaystyle\int_{\Omega_w}
  \left[
    \nabla d_{i} \cdot
    \left(
      \diffusion_{i}\nabla\concentration_{i}
      + \chargen_{i}\mobility_{i}\concentration_{i}\nabla\potential
      + \diffusion_{i} \vec{\beta_{i}} \concentration_{i}
      - \velocity \concentration_{i}
    \right)
  \right]
  \,d\Omega_w \notag \\
  & - \displaystyle\int_{\Gamma_\text{NP}}
  \left[
    d_{i}
    \left(
      \diffusion_{i} \nabla \concentration_{i}
      + \chargen_{i} \mobility_{i} \concentration_{i} \nabla \potential
      + \diffusion_{i} \vec{\beta_{i}} \concentration_{i}
      - \velocity \concentration_{i}
    \right)
    \cdot \vec{n}
  \right]
  \,d\Gamma_\text{NP} \notag
  \text{ .}
\end{align}
The integrals on the boundaries $\Gamma_\text{NP} = \Gamma_{w,c}+\Gamma_{w,t}+\Gamma_{p+m}$ are evaluated
using the Dirichlet \gls{bc} $\concentration_{i} = \cbulk$ for $\Gamma_{w,c}$ and $\Gamma_{w,t}$, and the no
flux \gls{bc} $\vec{n} \cdot \flux_{i} = 0$ for $\Gamma_{p+m}$

\subsection{Variable density and viscosity Navier-Stokes equation}

The steady-state, laminar fluid flow of an incompressible fluid with a variable density and viscosity is given
by the system of equations~\cite{Axelsson-2015}
\begin{align}
  \label{eq:ns_density_continuity}
  \velocity \cdot \nabla \density ={}& 0 \\
  \label{eq:ns_conservation}
  \left( \velocity \cdot \nabla \right) \left( \density\velocity \right)
  + \nabla \cdot \hydrostresstensor ={}& \volumeforce \text{\quad with }
  \hydrostresstensor =
    \pressure\identity - \viscosity\left[\nabla\velocity+\left(\nabla\velocity \right)^\mathsf{T}\right]
  \\
  \label{eq:ns_velocity_continuity}
  \nabla \cdot \left( \density\velocity \right) - \velocity \cdot \nabla \density ={}& 0
  \text { ,}
\end{align}
with fluid velocity $\velocity$, density $\density$, hydrodynamic stress tensor $\hydrostresstensor$,
viscosity $\viscosity$, pressure $\pressure$ and body force $\volumeforce$. The pressure test function $q$ is
used to derive the weak forms of \cref{eq:ns_density_continuity}
\begin{align}
  \displaystyle\int_{\Omega_w} \left[ \velocity \cdot \nabla \density \right] q \,d\Omega_w ={}&
  \displaystyle\int_{\Omega_w} \left[ q \velocity \cdot \nabla \density \right] \,d\Omega_w = 0 \text{ ,}
\end{align}
and \cref{eq:ns_velocity_continuity}
\begin{align}
  \displaystyle\int_{\Omega_w}
  &\left[\nabla \cdot \left( \density \velocity \right) - \velocity \cdot \nabla \density \right] q
  \,d\Omega_w \\ & =
  \displaystyle\int_{\Omega_w}
  \left[\nabla \cdot \left( \density \velocity \right) \right] q \,d\Omega_w
  -
  \displaystyle\int_{\Omega_w} 
  \left[ q \velocity \cdot \nabla \density \right] \,d\Omega_w \notag \\
  & =
  \displaystyle\int_{\Omega_w}
  \left[\nabla q \cdot \left( \density \velocity \right) \right] \,d\Omega_w
  -
  \displaystyle\int_{\Gamma_\text{NS}}
  \left[ q \left( \density \velocity \right) \cdot \vec{n} \right] \,d\Gamma_\text{NS}
  -
  \displaystyle\int_{\Omega_w}
  \left[ q \velocity \cdot \nabla \density \right] \,d\Omega_w \notag
  \text{ ,}
\end{align}
while for \cref{eq:ns_conservation} we use the velocity test function $\vec{v}=\left[v_r, v_\phi, v_z\right]$
\begin{align}
  \displaystyle\int_{\Omega_w}
  \left[
    \left( \velocity \cdot \nabla \right) \left( \density\velocity \right) + \nabla \cdot \hydrostresstensor
  \right]
  \cdot \vec{v} \,d\Omega_w
  =
  \displaystyle\int_{\Omega_w} \vec{\force} \cdot \vec{v} \,d\Omega_w 
\end{align}
which can be split into the domain and boundary integrals
\begin{align}
  \displaystyle\int_{\Omega_w} \vec{\force} \cdot \vec{v} \,d\Omega_w =
  \displaystyle\int_{\Omega_w} &
  \left[
    \left( \velocity \cdot \nabla \right) \left( \density\velocity \right) \cdot\vec{v}
  \right] 
  \,d\Omega_w 
  -
  \displaystyle\int_{\Omega_w}
  \left[
  \hydrostresstensor \cdot \nabla\vec{v}
  \right]
  \,d\Omega_w \notag \\
  & +
  \displaystyle\int_{\Gamma_\text{NS}}
  \left[
  \vec{v} \cdot
  \hydrostresstensor
  \cdot\vec{n}
  \right]
  \,d\Gamma_\text{NS}
  \text{ ,}
\end{align}
with boundaries $\Gamma_\text{NS} = \Gamma_{w,c}+\Gamma_{w,t}+\Gamma_{p+m}$. The no-slip Dirichlet \gls{bc}
$\velocity = 0$ is applied to $\Gamma_{p+m}$, and the no normal stress $\hydrostresstensor\vec{n}=0$ is used
for $\Gamma_{w,c}$ and $\Gamma_{w,t}$.

\cleardoublepage


\chapter[SI: Modeling the transport of ions and water through {ClyA}]%
        {Supplementary information: Modeling the transport of ions and water through {ClyA}}
\label{ch:transport_appendix}

\definecolor{shadecolor}{gray}{0.85}
\begin{shaded}
  \adapRSC{Willems-2020}
\newpage
\end{shaded}

\section{Comparison of the {ClyA} \lumen{} diameters}
\label{sec:transport_appendix:radius_comparision}

\begin{figure*}[t]
  \centering
  \includegraphics[scale=1]{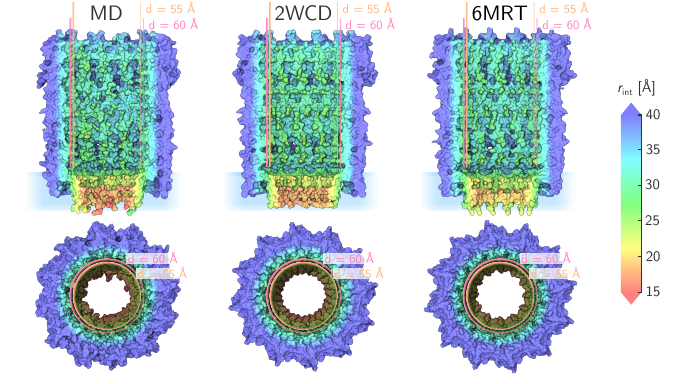}

  \caption[\textit{Lumen} diameters of {ClyA}.]%
  {%
    \textbf{\textit{Lumen} diameters of {ClyA}.}
    Side and top view of the molecular surface of the \gls{clya-as} equilibrated with \gls{md} (left), with
    the crystal \pdbid{2WCD}~\cite{Mueller-2009} (middle) and \gls{cryo-em} \pdbid{6MRT}~\cite{Peng-2019}
    (right) structures. Surfaces were colored according to $r_{\text{int}}$, the distance of each atom from
    the central axis of the pore, reduced with its van der Waals radius (see~\cref{eq:internal_radius}). The
    traditional diameter of \SI{5.5}{\nm} and the \SI{6.0}{\nm} are outlined in orange and pink respectively.
    Images were rendered using \gls{vmd}~\cite{Humphrey-1996}.
  }\label{fig:transport_radius_comparison_surface}
\end{figure*}

A comparison of the \gls{clya} structures equilibrated with molecular dynamics with the crystal
(\pdbid{2WCD}~\cite{Mueller-2009}) and \gls{cryo-em} (\pdbid{6MRT}~\cite{Peng-2019}) structures can be found
in \cref{fig:transport_radius_comparison_surface}. The molecular surface was colored according to the internal
radius $r_{\text{int}}$
\begin{align}\label{eq:internal_radius}
  r_{\text{int},i} = \sqrt{(x_i^2 + y_i^2)} - r_{\text{vdW},i}
\end{align}
with $x_i$, $y_i$ and $r_{\text{vdW},i}$ the $x$-coordinate, $y$-coordinate and van der Waals radius of atom
$i$. All structure were aligned to minimize their RSMD to the initial structure, which has its center of mass
at $(0,0,55)$.

Our comparison shows that the \gls{md} structure does not deviate significantly from the crystal and
\gls{cryo-em} structures in overall geometry. Moreover, the average nanopore diameter in the \cisi{} \lumen{}
is closer to \SI{6}{\nm} than the \SI{5.5}{\nm} reported earlier~\cite{Willems-Ruic-Biesemans-2019} The value
of \SI{5.5}{\nm} should be considered as a measure for the maximum size of a protein that can fit inside
dodecameric \gls{clya} without it touching the walls. The \SI{6}{\nm} diameter value given is the average
\lumen{} diameter, which is slightly larger than \SI{5.5}{\nm} given that it includes---rather than
excludes---all corrugations. 

\clearpage

\section{Tabulated ion selectivities}
\label{sec:transport_appendix:tab_ion_sel}

Tabulated data on the ion selectivity, in terms of cation transport number $\tna$ and cation permeability
ratio $\pna$ for various salt concentrations and voltages can be found in \cref{tab:ion_selectivities}.

\begin{table*}[b]
  \centering

  \captionsetup{width=10cm}
  \caption{Tabulated cation transport numbers and permeability ratios.}
  \label{tab:ion_selectivities}

  \renewcommand{\arraystretch}{1.2}
  \footnotesize

  \begin{tabularx}{10cm}{XXXXXXX}
    \toprule
     & \multicolumn{6}{c}{$\vbias$ [\si{\mV}]} \\
       \cmidrule{2-7}
     & \multicolumn{2}{c}{$+150$} & \multicolumn{2}{c}{$+100$} & \multicolumn{2}{c}{$+50$} \\
    \cmidrule(r){2-3} \cmidrule(r){4-5} \cmidrule(r){6-7}
    $\cbulk$ [\si{\Molar}]
     & $\tna$$^\text{\emph{a}}$ & $\pna$$^\text{\emph{b}}$
     & $\tna$$^\text{\emph{a}}$ & $\pna$$^\text{\emph{b}}$
     & $\tna$$^\text{\emph{a}}$ & $\pna$$^\text{\emph{b}}$ \\
    \midrule
    0.005 & 0.996 & 235 & 0.997 & 313 & 0.998 & 410 \\
    0.050 & 0.906 & 9.62 & 0.925 & 12.3 & 0.941 & 16.0 \\
    0.150 & 0.786 & 3.68 & 0.805 & 4.13 & 0.824 & 4.69 \\
    0.500 & 0.642 & 1.79 & 0.649 & 1.85 & 0.657 & 1.91 \\
    1.000 & 0.566 & 1.30 & 0.569 & 1.32 & 0.572 & 1.33 \\
    2.000 & 0.502 & 1.00 & 0.503 & 1.01 & 0.503 & 1.01 \\
    5.000 & 0.445 & 0.803 & 0.445 & 0.803 & 0.445 & 0.803 \\
    \midrule
    \midrule
     & \multicolumn{6}{c}{$\vbias$ [\si{\mV}]} \\
       \cmidrule{2-7}
     & \multicolumn{2}{c}{$-150$} & \multicolumn{2}{c}{$-100$} & \multicolumn{2}{c}{$-50$} \\
    \cmidrule(r){2-3} \cmidrule(r){4-5} \cmidrule(r){6-7}
    $\cbulk$ [\si{\Molar}]
      & $\tna$$^\text{\emph{a}}$ & $\pna$$^\text{\emph{b}}$
      & $\tna$$^\text{\emph{a}}$ & $\pna$$^\text{\emph{b}}$
      & $\tna$$^\text{\emph{a}}$ & $\pna$$^\text{\emph{b}}$ \\
    \midrule
    0.005 & 0.998 & 500 & 0.998 & 535 & 0.998 & 542 \\
    0.050 & 0.967 & 29.6 & 0.966 & 28.2 & 0.962 & 25.1 \\
    0.150 & 0.882 & 7.51 & 0.874 & 6.91 & 0.860 & 6.16 \\
    0.500 & 0.687 & 2.20 & 0.680 & 2.12 & 0.672 & 2.05 \\
    1.000 & 0.582 & 1.39 & 0.579 & 1.38 & 0.577 & 1.36 \\
    2.000 & 0.505 & 1.02 & 0.505 & 1.02 & 0.504 & 1.02 \\
    5.000 & 0.445 & 0.802 & 0.445 & 0.802 & 0.445 & 0.802 \\
    \bottomrule
  \end{tabularx}
  \begin{flushleft}
    $^\text{\emph{a}}$Cation transport number $\tna = \conductance_{\Na} / (\conductance_{\Na} +
    \conductance_{\Cl} )$;
    $^\text{\emph{b}}$Cation permeability ratio $\pna = \conductance_{\Na} / (\conductance_{\Cl} )$.
  \end{flushleft}
\end{table*}

\clearpage

\section{Peak values of the radial potential profiles inside {ClyA-AS}}

The peak values of the radial electrostatic potential $\radpot$ at the \cisi{} entry, middle of the lumen and
the \transi{} constriction for \SIlist{0.005;0.05;0.15;0.5;5}{\Molar} \ce{NaCl} are summarized in
\cref{tab:radial_potential}.

\begin{table}[b]
  \centering

  \captionsetup{width=8cm}
  \caption{Peak radial potential.}
  \label{tab:radial_potential}
  \footnotesize
  \begin{tabularx}{8cm}{SSSS}
    \toprule
    & \multicolumn{3}{c}{$\radpot$ [\si{\mV}]} \\
    \cmidrule{2-4}
    & {\cisi{}}      & {\lumen{}}    & {\transi{}}  \\
    {$\cbulk$ [\si{\Molar})]}
    & {$z\approx\mSI{10}{\nm}$} & {$z\approx\mSI{5}{\nm}$} & {$z\approx\mSI{0}{\nm}$} \\
    \midrule
    0.005          & -80         & -108       & -144   \\
    0.05           & -34         &  -50       &  -86   \\
    0.15           & -19         &  -29       &  -57   \\
    0.5            &  -9.3       &  -14       &  -30   \\
    5              &  -1.9       &   -1.7     &   -4.2 \\
    \bottomrule
  \end{tabularx}
\end{table}

\clearpage

\section{Surface integration to compute pore averaged values}

The average pore values for quantity of interest $X$ were computed by \cref{eq:pore_surface_integral}
\begin{align}\label{eq:pore_surface_integral}
  \left< X \right>_{\alpha} =
    \displaystyle\frac{\displaystyle\iint_{V_{\alpha}} \beta_{\alpha} X \,dr\,dz}
                      {\displaystyle\iint_{V_{\alpha}} \beta_{\alpha} \,dr\,dz}
  \text{ ,}
\end{align}
where
\begin{equation}
  \alpha=
  \begin{cases}
    \text{PT}, & d \ge 0  \text{~nm} \text{, average over the entire pore} \\
    \text{PB}, & d > 0.5  \text{~nm} \text{, average over the pore `bulk' }  \\
    \text{PS}, & d \le 0.5\text{~nm} \text{, average over the pore `surface' }
  \end{cases}
\end{equation}
and
\begin{align}
  \beta_{\text{PT}} &=
  \begin{cases}
    1, & \text{if}\ -1.85\le z \le 12.25  \text{ and } r \le r_\text{p}(z) \\
    0, & \text{otherwise}
  \end{cases} \\
  \beta_{\text{PB}} &=
  \begin{cases}
  1, & \text{if}\ -1.85\le z \le 12.25  \text{ and } r \le r_\text{p}(z) \text{ and } d > 0.5 \\
  0, & \text{otherwise}
  \end{cases} \\
  \beta_{\text{PS}} &=
  \begin{cases}
  1, & \text{if}\ -1.85\le z \le 12.25  \text{ and } r \le r_\text{p}(z) \text{ and } d \le 0.5 \\
  0, & \text{otherwise}
  \end{cases}
\end{align}
with $d$ the distance from the nanopore wall and $r_\text{p}(z)$ is the radius of the pore at height $z$.

\cleardoublepage


\backmatter

\includebibliography
\renewcommand*{\bibfont}{\small}
\printbibliography[title=\bibname]

@article{Wang-2018,
  doi = {10.1021/acssensors.7b00680},
  url = {https://doi.org/10.1021/acssensors.7b00680},
  year = {2018},
  month = jan,
  publisher = {American Chemical Society ({ACS})},
  volume = {3},
  number = {2},
  pages = {251--263},
  author = {Haiyan Wang and Jessica Ettedgui and Jacob Forstater and Joseph W. F. Robertson and Joseph E. Reiner and Huisheng Zhang and Siping Chen and John J. Kasianowicz},
  title = {{D}etermining the {P}hysical {P}roperties of {M}olecules with {N}anometer-{S}cale {P}ores},
  journal = {{ACS} Sens.}
}

@article{Howorka-2017,
  doi = {10.1038/nnano.2017.99},
  url = {https://doi.org/10.1038/nnano.2017.99},
  year = {2017},
  month = jul,
  publisher = {Springer Science and Business Media {LLC}},
  volume = {12},
  number = {7},
  pages = {619--630},
  author = {Stefan Howorka},
  title = {Building membrane nanopores},
  journal = {Nat. Nanotechnol.}
}

@article{Roderer-2017,
  doi = {10.1098/rstb.2016.0211},
  url = {https://doi.org/10.1098/rstb.2016.0211},
  year = {2017},
  month = jun,
  publisher = {The Royal Society},
  volume = {372},
  number = {1726},
  pages = {20160211},
  author = {Daniel Roderer and Rudi Glockshuber},
  title = {{A}ssembly mechanism of the {\textalpha}-pore{\textendash}forming toxin cytolysin A from
		            \textit{{E}scherichia coli}},
  journal = {Philos. Trans. Royal Soc. B}
}

@article{Eifler-2006,
  doi = {10.1038/sj.emboj.7601130},
  url = {https://doi.org/10.1038/sj.emboj.7601130},
  year = {2006},
  month = may,
  publisher = {Wiley},
  volume = {25},
  number = {11},
  pages = {2652--2661},
  author = {Nora Eifler and Michael Vetsch and Marco Gregorini and Philippe Ringler and Mohamed Chami and Ansgar Philippsen and Andrea Fritz and Shirley A M{\"{u}}ller and Rudi Glockshuber and Andreas Engel and Ulla Grauschopf},
  title = {{C}ytotoxin {ClyA} from Escherichia coli assembles to a 13-meric pore independent of its redox-state},
  journal = {{EMBO} J.}
}

@article{Peraro-2015,
  doi = {10.1038/nrmicro.2015.3},
  url = {https://doi.org/10.1038/nrmicro.2015.3},
  year = {2015},
  month = dec,
  publisher = {Springer Science and Business Media {LLC}},
  volume = {14},
  number = {2},
  pages = {77--92},
  author = {Matteo Dal Peraro and F. Gisou van der Goot},
  title = {Pore-forming toxins: ancient,  but never really out of fashion},
  journal = {Nat. Rev. Microbiol.}
}

@article{Kowalczyk-2011,
  doi = {10.1088/0957-4484/22/31/315101},
  url = {https://doi.org/10.1088/0957-4484/22/31/315101},
  year = {2011},
  month = jul,
  publisher = {{IOP} Publishing},
  volume = {22},
  number = {31},
  pages = {315101},
  author = {Stefan W Kowalczyk and Alexander Y Grosberg and Yitzhak Rabin and Cees Dekker},
  title = {{M}odeling the {C}onductance and {DNA} {B}lockade of {S}olid-{S}tate {N}anopores},
  journal = {Nanotechnology}
}

@article{Peng-2019,
  doi = {10.1371/journal.pone.0213423},
  url = {https://doi.org/10.1371/journal.pone.0213423},
  year = {2019},
  month = may,
  publisher = {Public Library of Science ({PLoS})},
  volume = {14},
  number = {5},
  pages = {e0213423},
  author = {Wei Peng and Marcela de Souza Santos and Yang Li and Diana R. Tomchick and Kim Orth},
  editor = {MD A. Motaleb},
  title = {{H}igh-resolution cryo-{EM} structures of the {E.} coli hemolysin {ClyA} oligomers},
  journal = {{PLOS} {ONE}}
}

@article{Mueller-2009,
  doi = {10.1038/nature08026},
  url = {https://doi.org/10.1038/nature08026},
  year = {2009},
  month = may,
  publisher = {Springer Nature},
  volume = {459},
  number = {7247},
  pages = {726--730},
  author = {Marcus Mueller and Ulla Grauschopf and Timm Maier and Rudi Glockshuber and Nenad Ban},
  title = {{T}he {S}tructure of a {C}ytolytic {\textalpha}-{H}elical {T}oxin {P}ore {R}eveals {I}ts {A}ssembly {M}echanism},
  journal = {Nature}
}

@article{Wallace-2000,
  doi = {10.1016/s0092-8674(00)81564-0},
  url = {https://doi.org/10.1016/s0092-8674(00)81564-0},
  year = {2000},
  month = jan,
  publisher = {Elsevier {BV}},
  volume = {100},
  number = {2},
  pages = {265--276},
  author = {Alistair J. Wallace and Timothy J. Stillman and Angela Atkins and Stuart J. Jamieson and Per A. Bullough and Jeffrey Green and Peter J. Artymiuk},
  title = {{E.} coli {H}emolysin ({HlyE},  {ClyA},  {SheA})},
  journal = {Cell}
}

@article{Mechaly-2011,
  doi = {10.1016/j.str.2010.11.013},
  url = {https://doi.org/10.1016/j.str.2010.11.013},
  year = {2011},
  month = feb,
  publisher = {Elsevier {BV}},
  volume = {19},
  number = {2},
  pages = {181--191},
  author = {Ariel E. Mechaly and Augusto Bellomio and David Gil-Cart{\'{o}}n and Koldo Morante and Mikel Valle and Juan Manuel Gonz{\'{a}}lez-Ma{\~{n}}as and Diego M.A. Gu{\'{e}}rin},
  title = {{S}tructural {I}nsights into the {O}ligomerization and {A}rchitecture of {E}ukaryotic {M}embrane {P}ore-{F}orming {T}oxins},
  journal = {Structure}
}

@article{Tanaka-2015,
  doi = {10.1038/ncomms7337},
  url = {https://doi.org/10.1038/ncomms7337},
  year = {2015},
  month = feb,
  publisher = {Springer Science and Business Media {LLC}},
  volume = {6},
  number = {1},
  author = {Koji Tanaka and Jose M.M. Caaveiro and Koldo Morante and Juan Manuel Gonz{\'{a}}lez-Ma{\~{n}}as and Kouhei Tsumoto},
  title = {Structural basis for self-assembly of a cytolytic pore lined by protein and lipid},
  journal = {Nat. Commun.}
}

@article{Locher-1998,
  doi = {10.1016/s0092-8674(00)81700-6},
  url = {https://doi.org/10.1016/s0092-8674(00)81700-6},
  year = {1998},
  month = dec,
  publisher = {Elsevier {BV}},
  volume = {95},
  number = {6},
  pages = {771--778},
  author = {Kaspar P Locher and Bernard Rees and Ralf Koebnik and Andr{\'{e}} Mitschler and Luc Moulinier and Jurg P Rosenbusch and Dino Moras},
  title = {{T}ransmembrane {S}ignaling across the {L}igand-{G}ated {FhuA} {R}eceptor},
  journal = {Cell}
}

@article{Xu-2019,
  doi = {10.1038/s41467-019-10272-3},
  url = {https://doi.org/10.1038/s41467-019-10272-3},
  year = {2019},
  month = may,
  publisher = {Springer Science and Business Media {LLC}},
  volume = {10},
  number = {1},
  author = {Jingwei Xu and Dianhong Wang and Miao Gui and Ye Xiang},
  title = {Structural assembly of the tailed bacteriophage phi29},
  journal = {Nat. Commun.}
}

@article{Wendell-2009,
  doi = {10.1038/nnano.2009.259},
  url = {https://doi.org/10.1038/nnano.2009.259},
  year = {2009},
  month = sep,
  publisher = {Springer Science and Business Media {LLC}},
  volume = {4},
  number = {11},
  pages = {765--772},
  author = {David Wendell and Peng Jing and Jia Geng and Varuni Subramaniam and Tae Jin Lee and Carlo Montemagno and Peixuan Guo},
  title = {{T}ranslocation of double-stranded {DNA} through membrane-adapted phi29 motor protein nanopores},
  journal = {Nat. Nanotechnol.}
}

@article{Lukoyanova-Kondos-2015,
  doi = {10.1371/journal.pbio.1002049},
  url = {https://doi.org/10.1371/journal.pbio.1002049},
  year = {2015},
  month = feb,
  publisher = {Public Library of Science ({PLoS})},
  volume = {13},
  number = {2},
  pages = {e1002049},
  author = {Natalya Lukoyanova and Stephanie C. Kondos and Irene Farabella and Ruby H. P. Law and Cyril F. Reboul and Tom T. Caradoc-Davies and Bradley A. Spicer and Oded Kleifeld and Daouda A. K. Traore and Susan M. Ekkel and Ilia Voskoboinik and Joseph A. Trapani and Tamas Hatfaludi and Katherine Oliver and Eileen M. Hotze and Rodney K. Tweten and James C. Whisstock and Maya Topf and Helen R. Saibil and Michelle A. Dunstone},
  editor = {Raimund Dutzler},
  title = {{C}onformational {C}hanges during {P}ore {F}ormation by the {P}erforin-{R}elated {P}rotein {P}leurotolysin},
  journal = {{PLOS} Biol.}
}

@article{Reboul-2016,
  doi = {10.1016/j.bbamem.2015.11.017},
  url = {https://doi.org/10.1016/j.bbamem.2015.11.017},
  year = {2016},
  month = mar,
  publisher = {Elsevier {BV}},
  volume = {1858},
  number = {3},
  pages = {475--486},
  author = {Cyril F. Reboul and James C. Whisstock and Michelle A. Dunstone},
  title = {{G}iant {MACPF}/{CDC} pore forming toxins: {A} class of their own},
  journal = {Biochim. Biophys. Acta Biomembr.}
}

@article{Stewart-2014,
  doi = {10.1074/jbc.m113.544890},
  url = {https://doi.org/10.1074/jbc.m113.544890},
  year = {2014},
  month = feb,
  publisher = {American Society for Biochemistry {\&} Molecular Biology ({ASBMB})},
  volume = {289},
  number = {13},
  pages = {9172--9181},
  author = {Sarah E. Stewart and Stephanie C. Kondos and Antony Y. Matthews and Michael E. D{\textquotesingle}Angelo and Michelle A. Dunstone and James C. Whisstock and Joseph A. Trapani and Phillip I. Bird},
  title = {{T}he {P}erforin {P}ore {F}acilitates the {D}elivery of {C}ationic {C}argos},
  journal = {J. Biol. Chem.}
}

@article{Song-1996,
  doi = {10.1126/science.274.5294.1859},
  url = {https://doi.org/10.1126/science.274.5294.1859},
  year = {1996},
  month = dec,
  publisher = {American Association for the Advancement of Science ({AAAS})},
  volume = {274},
  number = {5294},
  pages = {1859--1865},
  author = {L. Song and M. R. Hobaugh and C. Shustak and S. Cheley and H. Bayley and J. E. Gouaux},
  title = {{S}tructure of {S}taphylococcal alpha-{H}emolysin, a {H}eptameric {T}ransmembrane {P}ore},
  journal = {Science}
}

@article{Galdiero-2004,
  doi = {10.1110/ps.03561104},
  url = {https://doi.org/10.1110/ps.03561104},
  year = {2004},
  month = jun,
  publisher = {Wiley},
  volume = {13},
  number = {6},
  pages = {1503--1511},
  author = {Stefania Galdiero and Eric Gouaux},
  title = {{H}igh resolution crystallographic studies of {\textalpha}-hemolysin-phospholipid complexes define heptamer-lipid head group interactions: Implication for understanding protein-lipid interactions},
  journal = {Prot. Sci.}
}

@article{Sugawara-2015,
  doi = {10.1016/j.toxicon.2015.09.033},
  url = {https://doi.org/10.1016/j.toxicon.2015.09.033},
  year = {2015},
  month = dec,
  publisher = {Elsevier {BV}},
  volume = {108},
  pages = {226--231},
  author = {Takaki Sugawara and Daichi Yamashita and Koji Kato and Zhao Peng and Junki Ueda and Jun Kaneko and Yoshiyuki Kamio and Yoshikazu Tanaka and Min Yao},
  title = {{S}tructural basis for pore-forming mechanism of staphylococcal {\textalpha}-hemolysin},
  journal = {Toxicon}
}

@article{Iacovache-2016,
  doi = {10.1038/ncomms12062},
  url = {https://doi.org/10.1038/ncomms12062},
  year = {2016},
  month = jul,
  publisher = {Springer Science and Business Media {LLC}},
  volume = {7},
  number = {1},
  author = {Ioan Iacovache and Sacha De Carlo and Nuria Cirauqui and Matteo Dal Peraro and F. Gisou van der Goot and Beno{\^{\i}}t Zuber},
  title = {{C}ryo-{EM} structure of aerolysin variants reveals a novel protein fold and the pore-formation process},
  journal = {Nat. Commun.}
}

@article{Hildebrand-1991,
  author = {Hildebrand, A and Pohl, M and Bhakdi, S}, 
  title = {Staphylococcus aureus alpha-toxin. Dual mechanism of binding to target cells.},
  volume = {266}, 
  number = {26}, 
  pages = {17195-17200}, 
  year = {1991},
  URL = {http://www.jbc.org/content/266/26/17195.abstract}, 
  eprint = {http://www.jbc.org/content/266/26/17195.full.pdf+html}, 
  journal = {J. Biol. Chem.} 
}

@article {Bhakdi-1991,
	author = {Bhakdi, S and Tranum-Jensen, J},
	title = {{A}lpha-toxin of {S}taphylococcus aureus.},
	volume = {55},
	number = {4},
	pages = {733--751},
	year = {1991},
	publisher = {{A}merican {S}ociety for {M}icrobiology {J}ournals},
	issn = {0146-0749},
	URL = {https://mmbr.asm.org/content/55/4/733},
	eprint = {https://mmbr.asm.org/content/55/4/733.full.pdf},
	journal = {Microbiol. Mol. Biol. Rev.}
}

@article{Bezrukov-1993,
  doi = {10.1103/physrevlett.70.2352},
  url = {https://doi.org/10.1103/physrevlett.70.2352},
  year = {1993},
  month = apr,
  publisher = {American Physical Society ({APS})},
  volume = {70},
  number = {15},
  pages = {2352--2355},
  author = {Sergey M. Bezrukov and John J. Kasianowicz},
  title = {{C}urrent noise reveals protonation kinetics and number of ionizable sites in an open protein ion channel},
  journal = {Phys. Rev. Lett.}
}

@article{Kasianowicz-1995,
  doi = {10.1016/s0006-3495(95)79879-4},
  url = {https://doi.org/10.1016/s0006-3495(95)79879-4},
  year = {1995},
  month = jul,
  publisher = {Elsevier {BV}},
  volume = {69},
  number = {1},
  pages = {94--105},
  author = {J.J. Kasianowicz and S.M. Bezrukov},
  title = {{P}rotonation dynamics of the alpha-toxin ion channel from spectral analysis of {pH}-dependent current fluctuations},
  journal = {Biophys. J.}
}

@article{Kasianowicz-1999,
  doi = {10.1016/s0006-3495(99)77247-4},
  url = {https://doi.org/10.1016/s0006-3495(99)77247-4},
  year = {1999},
  month = feb,
  publisher = {Elsevier {BV}},
  volume = {76},
  number = {2},
  pages = {837--845},
  author = {John J. Kasianowicz and Daniel L. Burden and Linda C. Han and Stephen Cheley and Hagan Bayley},
  title = {{G}enetically {E}ngineered {M}etal {I}on {B}inding {S}ites on the {O}utside of a {C}hannel{\textquotesingle}s {T}ransmembrane {\textbeta}-{B}arrel},
  journal = {Biophys. J.}
}

@article{Roozbahani-2020,
  doi = {10.1002/smtd.202000266},
  url = {https://doi.org/10.1002/smtd.202000266},
  year = {2020},
  month = aug,
  publisher = {Wiley},
  pages = {2000266},
  author = {Golbarg Mohammadi Roozbahani and Xiaohan Chen and Youwen Zhang and Liang Wang and Xiyun Guan},
  title = {{N}anopore {D}etection of {M}etal {I}ons: {C}urrent {S}tatus and {F}uture {D}irections},
  journal = {Small Methods}
}

@article{Menestrina-2001,
  doi = {10.1016/s0041-0101(01)00153-2},
  url = {https://doi.org/10.1016/s0041-0101(01)00153-2},
  year = {2001},
  month = nov,
  publisher = {Elsevier {BV}},
  volume = {39},
  number = {11},
  pages = {1661--1672},
  author = {Gianfranco Menestrina and Mauro Dalla Serra and Gilles Pr{\'{e}}vost},
  title = {Mode of action of {\textbeta}-barrel pore-forming toxins of the staphylococcal {\textalpha}-hemolysin family},
  journal = {Toxicon}
}

@article{Menestrina-1986,
  doi = {10.1007/bf01869935},
  url = {https://doi.org/10.1007/bf01869935},
  year = {1986},
  month = jun,
  publisher = {Springer Science and Business Media {LLC}},
  volume = {90},
  number = {2},
  pages = {177--190},
  author = {Gianfranco Menestrina},
  title = {Ionic channels formed by Staphylococcus aureus alpha-toxin: {V}oltage-dependent inhibition by divalent and trivalent cations},
  journal = {J. Membr. Biol.}
}

@article{Valeva-2006,
  doi = {10.1074/jbc.m601960200},
  url = {https://doi.org/10.1074/jbc.m601960200},
  year = {2006},
  month = jul,
  publisher = {American Society for Biochemistry {\&} Molecular Biology ({ASBMB})},
  volume = {281},
  number = {36},
  pages = {26014--26021},
  author = {Angela Valeva and Nadja Hellmann and Iwan Walev and Dennis Strand and Markus Plate and Fatima Boukhallouk and Antje Brack and Kentaro Hanada and Heinz Decker and Sucharit Bhakdi},
  title = {{E}vidence {T}hat {C}lustered {P}hosphocholine {H}ead {G}roups {S}erve as {S}ites for {B}inding and {A}ssembly of an {O}ligomeric {P}rotein {P}ore},
  journal = {J. Biol. Chem.}
}

@article{Das-2007,
  doi = {10.1002/cbic.200600474},
  url = {https://doi.org/10.1002/cbic.200600474},
  year = {2007},
  month = jun,
  publisher = {Wiley},
  volume = {8},
  number = {9},
  pages = {994--999},
  author = {Somes K. Das and Manjula Darshi and Stephen Cheley and Mark I. Wallace and Hagan Bayley},
  title = {{M}embrane {P}rotein {S}toichiometry {D}etermined from the {S}tep-{W}ise {P}hotobleaching of {D}ye-{L}abelled {S}ubunits},
  journal = {{ChemBioChem}}
}

@article{Thompson-2011,
  doi = {10.1016/j.bpj.2011.09.054},
  url = {https://doi.org/10.1016/j.bpj.2011.09.054},
  year = {2011},
  month = dec,
  publisher = {Elsevier {BV}},
  volume = {101},
  number = {11},
  pages = {2679--2683},
  author = {James R. Thompson and Br{\'{\i}}d Cronin and Hagan Bayley and Mark I. Wallace},
  title = {{R}apid {A}ssembly of a {M}ultimeric {M}embrane {P}rotein {P}ore},
  journal = {Biophys. J.}
}

@article{Lee-2016b,
  doi = {10.1098/rsif.2015.0762},
  url = {https://doi.org/10.1098/rsif.2015.0762},
  year = {2016},
  month = jan,
  publisher = {The Royal Society},
  volume = {13},
  number = {114},
  pages = {20150762},
  author = {A. A. Lee and M. J. Senior and M. I. Wallace and T. E. Woolley and I. M. Griffiths},
  title = {{D}issecting the self-assembly kinetics of multimeric pore-forming toxins},
  journal = {J. Royal Soc. Interface}
}

@article{Bayoumi-2020,
  doi = {10.1021/acs.nanolett.0c04464},
  url = {https://doi.org/10.1021/acs.nanolett.0c04464},
  year = {2020},
  month = jan,
  publisher = {American Chemical Society ({ACS})},
  volume = {},
  number = {},
  pages = {},
  author = {Mariam Bayoumi and Stefanos K. Nomidis and Kherim Willems and Enrico Carlon and Giovanni Maglia},
  title = {{A}utonomous {A}nd {A}ctive {T}ransport {O}perated by an {E}ntropic {DNA} {P}iston},
  journal = {Accepted by Nano Lett.}
}

@article{Zernia-2020,
  doi = {10.1021/acsnano.9b09434},
  url = {https://doi.org/10.1021/acsnano.9b09434},
  year = {2020},
  month = jan,
  publisher = {American Chemical Society ({ACS})},
  volume = {14},
  number = {2},
  pages = {2296--2307},
  author = {Sarah Zernia and Nieck Jordy van der Heide and Nicole St{\'{e}}phanie Galenkamp and Giorgos Gouridis and Giovanni Maglia},
  title = {{C}urrent {B}lockades of {P}roteins inside {N}anopores for {R}eal-{T}ime {M}etabolome {A}nalysis},
  journal = {{ACS} Nano}
}

@article{Galenkamp-2020,
  doi = {10.1038/s41557-020-0437-0},
  url = {https://doi.org/10.1038/s41557-020-0437-0},
  year = {2020},
  month = apr,
  publisher = {Springer Science and Business Media {LLC}},
  volume = {12},
  number = {5},
  pages = {481--488},
  author = {Nicole St{\'{e}}phanie Galenkamp and Annemie Biesemans and Giovanni Maglia},
  title = {{D}irectional conformer exchange in dihydrofolate reductase revealed by single-molecule nanopore recordings},
  journal = {Nat. Chem.}
}

@article{Huang-2020,
  doi = {10.1021/acs.nanolett.0c00877},
  url = {https://doi.org/10.1021/acs.nanolett.0c00877},
  year = {2020},
  month = apr,
  publisher = {American Chemical Society ({ACS})},
  volume = {20},
  number = {5},
  pages = {3819--3827},
  author = {Gang Huang and Kherim Willems and Mart Bartelds and Pol van Dorpe and Misha Soskine and Giovanni Maglia},
  title = {{E}lectro-{O}smotic {V}ortices {P}romote the {C}apture of {F}olded {P}roteins by {PlyAB} {N}anopores},
  journal = {Nano Lett.}
}

@article{Willems-2020,
  doi = {10.1039/d0nr03114c},
  url = {https://doi.org/10.1039/d0nr03114c},
  year = {2020},
  publisher = {Royal Society of Chemistry ({RSC})},
  volume = {},
  number = {},
  pages = {},
  author = {Kherim Willems and Dino Rui{\'{c}} and Florian Lucas and Ujjal Barman and Niels Verellen and Johan Hofkens and Giovanni Maglia and Pol Van Dorpe},
  title = {{A}ccurate modeling of a biological nanopore with an extended continuum framework},
  journal = {Nanoscale}
}

@article{Willems-Ruic-Biesemans-2019,
  doi = {10.1021/acsnano.8b09137},
  url = {https://doi.org/10.1021/acsnano.8b09137},
  year = {2019},
  month = aug,
  publisher = {American Chemical Society ({ACS})},
  volume = {13},
  number = {9},
  pages = {9980--9992},
  author = {Kherim Willems and Dino Rui{\'{c}} and Annemie Biesemans and Nicole St{\'{e}}phanie Galenkamp and Pol Van Dorpe and Giovanni Maglia},
  title = {{E}ngineering and {M}odeling the {E}lectrophoretic {T}rapping of a {S}ingle {P}rotein {I}nside a {N}anopore},
  journal = {{ACS} Nano}
}

@article{Huang-2019,
  doi = {10.1038/s41467-019-08761-6},
  url = {https://doi.org/10.1038/s41467-019-08761-6},
  year = {2019},
  month = feb,
  publisher = {Springer Nature},
  volume = {10},
  number = {1},
  author = {Gang Huang and Arnout Voet and Giovanni Maglia},
  title = {{FraC} nanopores with adjustable diameter identify the mass of opposite-charge peptides with 44 dalton resolution},
  journal = {Nat. Commun.}
}

@article{Mutter-2018,
  doi = {10.1021/acschembio.8b00720},
  url = {https://doi.org/10.1021/acschembio.8b00720},
  year = {2018},
  month = oct,
  publisher = {American Chemical Society ({ACS})},
  volume = {13},
  number = {11},
  pages = {3153--3160},
  author = {Natalie L. Mutter and Misha Soskine and Gang Huang and In{\^{e}}s S. Albuquerque and Gon{\c{c}}alo J. L. Bernardes and Giovanni Maglia},
  title = {{M}odular {P}ore-{F}orming {I}mmunotoxins with {C}aged {C}ytotoxicity {T}ailored by {D}irected {E}volution},
  journal = {{ACS} Chem. Biol.}
}

@article{Galenkamp-2018,
  doi = {10.1038/s41467-018-06534-1},
  url = {https://doi.org/10.1038/s41467-018-06534-1},
  year = {2018},
  month = oct,
  publisher = {Springer Nature},
  volume = {9},
  number = {1},
  author = {Nicole St{\'{e}}phanie Galenkamp and Misha Soskine and Jos Hermans and Carsten Wloka and Giovanni Maglia},
  title = {{D}irect {E}lectrical {Q}uantification of {G}lucose and {A}sparagine from {B}odily {F}luids {U}sing {N}anopores},
  journal = {Nat. Commun.}
}

@article{Huang-2017,
  doi = {10.1038/s41467-017-01006-4},
  url = {https://doi.org/10.1038/s41467-017-01006-4},
  year = {2017},
  month = oct,
  publisher = {Springer Science and Business Media {LLC}},
  volume = {8},
  number = {1},
  author = {Gang Huang and Kherim Willems and Misha Soskine and Carsten Wloka and Giovanni Maglia},
  title = {{E}lectro-osmotic capture and ionic discrimination of peptide and protein biomarkers with {FraC} nanopores},
  journal = {Nat. Commun.}
}

@article{Wloka-2017,
  doi = {10.1021/acsnano.6b07760},
  url = {https://doi.org/10.1021/acsnano.6b07760},
  year = {2017},
  month = mar,
  publisher = {American Chemical Society ({ACS})},
  volume = {11},
  number = {5},
  pages = {4387--4394},
  author = {Carsten Wloka and Veerle Van Meervelt and Dewi van Gelder and Natasha Danda and Nienke Jager and Chris P. Williams and Giovanni Maglia},
  title = {{L}abel-{F}ree and {R}eal-{T}ime {D}etection of {P}rotein {U}biquitination with a {B}iological {N}anopore},
  journal = {{ACS} Nano}
}

@article{VanMeervelt-2017,
  doi = {10.1021/jacs.7b10106},
  url = {https://doi.org/10.1021/jacs.7b10106},
  year = {2017},
  month = dec,
  publisher = {American Chemical Society ({ACS})},
  volume = {139},
  number = {51},
  pages = {18640--18646},
  author = {Veerle Van Meervelt and Misha Soskine and Shubham Singh and Gea K. Schuurman-Wolters and Hein J. Wijma and Bert Poolman and Giovanni Maglia},
  title = {{R}eal-{T}ime {C}onformational {C}hanges and {C}ontrolled {O}rientation of {N}ative {P}roteins {I}nside a {P}rotein {N}anoreactor},
  journal = {J. Am. Chem. Soc.}
}

@article{Willems-VanMeervelt-2017,
  doi = {10.1098/rstb.2016.0230},
  url = {https://doi.org/10.1098/rstb.2016.0230},
  year = {2017},
  month = jun,
  publisher = {The Royal Society},
  volume = {372},
  number = {1726},
  pages = {20160230},
  author = {Kherim Willems and Veerle Van Meervelt and Carsten Wloka and Giovanni Maglia},
  title = {{S}ingle-{M}olecule {N}anopore {E}nzymology},
  journal = {Philos. Trans. Royal Soc. B}
}

@article{Wloka-2016,
  doi = {10.1002/anie.201606742},
  url = {https://doi.org/10.1002/anie.201606742},
  year = {2016},
  month = sep,
  publisher = {Wiley},
  volume = {55},
  number = {40},
  pages = {12494--12498},
  author = {Carsten Wloka and Natalie Lisa Mutter and Misha Soskine and Giovanni Maglia},
  title = {{A}lpha-{H}elical {F}ragaceatoxin{~}{C} {N}anopore {E}ngineered for {D}ouble-{S}tranded and {S}ingle-{S}tranded {N}ucleic {A}cid {A}nalysis},
  journal = {Angew. Chem. Int. Ed.}
}

@article{Franceschini-2016,
  doi = {10.1021/acsnano.6b03159},
  url = {https://doi.org/10.1021/acsnano.6b03159},
  year = {2016},
  month = aug,
  publisher = {American Chemical Society ({ACS})},
  volume = {10},
  number = {9},
  pages = {8394--8402},
  author = {Lorenzo Franceschini and Tine Brouns and Kherim Willems and Enrico Carlon and Giovanni Maglia},
  title = {{DNA} {T}ranslocation {T}hrough {N}anopores at {P}hysiological {I}onic {S}trengths {R}equires {P}recise {N}anoscale {E}ngineering},
  journal = {{ACS} Nano}
}

@article{Soskine-Biesemans-2015,
  doi = {10.1021/jacs.5b01520},
  url = {https://doi.org/10.1021/jacs.5b01520},
  year = {2015},
  month = apr,
  publisher = {American Chemical Society ({ACS})},
  volume = {137},
  number = {17},
  pages = {5793--5797},
  author = {Misha Soskine and Annemie Biesemans and Giovanni Maglia},
  title = {{S}ingle-{M}olecule {A}nalyte {R}ecognition with {ClyA} {N}anopores {E}quipped with {I}nternal {P}rotein {A}daptors},
  journal = {J. Am. Chem. Soc.}
}

@article{Biesemans-2015,
  doi = {10.1021/acs.nanolett.5b02309},
  url = {https://doi.org/10.1021/acs.nanolett.5b02309},
  year = {2015},
  month = aug,
  publisher = {American Chemical Society ({ACS})},
  volume = {15},
  number = {9},
  pages = {6076--6081},
  author = {Annemie Biesemans and Misha Soskine and Giovanni Maglia},
  title = {{A} {P}rotein {R}otaxane {C}ontrols the {T}ranslocation of {P}roteins {A}cross a {ClyA}
  {N}anopore},
  journal = {Nano Lett.}
}

@article{Ho-2015,
  doi = {10.1126/sciadv.1500905},
  url = {https://doi.org/10.1126/sciadv.1500905},
  year = {2015},
  month = dec,
  publisher = {American Association for the Advancement of Science ({AAAS})},
  volume = {1},
  number = {11},
  pages = {e1500905},
  author = {Ching-Wen Ho and Veerle Van Meervelt and Keng-Chang Tsai and Pieter-Jan De Temmerman and Jan Mast and Giovanni Maglia},
  title = {{E}ngineering a nanopore with co-chaperonin function},
  journal = {Sci. Adv.}
}

@article{VanMeervelt-2014,
  doi = {10.1021/nn506077e},
  url = {https://doi.org/10.1021/nn506077e},
  year = {2014},
  month = dec,
  publisher = {American Chemical Society ({ACS})},
  volume = {8},
  number = {12},
  pages = {12826--12835},
  author = {Veerle Van Meervelt and Misha Soskine and Giovanni Maglia},
  title = {{D}etection of {T}wo {I}someric {B}inding {C}onfigurations in a {P}rotein{\textendash}{A}ptamer {C}omplex with a {B}iological {N}anopore},
  journal = {{ACS} Nano}
}

@article{Soskine-2013,
  doi = {10.1021/ja4053398},
  url = {https://doi.org/10.1021/ja4053398},
  year = {2013},
  month = aug,
  publisher = {American Chemical Society ({ACS})},
  volume = {135},
  number = {36},
  pages = {13456--13463},
  author = {Misha Soskine and Annemie Biesemans and Marc De Maeyer and Giovanni Maglia},
  title = {{T}uning the {S}ize and {P}roperties of {ClyA} {N}anopores {A}ssisted by {D}irected {E}volution},
  journal = {J. Am. Chem. Soc.}
}

@article{Soskine-2012,
  doi = {10.1021/nl3024438},
  url = {https://doi.org/10.1021/nl3024438},
  year = {2012},
  month = aug,
  publisher = {American Chemical Society ({ACS})},
  volume = {12},
  number = {9},
  pages = {4895--4900},
  author = {Misha Soskine and Annemie Biesemans and Benjamien Moeyaert and Stephen Cheley and Hagan Bayley and Giovanni Maglia},
  title = {{A}n {E}ngineered {ClyA} {N}anopore {D}etects {F}olded {T}arget {P}roteins by {S}elective {E}xternal {A}ssociation and {P}ore {E}ntry},
  journal = {Nano Lett.}
}

@article{Franceschini-2013,
  doi = {10.1038/ncomms3415},
  url = {https://doi.org/10.1038/ncomms3415},
  year = {2013},
  month = sep,
  publisher = {Springer Science and Business Media {LLC}},
  volume = {4},
  number = {1},
  author = {Lorenzo Franceschini and Misha Soskine and Annemie Biesemans and Giovanni Maglia},
  title = {{A} nanopore machine promotes the vectorial transport of {DNA} across membranes},
  journal = {Nat. Commun.}
}

@article{Maglia-2010,
  author    = {Maglia, Giovanni and Heron, Andrew J. and Stoddart, David and Japrung,
  Deanpen and Bayley, Hagan},
  title     = {{A}nalysis of {S}ingle {N}ucleic {A}cid {M}olecules with {P}rotein {N}anopores},
  booktitle = {{S}ingle {M}olecule {T}ools, {P}art {B}:{S}uper-{R}esolution, {P}article {T}racking,
  {M}ultiparameter, and {F}orce {B}ased {M}ethods},
  series    = {Meth. Enzymol.},
  year      = {2010},
  volume    = {475},
  pages     = {591--623},
  doi       = {10.1016/S0076-6879(10)75022-9},
  month     = {1}
}

@article{Maglia-2008,
  doi = {10.1073/pnas.0808296105},
  url = {https://doi.org/10.1073/pnas.0808296105},
  year = {2008},
  month = dec,
  publisher = {Proceedings of the National Academy of Sciences},
  volume = {105},
  number = {50},
  pages = {19720--19725},
  author = {G. Maglia and M. R. Restrepo and E. Mikhailova and H. Bayley},
  title = {{E}nhanced translocation of single {DNA} molecules through {\textalpha}-hemolysin nanopores by manipulation of internal charge},
  journal = {Proc. Natl. Acad. Sci. U.S.A.}
}

@article{Li-2020,
  doi = {10.1021/acsnano.9b07385},
  url = {https://doi.org/10.1021/acsnano.9b07385},
  year = {2020},
  month = jan,
  publisher = {American Chemical Society ({ACS})},
  volume = {14},
  number = {2},
  pages = {1727--1737},
  author = {Xin Li and Kuo Hao Lee and Spencer Shorkey and Jianhan Chen and Min Chen},
  title = {{D}ifferent {A}nomeric {S}ugar {B}ound {S}tates of {M}altose {B}inding {P}rotein {R}esolved by a {C}ytolysin {A} {N}anopore {T}weezer},
  journal = {{ACS} Nano}
}

@article{Nomidis-2018,
  doi = {10.1088/1361-648x/aacc01},
  url = {https://doi.org/10.1088/1361-648x/aacc01},
  year = {2018},
  month = jun,
  publisher = {{IOP} Publishing},
  volume = {30},
  number = {30},
  pages = {304001},
  author = {Stefanos K Nomidis and Jef Hooyberghs and Giovanni Maglia and Enrico Carlon},
  title = {{DNA} capture into the {ClyA} nanopore: diffusion-limited versus reaction-limited processes},
  journal = {J. Phys. Condens. Matter}
}

@article{Benke-2015,
  doi = {10.1038/ncomms7198},
  url = {https://doi.org/10.1038/ncomms7198},
  year = {2015},
  month = feb,
  publisher = {Springer Science and Business Media {LLC}},
  volume = {6},
  number = {1},
  author = {Stephan Benke and Daniel Roderer and Bengt Wunderlich and Daniel Nettels and Rudi Glockshuber and Benjamin Schuler},
  title = {The assembly dynamics of the cytolytic pore toxin {ClyA}},
  journal = {Nat. Commun.}
}

@article{Vaidyanathan-2014,
  doi = {10.1039/c3ra45159c},
  url = {https://doi.org/10.1039/c3ra45159c},
  year = {2014},
  publisher = {Royal Society of Chemistry ({RSC})},
  volume = {4},
  number = {10},
  pages = {4930},
  author = {M. S. Vaidyanathan and Pradeep Sathyanarayana and Prabal K. Maiti and Sandhya S. Visweswariah and K. G. Ayappa},
  title = {{L}ysis dynamics and membrane oligomerization pathways for {C}ytolysin {A} ({ClyA}) pore-forming toxin},
  journal = {{RSC} Adv.}
}

@article{Rojko-2016,
  doi = {10.1016/j.bbamem.2015.09.007},
  url = {https://doi.org/10.1016/j.bbamem.2015.09.007},
  year = {2016},
  month = mar,
  publisher = {Elsevier {BV}},
  volume = {1858},
  number = {3},
  pages = {446--456},
  author = {Nejc Rojko and Mauro Dalla Serra and Peter Ma{\v{c}}ek and Gregor Anderluh},
  title = {Pore formation by actinoporins, cytolysins from sea anemones},
  journal = {Biochim. Biophys. Acta Biomembr.}
}

@article{Cosentino-2016,
  doi = {10.1016/j.bbamem.2015.09.013},
  url = {https://doi.org/10.1016/j.bbamem.2015.09.013},
  year = {2016},
  month = mar,
  publisher = {Elsevier {BV}},
  volume = {1858},
  number = {3},
  pages = {457--466},
  author = {Katia Cosentino and Uris Ros and Ana J. Garc{\'{i}}a-S{\'{a}}ez},
  title = {Assembling the puzzle: Oligomerization of {\textalpha}-pore forming proteins in membranes},
  journal = {Biochim. Biophys. Acta Biomembr.}
}

@article{Leung-2017,
  doi = {10.1038/nnano.2016.303},
  url = {https://doi.org/10.1038/nnano.2016.303},
  year = {2017},
  month = feb,
  publisher = {Springer Science and Business Media {LLC}},
  volume = {12},
  number = {5},
  pages = {467--473},
  author = {Carl Leung and Adrian W. Hodel and Amelia J. Brennan and Natalya Lukoyanova and Sharon Tran and Colin M. House and Stephanie C. Kondos and James C. Whisstock and Michelle A. Dunstone and Joseph A. Trapani and Ilia Voskoboinik and Helen R. Saibil and Bart W. Hoogenboom},
  title = {{R}eal-time visualization of perforin nanopore assembly},
  journal = {Nat. Nanotechnol.}
}

@article{Zhao-2019,
  doi = {10.1021/acsnano.8b09266},
  url = {https://doi.org/10.1021/acsnano.8b09266},
  year = {2019},
  month = feb,
  publisher = {American Chemical Society ({ACS})},
  author = {Shidi Zhao and Laura Restrepo-P{\'{e}}rez and Misha Soskine and Giovanni Maglia and Chirlmin Joo and Cees Dekker and Aleksei Aksimentiev},
  title = {{E}lectro-{M}echanical {C}onductance {M}odulation of a {N}anopore {U}sing a {R}emovable {G}ate},
  journal = {{ACS} Nano}
}

@article{Watanabe-2017,
  doi = {10.1021/acs.analchem.7b01550},
  url = {https://doi.org/10.1021/acs.analchem.7b01550},
  year = {2017},
  month = oct,
  publisher = {American Chemical Society ({ACS})},
  volume = {89},
  number = {21},
  pages = {11269--11277},
  author = {Hirokazu Watanabe and Alberto Gubbiotti and Mauro Chinappi and Natsumi Takai and Koji Tanaka and Kouhei Tsumoto and Ryuji Kawano},
  title = {{A}nalysis of {P}ore {F}ormation and {P}rotein {T}ranslocation {U}sing {L}arge {B}iological {N}anopores},
  journal = {Anal. Chem.}
}

@article{Garcia-Ortega-2011,
  doi = {10.1016/j.bbamem.2011.05.012},
  url = {https://doi.org/10.1016/j.bbamem.2011.05.012},
  year = {2011},
  month = sep,
  publisher = {Elsevier {BV}},
  volume = {1808},
  number = {9},
  pages = {2275--2288},
  author = {Luc{\'{i}}a Garc{\'{i}}a-Ortega and Jorge Alegre-Cebollada and Sara Garc{\'{i}}a-Linares and Marta Bruix and {\'{a}}lvaro Mart{\'{i}}nez-del-Pozo and Jos{\'{e}} G. Gavilanes},
  title = {{T}he behavior of sea anemone actinoporins at the water{\textendash}membrane interface},
  journal = {Biochim. Biophys. Acta Biomembr.}
}

@article{Wong-Ekkabut-2016b,
  doi = {10.1007/s10867-015-9396-x},
  url = {https://doi.org/10.1007/s10867-015-9396-x},
  year = {2016},
  month = aug,
  publisher = {Springer Science and Business Media {LLC}},
  volume = {42},
  number = {1},
  pages = {133--146},
  author = {Jirasak Wong-ekkabut and Mikko Karttunen},
  title = {{M}olecular dynamics simulation of water permeation through the alpha-hemolysin channel},
  journal = {J. Biol. Phys.}
}

@article{Aguilella-Arzo-2017,
  doi = {10.3390/e19030116},
  url = {https://doi.org/10.3390/e19030116},
  year = {2017},
  month = mar,
  publisher = {{MDPI} {AG}},
  volume = {19},
  number = {3},
  pages = {116},
  author = {Marcel Aguilella-Arzo and Mar{\'{i}}a Queralt-Mart{\'{i}}n and Mar{\'{i}}a-Lid{\'{o}}n Lopez and Antonio Alcaraz},
  title = {{F}luctuation-{D}riven {T}ransport in {B}iological {N}anopores. {A} {3D} {P}oisson{\textendash}{N}ernst{\textendash}{P}lanck {S}tudy},
  journal = {Entropy}
}

@article{Aguilella-Arzo-2020,
  doi = {10.1016/j.bioelechem.2019.107371},
  url = {https://doi.org/10.1016/j.bioelechem.2019.107371},
  year = {2020},
  month = feb,
  publisher = {Elsevier {BV}},
  volume = {131},
  pages = {107371},
  author = {Marcel Aguilella-Arzo and Vicente M. Aguilella},
  title = {{A}ccess resistance in protein nanopores. {A} structure-based computational approach},
  journal = {Bioelectrochemistry}
}

@article{Simakov-2018,
  doi = {10.1007/s00232-018-0013-3},
  url = {https://doi.org/10.1007/s00232-018-0013-3},
  year = {2018},
  month = jan,
  publisher = {Springer Science and Business Media {LLC}},
  volume = {251},
  number = {3},
  pages = {393--404},
  author = {Nikolay A. Simakov and Maria G. Kurnikova},
  title = {{M}embrane {P}osition {D}ependency of the {pKa} and {C}onductivity of the {P}rotein {I}on {C}hannel},
  journal = {J. Membr. Biol.}
}

@article{Howorka-2001,
  doi = {10.1038/90236},
  url = {https://doi.org/10.1038/90236},
  year = {2001},
  month = jul,
  publisher = {Springer Science and Business Media {LLC}},
  volume = {19},
  number = {7},
  pages = {636--639},
  author = {Stefan Howorka and Stephen Cheley and Hagan Bayley},
  title = {{S}equence-specific detection of individual {DNA} strands using engineered nanopores},
  journal = {Nat. Biotechnol.}
}

@article{Restrepo-Perez-2019a,
  doi = {10.1021/acsnano.9b05156},
  url = {https://doi.org/10.1021/acsnano.9b05156},
  year = {2019},
  month = sep,
  publisher = {American Chemical Society ({ACS})},
  author = {Laura Restrepo-P{\'{e}}rez and Gang Huang and Peggy R. Bohl{\"{a}}nder and Nathalie Worp and Rienk Eelkema and Giovanni Maglia and Chirlmin Joo and Cees Dekker},
  title = {{R}esolving {C}hemical {M}odifications to a {S}ingle {A}mino {A}cid within a {P}eptide {U}sing a {B}iological {N}anopore},
  journal = {{ACS} Nano}
}

@article{Stefureac-2006,
  doi = {10.1021/bi0604835},
  url = {https://doi.org/10.1021/bi0604835},
  year = {2006},
  month = aug,
  publisher = {American Chemical Society ({ACS})},
  volume = {45},
  number = {30},
  pages = {9172--9179},
  author = {Radu Stefureac and Yi-tao Long and Heinz-Bernhard Kraatz and Peter Howard and Jeremy S. Lee},
  title = {{T}ransport of {\textalpha}-{H}elical {P}eptides through {\textalpha}-{H}emolysin and {A}erolysin {P}ores},
  journal = {Biochemistry}
}

@article{Asandei-2015,
  doi = {10.1038/srep10419},
  url = {https://doi.org/10.1038/srep10419},
  year = {2015},
  month = jun,
  publisher = {Springer Nature},
  volume = {5},
  number = {1},
  author = {Alina Asandei and Mauro Chinappi and Jong-kook Lee and Chang Ho Seo and Loredana Mereuta and Yoonkyung Park and Tudor Luchian},
  title = {{P}lacement of {O}ppositely {C}harged {A}minoacids at a {P}olypeptide {T}ermini {D}etermines the {V}oltage-{C}ontrolled {B}raking of {P}olymer {T}ransport {T}hrough {N}anometer-{S}cale {P}ores},
  journal = {Sci. Rep.}
}

@article{Asandei-2015b,
  doi = {10.1021/acsami.5b04406},
  url = {https://doi.org/10.1021/acsami.5b04406},
  year = {2015},
  month = jul,
  publisher = {American Chemical Society ({ACS})},
  volume = {7},
  number = {30},
  pages = {16706--16714},
  author = {Alina Asandei and Mauro Chinappi and Hee-Kyoung Kang and Chang Ho Seo and Loredana Mereuta and Yoonkyung Park and Tudor Luchian},
  title = {{A}cidity-{M}ediated, {E}lectrostatic {T}uning of {A}symmetrically {C}harged {P}eptides {I}nteractions with {P}rotein {N}anopores},
  journal = {{ACS} Appl. Mater. Interfaces}
}

@article{Asandei-2016,
  doi = {10.1021/acsami.6b03697},
  url = {https://doi.org/10.1021/acsami.6b03697},
  year = {2016},
  month = may,
  publisher = {American Chemical Society ({ACS})},
  volume = {8},
  number = {20},
  pages = {13166--13179},
  author = {Alina Asandei and Irina Schiopu and Mauro Chinappi and Chang Ho Seo and Yoonkyung Park and Tudor Luchian},
  title = {{E}lectroosmotic {T}rap {A}gainst the {E}lectrophoretic {F}orce {N}ear a {P}rotein {N}anopore {R}eveals {P}eptide {D}ynamics {D}uring {C}apture and {T}ranslocation},
  journal = {{ACS} Appl. Mater. Interfaces}
}

@article{Li-2018,
  doi = {10.1002/celc.201800288},
  url = {https://doi.org/10.1002/celc.201800288},
  year = {2018},
  month = may,
  publisher = {Wiley},
  volume = {6},
  number = {1},
  pages = {126--129},
  author = {Shuang Li and Chan Cao and Jie Yang and Yi-Tao Long},
  title = {{D}etection of {P}eptides with {D}ifferent {C}harges and {L}engths by {U}sing the {A}erolysin {N}anopore},
  journal = {{ChemElectroChem}}
}

@article{Ouldali-2020,
  doi = {10.1038/s41587-019-0345-2},
  url = {https://doi.org/10.1038/s41587-019-0345-2},
  year = {2019},
  month = dec,
  publisher = {Springer Science and Business Media {LLC}},
  volume = {38},
  number = {2},
  pages = {176--181},
  author = {Hadjer Ouldali and Kumar Sarthak and Tobias Ensslen and Fabien Piguet and Philippe Manivet and Juan Pelta and Jan C. Behrends and Aleksei Aksimentiev and Abdelghani Oukhaled},
  title = {{E}lectrical recognition of the twenty proteinogenic amino acids using an aerolysin nanopore},
  journal = {Nat. Biotechnol.}
}

@article{Goyal-2014,
  doi = {10.1038/nature13768},
  url = {https://doi.org/10.1038/nature13768},
  year = {2014},
  month = sep,
  publisher = {Springer Science and Business Media {LLC}},
  volume = {516},
  number = {7530},
  pages = {250--253},
  author = {Parveen Goyal and Petya V. Krasteva and Nani Van Gerven and Francesca Gubellini and Imke Van den Broeck and Anastassia Troupiotis-Tsaïlaki and Wim Jonckheere and G{\'{e}}rard P{\'{e}}hau-Arnaudet and Jerome S. Pinkner and Matthew R. Chapman and Scott J. Hultgren and Stefan Howorka and R{\'{e}}mi Fronzes and Han Remaut},
  title = {{S}tructural and mechanistic insights into the bacterial amyloid secretion channel {CsgG}},
  journal = {Nature}
}

@article{Manrao-2012,
  doi = {10.1038/nbt.2171},
  url = {https://doi.org/10.1038/nbt.2171},
  year = {2012},
  month = mar,
  publisher = {Springer Science and Business Media {LLC}},
  volume = {30},
  number = {4},
  pages = {349--353},
  author = {Elizabeth A Manrao and Ian M Derrington and Andrew H Laszlo and Kyle W Langford and Matthew K Hopper and Nathaniel Gillgren and Mikhail Pavlenok and Michael Niederweis and Jens H Gundlach},
  title = {Reading {DNA} at single-nucleotide resolution with a mutant {MspA} nanopore and phi29 {DNA} polymerase},
  journal = {Nat. Biotechnol.}
}

@article{Kasianowicz-1996,
  doi = {10.1073/pnas.93.24.13770},
  url = {https://doi.org/10.1073/pnas.93.24.13770},
  year = {1996},
  month = nov,
  publisher = {Proceedings of the National Academy of Sciences},
  volume = {93},
  number = {24},
  pages = {13770--13773},
  author = {J. J. Kasianowicz and E. Brandin and D. Branton and D. W. Deamer},
  title = {{C}haracterization of individual polynucleotide molecules using a membrane channel},
  journal = {Proc. Natl. Acad. Sci. U.S.A.}
}

@article{Meller-2000,
  doi = {10.1073/pnas.97.3.1079},
  url = {https://doi.org/10.1073/pnas.97.3.1079},
  year = {2000},
  month = feb,
  publisher = {Proceedings of the National Academy of Sciences},
  volume = {97},
  number = {3},
  pages = {1079--1084},
  author = {A. Meller and L. Nivon and E. Brandin and J. Golovchenko and D. Branton},
  title = {{R}apid nanopore discrimination between single polynucleotide molecules},
  journal = {Proc. Natl. Acad. Sci. U.S.A.}
}

@article{Stoddart-2009,
  doi = {10.1073/pnas.0901054106},
  url = {https://doi.org/10.1073/pnas.0901054106},
  year = {2009},
  month = apr,
  publisher = {Proceedings of the National Academy of Sciences},
  volume = {106},
  number = {19},
  pages = {7702--7707},
  author = {D. Stoddart and A. J. Heron and E. Mikhailova and G. Maglia and H. Bayley},
  title = {{S}ingle-nucleotide discrimination in immobilized {DNA} oligonucleotides with a biological nanopore},
  journal = {Proc. Natl. Acad. Sci. U.S.A.}
}

@article{Firnkes-2010,
  doi = {10.1021/nl100861c},
  url = {https://doi.org/10.1021/nl100861c},
  year = {2010},
  month = jun,
  publisher = {American Chemical Society ({ACS})},
  volume = {10},
  number = {6},
  pages = {2162--2167},
  author = {Matthias Firnkes and Daniel Pedone and Jelena Knezevic and Markus D{\"{o}}blinger and Ulrich Rant},
  title = {{E}lectrically {F}acilitated {T}ranslocations of {P}roteins through {S}ilicon {N}itride {N}anopores: {C}onjoint and {C}ompetitive {A}ction of {D}iffusion, {E}lectrophoresis, and {E}lectroosmosis},
  journal = {Nano Lett.}
}

@article{Buchsbaum-2013,
  doi = {10.1021/nl401968r},
  url = {https://doi.org/10.1021/nl401968r},
  year = {2013},
  month = jul,
  publisher = {American Chemical Society ({ACS})},
  volume = {13},
  number = {8},
  pages = {3890--3896},
  author = {Steven F. Buchsbaum and Nick Mitchell and Hugh Martin and Matt Wiggin and Andre Marziali and Peter V. Coveney and Zuzanna Siwy and Stefan Howorka},
  title = {{D}isentangling {S}teric and {E}lectrostatic {F}actors in {N}anoscale {T}ransport {T}hrough {C}onfined {S}pace},
  journal = {Nano Lett.}
}

@article{Jain-2018,
  doi = {10.1038/nbt.4060},
  url = {https://doi.org/10.1038/nbt.4060},
  year = {2018},
  month = jan,
  publisher = {Springer Science and Business Media {LLC}},
  volume = {36},
  number = {4},
  pages = {338--345},
  author = {Miten Jain and Sergey Koren and Karen H Miga and Josh Quick and Arthur C Rand and Thomas A Sasani and John R Tyson and Andrew D Beggs and Alexander T Dilthey and Ian T Fiddes and Sunir Malla and Hannah Marriott and Tom Nieto and Justin O{\textquotesingle}Grady and Hugh E Olsen and Brent S Pedersen and Arang Rhie and Hollian Richardson and Aaron R Quinlan and Terrance P Snutch and Louise Tee and Benedict Paten and Adam M Phillippy and Jared T Simpson and Nicholas J Loman and Matthew Loose},
  title = {{N}anopore sequencing and assembly of a human genome with ultra-long reads},
  journal = {Nat. Biotechnol.}
}

@manual{ONT-2020,
    title = {Oxford Nanopore Technologies, https://nanoporetech.com},
    url = {https://nanoporetech.com},
    author = {},
    year = {2020 (accessed August 17, 2020)}
}

@article{Brown-2016,
  doi = {10.1038/nbt.3622},
  url = {https://doi.org/10.1038/nbt.3622},
  year = {2016},
  month = aug,
  publisher = {Springer Science and Business Media {LLC}},
  volume = {34},
  number = {8},
  pages = {810--811},
  author = {Clive G Brown and James Clarke},
  title = {{N}anopore development at Oxford Nanopore},
  journal = {Nat. Biotechnol.}
}

@article{Butler-2008,
  doi = {10.1073/pnas.0807514106},
  url = {https://doi.org/10.1073/pnas.0807514106},
  year = {2008},
  month = dec,
  publisher = {Proceedings of the National Academy of Sciences},
  volume = {105},
  number = {52},
  pages = {20647--20652},
  author = {T. Z. Butler and M. Pavlenok and I. M. Derrington and M. Niederweis and J. H. Gundlach},
  title = {Single-molecule {DNA} detection with an engineered {MspA} protein nanopore},
  journal = {Proc. Natl. Acad. Sci. U.S.A.}
}

@article{Wick-2019,
  doi = {10.1186/s13059-019-1727-y},
  url = {https://doi.org/10.1186/s13059-019-1727-y},
  year = {2019},
  month = jun,
  publisher = {Springer Science and Business Media {LLC}},
  volume = {20},
  number = {1},
  author = {Ryan R. Wick and Louise M. Judd and Kathryn E. Holt},
  title = {{P}erformance of neural network basecalling tools for {O}xford {N}anopore sequencing},
  journal = {Genome Biol.}
}

@article{Howorka-2020,
  doi = {10.1038/s41587-019-0401-y},
  url = {https://doi.org/10.1038/s41587-019-0401-y},
  year = {2020},
  month = jan,
  publisher = {Springer Science and Business Media {LLC}},
  volume = {38},
  number = {2},
  pages = {159--160},
  author = {Stefan Howorka and Zuzanna S. Siwy},
  title = {Reading amino acids in a nanopore},
  journal = {Nat. Biotechnol.}
}

@article{Rodriguez-Larrea-2013,
  doi = {10.1038/nnano.2013.22},
  url = {https://doi.org/10.1038/nnano.2013.22},
  year = {2013},
  month = mar,
  publisher = {Springer Science and Business Media {LLC}},
  volume = {8},
  number = {4},
  pages = {288--295},
  author = {David Rodriguez-Larrea and Hagan Bayley},
  title = {{M}ultistep protein unfolding during nanopore translocation},
  journal = {Nat. Nanotechnol.}
}

@article{Nivala-2013,
  doi = {10.1038/nbt.2503},
  url = {https://doi.org/10.1038/nbt.2503},
  year = {2013},
  month = feb,
  publisher = {Springer Science and Business Media {LLC}},
  volume = {31},
  number = {3},
  pages = {247--250},
  author = {Jeff Nivala and Douglas B Marks and Mark Akeson},
  title = {{U}nfoldase-mediated protein translocation through an {\textalpha}-hemolysin nanopore},
  journal = {Nat. Biotechnol.}
}

@article{Kennedy-2016,
  doi = {10.1038/nnano.2016.120},
  url = {https://doi.org/10.1038/nnano.2016.120},
  year = {2016},
  month = jul,
  publisher = {Springer Science and Business Media {LLC}},
  volume = {11},
  number = {11},
  pages = {968--976},
  author = {Eamonn Kennedy and Zhuxin Dong and Clare Tennant and Gregory Timp},
  title = {{R}eading the primary structure of a protein with 0.07 nm{\textsuperscript{3}} resolution using a subnanometre-diameter pore},
  journal = {Nat. Nanotechnol.}
}

@article{Spiering-2011,
  doi = {10.1021/nl201541y},
  url = {https://doi.org/10.1021/nl201541y},
  year = {2011},
  month = jul,
  publisher = {American Chemical Society ({ACS})},
  volume = {11},
  number = {7},
  pages = {2978--2982},
  author = {Andre Spiering and Sebastian Getfert and Andy Sischka and Peter Reimann and Dario Anselmetti},
  title = {{N}anopore {T}ranslocation {D}ynamics of a {S}ingle {DNA}-{B}ound {P}rotein},
  journal = {Nano Lett.}
}

@article{Fahie-2013,
  doi = {10.1074/jbc.m113.475350},
  url = {https://doi.org/10.1074/jbc.m113.475350},
  year = {2013},
  month = sep,
  publisher = {American Society for Biochemistry {\&} Molecular Biology ({ASBMB})},
  volume = {288},
  number = {43},
  pages = {31042--31051},
  author = {Monifa Fahie and Fabian B. Romano and Christina Chisholm and Alejandro P. Heuck and Mark Zbinden and Min Chen},
  title = {{A} {N}on-classical {A}ssembly {P}athway of {E}scherichia coli {P}ore-forming {T}oxin {C}ytolysin {A}},
  journal = {J. Biol. Chem.}
}

@article{Fahie-2015,
  doi = {10.1021/acs.analchem.5b03350},
  url = {https://doi.org/10.1021/acs.analchem.5b03350},
  year = {2015},
  month = oct,
  publisher = {American Chemical Society ({ACS})},
  volume = {87},
  number = {21},
  pages = {11143--11149},
  author = {Monifa A. Fahie and Bib Yang and Martin Mullis and Matthew A. Holden and Min Chen},
  title = {{S}elective {D}etection of {P}rotein {H}omologues in {S}erum {U}sing an {OmpG} {N}anopore},
  journal = {Anal. Chem.}
}

@article{Fahie-2015b,
  doi = {10.1021/acs.jpcb.5b06435},
  url = {https://doi.org/10.1021/acs.jpcb.5b06435},
  year = {2015},
  month = jul,
  publisher = {American Chemical Society ({ACS})},
  volume = {119},
  number = {32},
  pages = {10198--10206},
  author = {Monifa A. Fahie and Min Chen},
  title = {{E}lectrostatic {I}nteractions between {OmpG} {N}anopore and {A}nalyte {P}rotein {S}urface {C}an {D}istinguish between {G}lycosylated {I}soforms},
  journal = {J. Phys. Chem. B}
}

@article{Movileanu-2000,
  doi = {10.1038/80295},
  url = {https://doi.org/10.1038/80295},
  year = {2000},
  month = oct,
  publisher = {Springer Science and Business Media {LLC}},
  volume = {18},
  number = {10},
  pages = {1091--1095},
  author = {Liviu Movileanu and Stefan Howorka and Orit Braha and Hagan Bayley},
  title = {{D}etecting protein analytes that modulate transmembrane movement of a polymer chain within a single protein pore},
  journal = {Nat. Biotechn.}
}

@article{Niedzwiecki-2013,
  doi = {10.1021/nn400125c},
  url = {https://doi.org/10.1021/nn400125c},
  year = {2013},
  month = mar,
  publisher = {American Chemical Society ({ACS})},
  volume = {7},
  number = {4},
  pages = {3341--3350},
  author = {David J. Niedzwiecki and Raghuvaran Iyer and Philip N. Borer and Liviu Movileanu},
  title = {{S}ampling a {B}iomarker of the {H}uman {I}mmunodeficiency {V}irus across a {S}ynthetic {N}anopore},
  journal = {{ACS} Nano}
}

@article{Liu-2018,
  doi = {10.1002/anie.201803324},
  url = {https://doi.org/10.1002/anie.201803324},
  year = {2018},
  month = may,
  publisher = {Wiley},
  volume = {57},
  number = {37},
  pages = {11882--11887},
  author = {Lei Liu and Ting Li and Shouwen Zhang and Peng Song and Bingyuan Guo and Yuliang Zhao and Hai-Chen Wu},
  title = {{S}imultaneous {Q}uantification of {M}ultiple {C}ancer {B}iomarkers in {B}lood {S}amples through {DNA}-{A}ssisted {N}anopore {S}ensing},
  journal = {Angew. Chem. Int. Ed.}
}

@article{Uematsu-2018,
  doi = {10.1021/acs.jpcb.8b01975},
  url = {https://doi.org/10.1021/acs.jpcb.8b01975},
  year = {2018},
  month = feb,
  publisher = {American Chemical Society ({ACS})},
  volume = {122},
  number = {11},
  pages = {2992--2997},
  author = {Yuki Uematsu and Roland R. Netz and Lyd{\'{e}}ric Bocquet and Douwe Jan Bonthuis},
  title = {{C}rossover of the {P}ower-{L}aw {E}xponent for {C}arbon {N}anotube {C}onductivity as a {F}unction of {S}alinity},
  journal = {J. Phys. Chem. B}
}

@article{Grosberg-2010,
  doi = {10.1063/1.3495481},
  url = {https://doi.org/10.1063/1.3495481},
  year = {2010},
  month = oct,
  publisher = {{AIP} Publishing},
  volume = {133},
  number = {16},
  pages = {165102},
  author = {Alexander Y. Grosberg and Yitzhak Rabin},
  title = {{DNA} capture into a nanopore: {I}nterplay of diffusion and electrohydrodynamics},
  journal = {J. Chem. Phys.}
}

@article{Thompson-2003,
  doi = {10.1063/1.1609194},
  url = {https://doi.org/10.1063/1.1609194},
  year = {2003},
  month = oct,
  publisher = {{AIP} Publishing},
  volume = {119},
  number = {14},
  pages = {7503--7511},
  author = {Aidan P. Thompson},
  title = {{N}onequilibrium molecular dynamics simulation of electro-osmotic flow in a charged nanopore},
  journal = {J. Chem. Phys.}
}

@article{Qiao-Aluru-2003,
  doi = {10.1063/1.1543140},
  url = {https://doi.org/10.1063/1.1543140},
  year = {2003},
  month = mar,
  publisher = {{AIP} Publishing},
  volume = {118},
  number = {10},
  pages = {4692--4701},
  author = {R. Qiao and N. R. Aluru},
  title = {{I}on concentrations and velocity profiles in nanochannel electroosmotic flows},
  journal = {J. Chem. Phys.}
}

@article{Wong-2007,
  doi = {10.1063/1.2723088},
  url = {https://doi.org/10.1063/1.2723088},
  year = {2007},
  month = apr,
  publisher = {{AIP} Publishing},
  volume = {126},
  number = {16},
  pages = {164903},
  author = {C. T. A. Wong and M. Muthukumar},
  title = {{P}olymer capture by electro-osmotic flow of oppositely charged nanopores},
  journal = {J. Chem. Phys.}
}

@article{Mao-2014,
  doi = {10.1017/jfm.2014.214},
  url = {https://doi.org/10.1017/jfm.2014.214},
  year = {2014},
  month = may,
  publisher = {Cambridge University Press ({CUP})},
  volume = {749},
  pages = {167--183},
  author = {M. Mao and J.~D. Sherwood and S. Ghosal},
  title = {{E}lectro-osmotic flow through a nanopore},
  journal = {J. Fluid Mech.}
}

@article{Sherwood-2014,
  doi = {10.1021/la502349g},
  url = {https://doi.org/10.1021/la502349g},
  year = {2014},
  month = jul,
  publisher = {American Chemical Society ({ACS})},
  volume = {30},
  number = {31},
  pages = {9261--9272},
  author = {J. D. Sherwood and M. Mao and S. Ghosal},
  title = {{E}lectroosmosis in a {F}inite {C}ylindrical {P}ore: {S}imple {M}odels of {E}nd {E}ffects},
  journal = {Langmuir}
}

@article{Laohakunakorn-2015,
  doi = {10.1088/0957-4484/26/27/275202},
  url = {https://doi.org/10.1088/0957-4484/26/27/275202},
  year = {2015},
  month = jun,
  publisher = {{IOP} Publishing},
  volume = {26},
  number = {27},
  pages = {275202},
  author = {Nadanai Laohakunakorn and Ulrich F Keyser},
  title = {{E}lectroosmotic flow rectification in conical nanopores},
  journal = {Nanotechnology}
}

@article{Haywood-2014,
  doi = {10.1021/ac502596m},
  url = {https://doi.org/10.1021/ac502596m},
  year = {2014},
  month = nov,
  publisher = {American Chemical Society ({ACS})},
  volume = {86},
  number = {22},
  pages = {11174--11180},
  author = {Daniel G. Haywood and Zachary D. Harms and Stephen C. Jacobson},
  title = {{E}lectroosmotic {F}low in {N}anofluidic {C}hannels},
  journal = {Anal. Chem.}
}

@article{Tagliazucchi-2015,
  doi = {10.1016/j.mattod.2014.10.020},
  url = {https://doi.org/10.1016/j.mattod.2014.10.020},
  year = {2015},
  month = apr,
  publisher = {Elsevier {BV}},
  volume = {18},
  number = {3},
  pages = {131--142},
  author = {Mario Tagliazucchi and Igal Szleifer},
  title = {{T}ransport mechanisms in nanopores and nanochannels: can we mimic nature?},
  journal = {Mater. Today}
}

@article{Bonome-2017,
  doi = {10.1007/s10404-017-1928-1},
  url = {https://doi.org/10.1007/s10404-017-1928-1},
  year = {2017},
  month = may,
  publisher = {Springer Science and Business Media {LLC}},
  volume = {21},
  number = {5},
  author = {Emma Letizia Bonome and Fabio Cecconi and Mauro Chinappi},
  title = {{E}lectroosmotic flow through an {\textalpha}-hemolysin nanopore},
  journal = {Microfluid. Nanofluid.}
}

@article{Zhang-2014,
  doi = {10.1016/j.icheatmasstransfer.2014.10.024},
  url = {https://doi.org/10.1016/j.icheatmasstransfer.2014.10.024},
  year = {2014},
  month = dec,
  publisher = {Elsevier {BV}},
  volume = {59},
  pages = {101--105},
  author = {Chengbin Zhang and Pengfei Lu and Yongping Chen},
  title = {{M}olecular dynamics simulation of electroosmotic flow in rough nanochannels},
  journal = {Int. Commun. Heat Mass Transf.}
}

@article{Piguet-2014,
  doi = {10.1021/jz502360c},
  url = {https://doi.org/10.1021/jz502360c},
  year = {2014},
  month = dec,
  publisher = {American Chemical Society ({ACS})},
  volume = {5},
  number = {24},
  pages = {4362--4367},
  author = {Fabien Piguet and Francoise Discala and Marie-France Breton and Juan Pelta and Laurent Bacri and Abdelghani Oukhaled},
  title = {{E}lectroosmosis through {\textalpha}-{H}emolysin {T}hat {D}epends on {A}lkali {C}ation {T}ype},
  journal = {J. Phys. Chem. Lett.}
}

@article{Melnikov-2017,
  doi = {10.1103/physreve.95.063105},
  url = {https://doi.org/10.1103/physreve.95.063105},
  year = {2017},
  month = jun,
  publisher = {American Physical Society ({APS})},
  volume = {95},
  number = {6},
  author = {Dmitriy V. Melnikov and Zachery K. Hulings and Maria E. Gracheva},
  title = {{E}lectro-osmotic flow through nanopores in thin and ultrathin membranes},
  journal = {Phys. Rev. E}
}

@article{Belkin-2016,
  doi = {10.1021/acsami.6b00463},
  url = {https://doi.org/10.1021/acsami.6b00463},
  year = {2016},
  month = mar,
  publisher = {American Chemical Society ({ACS})},
  volume = {8},
  number = {20},
  pages = {12599--12608},
  author = {Maxim Belkin and Aleksei Aksimentiev},
  title = {{M}olecular {D}ynamics {S}imulation of {DNA} {C}apture and {T}ransport in {H}eated {N}anopores},
  journal = {ACS Appl. Mater. Interfaces}
}

@article{DiMarino-2015,
  doi = {10.1021/acs.jpclett.5b01077},
  url = {https://doi.org/10.1021/acs.jpclett.5b01077},
  year = {2015},
  month = jul,
  publisher = {American Chemical Society ({ACS})},
  volume = {6},
  number = {15},
  pages = {2963--2968},
  author = {Daniele Di Marino and Emma Letizia Bonome and Anna Tramontano and Mauro Chinappi},
  title = {{A}ll-{A}tom {M}olecular {D}ynamics {S}imulation of {P}rotein {T}ranslocation through an {\textalpha}-{H}emolysin {N}anopore},
  journal = {J. Phys. Chem. Lett.}
}

@article{Luan-2008,
  doi = {10.1103/physreve.78.021912},
  url = {https://doi.org/10.1103/physreve.78.021912},
  year = {2008},
  month = aug,
  publisher = {American Physical Society ({APS})},
  volume = {78},
  number = {2},
  author = {Binquan Luan and Aleksei Aksimentiev},
  title = {{E}lectro-osmotic screening of the {DNA} charge in a nanopore},
  journal = {Phys. Rev. E}
}

@article{Wilson-2019,
  doi = {10.1021/acssensors.8b01375},
  url = {https://doi.org/10.1021/acssensors.8b01375},
  year = {2019},
  month = mar,
  publisher = {American Chemical Society ({ACS})},
  volume = {4},
  number = {3},
  pages = {634--644},
  author = {James Wilson and Kumar Sarthak and Wei Si and Luyu Gao and Aleksei Aksimentiev},
  title = {{R}apid and {A}ccurate {D}etermination of {N}anopore {I}onic {C}urrent {U}sing a {S}teric {E}xclusion {M}odel},
  journal = {{ACS} Sens.}
}

@article{Schirmer-1999,
  doi = {10.1006/jmbi.1999.3326},
  url = {https://doi.org/10.1006/jmbi.1999.3326},
  year = {1999},
  month = dec,
  publisher = {Elsevier {BV}},
  volume = {294},
  number = {5},
  pages = {1159--1167},
  author = {Tilman Schirmer and Prashant S Phale},
  title = {{B}rownian dynamics simulation of ion flow through porin channels},
  journal = {J. Mol. Biol.}
}

@article{Millar-2008,
  doi = {10.1007/s10825-008-0230-6},
  url = {https://doi.org/10.1007/s10825-008-0230-6},
  year = {2008},
  month = feb,
  publisher = {Springer Science and Business Media {LLC}},
  volume = {7},
  number = {1},
  pages = {28--33},
  author = {C. Millar and R. Madathil and O. Beckstein and M. S. P. Sansom and S. Roy and A. Asenov},
  title = {{B}rownian simulation of charge transport in {\textalpha}-{H}aemolysin},
  journal = {J. Comput. Electron.}
}

@article{Egwolf-2010,
  doi = {10.1021/jp906791b},
  url = {https://doi.org/10.1021/jp906791b},
  year = {2010},
  month = mar,
  publisher = {American Chemical Society ({ACS})},
  volume = {114},
  number = {8},
  pages = {2901--2909},
  author = {Bernhard Egwolf and Yun Luo and D. Eric Walters and Beno{\^{\i}}t Roux},
  title = {{I}on {S}electivity of {\textalpha}-{H}emolysin with {\textbeta}-{C}yclodextrin {A}dapter. {II}. {M}ulti-{I}on {E}ffects {S}tudied with {G}rand {C}anonical {M}onte {C}arlo/{B}rownian {D}ynamics {S}imulations},
  journal = {J. Phys. Chem. B}
}

@article{DeBiase-2015,
  doi = {10.1002/jcc.23799},
  url = {https://doi.org/10.1002/jcc.23799},
  year = {2015},
  month = dec,
  publisher = {Wiley},
  volume = {36},
  number = {4},
  pages = {264--271},
  author = {Pablo M. De Biase and Suren Markosyan and Sergei Noskov},
  title = {{BROMOC} suite: {M}onte {C}arlo/{B}rownian dynamics suite for studies of ion permeation and {DNA} transport in biological and artificial pores with effective potentials},
  journal = {J. Computat. Chem.}
}

@article{DeBiase-2016,
  doi = {10.1039/c6nr00164e},
  url = {https://doi.org/10.1039/c6nr00164e},
  year = {2016},
  publisher = {Royal Society of Chemistry ({RSC})},
  volume = {8},
  number = {22},
  pages = {11571--11579},
  author = {Pablo M. De Biase and Eric N. Ervin and Prithwish Pal and Olga Samoylova and Suren Markosyan and Michael G. Keehan and Geoffrey A. Barrall and Sergei Yu. Noskov},
  title = {{W}hat controls open-pore and residual currents in the first sensing zone of alpha-hemolysin nanopore? {C}ombined experimental and theoretical study},
  journal = {Nanoscale}
}

@article{Basdevant-2019,
  doi = {10.1038/s41598-019-51942-y},
  url = {https://doi.org/10.1038/s41598-019-51942-y},
  year = {2019},
  month = oct,
  publisher = {Springer Science and Business Media {LLC}},
  volume = {9},
  number = {1},
  author = {Nathalie Basdevant and Delphine Dessaux and Rosa Ramirez},
  title = {{I}onic transport through a protein nanopore: a {C}oarse-{G}rained {M}olecular {D}ynamics {S}tudy},
  journal = {Sci. Rep.}
}

@article{Thakur-2019,
  doi = {10.1038/nbt.4316},
  url = {https://doi.org/10.1038/nbt.4316},
  year = {2019},
  month = dec,
  publisher = {Springer Science and Business Media {LLC}},
  volume = {37},
  number = {1},
  pages = {96--101},
  author = {Avinash Kumar Thakur and Liviu Movileanu},
  title = {{R}eal-time measurement of protein{\textendash}protein interactions at single-molecule resolution using a biological nanopore},
  journal = {Nat. Biotechnol.}
}

@article{Yusko-2017,
  doi = {10.1038/nnano.2016.267},
  url = {https://doi.org/10.1038/nnano.2016.267},
  year = {2017},
  month = dec,
  publisher = {Springer Science and Business Media {LLC}},
  volume = {12},
  number = {4},
  pages = {360--367},
  author = {Erik C. Yusko and Brandon R. Bruhn and Olivia M. Eggenberger and Jared Houghtaling and Ryan C. Rollings and Nathan C. Walsh and Santoshi Nandivada and Mariya Pindrus and Adam R. Hall and David Sept and Jiali Li and Devendra S. Kalonia and Michael Mayer},
  title = {{R}eal-time shape approximation and fingerprinting of single proteins using a nanopore},
  journal = {Nat. Nanotechnol.}
}

@article{Harrington-2019,
  doi = {10.1021/acsnano.8b07697},
  url = {https://doi.org/10.1021/acsnano.8b07697},
  year = {2019},
  month = dec,
  publisher = {American Chemical Society ({ACS})},
  volume = {13},
  number = {1},
  pages = {633--641},
  author = {Leon Harrington and Leila T. Alexander and Stefan Knapp and Hagan Bayley},
  title = {{S}ingle-{M}olecule {P}rotein {P}hosphorylation and {D}ephosphorylation by {N}anopore Enzymology},
  journal = {{ACS} Nano}
}

@incollection{Laszlo-2017,
  doi = {10.1016/bs.mie.2016.09.038},
  url = {https://doi.org/10.1016/bs.mie.2016.09.038},
  year = {2017},
  publisher = {Elsevier},
  pages = {387--414},
  author = {A.H. Laszlo and I.M. Derrrington and J.H. Gundlach},
  title = {{S}ubangstrom {M}easurements of {E}nzyme {F}unction {U}sing a {B}iological {N}anopore, {SPRNT}},
  booktitle = {Methods in Enzymology}
}

@article{Bayley-2001,
  doi = {10.1038/35093038},
  url = {https://doi.org/10.1038/35093038},
  year = {2001},
  month = sep,
  publisher = {Springer Science and Business Media {LLC}},
  volume = {413},
  number = {6852},
  pages = {226--230},
  author = {Hagan Bayley and Paul S. Cremer},
  title = {{S}tochastic sensors inspired by biology},
  journal = {Nature}
}

@article{Dekker-2007,
  doi = {10.1038/nnano.2007.27},
  url = {https://doi.org/10.1038/nnano.2007.27},
  year = {2007},
  month = mar,
  publisher = {Springer Science and Business Media {LLC}},
  volume = {2},
  number = {4},
  pages = {209--215},
  author = {Cees Dekker},
  title = {{S}olid-state nanopores},
  journal = {Nat. Nanotechnol.}
}

@article{Venkatesan-2011,
  doi = {10.1038/nnano.2011.129},
  url = {https://doi.org/10.1038/nnano.2011.129},
  year = {2011},
  month = sep,
  publisher = {Springer Science and Business Media {LLC}},
  volume = {6},
  number = {10},
  pages = {615--624},
  author = {Bala Murali Venkatesan and Rashid Bashir},
  title = {{N}anopore sensors for nucleic acid analysis},
  journal = {Nat. Nanotechnol.}
}

@article{Maffeo-2012,
  doi = {10.1021/cr3002609},
  url = {https://doi.org/10.1021/cr3002609},
  year = {2012},
  month = oct,
  publisher = {American Chemical Society ({ACS})},
  volume = {112},
  number = {12},
  pages = {6250--6284},
  author = {Christopher Maffeo and Swati Bhattacharya and Jejoong Yoo and David Wells and Aleksei Aksimentiev},
  title = {{M}odeling and {S}imulation of {I}on {C}hannels},
  journal = {Chem. Rev.}
}

@article{Thomas-2014,
  doi = {10.1002/smll.201302968},
  url = {https://doi.org/10.1002/smll.201302968},
  year = {2014},
  month = jan,
  publisher = {Wiley},
  volume = {10},
  number = {8},
  pages = {1453--1465},
  author = {Michael Thomas and Ben Corry and Tamsyn A. Hilder},
  title = {{W}hat {H}ave {W}e {L}earnt {A}bout the {M}echanisms of {R}apid {W}ater {T}ransport, {I}on {R}ejection and {S}electivity in {N}anopores from {M}olecular {S}imulation?},
  journal = {Small}
}

@article{Wang-2014,
  doi = {10.1039/c3cp51712h},
  url = {https://doi.org/10.1039/c3cp51712h},
  year = {2014},
  publisher = {Royal Society of Chemistry ({RSC})},
  volume = {16},
  number = {1},
  pages = {23--32},
  author = {Jingtao Wang and Minghui Zhang and Jin Zhai and Lei Jiang},
  title = {{T}heoretical simulation of the ion current rectification ({ICR}) in nanopores based on the {P}oisson{\textendash}{N}ernst{\textendash}{P}lanck ({PNP}) model},
  journal = {Phys. Chem. Chem. Phys.}
}

@article{Kim-2015,
  doi = {10.1016/j.bios.2015.02.020},
  url = {https://doi.org/10.1016/j.bios.2015.02.020},
  year = {2015},
  month = jul,
  publisher = {Elsevier {BV}},
  volume = {69},
  pages = {186--198},
  author = {Han Seul Kim and Yong-Hoon Kim},
  title = {{R}ecent progress in atomistic simulation of electrical current {DNA} sequencing},
  journal = {Biosens. Bioelectron.}
}

@article{Restrepo-Perez-2018,
  doi = {10.1038/s41565-018-0236-6},
  url = {https://doi.org/10.1038/s41565-018-0236-6},
  year = {2018},
  month = sep,
  publisher = {Springer Science and Business Media {LLC}},
  volume = {13},
  number = {9},
  pages = {786--796},
  author = {Laura Restrepo-P{\'{e}}rez and Chirlmin Joo and Cees Dekker},
  title = {{P}aving the way to single-molecule protein sequencing},
  journal = {Nat. Nanotechnol.}
}

@article{Chinappi-2020,
  doi = {10.1021/acsnano.0c06981},
  url = {https://doi.org/10.1021/acsnano.0c06981},
  year = {2020},
  month = nov,
  publisher = {American Chemical Society ({ACS})},
  volume = {14},
  number = {11},
  pages = {15816--15828},
  author = {Mauro Chinappi and Misa Yamaji and Ryuji Kawano and Fabio Cecconi},
  title = {{A}nalytical {M}odel for {P}article {C}apture in {N}anopores {E}lucidates {C}ompetition among {E}lectrophoresis, {E}lectroosmosis, and {D}ielectrophoresis},
  journal = {{ACS} Nano}
}

@article{Modi-2012,
  doi = {10.1039/c2nr31024d},
  url = {https://doi.org/10.1039/c2nr31024d},
  year = {2012},
  publisher = {Royal Society of Chemistry ({RSC})},
  volume = {4},
  number = {20},
  pages = {6166},
  author = {Niraj Modi and Mathias Winterhalter and Ulrich Kleinekath{\"{o}}fer},
  title = {{C}omputational modeling of ion transport through nanopores},
  journal = {Nanoscale}
}

@article{Bocquet-2010,
  doi = {10.1039/b909366b},
  url = {https://doi.org/10.1039/b909366b},
  year = {2010},
  publisher = {Royal Society of Chemistry ({RSC})},
  volume = {39},
  number = {3},
  pages = {1073--1095},
  author = {Lyd{\'{e}}ric Bocquet and Elisabeth Charlaix},
  title = {Nanofluidics, from bulk to interfaces},
  journal = {Chem. Soc. Rev.}
}

@article{Bocquet-2020,
  doi = {10.1038/s41563-020-0625-8},
  url = {https://doi.org/10.1038/s41563-020-0625-8},
  year = {2020},
  month = feb,
  publisher = {Springer Science and Business Media {LLC}},
  volume = {19},
  number = {3},
  pages = {254--256},
  author = {Lyd{\'{e}}ric Bocquet},
  title = {{N}anofluidics coming of age},
  journal = {Nat. Mat.}
}

@article{Sparreboom-2010,
  doi = {10.1088/1367-2630/12/1/015004},
  url = {https://doi.org/10.1088/1367-2630/12/1/015004},
  year = {2010},
  month = jan,
  publisher = {{IOP} Publishing},
  volume = {12},
  number = {1},
  pages = {015004},
  author = {W Sparreboom and A van den Berg and J C T Eijkel},
  title = {{T}ransport in nanofluidic systems: a review of theory and applications},
  journal = {New J. Phys.}
}

@article{Wong-ekkabut-2016a,
  doi = {10.1016/j.bbamem.2016.02.004},
  url = {https://doi.org/10.1016/j.bbamem.2016.02.004},
  year = {2016},
  month = oct,
  publisher = {Elsevier {BV}},
  volume = {1858},
  number = {10},
  pages = {2529--2538},
  author = {Jirasak Wong-ekkabut and Mikko Karttunen},
  title = {{T}he good, the bad and the user in soft matter simulations},
  journal = {Biochim. Biophys. Acta Biomembr.}
}

@article{Allen-1999,
  doi = {10.1063/1.480132},
  url = {https://doi.org/10.1063/1.480132},
  year = {1999},
  month = nov,
  publisher = {{AIP} Publishing},
  volume = {111},
  number = {17},
  pages = {7985--7999},
  author = {T. W. Allen and S. Kuyucak and S.-H. Chung},
  title = {{T}he effect of hydrophobic and hydrophilic channel walls on the structure and diffusion of water and ions},
  journal = {J. Chem. Phys.}
}

@article{Lynden-Bell-1996,
  doi = {10.1063/1.472757},
  url = {https://doi.org/10.1063/1.472757},
  year = {1996},
  month = nov,
  publisher = {{AIP} Publishing},
  volume = {105},
  number = {20},
  pages = {9266--9280},
  author = {R. M. Lynden-Bell and Jayendran C. Rasaiah},
  title = {Mobility and solvation of ions in channels},
  journal = {J. Chem. Phys.}
}

@article{Gramse-2013,
  doi = {10.1016/j.bpj.2013.02.011},
  url = {https://doi.org/10.1016/j.bpj.2013.02.011},
  year = {2013},
  month = mar,
  publisher = {Elsevier {BV}},
  volume = {104},
  number = {6},
  pages = {1257--1262},
  author = {G. Gramse and A. Dols-Perez and M.A. Edwards and L. Fumagalli and G. Gomila},
  title = {{N}anoscale {M}easurement of the {D}ielectric {C}onstant of {S}upported {L}ipid {B}ilayers in {A}queous {S}olutions with {E}lectrostatic {F}orce {M}icroscopy},
  journal = {Biophys. J.}
}

@article{Hoogerheide-2014,
  doi = {10.1021/nn5025829},
  url = {https://doi.org/10.1021/nn5025829},
  year = {2014},
  month = jun,
  publisher = {American Chemical Society ({ACS})},
  volume = {8},
  number = {7},
  pages = {7384--7391},
  author = {David P. Hoogerheide and Bo Lu and Jene A. Golovchenko},
  title = {{P}ressure{\textendash}{V}oltage {T}rap for {DNA} near a {S}olid-{S}tate {N}anopore},
  journal = {{ACS} Nano}
}

@article{Hsu-2017,
  doi = {10.1021/acs.jpcc.7b06798},
  url = {https://doi.org/10.1021/acs.jpcc.7b06798},
  year = {2017},
  month = sep,
  publisher = {American Chemical Society ({ACS})},
  volume = {121},
  number = {37},
  pages = {20517--20523},
  author = {Wei-Lun Hsu and Dalton J. E. Harvie and Malcolm R. Davidson and Dave E. Dunstan and Junho Hwang and Hirofumi Daiguji},
  title = {{V}iscoelectric {E}ffects in {N}anochannel {E}lectrokinetics},
  journal = {J. Phys. Chem. C}
}

@article{Hsu-2017b,
  doi = {10.1021/acs.jpcc.6b09907},
  url = {https://doi.org/10.1021/acs.jpcc.6b09907},
  year = {2017},
  month = feb,
  publisher = {American Chemical Society ({ACS})},
  volume = {121},
  number = {8},
  pages = {4576--4582},
  author = {Jyh-Ping Hsu and Shu-Tuan Yang and Chih-Yuan Lin and Shiojenn Tseng},
  title = {{I}onic {C}urrent {R}ectification in a {C}onical {N}anopore: {I}nfluences of {E}lectroosmotic {F}low and {T}ype of {S}alt},
  journal = {J. Phys. Chem. C}
}

@article{Axelsson-2015,
  doi = {10.3846/13926292.2015.1021395},
  url = {https://doi.org/10.3846/13926292.2015.1021395},
  year = {2015},
  month = mar,
  publisher = {Vilnius Gediminas Technical University},
  volume = {20},
  number = {2},
  pages = {232--260},
  author = {Owe Axelsson and Xin He and Maya Neytcheva},
  title = {{N}umerical {S}olution of the {T}ime-{D}ependent {N}avier--{S}tokes {E}equation for {V}ariable {D}ensity--{V}ariable {V}iscosity. {P}art {I}},
  journal = {Math. Model. Anal.}
}

@article{White-2008,
  doi = {10.1021/la702955k},
  url = {https://doi.org/10.1021/la702955k},
  year = {2008},
  month = mar,
  publisher = {American Chemical Society ({ACS})},
  volume = {24},
  number = {5},
  pages = {2212--2218},
  author = {Henry S. White and Andreas Bund},
  title = {{I}on {C}urrent {R}ectification at {N}anopores in {G}lass {M}embranes},
  journal = {Langmuir}
}

@article{Baldessari-2008-1,
  doi = {10.1016/j.jcis.2008.06.007},
  url = {https://doi.org/10.1016/j.jcis.2008.06.007},
  year = {2008},
  month = sep,
  publisher = {Elsevier {BV}},
  volume = {325},
  number = {2},
  pages = {526--538},
  author = {Fabio Baldessari},
  title = {{E}lectrokinetics in nanochannels: part {I}. {E}lectric double layer overlap and
  channel-to-well equilibrium},
  journal = {J. Colloid Interface Sci.}
}

@article{Baldessari-2008-2,
  doi = {10.1016/j.jcis.2008.06.008},
  url = {https://doi.org/10.1016/j.jcis.2008.06.008},
  year = {2008},
  month = sep,
  publisher = {Elsevier {BV}},
  volume = {325},
  number = {2},
  pages = {539--546},
  author = {Fabio Baldessari},
  title = {Electrokinetics in nanochannels: part {II}. Mobility dependence on ion density and ionic
  current measurements},
  journal = {J. Colloid Interface Sci.}
}

@article{Constantin-2007,
  doi = {10.1103/physreve.76.041202},
  url = {https://doi.org/10.1103/physreve.76.041202},
  year = {2007},
  month = oct,
  publisher = {American Physical Society ({APS})},
  volume = {76},
  number = {4},
  author = {Drago{\c{s}} Constantin and Zuzanna S. Siwy},
  title = {{P}oisson-{N}ernst-{P}lanck model of ion current rectification through a nanofluidic diode},
  journal = {Phys. Rev. E}
}

@article{Pederson-2015,
  doi = {10.1021/acs.jpcb.5b04955},
  url = {https://doi.org/10.1021/acs.jpcb.5b04955},
  year = {2015},
  month = aug,
  publisher = {American Chemical Society ({ACS})},
  volume = {119},
  number = {33},
  pages = {10448--10455},
  author = {Emmanuel D. Pederson and Jonathan Barbalas and Bryon S. Drown and Michael J. Culbertson and Lisa M. Keranen Burden and John J. Kasianowicz and Daniel L. Burden},
  title = {{P}roximal {C}apture {D}ynamics for a {S}ingle {B}iological {N}anopore {S}ensor},
  journal = {J. Phys. Chem. B}
}

@article{Simakov-2010,
  doi = {10.1021/jp1046062},
  url = {https://doi.org/10.1021/jp1046062},
  year = {2010},
  month = nov,
  publisher = {American Chemical Society ({ACS})},
  volume = {114},
  number = {46},
  pages = {15180--15190},
  author = {Nikolay A. Simakov and Maria G. Kurnikova},
  title = {{S}oft {W}all {I}on {C}hannel in {C}ontinuum {R}epresentation with {A}pplication to {M}odeling {I}on {C}urrents in {\textalpha}-{H}emolysin},
  journal = {J. Phys. Chem. B}
}

@article{Chaudhry-2014,
  doi = {10.4208/cicp.101112.100413a},
  url = {https://doi.org/10.4208/cicp.101112.100413a},
  year = {2014},
  month = jan,
  publisher = {Global Science Press},
  volume = {15},
  number = {1},
  pages = {93--125},
  author = {Jehanzeb Hameed Chaudhry and Jeffrey Comer and Aleksei Aksimentiev and Luke N. Olson},
  title = {{A} {S}tabilized {F}inite {E}lement {M}ethod for {M}odified {P}oisson-{N}ernst-{P}lanck {E}quations to {D}etermine {I}on {F}low {T}hrough a {N}anopore},
  journal = {Commun. Comput. Phys.}
}

@article{Gillespie-2002,
  doi = {10.1088/0953-8984/14/46/317},
  url = {https://doi.org/10.1088/0953-8984/14/46/317},
  year = {2002},
  month = nov,
  publisher = {{IOP} Publishing},
  volume = {14},
  number = {46},
  pages = {12129--12145},
  author = {Dirk Gillespie and Wolfgang Nonner and Robert S Eisenberg},
  title = {{C}oupling {P}oisson-{N}ernst-{P}lanck and density functional theory to calculate ion flux},
  journal = {J. Phys. Condens. Matter}
}

@article{Cervera-2005,
  doi = {10.1209/epl/i2005-10054-x},
  url = {https://doi.org/10.1209/epl/i2005-10054-x},
  year = {2005},
  month = jul,
  publisher = {{IOP} Publishing},
  volume = {71},
  number = {1},
  pages = {35--41},
  author = {J Cervera and B Schiedt and P Ram{\'{i}}rez},
  title = {{A} {P}oisson/{N}ernst-{P}lanck model for ionic transport through synthetic conical nanopores},
  journal = {{EPL}}
}

@article{Ramirez-2007,
  doi = {10.1063/1.2735608},
  url = {https://doi.org/10.1063/1.2735608},
  year = {2007},
  month = may,
  publisher = {{AIP} Publishing},
  volume = {126},
  number = {19},
  pages = {194703},
  author = {Patricio Ram{\'{i}}rez and Vicente G{\'{o}}mez and Javier Cervera and Birgitta Schiedt and Salvador Maf{\'{e}}},
  title = {{I}on transport and selectivity in nanopores with spatially inhomogeneous fixed charge distributions},
  journal = {J. Chem. Phys.}
}

@article{Lin-2015,
  doi = {10.1021/acs.analchem.5b03074},
  url = {https://doi.org/10.1021/acs.analchem.5b03074},
  year = {2015},
  month = dec,
  publisher = {American Chemical Society ({ACS})},
  volume = {88},
  number = {2},
  pages = {1176--1187},
  author = {Jeng-Yang Lin and Chih-Yuan Lin and Jyh-Ping Hsu and Shiojenn Tseng},
  title = {{I}onic {C}urrent {R}ectification in a {pH}-{T}unable {P}olyelectrolyte {B}rushes {F}unctionalized {C}onical {N}anopore: {E}ffect of {S}alt {G}radient},
  journal = {Anal. Chem.}
}

@article{Burger-2012,
  doi = {10.1088/0951-7715/25/4/961},
  url = {https://doi.org/10.1088/0951-7715/25/4/961},
  year = {2012},
  month = mar,
  publisher = {{IOP} Publishing},
  volume = {25},
  number = {4},
  pages = {961--990},
  author = {M Burger and B Schlake and M-T Wolfram},
  title = {{N}onlinear {P}oisson{\textendash}{N}ernst{\textendash}{P}lanck equations for ion flux through confined geometries},
  journal = {Nonlinearity}
}

@article{Hulings-2018,
  doi = {10.1088/1361-6528/aada64},
  url = {https://doi.org/10.1088/1361-6528/aada64},
  year = {2018},
  month = sep,
  publisher = {{IOP} Publishing},
  volume = {29},
  number = {44},
  pages = {445204},
  author = {Zachery K Hulings and Dmitriy V Melnikov and Maria E Gracheva},
  title = {{B}rownian dynamics simulations of the ionic current traces for a neutral nanoparticle translocating through a nanopore},
  journal = {Nanotechnology}
}

@article{Valisko-2019,
  doi = {10.1063/1.5091789},
  url = {https://doi.org/10.1063/1.5091789},
  year = {2019},
  month = apr,
  publisher = {{AIP} Publishing},
  volume = {150},
  number = {14},
  pages = {144703},
  author = {M{\'{o}}nika Valisk{\'{o}} and Bart{\l}omiej Matejczyk and Zolt{\'{a}}n Hat{\'{o}} and Tam{\'{a}}s Krist{\'{o}}f and Eszter M{\'{a}}dai and D{\'{a}}vid Fertig and Dirk Gillespie and Dezs{\H{o}} Boda},
  title = {{M}ultiscale analysis of the effect of surface charge pattern on a nanopore's rectification and selectivity properties: {F}rom all-atom model to {P}oisson-{N}ernst-{P}lanck},
  journal = {J. Chem. Phys.}
}

@article{Tian-2003,
  doi = {10.1063/1.1621614},
  url = {https://doi.org/10.1063/1.1621614},
  year = {2003},
  month = dec,
  publisher = {{AIP} Publishing},
  volume = {119},
  number = {21},
  pages = {11475--11483},
  author = {Pu Tian and Grant D. Smith},
  title = {{T}ranslocation of a polymer chain across a nanopore: {A} {B}rownian dynamics simulation study},
  journal = {J. Chem. Phys.}
}

@article{Noskov-2004,
  doi = {10.1529/biophysj.104.044008},
  url = {https://doi.org/10.1529/biophysj.104.044008},
  year = {2004},
  month = oct,
  publisher = {Elsevier {BV}},
  volume = {87},
  number = {4},
  pages = {2299--2309},
  author = {Sergei Yu. Noskov and Wonpil Im and Beno{\^{\i}}t Roux},
  title = {{I}on {P}ermeation through the {\textalpha}-{H}emolysin {C}hannel: {T}heoretical {S}tudies {B}ased on {B}rownian {D}ynamics and {P}oisson-{N}ernst-{P}lank {E}lectrodiffusion {T}heory},
  journal = {Biophys. J.}
}

@article{Chen-2016,
  doi = {10.1007/s11538-016-0196-7},
  url = {https://doi.org/10.1007/s11538-016-0196-7},
  year = {2016},
  month = aug,
  publisher = {Springer Science and Business Media {LLC}},
  volume = {78},
  number = {8},
  pages = {1703--1726},
  author = {Duan Chen},
  title = {{A} {N}ew {P}oisson-{N}ernst-{P}lanck {M}odel with {I}on-{W}ater {I}nteractions for {C}harge {T}ransport in {I}on {C}hannels.},
  journal = {Bull. Math. Biol.}
}

@article{Kilic-2007,
  doi = {10.1103/physreve.75.021503},
  url = {https://doi.org/10.1103/physreve.75.021503},
  year = {2007},
  month = feb,
  publisher = {American Physical Society ({APS})},
  volume = {75},
  number = {2},
  author = {Mustafa Sabri Kilic and Martin Z. Bazant and Armand Ajdari},
  title = {{S}teric {E}ffects in the {D}ynamics of {E}lectrolytes at {L}arge {A}pplied {V}oltages. {II}. {M}odified {P}oisson-{N}ernst-{P}lanck {E}quations},
  journal = {Phys. Rev. E}
}

@article{Bazant-2009,
  doi = {10.1016/j.cis.2009.10.001},
  url = {https://doi.org/10.1016/j.cis.2009.10.001},
  year = {2009},
  month = nov,
  publisher = {Elsevier {BV}},
  volume = {152},
  number = {1-2},
  pages = {48--88},
  author = {Martin Z. Bazant and Mustafa Sabri Kilic and Brian D. Storey and Armand Ajdari},
  title = {{T}owards an understanding of induced-charge electrokinetics at large applied voltages in concentrated solutions},
  journal = {Adv. Colloid Interface Sci.}
}

@article{Lu-2011,
  doi = {10.1016/j.bpj.2011.03.059},
  url = {https://doi.org/10.1016/j.bpj.2011.03.059},
  year = {2011},
  month = may,
  publisher = {Elsevier {BV}},
  volume = {100},
  number = {10},
  pages = {2475--2485},
  author = {Benzhuo Lu and Y.C. Zhou},
  title = {{P}oisson-{N}ernst-{P}lanck {E}quations for {S}imulating {B}iomolecular {D}iffusion-{R}eaction {P}rocesses {II}: {S}ize {E}ffects on {I}onic {D}istributions and {D}iffusion-{R}eaction {R}ates},
  journal = {Biophys. J.}
}

@article{Borukhov-1997,
  doi = {10.1103/physrevlett.79.435},
  url = {https://doi.org/10.1103/physrevlett.79.435},
  year = {1997},
  month = jul,
  publisher = {American Physical Society ({APS})},
  volume = {79},
  number = {3},
  pages = {435--438},
  author = {Itamar Borukhov and David Andelman and Henri Orland},
  title = {{S}teric {E}ffects in {E}lectrolytes: {A} {M}odified {P}oisson-{B}oltzmann {E}quation},
  journal = {Phys. Rev. Lett.}
}

@article{Daiguji-2004,
  doi = {10.1021/nl0348185},
  url = {https://doi.org/10.1021/nl0348185},
  year = {2004},
  month = jan,
  publisher = {American Chemical Society ({ACS})},
  volume = {4},
  number = {1},
  pages = {137--142},
  author = {Hirofumi Daiguji and Peidong Yang and Arun Majumdar},
  title = {{I}on {T}ransport in {N}anofluidic {C}hannels},
  journal = {Nano Lett.}
}

@article{Eisenberg-1996,
  doi = {10.1007/s002329900026},
  url = {https://doi.org/10.1007/s002329900026},
  year = {1996},
  month = mar,
  publisher = {Springer Science and Business Media {LLC}},
  volume = {150},
  number = {1},
  pages = {1--25},
  author = {R.S. Eisenberg},
  title = {{C}omputing the {F}ield in {P}roteins and {C}hannels},
  journal = {J. Membr. Biol.}
}

@article{Chen-1997,
  doi = {10.1016/s0006-3495(97)78650-8},
  url = {https://doi.org/10.1016/s0006-3495(97)78650-8},
  year = {1997},
  month = jan,
  publisher = {Elsevier {BV}},
  volume = {72},
  number = {1},
  pages = {97--116},
  author = {D. Chen and J. Lear and B. Eisenberg},
  title = {{P}ermeation through an open channel: {P}oisson-{N}ernst-{P}lanck theory of a synthetic ionic channel},
  journal = {Biophys. J.}
}

@article{Im-2002,
  doi = {10.1016/s0022-2836(02)00778-7},
  url = {https://doi.org/10.1016/s0022-2836(02)00778-7},
  year = {2002},
  month = sep,
  publisher = {Elsevier {BV}},
  volume = {322},
  number = {4},
  pages = {851--869},
  author = {Wonpil Im and Beno{\^{\i}}t Roux},
  title = {{I}on {P}ermeation and {S}electivity of {OmpF} {P}orin: {A} {T}heoretical {S}tudy {B}ased on {M}olecular {D}ynamics, {B}rownian {D}ynamics, and {C}ontinuum {E}lectrodiffusion {T}heory},
  journal = {J. Mol. Biol.}
}

@article{Melnikov-2020,
  doi = {10.1021/acs.jpcc.0c04829},
  url = {https://doi.org/10.1021/acs.jpcc.0c04829},
  year = {2020},
  month = aug,
  publisher = {American Chemical Society ({ACS})},
  volume = {124},
  number = {36},
  pages = {19802--19808},
  author = {Dmitriy V. Melnikov and Zachery K. Hulings and Maria E. Gracheva},
  title = {{C}oncentration {P}olarization, {S}urface {C}harge, and {I}onic {C}urrent {B}lockade in {N}anopores},
  journal = {J. Phys. Chem. C}
}

@article{Corry-2000,
  doi = {10.1016/s0009-2614(00)00206-2},
  url = {https://doi.org/10.1016/s0009-2614(00)00206-2},
  year = {2000},
  month = mar,
  publisher = {Elsevier {BV}},
  volume = {320},
  number = {1-2},
  pages = {35--41},
  author = {B. Corry and S. Kuyucak and S.H. Chung},
  title = {{I}nvalidity of continuum theories of electrolytes in nanopores},
  journal = {Chem. Phys. Lett.}
}

@article{Collins-2012,
  doi = {10.1016/j.bpc.2012.04.002},
  url = {https://doi.org/10.1016/j.bpc.2012.04.002},
  year = {2012},
  month = jun,
  publisher = {Elsevier {BV}},
  volume = {167},
  pages = {43--59},
  author = {Kim D. Collins},
  title = {{W}hy continuum electrostatics theories cannot explain biological structure, polyelectrolytes or ionic strength effects in ion{\textendash}protein interactions},
  journal = {Biophys. Chem.}
}

@article{Zhou-2020,
  doi = {10.1021/acs.jpcb.9b10702},
  url = {https://doi.org/10.1021/acs.jpcb.9b10702},
  year = {2020},
  month = feb,
  publisher = {American Chemical Society ({ACS})},
  author = {Wanqi Zhou and Hu Qiu and Yufeng Guo and Wanlin Guo},
  title = {{M}olecular {I}nsights into {D}istinct {D}etection {P}roperties of {\textalpha}-{H}emolysin, {MspA}, {CsgG}, and {A}erolysin {N}anopore {S}ensors},
  journal = {J. Phys. Chem. B}
}

@article{Aksimentiev-2005,
  doi = {10.1529/biophysj.104.058727},
  url = {https://doi.org/10.1529/biophysj.104.058727},
  year = {2005},
  month = jun,
  publisher = {Elsevier {BV}},
  volume = {88},
  number = {6},
  pages = {3745--3761},
  author = {Aleksij Aksimentiev and Klaus Schulten},
  title = {{I}maging {\textalpha}-{H}emolysin with {M}olecular {D}ynamics: {I}onic {C}onductance, {O}smotic {P}ermeability, and the {E}lectrostatic {P}otential {M}ap},
  journal = {Biophys. J.}
}

@article{Bhattacharya-2011,
  doi = {10.1021/jp111441p},
  url = {https://doi.org/10.1021/jp111441p},
  year = {2011},
  month = feb,
  publisher = {American Chemical Society ({ACS})},
  volume = {115},
  number = {10},
  pages = {4255--4264},
  author = {Swati Bhattacharya and Julien Muzard and Linda Payet and Jerome Math{\'{e}} and Ulrich Bockelmann and Aleksei Aksimentiev and Virgile Viasnoff},
  title = {{R}ectification of the {C}urrent in {\textalpha}-{H}emolysin {P}ore {D}epends on the {C}ation {T}ype: {T}he {A}lkali {S}eries {P}robed by {M}olecular {D}ynamics {S}imulations and {E}xperiments},
  journal = {J. Phys. Chem. C}
}

@article{Bhattacharya-2012,
  doi = {10.1021/nn3019943},
  url = {https://doi.org/10.1021/nn3019943},
  year = {2012},
  month = jul,
  publisher = {American Chemical Society ({ACS})},
  volume = {6},
  number = {8},
  pages = {6960--6968},
  author = {Swati Bhattacharya and Ian M. Derrington and Mikhail Pavlenok and Michael Niederweis and Jens H. Gundlach and Aleksei Aksimentiev},
  title = {{M}olecular {D}ynamics {S}tudy of {MspA} {A}rginine {M}utants {P}redicts {S}low {DNA} {T}ranslocations and {I}on {C}urrent {B}lockades {I}ndicative of {DNA} {S}equence},
  journal = {{ACS} Nano}
}

@article{Bhattacharya-2016,
  doi = {10.1021/acsnano.6b00940},
  url = {https://doi.org/10.1021/acsnano.6b00940},
  year = {2016},
  month = apr,
  publisher = {American Chemical Society ({ACS})},
  volume = {10},
  number = {4},
  pages = {4644--4651},
  author = {Swati Bhattacharya and Jejoong Yoo and Aleksei Aksimentiev},
  title = {{W}ater {M}ediates {R}ecognition of {DNA} {S}equence via {I}onic {C}urrent {B}lockade in a {B}iological {N}anopore},
  journal = {{ACS} Nano}
}

@article{Manara-2015,
  doi = {10.1038/srep12783},
  url = {https://doi.org/10.1038/srep12783},
  year = {2015},
  month = aug,
  publisher = {Springer Science and Business Media {LLC}},
  volume = {5},
  number = {1},
  author = {Richard M.A. Manara and E. Jayne Wallace and Syma Khalid},
  title = {{DNA} sequencing with {MspA}: {M}olecular {D}ynamics simulations reveal free-energy differences between sequencing and non-sequencing mutants},
  journal = {Sci. Rep.}
}

@article{Cao-2018,
  doi = {10.1038/s41467-018-05108-5},
  url = {https://doi.org/10.1038/s41467-018-05108-5},
  year = {2018},
  month = jul,
  publisher = {Springer Science and Business Media {LLC}},
  volume = {9},
  number = {1},
  author = {Chan Cao and Meng-Yin Li and Nuria Cirauqui and Ya-Qian Wang and Matteo Dal Peraro and He Tian and Yi-Tao Long},
  title = {{M}apping the sensing spots of aerolysin for single oligonucleotides analysis},
  journal = {Nat. Commun.}
}

@article{Cao-2019,
  doi = {10.1038/s41467-019-12690-9},
  url = {https://doi.org/10.1038/s41467-019-12690-9},
  year = {2019},
  month = oct,
  publisher = {Springer Science and Business Media {LLC}},
  volume = {10},
  number = {1},
  pages = {4918},
  author = {Chan Cao and Nuria Cirauqui and Maria Jose Marcaida and Elena Buglakova and Alice Duperrex and Aleksandra Radenovic and Matteo Dal Peraro},
  title = {{S}ingle-molecule sensing of peptides and nucleic acids by engineered aerolysin nanopores},
  journal = {Nat. Commun.}
}

@article{Renkin-1954,
  doi = {10.1085/jgp.38.2.225},
  url = {https://doi.org/10.1085/jgp.38.2.225},
  year = {1954},
  month = {11},
  volume    = {38},
  number    = {2},
  pages     = {225--43},
  author    = {Renkin, Eugene M.},
  title     = {{F}iltration, diffusion, and molecular sieving through porous cellulose membranes},
  journal   = {J. Gen. Physiol.}
}

@article{Deen-1987,
  doi = {10.1002/aic.690330902},
  url = {https://doi.org/10.1002/aic.690330902},
  year = {1987},
  month = sep,
  publisher = {Wiley},
  volume = {33},
  number = {9},
  pages = {1409--1425},
  author = {W. M. Deen},
  title = {{H}indered transport of large molecules in liquid-filled pores},
  journal = {{AIChE} J.}
}

@article{Dechadilok-2006,
  doi = {10.1021/ie051387n},
  url = {https://doi.org/10.1021/ie051387n},
  year = {2006},
  month = oct,
  publisher = {American Chemical Society ({ACS})},
  volume = {45},
  number = {21},
  pages = {6953--6959},
  author = {Panadda Dechadilok and William M. Deen},
  title = {{H}indrance {F}actors for {D}iffusion and {C}onvection in {P}ores},
  journal = {Ind. Eng. Chem. Res.}
}

@article{Anderson-1972,
  doi = {10.1039/f19726800744},
  url = {https://doi.org/10.1039/f19726800744},
  year = {1972},
  publisher = {Royal Society of Chemistry ({RSC})},
  volume = {68},
  number = {0},
  pages = {744},
  author = {John L. Anderson and John A. Quinn},
  title = {{I}onic mobility in microcapillaries. {A} test for anomalous water structures},
  journal = {J. Chem. Soc. Faraday Trans.}
}

@article{Muthukumar-2010,
  doi = {10.1063/1.3429882},
  url = {https://doi.org/10.1063/1.3429882},
  year = {2010},
  month = may,
  publisher = {{AIP} Publishing},
  volume = {132},
  number = {19},
  pages = {195101},
  author = {M. Muthukumar},
  title = {{T}heory of capture rate in polymer translocation},
  journal = {J. Chem. Phys.}
}

@article{Makarov-1998,
  doi = {10.1016/s0006-3495(98)77502-2},
  url = {https://doi.org/10.1016/s0006-3495(98)77502-2},
  year = {1998},
  month = jul,
  publisher = {Elsevier {BV}},
  volume = {75},
  number = {1},
  pages = {150--158},
  author = {Vladimir A. Makarov and Michael Feig and B. Kim Andrews and B. Montgomery Pettitt},
  title = {{D}iffusion of {S}olvent around {B}iomolecular {S}olutes: {A} {M}olecular {D}ynamics {S}imulation {S}tudy},
  journal = {Biophys. J.}
}

@article{Pronk-2014,
  doi = {10.1038/ncomms4034},
  url = {https://doi.org/10.1038/ncomms4034},
  year = {2014},
  month = jan,
  publisher = {Springer Science and Business Media {LLC}},
  volume = {5},
  number = {1},
  author = {Sander Pronk and Erik Lindahl and Peter M. Kasson},
  title = {{D}ynamic heterogeneity controls diffusion and viscosity near biological interfaces},
  journal = {Nat. Commun.}
}

@article{Ye-2011,
  doi = {10.1186/1556-276x-6-87},
  url = {https://doi.org/10.1186/1556-276x-6-87},
  year = {2011},
  publisher = {Springer Science and Business Media {LLC}},
  volume = {6},
  number = {1},
  pages = {87},
  author = {Hongfei Ye and Hongwu Zhang and Zhongqiang Zhang and Yonggang Zheng},
  title = {{S}ize and temperature effects on the viscosity of water inside carbon nanotubes},
  journal = {Nanoscale Res. Lett.}
}

@article{Manghi-2018,
  doi = {10.1103/physreve.98.012605},
  url = {https://doi.org/10.1103/physreve.98.012605},
  year = {2018},
  month = jul,
  publisher = {American Physical Society ({APS})},
  volume = {98},
  number = {1},
  author = {Manoel Manghi and John Palmeri and Khadija Yazda and Fran{\c{c}}ois Henn and Vincent Jourdain},
  title = {{R}ole of charge regulation and flow slip in the ionic conductance of nanopores: {A}n analytical approach},
  journal = {Phys. Rev. E}
}

@article{Kannam-2017,
  doi = {10.1063/1.4975161},
  url = {https://doi.org/10.1063/1.4975161},
  year = {2017},
  month = feb,
  publisher = {{AIP} Publishing},
  volume = {146},
  number = {5},
  pages = {054108},
  author = {Sridhar Kumar Kannam and Matthew T. Downton},
  title = {{T}ranslational diffusion of proteins in nanochannels},
  journal = {J. Chem. Phys.}
}

@article{McMullen-2017,
  doi = {10.1021/acsnano.7b06767},
  url = {https://doi.org/10.1021/acsnano.7b06767},
  year = {2017},
  month = nov,
  publisher = {American Chemical Society ({ACS})},
  volume = {11},
  number = {11},
  pages = {11669--11677},
  author = {Angus J. McMullen and Jay X. Tang and Derek Stein},
  title = {{N}anopore {M}easurements of {F}ilamentous {V}iruses {R}eveal a {S}ub-nanometer-{S}cale {S}tagnant {F}luid {L}ayer},
  journal = {{ACS} Nano}
}

@article{Duan-2010,
  doi = {10.1038/nnano.2010.233},
  url = {https://doi.org/10.1038/nnano.2010.233},
  year = {2010},
  month = nov,
  publisher = {Springer Science and Business Media {LLC}},
  volume = {5},
  number = {12},
  pages = {848--852},
  author = {Chuanhua Duan and Arun Majumdar},
  title = {{A}nomalous ion transport in 2-nm hydrophilic nanochannels},
  journal = {Nat. Nanotechnol.}
}

@article{Cozmuta-2005,
  doi = {10.1080/08927020412331308467},
  url = {https://doi.org/10.1080/08927020412331308467},
  year = {2005},
  month = feb,
  publisher = {Informa {UK} Limited},
  volume = {31},
  number = {2-3},
  pages = {79--93},
  author = {Ioana Cozmuta and James T. O{\textquotesingle}Keeffe and Deepak Bose and Viktor Stolc},
  title = {{H}ybrid {MD}-{N}ernst {P}lanck model of {\textalpha}-hemolysin conductance properties},
  journal = {Mol. Simulat.}
}

@article{OKeeffe-2007,
  doi = {10.1016/j.chemphys.2007.09.013},
  url = {https://doi.org/10.1016/j.chemphys.2007.09.013},
  year = {2007},
  month = dec,
  publisher = {Elsevier {BV}},
  volume = {342},
  number = {1-3},
  pages = {25--32},
  author = {James O'Keeffe and Ioana Cozmuta and Deepak Bose and Viktor Stolc},
  title = {{A} predictive {MD}-{N}ernst{\textendash}{P}lanck model for transport in alpha-hemolysin: {M}odeling anisotropic ion currents},
  journal = {Chem. Phys.}
}

@article{Liu-2013,
  doi = {10.1021/jp408330f},
  url = {https://doi.org/10.1021/jp408330f},
  year = {2013},
  month = sep,
  publisher = {American Chemical Society ({ACS})},
  volume = {117},
  number = {40},
  pages = {12051--12058},
  author = {Jinn-Liang Liu and Bob Eisenberg},
  title = {{C}orrelated {I}ons in a {C}alcium {C}hannel {M}odel: {A} {P}oisson{\textendash}{F}ermi {T}heory},
  journal = {J. Phys. Chem. B}
}

@article{Liu-2020,
  doi = {10.3390/e22050550},
  url = {https://doi.org/10.3390/e22050550},
  year = {2020},
  month = may,
  publisher = {{MDPI} {AG}},
  volume = {22},
  number = {5},
  pages = {550},
  author = {Jinn-Liang Liu and Bob Eisenberg},
  title = {{M}olecular {M}ean-{F}ield {T}heory of {I}onic {S}olutions: {A} {P}oisson-{N}ernst-{P}lanck-{B}ikerman {M}odel},
  journal = {Entropy}
}

@article{Vo-2016,
  doi = {10.1038/srep33881},
  url = {https://doi.org/10.1038/srep33881},
  year = {2016},
  month = sep,
  publisher = {Springer Science and Business Media {LLC}},
  volume = {6},
  number = {1},
  author = {Truong Quoc Vo and BoHung Kim},
  title = {{T}ransport {P}henomena of {W}ater in {M}olecular {F}luidic {C}hannels},
  journal = {Sci. Rep.}
}

@article{Furini-2006,
  doi = {10.1529/biophysj.105.078741},
  url = {https://doi.org/10.1529/biophysj.105.078741},
  year = {2006},
  month = nov,
  publisher = {Elsevier {BV}},
  volume = {91},
  number = {9},
  pages = {3162--3169},
  author = {Simone Furini and Francesco Zerbetto and Silvio Cavalcanti},
  title = {{A}pplication of the {P}oisson-{N}ernst-{P}lanck {T}heory with {S}pace-{D}ependent {D}iffusion {C}oefficients to {KcsA}},
  journal = {Biophys. J.}
}

@article{Liu-2015,
  doi = {10.1103/physreve.92.012711},
  url = {https://doi.org/10.1103/physreve.92.012711},
  year = {2015},
  month = jul,
  publisher = {American Physical Society ({APS})},
  volume = {92},
  number = {1},
  author = {Jinn-Liang Liu and Bob Eisenberg},
  title = {{N}umerical methods for a {P}oisson-{N}ernst-{P}lanck-{F}ermi model of biological ion channels},
  journal = {Phys. Rev. E}
}

@article{Bhadauria-2017,
  doi = {10.1063/1.4982731},
  url = {https://doi.org/10.1063/1.4982731},
  year = {2017},
  month = may,
  publisher = {{AIP} Publishing},
  volume = {146},
  number = {18},
  pages = {184106},
  author = {Ravi Bhadauria and N. R. Aluru},
  title = {{M}ultiscale modeling of electroosmotic flow: {E}ffects of discrete ion, enhanced viscosity, and surface friction},
  journal = {J. Chem. Phys.}
}

@article{Buchner-1999,
  doi = {10.1021/jp982977k},
  url = {https://doi.org/10.1021/jp982977k},
  year = {1999},
  month = jan,
  publisher = {American Chemical Society ({ACS})},
  volume = {103},
  number = {1},
  pages = {1--9},
  author = {Richard Buchner and Glenn T. Hefter and Peter M. May},
  title = {{D}ielectric {R}elaxation of {A}queous {NaCl} {S}olutions},
  journal = {J. Phys. Chem. A}
}

@article{Gavish-2016,
  doi = {10.1103/physreve.94.012611},
  url = {https://doi.org/10.1103/physreve.94.012611},
  year = {2016},
  month = jul,
  publisher = {American Physical Society ({APS})},
  volume = {94},
  number = {1},
  author = {Nir Gavish and Keith Promislow},
  title = {{D}ependence of the dielectric constant of electrolyte solutions on ionic concentration: A microfield approach},
  journal = {Phys. Rev. E}
}

@article{Hai-Lang-1996,
  doi = {10.1021/je9501402},
  url = {https://doi.org/10.1021/je9501402},
  year = {1996},
  month = jan,
  publisher = {American Chemical Society ({ACS})},
  volume = {41},
  number = {3},
  pages = {516--520},
  author = {Hai-Lang, Zhang and Shi-Jun, Han},
  title = {{V}iscosity and {D}ensity of {W}ater + {S}odium {C}hloride + {P}otassium {C}hloride {S}olutions at 298.15 {K}},
  journal = {J. Chem. Eng. Data}
}

@book{Mills-1989,
  author    = {Mills, R. and Lobo, V.M.M.},
  title     = {{S}elf-diffusion in electrolyte solutions: a critical examination of data compiled from the
  literature},
  number    = {v. 36},
  isbn      = {9780444416896},
  lccn      = {lc88037012},
  series    = {Phys. Sci. Data},
  url       = {https://books.google.be/books?id=LfeFAAAAIAAJ},
  year      = {1989},
  publisher = {Elsevier}
}

@article{Panopoulos-1986,
  doi = {10.1007/bf00646692},
  url = {https://doi.org/10.1007/bf00646692},
  year = {1986},
  month = mar,
  publisher = {Springer Science and Business Media {LLC}},
  volume = {15},
  number = {3},
  pages = {243--252},
  author = {Demetrios K. Panopoulos and Hiroyuki Kaneko and Michael Spiro},
  title = {{T}ransference numbers of sodium chloride in concentrated aqueous solutions and chloride conductances in several concentrated electrolyte solutions},
  journal = {J. Sol. Chem.}
}

@article{Currie-1960,
  doi = {10.1021/j100840a037},
  url = {https://doi.org/10.1021/j100840a037},
  year = {1960},
  month = nov,
  publisher = {American Chemical Society ({ACS})},
  volume = {64},
  number = {11},
  pages = {1751--1753},
  author = {D. J. Currie and A. R. Gordon},
  title = {{TRANSFERENCE} {NUMBERS} {FOR} {CONCENTRATED} {AQUEOUS} {SODIUM} {CHLORIDE} {SOLUTIONS}, {AND} {THE} {IONIC} {CONDUCTANCES} {FOR} {POTASSIUM} {AND} {SODIUM} {CHLORIDES}},
  journal = {J. Phys. Chem.}
}

@article{Schonert-2013,
  doi = {10.1007/s10953-012-9944-y},
  url = {https://doi.org/10.1007/s10953-012-9944-y},
  year = {2013},
  month = jan,
  publisher = {Springer Science and Business Media {LLC}},
  volume = {43},
  number = {1},
  pages = {26--39},
  author = {Hansj{\"{u}}rgen Sch{\"{o}}nert},
  title = {{T}ransference {N}umbers in the {H2O}~+~{NaCl}~+~{MgCl}2 {S}ystem at {T}~=~298.15~{K}, {D}etermined in a {F}ive-{C}ompartment {H}ittorf {C}ell},
  journal = {J. Sol. Chem.}
}

@article{DellaMonica-1979,
  doi = {10.1016/0013-4686(79)87099-1},
  url = {https://doi.org/10.1016/0013-4686(79)87099-1},
  year = {1979},
  month = sep,
  publisher = {Elsevier {BV}},
  volume = {24},
  number = {9},
  pages = {1013--1017},
  author = {Mario Della Monica and G. Petrella and A. Sacco and S. Bufo},
  title = {{T}ransference numbers in concentrated sodium chloride solutions},
  journal = {Electrochim. Acta}
}

@article{Bianchi-1989,
  doi = {10.1007/bf00657336},
  url = {https://doi.org/10.1007/bf00657336},
  year = {1989},
  month = may,
  publisher = {Springer Science and Business Media {LLC}},
  volume = {18},
  number = {5},
  pages = {485--491},
  author = {Hugo Bianchi and Horacio R. Corti and Roberto Fern{\'{a}}ndez-Prini},
  title = {{T}he conductivity of concentrated aqueous mixtures of {NaCl} and {MgCl\textsubscript{2}} at 25{\textdegree}{C}},
  journal = {J. Sol. Chem.}
}

@article{Goldsack-1976,
  doi = {10.1139/v76-418},
  url = {https://doi.org/10.1139/v76-418},
  year = {1976},
  month = sep,
  publisher = {Canadian Science Publishing},
  volume = {54},
  number = {18},
  pages = {2953--2966},
  author = {Douglas E. Goldsack and Raymond Franchetto and Arlene (Anttila) Franchetto},
  title = {{S}olvation effects on the conductivity of concentrated electrolyte solutions},
  journal = {Can. J. Chem.}
}

@article{ContrerasAburto-2013-1,
  doi = {10.1063/1.4822297},
  url = {https://doi.org/10.1063/1.4822297},
  year = {2013},
  month = oct,
  publisher = {{AIP} Publishing},
  volume = {139},
  number = {13},
  pages = {134109},
  author = {Claudio Contreras Aburto and Gerhard N{\"{a}}gele},
  title = {{A} unifying mode-coupling theory for transport properties of electrolyte solutions. {I}. {G}eneral scheme and limiting laws},
  journal = {J. Chem. Phys.}
}

@article{ContrerasAburto-2013-2,
  doi = {10.1063/1.4822298},
  url = {https://doi.org/10.1063/1.4822298},
  year = {2013},
  month = oct,
  publisher = {{AIP} Publishing},
  volume = {139},
  number = {13},
  pages = {134110},
  author = {Claudio Contreras Aburto and Gerhard N{\"{a}}gele},
  title = {A unifying mode-coupling theory for transport properties of electrolyte solutions. {II}. {R}esults for equal-sized ions electrolytes},
  journal = {J. Chem. Phys.}
}

@misc{Haynes-2017,
  title = {{CRC} {H}andbook of {C}hemistry and {P}hysics, 97th edn, {V}ol. 2016--2017},
  author = {Haynes, WM and Lide, DR and Bruno, TJ},
  year = {2017},
  publisher = {CRC Press}
}

@article{Esteso-1976,
  doi = {10.1039/f19767201425},
  url = {https://doi.org/10.1039/f19767201425},
  year = {1976},
  publisher = {Royal Society of Chemistry ({RSC})},
  volume = {72},
  number = {0},
  pages = {1425},
  author = {Miguel A. Esteso and Chee-Yan Chan and Michael Spiro},
  title = {{E}xamination of a transference number anomaly: {0.02 mol~dm$^{-3}$} aqueous sodium chloride at 25{\textdegree}C},
  journal = {J. Chem. Soc. Faraday Trans.}
}

@article{Jurrus-2018,
  doi = {10.1002/pro.3280},
  url = {https://doi.org/10.1002/pro.3280},
  year = {2018},
  month = oct,
  publisher = {Wiley},
  volume = {27},
  number = {1},
  pages = {112--128},
  author = {Elizabeth Jurrus and Dave Engel and Keith Star and Kyle Monson and Juan Brandi and Lisa E. Felberg and David H. Brookes and Leighton Wilson and Jiahui Chen and Karina Liles and Minju Chun and Peter Li and David W. Gohara and Todd Dolinsky and Robert Konecny and David R. Koes and Jens Erik Nielsen and Teresa Head-Gordon and Weihua Geng and Robert Krasny and Guo-Wei Wei and Michael J. Holst and J. Andrew McCammon and Nathan A. Baker},
  title = {{I}mprovements to the {APBS} biomolecular solvation software suite},
  journal = {Protein Sci.}
}

@article{Sondergaard-2011,
  doi = {10.1021/ct200133y},
  url = {https://doi.org/10.1021/ct200133y},
  year = {2011},
  month = jun,
  publisher = {American Chemical Society ({ACS})},
  volume = {7},
  number = {7},
  pages = {2284--2295},
  author = {Chresten R. S{\o}ndergaard and Mats H. M. Olsson and Micha{\l} Rostkowski and Jan H. Jensen},
  title = {Improved Treatment of Ligands and Coupling Effects in Empirical Calculation and Rationalization of {pKa} Values},
  journal = {J. Chem. Theory Comput.}
}

@article{Olsson-2011,
  doi = {10.1021/ct100578z},
  url = {https://doi.org/10.1021/ct100578z},
  year = {2011},
  month = jan,
  publisher = {American Chemical Society ({ACS})},
  volume = {7},
  number = {2},
  pages = {525--537},
  author = {Mats H. M. Olsson and Chresten R. S{\o}ndergaard and Michal Rostkowski and Jan H. Jensen},
  title = {{PROPKA}3: Consistent Treatment of Internal and Surface Residues in Empirical {pKa} Predictions},
  journal = {J. Chem. Theory Comput.}
}

@article{Wang-2015,
  doi = {10.1002/prot.24935},
  url = {https://doi.org/10.1002/prot.24935},
  year = {2015},
  month = oct,
  publisher = {Wiley},
  volume = {83},
  number = {12},
  pages = {2186--2197},
  author = {Lin Wang and Lin Li and Emil Alexov},
  title = {{pKa} predictions for proteins, {RNAs}, and {DNAs} with the {G}aussian dielectric function using {DelPhi} {pKa}},
  journal = {Proteins}
}

@article{Best-2012,
  doi = {10.1021/ct300400x},
  url = {https://doi.org/10.1021/ct300400x},
  year = {2012},
  month = aug,
  publisher = {American Chemical Society ({ACS})},
  volume = {8},
  number = {9},
  pages = {3257--3273},
  author = {Robert B. Best and Xiao Zhu and Jihyun Shim and Pedro E. M. Lopes and Jeetain Mittal and Michael Feig and Alexander D. MacKerell},
  title = {{O}ptimization of the {A}dditive {CHARMM} {A}ll-{A}tom {P}rotein {F}orce {F}ield {T}argeting {I}mproved {S}ampling of the {B}ackbone phi, {\textpsi} and {S}ide-{C}hain {\textchi}(1) and {\textchi}(2) {D}ihedral {A}ngles.},
  journal = {J. Chem. Theory Comput.}
}

@article{Huang-2016,
  doi = {10.1038/nmeth.4067},
  url = {https://doi.org/10.1038/nmeth.4067},
  year = {2016},
  month = nov,
  publisher = {Springer Science and Business Media {LLC}},
  volume = {14},
  number = {1},
  pages = {71--73},
  author = {Jing Huang and Sarah Rauscher and Grzegorz Nawrocki and Ting Ran and Michael Feig and Bert L de Groot and Helmut Grubm{\"{u}}ller and Alexander D MacKerell},
  title = {{CHARMM}36m: an improved force field for folded and intrinsically disordered proteins},
  journal = {Nat. Meth.}
}

@article{Oostenbrink-2004,
  doi = {10.1002/jcc.20090},
  url = {https://doi.org/10.1002/jcc.20090},
  year = {2004},
  publisher = {Wiley},
  volume = {25},
  number = {13},
  pages = {1656--1676},
  author = {Chris Oostenbrink and Alessandra Villa and Alan E. Mark and Wilfred F. Van Gunsteren},
  title = {{A} biomolecular force field based on the free enthalpy of hydration and solvation: The {GROMOS} force-field parameter sets {53A5} and {53A6}},
  journal = {J. Computat. Chem.}
}

@article{Daiguji-2010,
  doi = {10.1039/b820556f},
  url = {https://doi.org/10.1039/b820556f},
  year = {2010},
  publisher = {Royal Society of Chemistry ({RSC})},
  volume = {39},
  number = {3},
  pages = {901--911},
  author = {Hirofumi Daiguji},
  title = {{I}on transport in nanofluidic channels},
  journal = {Chem. Soc. Rev.}
}

@article{Zhang-2016,
  doi = {10.1021/acs.chemrev.5b00608},
  url = {https://doi.org/10.1021/acs.chemrev.5b00608},
  year = {2016},
  month = dec,
  publisher = {American Chemical Society ({ACS})},
  volume = {116},
  number = {1},
  pages = {215--257},
  author = {Anqi Zhang and Charles M. Lieber},
  title = {Nano-Bioelectronics},
  journal = {Chem. Rev.}
}

@article{Chen-2013,
  doi = {10.1039/c2ra22351a},
  url = {https://doi.org/10.1039/c2ra22351a},
  year = {2013},
  publisher = {Royal Society of Chemistry ({RSC})},
  volume = {3},
  number = {14},
  pages = {4473},
  author = {Chao Chen and Qingji Xie and Dawei Yang and Hualing Xiao and Yingchun Fu and Yueming Tan and Shouzhuo Yao},
  title = {{R}ecent advances in electrochemical glucose biosensors: a review},
  journal = {{RSC} Adv.}
}

@article{Deamer-2016,
  doi = {10.1038/nbt.3423},
  url = {https://doi.org/10.1038/nbt.3423},
  year = {2016},
  month = may,
  publisher = {Springer Science and Business Media {LLC}},
  volume = {34},
  number = {5},
  pages = {518--524},
  author = {David Deamer and Mark Akeson and Daniel Branton},
  title = {{T}hree decades of nanopore sequencing},
  journal = {Nat. Biotechnol.}
}

@article{Schoch-2008,
  doi = {10.1103/revmodphys.80.839},
  url = {https://doi.org/10.1103/revmodphys.80.839},
  year = {2008},
  month = jul,
  publisher = {American Physical Society ({APS})},
  volume = {80},
  number = {3},
  pages = {839--883},
  author = {Reto B. Schoch and Jongyoon Han and Philippe Renaud},
  title = {{T}ransport phenomena in nanofluidics},
  journal = {Rev. Mod. Phys.}
}

@article{Grochowski-2008,
  doi = {10.1002/bip.20877},
  url = {https://doi.org/10.1002/bip.20877},
  year = {2008},
  publisher = {Wiley},
  volume = {89},
  number = {2},
  pages = {93--113},
  author = {Pawe{\l} Grochowski and Joanna Trylska},
  title = {{C}ontinuum molecular electrostatics, salt effects, and counterion binding{\textemdash}{A} review of the {P}oisson{\textendash}{B}oltzmann theory and its modifications},
  journal = {Biopolymers}
}

@article{Neuman-2004,
  doi       = {10.1063/1.1785844},
  url       = {https://doi.org/10.1063/1.1785844},
  year      = {2004},
  publisher = {{AIP} Publishing},
  volume    = {75},
  number    = {9},
  pages     = {2787--2809},
  author    = {Keir C. Neuman and Steven M. Block},
  title     = {{O}ptical {T}rapping},
  journal   = {Rev. Sci. Instrum.}
}

@article{Bruckbauer-2007,
  doi = {10.1529/biophysj.107.104737},
  url = {https://doi.org/10.1529/biophysj.107.104737},
  year = {2007},
  publisher = {Elsevier {BV}},
  volume = {93},
  number = {9},
  pages = {3120--3131},
  author = {Andreas Bruckbauer and Peter James and Dejian Zhou and Ji Won Yoon and David Excell and Yuri Korchev and Roy Jones and David Klenerman},
  title = {{N}anopipette {D}elivery of {I}ndividual {M}olecules to {C}ellular {C}ompartments for {S}ingle-{M}olecule {F}luorescence {T}racking},
  journal = {Biophys. J.}
}

@article{Yang-2009,
  doi = {10.1038/nature07593},
  url = {https://doi.org/10.1038/nature07593},
  year = {2009},
  publisher = {Springer Nature},
  volume = {457},
  number = {7225},
  pages = {71--75},
  author = {Allen H. J. Yang and Sean D. Moore and Bradley S. Schmidt and Matthew Klug and Michal Lipson and David Erickson},
  title = {{O}ptical {M}anipulation of {N}anoparticles and {B}iomolecules in {S}ub-{W}avelength {S}lot {W}aveguides},
  journal = {Nature}
}

@article{Bergeron-2013,
  doi = {10.3791/4424},
  url = {https://doi.org/10.3791/4424},
  year = {2013},
  month = jan,
  publisher = {{MyJove} Corporation},
  number = {71},
  author = {Jarrah Bergeron and Ana Zehtabi-Oskuie and Saeedeh Ghaffari and Yuanjie Pang and Reuven Gordon},
  title = {{O}ptical {T}rapping of {N}anoparticles},
  journal = {J. Vis. Exp.}
}

@article{Mandal-2010,
  doi = {10.1021/nl9029225},
  url = {https://doi.org/10.1021/nl9029225},
  year = {2010},
  month = jan,
  publisher = {American Chemical Society ({ACS})},
  volume = {10},
  number = {1},
  pages = {99--104},
  author = {Sudeep Mandal and Xavier Serey and David Erickson},
  title = {{N}anomanipulation {U}sing {S}ilicon {P}hotonic {C}rystal {R}esonators},
  journal = {Nano Lett.}
}

@article{Juan-2009,
  doi = {10.1038/nphys1422},
  url = {https://doi.org/10.1038/nphys1422},
  year = {2009},
  month = oct,
  publisher = {Springer Nature},
  volume = {5},
  number = {12},
  pages = {915--919},
  author = {Mathieu L. Juan and Reuven Gordon and Yuanjie Pang and Fatima Eftekhari and Romain Quidant},
  title = {{S}elf-induced back-action optical trapping of dielectric nanoparticles},
  journal = {Nat. Phys.}
}

@article{Kerman-2015,
  doi = {10.1039/c5nr05341b},
  url = {https://doi.org/10.1039/c5nr05341b},
  year = {2015},
  publisher = {Royal Society of Chemistry ({RSC})},
  volume = {7},
  number = {44},
  pages = {18612--18618},
  author = {Sarp Kerman and Chang Chen and Yi Li and Wim Van Roy and Liesbet Lagae and Pol Van Dorpe},
  title = {{R}aman {F}ingerprinting of {S}ingle {D}ielectric {N}anoparticles in {P}lasmonic {N}anopores},
  journal = {Nanoscale}
}

@article{Chen-2011,
  doi = {10.1021/nl2031458},
  url = {https://doi.org/10.1021/nl2031458},
  year = {2011},
  month = dec,
  publisher = {American Chemical Society ({ACS})},
  volume = {12},
  number = {1},
  pages = {125--132},
  author = {Chang Chen and Mathieu L. Juan and Yi Li and Guido Maes and Gustaaf Borghs and Pol Van Dorpe and Romain Quidant},
  title = {{E}nhanced {O}ptical {T}rapping and {A}rrangement of {N}ano-{O}bjects in a {P}lasmonic {N}anocavity},
  journal = {Nano Lett.}
}

@article{Pang-2011,
  doi = {10.1021/nl203719v},
  url = {https://doi.org/10.1021/nl203719v},
  year = {2011},
  month = dec,
  publisher = {American Chemical Society ({ACS})},
  volume = {12},
  number = {1},
  pages = {402--406},
  author = {Yuanjie Pang and Reuven Gordon},
  title = {{O}ptical {T}rapping of a {S}ingle {P}rotein},
  journal = {Nano Lett.}
}

@article{Cohen-2005,
  doi = {10.1063/1.1872220},
  url = {https://doi.org/10.1063/1.1872220},
  year = {2005},
  month = feb,
  publisher = {{AIP} Publishing},
  volume = {86},
  number = {9},
  pages = {093109},
  author = {Adam E. Cohen and W. E. Moerner},
  title = {{M}ethod for {T}rapping and {M}anipulating {N}anoscale {O}bjects in {S}olution},
  journal = {Appl. Phys. Lett.}
}

@article{Goldsmith-2011,
  doi = {10.1073/pnas.1113572108},
  url = {https://doi.org/10.1073/pnas.1113572108},
  year = {2011},
  month = oct,
  publisher = {Proceedings of the National Academy of Sciences},
  volume = {108},
  number = {42},
  pages = {17269--17274},
  author = {R. H. Goldsmith and L. C. Tabares and D. Kostrz and C. Dennison and T. J. Aartsma and G. W. Canters and W. E. Moerner},
  title = {{R}edox {C}ycling and {K}inetic {A}nalysis of {S}ingle {M}olecules of {S}olution-{P}hase {N}itrite {R}eductase},
  journal = {Proc. Natl. Acad. Sci. U. S. A.}
}

@article{Cohen-2006,
  doi = {10.1073/pnas.0509976103},
  url = {https://doi.org/10.1073/pnas.0509976103},
  year = {2006},
  month = mar,
  publisher = {Proceedings of the National Academy of Sciences},
  volume = {103},
  number = {12},
  pages = {4362--4365},
  author = {A. E. Cohen and W. E. Moerner},
  title = {{S}uppressing {B}rownian {M}otion of {I}ndividual {B}iomolecules in {S}olution},
  journal = {Proc. Natl. Acad. Sci. U. S. A.}
}

@article{Fields-2011,
  doi = {10.1073/pnas.1103554108},
  url = {https://doi.org/10.1073/pnas.1103554108},
  year = {2011},
  month = may,
  publisher = {Proceedings of the National Academy of Sciences},
  volume = {108},
  number = {22},
  pages = {8937--8942},
  author = {A. P. Fields and A. E. Cohen},
  title = {{E}lectrokinetic {T}rapping at the {O}ne {N}anometer {L}imit},
  journal = {Proc. Natl. Acad. Sci. U. S. A.}
}

@article{Goldsmith-2010,
  doi = {10.1038/nchem.545},
  url = {https://doi.org/10.1038/nchem.545},
  year = {2010},
  month = jan,
  publisher = {Springer Science and Business Media {LLC}},
  volume = {2},
  number = {3},
  pages = {179--186},
  author = {Randall H. Goldsmith and W. E. Moerner},
  title = {{W}atching {C}onformational- and {P}hotodynamics of {S}ingle {F}luorescent {P}roteins in {S}olution},
  journal = {Nat. Chem.}
}

@article{Krishnan-2010,
  doi = {10.1038/nature09404},
  url = {https://doi.org/10.1038/nature09404},
  year = {2010},
  month = oct,
  publisher = {Springer Nature},
  volume = {467},
  number = {7316},
  pages = {692--695},
  author = {Madhavi Krishnan and Nassiredin Mojarad and Philipp Kukura and Vahid Sandoghdar},
  title = {{G}eometry-{I}nduced {E}lectrostatic {T}rapping of {N}anometric {O}bjects in a {F}luid},
  journal = {Nature}
}

@article{Myers-2015,
  doi = {10.1038/nnano.2015.173},
  url = {https://doi.org/10.1038/nnano.2015.173},
  year = {2015},
  month = aug,
  publisher = {Springer Nature},
  volume = {10},
  number = {10},
  pages = {886--891},
  author = {Christopher J. Myers and Michele Celebrano and Madhavi Krishnan},
  title = {{I}nformation {S}torage and {R}etrieval in a {S}ingle {L}evitating {C}olloidal {P}article},
  journal = {Nat. Nanotechnol.}
}

@article{RinconRestrepo-2011,
  doi = {10.1021/nl1038874},
  url = {https://doi.org/10.1021/nl1038874},
  year = {2011},
  month = feb,
  publisher = {American Chemical Society ({ACS})},
  volume = {11},
  number = {2},
  pages = {746--750},
  author = {Marcela Rincon-Restrepo and Ellina Mikhailova and Hagan Bayley and Giovanni Maglia},
  title = {{C}ontrolled {T}ranslocation of {I}ndividual {DNA} {M}olecules through {P}rotein {N}anopores with {E}ngineered {M}olecular {B}rakes},
  journal = {Nano Lett.}
}

@article{Squires-2015,
  doi = {10.1038/srep11643},
  url = {https://doi.org/10.1038/srep11643},
  year = {2015},
  month = jun,
  publisher = {Springer Nature},
  volume = {5},
  number = {1},
  author = {Allison Squires and Evrim Atas and Amit Meller},
  title = {{N}anopore {S}ensing of {I}ndividual {T}ranscription {F}actors {B}ound to {DNA}},
  journal = {Sci. Rep.}
}

@article{Yang-2018,
  doi = {10.1021/acs.nanolett.8b02968},
  url = {https://doi.org/10.1021/acs.nanolett.8b02968},
  year = {2018},
  month = sep,
  publisher = {American Chemical Society ({ACS})},
  volume = {18},
  number = {10},
  pages = {6469--6474},
  author = {Wayne Yang and Laura Restrepo-P{\'{e}}rez and Michel Bengtson and Stephanie J. Heerema and Anthony Birnie and Jaco van der Torre and Cees Dekker},
  title = {{D}etection of {CRISPR}-{dCas}9 on {DNA} with {S}olid-{S}tate {N}anopores},
  journal = {Nano Lett.}
}

@article{Yusko-2011,
  doi = {10.1038/nnano.2011.12},
  url = {https://doi.org/10.1038/nnano.2011.12},
  year = {2011},
  month = feb,
  publisher = {Springer Nature},
  volume = {6},
  number = {4},
  pages = {253--260},
  author = {Erik C. Yusko and Jay M. Johnson and Sheereen Majd and Panchika Prangkio and Ryan C. Rollings and Jiali Li and Jerry Yang and Michael Mayer},
  title = {{C}ontrolling {P}rotein {T}ranslocation {T}hrough {N}anopores with {B}io-{I}nspired {F}luid {W}alls},
  journal = {Nat. Nanotechnol.}
}

@article{Keyser-2006,
  doi = {10.1038/nphys344},
  url = {https://doi.org/10.1038/nphys344},
  year = {2006},
  month = jul,
  publisher = {Springer Nature},
  volume = {2},
  number = {7},
  pages = {473--477},
  author = {Ulrich F. Keyser and Bernard N. Koeleman and Stijn van Dorp and Diego Krapf and Ralph M. M. Smeets and Serge G. Lemay and Nynke H. Dekker and Cees Dekker},
  title = {{D}irect {F}orce {M}easurements on {DNA} in a {S}olid-{S}tate {N}anopore},
  journal = {Nat. Phys.}
}

@article{Lieberman-2010,
  doi = {10.1021/ja1087612},
  url = {https://doi.org/10.1021/ja1087612},
  year = {2010},
  month = dec,
  publisher = {American Chemical Society ({ACS})},
  volume = {132},
  number = {50},
  pages = {17961--17972},
  author = {Kate R. Lieberman and Gerald M. Cherf and Michael J. Doody and Felix Olasagasti and Yvette Kolodji and Mark Akeson},
  title = {{P}rocessive {R}eplication of {S}ingle {DNA} {M}olecules in a {N}anopore {C}atalyzed by {P}hi29 {DNA} {P}olymerase},
  journal = {J. Am. Chem. Soc.}
}

@article{Plesa-2013,
  doi = {10.1021/nl3042678},
  url = {https://doi.org/10.1021/nl3042678},
  year = {2013},
  month = jan,
  publisher = {American Chemical Society ({ACS})},
  volume = {13},
  number = {2},
  pages = {658--663},
  author = {Calin Plesa and Stefan W. Kowalczyk and Ruben Zinsmeester and Alexander Y. Grosberg and Yitzhak Rabin and Cees Dekker},
  title = {{F}ast {T}ranslocation of {P}roteins {T}hrough {S}olid {S}tate {N}anopores},
  journal = {Nano Lett.}
}

@article{Hu-2018,
  doi = {10.1021/acsnano.8b00734},
  url = {https://doi.org/10.1021/acsnano.8b00734},
  year = {2018},
  month = apr,
  publisher = {American Chemical Society ({ACS})},
  volume = {12},
  number = {5},
  pages = {4494--4502},
  author = {Rui Hu and Jo{\~{a}}o V. Rodrigues and Pradeep Waduge and Hirohito Yamazaki and Benjamin Cressiot and Yasmin Chishti and Lee Makowski and Dapeng Yu and Eugene Shakhnovich and Qing Zhao and Meni Wanunu},
  title = {{D}ifferential {E}nzyme {F}lexibility {P}robed {U}sing {S}olid-{S}tate {N}anopores},
  journal = {{ACS} Nano}
}

@article{Waduge-2017,
  doi = {10.1021/acsnano.7b01212},
  url = {https://doi.org/10.1021/acsnano.7b01212},
  year = {2017},
  month = may,
  publisher = {American Chemical Society ({ACS})},
  volume = {11},
  number = {6},
  pages = {5706--5716},
  author = {Pradeep Waduge and Rui Hu and Prasad Bandarkar and Hirohito Yamazaki and Benjamin Cressiot and Qing Zhao and Paul C. Whitford and Meni Wanunu},
  title = {{N}anopore-{B}ased {M}easurements of {P}rotein {S}ize, {F}luctuations, and {C}onformational {C}hanges},
  journal = {{ACS} Nano}
}

@article{Larkin-2014,
  doi = {10.1016/j.bpj.2013.12.025},
  url = {https://doi.org/10.1016/j.bpj.2013.12.025},
  year = {2014},
  month = feb,
  publisher = {Elsevier {BV}},
  volume = {106},
  number = {3},
  pages = {696--704},
  author = {Joseph Larkin and Robert Y. Henley and Murugappan Muthukumar and Jacob~K. Rosenstein and Meni Wanunu},
  title = {{H}igh-{B}andwidth {P}rotein {A}nalysis {U}sing {S}olid-{S}tate {N}anopores},
  journal = {Biophys. J.}
}

@article{Rosenstein-2012,
  doi = {10.1038/nmeth.1932},
  url = {https://doi.org/10.1038/nmeth.1932},
  year = {2012},
  month = mar,
  publisher = {Springer Science and Business Media {LLC}},
  volume = {9},
  number = {5},
  pages = {487--492},
  author = {Jacob K Rosenstein and Meni Wanunu and Christopher A Merchant and Marija Drndic and Kenneth L Shepard},
  title = {{I}ntegrated nanopore sensing platform with sub-microsecond temporal resolution},
  journal = {Nat. Methods}
}

@article{McNally-2010,
  doi = {10.1021/nl1012147},
  url = {https://doi.org/10.1021/nl1012147},
  year = {2010},
  month = jun,
  publisher = {American Chemical Society ({ACS})},
  volume = {10},
  number = {6},
  pages = {2237--2244},
  author = {Ben McNally and Alon Singer and Zhiliang Yu and Yingjie Sun and Zhiping Weng and Amit Meller},
  title = {{O}ptical {R}ecognition of {C}onverted {DNA} {N}ucleotides for {S}ingle-{M}olecule {DNA} {S}equencing {U}sing {N}anopore {A}rrays},
  journal = {Nano Lett.}
}

@article{Mohammad-2008,
  doi = {10.1021/ja710787a},
  url = {https://doi.org/10.1021/ja710787a},
  year = {2008},
  month = mar,
  publisher = {American Chemical Society ({ACS})},
  volume = {130},
  number = {12},
  pages = {4081--4088},
  author = {Mohammad and Sumit Prakash and Andreas Matouschek and Liviu Movileanu},
  title = {{C}ontrolling a {S}ingle {P}rotein in a {N}anopore {T}hrough {E}lectrostatic {T}raps},
  journal = {J. Am. Chem. Soc.}
}

@article{RodriguezLarrea-2013,
  doi = {10.1038/nnano.2013.22},
  url = {https://doi.org/10.1038/nnano.2013.22},
  year = {2013},
  month = mar,
  publisher = {Springer Nature},
  volume = {8},
  number = {4},
  pages = {288--295},
  author = {David Rodriguez-Larrea and Hagan Bayley},
  title = {{M}ultistep {P}rotein {U}nfolding {D}uring {N}anopore {T}ranslocation},
  journal = {Nat. Nanotechnol.}
}

@article{Mereuta-2014,
  doi = {10.1038/srep03885},
  url = {https://doi.org/10.1038/srep03885},
  year = {2014},
  month = jan,
  publisher = {Springer Nature},
  volume = {4},
  number = {1},
  author = {Loredana Mereuta and Mahua Roy and Alina Asandei and Jong Kook Lee and Yoonkyung Park and Ioan Andricioaei and Tudor Luchian},
  title = {{S}lowing {D}own {S}ingle-{M}olecule {T}rafficking {T}hrough a {P}rotein {N}anopore {R}eveals {I}ntermediates for {P}eptide {T}ranslocation},
  journal = {Sci. Rep.}
}

@article{Chinappi-2015,
  doi = {10.1103/physreve.92.032714},
  url = {https://doi.org/10.1103/physreve.92.032714},
  year = {2015},
  month = sep,
  publisher = {American Physical Society ({APS})},
  volume = {92},
  number = {3},
  author = {Mauro Chinappi and Tudor Luchian and Fabio Cecconi},
  title = {{N}anopore {T}weezers: {V}oltage-{C}ontrolled {T}rapping and {R}eleasing of {A}nalytes},
  journal = {Phys. Rev. E}
}

@article{Talaga-2009,
  doi = {10.1021/ja901088b},
  url = {https://doi.org/10.1021/ja901088b},
  year = {2009},
  month = jul,
  publisher = {American Chemical Society ({ACS})},
  volume = {131},
  number = {26},
  pages = {9287--9297},
  author = {David S. Talaga and Jiali Li},
  title = {{S}ingle-{M}olecule {P}rotein {U}nfolding in {S}olid {S}tate {N}anopores},
  journal = {J. Am. Chem. Soc.}
}

@article{Ainavarapu-2005,
  doi = {10.1529/biophysj.105.062034},
  url = {https://doi.org/10.1529/biophysj.105.062034},
  year = {2005},
  month = nov,
  publisher = {Elsevier {BV}},
  volume = {89},
  number = {5},
  pages = {3337--3344},
  author = {Sri Rama Koti Ainavarapu and Lewyn Li and Carmen L. Badilla and Julio M. Fernandez},
  title = {{L}igand {B}inding {M}odulates the {M}echanical {S}tability of {D}ihydrofolate {R}eductase},
  journal = {Biophys. J.}
}

@article{Babakinejad-2013,
  doi = {10.1021/ac4021769},
  url = {https://doi.org/10.1021/ac4021769},
  year = {2013},
  month = sep,
  publisher = {American Chemical Society ({ACS})},
  volume = {85},
  number = {19},
  pages = {9333--9342},
  author = {Babak Babakinejad and Peter J{\"{o}}nsson and Ainara L{\'{o}}pez C{\'{o}}rdoba and Paolo Actis and Pavel Novak and Yasufumi Takahashi and
  Andrew Shevchuk and Uma Anand and Praveen Anand and Anna Drews and Antonio Ferrer-Montiel and David Klenerman and Yuri E. Korchev},
  title = {{L}ocal {D}elivery of {M}olecules from a {N}anopipette for {Q}uantitative {R}eceptor {M}apping on {L}ive {C}ells},
  journal = {Anal. Chem.}
}

@article{Dolinsky-2004,
  doi = {10.1093/nar/gkh381},
  url = {https://doi.org/10.1093/nar/gkh381},
  year = {2004},
  month = jul,
  publisher = {Oxford University Press ({OUP})},
  volume = {32},
  number = {Web Server},
  pages = {W665--W667},
  author = {T. J. Dolinsky and J. E. Nielsen and J. A. McCammon and N. A. Baker},
  title = {{PDB2PQR}: an {A}utomated {P}ipeline for the {S}etup of {P}oisson-{B}oltzmann {E}lectrostatics {C}alculations},
  journal = {Nucleic Acids Res.}
}

@article{Dolinsky-2007,
  doi = {10.1093/nar/gkm276},
  url = {https://doi.org/10.1093/nar/gkm276},
  year = {2007},
  month = may,
  publisher = {Oxford University Press ({OUP})},
  volume = {35},
  number = {Web Server},
  pages = {W522--W525},
  author = {T. J. Dolinsky and P. Czodrowski and H. Li and J. E. Nielsen and J. H. Jensen and G. Klebe and N. A. Baker},
  title = {{PDB2PQR}: {E}xpanding and {U}pgrading {A}utomated {P}reparation of {B}iomolecular {S}tructures for {M}olecular {S}imulations},
  journal = {Nucleic Acids Res.}
}

@article{Li-2005,
  doi = {10.1002/prot.20660},
  url = {https://doi.org/10.1002/prot.20660},
  year = {2005},
  month = oct,
  publisher = {Wiley},
  volume = {61},
  number = {4},
  pages = {704--721},
  author = {Hui Li and Andrew D. Robertson and Jan H. Jensen},
  title = {{V}ery {F}ast {E}mpirical {P}rediction and {R}ationalization of {P}rotein {pKa} {V}alues},
  journal = {Proteins}
}

@article{Baker-2001,
  doi = {10.1073/pnas.181342398},
  url = {https://doi.org/10.1073/pnas.181342398},
  year = {2001},
  month = aug,
  publisher = {Proceedings of the National Academy of Sciences},
  volume = {98},
  number = {18},
  pages = {10037--10041},
  author = {N. A. Baker and D. Sept and S. Joseph and M. J. Holst and J. A. McCammon},
  title = {{E}lectrostatics of {N}anosystems: {A}pplication to {M}icrotubules and the {R}ibosome},
  journal = {Proc. Natl. Acad. Sci. U. S. A.}
}

@incollection{Baker-2005,
  doi = {10.1002/0471720895.ch5},
  url = {https://doi.org/10.1002/0471720895.ch5},
  year = {2005},
  month = jan,
  publisher = {John Wiley {\&} Sons,  Inc.},
  pages = {349--379},
  author = {Nathan A. Baker},
  title = {{B}iomolecular {A}pplications of {P}oisson-{B}oltzmann {M}ethods},
  booktitle = {Rev. Comput. Chem.}
}

@article{Stone-2010,
  doi = {10.1109/mcse.2010.69},
  url = {https://doi.org/10.1109/mcse.2010.69},
  year = {2010},
  month = may,
  publisher = {Institute of Electrical and Electronics Engineers ({IEEE})},
  volume = {12},
  number = {3},
  pages = {66--73},
  author = {John E. Stone and David Gohara and Guochun Shi},
  title = {{OpenCL}: {A} {P}arallel {P}rogramming {S}tandard for {H}eterogeneous {C}omputing {S}ystems},
  journal = {Comput. Sci. Eng.}
}

@article{Miles-2001,
  doi = {10.1021/bi010454o},
  url = {https://doi.org/10.1021/bi010454o},
  year = {2001},
  month = jul,
  publisher = {American Chemical Society ({ACS})},
  volume = {40},
  number = {29},
  pages = {8514--8522},
  author = {George Miles and Stephen Cheley and Orit Braha and Hagan Bayley},
  title = {{T}he {S}taphylococcal {L}eukocidin {B}icomponent {T}oxin {F}orms {L}arge {I}onic {C}hannels},
  journal = {Biochemistry}
}

@incollection{Miyazaki-2011,
  doi = {10.1016/b978-0-12-385120-8.00017-6},
  url = {https://doi.org/10.1016/b978-0-12-385120-8.00017-6},
  year = {2011},
  publisher = {Elsevier},
  pages = {399--406},
  author = {Kentaro Miyazaki},
  title = {{MEGAWHOP} {C}loning},
  booktitle = {Methods Enzymol.}
}

@article{Movileanu-2005,
  doi = {10.1529/biophysj.104.057406},
  url = {https://doi.org/10.1529/biophysj.104.057406},
  year = {2005},
  month = aug,
  publisher = {Elsevier {BV}},
  volume = {89},
  number = {2},
  pages = {1030--1045},
  author = {Liviu Movileanu and Jason P. Schmittschmitt and J. Martin Scholtz and Hagan Bayley},
  title = {{I}nteractions of {P}eptides with a {P}rotein {P}ore},
  journal = {Biophys. J.}
}

@article{Thevenet-2012,
  doi = {10.1093/nar/gks419},
  url = {https://doi.org/10.1093/nar/gks419},
  year = {2012},
  month = may,
  publisher = {Oxford University Press ({OUP})},
  volume = {40},
  number = {W1},
  pages = {W288--W293},
  author = {P. Thevenet and Y. Shen and J. Maupetit and F. Guyon and P. Derreumaux and P. Tuffery},
  title = {{PEP}-{FOLD}: an {U}pdated \textit{de novo} {S}tructure {P}rediction {S}erver for {B}oth {L}inear and {D}isulfide {B}onded {C}yclic {P}eptides},
  journal = {Nucleic Acids Res.}
}

@article{Shen-2014,
  doi = {10.1021/ct500592m},
  url = {https://doi.org/10.1021/ct500592m},
  year = {2014},
  month = sep,
  publisher = {American Chemical Society ({ACS})},
  volume = {10},
  number = {10},
  pages = {4745--4758},
  author = {Yimin Shen and Julien Maupetit and Philippe Derreumaux and Pierre Tuff{\'{e}}ry},
  title = {{I}mproved {PEP}-{FOLD} {A}pproach for {P}eptide and {M}iniprotein {S}tructure {P}rediction},
  journal = {J. Chem. Theory Comput.}
}

@article{Boukhet-2016,
  doi = {10.1039/c6nr06936c},
  url = {https://doi.org/10.1039/c6nr06936c},
  year = {2016},
  publisher = {Royal Society of Chemistry ({RSC})},
  volume = {8},
  number = {43},
  pages = {18352--18359},
  author = {Mordjane Boukhet and Fabien Piguet and Hadjer Ouldali and Manuela Pastoriza-Gallego and Juan Pelta and Abdelghani Oukhaled},
  title = {{P}robing {D}riving {F}orces in {A}erolysin and {\textalpha}-{H}emolysin {B}iological {N}anopores: {E}lectrophoresis \textit{versus} {E}lectroosmosis},
  journal = {Nanoscale}
}

@article{Lu-2012,
  doi = {10.1103/physreve.86.011921},
  url = {https://doi.org/10.1103/physreve.86.011921},
  year = {2012},
  month = jul,
  publisher = {American Physical Society ({APS})},
  volume = {86},
  number = {1},
  author = {Bo Lu and David P. Hoogerheide and Qing Zhao and Dapeng Yu},
  title = {{E}ffective {D}riving {F}orce {A}pplied on {DNA} {I}nside a {S}olid-{S}tate {N}anopore},
  journal = {Phys. Rev. E}
}

@article{Ai-2011,
  doi = {10.1039/c0cp02267e},
  url = {https://doi.org/10.1039/c0cp02267e},
  year = {2011},
  publisher = {Royal Society of Chemistry ({RSC})},
  volume = {13},
  number = {9},
  pages = {4060},
  author = {Ye Ai and Shizhi Qian},
  title = {{E}lectrokinetic particle translocation through a nanopore},
  journal = {Phys. Chem. Chem. Phys.}
}

@article{vanDorp-2009,
  doi = {10.1038/nphys1230},
  url = {https://doi.org/10.1038/nphys1230},
  year = {2009},
  month = mar,
  publisher = {Springer Nature},
  volume = {5},
  number = {5},
  pages = {347--351},
  author = {Stijn van Dorp and Ulrich F. Keyser and Nynke H. Dekker and Cees Dekker and Serge G. Lemay},
  title = {{O}rigin of the {E}lectrophoretic {F}orce on {DNA} in {S}olid-{S}tate {N}anopores},
  journal = {Nat. Phys.}
}

@article{Wanunu-2009,
  doi = {10.1038/nnano.2009.379},
  url = {https://doi.org/10.1038/nnano.2009.379},
  year = {2009},
  month = dec,
  publisher = {Springer Science and Business Media {LLC}},
  volume = {5},
  number = {2},
  pages = {160--165},
  author = {Meni Wanunu and Will Morrison and Yitzhak Rabin and Alexander Y. Grosberg and Amit Meller},
  title = {{E}lectrostatic focusing of unlabelled {DNA} into nanoscale pores using a salt gradient},
  journal = {Nat. Nanotechnol.}
}

@article{Oukhaled-2011,
  doi = {10.1021/nn1034795},
  url = {https://doi.org/10.1021/nn1034795},
  year = {2011},
  month = apr,
  publisher = {American Chemical Society ({ACS})},
  volume = {5},
  number = {5},
  pages = {3628--3638},
  author = {Abdelghani Oukhaled and Benjamin Cressiot and Laurent Bacri and Manuela Pastoriza-Gallego and Jean-Michel Betton and Eric Bourhis and Ralf Jede and Jacques Gierak and Loïc Auvray and Juan Pelta},
  title = {{D}ynamics of {C}ompletely {U}nfolded and {N}ative {P}roteins through {S}olid-{S}tate {N}anopores as a {F}unction of {E}lectric {D}riving {F}orce},
  journal = {{ACS} Nano}
}

@article{Cressiot-2012,
  doi = {10.1021/nn301672g},
  url = {https://doi.org/10.1021/nn301672g},
  year = {2012},
  month = jun,
  publisher = {American Chemical Society ({ACS})},
  volume = {6},
  number = {7},
  pages = {6236--6243},
  author = {Benjamin Cressiot and Abdelghani Oukhaled and Gilles Patriarche and Manuela Pastoriza-Gallego and Jean-Michel Betton and Lo{\"{\i}}c Auvray and Murugappan Muthukumar and Laurent Bacri and Juan Pelta},
  title = {{P}rotein {T}ransport through a {N}arrow {S}olid-{S}tate {N}anopore at {H}igh {V}oltage: {E}xperiments and {T}heory},
  journal = {{ACS} Nano}
}

@article{Cressiot-2015,
  doi = {10.1021/acsnano.5b03053},
  url = {https://doi.org/10.1021/acsnano.5b03053},
  year = {2015},
  month = aug,
  publisher = {American Chemical Society ({ACS})},
  volume = {9},
  number = {9},
  pages = {9050--9061},
  author = {Benjamin Cressiot and Esther Braselmann and Abdelghani Oukhaled and Adrian H. Elcock and Juan Pelta and Patricia L. Clark},
  title = {{D}ynamics and {E}nergy {C}ontributions for {T}ransport of {U}nfolded {P}ertactin through a {P}rotein {N}anopore},
  journal = {{ACS} Nano}
}

@article{Verschueren-2019,
  doi = {10.1002/smtd.201800465},
  url = {https://doi.org/10.1002/smtd.201800465},
  year = {2019},
  month = jan,
  publisher = {Wiley},
  pages = {1800465},
  author = {Daniel Verschueren and Xin Shi and Cees Dekker},
  title = {{N}ano-{O}ptical {T}weezing of {S}ingle {P}roteins in {P}lasmonic {N}anopores},
  journal = {Small Methods}
}

@article{Gooding-2016,
  doi = {10.1002/anie.201600495},
  url = {https://doi.org/10.1002/anie.201600495},
  year = {2016},
  month = jul,
  publisher = {Wiley},
  volume = {55},
  number = {38},
  pages = {11354--11366},
  author = {J. Justin Gooding and Katharina Gaus},
  title = {{S}ingle-{M}olecule {S}ensors: {C}hallenges and {O}pportunities for {Q}uantitative {A}nalysis},
  journal = {Angew. Chem., Int. Ed.}
}

@article{Chen-2018,
  doi = {10.1038/s41467-018-04118-7},
  url = {https://doi.org/10.1038/s41467-018-04118-7},
  year = {2018},
  month = apr,
  publisher = {Springer Science and Business Media {LLC}},
  volume = {9},
  number = {1},
  author = {Chang Chen and Yi Li and Sarp Kerman and Pieter Neutens and Kherim Willems and Sven Cornelissen and Liesbet Lagae and Tim Stakenborg and Pol Van Dorpe},
  title = {{H}igh {S}patial {R}esolution {N}anoslit {SERS} for {S}ingle-{M}olecule {N}ucleobase {S}ensing},
  journal = {Nat. Commun.}
}

@article{Bradac-2018,
  doi = {10.1002/adom.201800005},
  url = {https://doi.org/10.1002/adom.201800005},
  year = {2018},
  month = apr,
  publisher = {Wiley},
  volume = {6},
  number = {12},
  pages = {1800005},
  author = {Carlo Bradac},
  title = {{N}anoscale {O}ptical {T}rapping: {A} {R}eview},
  journal = {Adv. Opt. Mater.}
}

@proceeding{Kotnala-2014,
  doi = {10.1117/12.2062051},
  url = {https://doi.org/10.1117/12.2062051},
  year = {2014},
  month = sep,
  volume = {9164},
  publisher = {{SPIE}},
  author = {Abhay Kotnala and Ahmed A. Al-Balushi and Reuven Gordon},
  editor = {Kishan Dholakia and Gabriel C. Spalding},
  title = {{O}ptical {T}weezers for {F}ree-{S}olution {L}abel-{F}ree {S}ingle {B}io-{M}olecule {S}tudies},
  booktitle = {{O}ptical {T}rapping and {O}ptical {M}icromanipulation {XI}}
}

@article{Xie-2001,
  doi = {10.1002/1438-5171(200112)2:4<229::aid-simo229>3.0.co;2-9},
  url = {https://doi.org/10.1002/1438-5171(200112)2:4<229::aid-simo229>3.0.co;2-9},
  year = {2001},
  month = dec,
  publisher = {Wiley},
  volume = {2},
  number = {4},
  pages = {229--236},
  author = {Sunney Xie},
  title = {{S}ingle-{M}olecule {A}pproach to {E}nzymology},
  journal = {Single Mol.}
}

@article{Baker-2017,
  doi = {10.1002/wnan.1477},
  url = {https://doi.org/10.1002/wnan.1477},
  year = {2017},
  month = apr,
  publisher = {Wiley},
  volume = {10},
  number = {1},
  pages = {e1477},
  author = {James E. Baker and Ryan P. Badman and Michelle D. Wang},
  title = {{N}anophotonic {T}rapping: {P}recise {M}anipulation and {M}easurement of {B}iomolecular {A}rrays},
  journal = {Wiley Interdiscip. Rev.: Nanomed. Nanobiotechnol.}
}

@article{Kohen-2015,
  doi = {10.12688/f1000research.6968.1},
  url = {https://doi.org/10.12688/f1000research.6968.1},
  year = {2015},
  month = dec,
  publisher = {F1000 (Faculty of 1000 Ltd)},
  volume = {4},
  pages = {1464},
  author = {Amnon Kohen},
  title = {{D}ihydrofolate {R}eductase {A}s a {M}odel for {S}tudies of {E}nzyme {D}ynamics and {C}atalysis},
  journal = {F1000Research}
}

@article{Levene-2003,
  doi = {10.1126/science.1079700},
  url = {https://doi.org/10.1126/science.1079700},
  year = {2003},
  month = jan,
  publisher = {American Association for the Advancement of Science ({AAAS})},
  volume = {299},
  number = {5607},
  pages = {682--686},
  author = {M. J. Levene},
  title = {{Z}ero-{M}ode {W}aveguides for {S}ingle-{M}olecule {A}nalysis at {H}igh {C}oncentrations},
  journal = {Science}
}

@article{Derrington-2015,
  doi = {10.1038/nbt.3357},
  url = {https://doi.org/10.1038/nbt.3357},
  year = {2015},
  month = sep,
  publisher = {Springer Nature},
  volume = {33},
  number = {10},
  pages = {1073--1075},
  author = {Ian M Derrington and Jonathan M Craig and Eric Stava and Andrew H Laszlo and Brian C Ross and Henry Brinkerhoff and Ian C Nova and Kenji Doering and Benjamin I Tickman and Mostafa Ronaghi and Jeffrey G Mandell and Kevin L Gunderson and Jens H Gundlach},
  title = {{S}ubangstrom {S}ingle-{M}olecule {M}easurements of {M}otor {P}roteins {U}sing a {N}anopore},
  journal = {Nat. Biotechnol.}
}

@article{Laszlo-2016,
  doi = {10.1016/j.ymeth.2016.03.026},
  url = {https://doi.org/10.1016/j.ymeth.2016.03.026},
  year = {2016},
  month = aug,
  publisher = {Elsevier {BV}},
  volume = {105},
  pages = {75--89},
  author = {Andrew H. Laszlo and Ian M. Derrington and Jens H. Gundlach},
  title = {{MspA} {N}anopore {A}s a {S}ingle-{M}olecule {T}ool: {F}rom {S}equencing to {SPRNT}},
  journal = {Methods}
}

@article{Ma-2009,
  doi = {10.1002/cbic.200900526},
  url = {https://doi.org/10.1002/cbic.200900526},
  year = {2009},
  month = nov,
  publisher = {Wiley},
  volume = {11},
  number = {1},
  pages = {25--34},
  author = {Long Ma and Scott L. Cockroft},
  title = {{B}iological {N}anopores for {S}ingle-{M}olecule {B}iophysics},
  journal = {{ChemBioChem}}
}

\includecv{curriculum}

\includepublications{publications}

\makebackcoverXII

\end{document}